\documentclass[12pt,oneside,letterpaper]{book}
\usepackage{flafter}
\usepackage{float}
\usepackage[utf8]{inputenc}
\usepackage[table]{xcolor} 
\usepackage{placeins}
\usepackage{url}
\usepackage{geometry}
\usepackage{setspace}
\usepackage{adjustbox}
\usepackage{graphicx}
\usepackage{changepage}
\usepackage{tabularx} 
\usepackage{tikz} 
\usetikzlibrary{decorations.pathreplacing}
\usetikzlibrary{positioning, arrows.meta, shapes.geometric, automata}
\usetikzlibrary{shapes.geometric, arrows.meta, positioning}
\usetikzlibrary{fit}
\usetikzlibrary{calc}
\usepackage{pgfplots}
\pgfplotsset{compat=1.17}
\usepackage{subcaption}
\usepackage{tocbibind}
\usepackage{tocloft}
\usepackage{appendix}
\usepackage{amsmath}
\usepackage{amssymb}
\usepackage{longtable}
\usepackage{array}

\usepackage{hyperref}
\usepackage{xcolor} 
\usepackage{enumitem}
\usepackage{listings}
\usepackage{lipsum}
\usepackage{geometry} 

\usepackage{setspace} 
\usepackage{color}

\usepackage[linesnumbered,ruled,vlined, noend]{algorithm2e}

\usepackage{booktabs} 
\usepackage[table]{xcolor} 
\usepackage{colortbl} 
\definecolor{lightgray}{gray}{0.9} 

\DeclareUnicodeCharacter{0306}{\u}

\begin{document}
\doublespacing 

\frontmatter

\begin{singlespacing}
\begin{titlepage}
   \centering
   {Choreographing the Rhythms of Observation:\par}
   {Dynamics for Ranged Observer Bipartite-Unipartite SpatioTemporal (ROBUST) Networks\par}
   \mbox{} \\
   \mbox{} \\
   \mbox{} \\
   \mbox{} \\
   \mbox{} \\
   \mbox{} \\
   {A Dissertation \par}
   \mbox{} \\
   \mbox{} \\
   \mbox{} \\
   \mbox{} \\
   \mbox{} \\
   \mbox{} \\
   {Submitted to the Graduate Faculty of \\
   the University of New Orleans \\
   in partial fulfillment of the \\
   requirements for the degree of \\
   \mbox{} \\
   \mbox{} \\
   \mbox{} \\
   \mbox{} \\
   \mbox{} \\
   \mbox{} \\
   Doctor of Philosophy \\
   in \\
   Engineering and Applied Science \par }
   \mbox{} \\
   \mbox{} \\
   \mbox{} \\
   \mbox{} \\
   \mbox{} \\
   \mbox{} \\
   {by\\
   \mbox{} \\
   (Ted) Edward Holmberg\\
   \mbox{} \\
   B.A. University of New Orleans, 2007\\
   M.Sc. University of New Orleans, 2014\\}
   \mbox{} \\
    {May, 2024} 
\end{titlepage}
\end{singlespacing}

\newpage

\setcounter{page}{2}  

\tableofcontents
\newpage
\listoffigures
\newpage
\listoftables
\newpage
\listofalgorithms
\newpage
\newpage

\begin{spacing}{1} 
\chapter*{\centering Abstract} 
\addcontentsline{toc}{chapter}{Abstract}

Existing network analysis methods struggle to optimize observer placements in dynamic environments with limited visibility. This dissertation introduces the novel ROBUST (Ranged Observer Bipartite-Unipartite SpatioTemporal) framework, offering a significant advancement in modeling, analyzing, and optimizing observer networks within complex spatiotemporal domains. ROBUST leverages a unique bipartite-unipartite approach, distinguishing between observer and observable entities while incorporating spatial constraints and temporal dynamics. \\

This research extends spatiotemporal network theory by introducing novel graph-based measures, including myopic degree, spatial closeness centrality, and edge length proportion. These measures, coupled with advanced clustering techniques like Proximal Recurrence, provide insights into network structure, resilience, and the effectiveness of observer placements.
The ROBUST framework demonstrates superior resource allocation and strategic responsiveness compared to conventional models. Case studies in oceanographic monitoring, urban safety networks, and multi-agent path planning showcases its practical applicability and adaptability. Results demonstrate significant improvements in coverage, response times, and overall network efficiency. \\

This work paves the way for future research in incorporating imperfect knowledge, refining temporal pathing methodologies, and expanding the scope of applications. By bridging theoretical advancements with practical solutions, ROBUST stands as a significant contribution to the field, promising to inform and inspire ongoing and future endeavors in network optimization and multi-agent system planning.
\end{spacing}

\newpage

\doublespacing 
\mainmatter
\chapter{Introduction}
\label{chap:intro}
\section{Background}
Real-time spatial data is often generated at a pace and volume that far exceeds our capacity to collect it \cite{Li2022RealTimeGIS}. This poses a significant challenge to observers such as sensors, agents, and cameras, which must locate the most crucial data from an overwhelmingly large pool of possibilities \cite{Ladner2002, Chung2001, wilson2003}.  Continually determining how to acquire current and future real-time data can be likened to a 'dance.' The development of Responsive Observer Bipartite-Unipartite SpatioTemporal (ROBUST) networks was motivated by the need to systematically structure these decisions and the resulting dynamics – essentially choreographing the observations – into a conceptual and mathematical model. ROBUST networks are designed to navigate, analyze, and optimize the intricacies of these observational processes. 

This work focuses on using  ROBUST networks to optimize resource allocation, enhance prediction accuracy, and bolster strategic responsiveness. This entails a thorough examination of the relationships between observers and observables across spatial and temporal dimensions. Effective positioning of observers for optimal data collection is a pivotal element of this research.

Additionally, this work aims to develop scalable, adaptable, and robust solutions for managing real-time spatiotemporal data observations. ROBUST networks are explored in detail within this research demonstrating their effectiveness through examples and case studies in subsequent chapters.

\section{Problem Statement}
Current network models are not effective enough to deal with the complex and varied interactions in spatiotemporal bipartite networks, especially when precise and optimal network configurations are needed \cite{Ferreira2020}. These models struggle to cope with the intricate timing and spatial aspects of these networks \cite{Hamdi2022}.

Several issues make this problem more difficult. External factors often make network behavior unpredictable, there's a need to balance different goals within the network \cite{Shekhar2015}, and there are potential threats to the network's security and reliability. Furthermore, there's a lack of effective methods to test and confirm that these networks are functioning correctly.

The aim of this dissertation is to develop new methods and tools to better understand and improve ROBUST networks, making them more effective for practical use in real-world situations.

\section{Objectives of the Study}
The main goal of this study is to create, refine, and test a system for managing ROBUST networks. ROBUST networks are expected to be more effective and provide clearer insights than current network models due to the novel approach of coupling a Bipartite-Unipartite workflow. The focus is on understanding and utilizing patterns in space and time to improve how these networks work and to make better decisions about where to place observation tools.

Key goals of the study include:

\begin{enumerate}
    \item \textbf{Developing an Improved Network Model:} By adopting spatiotemporal graph theory over traditional static graph approaches, this model emphasizes strict spatial properties in nodes and adapts to evolutionary changes over time. It aims to enhance accuracy and practicality in real-world applications, utilizing dynamic spatial and temporal data for refined network modeling and analysis.

    \item \textbf{Analyzing Spatial-Temporal Patterns:} Utilizing advanced graph analysis methods to scrutinize space-time patterns within ROBUST networks. This analysis is geared towards identifying the most influential nodes, assessing network robustness, pinpointing areas of brittleness, and predicting potential evolutions within the network. These insights are crucial for enhancing network functionality, guiding strategic placement and adjustments of observation nodes, and preparing for future changes. 

    \item \textbf{In-depth Analysis of Decision-Making:} This goal involves adopting a comprehensive analytical approach to understand decision-making in uncertain environments. Key to this approach is the incorporation of concepts like the `chain of regret,' a tool used to evaluate the impact of past decisions on current and future network states. This analysis aims to distinguish between outcomes influenced by inherent randomness in the network and those arising from its inefficiencies.

    \item \textbf{Accurate Representation of Problem Domains:} This objective focuses on ensuring that ROBUST networks are tailored to accurately represent specific real-world scenarios across different domains. It involves developing strategies to apply ROBUST networks effectively in varied contexts, determining which spatiotemporal properties are integral to each scenario, and deciding the appropriate granularity of analysis. Central to this process is the precise identification and modeling of the two primary entities in these networks: observers and observables. 

\end{enumerate}

\section{Significance of the Research}
This research aims to advance the field of network analysis, with particular focus on spatiotemporal bipartite networks. It addresses critical gaps in existing methodologies, offering both theoretical and practical contributions to the understanding and optimization of complex systems. Overall, the research expands the scope and application of network analysis, leading to systems that are more robust, adaptable, and aligned with the nuanced requirements of spatiotemporal dynamics. The implications of this research span a wide range of disciplines and industries, underlining the interconnectedness of modern observational systems and the need for advanced analytical approaches.

\subsubsection*{Key Significance Highlights:}
\begin{itemize}
\item \textbf{Efficient Resource Use:} By reducing the observer node set in ROBUST networks, the research enhances resource efficiency, balancing operational costs with effective data coverage.
\item \textbf{Improved Response Times:} Minimizing distances between nodes in ROBUST networks shortens response times, which is vital in dynamic real-world scenarios.
\item \textbf{Enhanced Outcomes:} Utilizing temporal centrality measures, the research leads to optimized network configurations and strategic observer node placements, improving overall network performance.
\item \textbf{Temporal Network Pathing:} The introduction of Weighted Aggregate Inter-Temporal Rewards (WAITR) planning. This approach employs the principles of connectivity and flow to predict network states over time, enabling networks to dynamically adapt to environmental changes and emerging data patterns.
\item \textbf{Real-time Network Evaluation:} Continuous reconfiguration and real-time assessment are emphasized, allowing for agile adaptations and performance optimization of the networks.
\item \textbf{Addressing Stochastic Challenges:} The `chain of regret' measure is incorporated to manage the inherent unpredictability in networks, enhancing decision-making in uncertain environments.
\item \textbf{Clustering and Spatiotemporal Coherence:} The Proximal Recurrence (PR) methodology ensures accurate clustering and coherent representation of spatial-temporal aspects in network analysis.
\end{itemize}

\section{Research Questions}
The research questions at the heart of this study are designed to address the complexities of ROBUST networks. The questions aim to probe the depths of network dynamics, uncovering how they can be optimized, managed, and leveraged to improve both theoretical understanding and practical applications. Specifically, the study seeks to answer the following critical questions:
\begin{enumerate}
    \item How can the physical location and spatial relationship between observer and observable entities within a bipartite network significantly affect its dynamics and the overall network analysis?

    \item How do the temporal dynamics of interactions or behaviors of entities within these networks evolve, and how might past interactions influence future ones?

    \item Given the inherent randomness and uncertainty within these domains, how can probabilistic approaches be more appropriately applied to predict future states or interactions within the network?

    \item What are the most effective strategies for ensuring that observers only interact with observables, maintaining the integrity of intra-type interactions within the network?

    \item How can clear objectives such as optimizing network performance, maximizing coverage, or minimizing costs be defined and achieved within the constraints of a ROBUST network?

\end{enumerate}

By addressing these questions, this research intends to contribute to a richer understanding and more sophisticated handling of complex networks. The ultimate goal is to articulate and apply novel strategies that not only meet the theoretical challenges posed by spatiotemporal and bipartite dynamics but also address practical considerations in deploying these networks across various domains.

\section{Research Hypotheses}
These hypotheses are central to the research objectives, guiding the investigative approach and laying the groundwork for the subsequent methodology, analysis, and discussion. They are designed to be tested through a series of controlled experiments and comparative analyses, aiming to demonstrate the validity and utility of ROBUST networks in real-world scenarios.

\begin{enumerate}
    \item[\textbf{H1:}] \textit{Utilizing ROBUST networks for dataset analysis will result in more precise and actionable insights than those achievable with conventional models, such as heuristic algorithms, Integer Linear Programming (ILP), Greedy Algorithms, Combinatorial Optimization.} The effectiveness of ROBUST networks will be quantified using coverage and robustness measures, which will assess their proficiency in dynamic spatiotemporal data analysis. This hypothesis focuses on the distinctive strength of ROBUST networks to adhere to spatiotemporal constraints governing node behaviors, providing enhanced precision in analyses that factor in both spatial and temporal dynamics of the network.

    \item[\textbf{H2:}] \textit{ROBUST networks will demonstrate superior resource allocation and strategic responsiveness compared to the alternative models cited in Chapter~\ref{chap:lit_review} for observational networks.} This hypothesis will be tested by determining the optimal balance between minimizing node insertions and maximizing event capture. The focus is on demonstrating the efficiency and responsiveness of ROBUST networks relative to established models

    \item[\textbf{H3:}] \textit{The integration of spatial-temporal graph analysis in ROBUST networks will lead to improved optimization, evidenced by more effective placement of observer nodes and enhanced network efficiency.} This hypothesis posits that spatial-temporal graph methodology intrinsic to ROBUST networks will yield superior network optimization outcomes. The assessment will involve evaluating the overall network centrality distributions and nodal centrality to quantify the improvements in observer node placement and overall network efficiency.
\end{enumerate}

Each hypothesis is tailored to be empirically testable, providing a clear framework for examining the advancements offered by ROBUST networks in the realm of spatiotemporal data analysis.

\section{Dissertation Structure}
The structure of this dissertation is meticulously designed to guide readers through the complex choreography of observation within Responsive Observer Bipartite-Unipartite SpatioTemporal (ROBUST) Networks. 

\subsubsection*{The dissertation is organized into the following chapters:}

\subsubsection*{Chapter 1: Introduction}
\vspace{-2mm}
 Introduces the concept and significance of ROBUST networks, highlighting the challenges in handling real-time spatial data and setting the stage for exploring advanced network analysis methods to address these issues.

\subsubsection*{Chapter 2: Literature Review}
\vspace{-2mm}
Delves into the current state of research surrounding the analysis of sensor networks in a spatiotemporal environment.

\subsubsection*{Chapter 3: ROBUST Network Theory}
\vspace{-2mm}
Introduces the Ranged Observer Bipartite-Unipartite SpatioTemporal (ROBUST) Network, a novel framework for modeling the complex dynamics between spatially-sensitive observers and observable entities, expanding upon earlier theoretical concepts and aiming to optimize the performance of observational systems through strategic placement and dynamic reconfiguration of observers.

\subsubsection*{Chapter 4: ROBUST Measures and Analysis}
\vspace{-2mm}
This chapter includes some novel graph-based measures designed for ROBUST network
analysis.

\subsubsection*{Chapter 5: ROBUST and Static Observers}
\vspace{-2mm}
This chapter delves into optimizing observer node placements in networks, employing bipartite and unipartite analyses to enhance network coverage and resilience, and strategically positioning observer nodes to fill coverage gaps and improve the network's overall efficiency.

\subsubsection*{Chapter 6: ROBUST and Mobile Observers}
\vspace{-2mm}
This chapter expands on the foundational principles of the ROBUST Network by examining the application of mobile sensor placements, delineating a three-phase process involving the PREP Mapper, TED Predictor, and WAITR Planner to optimize sensor positions for real-time adaptability and responsiveness, thereby enhancing the network's effectiveness in diverse operational environments.

\subsubsection*{Chapter 7: Case Study I: Oceanographic Monitoring}
\vspace{-2mm}
Presents real-world applications and  practical implementation of ROBUST networks in environmental modeling.

\subsubsection*{Chapter 8: Case Study II: Urban Safety Network}
\vspace{-2mm}
Presents real-world applications and the practical implementation of ROBUST networks in urban safety.

\subsubsection*{Chapter 9: Case Study III: Multiagent Planning}
\vspace{-2mm}
This chapter presents a case study on multiagent spatiotemporal planning strategies for Autonomous Underwater Vehicles (AUVs) using the ROBUST Network, focusing on optimizing AUV fleet paths in the Gulf of Mexico to enhance data collection efficiency by leveraging oceanographic variabilities.

\subsubsection*{Chapter 10: Benchmarking Approach}
\vspace{-2mm}
This chapter introduces a benchmarking framework using a synthetic dataset to evaluate static and mobile sensor placement methods, analyzes the results under the assumption of perfect knowledge, and establishes a baseline for comparison with real-world scenarios.

\subsubsection*{Chapter 11: Benchmarking Results}
\vspace{-2mm}
Presents experimental results and visualizations comparing the performance of literature-based and proposed sensor placement methods (both static and mobile) on synthetic benchmark datasets.

\subsubsection*{Chapter 12: Benchmarking Analysis}
\vspace{-2mm}
This chapter includes an analysis and discussion of the research findings and their implications from the Benchmarking results.

\subsubsection*{Chapter 13: Conclusions and Future Work}
\vspace{-2mm}
Summarizes the findings, discusses the contributions to the field, and outlines the directions for future research, thereby encapsulating the research journey and its broader implications.

\subsubsection*{Appendix A: Theoretical Foundations}
\vspace{-2mm}
The theoretical foundations for the study are established by exploring graph theory and analysis, spatial graphs, dynamic graphs, and temporal graphs, ultimately leading to the development of spatiotemporal graphs.

\subsubsection*{Appendix B: ROBUST Dynamics}
\vspace{-2mm}
This chapter explores the nuanced dynamics within the ROBUST Network, focusing on the complex interplay between cooperative and competitive relationships among observer and observable nodes, crucial for designing and managing multiagent systems effectively.

\subsubsection*{Appendix C: Collaborative Bipartite Dynamics}
\vspace{-2mm}
This chapter explores collaborative dynamics in specialized observer-observable relationships within the ROBUST network, detailing how different bipartite interactions like Server-Client, Waiter-Client, Manager-Worker, and Guard-Citizen contribute to optimizing network performance and efficiency through strategic role-based interactions.

\subsubsection*{Appendix D: Competitive Bipartite Dynamics}
\vspace{-2mm}
This chapter analyzes competitive dynamics within the ROBUST Network, focusing on various adversarial relationships like Invader-Defender and Predator-Prey, detailing their strategic interactions and the implications for network resilience and optimization

\subsubsection*{Appendix E: Potential Cases for ROBUST Networks}
\vspace{-2mm}
Presents possible real-world applications and the practical implementation of ROBUST network across various domains to showcase its vast and wide-ranging usability opportunities.

\subsubsection*{Appendix F: Benchmarking Environment}
\vspace{-2mm}
Proposes an approach to dissecting and evaluating distinct spatiotemporal data behaviors in two primary case studies (Oceanography Data and Crime Data), with the goal of extrapolating insights and creating synthetic datasets to generalize these behaviors. This approach allows for testing and evaluation of analytical methods across various spatiotemporal scenarios, thus enhancing the understanding and optimization of data analysis techniques.

\section{Scope of the Study}
The scope of this study encompasses the development and application of ROBUST networks within stochastic spatiotemporal environments. It focuses on optimizing sensor placement strategies, enhancing decision-making processes through the innovative ``WAITR" metric, and assessing network performance in unpredictable settings. The study critically examines the application of these networks in real-world scenarios, such as environmental monitoring and urban safety, utilizing spatial-temporal data to inform network behavior and efficiency.

Moreover, the research addresses the challenges of managing uncertainty within these networks, applying stochastic modeling techniques to provide actionable insights. Temporal analysis, including forecasting, nowcasting, and hindcasting, plays a supportive role in refining the network models and enabling robust predictions and evaluations of network efficacy.

\chapter{Literature Review }
\label{chap:lit_review}
\section{Introduction}
\subsection{Overview}
There are several potential approaches for addressing the problem of sensor placements in spatiotemporal environments. This chapter covers several popular approaches and some new approaches. These approaches  may be contrasted and compared to the approach proposed in this research.  

\subsection{Case Studies and Applications}
In evaluating the performance and generalizability of the Greedy Algorithm for sensor placement, this study leverages customizable synthetic datasets, simulating dynamic heatmaps. This approach ensures that the algorithm's effectiveness is not overfitted to a specific case. It also allows us to explore a wide range of scenarios, providing insights into how the algorithm adapts to various conditions and challenges.

\section{Frequency Clustering}

\subsection{Overview}
Frequency clustering (i.e., mode-based clustering) focuses on identifying the dominant occurrence within a dataset.  This simple approach serves as a frequentist baseline for sensor placement, highlighting the areas with the highest concentration of events. It can be a valuable starting point (i.e. base case) for analysis, offering insights into dominant recurrent patterns before applying more complex optimization strategies. \cite{Neyman1977}

\begin{figure}[htbp]
  \begin{adjustwidth}{-1cm}{-1cm} 
    \centering
    \includegraphics[width=\linewidth]{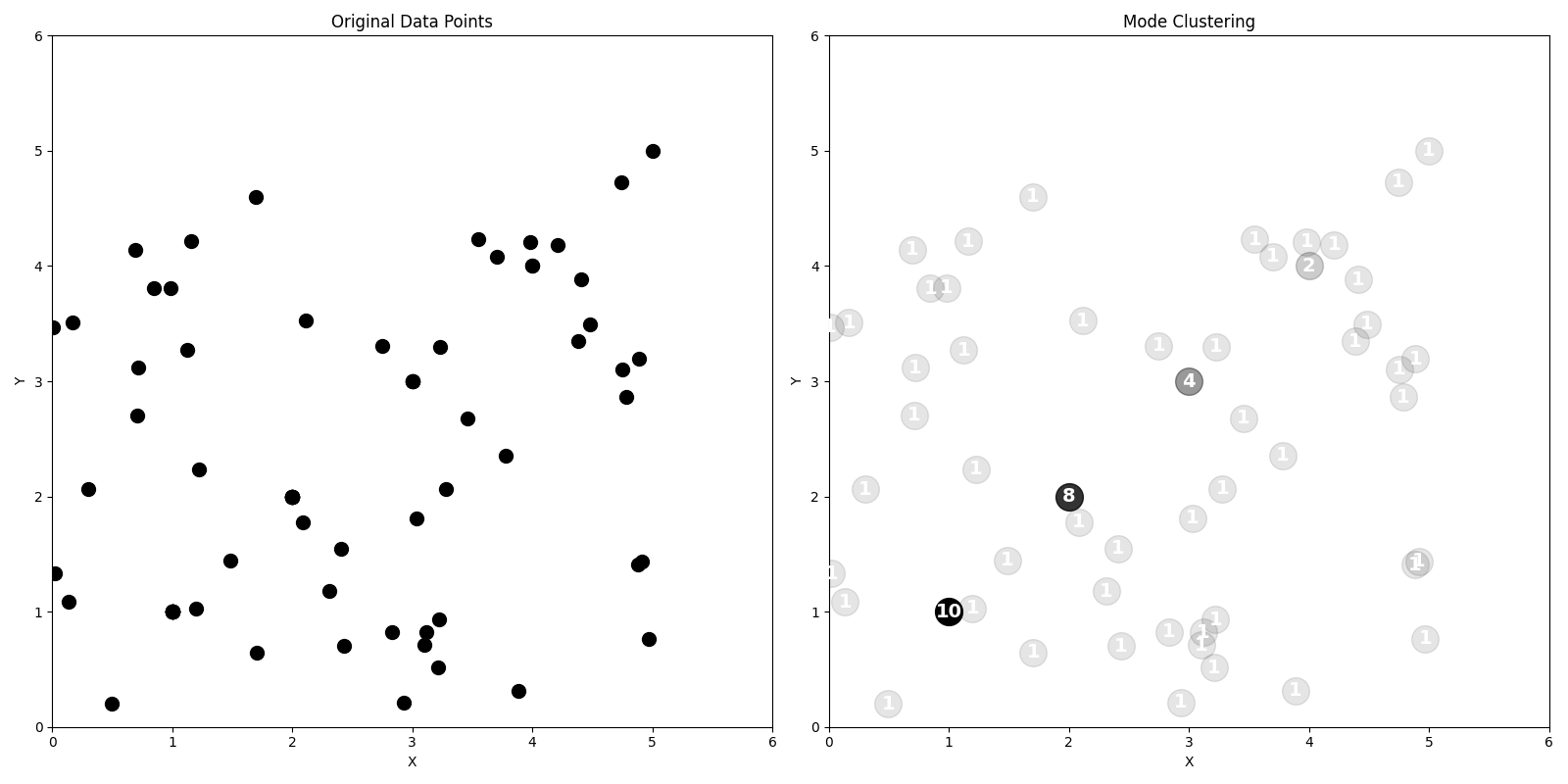} 
    \caption{Example: Frequency clustering}
    \label{fig:mode-lit-review-demo}
  \end{adjustwidth}
\end{figure}

Figure~\ref{fig:mode-lit-review-demo} contrasts traditional data display with frequency clustering. The left panel illustrates the original dataset with uniformly black dots. The right panel employs mode clustering, where points are visualized with varying levels of opacity to reflect their frequency; numbers indicate the count of occurrences.

\subsection{Methodology}
Frequency Clustering identifies the most frequently occurring data point in the dataset, which represents the statistical mode. This process emphasizes the most common values and disregards spatial relationships among data points.

\subsubsection{Identification of Dominant Occurrences}
Frequency Clustering determines the most common occurrences within a dataset, offering insights into the prevalent events or values.

\begin{algorithm}[H]
\setstretch{1}
\SetAlgoLined
\LinesNumbered  
\KwIn{Dataset \( D \)}
\KwOut{Sorted map \( M \) of data points by frequency}
\Begin{
    Initialize an empty map \( M \) to store the frequency of each point\;
    \ForEach{point in \( D \)}{
        \eIf{point is in \( M \)}{
            Increment the count of point in \( M \)\;
        }{
            Add point to \( M \) with a count of 1\;
        }
    }
    Sort \( M \) by value in descending order\;
    \KwRet{\( M \)}\;
}
\caption{Frequency Clustering Algorithm}
\end{algorithm}

\subsection{Application in Sensor Placement}
Frequency Clustering can be applied to sensor placement by identifying the most common events captured by sensors. However, its limitations become apparent in complex bipartite spatiotemporal networks.

\subsubsection{Inadequacy in Spatial-Temporal Contexts}
The technique's focus on frequency without considering spatial dimensions may miss important spatial clusters, leading to suboptimal node insertion strategies in sensor networks.

\subsubsection{Potential Overlooking of Crucial Hubs}
Frequency Clustering might bypass less frequent but strategically important spatial hotspots or hubs in a network, which could be vital for understanding and optimizing sensor placement.

\subsubsection{Neglect of Spatial Relationships}
While identifying frequent occurrences, Frequency Clustering overlooks the spatial distribution and relationships of data points, which can be crucial in many contexts.

\subsection{Advantages}
\begin{itemize}
    \item \textbf{Simplicity:} Frequency Clustering is straightforward and easy to implement.
    \item \textbf{Highlighting Dominant Trends:} It effectively brings out the most common occurrences in a dataset.
\end{itemize}

\subsection{Limitations and Concerns}
\begin{itemize}
    \item \textbf{Neglect of Spatial Relationships:} The method does not consider the spatial positioning of data points, crucial in sensor networks.
    \item \textbf{Potential for Overlooking Important Patterns:} It may miss crucial spatial-temporal patterns that are not the most frequent but are strategically significant.
\end{itemize}

\subsection{Contextual Comparison}
Compared to more sophisticated methods like Genetic Algorithms, Frequency Clustering offers a simpler approach focused on frequency analysis. However, its neglect of spatial and temporal dynamics makes it less suitable for complex sensor placement problems where spatial relationships and diverse temporal patterns play a significant role. While it can provide insights into dominant trends, it may need to be paired with other methods for a comprehensive analysis of sensor data.

\section{K-means Clustering}

\subsection{Overview}
K-means clustering is a popular algorithm that partitions datasets into \( k \) distinct clusters by minimizing intra-cluster variances. It's a fundamental method for identifying centroids across spatially diverse data points. A centroid is the central point of each cluster, calculated as the mean position of all the points in the cluster \cite{MacQueen1967}.

\begin{figure}[htbp]
  \begin{adjustwidth}{-1cm}{-1cm} 
    \centering
    \includegraphics[width=\linewidth]{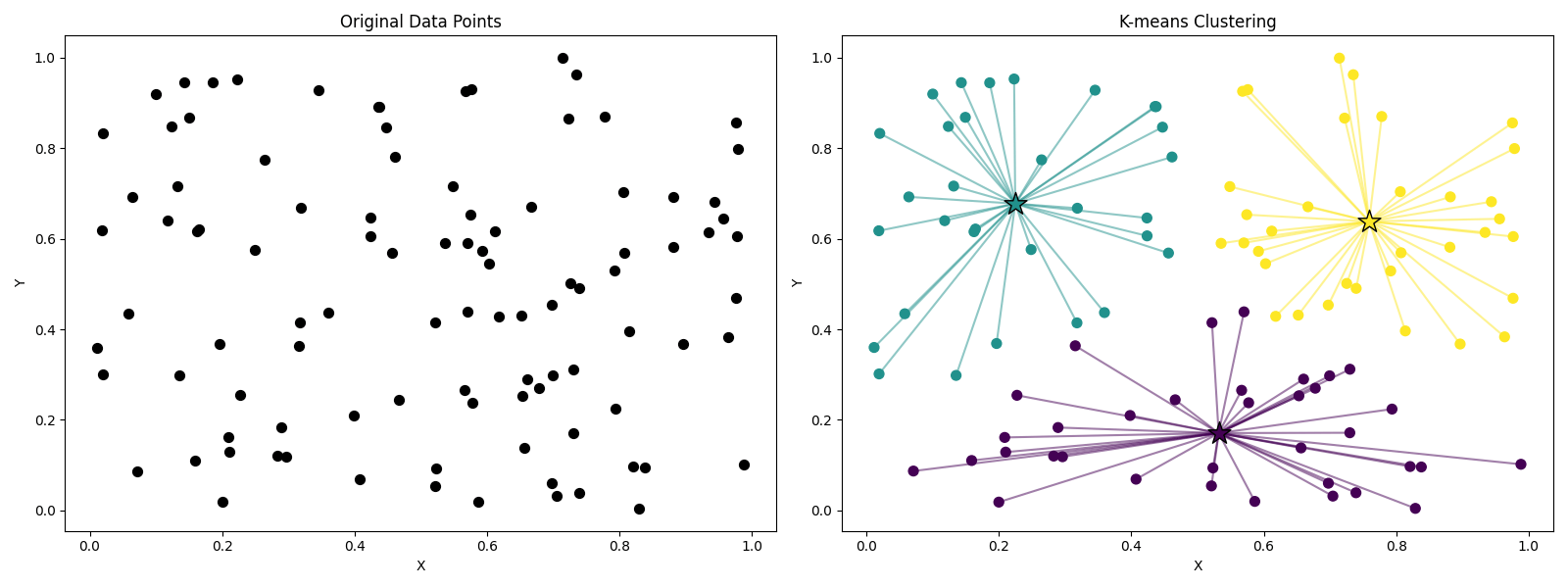} 
    \caption{Example: K-means clustering}
    \label{fig:kmeans-lit-review-demo}
  \end{adjustwidth}
\end{figure}

Figure~\ref{fig:kmeans-lit-review-demo} displays K-means clustering: the left panel shows original data points in black, and the right panel shows the clustered data with centroids marked by stars and linked to their associated points.

\subsection{Methodology}
The algorithm assigns data points to the nearest cluster center and recalculates the center until convergence. It requires the number of clusters, \( k \), to be specified in advance. K-means iteratively updates the positions of cluster centroids to minimize the total intra-cluster variance. \cite{Jain2010}

\begin{algorithm}[H]
\setstretch{1}
\SetAlgoLined
\LinesNumbered  
\KwResult{Partition dataset into \( k \) clusters}
\Begin{
    Initialize centroids randomly\;
    \Repeat{centroids do not change significantly}{
        \ForEach{data point}{
            Assign the data point to the cluster with the closest centroid\;
        }
        \ForEach{cluster}{
            Recalculate the centroid as the mean of all points in the cluster\;
        }
    }
    \KwRet{clusters and their centroids}\;
}
\caption{K-means Clustering Algorithm}
\end{algorithm}

\subsection{Spatiotemporal Adaptations}
Researchers have extended K-means to address the complexities of spatiotemporal data. For example, Dorabiala et al. (2022) proposed Spatiotemporal K-means (STKM), a specialized method designed to identify moving clusters within spatiotemporal datasets \cite{dorabiala2022l}.

\subsection{Application in Sensor Placement}
K-means clustering has found applications in sensor placement by grouping sensors or observed events based on their spatial and sometimes temporal characteristics. Here are a few recent examples:

\begin{itemize}
        \item \textbf{Water Distribution Networks:} Gautam et al. (2022) employed K-means clustering to optimize sensor placement in water distribution networks, aiming to maximize the detection of contamination events \cite{Gautam2022}.
    
        \item \textbf{Environmental Monitoring:} Li et al. (2024) proposed a dynamic sensor placement strategy for pig houses, leveraging a three-way K-means model to adapt sensor positions based on changing environmental conditions \cite{Li2024}.

        \item \textbf{Smart Home Networks:} Simonsson et al. (2023) employed K-means clustering to analyze resident movement patterns captured by 3D depth cameras within smart homes. By clustering locations based on similarity, they were able to identify optimal sensor positions and fields of view for activity monitoring \cite{Simonsson2023}.
\end{itemize}

\subsubsection{Limitations in Sensor Placements}
Its straightforward approach can oversimplify complex spatiotemporal patterns in sensor data. In scenarios with complex spatial-temporal dynamics, K-means may fail to capture subtle yet crucial patterns due to its rigid clustering approach.

\subsubsection{Inadequate for Observational Range Considerations}
K-means does not account for the observational range of entities, like surveillance cameras, which is a critical aspect in strategic sensor placement.

\subsection{Advantages}
\begin{itemize}
    \item \textbf{Simplicity:} K-means is easy to understand and implement.
    \item \textbf{Efficiency:} It is computationally less intensive, making it suitable for large datasets.
\end{itemize}

\subsection{Limitations and Concerns}
\begin{itemize}
    \item \textbf{Predefined Cluster Number:} The need to specify \( k \) in advance can be limiting in dynamic real-world scenarios.
    \item \textbf{Rigidity in Assignment:} The algorithm's insistence on assigning every data point to a cluster can obscure important patterns.
    \item \textbf{Lack of Spatial Consideration:} K-means does not inherently consider the spatial reach of the entities involved, such as sensors or cameras.
\end{itemize}

\subsection{Contextual Comparison}
While K-means provides a straightforward approach to clustering, it lacks the sophistication needed for complex sensor placement scenarios, especially when compared to methods like Genetic Algorithms. The latter offers a more nuanced understanding of spatial and temporal constraints, making it better suited for such applications. K-means, however, can be a good starting point or a supplementary method in simpler situations where high-level grouping is sufficient.

\section{DBSCAN Clustering}

\subsection{Overview}
Density-Based Spatial Clustering of Applications with Noise (DBSCAN) is a clustering algorithm that forms clusters based on density proximity. Unlike traditional clustering methods, it does not require pre-specification of the number of clusters and can identify outliers as noise. 

\begin{figure}[htbp]
  \begin{adjustwidth}{-1cm}{-1cm} 
    \centering
    \includegraphics[width=\linewidth]{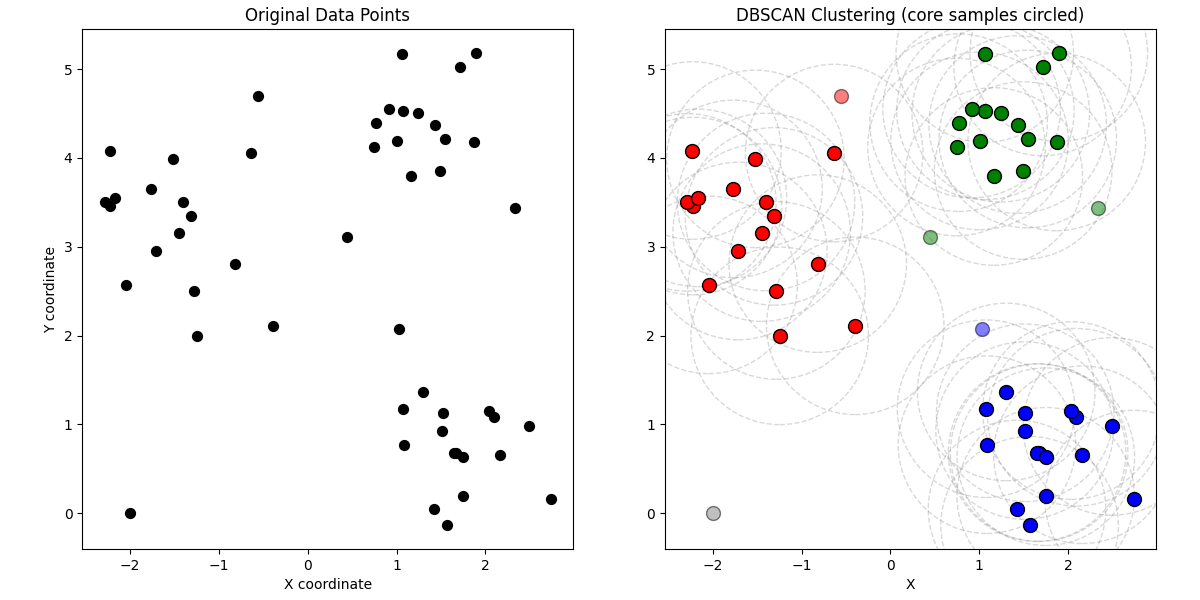} 
    \caption{Example: DBSCAN clustering}
    \label{fig:dbcan-lit-review-demo}
  \end{adjustwidth}
\end{figure}

Figure~\ref{fig:dbcan-lit-review-demo} contrasts unclustered data (left panel) with DBSCAN results (right panel), where vibrant dots indicate core points within their gray radius, muted dots mark boundary points, and gray dots denote noise.

\subsection{Methodology}
DBSCAN starts from an arbitrary point and expands clusters by exploring neighboring points that satisfy a predefined density condition, thereby connecting points directly or indirectly within a density reach. \cite{Ester1996}
\subsubsection{Handling Noise}
Points that do not meet the density criteria are marked as noise, allowing DBSCAN to focus on significant clusters while ignoring outliers.

\begin{algorithm}[H]
\setstretch{1}
\SetAlgoLined
\LinesNumbered  
\KwIn{Dataset $D$, radius $\varepsilon$, minimum number of points $minPts$}
\KwOut{Clusters based on density and noise identification}
\Begin{
    \ForEach{point in $D$}{
        \If{point is not visited}{
            mark point as visited\;
            NeighborPts $\leftarrow$ regionQuery(point, $\varepsilon$)\;
            \If{sizeof(NeighborPts) $<$ $minPts$}{
                mark point as NOISE\;
            }
            \Else{
                increment cluster count\;
                expandCluster(point, NeighborPts, cluster, $\varepsilon$, $minPts$)\;
            }
        }
    }
}
\caption{DBSCAN Clustering Algorithm}
\end{algorithm}

\subsubsection{Expansion of Clusters}
Once a dense point is found, DBSCAN expands the cluster by including all density-reachable points, effectively growing clusters based on local density criteria.

\begin{algorithm}[H]
\setstretch{1}
\SetAlgoLined
\SetKwFunction{FMain}{expandCluster}
\SetKwProg{Fn}{Function}{:}{}
\Fn{\FMain{point, NeighborPts, cluster, eps, minPts}}{
    add point to cluster\;
    \While{NeighborPts is not empty}{
        P $\leftarrow$ NeighborPts.pop()\;
        \If{P is not visited}{
            mark P as visited\;
            NeighborPts' $\leftarrow$ regionQuery(P, eps)\;
            \If{sizeof(NeighborPts') $\geq$ minPts}{
                NeighborPts $\leftarrow$ NeighborPts joined with NeighborPts'\;
            }
        }
        \If{P is not yet member of any cluster}{
            add P to cluster\;
        }
    }
}
\caption{expandCluster function used in DBSCAN}
\end{algorithm}

\subsubsection{Density Reach}
The algorithm defines a minimum number of points within a given radius to identify dense regions, forming the basis of a cluster.

\begin{algorithm}[H]
\setstretch{1}
\SetAlgoLined
\SetKwFunction{FMain}{regionQuery}
\SetKwProg{Fn}{Function}{:}{}
\Fn{\FMain{source\_point, eps, Dataset}}{
    neighbors $\leftarrow$ empty list\;
    \ForEach{target\_point in Dataset}{
        \If{distance(source\_point, target\_point') $\leq$ eps}{
            append target\_point to neighbors\;
        }
    }
    \KwRet{neighbors}\;
}
\caption{regionQuery function used in DBSCAN}
\end{algorithm}

\subsection{Spatiotemporal Adaptations}
Researchers have extended DBSCAN to address spatiotemporal data. For example, Birant and Kut (2007) proposed Spatiotemporal DBSCAN (ST-DBSCAN), a specialized method designed to identify clusters that evolve over time and space \cite{Birant2007}

\subsection{Application in Sensor Placement}
DBSCAN's approach to clustering can be applied to sensor placement, especially when analyzing spatial-temporal data. Here are a few recent examples:

\begin{itemize}

        \item \textbf{Energy Management Network} Yoganathan et al. (2018) proposed a data-driven approach to optimizing sensor placement for energy management in buildings. They combined clustering algorithms, information loss analysis, and the Pareto principle to determine the ideal number and locations of sensors to maximize indoor environment monitoring while minimizing cost and redundancy \cite{Yoganathan2018}

        \item \textbf{Methan Detection Network} Wang et al. (2021) developed an unsupervised machine learning framework to optimize sensor placement for methane leak detection in oil and gas facilities. Their approach integrates facility data, historical leak rates, meteorological data, and atmospheric dispersion models. Sensor locations are optimized to maximize leak detection with a limited budget, and DBSCAN is used for spatial clustering to reduce redundancy. \cite{wang2021unsupervised}
    
\end{itemize}

\subsubsection{Limitations in Sensor Placement}
However, its specific characteristics introduce limitations in the context for ranged-based sensor placements.

\paragraph{Merging of Adjacent Clusters}
DBSCAN tends to merge close clusters, which can lead to larger, less meaningful clusters in sensor networks and obscure true spatial-temporal patterns.

\paragraph{Lack of Observational Range Consideration}
The algorithm does not inherently consider the observational range of sensors, which is critical in many spatial network analyses.

\paragraph{No Centroid Generation}
DBSCAN does not produce centroid points for clusters, which can be a drawback in strategizing node insertions where centroids are valuable.

\subsection{Advantages}
\begin{itemize}
    \item \textbf{Flexibility in Number of Clusters:} DBSCAN automatically determines the number of clusters based on data density.
    \item \textbf{Outlier Detection:} It effectively identifies and excludes outliers, focusing on significant clusters.
\end{itemize}

\subsection{Limitations and Concerns}
\begin{itemize}
    \item \textbf{Sensitivity to Parameters:} The results are highly sensitive to the choice of density parameters.
    \item \textbf{Difficulty with Varying Densities:} DBSCAN struggles with data sets where clusters have varying densities.
    \item \textbf{Inadequacy for Complex Spatial Networks:} It is less effective in scenarios requiring detailed consideration of spatial influence and observational range.
\end{itemize}

\subsection{Contextual Comparison}
Compared to other clustering methods like K-means or sophisticated approaches like Genetic Algorithms, DBSCAN offers a unique perspective by focusing on density-based clustering. However, its limitations in handling complex spatial-temporal data and observational ranges make it less suited for intricate sensor placement problems where these factors are crucial. It excels in scenarios where automatic cluster determination and outlier exclusion are beneficial but may require supplementation with other methods for more comprehensive analysis.

\section{Greedy Algorithm}
\subsection{Overview}
The Greedy Algorithm represents a straightforward approach in sensor placement tasks. It iteratively makes the locally optimal choice at each stage with the intent of finding a global optimum. \cite{ Jungnickel1999} In the context of sensor placement, this algorithm relocates each sensor within a predefined radius to maximize the coverage or reading quality.

\begin{figure}[htbp]
  \begin{adjustwidth}{-1cm}{-1cm} 
    \centering
    \includegraphics[width=0.6\linewidth]{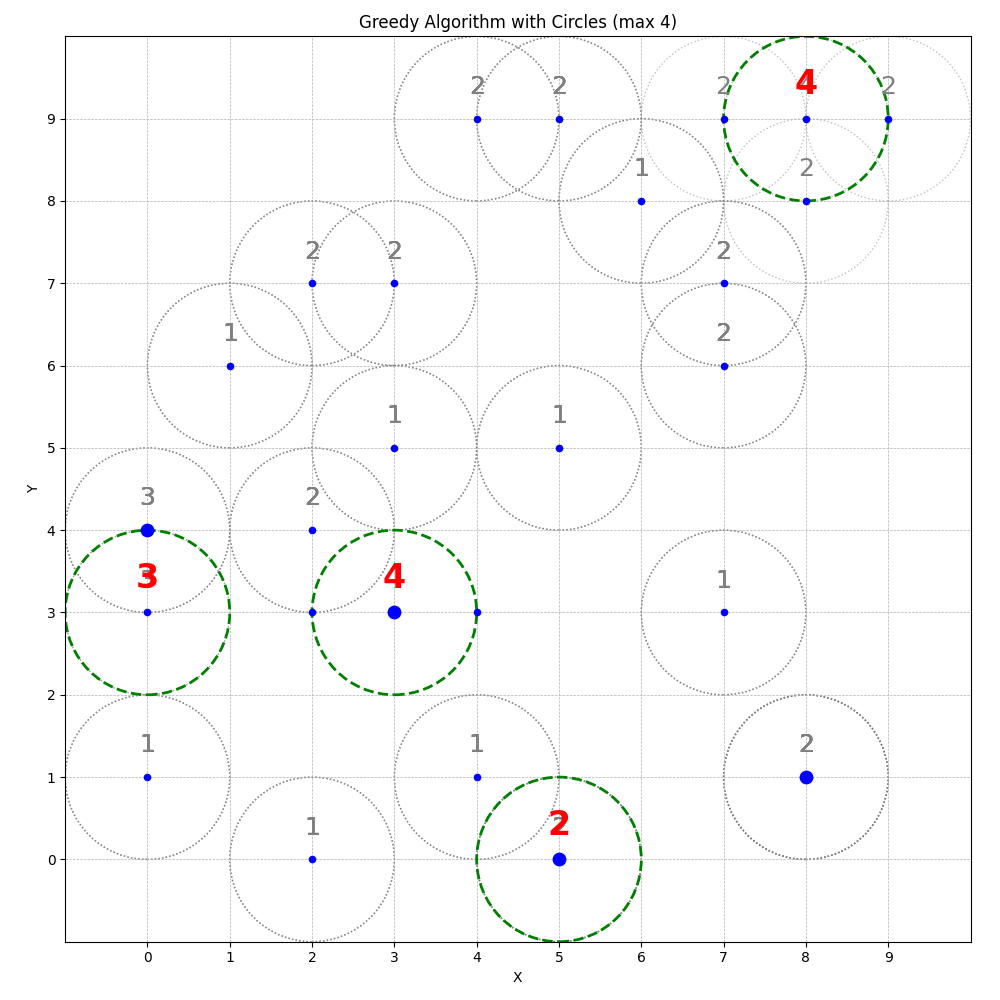} 
    \caption{Example: Greedy clustering}
    \label{fig:greedy-lit-review-demo}
  \end{adjustwidth}
\end{figure}

Figure~\ref{fig:greedy-lit-review-demo} visualizes the greedy algorithm applied to point-selected clustering. Each blue dot is a point; larger dots represent overlapping points. The algorithm selects the center point (shown as green dashed circles) that covers the maximum number of other points within their radius. Numbers indicate the count of points each selected sensor covers. Grey dashed circles illustrate alternative placements considered by the algorithm but ultimately not chosen due to the limited number of sensors allowed, showcasing the algorithm's local optimization behavior.

\subsection{Methodology}
This algorithm evaluates each point independently, selecting positions that yield the maximum sum of values within a specified region around the point. The process iteratively updates circle positions without backtracking, a characteristic hallmark of greedy algorithms \cite{Cormen2009}.

\vspace{8pt}
\begin{algorithm}[H]
\setstretch{1}
\SetAlgoLined
\LinesNumbered  
\KwIn{Set of points $P$, coverage radius $r$, maximum number of circles $n_{max}$}
\KwOut{Optimal placement of circles}
\Begin{
    Initialize an empty list of circles $S$\;
    Initialize all points as uncovered $U \leftarrow P$\;
    \While{$\left|S\right| < n_{max}$ and $U \neq \emptyset$}{
        $p_{best} \leftarrow$ null\;
        $max\_coverage \leftarrow 0$\;
        \ForEach{$p$ in $U$}{
            $coverage \leftarrow \left| \{ u \in U : \text{distance}(u, p) \leq r \} \right|$\;
            \If{$coverage > max\_coverage$}{
                $max\_coverage \leftarrow coverage$\;
                $p_{best} \leftarrow p$\;
            }
        }
        \If{$p_{best}$ is null}{
            break\;
        }
        Add $p_{best}$ to $S$\;
        Update $U$ to remove points covered by $p_{best}$\;
    }
}
\caption{Greedy Placement Algorithm}
\end{algorithm}

\subsection{Applications in Sensor Placement}
The Greedy Algorithm has practical applications for sensor placement due to its ability to make quick, localized decisions and facilitate the efficient management of sensor networks, especially in dynamic environments. Here are a some recent examples:

\begin{itemize}
    \item \textbf{Salinity Monitoring:} Aydin et al. (2019) combined a PCA model with a greedy algorithm to optimize sensor placement for salinity estimation in drainage monitoring networks. Their approach aimed to minimize salinity reconstruction errors for improved water resource management. \cite{Aydin2019}

    \item \textbf{Maritime Surveillance Optimization:} Nguyen et al. (2023) propose a multi-agent approach for sensor allocation and path planning for mobile sensors with limited field of view in maritime environments. Their method employs a greedy sensor assignment algorithm and regret-matching learning to enhance situational awareness with limited resources (i.e. sensors). \cite{Nguyen2023}
\end{itemize}

\subsubsection{Adaptation to Environmental Changes}
Greedy algorithms excel in adapting to temporal and spatial changes in the environment. This feature is particularly useful in spatiotemporal sensor networks, where sensor positions need constant adjustment based on the evolving data landscape.

\subsubsection{Handling of Occupied Positions}
A significant aspect of the Greedy Algorithm in sensor placement is its approach to managing already occupied positions. This is especially pertinent in dense sensor networks where space for optimal sensor placement may be limited. The algorithm's strategy to circumvent this challenge involves checking for position availability and, if necessary, adjusting to the next best available position. This mechanism ensures continuous operation of the sensor network even in scenarios with limited spatial options, reflecting a practical adaptation of the algorithm to real-world constraints.

\subsubsection{Scalability and Performance}
While the Greedy Algorithm provides a fast solution for smaller networks, scalability remains a challenge in larger networks. The computational efficiency drops as the number of sensors increases, highlighting the need for hybrid approaches in extensive networks.

\subsubsection{Integration with Predictive Models}
The algorithm's effectiveness can be enhanced by integrating it with predictive models. These models can provide foresight into future environmental changes, allowing the Greedy Algorithm to make more informed decisions and somewhat mitigating its inherent short-sightedness.

\subsubsection{Sequential Decision-Making}
In sequential decision-making scenarios, the algorithm's inability to backtrack can lead to inconsistencies in sensor placements over time. Advanced sensor management strategies can be employed to periodically re-evaluate and adjust the network, ensuring continuous optimization despite the algorithm's limitations.

\subsection{Advantages}
The primary advantages of the Greedy Algorithm include its simplicity and computational efficiency. It is particularly effective in scenarios where a near-optimal solution is acceptable and rapid decision-making is crucial.

\subsection{Limitations and Concerns}
Despite its advantages, the Greedy Algorithm has inherent limitations, especially in complex sensor networks:
\begin{itemize}
    \item \textbf{Local Optima:} The algorithm's focus on immediate gain often leads to suboptimal global solutions, especially in complex environments with multiple interacting variables.
    \item \textbf{Lack of Future Planning:} The algorithm does not account for future states or possibilities, potentially leading to less effective overall network performance over time.
    \item \textbf{Sequential Inconsistencies:} In the context of sequential decision-making, the lack of backtracking or consideration of future states can result in disjointed or inconsistent sensor placements over time.
\end{itemize}

\subsection{Contextual Comparison}
When compared to other algorithms, such as dynamic programming or backtracking algorithms, the Greedy Algorithm offers a trade-off between computational speed and the quality of the solution. In scenarios where rapid decision-making is prioritized over the optimality of results, the Greedy Algorithm remains a viable option. However, for applications demanding high precision and long-term planning, alternative methods may provide more effective solutions.

\section{Genetic Algorithm}
\subsection{Overview}
The Genetic Algorithm (GA) represents a sophisticated approach to sensor placement problems. Unlike simpler methods like the Greedy Algorithm, GA is based on the principles of natural selection and genetics \cite{Holland1975}. It is particularly effective in exploring a large search space to find optimal or near-optimal solutions for complex problems such as sensor placement, where multiple variables and constraints are involved.

\begin{figure}[htbp]
  \begin{adjustwidth}{-1cm}{-1cm} 
    \centering
    \includegraphics[width=0.6\linewidth]{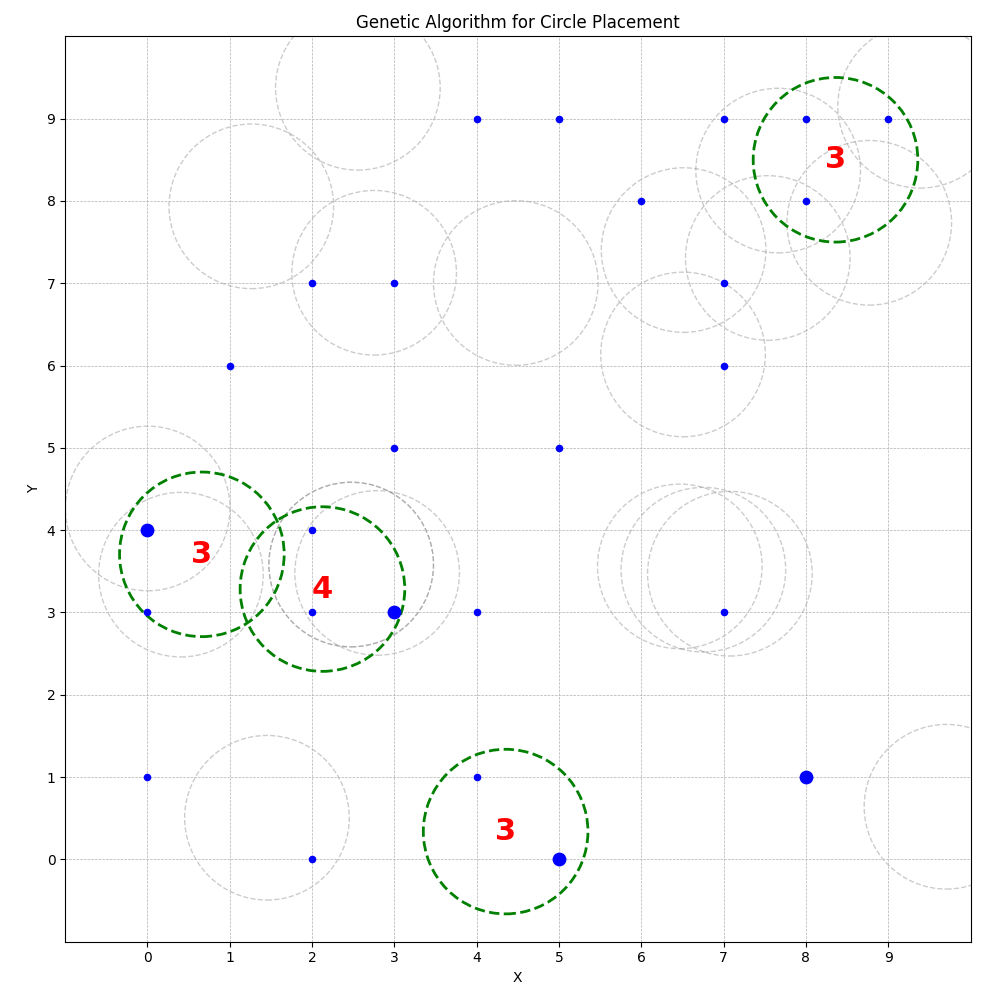} 
    \caption{Example: Genetic clustering}
    \label{fig:genetic-lit-review-demo}
  \end{adjustwidth}
\end{figure}

Figure~\ref{fig:genetic-lit-review-demo} shows the results of a Genetic Algorithm (GA). Each blue dot indicates a point to be covered, with larger dots signifying the presence of multiple overlapping points. The final cluster positions are denoted by green dashed circles, where each red numeral within a circle reflects the count of points it encloses. Grey dashed circles represent previous leading configurations throughout the GA's iterations, providing insight into the evolutionary progression and the exploration of the solution space.

\subsection{Methodology}
GAs work by maintaining a population of potential solutions, known as individuals. These individuals evolve over successive generations. In each generation, individuals are evaluated using a fitness function, selected based on their fitness, and then combined and mutated to produce a new generation of individuals. This process continues until a satisfactory solution is found or a predefined number of generations have been completed. \cite{Whitley1994, Holland1975}  \\

\begin{algorithm}[H]
\setstretch{1}
\SetAlgoLined
\LinesNumbered
\KwIn{Initial population $P$, fitness function $f$, crossover rate $c_r$, mutation rate $m_r$, number of generations $g$}
\KwOut{Optimal or near-optimal solution}
\Begin{
    \For{$i \leftarrow 1$ \KwTo $g$}{
        Evaluate each individual in $P$ using $f$\;
        Select a subset of $P$ based on fitness to create a mating pool\;
        \While{new population size $< |P|$}{
            Select parents from the mating pool\;
            Perform crossover on parents with probability $c_r$ to create offspring\;
            Perform mutation on offspring with probability $m_r$\;
            Add offspring to the new population\;
        }
        Replace $P$ with the new population\;
    }
    \KwRet{The best individual from $P$ according to $f$}\;
}
\caption{Genetic Algorithm for Sensor Placement}
\end{algorithm}

\subsubsection{Population Diversity}
In GAs, population diversity is crucial. It prevents the algorithm from converging prematurely on suboptimal solutions. By maintaining a diverse set of solutions, the algorithm can explore various parts of the search space and enhance its ability to find the global optimum. \cite{Kramer2017} \\

\begin{algorithm}[H]
\setstretch{1}
\SetAlgoLined
\KwIn{Population $P$, diversity threshold $\delta$}
\KwOut{Diverse population}
\Begin{
    Calculate diversity measure for $P$\;
    \If{diversity measure $< \delta$}{
        Introduce new random individuals into $P$\;
        Re-evaluate diversity measure\;
    }
    \KwRet{$P$}\;
}
\caption{Maintaining Population Diversity in Genetic Algorithm}
\end{algorithm}

\subsubsection{Crossover and Mutation}
Crossover combines parts of two or more parent solutions to create new offspring, while mutation introduces random changes to individual solutions. These operators help in exploring new areas of the search space and avoiding local optima. \cite{Kramer2017} \\

\begin{algorithm}[H]
\setstretch{1}
\SetAlgoLined
\KwIn{Parents $p_1$, $p_2$, crossover rate $c_r$, mutation rate $m_r$}
\KwOut{Offspring $o_1$, $o_2$}
\Begin{
    \eIf{random number $< c_r$}{
        Perform crossover on $p_1$ and $p_2$ to produce $o_1$ and $o_2$\;
    }{
        Set $o_1 \leftarrow p_1$ and $o_2 \leftarrow p_2$\;
    }
    Mutate $o_1$ and $o_2$ with probability $m_r$\;
    \KwRet{$o_1$, $o_2$}\;
}
\caption{Crossover and Mutation in Genetic Algorithm}
\end{algorithm}

\subsubsection{Generational Evolution}
Through generational evolution, GAs refine their solutions over time. \cite{Kramer2017} \\

\begin{algorithm}[H]
\setstretch{1}
\SetAlgoLined
\KwIn{Current generation population $P_{curr}$, fitness function $f$}
\KwOut{Next generation population $P_{next}$}
\Begin{
    Evaluate $P_{curr}$ using $f$\;
    Select the fittest individuals from $P_{curr}$\;
    Generate $P_{next}$ using crossover and mutation on the selected individuals\;
    \KwRet{$P_{next}$}\;
}
\caption{Generational Evolution in Genetic Algorithm}
\end{algorithm}

\subsection{Applications in Sensor Placement}
In the context of sensor placement, GAs offer several unique advantages. For instance, they can handle multiple constraints and objectives, making them well-suited for complex scenarios where traditional algorithms might struggle. Here are some recent examples:

\begin{itemize}
    \item \textbf{Police Patrol Networks:}  Jiang et al. (2022) propose a genetic algorithm-based framework to optimize patrol routes for city inspectors (i.e. mobile sensors) in smart city management. Their approach classifies road segments by event frequency and employs a modified genetic algorithm (DP-MOGA) to minimize response time and the number of inspectors needed. The model is tested using real-world patrol data from Zhengzhou, China. \cite{Jiang2022}

    \item \textbf{Multi-Agent Coverage Optimization:}  Sadek et al. (2021) introduce a multi-agent approach for optimizing coverage path planning in unknown environments. Their solution leverages dynamic programming and a genetic algorithm to achieve faster coverage times, minimize redundant coverage, and reduce communication overhead. \cite{Sadek2021}
\end{itemize}

\subsubsection{Handling of Constraints}
In addition, The the flexibility of GAs is evident in their ability to manage various constraints, such as movement limits and speed restrictions of sensors. By customizing crossover and mutation operations, GAs can ensure that these constraints are respected throughout the optimization process. \cite{Holland1975}

\subsubsection{Adaptation to Diverse Scenarios}
GAs are adaptable to a wide range of sensor placement scenarios. Whether it's a static environment or a dynamic one with temporal and spatial variations, GAs can evolve solutions that are well-suited to the specific characteristics of the environment. \cite{Whitley1994}

\subsubsection{Integration with Sensor Networks}
In sensor networks, GAs can be used to optimize sensor positions over time, taking into account factors like coverage, connectivity, and energy efficiency. This makes them a valuable tool in the design and management of efficient and effective sensor networks.

\subsubsection{Design of Fitness Function}
The fitness function in GAs for sensor placement is crucial as it guides the evolution of solutions. In this implementation, the fitness function is designed to ensure that spatial constraints are adhered to, preventing sensor overlap and maximizing sensor coverage across all time steps. It evaluates not just the immediate sensor placement but also its impact over time, favoring solutions that maintain effective coverage throughout the sensor network's operational duration.

\subsubsection{Crossover and Mutation with Spatiotemporal Constraints}
In this case, the crossover and mutation would be tailored to respect the spatiotemporal constraints inherent in sensor placement. The crossover operation involves swapping entire temporal paths of sensors between two parent solutions. This ensures that each sensor's movement history is preserved, adhering to speed limits and other movement constraints. Mutation, on the other hand, is implemented with careful consideration of future positions. Any mutation at a given timestep influences the subsequent positions of the sensor, maintaining the continuity and feasibility of its path.

\subsubsection{Avoidance of Sensor Overlap}
A key consideration in our GA is the avoidance of sensor overlap. This is achieved through a careful design of the fitness function and mutation operations. The fitness function penalizes configurations where sensors overlap in their coverage, thus promoting a spread of sensors across the monitored area. The mutation operation also ensures that any changes in sensor position do not lead to overlaps, respecting the spatial exclusivity required for effective sensor deployment.

\subsection{Advantages}
\begin{itemize}
    \item \textbf{Exploration of Large Search Space:} GAs are capable of exploring a vast search space more efficiently than traditional methods, increasing the likelihood of finding superior solutions.
    \item \textbf{Handling of Multiple Objectives and Constraints:} They can simultaneously consider multiple objectives and constraints, making them highly versatile.
    \item \textbf{Adaptability:} GAs can adapt to changing environments and requirements, making them suitable for dynamic scenarios.
    \item \textbf{Temporal Optimization:} Unlike algorithms that focus on immediate gains, GAs consider the entire timeline, optimizing sensor positions for the duration of their operation.

\end{itemize}

\subsection{Limitations and Concerns}
Despite their advantages, GAs have certain limitations:
\begin{itemize}
    \item \textbf{Computational Intensity:} They can be computationally intensive, especially for large populations or many generations.
    \item \textbf{No Guarantee of Optimal Solution:} GAs do not always guarantee an optimal solution and may sometimes converge on suboptimal solutions.
    \item \textbf{Requirement for Parameter Tuning:} The effectiveness of a GA depends significantly on the tuning of its parameters, which can be a complex and time-consuming process.
    \item \textbf{Complexity in Handling Spatiotemporal Constraints:} The need to respect multiple spatiotemporal constraints adds complexity to the algorithm, requiring sophisticated crossover and mutation strategies.
\end{itemize}

\subsection{Contextual Comparison}
Compared to other algorithms like the Greedy Algorithm, GAs offer a more robust and flexible approach for complex sensor placement problems. They are better suited for scenarios where the exploration of a large search space and handling of multiple constraints are critical. However, their computational intensity and the need for careful parameter tuning can be seen as trade-offs compared to simpler, more straightforward algorithms.

\section{Integer Linear Programming (ILP)}
\subsection{Overview}
Integer Linear Programming (ILP) is a mathematical optimization method, highly effective for binary variable problems such as point placements. It stands out for providing optimal solutions within specific linear constraints and objectives, offering a contrast to heuristic methods that yield approximate solutions. \cite{Nemhauser1988}

\begin{figure}[htbp]
  \begin{adjustwidth}{-1cm}{-1cm} 
    \centering
    \includegraphics[width=0.6\linewidth]{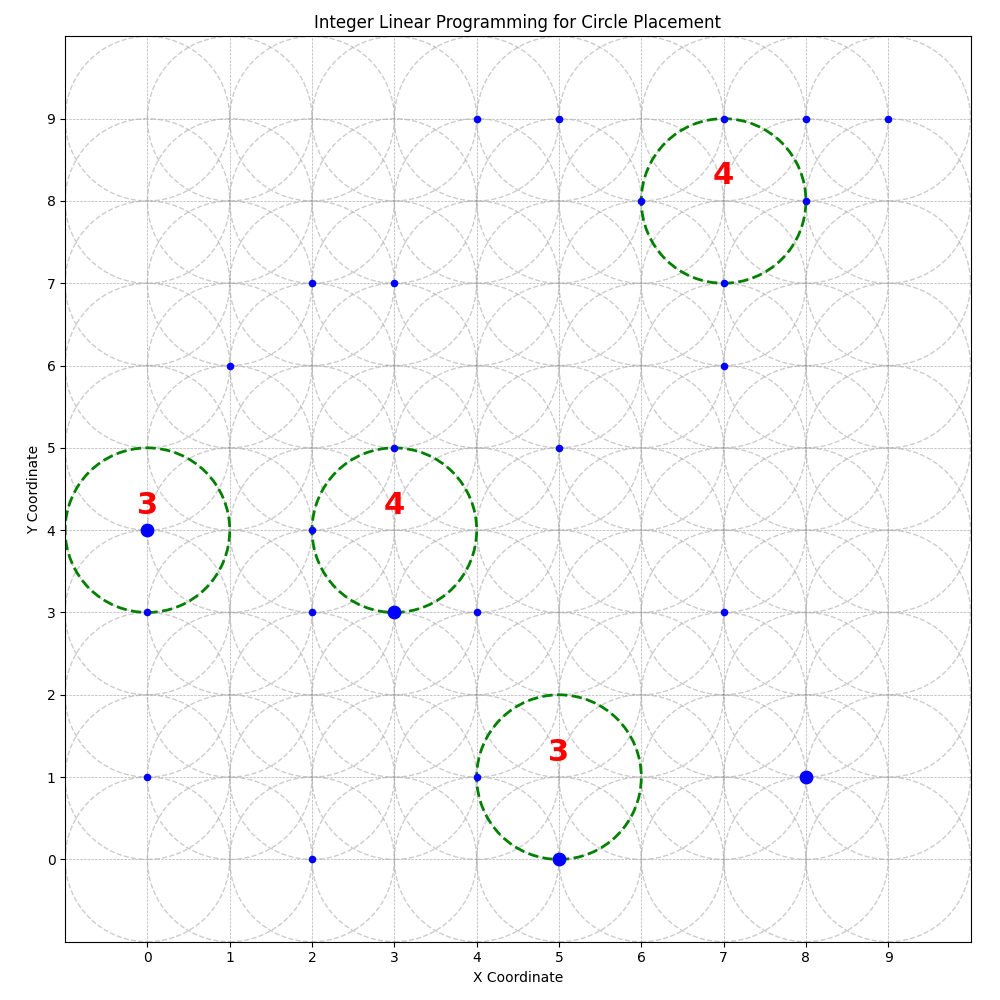} 
    \caption{Example: Integer Linear Programming (ILP) clustering}
    \label{fig:ilp-lit-review-demo}
  \end{adjustwidth}
\end{figure}

Figure~\ref{fig:ilp-lit-review-demo} illustrates the application of Integer Linear Programming (ILP) for optimal circle placement. Each blue dot represents a point to be covered, with larger dots signifying multiple overlapping points at that location. The green dashed circles highlight the optimal solution determined by ILP, with the red numerals inside each circle indicating the total number of points encompassed by that circle. The grey circles represent all possible circle positions evaluated during the ILP process, showcasing the exhaustive search within the solution space to arrive at the optimal configuration.

\subsection{Methodology}
ILP tackles the problem of sensor placement by setting it up as an optimization problem with integer constraints. This involves the definition of decision variables, an objective function, and constraints, all expressed in linear terms. \cite{Schrijver1998}

\subsubsection{Decision Variables}
Decision variables in ILP are binary, denoting whether a sensor is placed (1) or not (0) at a potential location. \cite{Nemhauser1988} \\

\begin{algorithm}[H]
\setstretch{1}
\SetAlgoLined
\LinesNumbered  
\KwIn{Set of potential sensor locations $L$}
\KwOut{Decision variables $x_l$ for each location $l \in L$}
\Begin{
    \ForEach{location $l \in L$}{
        $x_l \gets$ Binary variable, $x_l \in \{0,1\}$\;
    }
}
\caption{ILP Definition of Decision Variables}
\end{algorithm}

\subsubsection{Objective Function}
The objective function aims to maximize the coverage of points, represented by the sum of the values within the range of each placed point. \cite{Nemhauser1988} \\

\begin{algorithm}[H]
\setstretch{1}
\SetAlgoLined
\LinesNumbered  
\KwIn{Decision variables $x_l$, coverage values $c_{l,p}$ for each location $l$ and point $p$}
\KwOut{Maximized total coverage}
\Begin{
    Maximize $\sum_{l \in L} \sum_{p \in P} c_{l,p} x_l$\;
}
\caption{ILP Objective Function for Point Coverage}
\end{algorithm}

\subsubsection{Constraints}
Constraints ensure the practical feasibility of the circle placement, such as limiting the total number of circles and avoiding overlapping coverage areas.  \cite{Nemhauser1988} \\

\begin{algorithm}[H]
\setstretch{1}
\SetAlgoLined
\LinesNumbered  
\KwIn{Decision variables $x_l$, maximum number of circles $M$, overlap matrix $o_{l,m}$}
\KwOut{Feasible solution satisfying all constraints}
\Begin{
    \tcp{Limit the number of circles}
    $\sum_{l \in L} x_l \leq M$\;
    \ForEach{location $l \in L$}{
        \ForEach{location $m \in L \setminus \{l\}$}{
            \tcp{Prevent overlap in point coverage}
            $x_l + x_m \leq 1$ if $o_{l,m} = 1$\;
        }
    }
}
\caption{ILP Constraints for Circle Placement}
\end{algorithm}

\subsection{Applications in Sensor Placement}
ILP is adept at managing complex sensor placement scenarios, particularly with dynamic environmental changes and multiple constraints. Here are some recent examples:

\begin{itemize}
    \item \textbf{Smart Home Target Tracking:} Gholizadeh-Tayyar et al. (2020) introduce an ILP model to optimize sensor placement for indoor target tracking in smart homes. Their approach uniquely integrates target tracking methods with sensor deployment, accounting for smart home layouts, sensor parameters, and system reliability. \cite{Gholizadeh-Tayyar2020}   
\end{itemize}

\subsubsection{Constraint Management}
ILP rigorously manages constraints, including those on sensor movement limits and non-overlapping coverage areas. Specifically, it mirrors the constraints in the code by limiting potential sensor positions based on their previous locations and a defined maximum movement speed. This approach ensures that sensor repositioning between time steps is both realistic and feasible. \cite{Schrijver1998}

\subsubsection{Movement and Coverage Optimization}
The ILP model may include constraints such as movement speed while also optimizing sensor coverage. To avoid overlap, a constraint is applied similar to the non-overlapping constraint in the code, where sensor positions are chosen such that their coverage areas do not intersect within a specified radius. This ensures effective utilization of each sensor's coverage capacity without redundancy.

\subsubsection{Adaptation to Temporal Variations}
In dynamic environments, ILP adapts sensor positions in response to changing scenarios over time. This adaptation is akin to the time-step adjustments in the model, where sensor positions are re-calibrated for each time step, accounting for both the movement constraints and the need to avoid overlapping coverage areas.

\subsection{Advantages}
\begin{itemize}
    \item \textbf{Optimality:} ILP ensures optimal solutions, a notable strength over heuristic methods.
    \item \textbf{Precise Constraint Handling:} Its explicit incorporation of various constraints makes it highly effective.
    \item \textbf{Adaptability:} Suitable for environments with temporal variations, ILP adapts well to changing conditions.
    \item \textbf{Efficient Coverage:} Focusing on maximizing coverage, ILP ensures efficient sensor network utilization.
\end{itemize}

\subsection{Limitations and Concerns}
The limitations of ILP include:
\begin{itemize}
    \item \textbf{Computational Demand:} Particularly intense for large-scale problems.
    \item \textbf{Restriction to Linear Formulations:} Requires linear constraints and objective functions.
    \item \textbf{Binary Focus:} Mainly handles binary decision variables.
    \item \textbf{Time-Consuming:} Solution acquisition can be slow for complex problems.
    \item \textbf{Parameter Sensitivity:} Minor changes can significantly alter solutions.
\end{itemize}

\subsection{Contextual Comparison}
ILP's deterministic and optimal solutions are advantageous for maximum coverage and precision compared to heuristic methods like Genetic Algorithms and simpler methods like the Greedy Algorithm. However, its slower solution times and computational intensity, coupled with the need for linear formulations, make it less flexible than these alternatives.

\section{Mixed Integer Linear Programming (MILP)}
\subsection{Overview}
Mixed Integer Linear Programming (MILP) extends the capabilities of Integer Linear Programming (ILP) by introducing continuous variables alongside binary decision variables. This expansion enables MILP to model more complex and dynamic scenarios, such as circle (i.e. sensor) placement with varied movement patterns and speeds. 

\begin{figure}[htbp]
  \begin{adjustwidth}{-1cm}{-1cm} 
    \centering
    \includegraphics[width=0.6\linewidth]{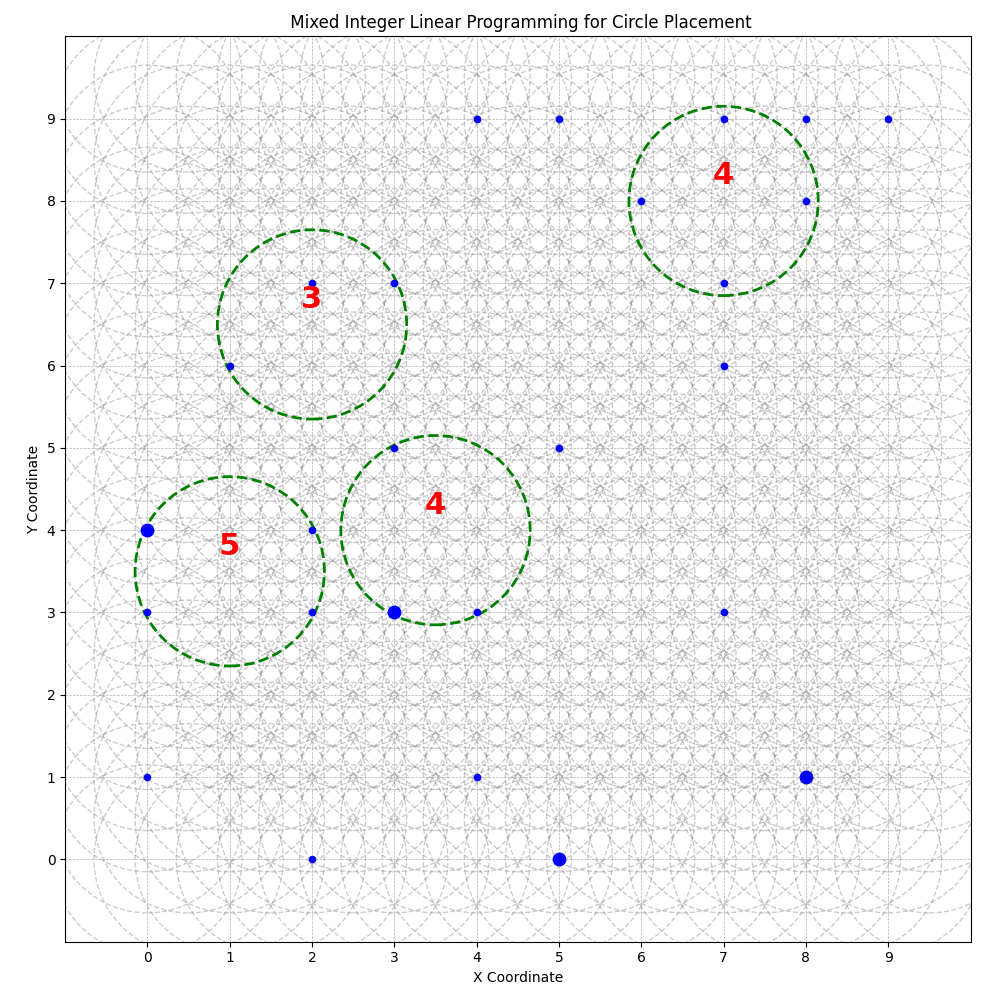} 
    \caption{Example: Mixed Integer Linear Programming (MILP) clustering}
    \label{fig:milp-lit-review-demo}
  \end{adjustwidth}
\end{figure}

Figure~\ref{fig:milp-lit-review-demo} illustrates the application of Mixed Integer Linear Programming (MILP) for optimal circle placement. Each blue dot represents a point to be covered, with larger dots indicating multiple overlapping points at the same location. The green dashed circles highlight the optimal solution determined by MILP, marked by red numerals inside each circle that indicate the total number of points encompassed. The grey circles represent all potential circle positions, including fractional coordinates, evaluated during the MILP process. This demonstrates the comprehensive search within an expanded solution space, encompassing both integer and fractional positions, to determine the optimal configuration.

\subsection{Methodology}
MILP approaches sensor placement by incorporating continuous variables and modified constraints, thereby allowing a more nuanced representation of sensor behavior over time. 

\subsubsection{Decision Variables}
In contrast to ILP's binary-only approach, MILP uses both binary and continuous variables. Binary variables (has\_circle) indicate the presence or absence of a circle at a location, while continuous variables (movement\_step) represent the circle's movement. \cite{LeeLeyffer2012} \\

\begin{algorithm}[H]
\setstretch{1}
\SetAlgoLined
\KwIn{Set of potential sensor locations $L$, set of time steps $T$}
\KwOut{Binary sensor placement variables and continuous movement variables}
\Begin{
    \ForEach{location $l$ in $L$}{
        sensor\_vars[$l$] $\leftarrow$ binary variable indicating sensor presence at location $l$\;
    }
    \ForEach{location $l$ in $L$, time $t$ in $T$}{
        movement\_vars[$l$, $t$] $\leftarrow$ continuous variable indicating sensor movement at location $l$ at time $t$\;
    }
}
\caption{Definition of Decision Variables in MILP}
\end{algorithm}

\subsubsection{Objective Function}
MILP's objective function is dual-faceted: maximizing total circle coverage and minimizing movement. This dual focus aligns with temporal scenarios where both coverage and movement efficiency are vital. \cite{LeeLeyffer2012} \\

\begin{algorithm}[H]
\setstretch{1}
\SetAlgoLined
\KwIn{sensor\_vars, movement\_vars}
\KwOut{Optimized objective value}
\Begin{
    coverage\_sum $\leftarrow$ sum of sensor\_vars indicating total coverage\;
    movement\_sum $\leftarrow$ sum of movement\_vars indicating total movement\;
    Objective $\leftarrow$ maximize(coverage\_sum) - minimize(movement\_sum)\;
    Solve MILP with Objective\;
}
\caption{MILP Objective Function for Circle Placement}
\end{algorithm}

\subsubsection{Enhanced Constraints}
MILP introduces advanced constraints that govern circle movement, limiting it based on previous positions and maximum speed. These constraints offer a realistic portrayal of circle (i.e. sensor) dynamics, especially in variable environments. \cite{LeeLeyffer2012} \\

\begin{algorithm}[H]
\setstretch{1}
\SetAlgoLined
\KwIn{sensor\_vars, movement\_vars, maximum speed $max\_speed$}
\Begin{
    \ForEach{location $l$ in $L$, time $t$ and $t+1$ in $T$}{
        \eIf{sensor\_vars[$l$] at time $t$ and $t+1$ is active}{
            Constrain movement\_vars[$l$, $t$] $\leq$ $max\_speed$\;
        }{
            Constrain movement\_vars[$l$, $t$] $=$ 0\;
        }
    }
}
\caption{Enhanced Constraints for Sensor Movement in MILP}
\end{algorithm}

\subsection{Applications in Sensor Placement}
MILP's application in sensor placement is distinguished by its ability to handle more complex and dynamic situations than ILP. Here are some recent examples:

\begin{itemize}
    \item \textbf{Sensor Network Optimization for Methane Emissions:} Klise et al. (2020) present a MILP approach for optimizing sensor placement in methane emission monitoring scenarios. They develop the open-source Chama package to determine ideal sensor positions and detection thresholds for maximizing leak detection efficacy. Their model accounts for uncertainties in wind conditions and emission patterns. \cite{Klise2020}
\end{itemize}

\subsubsection{Dynamic Movement Modeling}
Unlike ILP, MILP models sensor movements continuously, reflecting realistic sensor behavior. The inclusion of continuous variables in the code for sensor movements allows the model to account for varying speeds and gradual repositioning.

\subsubsection{Comprehensive Coverage and Movement Optimization}
MILP not only considers total coverage but also integrates movement costs into its optimization process. This is exemplified in the code's objective function, which includes the total movement variable, striking a balance between coverage and movement efficiency.

\subsubsection{Adaptation to Complex Scenarios}
MILP's advanced constraints and continuous variables make it adept at handling more intricate scenarios. The model can simulate scenarios where sensor movement and its costs are significant factors, as reflected in the movement constraints and continuous variables in the code.

\subsection{Advantages}
\begin{itemize}
    \item \textbf{Enhanced Realism:} MILP's use of continuous variables offers a more realistic depiction of sensor movements.
    \item \textbf{Balanced Optimization:} It considers both coverage maximization and movement minimization, aligning with practical needs.
    \item \textbf{Flexibility:} More suited for complex and dynamic environments compared to ILP.
    \item \textbf{Comprehensive Modeling:} MILP's ability to incorporate varied constraints and variables allows for detailed scenario simulation.
\end{itemize}

\subsection{Limitations and Concerns}
The limitations of MILP include:
\begin{itemize}
    \item \textbf{Computational Complexity:} More demanding than ILP due to the inclusion of continuous variables.
    \item \textbf{Solution Difficulty:} Finding optimal solutions can be more challenging and time-consuming.
    \item \textbf{Model Complexity:} Requires careful formulation and understanding of constraints and variables.
\end{itemize}

\subsection{Contextual Comparison}
MILP, with its inclusion of continuous variables and dual-objective focus, offers a more nuanced approach than ILP, particularly in scenarios where sensor movement and its implications are critical. While it maintains the strengths of ILP in terms of constraint handling and coverage optimization, MILP stands out in its ability to model sensor dynamics more realistically. However, this comes at the cost of increased computational complexity and solution difficulty.

\section{Graph Signal Sampling}

Graph Signal Sampling is an advanced technique in signal processing that applies the principles of graph theory to sample and reconstruct signals on irregular domains. It extends traditional signal processing, which typically deals with time or spatially-regular signals, to complex networks or graphs. 

\subsection{Graph Sampling Theory}
\subsubsection{Theoretical Framework}
Graph sampling theory establishes a method for selecting a subset of nodes (vertices) in a graph to reconstruct a signal defined over the entire graph. This theory is grounded in understanding how the structure of a graph (how nodes are interconnected) affects the signal (information or data) residing on the graph. The main objective is to identify pivotal nodes that can represent the entire signal on the graph, facilitating efficient signal processing tasks like compression, reconstruction, and analysis. \cite{Nomura2022}

\subsubsection{Practical Implications}
In real-world applications, such as sensor networks, this theory provides a systematic approach for deploying a limited number of sensors to capture essential information. It's especially beneficial in environments where deploying sensors at every possible location is neither feasible nor cost-effective. Graph sampling theory helps in making informed decisions about where to place sensors for maximum coverage and efficiency.

\subsection{Sampling Strategy}
\subsubsection{Node Selection Criteria}
The strategy for sampling on a graph involves selecting nodes that best represent the overall signal. This decision-making process is informed by various graph attributes, including node centrality, connectivity, and the distribution of signal values. The goal is to choose nodes that, together, can reconstruct the full graph signal as accurately as possible. \cite{Nomura2022}

\paragraph{Voronoi Regions}
Voronoi regions in graph signal sampling divide the graph into distinct areas, each centered around a selected node or sensor. These regions help in determining the influence or coverage area of each sensor, which ensures that the entire graph is effectively monitored. Using Voronoi regions helps in evenly distributing the sensors across the network for balanced signal representation. \cite{Nomura2022}

\paragraph{P-Hop Neighbors}
P-hop neighbor analysis in graph sampling focuses on the proximity of nodes within a specific number of hops (steps) in the graph. By considering nodes that are within a certain hop distance, the strategy ensures that the sampled nodes capture local signal variations effectively and leads to more accurate signal reconstruction from limited observations. \cite{Nomura2022}

\subsection{Online Dictionary Learning}
\subsubsection{Adaptive Dictionary Construction}
Online Dictionary Learning in the context of graph signal sampling refers to the continuous development and refinement of a basis set (dictionary) that can efficiently represent the graph signals. These dictionaries are collections of basis vectors that capture significant patterns or features in the signal, allowing for efficient signal approximation and analysis. \cite{Nomura2022}

\paragraph{Iterative Updates}
This process is iterative, meaning the dictionary is constantly updated as new data is observed. This iterative nature allows the dictionary to evolve and adapt to changes in the signal, ensuring it remains an accurate representation of the current state of the graph signal. \cite{Nomura2022}

\paragraph{Sparse Regularization}
Sparse regularization in dictionary learning is a method that emphasizes the use of a small number of significant basis vectors for signal representation. It aims to achieve a sparse representation where most of the coefficients are zero or close to zero, retaining only those that are most relevant for capturing the key features of the data. This is particularly beneficial in handling high-dimensional data because it simplifies the model, making it both computationally efficient and easier to interpret. The process typically involves a technique like soft-thresholding, which reduces the influence of less significant data elements by honing in on the most crucial aspects of the signal. This selective focus on important features leads to a more meaningful and manageable representation of complex data. \cite{Nomura2022}

\paragraph{Real-Time Data Adaptation}
Adapting to real-time data is a critical aspect of this method, allowing for responsive signal processing in dynamic environments. However, this adaptability introduces challenges, such as managing incomplete or noisy data and ensuring computational efficiency, particularly when the graph or the signal changes rapidly.

\subsection{Dynamic Sensor Placement}
\subsubsection{Sensor Positioning Strategies}
Dynamic sensor placement is a strategy in graph signal sampling that involves the continuous adjustment of sensor locations based on the evolving nature of the graph signal. Unlike static placement, where sensors remain fixed, dynamic placement allows for a more responsive and flexible monitoring system.

\paragraph{Utility-Based Placement}
A utility function, often based on real-time data like heatmaps, guides the placement of sensors. This function evaluates the importance or relevance of different locations on the graph, allowing sensors to be positioned where they are most needed. This method ensures that sensors are always optimally located to capture crucial signal information.

\paragraph{Dynamic Adaptation Challenges}
 Although it is advantageous in adapting to changing conditions, dynamic sensor placement also poses several challenges. These include computational complexity due to constant recalculations, the necessity for accurate and timely data processing, and logistical considerations in physically moving sensors within the network, if applicable.

\chapter{ROBUST Network Theory}
\label{chap:robust_network}

\section{Overview}
The Ranged Observer Bipartite-Unipartite SpatioTemporal (ROBUST) Network is a novel framework for modeling the intricate dynamics between spatially-sensitive observers and observable entities. It addresses the challenges identified in Chapters \ref{chap:intro} - \ref{chap:lit_review} and expands upon the theoretical concepts presented in Chapter \ref{chap:theoretical_foundations}.

\vspace{2mm}
\begin{figure}[H]
    \centering
    \begin{tikzpicture}
      \node[draw, thick, rounded corners] (observer) at (0, 0) {observer};
    
      \node[draw, thick, rounded corners] (box1) at (-2, -1.5) {observable};
      \node[draw, thick, rounded corners] (box3) at (2, -1.5) {observable};
      \node[draw, thick, rounded corners] (box4) at (-2, 1.5) {observable};
      \node[draw, thick, rounded corners] (box6) at (2, 1.5) {observable};
      \node[draw, thick, rounded corners] (box7) at (-3, 0) { observable };
      \node[draw, thick, rounded corners] (box8) at (3, 0) { observable };

      \draw[->] (box4) -- (observer);
      \draw[->] (box6) -- (observer);
      \draw[->] (box1) -- (observer);
      \draw[->] (box3) -- (observer);
      \draw[->] (box7) -- (observer);
      \draw[->] (box8) -- (observer);
    \end{tikzpicture}
    \caption{Example Observer-Observable Network}
    \label{fig:robust-intro}
\end{figure}
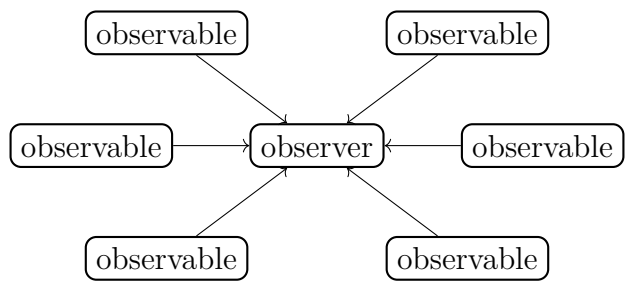

\subsection{Approach}
ROBUST is a mathematical and conceptual framework that represents the spatial and temporal relationships between observers and observable events. \cite{Holmberg2022} It consists of a set of nodes (representing observers and observables) and a set of edges (representing the spatiotemporal interactions between them). \cite{Holmberg2023}

\subsection{Components}

\paragraph{Observer Nodes:} Controllable elements in the network, such as sensors or cameras, that can observe events.

\paragraph{Observable Nodes:} Events or phenomena that are being observed.

\paragraph{Ranged Edges:} Edges in ROBUST have both a spatial and temporal component. \cite{Holmberg2022} The spatial component defines the distance between the observer and observable nodes, while the temporal component defines the time interval over which the observation takes place. \cite{Holmberg2023}

\subsection{Analysis Techniques}

ROBUST can be analyzed using both bipartite and unipartite techniques.

\paragraph{Bipartite Analysis:} Focuses on the interactions between observer and observable nodes.

\paragraph{Unipartite Analysis:} Focuses on the structure and dynamics of the network as a whole.

\subsection{Objective}
The ultimate objective of ROBUST is to develop a sophisticated framework for optimizing the performance of observational systems. This framework will allow users to:
\begin{itemize}
    \item Identify optimal locations for static observers to ensure coverage and efficiency.
    \item Plan and reconfigure paths for dynamic observers to adapt to changing environments and targets.
    \item Integrate continuous observer models for ongoing, real-time environmental monitoring.
    \item Expand the observational reach of existing systems to cover larger or more complex areas.
\end{itemize}

\section{Definition of the ROBUST Network}
The ROBUST Network is a novel graph-theoretic framework designed to model complex observational systems. It consists of two distinct sets of entities: observer nodes and observable nodes.

\subsubsection{Observer Nodes}
Observer Nodes represent entities that can observe events, such as sensors or cameras. Each observer node has a defined myopic range, which limits the distance over which it can observe events. \cite{Holmberg2022,Holmberg2023}

\subsubsection{Observerable Nodes}
Observable Nodes represent events or phenomena that can be observed. These nodes can be further classified into two categories:

\begin{itemize}

    \item  \textbf{Observed Events:} Events that are within the myopic range of at least one observer node.
    \item  \textbf{Unobserved Events:} Events that are not within the myopic range of any observer node.
The ROBUST Network is a bipartite graph, meaning that edges can only exist between observer nodes and observable nodes. An edge is created between an observer node and an observable node if the observer node is within the myopic range of the observable node.
\end{itemize}

The ROBUST Network also incorporates temporality, allowing events to occur at different points in time. This enables the network to capture the dynamic nature of observation.

In addition to the bipartite structure, the ROBUST Network also includes a unipartite grouping of unobserved events. This grouping is formed by clustering unobserved event nodes that are within the myopic range of each other. This unipartite grouping provides insights into the spatial distribution and potential significance of unobserved events.

\subsection{Extending Spatiotemporal Networks to ROBUST Networks}

Building upon the foundational concepts of spatiotemporal networks introduced in Chapter \ref{chap:theoretical_foundations}, we further elaborate the model to accommodate the specificities of the ROBUST Network. The ROBUST Network integrates the dynamic nature of spatiotemporal networks with a bipartite structure by distinguishing between observer and observable entities, and introduces the concept of myopia to model spatial constraints effectively.

\subsubsection{Notation and Terminology}

\begin{itemize}
\item \textbf{Sets:}
    \begin{itemize}
    \item $\mathbb{R}^n$: Represents an n-dimensional, real-valued space. In our model, this space describes the spatial coordinates of nodes.
    \item $\mathbb{A}$: This denotes a general set of possible attributes for nodes. Examples of node attributes could include sensor battery level, signal strength, or event intensity. 
    \item $\mathbb{B}$: Similar to $\mathbb{A}$, this represents a set of attributes for edges. Edge attributes  could include observation quality or signal strength  between nodes. 
    \end{itemize}

\item \textbf{Mathematical Concepts:}
    \begin{itemize}
    \item \textbf{Bipartite Graph:} A graph where nodes are divided into two distinct sets, and edges can only exist between nodes in different sets (never between nodes of the same set).
    \item \textbf{Myopic Range:} The maximum distance within which an observer node can sense or observe an event.
    \end{itemize}

\item \textbf{Specific to the ROBUST Network Model:}
    \begin{itemize}
    \item $V_O$: The set of observer nodes.
    \item $V_E$: The set of observable nodes.
    \item $V_E^{obs}$: The subset of observable nodes that are currently within the myopic range of at least one observer.
    \item $V_E^{unobs}$: The subset of observable nodes that are not currently within the myopic range of any observer.
    \end{itemize}
\end{itemize}

\subsubsection{Mathematical Formulation of ROBUST Networks}
Let the ROBUST Network be denoted as \(G_{st} = (V, E, P, T, A_V, A_E)\), where:

\begin{itemize}
    \item Nodes \((V)\): \(V\) represents the collection of all nodes in the network. These nodes are divided into two sets:
    \begin{itemize}
        \item Observer Nodes \((V_O)\) represent entities that can observe events, such as sensors or cameras. Each observer node has a defined myopic range, which limits the distance over which it can observe events.
        \item Observable Nodes \((V_E)\) represent events or phenomena that can be observed. These nodes can be further classified into two categories:
        \begin{itemize}
            \item Observed Events \((V_{E_{obs}})\): Events that are within the myopic range of at least one observer node.
            \item Unobserved Events \((V_{E_{unobs}})\): Events that are not within the myopic range of any observer node.
        \end{itemize}
    \end{itemize}
    \item Edges \((E)\):  \(E\) represents the connections between observer nodes and observable nodes. An edge only exists between an observer node and an observable node if the observable node is within the myopic range of the observer node, adhering to a bipartite graph configuration.
    \item Spatial Positioning \((P)\): \(P: V \times T \rightarrow \mathbb{R}^n\) captures the spatial location of each node (observer or observable) at a specific point in time \((t)\). This function incorporates the myopic constraints that influence observer capabilities.
    \item Temporal Domain \((T)\): \(T\) represents the time domain, allowing the model to capture the dynamic nature of observation.
    \item Time-Variant Attributes \((A_V \text{ and } A_E)\):
    \begin{itemize}
        \item \(A_V: V \times T \rightarrow \mathbb{A}\) represents time-variant attributes associated with each node in the network (observer or observable) over time \((t)\).
        \item \(A_E: E \times T \rightarrow \mathbb{B}\) represents time-variant attributes associated with the edges between nodes over time \((t)\). These attributes could include the quality of observation or the strength of the signal between an observer and an observable node.
    \end{itemize}
\end{itemize}

The functions  $P$, $A_V$ , and $A_E$ introduce dimensions of space, attributes of nodes, and attributes of edges that are vital for the temporal-spatial analysis of networks under the constraints of myopic observation.

\section{Observer Nodes}
\label{sec:observer_nodes}

Observer nodes are configurable in the ROBUST Network, acting as the system's controllable elements capable of monitoring and collecting data from the environment. In practice, these nodes have severe spatiotemporal constraints, including a limited range of observation and a limited ability to observe events that occur outside of a certain time frame. 

\subsection{Observer Role in ROBUST}
Observer nodes form the backbone of the ROBUST Network, endowed with the autonomy to monitor, interact with their environment, and make decisions under direct operational control. Their primary role extends beyond comprehensive surveillance across the network's spatiotemporal dimensions; they effectively become the eyes, operational arms, and decision-making brains of the system. Tasked with discerning and responding to the dynamic landscape of observable events, their deployment and actions are strategically informed to optimize coverage and efficiency. This involves not only the physical positioning for optimal data acquisition but also the adaptability to shift focus based on evolving observational priorities, thus highlighting regions of high interest or activity. As the principal decision-making agents, observer nodes dynamically interpret data to inform their actions, ensuring the network's adaptability and responsiveness. This foundational role underpins the subsequent discussions on deployment strategies and bipartite dynamics, providing a clear delineation of observer nodes as distinct, proactive, and decision-making agents within the network's architecture.

\subsection{Observer View and Myopia}
Observer nodes in the ROBUST Network can represent either mechanical, biological, or virtual entities. Their ability to sense the environment in the world is based on their specific embodiment. Observer nodes in the ROBUST Network are equipped with a variety of sensors that allow them to interact with their environment and collect data. The type of sensor used determines the observer node's view, which includes the range, resolution, and type of data that can be collected.

\subsubsection{Type of Sensors}
Observer nodes in the ROBUST Network can use a wide range of sensors, including:  \cite{Ladner2002, Chung2001, wilson2003}

\begin{itemize}
    \item \textbf{Mechanical sensors:} Sensors that use mechanical components to detect and measure physical phenomena. Examples of mechanical sensors include cameras, microphones, and accelerometers.
    \item \textbf{Biological sensors:} Sensors that use biological components to detect and measure chemical or biological phenomena. Examples of biological sensors include biosensors and chemical sensors.
    \item \textbf{Virtual sensors:} Software engineering tools that can be used to detect and measure changes in the state of a system. Examples of virtual sensors include differencing and hashing between two states of code.
\end{itemize}

\subsubsection{Generalized Sensor Categories in Observer Nodes}
Observer nodes within the ROBUST Network utilize a broad spectrum of sensors, each designed to detect specific changes in the environment. These changes can be physical or digital, illustrating the flexibility and depth of environmental monitoring. The following categories encapsulate the range of changes these sensors are designed to detect:

\begin{itemize}
    \item \textbf{Light Sensors:} Detect changes in light intensity, color spectrum, and patterns, applicable in both natural and artificial lighting conditions.
    
    \item \textbf{Sound Sensors:} Capture variations in sound waves, including frequency, amplitude, and direction, crucial for audio analysis and surveillance.
    
    \item \textbf{Pressure Sensors:} Measure changes in pressure, including atmospheric, liquid, or gas pressures, relevant for environmental monitoring and industrial processes.
    
    \item \textbf{Chemical Sensors:} Identify changes in chemical compositions, detecting specific substances or changes in air and water quality.
    
    \item \textbf{Temperature Sensors:} Monitor fluctuations in temperature, essential for environmental control, weather monitoring, and industrial processes.
    
    \item \textbf{Motion Sensors:} Detect movement or displacement, applicable in security systems, wildlife tracking, and automated systems.
    
    \item \textbf{Dimension Sensors:} Measure changes in size or volume, including physical dimensions like length, width, height, or digital dimensions such as file size, useful in logistics, manufacturing, and digital storage management.
    
    \item \textbf{Weight Sensors:} Gauge changes in weight or mass, crucial for industrial weighing systems, health monitoring, and inventory control.
    
    \item \textbf{Electromagnetic Sensors:} Detect electromagnetic fields or waves, applicable in navigation, communication, and scientific research.
    
    \item \textbf{Environmental Sensors:} Comprehensive category covering sensors that monitor humidity, pH levels, salinity, and more, providing a holistic view of environmental conditions.
\end{itemize}

This classification not only underscores the diversity of sensors integrated into observer nodes but also reflects the wide-ranging nature of environmental changes they are tasked with detecting.  \cite{Ladner2002, Chung2001, wilson2003}

\subsubsection{Sensor Capabilities}
Each type of sensor has its own unique set of capabilities, including:
\begin{itemize}
    \item \textbf{Range:} The distance over which the sensor can collect data.
    \item \textbf{Resolution:} The level of detail that the sensor can capture.
    \item \textbf{Temporal resolution:} The frequency at which the sensor can collect data.
\end{itemize}

\subsubsection{Limitations of Sensors}
Some sensors have inherent limitations, such as line-of-sight or field of view. For example, visual sensors can only see objects that are within their line of sight. While audio sensors can also be affected by line-of-sight, sound waves can diffract around obstacles to some extent, allowing them to be heard even if the source is not directly visible. Other sensors have no such limitations and can detect objects or phenomena that are not directly visible. For example, GPS sensors can track the location of an object even in conditions of poor visibility, such as darkness or fog.

These limitations are important to consider when selecting sensors for observer nodes in addition to where to place and in what orientation. The specific limitations of a sensor will determine how well it can be used to collect data in a particular environment. 

\subsubsection{Myopia}
The term "myopia" in the context of the ROBUST Network refers to the inherent limitations in a sensor's range and field of view. This affects how and what observers can detect. For example, a sensor with a short range may not be able to detect objects that are far away. Similarly, a sensor with a narrow field of view may not be able to detect objects that are not directly in front of it.

\subsubsection{Importance of Observer View and Myopia}
Understanding the interplay between sensor capabilities and observational needs is crucial for optimizing the deployment and operation of observer nodes. The network's design must account for these factors, ensuring that nodes are not only well-equipped for their intended roles but also placed and scheduled to overcome or mitigate their myopic constraints. This section sets the groundwork for discussing how strategic placement, technology selection, and coordination enhance the network's overall observational capacity.

\subsubsection{Additional Considerations}
In addition to the factors discussed above, there are a number of other considerations that can affect observer view and myopia, including:

\begin{itemize}
    \item \textbf{Orientation:} The direction in which the sensor is pointing.
    \item \textbf{Obstructions:} Objects that can block the sensor's view.
    \item \textbf{Environmental conditions:} Factors such as lighting, noise, and weather can affect the sensor's performance.
\end{itemize}

By taking all of these factors into account, it is possible to design and deploy observer nodes that can effectively collect the data needed to meet the objectives of the ROBUST Network.

\subsubsection{Abstract Sensors for Existential Observations}
In addition to the physical and digital sensors outlined previously, observer nodes may also integrate abstract sensors designed to detect and qualify existential observables. These abstract sensors represent logical conditions, states, or events that, while not directly measurable through physical means, are crucial for the comprehensive monitoring and analysis capabilities of the network.

\paragraph{Definition and Examples:}
Abstract sensors operate by interpreting data, signals, or inputs from various sources to determine the presence, absence, or state of a specific observable. Examples include:
\begin{itemize}
    \item \textbf{Presence Detection:} Logical conditions assessing the presence or absence of an entity within a defined space, akin to the waiter-client model where the primary observable is the arrival or presence of a client.
    \item \textbf{State Change Detection:} Sensors that monitor for changes in system states or conditions, signaling transitions that are indicative of significant events or actions.
    \item \textbf{Event Triggering:} Conditions or algorithms designed to detect specific patterns or sequences of actions that signify an event of interest has occurred or is imminent.
\end{itemize}

\paragraph{Importance in Observer Nodes:}
The inclusion of abstract sensors expands the observational scope of the ROBUST Network, allowing for the detection of nuanced or complex phenomena that traditional sensors may not directly capture. This capability is particularly important in contexts where the observable is conceptual or based on the aggregation of multiple data points or signals.

\paragraph{Conclusion}
The observer view and myopia are important factors to consider when designing and deploying observer nodes in the ROBUST Network. By understanding the capabilities and limitations of different types of sensors, and by taking into account the specific environmental conditions in which the nodes will be deployed, it is possible to optimize the network's observational capacity and achieve the desired objectives.

\subsection{Observer Movement}
The mobility of observer nodes within the ROBUST Network significantly impacts their observational capabilities and the network's overall effectiveness. Observers can be categorized based on their movement patterns, which dictate how they navigate the environment and adapt to dynamic conditions. 
\paragraph{Static Observers:} These nodes remain fixed in predetermined locations, providing consistent coverage in areas of strategic importance. Their stationary nature simplifies deployment but requires careful planning to ensure comprehensive coverage.

\paragraph{Dynamic Observers:} Dynamic observers are capable of movement, enhancing the network's adaptability and coverage. They can be further classified based on their movement patterns:

\begin{itemize}
    \item \textbf{Discrete Dynamic Observers:} These nodes move in predetermined intervals or in response to specific triggers, allowing for flexible coverage that adapts to changing conditions or requirements.
    \item \textbf{Continuous Dynamic Observers:} With the ability to move continuously, these observers can track dynamic phenomena or adjust their positioning in real-time, offering the highest level of adaptability and coverage optimization.
\end{itemize}

\paragraph{Environmental and Terrain Considerations:}
\begin{itemize}
    \item \textbf{Obstacle Avoidance:} Essential for navigating complex physical environments, from urban landscapes to natural terrains, ensuring uninterrupted data collection.
    \item \textbf{Terrain Adaptation:} Observers must be equipped to handle diverse environments, such as aquatic settings for Autonomous Underwater Vehicles (AUVs) or aerial conditions for drones.
    \item \textbf{Energy Consumption and Power Requirements:} For dynamic observers, it is crucial to consider their energy consumption and power requirements, especially in situations where they need to move frequently. 
\end{itemize}

\paragraph{Abstract Movement in Virtual Environments:}
In digital or virtual settings, the concept of movement extends to navigating through data structures, networks, or virtual spaces. Here, movement can be interpreted as the process of exploring digital domains, such as traversing directories, databases, or virtual worlds, where spatial measures are represented by complexity levels, data access paths, or interaction within software environments. For example, a web crawler can be used to navigate and collect data from a website.

\paragraph{Implications for Network Design:}
The choice between static and dynamic observers influences the network's design and operational strategies. While static observers provide reliable coverage in key areas, dynamic observers offer flexibility to respond to unexpected changes or events. The integration of both types within the network can provide a balanced approach, combining stability with adaptability.

\subsection{Multi-Observer Coordination}
Effective coordination among observer nodes is pivotal in the ROBUST Network, ensuring that collective observation efforts are harmonized for maximal coverage and efficiency. This subsection delves into the mechanisms of multi-observer coordination, shedding light on how these nodes collaborate and, at times, compete within the same operational space to fulfill the network's objectives.

\paragraph{Collaborative Dynamics:}
Observer nodes, by design, share an action space where their operations and data collection efforts influence one another. Through strategic collaboration, these nodes can enhance the network's capacity to monitor and analyze the environment comprehensively. Coordination mechanisms, such as consensus algorithms and coordinated scheduling, play a crucial role in synchronizing observer activities, ensuring that data collection is optimized across spatial and temporal dimensions without excessive overlap or redundancy.

\paragraph{Adversarial and Competitive Interactions:}
Not all interactions within the network are cooperative. In certain scenarios, as explored in Chapter \ref{chap:robust-dynamics}: \textit{ROBUST Dynamics}, observer nodes may find themselves in competitive or even adversarial relationships. These situations arise when nodes compete for resources or observation priorities, necessitating strategies that balance individual node objectives with overarching network goals. This dynamic introduces a layer of complexity in coordination, where the network must navigate between collaboration and competition to achieve optimal outcomes.

\paragraph{Strategies for Effective Coordination:}
Achieving effective multi-observer coordination requires a multifaceted approach, incorporating both algorithmic solutions and operational strategies:
\begin{itemize}
    \item \textbf{Algorithmic Coordination:} Utilization of advanced algorithms to dynamically allocate tasks, manage resources, and schedule observations among nodes, ensuring that each node's contributions are aligned with network-wide goals.
    \item \textbf{Operational Strategies:} Development of protocols for communication, data sharing, and conflict resolution among nodes, facilitating a cooperative environment even in the presence of potential competitive dynamics.
\end{itemize}

\paragraph{Impact on Network Performance:}
The manner in which observer nodes coordinate their actions directly influences the ROBUST Network's efficiency and efficacy. Properly orchestrated, multi-observer coordination enhances the network's ability to adapt to environmental changes, detect and analyze events of interest, and utilize resources efficiently, thereby maximizing the collective observational power of the network.

\paragraph{Conclusion:}
Multi-observer coordination is a cornerstone of the ROBUST Network's operational paradigm, ensuring that the collective capabilities of observer nodes are leveraged to their fullest potential. By fostering an environment where nodes can effectively collaborate and navigate competitive dynamics, the network can achieve its objectives more effectively.

\subsection{Strategic Deployment}
Optimizing the ROBUST Network's performance hinges on the strategic deployment of observer nodes. This critical process involves meticulously planning the placement of nodes to achieve comprehensive environmental coverage and efficient data collection, tailored to the network's dynamic needs.

\paragraph{Adaptive Coverage Strategy:}
Strategic deployment is characterized by an adaptive coverage strategy. This approach involves identifying and prioritizing regions of interest while considering the potential for areas devoid of significant activity to be strategically overlooked. Such selectivity enhances network efficiency, allowing for the reallocation of resources to areas with higher observational value. This adaptive strategy ensures that the deployment of observer nodes is both precise and flexible, meeting the network's objectives within the confines of available resources and operational capabilities.

\paragraph{Placement Algorithms and Models:}
\begin{itemize}
    \item \textbf{Optimization Algorithms:} The deployment strategy incorporates sophisticated algorithms designed to analyze environmental data and predict areas of high interest. These algorithms assist in determining the most effective locations for observer node placement, ensuring optimal coverage and data collection efficiency.
    \item \textbf{Simulation Models:} Before actual deployment, simulation models are utilized to evaluate the potential impact of various placement strategies. These models help in understanding how different configurations might affect network performance, allowing for adjustments that enhance coverage and efficiency.
    \begin{itemize}
        \item \textbf{Monte Carlo Simulations:} Utilizing random sampling and probabilistic analysis, Monte Carlo simulations offer insights into the behavior of the network under a wide range of scenarios. This method is particularly valuable for identifying robust deployment strategies that maintain coverage and efficiency amidst uncertainty and environmental variability.
    \end{itemize}

\end{itemize}

\paragraph{Dynamic Reconfiguration:}
\begin{itemize}
    \item \textbf{Responsive Adjustment:} Observer nodes are equipped with the capability for dynamic reconfiguration, allowing them to adapt to environmental changes or new observational requirements. 
    \item \textbf{Autonomous and Directed Movements:} Whether through autonomous decision-making or directives from a central control, observer nodes can adjust their positions, operational parameters, or focus areas. This movement strategy is integral to the network's ability to dynamically realign its resources and maintain effective surveillance across varying conditions.
\end{itemize}

\paragraph{Conclusion:}
The strategic deployment and dynamic reconfiguration of observer nodes are cornerstone practices within the ROBUST Network, ensuring that its observational capabilities are maximized. Through careful planning, adaptive strategies, and the use of advanced algorithms and models, the network achieves a delicate balance between comprehensive coverage and resource efficiency. These practices underscore the network's adaptability and precision in capturing and analyzing environmental data, foundational to its success.

\section{Observable Nodes}
\label{sec:observable_nodes}

Observable nodes within the ROBUST Network represent the entities, events, or conditions that are subject to observation by observer nodes. These nodes encapsulate a wide array of phenomena, ranging from environmental variables and physical objects to digital events and states. This section delves into the nature of observable nodes, their classification based on observability, their interaction dynamics with observer nodes, and their broader implications for network design and operational efficiency.  

\subsection{Observable Role in ROBUST}
Observable nodes are critical elements within the ROBUST Network, serving as the focal points of observation and analysis by observer nodes. Unlike observer nodes, which actively monitor and interact with the environment, observable nodes represent the dynamic and diverse set of phenomena that populate the network's environment, from physical entities and environmental conditions to digital signals and states. Their role is passive yet pivotal; they constitute the subjects of the network's surveillance efforts, providing the raw data and insights upon which decisions and actions are based.

Observable nodes, by their very nature, introduce variability and complexity into the network's operational landscape. They do not possess autonomy in the same way observer nodes do; rather, their existence, behavior, and changes are what the network seeks to detect, understand, and respond to. These nodes embody the network's external interface, the bridge between the digital and physical worlds, or the interplay of various data layers within a purely digital realm.

Their presence and behavior are key determinants of the network's observational strategies and priorities. The detection, classification, and analysis of observable nodes drive the adaptive responses of the network, shaping the deployment of observer nodes and the allocation of resources. Observable nodes, therefore, play a central role in defining the network's objectives, influencing its design and evolution over time. They are the catalysts for the network's actions, prompting adjustments in observational focus and strategies to ensure that the network remains aligned with its overarching goals of comprehensiveness, efficiency, and adaptability.

In essence, observable nodes are at the heart of the ROBUST Network's purpose and functionality. They are the raison d'être for the network's existence, challenging it to continually refine and adapt its observational capabilities to meet the demands of an ever-changing environment. Through the lens of observable nodes, the network navigates the complexity of its operational domain, striving to achieve a detailed understanding and effective management of the phenomena it seeks to monitor.

\subsection{Behavior Variability of Observable Nodes}
Observable nodes within the ROBUST Network exhibit a wide spectrum of behaviors, reflecting the diverse nature of the phenomena these nodes represent. This variability is crucial for understanding the dynamic interaction between observable nodes and observer nodes, as it directly impacts the network's observational strategies and data collection methodologies.

\paragraph{Spectrum of Responses:}
Observable nodes' responses to observation efforts can be categorized into several distinct behaviors:
\begin{itemize}
    \item \textbf{Disregard for Observers:} Some observable nodes may exhibit behaviors that are indifferent to the presence of observer nodes. This category typically includes passive environmental phenomena or digital states that remain unchanged regardless of observation.
    \item \textbf{Seeking Observers:} In contrast, certain observable nodes might actively seek to be observed. This behavior is often seen in systems where feedback from observation can trigger desired actions or responses, such as in automated monitoring systems that rely on observer input to initiate corrective measures.
    \item \textbf{Evading Observation:} Some nodes may exhibit behaviors aimed at evading detection or observation. This is particularly relevant in contexts involving privacy concerns, security measures, or competitive dynamics where visibility to observer nodes might lead to undesired outcomes.
\end{itemize}

\subsection{Movement and Dynamics of Observable Nodes}
The dynamics and movement patterns of observable nodes within the ROBUST Network are essential factors influencing their detectability and the overall effectiveness of the observation strategy. These patterns can range from entirely random to highly predictable, based on inherent or environmental factors.

\paragraph{Classification of Dynamics:}
Observable nodes exhibit a variety of dynamic behaviors that can be broadly classified into three categories:
\begin{itemize}
    \item \textbf{Random Dynamics:} Some observable nodes may move or change in unpredictable ways, making it challenging for observer nodes to track them consistently. This randomness requires flexible and adaptive observation strategies to ensure effective coverage.
    \item \textbf{Stochastic (Probabilistic) Dynamics:} Other nodes display stochastic behaviors, where their movements or changes follow known probability distributions. This allows for more strategic planning of observation efforts, utilizing statistical models to predict future states or positions.
    \item \textbf{Evolutionary Dynamics:} Certain observable nodes may evolve over time, following predictable patterns that can be modeled and anticipated. Understanding these evolutionary trends enables observer nodes to adapt their observation tactics to stay aligned with the nodes' trajectories.
\end{itemize}

\subsection{Temporal Existence of Observable Nodes}
The temporal existence of observable nodes within the ROBUST Network significantly impacts the network's observation strategies and data analysis processes. This existence varies widely, ranging from ephemeral phenomena that appear and disappear within a single observational frame to persistent entities that remain detectable across multiple frames or even indefinitely. Understanding these temporal patterns is crucial for tailoring the network's approach to monitoring and analyzing these nodes effectively.

\paragraph{Transient Observable Nodes:}
Transient observable nodes are those that appear briefly and then vanish, often within a single observational frame. Their fleeting nature poses unique challenges for observation, requiring rapid detection and analysis methods to capture and interpret the data they provide before they disappear. Such nodes often represent short-lived events or conditions that can offer valuable insights into the dynamic nature of the network's environment.

\paragraph{Persistent Observable Nodes:}
In contrast, persistent observable nodes maintain their presence over extended periods, ranging from several frames to a constant state throughout the network's operational timeline. Their enduring existence allows for more comprehensive observation and analysis, facilitating long-term studies and trend analysis. Persistent nodes typically include stable environmental conditions, ongoing physical processes, or continuous digital activities.

\paragraph{Influence on Observation Planning and Analysis:}
The temporal existence of observable nodes directly influences the ROBUST Network's observation planning and data analysis strategies. For transient nodes, the network must deploy rapid-response observation techniques capable of quickly capturing and processing the fleeting data. This might involve high-frequency monitoring or the use of real-time analytics to interpret the data as it is collected.

For persistent nodes, the strategy shifts towards sustained observation and in-depth analysis, leveraging the accumulated data over time to uncover trends, patterns, and long-term changes. This approach may utilize slower, more methodical observation methods and complex analytical models to extract meaningful insights from the extended dataset.

\paragraph{Adapting to Temporal Variability:}
Adapting to the temporal variability of observable nodes requires the ROBUST Network to be versatile and responsive in its observation capabilities. By recognizing and accommodating the different temporal existences of these nodes, the network can optimize its observation and analysis efforts, ensuring that both transient and persistent phenomena are effectively monitored and understood. This adaptability enhances the network's overall effectiveness in capturing the full spectrum of observable phenomena within its domain.

\subsection{Mobility of Observable Nodes}
Observable nodes within the ROBUST Network exhibit a range of mobility characteristics, from stationary entities to those that demonstrate various degrees of movement. This mobility plays a significant role in determining the approach and techniques employed by the network to observe and analyze these nodes effectively.

\paragraph{Stationary Observable Nodes:}
Stationary observable nodes are those that remain fixed in location over time. Their immobility simplifies the process of observation, as their position relative to observer nodes does not change. Stationary nodes typically include physical structures, environmental landmarks, or fixed digital entities. The stable nature of these nodes allows for consistent observation and facilitates longitudinal studies to monitor changes or trends over time.

\paragraph{Mobile Observable Nodes:}
In contrast, mobile observable nodes are characterized by their ability to move or change position within the network's environment. This mobility can vary widely, from predictable patterns to erratic movements, and poses additional challenges for observation. Mobile nodes require dynamic observation strategies that can adapt to their changing positions, ensuring effective coverage and data collection. Examples of mobile observable nodes include moving physical objects, dynamic environmental phenomena, or shifting digital states.

\paragraph{Observation Techniques for Mobile Nodes:}
To effectively track and observe mobile observable nodes, the ROBUST Network employs a variety of techniques and technologies, including:
\begin{itemize}
    \item \textbf{Real-time Tracking:} Utilizing sensors and algorithms that can monitor the position and movement of nodes in real time, enabling observer nodes to adjust their focus and maintain coverage of mobile targets.
    \item \textbf{Predictive Modeling:} Implementing models that predict the future movements of observable nodes based on historical data and observed patterns, facilitating anticipatory adjustments in observation strategies.
    \item \textbf{Adaptive Observation Methods:} Employing flexible observation methods that can quickly respond to changes in the mobility of observable nodes, such as adjusting sensor configurations or deploying mobile observer nodes to follow the targets.
\end{itemize}

\paragraph{Impact on Network Design and Operation:}
The mobility of observable nodes significantly influences the design and operational strategies of the ROBUST Network. Accounting for both stationary and mobile nodes ensures that the network can effectively capture a comprehensive dataset, reflecting the full spectrum of dynamics within its observation domain. This consideration is crucial for the network's ability to provide accurate, timely, and relevant data for analysis and decision-making processes.

\subsection{Interest Levels of Observable Nodes}
The significance of observable nodes within the ROBUST Network is a critical factor in prioritizing observation efforts and allocating resources. This significance, or level of interest, varies widely among observable nodes, influencing how the network manages its observational capabilities to effectively monitor and analyze these entities.

\paragraph{Binary and Continuous Classification:}
To address the diverse range of observable nodes, the ROBUST Network employs a classification system that categorizes nodes based on their level of interest. This system is designed to facilitate decision-making processes regarding observation priorities and resource allocation:

\begin{itemize}
    \item \textbf{Binary Classification:} This approach simplifies the classification by categorizing observable nodes into two distinct groups: those of interest and those not of interest. It enables a clear distinction, directing observation efforts towards nodes deemed significant for the network's objectives.
    \item \textbf{Continuous Classification:} In contrast, a continuous classification system offers a more nuanced understanding of significance, rating observable nodes on a scale (e.g., from 0 to 1). This method allows for a gradation of interest levels, providing a more detailed framework for prioritizing observation and analysis.
\end{itemize}

\paragraph{Determining Interest Levels:}
Interest levels are determined based on several factors, including the potential impact of the observable node on network objectives, the novelty or rarity of the phenomena, and the relevance of the data it can provide. This assessment is crucial for optimizing the network's observation strategies, ensuring that resources are focused on nodes that offer the greatest value or insight.

\paragraph{Implications for Observation Strategy:}
The classification of observable nodes by interest levels directly influences the ROBUST Network's observation strategy. Nodes classified as highly significant may warrant more intensive observation efforts, including higher frequency monitoring, dedicated resources, and specialized analysis techniques. Conversely, nodes deemed of lesser interest may be monitored less frequently or with standard methods, allowing the network to allocate its resources more efficiently.

\paragraph{Adapting to Changes in Interest Levels:}
Interest levels of observable nodes may change over time, necessitating flexibility in the network's observation strategy. The ROBUST Network is designed to adapt to these changes, reevaluating the significance of nodes as new information becomes available or as the network's objectives evolve. This dynamic approach ensures that the network remains focused on the most relevant and impactful observable nodes, enhancing its overall efficiency and effectiveness.

This section highlights the importance of classifying observable nodes by their level of interest, providing a foundation for targeted observation and analysis within the ROBUST Network. By prioritizing nodes based on their significance, the network can more effectively manage its resources, adapt to changing conditions, and achieve its operational goals.

\subsection{Observational Myopia in Observable Nodes}
Observational myopia within the ROBUST Network refers to the potential perceptual limitations of observable nodes, contingent on their capacity to sense and respond to their environment. This concept is particularly relevant in scenarios where observable nodes possess some form of "sight" or sensing mechanism, allowing them to perceive and consequently react to the presence or actions of observer nodes or other environmental stimuli.

\paragraph{Conditional Sensing and Response:}
Not all observable nodes are endowed with sensing capabilities; however, for those that are, their interaction with the environment and observer nodes may be influenced by their sensory range and acuity. This sensory perception, akin to myopia, can dictate:
\begin{itemize}
    \item \textbf{Behavioral Adaptations:} Observable nodes with sensory capabilities might adapt their behavior based on their perception of the environment, including the presence of observer nodes. These adaptations could range from evasive maneuvers to mitigate detection to alterations in their operational state in response to perceived stimuli.
    \item \textbf{Perceptual Limitations:} The extent and clarity of these nodes' sensory perception can significantly impact their detectability and the strategies observer nodes must employ to monitor them effectively. Knowing the myopic range of observable nodes helps in strategizing observation to avoid detection or to ensure visibility without eliciting reactive behaviors.
\end{itemize}

\paragraph{Strategic Implications for Observer Nodes:}
Observerable nodes may also have observational properties that the observers nodes should consider in certain scenarios. Understanding the observational myopia of observable nodes is pivotal for the ROBUST Network's strategy in two key areas:
\begin{itemize}
    \item \textbf{Stealth and Detection Avoidance:} In cases where it is crucial for observer nodes to remain undetected to avoid influencing the behavior of observable nodes, knowing their myopic limitations allows observer nodes to maintain surveillance from beyond the ``observational threshold" of the observable nodes.
    \item \textbf{Predictive Behavior Modeling:} By anticipating the reactions of observable nodes to their environment or direct observation, the network can refine its predictive models to account for potential behavioral changes. This enhances the accuracy of data analysis and the efficacy of the network's adaptive responses.
\end{itemize}

This nuanced understanding of observational myopia in observable nodes enriches the ROBUST Network's operational dynamics, facilitating a more informed and strategic approach to observation and interaction within its domain.

\subsection{Observation Status of Observable Nodes}
The observation status of observable nodes within the ROBUST Network plays a crucial role in the network's data collection and analysis efforts. This status, which indicates whether nodes are currently under observation or remain unobserved, directly impacts the network's ability to gather comprehensive and accurate data.

\paragraph{Actively Observed Nodes:}
Observable nodes that are actively observed by the network's observer nodes contribute significantly to the network's data repository. These nodes are monitored through various observation techniques, allowing the network to capture real-time or near-real-time data about their states, behaviors, and interactions. The continuous or periodic observation of these nodes facilitates a deeper understanding of the network's operational environment, enabling more informed decision-making and strategy development.

\paragraph{Unobserved Nodes:}
Conversely, nodes that remain unobserved pose a challenge to the network's objective of achieving comprehensive surveillance. Unobserved nodes can result from various factors, including limitations in the network's observational capabilities, strategic decisions to prioritize certain nodes over others, or the transient nature of some observable nodes that eludes detection. The absence of data from these nodes creates gaps in the network's understanding, potentially impacting the accuracy and effectiveness of its analyses and responses.

\paragraph{Implications for Network Strategy:}
The observation status of observable nodes necessitates strategic planning and resource allocation within the ROBUST Network to optimize observation coverage and data collection:
\begin{itemize}
    \item \textbf{Enhancing Observational Capabilities:} To minimize the number of unobserved nodes, the network may invest in expanding its observational infrastructure, deploying additional observer nodes, or integrating advanced sensing technologies to extend its reach and sensitivity.
    \item \textbf{Dynamic Prioritization:} The network can employ dynamic prioritization algorithms to adjust its observational focus in real-time, based on changing conditions, emerging priorities, or newly detected nodes, ensuring that critical data is not overlooked.
    \item \textbf{Data Analysis and Modeling:} Leveraging advanced data analysis and predictive modeling techniques can help compensate for the absence of direct observations, inferring the states and behaviors of unobserved nodes from the available data and known patterns.
\end{itemize}

\paragraph{Enhancing Observational Coverage:}
Addressing the challenge of unobserved nodes and maximizing the comprehensiveness of data collection are essential for the ROBUST Network's effectiveness. By continuously assessing and adjusting its observation status across the spectrum of observable nodes, the network can enhance its operational intelligence, adaptability, and overall performance.

\subsection{Inter-node Communication Dynamics among Observable Nodes}
The dynamics of communication between observable nodes within the ROBUST Network are pivotal in shaping the network's understanding of collective behaviors and interaction outcomes. This communication, inherently dependent on the proximity and connectivity between nodes, dictates the flow of information, and enables nodes to potentially coordinate actions, share state information, or collectively respond to environmental stimuli.

\paragraph{Communication Range and Capabilities:}
Observable nodes' ability to communicate is fundamentally influenced by their range of communication—determined by their physical or digital proximity and the medium of interaction. This range defines the nodes' capacity to establish connections with one another, directly affecting the network's insight into their collective dynamics.

\begin{itemize}
    \item \textbf{Direct Communication:} In scenarios where nodes are within direct communication range of each other, they can exchange information seamlessly, leading to synchronized or coordinated behaviors that might influence the network's operational environment.
    \item \textbf{Indirect Communication:} Observable nodes might also engage in indirect communication, mediated through environmental modifications or by influencing intermediary nodes, extending their influence beyond immediate proximity.
\end{itemize}

\paragraph{Influence on Collective Behaviors:}
The inter-node communication capabilities of observable nodes play a critical role in the emergence of collective behaviors. These behaviors, ranging from group movements to distributed processing tasks, can significantly impact the network's data collection and analysis strategies by altering the nodes' observable characteristics.

\paragraph{Observational Myopia and Communication:}
Observational myopia, or the limited perceptual range of observable nodes, may also affect their communication dynamics. Nodes with restricted sensory capabilities may only communicate with or respond to neighboring nodes within their perceptual threshold, potentially leading to localized rather than global collective behaviors.

\paragraph{Strategic Implications for the Network:}
Understanding the communication dynamics among observable nodes is crucial for the ROBUST Network to accurately interpret collective behaviors and adapt its observation strategies accordingly. By analyzing these dynamics, the network can anticipate changes in the operational environment, optimize its observation efforts, and potentially influence node behaviors through strategic interventions.

This focus on inter-node communication dynamics underscores the importance of considering not only the individual characteristics of observable nodes but also the complex interactions that occur among them. Such an approach enhances the network's ability to navigate its operational domain effectively, ensuring a comprehensive and nuanced understanding of observable phenomena.

\subsection{Adaptability to Environmental Changes in Observable Nodes}
The adaptability of observable nodes to environmental changes within the ROBUST Network is a critical aspect of their role and functionality. This adaptability reflects the capacity of these nodes to modify their behavior or state in response to external stimuli or shifts in their operational environment. Such changes can significantly influence both the observability of these nodes and the overall adaptability of the network.

\paragraph{Natural Responses to Environmental Shifts:}
Observable nodes may exhibit natural responses to environmental changes, altering their behavior or physical states in ways that can either facilitate or hinder their detection by observer nodes. These responses might include:
\begin{itemize}
    \item \textbf{Behavioral Adjustments:} Changes in activity patterns, movement behaviors, or interaction with other nodes in response to environmental stimuli.
    \item \textbf{State Transitions:} Modifications in physical or digital states that reflect adaptation to new environmental conditions.
\end{itemize}

\paragraph{Sensory-Driven Adaptations:}
For observable nodes equipped with sensory capabilities, their adaptability can also be influenced by their sensory perceptions, which are subject to observational myopia. This myopia can limit the nodes' ability to fully perceive their surroundings, potentially affecting their adaptive responses:
\begin{itemize}
    \item \textbf{Limited Sensory Range:} Nodes with restricted sensory perception may react to a narrower set of environmental cues, leading to adaptations that are based on incomplete information about the environment.
    \item \textbf{Response to Perceived Stimuli:} Adaptive behaviors or state changes in response to perceived stimuli, influenced by the nodes' myopic view of their surroundings.
\end{itemize}

\paragraph{Implications for Network Observability and Adaptability:}
The adaptive responses of observable nodes to environmental changes pose both challenges and opportunities for the ROBUST Network:
\begin{itemize}
    \item \textbf{Enhanced Observability:} Adaptive behaviors may make certain nodes more detectable or provide new data points for observation, enhancing the network's ability to monitor and analyze the operational environment.
    \item \textbf{Adaptability Challenges:} Conversely, rapid or unpredictable adaptations can challenge the network's current observational strategies, requiring adjustments to maintain effective coverage and data accuracy.
\end{itemize}

\paragraph{Strategic Considerations for Observational Strategies:}
Understanding and anticipating the adaptability of observable nodes to environmental changes is essential for developing robust observational strategies. This includes employing adaptive observation technologies and methodologies capable of capturing the dynamic nature of observable nodes, as well as integrating predictive models to forecast potential adaptations and their impacts on observability.

This subsection underscores the importance of considering the adaptability of observable nodes within the context of the ROBUST Network's operational planning and strategy development. By accounting for the dynamic interplay between observable nodes and their environment, the network enhances its resilience and effectiveness in achieving comprehensive situational awareness.

\section{Ranged Edges and Spatial-Temporal Constraints in ROBUST Network}

\subsection{Introduction to Ranged Edges}
Ranged edges within the ROBUST Network signify the sensor view connections between observer and observable nodes, influenced significantly by the observer nodes' myopia. These edges are not only spatial but also temporal, bridging the gap between observer and observable nodes across different time frames. Unlike traditional network edges designed for pathfinding or direct communication, ranged edges facilitate sensor view clustering or grouping, playing a pivotal role in the network's ability to dynamically adapt to both spatial constraints and temporal changes.

\subsection{Proximal Recurrence: A Spatial-Temporal Clustering Technique}
The ROBUST Network implements Proximal Recurrence, an innovative clustering technique that dynamically adjusts to the spatial constraints and myopia of observer nodes while also considering the temporal dynamics of the network. This technique ensures the optimal placement of observer nodes relative to observable nodes across both space and time, thereby enhancing the network's efficacy in monitoring and data collection. Proximal Recurrence leverages both p-hops and distances to define the spatial division and movement, and incorporates temporal adjustments to accommodate the unique needs of the network's operational environment.

\subsection{Influence of Observer Myopia}
The effective range of these edges is predominantly determined by the myopia inherent to the observer nodes. Myopia, in this context, refers to the limited sensing range of observer nodes, affecting their ability to detect and interact with observable nodes within the environment. The ROBUST Network's architecture and algorithms take into consideration the observer myopia to dynamically adjust the sensor view clustering, ensuring comprehensive coverage and responsiveness despite the spatial, perceptual, and temporal limitations.

\subsection{Strategic Importance of Ranged Edges}
Ranged edges and the Proximal Recurrence technique are fundamental to the ROBUST Network's strategy for overcoming spatial constraints and optimizing observer node deployment. By accounting for observer myopia, leveraging dynamic clustering, and adapting to temporal changes, the network achieves a balanced and effective distribution of observational capabilities. This approach not only maximizes the efficiency of the network's resource allocation but also enhances its adaptability and performance in varied and challenging operational contexts, ensuring that critical areas within the environment are adequately monitored across time.

\section{Bipartite Analysis in ROBUST Network}

\subsection{Introduction}
Introduced in Chapter \ref{chap:theoretical_foundations}, Bipartite Analysis leverages the theoretical foundation of bipartite networks to evaluate the performance and interaction between observer nodes (\(O\)) and observable nodes (\(E\)) within the ROBUST Network. This approach models the two distinct sets of nodes and their connections, focusing on assessing the coverage of observable events by observer nodes. 

\subsection{Objectives of Bipartite Analysis}
The primary goals of Bipartite Analysis include:
\begin{enumerate}
    \item Optimizing the current network's observability by improving the placement and effectiveness of observer nodes.
    \item Assessing the network's configuration to distinguish between observed and unobserved events, thereby identifying coverage gaps.
    \item Informing strategic decisions to enhance network performance and coverage through data-driven insights.
\end{enumerate}

\subsection{Methodology}
The methodology encompasses constructing a bipartite graph, wherein edges denote potential observational links between observer and observable nodes. This facilitates:
\begin{itemize}
    \item \textbf{Distance Matrix Computation:} Assessing spatial and temporal distances to understand potential interactions.
    \item \textbf{Observability Assessment:} Evaluating observer node centrality and classifying events by observability.
    \item \textbf{Strategic Network Optimization:} Applying analysis results to guide observer node deployment and network adjustments.
\end{itemize}

\subsection{Rationale}
The Bipartite Analysis is named for its foundational structure, which distinctly separates observer nodes and observable events into two interconnected sets. This separation allows for a focused analysis of the observational relationships and the efficiency of the network's design in capturing observable phenomena.

\section{Unipartite Analysis in ROBUST Network}

\subsection{Introduction}
Building on the concepts introduced in Chapter \ref{chap:theoretical_foundations} and expanding upon the Bipartite Analysis, Unipartite Analysis focuses on the network's expansion by targeting unobserved events (\(E_{unobserved}\)). This analysis specifically addresses clusters of these events, aiming to strategically place new observer nodes to mitigate observable coverage gaps previously identified.

\subsection{Objectives}
The Unipartite Analysis pursues several critical objectives to enhance the ROBUST Network's efficiency and coverage:
\begin{enumerate}
    \item To identify clusters of unobserved events within the network, as delineated in Chapter \ref{chap:theoretical_foundations}'s discussion on network structures.
    \item To develop strategic placement plans for new observer nodes, enhancing coverage and observability of the network.
    \item To align network expansions with the overarching goal of maximizing the network's observability and performance.
\end{enumerate}

\subsection{Methodology}
Unipartite Analysis employs spatial clustering techniques to group unobserved events based on their proximity and the likelihood of being collectively observed by new observer placements. This process involves:
\begin{itemize}
    \item \textbf{Spatial Clustering:} Identifying dense clusters of unobserved events using spatial analysis techniques.
    \item \textbf{Observer Placement Strategy:} Developing strategies for the placement of new observer nodes based on cluster locations to maximize their observational impact on unobserved events.
    \item \textbf{Performance Forecasting:} Estimating the potential increase in network observability and efficiency from the proposed expansions.
\end{itemize}

\subsection{Rationale}
The Unipartite Analysis directly addresses the need for network expansion by focusing solely on the set of unobserved events, unlike the Bipartite Analysis, which considers the relationships between observer nodes and all events. This targeted approach ensures that network expansions are strategically aligned with enhancing overall network performance.

\section{Novelty of ROBUST Network}

\subsection{Innovative Approach to Network Analysis}
The ROBUST Network embodies a groundbreaking approach to network analysis, uniquely combining bipartite and unipartite analyses. This dual methodology provides a holistic assessment of the network's performance, focusing on optimizing the deployment and effectiveness of observer nodes vis-à-vis observable events. The innovation lies in dynamically transitioning between bipartite and unipartite groupings, enabling detailed insights into both the current state of the observer network and potential for spatiotemporal enhancements.

\subsection{Alignment with Research Hypotheses}
This novel approach directly addresses the research hypotheses established in Chapter \ref{chap:intro}:
\begin{enumerate}
    \item \textbf{H1:} The dual analysis framework enhances precision and actionable insights beyond conventional models, fulfilling the hypothesis of ROBUST networks' superior performance in spatiotemporal data analysis.
    
    \item \textbf{H2:} By optimizing resource allocation and strategic responsiveness through targeted observer node placements, the ROBUST network demonstrates the efficiency and adaptability hypothesized to surpass alternative models.
    
    \item \textbf{H3:} The integration of spatial-temporal graph analysis, intrinsic to the ROBUST framework, supports the hypothesis of improved optimization and network efficiency, enabling effective observer node placement and network expansion.
\end{enumerate}

\subsection{Strategic Implications}
The incorporation of bipartite and unipartite analyses into ROBUST's analytical framework offers:
\begin{itemize}
    \item A comprehensive evaluation of the observer network's current state, examining both observed and unobserved events to determine effectiveness and identify coverage gaps.
    
    \item Data-driven identification of strategic expansion locations, informed by unipartite analysis of unobserved event clusters.
    
    \item The formulation of strategic observer node placements, maximizing performance and enhancing network observability based on bipartite analysis insights.
\end{itemize}

\subsection{Conclusion}
The novel application of both bipartite and unipartite analyses within the ROBUST Network signifies a significant advancement in network analysis methodologies. By leveraging these dual perspectives, ROBUST effectively evaluates and adapts to the dynamic requirements of its operational environment, ensuring targeted and efficient network expansions. This approach not only maximizes observational capabilities but also aligns with the foundational research hypotheses, demonstrating the network's innovative potential in spatiotemporal data analysis.

\chapter{ROBUST Measures and Analysis}
In this chapter, we introduce novel graph-based measures for ROBUST network analysis, focusing on spatial metrics and structural insights that highlight the unique interplay between nodes. These measures, emerging from the inherent ROBUST nature of network connections, offer a fresh perspective on network resilience, connectivity, and spatial properties. Our approach enriches the understanding of network dynamics, paving the way for optimized strategies in network design and management.

\section{Spatial Metrics for Node Analysis}

Adapting traditional network metrics to incorporate spatial elements is essential in analyzing spatial networks. The following are key aspects of this adaptation, specifically focused on individual nodes:

\subsection{ \textit{Myopic Degree:} } 
Measures a node's connectivity within a specific spatial range, focusing on the local neighborhood density of a node. This metric quantifies the number of connections a node has within a defined range, emphasizing the immediate spatial neighborhood. This metric extends that of the traditional concept of degree centrality \cite{Newman2018, Wasserman1994}, which quantifies the number of connections a node has to measure its influence. Unlike traditional degree centrality, which considers all connections regardless of distance, Myopic Degree focuses on connections within a defined spatial range. This is particularly important in spatial networks where the proximity of nodes significantly impacts their interactions and influence.
\vspace{5mm}

The mathematical representation is:
   \begin{equation}
   \text{Spatial Degree}(v_i) = |\{v_j \in V : d(P(v_i), P(v_j)) \leq \theta \}|
   \end{equation}

where \( \theta \) is a threshold distance for considering an edge to exist. 
   
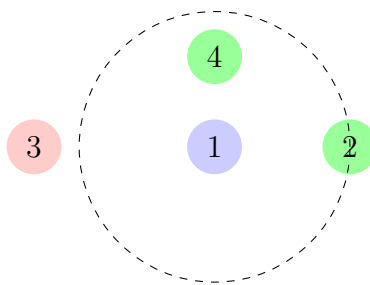
\begin{figure}[H]
    \centering
    \begin{tikzpicture}
      [scale=1.2,auto=left]
    \node[shape=circle,fill=blue!20] (c1) at (2,2) {1};
      \node[shape=circle,fill=green!40] (n1) at (3.5,2) {2};
      \node[shape=circle,fill=red!20] (n2) at (0,2) {3};
      \node[shape=circle,fill=green!40] (n3) at (2,3) {4};

      \draw[dashed] (2,2) circle (1.5cm);
    \end{tikzpicture}
    \caption{Spatial centrality of a node in a network.} 
    \label{fig:spatial-centrality}
\end{figure}

In figure \ref{fig:spatial-centrality} Node 1 has a spatial centrality of 2, being close to Nodes 2 and 4 within a certain range.

\subsection{ \textit{Spatial Closeness Centrality:} } 
Assesses a node's centrality based on its spatial distance to all other nodes in the network, similar to the concept of closeness centrality in traditional network analysis \cite{Freeman1977, Sabidussi1966}. It calculates how central a node is in the network, considering the spatial distances. The formula is:
   \begin{equation}
   C_{\text{closeness}}(v_i) = \left(\sum_{v_j \in V, v_j \neq v_i} d(P(v_i), P(v_j))\right)^{-1}
   \end{equation}

This measure provides a global view of the node's position and influence within the network's spatial structure, taking into account its average distance to all other nodes.  

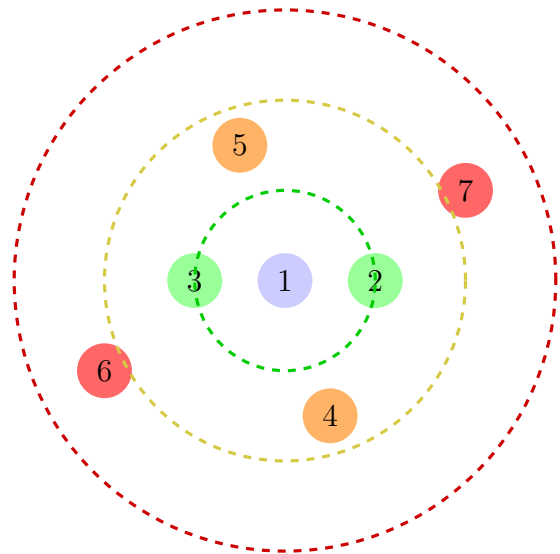
\begin{figure}[H]
    \centering
    \begin{tikzpicture}
      [scale=1.2,auto=left]
      \node[shape=circle,fill=blue!20] (c1) at (2,2) {1};

      \node[shape=circle,fill=green!40] (n1) at (3,2) {2};
      \node[shape=circle,fill=green!40] (n2) at (1,2) {3};

      \node[shape=circle,fill=orange!60] (n3) at (2.5,0.5) {4};
      \node[shape=circle,fill=orange!60] (n4) at (1.5,3.5) {5};

      \node[shape=circle,fill=red!60] (n5) at (0,1) {6};
      \node[shape=circle,fill=red!60] (n6) at (4,3) {7};

      \draw[very thick,dashed, green!80!black] (2,2) circle (1cm); 
      \draw[very thick,dashed, yellow!80!black] (2,2) circle (2cm); 
      \draw[very thick,dashed, red!80!black] (2,2) circle (3cm); 

    \end{tikzpicture}
    \caption{Illustration of spatial closeness centrality.}
    \label{fig:spatial-closeness-centrality}
\end{figure}

In Figure \ref{fig:spatial-closeness-centrality} illustrates the spatial closeness centrality of Node 1 in a network. Nodes within varying orbits indicate their distance to the central node: green (close), gold (medium distance), and red (far).

\section{Edge-Based Spatial Metrics}

\subsection{Spatial Edge Density}
 Spatial edge density is a global-based measure that evaluates how close a network is to being maximally connected within its spatial constraints. It assesses the overall concentration of edges across the entire network, considering the spatial arrangement and constraints such as physical distance, geographical barriers, or range-based limitations. Unlike traditional edge density, which simply counts the ratio of existing edges to the maximum possible in a complete graph \cite{DiestelGraphTheory2017, Bollobas1998}, this metric adapts the denominator to reflect the maximum feasible edges under spatial constraints. 

The formula can be expressed as:

\begin{equation}
\text{Spatial Edge Density} = \frac{\text{Number of Actual Edges}}{\text{Maximum Feasible Edges under Spatial Constraints}}
\end{equation}

This measure differs from individual edge metrics, as it does not focus on the properties or importance of specific edges but rather evaluates the network's connectivity as a whole.

\begin{figure}[htbp]
    \centering
    \begin{minipage}[b]{0.45\textwidth}
        \centering
        \begin{tikzpicture}
            \draw[dashed,green!80!black, very thick] (0,0) circle (1.5cm);
            \draw[dashed,red!80!black, very thick] (1,1) circle (1.5cm);
            \draw[dashed,blue!80!black, very thick] (2,0) circle (1.5cm);

            \draw[line width=3] (0,0) -- (1,1);
            \draw[line width=3] (1,1) -- (2,0);

            \node[shape=circle,draw,fill=green!40] (A) at (0,0) {A};
            \node[shape=circle,draw, fill=red!20] (B) at (1,1) {B};
            \node[shape=circle,draw, fill=blue!20] (C) at (2,0) {C};
        \end{tikzpicture}
        \caption*{Maximally Connected Network}
    \end{minipage}
    \hfill
    \begin{minipage}[b]{0.45\textwidth}
        \centering
        \begin{tikzpicture}
            \draw[dashed,green!80!black, very thick] (0,0) circle (1.5cm);
            \draw[dashed,red!80!black, very thick] (1,1) circle (1.5cm);
            \draw[dashed,blue!80!black, very thick] (2,0) circle (1.5cm);

            \draw[line width=3] (0,0) -- (1,1);

            \node[shape=circle,draw,fill=green!40] (A) at (0,0) {A};
            \node[shape=circle,draw, fill=red!20] (B) at (1,1) {B};
            \node[shape=circle,draw, fill=blue!20] (C) at (2,0) {C};
        \end{tikzpicture}
        \caption*{Less Connected Network}
    \end{minipage}
    \caption{Comparison of Network Connectivity}
    \label{fig:spatial-edge-density}
\end{figure}

Figure \ref{fig:spatial-edge-density} contrasts two network configurations. On the left, the Maximally Connected Network includes edges A-B and B-C, showing nodes A, B, and C are within range of one another, indicated by dashed circles. Node A and C are not connected, highlighting their out-of-range position. The right side illustrates a Less Connected Network with the same nodes and range constraints but fewer edges, emphasizing the network's reduced connectivity. The edge density is computed as the actual to possible edges ratio, here (1 actual edge / 2 possible edges), resulting in a density of 0.5.

\subsection{Edge Length Proportion}
The Edge Length Proportion is a spatial metric that quantifies the proportion of an individual edge's length relative to the total length of all edges in the network. For a given edge \(e\), it is calculated as the edge's length divided by the sum of the lengths of all edges within the network. This can be mathematically represented as:

\begin{equation}
\text{Edge Length Proportion (of edge e)} = \frac{\text{Length of edge e}}{\sum_{\text{all edges } e' \in E} \text{Length of edge } e'}
\end{equation}

This ratio is particularly useful in spatial networks for evaluating the scale of an edge's contribution to the network's total length, offering insights into infrastructure planning, network robustness and resilience, as discussed in research on network vulnerability and disruption \cite{Ellens2013, Schneider2013, Gao2016, Sydney2008}

This definition makes it clear that the measure is an edge-specific metric that relates its length to the aggregate of all edge lengths in the network, thus providing a perspective on the edge's contribution to the overall structure and connectivity of the spatial network.

\begin{figure}[htbp]
    \centering
    \begin{tikzpicture}[auto,node distance=2cm,
        thick,main node/.style={circle,draw,font=\sffamily\Large\bfseries}]

        \node[shape=circle,draw,fill=blue!20] (A) at (0,0) {A};
        \node[shape=circle,draw, fill=blue!20] (B) at (1,1) {B};
        \node[shape=circle,draw, fill=blue!20] (C) at (2,0) {C};
        \node[shape=circle,draw, fill=blue!20] (D) at (6,1) {D};
        \draw[very thick] (A) -- (B) node[midway] {1};
        \draw[very thick] (A) -- (C) node[midway] {2};
        \draw[very thick] (B) -- (C) node[midway] {1};
        \draw[very thick, red] (C) -- (D) node[midway, black] {5};
      
    \end{tikzpicture}
    \caption{Illustration of Edge Length Proportion in a Spatial Network}
    \label{fig:edge-length-proportion}
\end{figure}
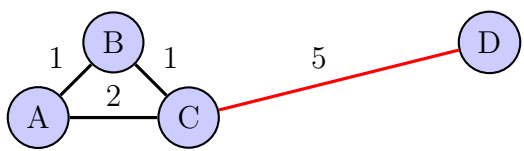

Figure \ref{fig:edge-length-proportion} demonstrates the concept of Edge Length Proportion in a spatial network. The network consists of four nodes labeled A, B, C, and D. The edges between these nodes are marked with their respective lengths. Specifically, edge A-B has a length of 1, A-C is 2, B-C is 1, and C-D is the longest edge with a length of 5. To compute the Edge Length Proportion for the edge C-D, we calculate it as the length of edge C-D divided by the total length of all edges in the network. The total length of all edges is \(1 + 2 + 1 + 5 = 9\), and thus the Edge Length Proportion for edge C-D is \( \frac{5}{9} \approx 0.56 \), indicating that this edge constitutes approximately 56\% of the total length of all edges in this network.

\section{Spatial Properties and Network Structures}
This subsection examines the impact of spatial characteristics on the structure of networks. It focuses on how the physical positioning of nodes and the distances between them shape the network's topology, leading to unique emergent patterns and dynamics.

\subsection{Spatial Clustering Coefficient}
An adaptation of the clustering coefficient, a fundamental concept in network theory \cite{Watts1998, Newman2003}, for spatial networks would consider not only the number of closed triplets but also the compactness of these triplets in space. This Spatial Clustering Coefficient provides insight into the tendency of nodes to form tightly-knit communities that are also geographically proximal.

\begin{equation}
\text{Spatial Clustering Coefficient}(v) = \frac{\text{Number of closed triplets involving } v}{\text{Number of all possible spatially close triplets involving } v}
\end{equation}

\begin{figure}[htbp]
    \centering
    \includegraphics[width=\textwidth]{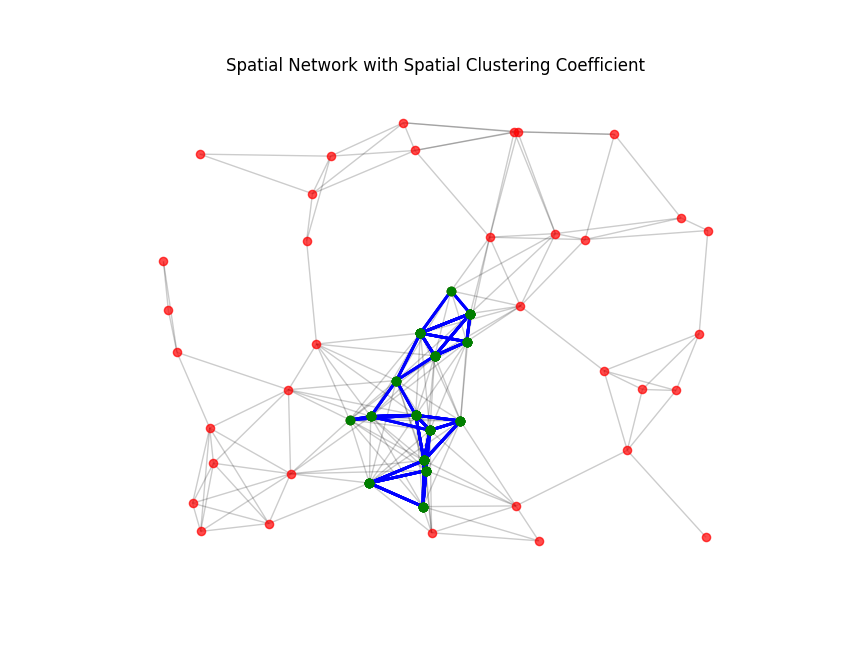}
    \caption{Illustration of Pearson Correlation in Spatial Networks}
    \label{fig:spatial-cluster-coef}
\end{figure}

Figure \ref{fig:spatial-cluster-coef} showcases a spatial network graph emphasizing the spatial clustering coefficient concept. Nodes forming triangles within a given radius are colored green, and the edges connecting these nodes are thickened and colored blue, signifying their contribution to the spatial clustering coefficient. Nodes that form triangular connections but lie outside the specified proximity threshold are colored red with standard edges, illustrating that they do not contribute to the coefficient. This figure visually underscores the significance of geographical proximity in the formation of tightly-knit node communities in spatial networks.

\vspace{6mm}
\begin{algorithm}[H]
\setstretch{1}
\SetAlgoLined
\LinesNumbered
\KwIn{Number of nodes $num\_nodes$, connection radius $radius$, threshold distance $threshold\_distance$}
\KwOut{Spatial network graph $G$, positions $pos$, spatial triangles $spatial\_triangles$}
\Begin{
    $G \leftarrow$ random geometric graph with $num\_nodes$ and $radius$\;
    $pos \leftarrow$ get node positions from $G$\;
    $spatial\_triangles \leftarrow$ empty list\;
    \For{each node $node$ in $G$}{
        $neighbors \leftarrow$ list of neighbors of $node$\;
        \For{each pair $(u, v)$ in combinations of $neighbors$}{
            \If{$G$ has edge $(u, v)$}{
                \If{distance between $node$ and $u$ $<$ $threshold\_distance$ \textbf{and} distance between $node$ and $v$ $<$ $threshold\_distance$ \textbf{and} distance between $u$ and $v$ $<$ $threshold\_distance$}{
                    append $(node, u, v)$ to $spatial\_triangles$\;
                }
            }
        }
    }
    \KwRet{$G$, $pos$, $spatial\_triangles$}\;
}
\caption{Creation of a Spatial Network with Spatial Clustering Coefficient}
\end{algorithm}
\vspace{6mm}

The algorithm delineates the construction of a spatial network and the identification of spatial triangles that contribute to the spatial clustering coefficient. It begins by generating a random geometric graph $G$ representing the spatial network, with nodes distributed within a given connection radius. Positions $pos$ of these nodes are then recorded. The algorithm proceeds to identify all triangles in the network where each edge is below the threshold distance, indicative of spatial proximity. These triangles are stored in a list $spatial\_triangles$. The output is the spatial network graph $G$, the positions of the nodes $pos$, and the list of spatial triangles $spatial\_triangles$, which are used to compute the spatial clustering coefficient.

\subsection{ Resilience in Spatial Networks}

Resilience in spatial networks goes beyond the traditional concept of maintaining connectivity, aligning with broader research on network robustness and resilience \cite{Ellens2013, Schneider2013, Gao2016}. It involves preserving the integrity of spatial segments—distinct areas that are effectively controlled or influenced by nodes. This approach is analogous to ensuring service coverage or operational influence in real-world spatial systems.

The spatial resilience of a network, denoted by \(R_s\), is quantified by the proportion of the network's spatial segments that remain intact after disruptions:

\begin{equation}
R_s = \frac{1}{|V|} \sum_{v_i \in V} \mathbb{1}_{\text{intact}}(S(v_i))
\end{equation}

In this equation, \(|V|\) is the total number of nodes, \(S(v_i)\) represents the spatial segment associated with node \(v_i\), and \(\mathbb{1}_{\text{intact}}\) is the indicator function that returns 1 if the segment is intact and 0 otherwise after a disruption. This measure captures the robustness of a network in terms of spatial coverage and influence, which is critical in applications like territorial management, emergency services deployment, and infrastructure maintenance.

\begin{figure}[htbp]
    \centering
    \includegraphics[width=\textwidth]{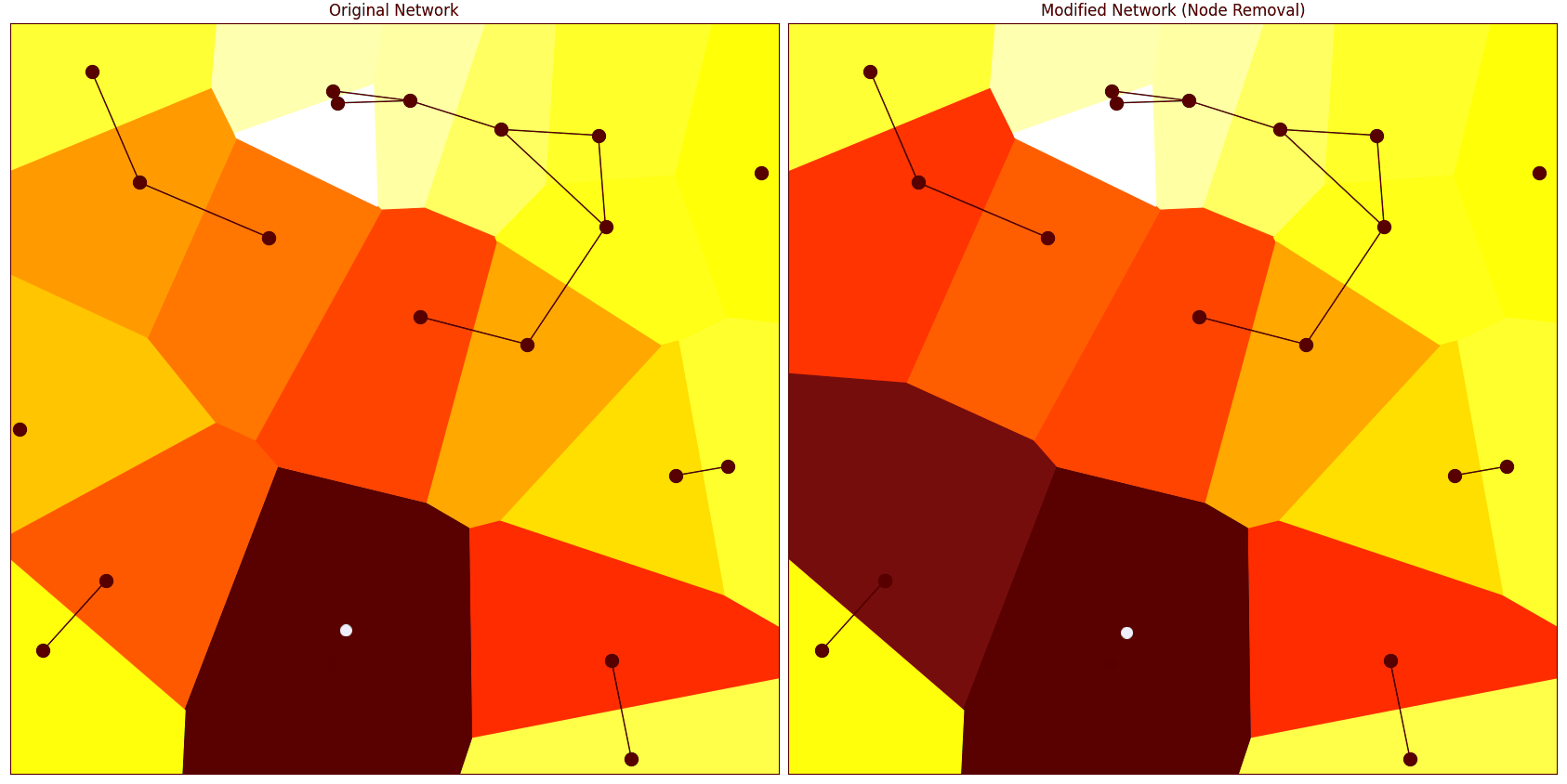}
    \caption{Spatial Resilience}
    \label{fig:spatial-resilience}
\end{figure}

Figure \ref{fig:spatial-resilience} visualizes the concept of spatial resilience in network structures, portraying how space is segmented and influenced by nodes before and after a disruption. The color-coded heatmap emphasizes the size of each segment: smaller segments are marked in yellow, denoting a preferable, tighter scope of influence, while larger segments are in red, indicating a broader, less ideal influence. Notably, the largest segments are depicted in dark red, signifying the most substantial loss of control. The left panel shows the original network with its segmented spatial domain, while the right panel reveals the network's altered spatial configuration following the removal of a node. The deepening of colors towards dark red in the right panel underscores an increased strain on the remaining nodes, illustrating the network's diminished spatial resilience.

\chapter{ROBUST and Static Observers}
\label{chap:robust_static}

This chapter details how to analyze and optimize static observer networks through bipartite and unipartite analysis within ROBUST networks. The first section, Bipartite Dynamics, evaluates the current observer nodes' performance in monitoring the network's critical events. It identifies well-covered areas, coverage gaps, and weaknesses in observer placement.

Following this, Unipartite Dynamics uses these insights to optimize observer locations. It addresses uncovered events and aims to fill in data gaps by strategically positioning new observers.

The primary objectives are to enhance network coverage with existing resources, utilize spatial and range analysis for better scalability and performance, and apply temporality to improve Proximal Recurrence (PR) clustering. These efforts aim to fortify the static observer network against evolving challenges.

\section*{I. Bipartite Dynamics: Observer Network Analysis}

This section presents a comprehensive analysis of observer network efficacy and coverage. A Bipartite pairing is employed to systematically examine the spatial relationships between stationary observer nodes and events over time. Temporal aggregation is particularly pertinent, as it allows for identifying spatial densities and patterns from accumulated data, taking advantage of the static nature of observer nodes. The approach incorporates a sequence of computational steps outlined in the algorithm below, which collectively address the challenges of scalable spatiotemporal data assessment.

\begin{figure}[h!]
  \centering
  \includegraphics[width=0.5\linewidth]{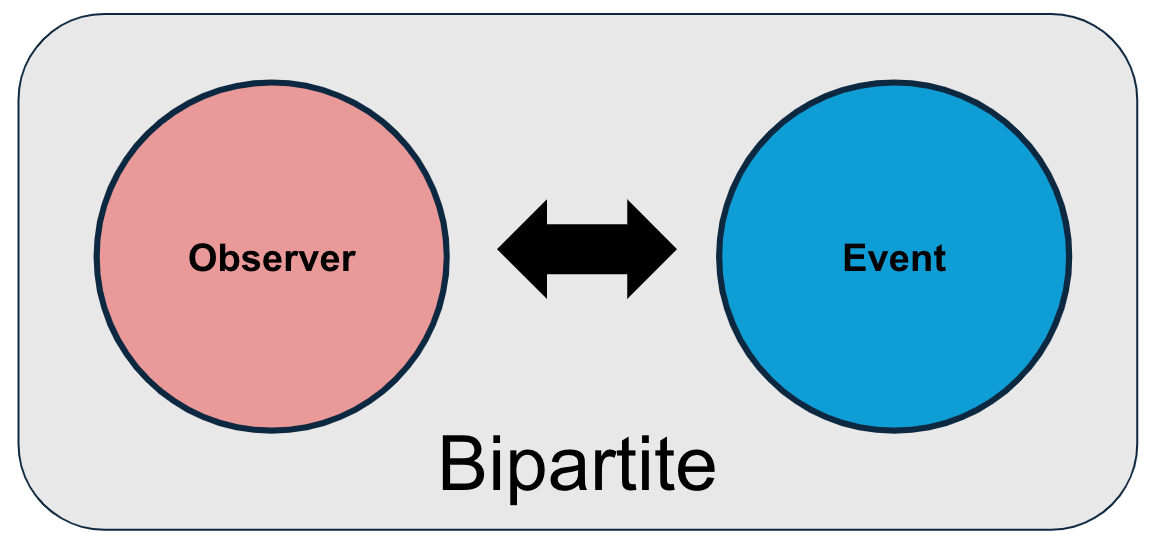}
  \caption{Bipartite Dynamics between Observers and Events}
  \label{fig:bipartite-dynamics}
\end{figure}

\subsubsection*{Algorithm Overview:}
The algorithm proceeds through the following steps to analyze observer networks:

\begin{figure}[h!]
  \centering
  \includegraphics[width=\linewidth]{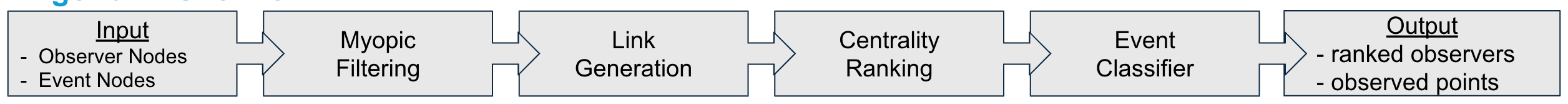}
  \caption{Flowchart of Analysis using Bipartite Dynamics}
  \label{fig:bipartite-flowchart}
\end{figure}

\begin{enumerate}
    \item \textbf{Myopic Filtering:} Confines analysis to the immediate range of each observer, minimizing complexity.
    \item \textbf{Link Generation:} Connects observers to events within a defined proximity, establishing a network graph.
    \item \textbf{Centrality Ranking:} Assesses observer influence within the network to locate pivotal nodes and highlight constraints.
    \item \textbf{Event Classification:} Separates events into observed and non-observed categories, based on their links to observer nodes.
\end{enumerate}

\section{Myopic Filtering}

This section introduces Myopic Filtering, an approach to enhance computational efficiency in analyzing static observer networks. Instead of evaluating the entire network, the algorithm employs a radius-based method focusing on the most relevant spatial segments around each observer node.

\subsection{Vectorized Approach}
Using broadcasting in array operations, the algorithm calculates distances between observer nodes and event points, but only within a specified coverage radius. This approach significantly lowers the computational load by avoiding the evaluation of events outside this radius.

\subsubsection{Spatial Indexing for Efficient Distance Computations}

The algorithm incorporates spatial indexing to streamline the distance computation process between observers and events:

\begin{itemize}
    \item \textbf{Spatial Data Structure}: A spatial index, like an R-tree \cite{Guttman1984} or a grid-based hash \cite{Samet2006}, organizes event locations for efficient access.
    
    \item \textbf{Event Indexing}: All events are indexed in this structure, facilitating quick retrieval based on their spatial coordinates.
    
    \item \textbf{Efficient Querying}: Observers query the spatial index to identify nearby events within their coverage radius, bypassing distant events to only compute distances for relevant, proximate events.
    
    \item \textbf{Selective Distance Calculation}: Distances are computed solely between each observer and the nearby events identified, reducing the volume of distance calculations.
\end{itemize}

\subsubsection{GPU-Accelerated Computation}
GPU acceleration, leveraging technologies like CUDA and parallel processing architectures \cite{Nickolls2008, Kirk2010}, enables the algorithm to process extensive datasets rapidly. By harnessing the parallel processing capabilities of GPUs, Myopic Filtering can handle a substantial number of observer nodes and event points with increased speed and efficiency.

\subsection{Visualization of Myopic Filtering}
The visualization illustrates the concept of Myopic Filtering by depicting observer nodes within their coverage areas and the distribution of event points.

\begin{figure}[h!]
  \centering
  \includegraphics[width=0.75\linewidth]{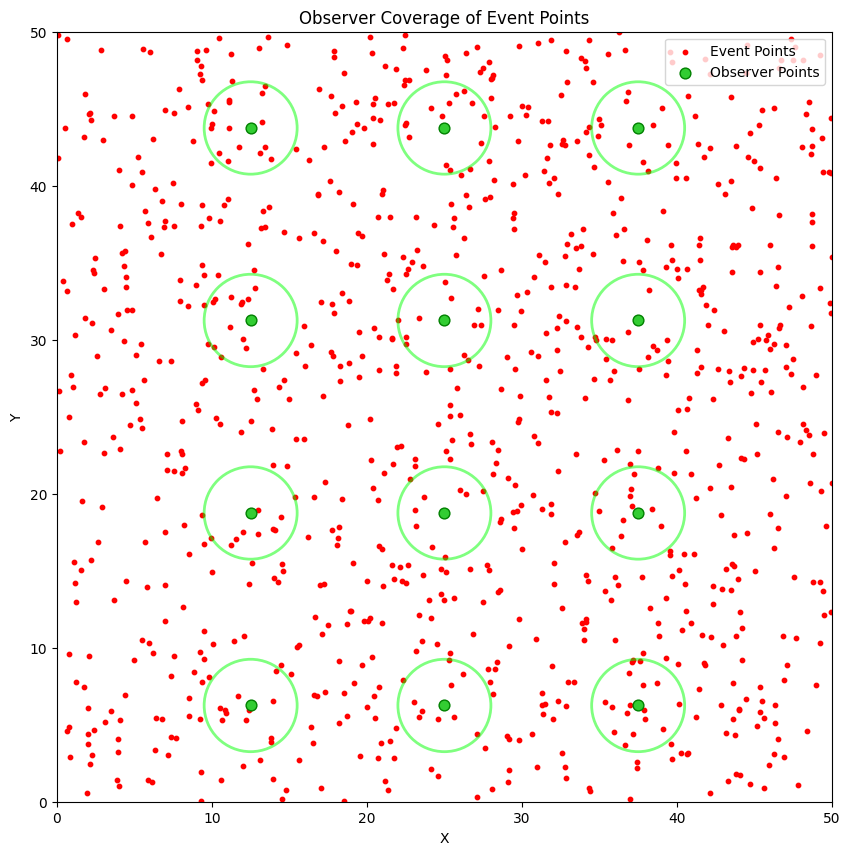}
  \caption{Myopic Filtering using Bipartite Dynamics}
  \label{fig:bipartite-myopic-filter}
\end{figure}

\section{Link Generation}

Following the myoptic filtering of event points within observer coverage areas through Myopic Filtering, the next step is Link Generation. This process involves creating connections, or links, between observer nodes and event nodes to enable detailed network analysis.

\subsection{Unweighted Links}
Initially, links are established on a binary basis: an unweighted link is formed if an event falls within an observer's coverage radius. This preliminary step lays the groundwork for introducing more sophisticated, weighted links.

\subsection{Weighted Links}
The transition from unweighted to weighted links is guided by the following criteria, aligning with principles of weighted network analysis \cite{Barrat2004, Opsahl2010}, which add depth and specificity to the network analysis:

\begin{itemize}
    \item \textbf{Spatial Proximity}: Weights are inversely proportional to the distance between observers and events, reflecting the concept of spatial decay for sensing capabilities \cite{Barthelemy2022}. Closer interactions are emphasized, implying that events within a shorter distance to an observer node are of greater value.

    \item \textbf{Temporal Relevance}: Time is integrated into the weighting, with more recent events potentially receiving higher priority, similar to temporal weighting schemes used in dynamic network analysis \cite{Holme2012}. This temporal weighting can be critical for dynamic environments where recent data may be more indicative of the current state of the network.

    \item \textbf{Imperfect Knowledge}: Acknowledging the inherent uncertainty in data, a probabilistic weight is applied to each event, incorporating aspects of uncertainty modeling into the network representation \cite{Berger1985}. This factor considers the likelihood of occurrence or detection accuracy, thus incorporating uncertainty into the network model.
\end{itemize}

\subsection{Visualization of Link Generation}
The visualization demonstrates the Link Generation process, showcasing how connections are established between observer nodes and event points within their coverage areas. Blue lines represent the links from observers to events, distinguishing between viewed events within the coverage radius and hidden events outside of it.

\begin{figure}[h!]
  \centering
  \includegraphics[width=0.75\linewidth]{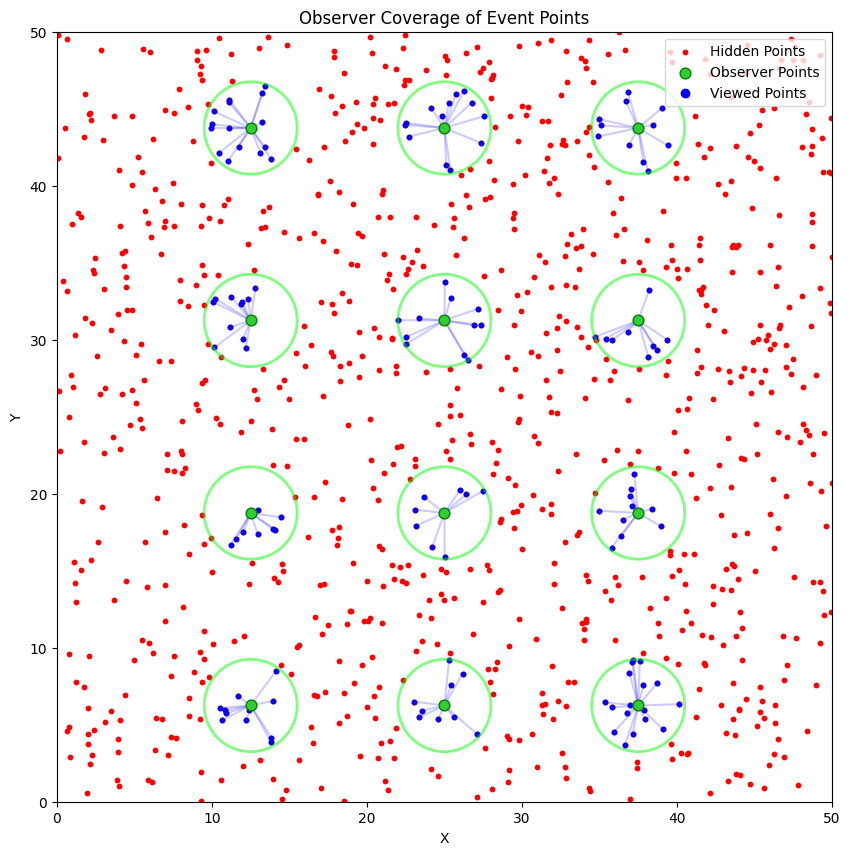}
  \caption{Link Generation using Bipartite Dynamics}
  \label{fig:bipartite-link-generation}
\end{figure}

\section{Centrality Ranking}

After establishing links between observer nodes and event nodes through Myopic Filtering and Link Generation, the subsequent analytic phase is Centrality Ranking.  This process quantifies the importance or `centrality' of each observer within the network, aligning with established concepts in network analysis for assessing node influence and importance \cite{Freeman1977, Bonacich1987}. Centrality is based on their connectivity to event nodes.

\subsection{Unweighted Centrality}
Initially, centrality is measured in an unweighted manner, utilizing the concept of degree centrality \cite{Opsahl2010} to account for the sheer number of direct connections an observer node maintains with event nodes.

\begin{itemize}
    \item \textbf{Centrality Definition}: Unweighted centrality is defined as the count of an observer's unique links to events, reflecting the scope of their monitoring capabilities.
    
    \item \textbf{Coverage Breadth}: This metric assesses an observer's range of coverage, seeking to ensure that event observation is widespread and unique, avoiding redundant coverage which could skew the centrality measure.
\end{itemize}

\subsection{Weighted Centrality}
To capture the nuanced dynamics of observer influence, centrality metrics are extended beyond mere counts to include the weight of each link, similar to weighted centrality measures used in network analysis \cite{Brandes2001, Newman2005}.

\begin{itemize}
    \item \textbf{Weighted Centrality Definition}: Incorporates the strength or significance of each observer-event link, which is calculated by summing the weights of all connecting edges.
    
    \item \textbf{Application of Weighted Centrality}: This measure evaluates an observer's influence by considering not just the count of connections but also their quality—determined by spatial proximity, temporal relevance, and the certainty associated with each observed event.
\end{itemize}

\subsection{Considerations for Centrality}
The computation of centrality involves critical considerations to refine its accuracy and applicability:

\begin{itemize}
    \item \textbf{Unique Assignment of Events}: Where feasible, events are preferentially assigned to a single observer to maintain the integrity of centrality scores by minimizing overlap and ensuring clear responsibility and observation demarcation.
    
    \item \textbf{Quantitative Analysis}: The centrality score serves as a quantitative indicator of an observer's prominence in the network, guiding resource allocation and strategic planning within the observational framework.
\end{itemize}

\subsection{Visualization of Centrality Ranking}
The visualization displays centrality rankings within a network by numbering observer nodes—connected by blue lines to events they cover.

\begin{figure}[h!]
  \centering
  \includegraphics[width=0.75\linewidth]{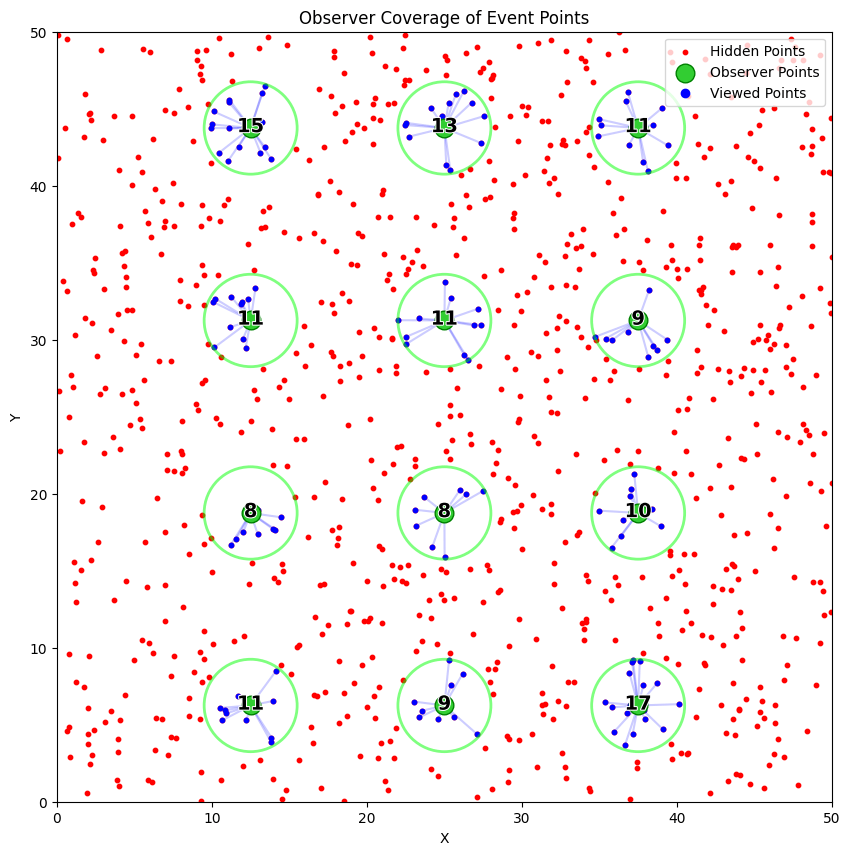}
  \caption{Centrality Ranking using Bipartite Dynamics}
  \label{fig:bipartite-centrality-ranking}
\end{figure}

\clearpage
\section{Event Classifier}

The Event Classifier systematically categorizes events into distinct states based on their detection by observer nodes, aligning with research on event detection and classification techniques in temporal networks \cite{Kovanen2011}.

\subsection{Binary Classification}
Initially, the classifier employs a binary distinction:

\begin{itemize}
    \item \textbf{Viewed Events}: These events are directly linked to at least one observer node, indicating successful observation and coverage.
    \item \textbf{Hidden Events}: Events lacking direct connections to observer nodes, revealing potential coverage gaps within the network.
\end{itemize}

\subsection{Weighted Classification}
The weighted classifier leverages a more diverse set, assessing events not simply as `viewed' or `hidden' but across a continuum based on detailed quality and contextual observations, similar to weighted classification schemes used in various data analysis applications \cite{Aggarwal2014}.

\begin{itemize}
    \item \textbf{Criteria for Weight Assignment}: Weights may be dynamically determined by factors including:
    \begin{itemize}
        \item \textit{Spatial Proximity:} This considers the physical closeness of events to observers, often modeled using spatial decay functions that capture the diminishing importance of events as distance increases \cite{Barthelemy2022}.
        \item \textit{Temporal Relevance:} This distinguishes events based on their recency, reflecting the idea that recent events may hold greater significance in dynamic networks \cite{Holme2012}.
        \item \textit{Probability of Event Occurrence:} This incorporates the likelihood or reliability of events, aligning with methods of uncertainty modeling in data analysis to account for potential errors or incomplete information \cite{Berger1985}.
    \end{itemize}

    \item \textbf{Event Valuation Spectrum}: 
    \begin{itemize}
        \item \textit{Likely vs. Unlikely}: Events are assessed on a spectrum, with those detected within close proximity and with high probability metrics considered 'likely' and assigned greater significance.
        \item \textit{Recent vs. Historical}: Events are also classified as 'recent' or 'historical' to reflect their temporal significance, with recent events prioritized based on their immediate relevance to the network's current status.
         \item \textit{Near vs. Far}: Events are also classified as 'near' or 'far' to reflect their spatial significance, with near events prioritized based on their closeness.
    \end{itemize}

    \item \textbf{Multiple Case Classification}: This approach allows for a nuanced evaluation of events, enabling the network to prioritize events based on a combination of observed characteristics and their potential impact on network operations.

\end{itemize}

\subsection{Visualization of Event Classification}
The visualization displays event classification within a network where blue nodes are observed and red are hidden.

\begin{figure}[h!]
  \centering
  \includegraphics[width=0.75\linewidth]{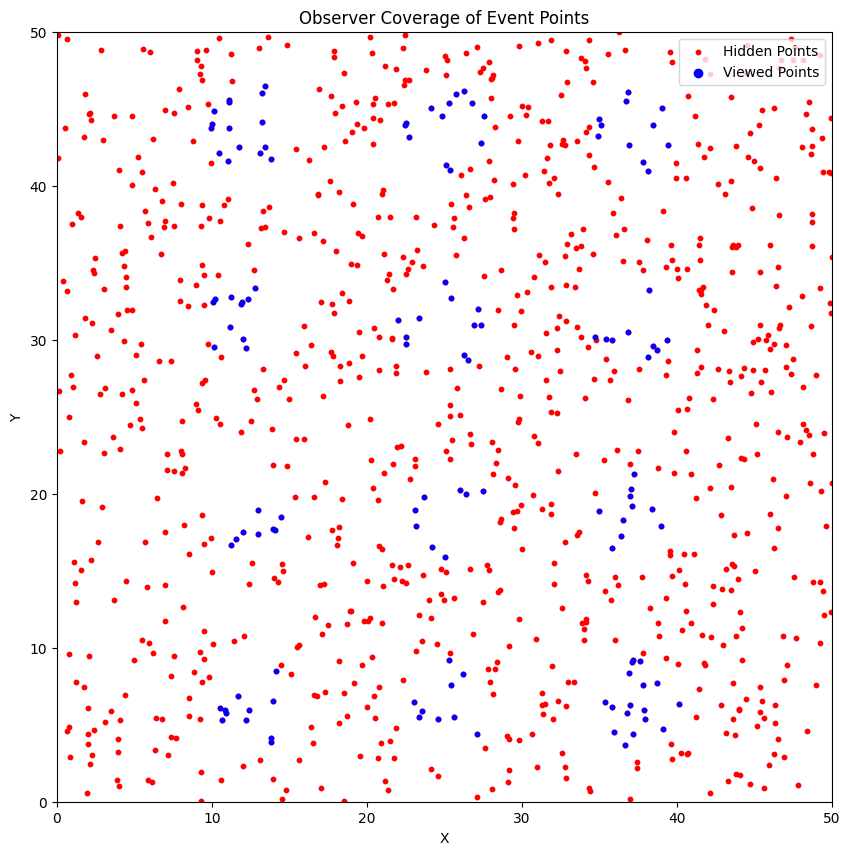}
  \caption{Event Classifier using Bipartite Dynamics}
  \label{fig:bipartite-event-classifier}
\end{figure}

\section{Coverage Analysis}
This section delves into quantitatively assessing the effectiveness of observer nodes in monitoring event nodes throughout the network, aligning with research on network coverage problems and optimization techniques \cite{Meguerdichian2001}. This analysis quantifies the reach and limitations of the current observer deployment and for guiding improvements in network design.

\subsection{Overview of Coverage Metrics}
Coverage analysis hinges on several key metrics that collectively describe the performance and efficiency of the observer network:

\begin{enumerate}
    \item \textbf{Total Coverage}: This metric quantifies the proportion of event nodes that are observed by at least one observer across the entire network. For instance, a Total Coverage Score of 13.1\% would indicate that 13.1\% of all events are monitored by at least one observer, reflecting the overall effectiveness of the network, a key consideration in sensor placement optimization \cite{Huang2005}.
    
    \item \textbf{Average Coverage per Observer}: Reflecting the efficiency and value of each observer node, this score averages the number of event nodes each observer covers. An example score of 10.92 suggests that, on average, each observer monitors nearly 11 events, highlighting the individual contribution of observers to network coverage and the importance of resource allocation in sensor networks \cite{Zou2004}.
    
    \item \textbf{Distribution of Degrees}: This analysis examines the number of connections each observer has with event nodes, shedding light on the network's connectivity and its impact on information flow and resilience \cite{Albert2002}. By identifying observers with high connectivity (hubs) and those with few links, this insight assists in strategic network adjustments.
\end{enumerate}

\subsection{Strategic Implications}
The insights gained from the coverage analysis play a critical role in network optimization:

\begin{itemize}
    \item \textbf{Optimizing Observer Placement}: By understanding the existing coverage and connectivity, strategic decisions can be made to either reposition existing observers or to add new ones in areas where event nodes are currently under-monitored, aligning with principles of sensor placement optimization \cite{Krause2008}.
    
    \item \textbf{Enhancing Network Efficiency}: Adjustments based on the Average Coverage per Observer can lead to improved efficiency, ensuring that resources are allocated where they yield the highest coverage impact, contributing to overall network optimization and performance enhancement \cite{Abhishek2018}.
\end{itemize}

\subsection{Visualization of Coverage Metrics}
This visualization highlights the distribution of observer effectiveness and identifies both highly effective and underperforming areas of the network.

\begin{figure}[h!]
  \centering
  \includegraphics[width=0.75\linewidth]{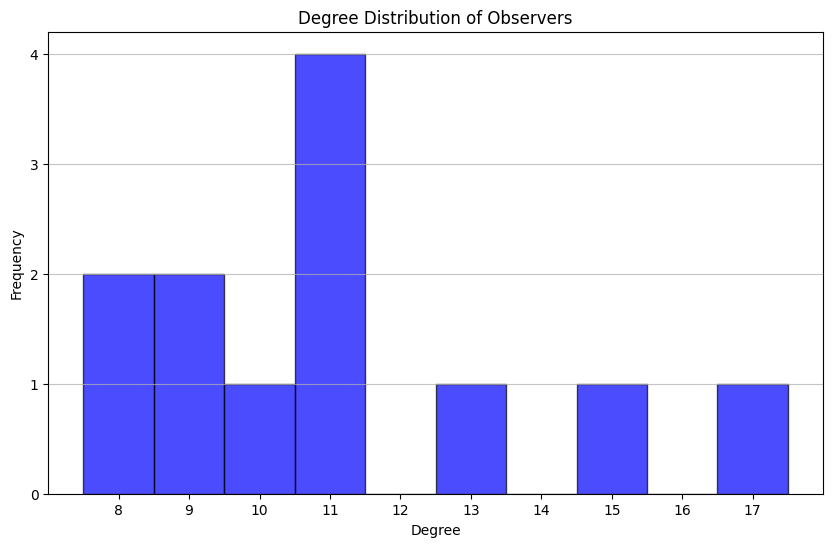}
  \caption{Centrality Analysis using Bipartite Dynamics}
  \label{fig:bipartite-centrality-analysis}
\end{figure}

\clearpage  
\section*{II. Unipartite Dynamics: Optimize Node Placements  }

Building upon the insights gained from the Bipartite Dynamics section, this section shifts focus towards optimizing node placements within the observer network. The Unipartite Dynamics approach applies advanced spatial analysis techniques to maximize coverage efficiently with a finite set of resources.

\begin{figure}[h!]
  \centering
  \includegraphics[width=0.45\linewidth]{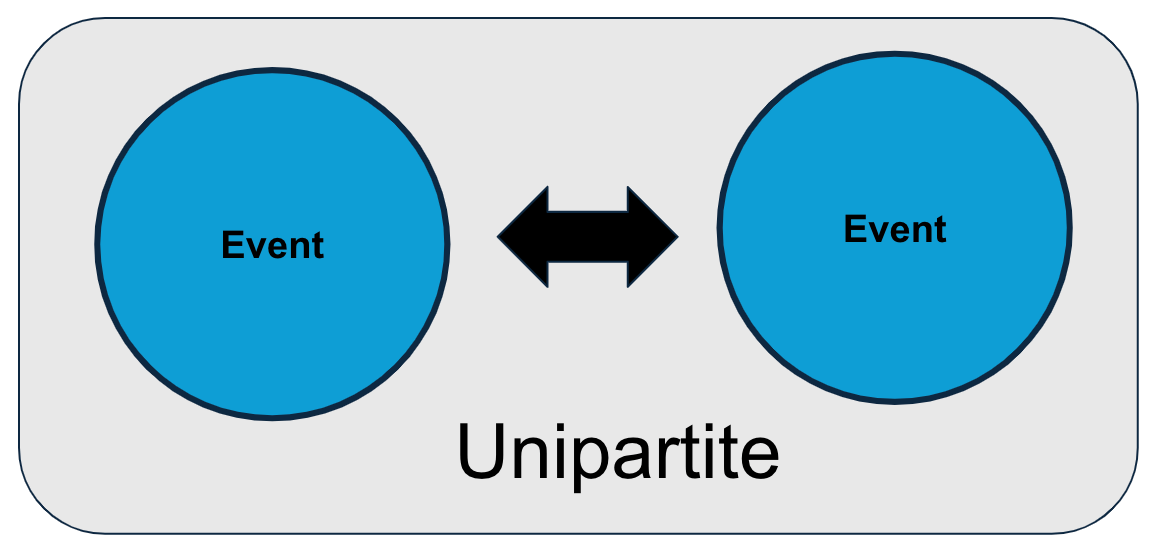}
  \caption{Unipartite Dynamics between Unobserved Events}
  \label{fig:unipartite-dynamics}
\end{figure}

\subsection{Unipartite Dynamics: Parameters and Outputs}

\begin{figure}[h!]
  \centering
  \includegraphics[width=0.8\linewidth]{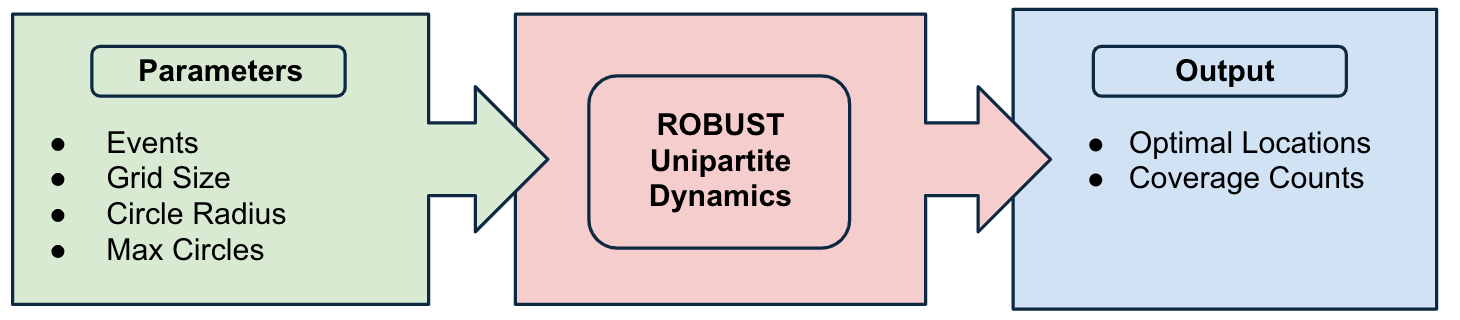}
  \caption{Unipartite Input Parameters and Outputs}
  \label{fig:unipartite-param-return}
\end{figure}

\subsubsection*{Input Parameters}
\begin{itemize}
    \item \textbf{Events:} The list of spatial events that are yet to be observed by the network, serving as the target points for coverage optimization.
    \item \textbf{Grid Size:} The scale of the computational matrix which defines the spatial resolution of analysis—larger grids allow for a more granular examination.
    \item \textbf{Circle Radius:} Represents the reach or the effective coverage range of each observer node, crucial for calculating influence over the event points.
    \item \textbf{Max Circles:} This constraint sets the upper limit on the number of observer nodes to deploy, aligning with resource availability and strategic considerations.
\end{itemize}

\subsubsection*{Expected Outputs}
\begin{itemize}
    \item \textbf{Optimal Locations:} A set of coordinates pinpointing the strategic placement of observer nodes to maximize event coverage and network efficiency.
    \item \textbf{Coverage Counts:} A tally of events within the radius of influence of each placed observer node, providing a quantitative measure of the network's improved observation capacity.
\end{itemize}

\subsubsection*{Algorithm Overview:}
 This step-by-step process designed to optimize the placement of resources based on advanced spatial analysis:

 \begin{figure}[h!]
  \centering
  \includegraphics[height=0.4\linewidth]{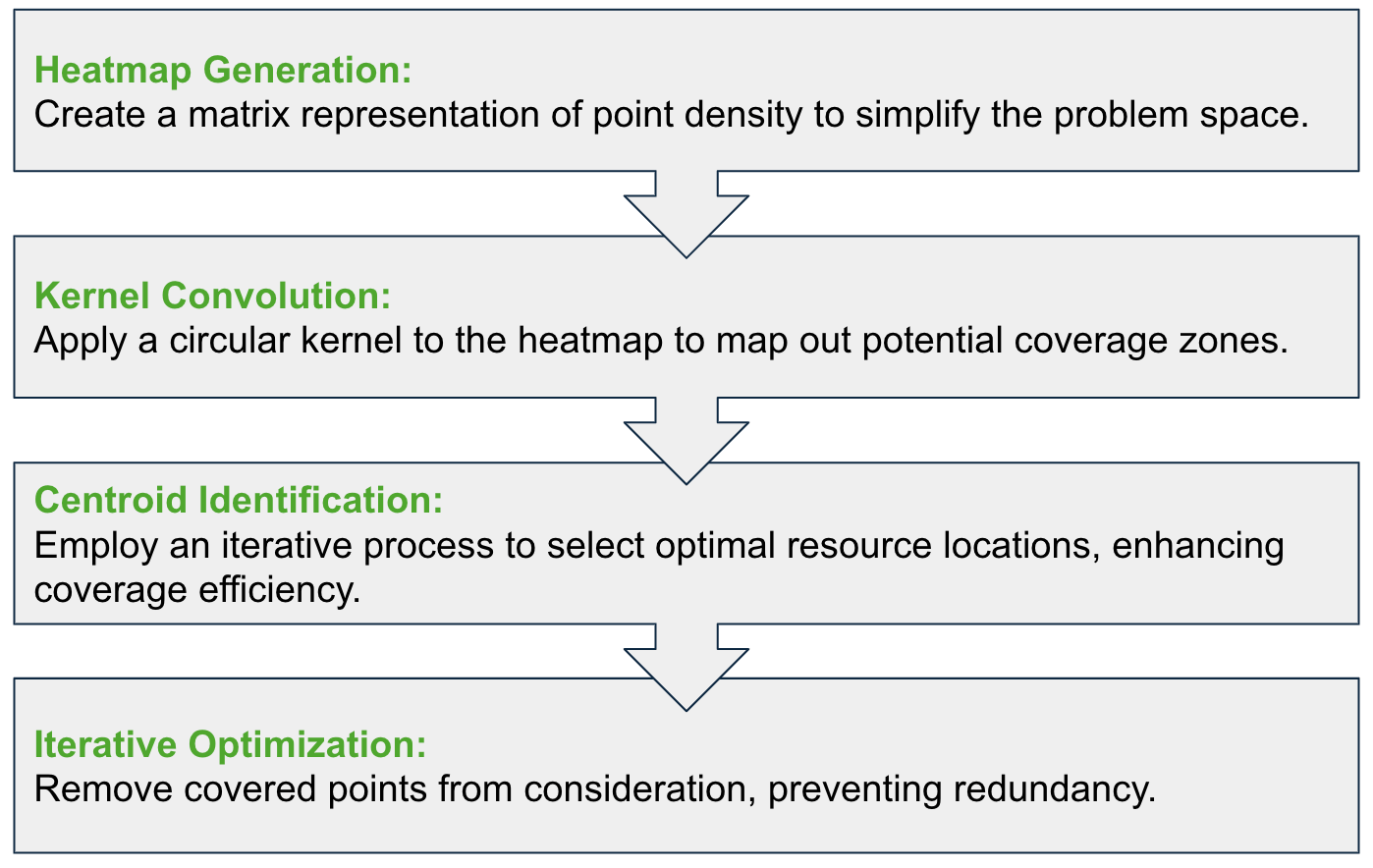}
  \caption{Flowchart of Analysis using Unipartite Dynamics}
  \label{fig:unipartite-dynamics-flowchart}
\end{figure}

\begin{enumerate}
    \item \textbf{Heatmap Generation}: Initially, create a matrix representing the density of event points across the network. This heatmap simplifies the problem space, providing a visual representation of areas that require increased coverage.
    
    \item \textbf{Kernel Convolution}: Apply a circular kernel to the heatmap to effectively map out potential coverage zones. This technique uses spatial convolution to simulate the range of coverage each observer node could provide, optimizing placements based on the spatial overlap of these zones.
    
    \item \textbf{Centroid Identification}: Utilize an iterative process to identify the centroids of these coverage zones. By strategically placing observer nodes at these centroids, the network can maximize the area each node covers.
    
    \item \textbf{Iterative Optimization}: Continuously refine node placements by removing points already covered from further consideration. This step helps to eliminate redundancy and ensures each observer is utilized to its fullest potential.
\end{enumerate}

\subsection{Visualization of Initial Unobserved Events }
This visualization represents the aggregated positions of the initial set of unobserved events. The following sections will apply the defined methods to this demo to illustrate the applicability. 

\begin{figure}[h!]
  \centering
  \includegraphics[width=0.75\linewidth]{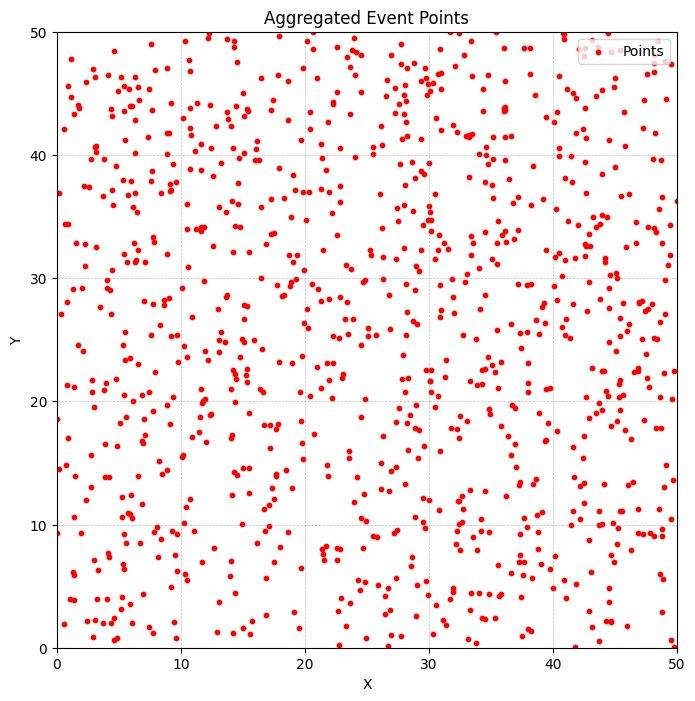}
  \caption{Initial set of Aggregated Unobserved Events}
  \label{fig:unipartite-initial-points}
\end{figure}

\clearpage
\section{Point Density Heatmap}

\subsection{Introduction to Heatmap Utilization}
A point density heatmap is a matrix representation that leverages the principles of kernel density estimation \cite{Silverman1986, Wand1995} to capture the frequency of events in a location. This section delves into the construction and application of heatmaps to distill complex spatial relationships into actionable data.

\subsection{Constructing the Point Density Matrix}
The construction of a point density heatmap involves categorizing spatial data into a grid using spatial binning techniques \cite{Bailey1995, Cressie2012}, with each cell representing a specific area's event density. This process simplifies the computational task from analyzing individual points to evaluating discrete grid areas.

\begin{enumerate}
    \item \textbf{Spatial Binning Process:} Points are assigned to bins based on their coordinates, transforming the continuous spatial field into a discrete set of bins. Each bin aggregates the density of points within its bounds, reducing the granularity but preserving essential spatial information. The bins represent the boundaries for defining if points are at the ``same" position. 
    
    \item \textbf{Matrix Representation:} The resulting bins are then arranged into a matrix, where each cell's value corresponds to the density of points within that bin's area. This matrix acts as a heatmap,representing areas of high and low event concentrations.
\end{enumerate}

\subsection{Methodological Approach}
The methodology for creating a point density heatmap emphasizes computational efficiency, particularly for large datasets where direct point-to-point comparisons are impractical due to their quadratic complexity.

\begin{enumerate}
    \item \textbf{Algorithmic Efficiency:} By aggregating points into bins, the computational complexity is dramatically reduced. This efficiency gain allows for rapid assessment of point distribution and density without the need for pairwise distance calculations.
    
    \item \textbf{Computational Acceleration:} Modern computational tools such as CUDA \cite{Nickolls2008} provide the necessary acceleration to handle extensive spatial datasets. This utilization of advanced computational resources ensures that the heatmap generation process remains scalable and time-efficient.
\end{enumerate}

\subsection{Visualization}
This a visualization of applying a point density heatmap with resolution as 50x50 to the initial set of unobserved nodes.  Cells that are brighter have a higher density of events.

\begin{figure}[h!]
  \centering
  \includegraphics[width=0.75\linewidth]{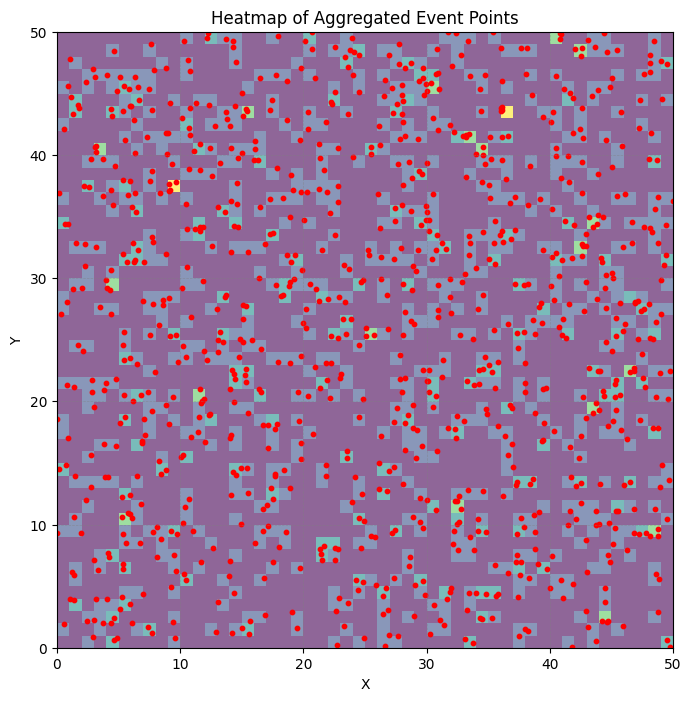}
  \caption{Point Density Heatmap using Unipartite Dynamics}
  \label{fig:unipartite-density-heatmap}
\end{figure}

\clearpage
\section{Kernel Convolution}

\subsection{Theoretical Foundation of Kernel Convolution}
Kernel convolution, an operation adapted from image processing and computer vision \cite{Gonzalez2018, Szeliski2010}, involves superimposing a defined kernel shape, such as a circle or square, over a matrix to identify regions of interest within the data. In the context of coverage mapping, this technique is employed to determine the areas with the highest density of events that fall within a given range or influence zone of an observer node.

\subsection{Operationalizing Convolution in Spatial Analysis}
The kernel convolution process is articulated through the following steps:

\begin{enumerate}
    \item \textbf{Kernel Overlay:} A kernel, representing the coverage radius of a potential observer node, is applied to each point on the heatmap. This step simulates the coverage area of the observer.
    
    \item \textbf{Convolution Mechanics:} 
    As the kernel moves across the heatmap, it aggregates the values of the points that fall under it, performing a convolution operation \cite{Gonzalez2018} that is akin to a sliding window that computes the sum of covered event densities.
    
    \item \textbf{Coverage Potential Identification:} Through this process, the algorithm highlights regions where the kernel encompasses a high concentration of neighboring events.
\end{enumerate}

\subsection{Convolution Process in Detail}
The convolution mechanism can be mathematically formalized as follows:

\begin{itemize}
    \item Let \( H \) be the heatmap matrix where \( H_{ij} \) represents the density of events at grid cell \( (i, j) \).
    \item Let \( K \) be the kernel matrix with a predefined shape and size, where \( K_{ab} \) represents the weight or influence of the kernel at position \( (a, b) \) relative to its center.
    \item The convolution \( C \) at a particular point \( (i, j) \) on the heatmap is calculated as:
    \[
    C_{ij} = \sum_{a,b} H_{i+a,j+b} \cdot K_{ab}
    \]
    \item The points \( (i, j) \) where \( C_{ij} \) is maximized are identified as candidate locations for observer placement.
\end{itemize}

\subsection{Scalability and Computational Efficiency}
The scalable nature of this approach allows it to efficiently handle large datasets:

\begin{itemize}
\item The reduction from a direct point-to-point analysis to a manageable convolution operation significantly lowers computational time \cite{Cormen2009}.

\item Utilization of computational accelerators like GPUs, using libraries such as CUDA \cite{Kirk2010},  enables the handling of extensive data grids, enhancing the method's applicability to larger networks.
\end{itemize}

\subsection{Visualization}
This a visualization of applying a kernel convolution on the point density heatmap.  Cells that are brighter have a higher density while also considering neighboring influences.

\begin{figure}[h!]
  \centering
  \includegraphics[width=0.75\linewidth]{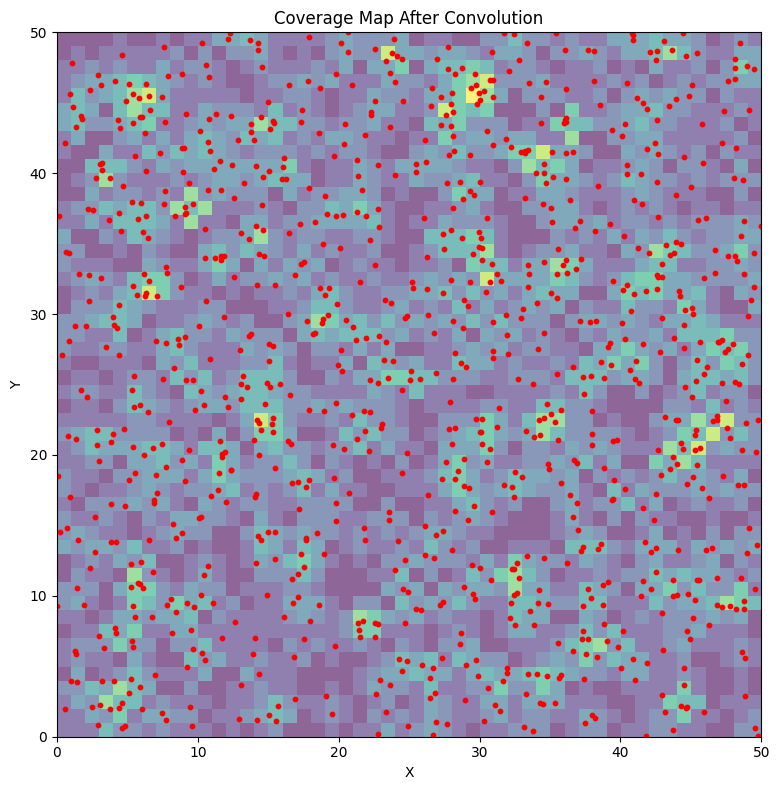}
  \caption{Kernel Convolution using Unipartite Dynamics}
  \label{fig:unipartite-kernel-convolution}
\end{figure}

\clearpage
\section{Centroid Identification and Iterative Optimization}

\subsection{Objective and Methodology} 
The objective is to locate the focal points—centroids—of each coverage area to place observer nodes most effectively. The process uses the output from kernel convolution to guide the placement of these centroids.

\subsection{Iterative Centroid Selection Process}
This process involves a series of steps to ensure each centroid's location maximizes the coverage of event points:

\begin{enumerate}
\item \textbf{Initial Centroid Location:} The algorithm begins by identifying the peak value within the coverage map, which indicates the area with the highest cumulative event density and potential for coverage. This peak serves as the first centroid.

\item \textbf{Coverage Area Neutralization:} Once the first centroid is placed, its coverage radius is projected onto the heatmap, and the densities of all points within this radius are reduced to zero. This prevents these points from influencing the location of subsequent centroids.

\item \textbf{Subsequent Centroids:} The selection process repeats, each time identifying the next highest value on the updated coverage map and neutralizing the corresponding coverage area. This iterative approach continues until all centroids are placed or the remaining points on the heatmap do not justify additional coverage.

\end{enumerate}

\subsection{Resultant Coverage Optimization}
The outcome of this process is a set of strategically placed centroids that collectively maximize the coverage of event points. Each centroid's placement is optimized to extend the observer network's reach while avoiding redundancy, ensuring that each resource contributes effectively to the overall network performance.

\subsection{Visualization}
This shows where the top ten positions for observers should be placed.

\begin{figure}[h!]
  \centering
  \includegraphics[width=0.75\linewidth]{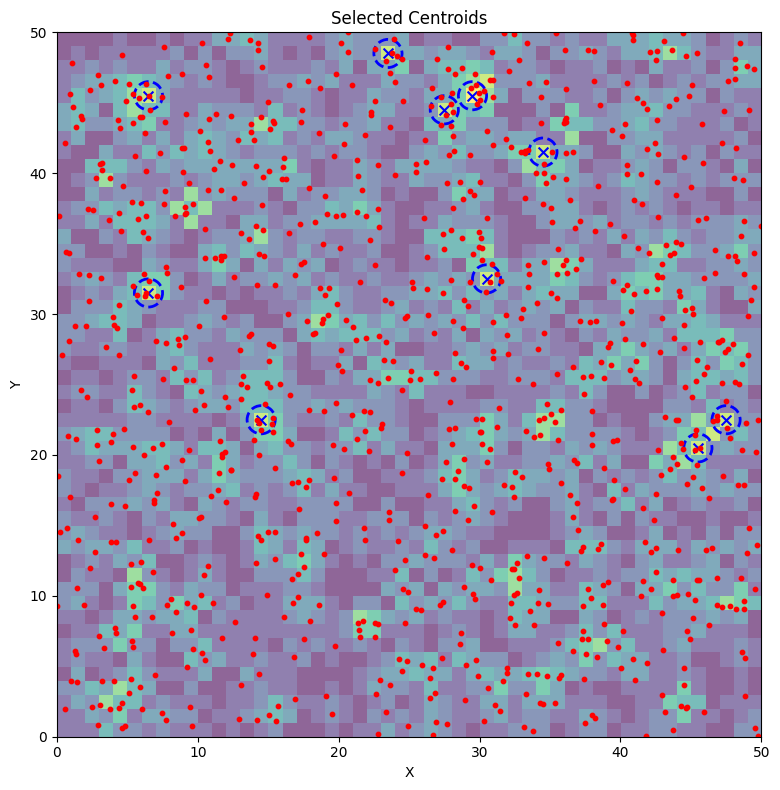}
  \caption{Centroid Identification using Unipartite Dynamics}
  \label{fig:unipartite-centroid-identification}
\end{figure}

\clearpage

\chapter{ROBUST and Mobile Observers}
\label{chap:robust_mobile}

\section{Introduction}

Building on the foundational principles of the ROBUST Network, this chapter delves into the application of dynamic sensor placements, a critical aspect in scenarios demanding real-time adaptability and responsiveness. The process for dynamically placing sensors in the ROBUST Network is divided into three distinct phases.  Each phase plays a vital role in ensuring the network's effectiveness.

The first phase involves the Proximal Recurrent Event Partition (PREP) Mapper which extends from the ROBUST Network. This phase focuses on mapping a network of potential waypoints, which are crucial positions or areas the sensors might occupy. The Mapper's role is to lay out these strategic points.

In the second phase, the TED (Temporal Event Dynamics) Predictor comes into play. This phase is dedicated to predicting which waypoints will be active at any given time frame. For the purpose of this research, the TED Predictor operates under the assumption of perfect knowledge, enabling it to make accurate and informed predictions about the network's dynamics.

The final phase involves the WAITR (Weighted Aggregate Intra-Temporal Reward) Planner. This component is responsible for computing the optimal paths for sensor movements by ensuring that sensors are positioned effectively for each time frame. The WAITR Planner takes into account various factors to determine the most advantageous routes for the sensors, maximizing observational efficiency and coverage.

Through these three phases – the PREP Mapper, the TED Predictor, and the WAITR Planner – the ROBUST Network achieves a high level of dynamism and adaptability, essential for real-time monitoring and response in complex environments. 

The algorithms presented in this chapter refers to the implementation details used for benchmarking in Chapter \ref{chap:benchmark_results}: \textit{Benchmarking Results}.

\section{Phase 1: PREP Mapper}

The Proximal Recurrent Event Partition (PREP) Mapper forms the first phase in the dynamic sensor placement process within the ROBUST Network. Its primary function is to map out potential waypoints based on points of interest. These waypoints are strategically vital positions or areas that sensors might occupy for optimal observational coverage. 

\vspace{8mm}

\resizebox{\textwidth}{!}{%
\begin{tikzpicture}[
    scale=0.75,
    node distance=1cm,
    auto,
    thick,
    main/.style={rectangle, draw, text width=4cm, minimum height=1.5cm, align=center, font=\sffamily\small\linespread{0.8}\selectfont},
    to/.style={->, shorten >=1pt, semithick, font=\sffamily\footnotesize},
    every node/.append style={font=\sffamily\small}
]

    \node[main] (input) {Input Data};
    \node[main] (nodeGen) [right=of input] {Node Generation};
    \node[main] (linkGen) [right=of nodeGen] {Link Generation};
    \node[main] (waypoint) [right=of linkGen] {Waypoint Network};
    \node[main] (getObsv) [below=of nodeGen] {Get Observables};
    \node[main] (getObs) [below=of getObsv] {Get Observers};
    \node[main] (filterUnobs) [below=of getObs] {Filter for Unobserved};
    \node[main] (unobsDist) [below=of linkGen] {Unobserved Distance Matrix};
    \node[main] (linkClosest) [below=of unobsDist] {Link Closest Nodes by Move Threshold};
    \node[main] (addInter) [below=of linkClosest] {Add Bridge Nodes if needed};
    
    \draw[to] (input) -- (nodeGen);
    \draw[to] (nodeGen) -- (linkGen);
    \draw[to] (linkGen) -- (waypoint);
    \draw[to] (nodeGen) -- (getObsv);
    \draw[to] (getObsv) -- (getObs);
    \draw[to] (getObs) -- (filterUnobs);
    \draw[to] (linkGen) -- (unobsDist);
    \draw[to] (unobsDist) -- (linkClosest);
    \draw[to] (linkClosest) -- (addInter) ;
\end{tikzpicture}
}

\subsection{Node Generation}

\subsubsection{Algorithm Overview:}
The Node Generation algorithm in the PREP Mapper is designed to identify waypoints based on heatmap data representing event occurrences. The process involves analyzing the spatial distribution of events and determining key locations that would serve as effective waypoints for sensor deployment.  

\textbf{Notations:}
\begin{itemize}
    \item \( H \): Heatmap matrix representing event intensity over the monitored area.
    \item \( WP \): Set of Waypoint nodes, initially empty.
    \item \( r \): Radius threshold for waypoint effectiveness.
\end{itemize}

\begin{enumerate}
    \item Extract event locations.
    \item Filter observed events.
    \item Select waypoints based on effectiveness and spatial distribution.
\end{enumerate}

\textbf{1. Extract Event Locations:}
\[
E = \text{get\_events}(H, \text{threshold})
\]
Where \( E \) represents the set of event locations extracted from the heatmap data \( H \).

\subsubsection{Get Events \texttt{(get\_events)} } 
To express the \texttt{get\_events} function in terms of matrices and vector operations, we represent it using mathematical notations that correspond to the operations performed in the code. The function identifies cells in a heatmap where the event intensity exceeds a specified threshold.

\textbf{Heatmap Matrix \( H \)}: Represent the heatmap as a matrix \( H \) where each element \( H_{ij} \) corresponds to the intensity value at the cell located at row \( i \) and column \( j \).

\textbf{Thresholding Operation}: Apply a thresholding operation to identify cells where the intensity exceeds the threshold. This can be represented as a function \( T(H, \theta) \) that returns a binary matrix \( M \) of the same dimensions as \( H \), where each element is determined by:
\[
M_{ij} = 
\begin{cases} 
1 & \text{if } H_{ij} \geq \theta \\
0 & \text{otherwise}
\end{cases}
\]
Here, \( \theta \) is the threshold (e.g., 0.9 in your code).

\textbf{Reduction Across Time Dimension}: The event occurrence is determined by applying a maximum operation across the time dimension of the heatmap. This operation identifies if an event has occurred at least once at each spatial location throughout the dataset.

\textbf{Event Locations Extraction}: The function \texttt{get\_events} identifies the indices of cells where events occurred, which corresponds to finding the row and column indices of the matrix \( M \) where the value is 1.

\textbf{Event Locations Matrix \( E \)}: Represent the set of event locations as a matrix \( E \), where each row corresponds to the coordinates of an event. The coordinates are extracted from the matrix \( M \) as follows:
\[
E = \{ (i, j) \mid M_{ij} = 1 \}
\]
The coordinates \( (i, j) \) represent the spatial locations of the events.

Bringing it all together, the mathematical expression corresponding to the \texttt{get\_events} function can be:

\textbf{Extract Event Locations:}
\[
E = \{ (i, j) \mid M_{ij} = 1 \} \quad \text{where} \quad M = T(H, \theta)
\]

\vspace{12pt}
\textbf{2. Filter Observed Events (\texttt{init\_robust\_net}):} \\
The \texttt{init\_robust\_net} function filters observed events from unobserved ones by evaluating the proximity of sensor nodes to event nodes. This is crucial for identifying areas where events are effectively observed and areas needing additional sensor coverage.

\textbf{Function Overview:}
\[
Observed, Unobserved = \text{init\_robust\_net}(S, E, r)
\]
This function classifies events as observed or unobserved based on their proximity to sensor nodes, using a distance matrix and a radius threshold.

\textbf{Notations:}
\begin{itemize}
    \item \( S \): Matrix representing sensor nodes, where each row is a node with coordinates.
    \item \( E \): Matrix representing event nodes, where each row is a node with coordinates.
    \item \( r \): Radius threshold for determining observation effectiveness.
    \item \( DM \): Distance matrix between sensor nodes and event nodes.
\end{itemize}

\textbf{Distance Calculation Between Sensors and Events}:
The first step in the function is to calculate the distances between each sensor node and event node.
\[
DM_{i,j} = \sqrt{(S_{i,x} - E_{j,x})^2 + (S_{i,y} - E_{j,y})^2}
\]
for each pair of sensor node \(i\) and event node \(j\), where \( S_{i,x} \) and \( S_{i,y} \) are the coordinates of the sensor node, and \( E_{j,x} \) and \( E_{j,y} \) are the coordinates of the event node.

\textbf{Classification of Observed and Unobserved Events}:
Based on the distances calculated in the matrix \( DM \) and the specified radius \( r \), the function then classifies events as observed or unobserved. Events within the radius of any sensor node are considered observed, while those outside are considered unobserved.

\vspace{12pt}

\textbf{3. Select Waypoints (\texttt{next\_observer\_nodes}):} \\
The \texttt{next\_observer\_nodes} function is designed to identify optimal waypoints based on the spatial distribution and density of unobserved events. This involves analyzing distances among unobserved event nodes and clustering them to determine the most effective locations for sensor placement.

\textbf{Function Overview:}
\[
WP = \text{next\_observer\_nodes}(DM, r)
\]
This function clusters unobserved event nodes based on their proximity, using a precomputed distance matrix and a radius threshold, to identify dense clusters that are optimal for waypoint selection.

\textbf{Notations:}
\begin{itemize}
    \item \( U \): Matrix representing unobserved event nodes.
    \item \( r \): Radius threshold for clustering.
    \item \( DM \): Distance matrix among unobserved event nodes.
    \item \( WP \): Set of Waypoint nodes, determined by clustering.
\end{itemize}

\textbf{Distance Calculation Among Unobserved Events}:
The first step in the function is to calculate the distances between each pair of unobserved event nodes.
\[
DM_{i,j} = \sqrt{(U_{i,x} - U_{j,x})^2 + (U_{i,y} - U_{j,y})^2}
\]
for each pair of unobserved event node \(i\) and event node \(j\), where \( U_{i,x} \) and \( U_{i,y} \) are the coordinates of the unobserved event nodes.

\textbf{Clustering Unobserved Events}:
Using the distance matrix \( DM \), the function clusters unobserved event nodes based on their proximity, defined by the radius \( r \). This step involves identifying all unobserved nodes within radius \( r \) of each centroid node and grouping them into clusters.

\textbf{Selecting Densest Clusters}:
Clusters are sorted based on their density, and the densest clusters are selected as waypoints. This is achieved by:
\[
WP = \text{get\_densest\_clusters}(DM, U, r)
\]
where the function \(\text{get\_densest\_clusters}\) identifies clusters of unobserved event nodes within radius \( r \) and selects the densest ones as waypoints.

\begin{figure}[htbp]
  \begin{adjustwidth}{-2cm}{-2cm} 
    \centering
    \begin{minipage}[b]{0.6\textwidth} 
      \centering
      \includegraphics[width=\textwidth]{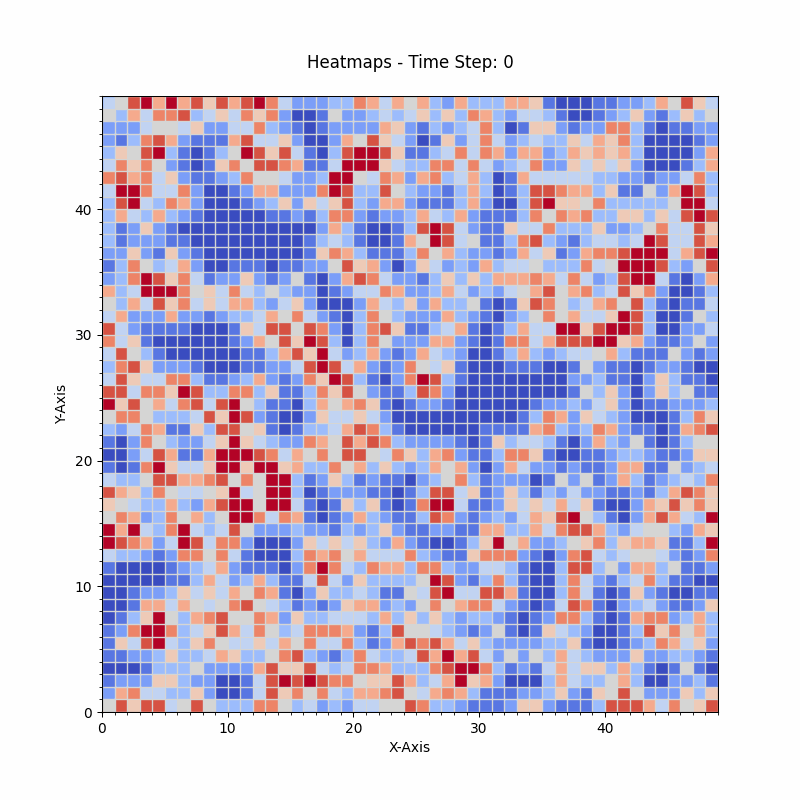}
      \caption{Grid with Heat-mapped Events}
      \label{fig:first}
    \end{minipage}
    \hfill
    \begin{minipage}[b]{0.6\textwidth} 
      \centering
      \includegraphics[width=\textwidth]{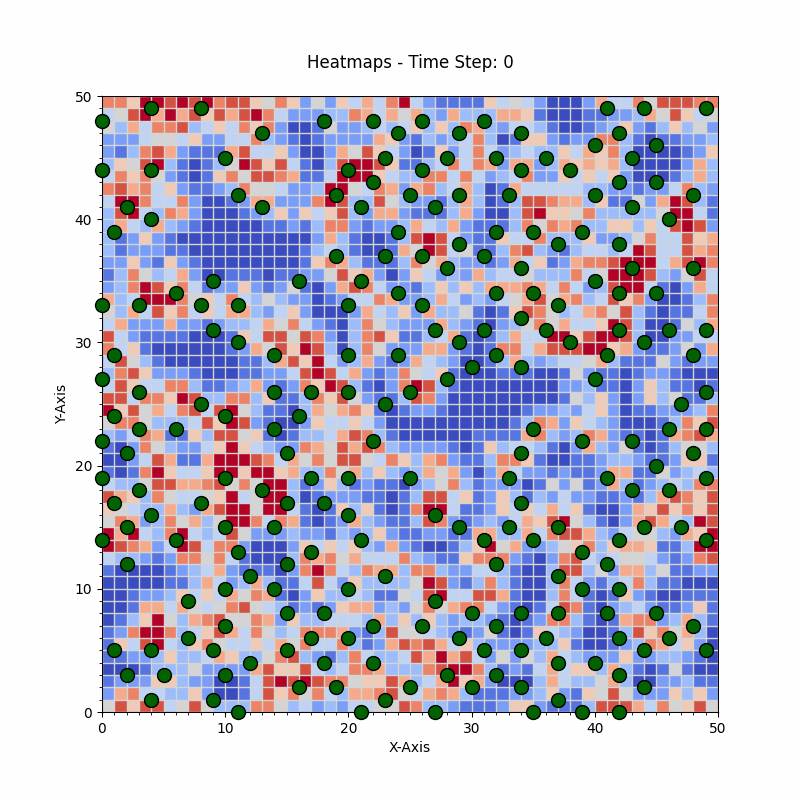}
      \caption{Waypoint Nodes with PREP Mapper}
      \label{fig:second}
    \end{minipage}
  \end{adjustwidth}
\end{figure}

\subsubsection{Node Generation Visualization}
The figure consists of two panels, side-by-side, that illustrate the initial steps in developing a waypoint network for mobile sensors. The left panel (Figure~\ref{fig:first}) shows am example heatmap visualization, where each cell's color represents the level of interest within that region.  Bluer colors indicate less interesting areas, while red colors represent "hot zones" with a higher concentration of events of interest.

The right panel (Figure~\ref{fig:second}) depicts the placement of waypoint nodes which represent potential locations for mobile sensors to visit. The goal is to optimize the placement of these waypoints to maximize mobile sensor coverage of the most important areas (hot zones) identified in the heatmap.

\subsection{Link Generation}

The Link Generation phase in the PREP Mapper aims to establish connections between identified waypoints. This involves calculating distances between waypoints and generating links based on proximity.

\textbf{Notations:}
\begin{itemize}
    \item \( WP \): Matrix representing Waypoint nodes, where each row is a node with coordinates.
    \item \( DM \): Distance matrix among waypoint nodes.
    \item \( L \): Set of Links between waypoints.
\end{itemize}

\textbf{Distance Calculation Among Waypoints}:
To calculate the distances among waypoint nodes, we represent the waypoints as a matrix \( WP \) and compute the distance matrix \( DM \) as follows:
\[
DM_{i,j} = \sqrt{(WP_{i,x} - WP_{j,x})^2 + (WP_{i,y} - WP_{j,y})^2}
\]
for each pair of waypoint nodes \(i\) and \(j\), where \( WP_{i,x} \) and \( WP_{i,y} \) are the coordinates of the waypoint nodes.

\textbf{Generating Proximity-Based Links}:
The process of creating links between waypoints is based proximity. This involves identifying the nodes closest to each waypoint and forming links accordingly. The mathematical operations are as follows:

\textbf{Notations:}
\begin{itemize}
    \item \( DM \): Distance matrix among waypoint nodes.
    \item \( n \): Number of links to be generated for each waypoint.
    \item \( L \): Set of proximity-based links between waypoints.
\end{itemize}

\textbf{Exclude Self-Links}:
First, self-links are excluded by setting the diagonal elements of the distance matrix \( DM \) to infinity:
\[
DM_{ii} = \infty \quad \text{for all } i
\]

\textbf{Sort Distances and Identify Closest Nodes}:
For each waypoint node, identify the \( n \) closest other nodes based on the distances in \( DM \). This is done by sorting the distances for each node and selecting the top \( n \) indices:
\[
\text{Indices}_{i} = \text{argsort}(DM_{i, :})[0:n]
\]

\textbf{Create Link Tuples}:
Links are then formed as tuples of node indices and their corresponding distances. Each link connects a waypoint to one of its \( n \) closest nodes:
\[
L = \bigcup_{i=1}^{\text{num\_nodes}} \{ (i, \text{Indices}_{i,j}, DM_{i, \text{Indices}_{i,j}}) : j = 1, \ldots, n \}
\]
where \( \text{num\_nodes} \) is the total number of waypoint nodes and \( \text{Indices}_{i,j} \) is the \( j \)-th closest node to waypoint \( i \).

\textbf{Unique Edge Identification}:
After generating the links, it's crucial to ensure the uniqueness of these links and avoid redundancy. This process involves organizing the edges, sorting them lexicographically, and filtering out duplicates. The mathematical operations for this process are as follows:

\textbf{Notations:}
\begin{itemize}
    \item \( L \): Initial set of links (edges) between waypoints.
    \item \( L_{unique} \): Final set of unique links between waypoints.
\end{itemize}

\textbf{Organizing Edge Columns}:
First, reorganize each edge such that the smaller node index appears first. This is done by comparing the node indices in each edge and swapping them if necessary:
\[
\text{for each } (i, j, d) \in L, \quad (i', j') = 
\begin{cases} 
(i, j) & \text{if } i < j \\
(j, i) & \text{if } i > j
\end{cases}
\]
where \( d \) is the distance between nodes \( i \) and \( j \), and \( (i', j') \) are the reorganized node indices.

\textbf{Lexicographical Sorting}:
Sort the edges lexicographically based on the organized node indices. This can be represented as:
\[
L_{sorted} = \text{sort\_lexicographically}(L)
\]
where \( L_{sorted} \) is the sorted set of edges.

\textbf{Filtering Unique Edges}:
Finally, filter out duplicate edges to retain only unique connections. This involves comparing adjacent edges in the sorted list and selecting edges that are not duplicates:
\[
L_{unique} = \text{unique}(L_{sorted})
\]
where \( L_{unique} \) represents the set of unique edges.

\subsubsection{Link Generation Visualization}
\begin{figure}[htp]
  \centering
  \includegraphics[width=\linewidth]{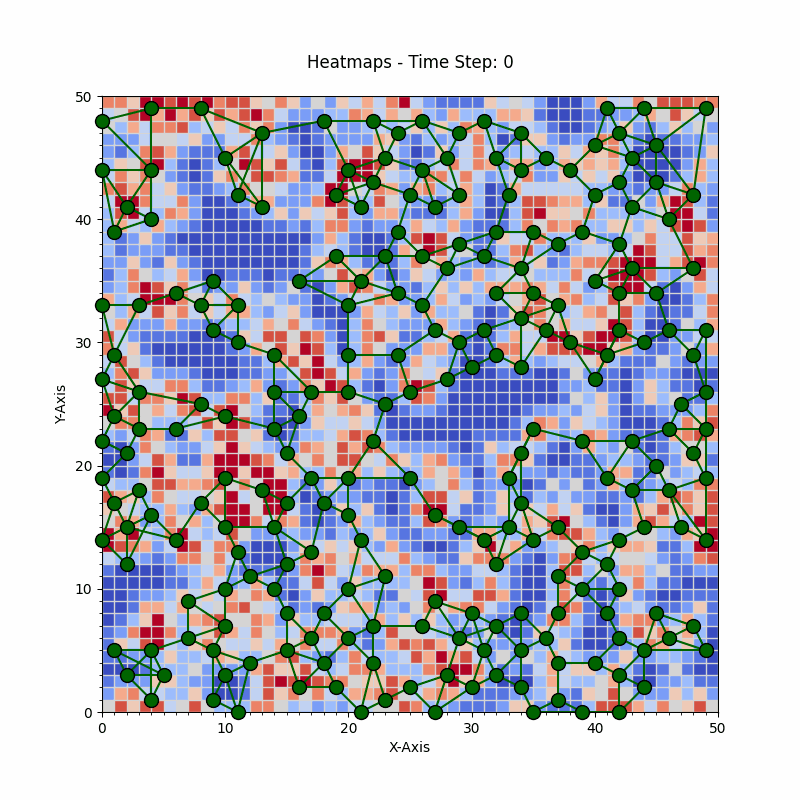} 
  \caption{Waypoint Links with PREP mapper}
  \label{fig:links-prep-mapper}
\end{figure}

This Figure \ref{fig:links-prep-mapper} illustrates link generation, building upon the waypoint nodes. Edges are formed between waypoint nodes based on proximity-based thresholds. Waypoints within a defined radius of each other are connected by edges. These edges serve as potential navigation routes for the mobile sensors, ensuring connectivity within the network. The goal is to optimize the generation of edges to ensure effective routes that enable comprehensive coverage by the mobile sensors.

\clearpage

\section{Phase 2: TED Predictor}

\paragraph{Overview}
The TED (Temporal Event Dynamics) Predictor represents a pivotal element in the ROBUST Network, tasked with the critical function of determining active waypoints in dynamic environments. It operates by analyzing spatiotemporal data, primarily focusing on the intensity and distribution of events as captured by heatmaps. This phase is central to the network's adaptability, enabling it to respond effectively to changing conditions and requirements.

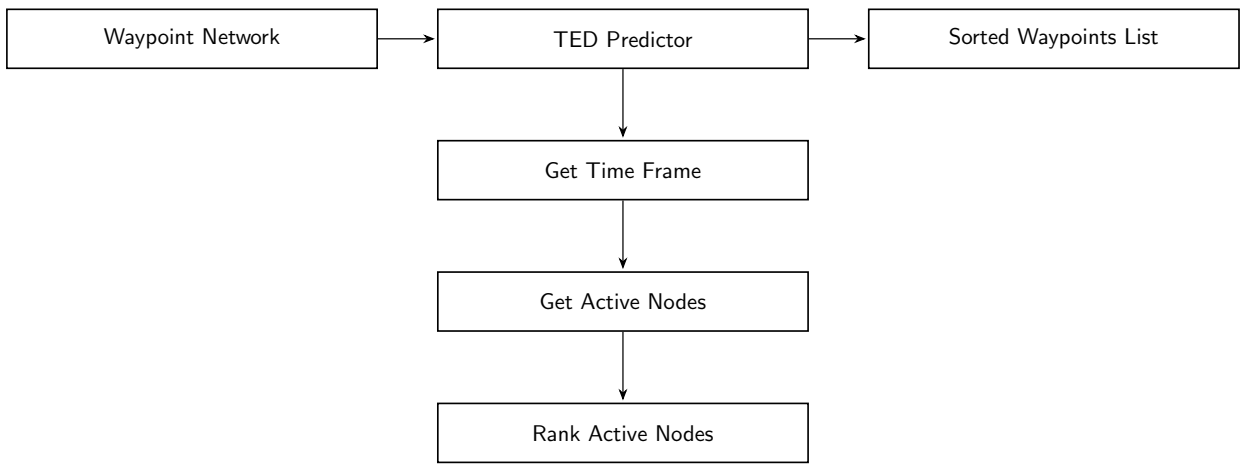
\begin{figure}[htbp]
    \centering
    \resizebox{\textwidth}{!}{%
    \begin{tikzpicture}[
        node distance=1.2cm and 1cm,
        auto,
        thick,
        main/.style={rectangle, draw, text width=6cm, minimum height=1cm, align=center, font=\sffamily\small},
        to/.style={-Stealth, shorten >=1pt, semithick, font=\sffamily\footnotesize}
    ]
    
        \node[main] (waypoint) {Waypoint Network};
        \node[main] (ted) [right=of waypoint] {TED Predictor};
        \node[main] (sorted) [right=of ted] {Sorted Waypoints List};
        \node[main] (timeframe) [below=of ted] {Get Time Frame};
        \node[main] (activenodes) [below=of timeframe] {Get Active Nodes};
        \node[main] (ranknodes) [below=of activenodes] {Rank Active Nodes};
    
        \draw[to] (waypoint) -- (ted);
        \draw[to] (ted) -- (sorted);
        \draw[to] (ted) -- (timeframe);
        \draw[to] (timeframe) -- (activenodes);
        \draw[to] (activenodes) -- (ranknodes);
    
    \end{tikzpicture}
    }
    \caption{Process flow from Waypoint Network to Ranked Active Nodes.}
    \label{fig:process-flow}
\end{figure}

\subsection{Perfect Knowledge vs Imperfect Knowledge}

The TED Predictor operates under two distinct scenarios: Perfect Knowledge and Imperfect Knowledge, each with its implications for the network's functionality and accuracy.

\subsubsection{Perfect Knowledge Scenario}
In Perfect Knowledge, active waypoints are identified directly from heatmaps, eliminating forecasting errors. This process involves analyzing the concentration and positioning of high-intensity events in the heatmaps.

\subsubsection{Imperfect Knowledge Scenario}
Imperfect Knowledge involves making predictions about event occurrences, potentially leading to discrepancies with actual data. This is significant in real-world scenarios where future events are forecasted based on historical or probabilistic data.

\subsubsection{Error Computation}
Error measurement quantifies the discrepancy between outcomes in Perfect and Imperfect Knowledge scenarios. This evaluates the precision of predictions and guides future planning strategies.

\paragraph{Mathematical Expressions}
\begin{itemize}
    \item \textbf{Heatmap Data Representation:} $H_t$: Heatmap matrix at time $t$, showing event intensities. $\theta$: Threshold for significant event intensity.
    \item \textbf{Identifying Active Waypoints:} For each timeframe $t$, active waypoints are identified using $H_t$ and densest clusters. $A_t$: Set of active waypoints at time $t$.
    \[ A_t = \text{get\_active\_nodes\_using\_clusters}(H_t, \text{Densest\_Clusters}, \theta) \]
    \item \textbf{Sorting Active Waypoints:} Waypoints are sorted based on the number of unique high-intensity events they observe. $A_{t, \text{sorted}}$: Organized array of active waypoints at time $t$.
    \item \textbf{Perfect vs Imperfect Knowledge:} Compares active waypoints determined under Perfect Knowledge with forecasts under Imperfect Knowledge. Error: Variation in number or location of active waypoints between scenarios.
    \[ \text{Error} = |A_{t, \text{perfect}} - A_{t, \text{imperfect}}| \]
\end{itemize}

\subsubsection{Analytical Breakdown of Waypoint Activation Process}
\begin{itemize}
    \item \textbf{Event Intensity Thresholding:} Implementing a threshold to the heatmap to distinguish significant events.
    \[ M_{ij} = \begin{cases} 1 & \text{if } H_{t,ij} \geq \theta \\ 0 & \text{otherwise} \end{cases} \]
    \item \textbf{Activation Assessment for Waypoints:} Evaluating whether waypoints cover high-intensity events and marking them as active for the corresponding timeframe.
    \item \textbf{Prioritization of Waypoints:} Waypoints are ranked based on the count of unique events they cover, focusing on those with a wider event coverage.
\end{itemize}

\paragraph{Concluding Note}
Errors in the prediction phase can lead to 'regret' in the planning phase. This regret represents the discrepancy in effectiveness of sensor placements evaluated against the outcomes from the prediction phase. This understanding is vital for enhancing both prediction and planning approaches within the ROBUST Network.

This section details the TED Predictor's methodology, aligning with the principles of selecting and appraising active waypoints in a comprehensive and mathematical manner.

\subsubsection{Visualization of TED Predictors}
This figure depicts the activity of a waypoint network across two timesteps. The left panel shows the network at timestep 0 (\textit{Heatmaps-Timestep: 0}) and the right panel shows the network at timestep 1 (\textit{Heatmaps-Timestep: 1}).  The waypoints represent potential locations for mobile sensors, and their colors indicate their activity levels.

\paragraph{Node Colors:}
\begin{itemize}
    \item \textbf{Gray:} Inactive waypoints at the current timestep.
    \item \textbf{Lime Green:} Active waypoints at the current timestep.
    \item \textbf{Gold:} Highly active waypoints, in range of the most significant event at the current timestep (Top 5).
\end{itemize}
    
\begin{figure}[htbp]
    \centering
    \vspace{-0.5cm}
    \includegraphics[width=0.7\textwidth]{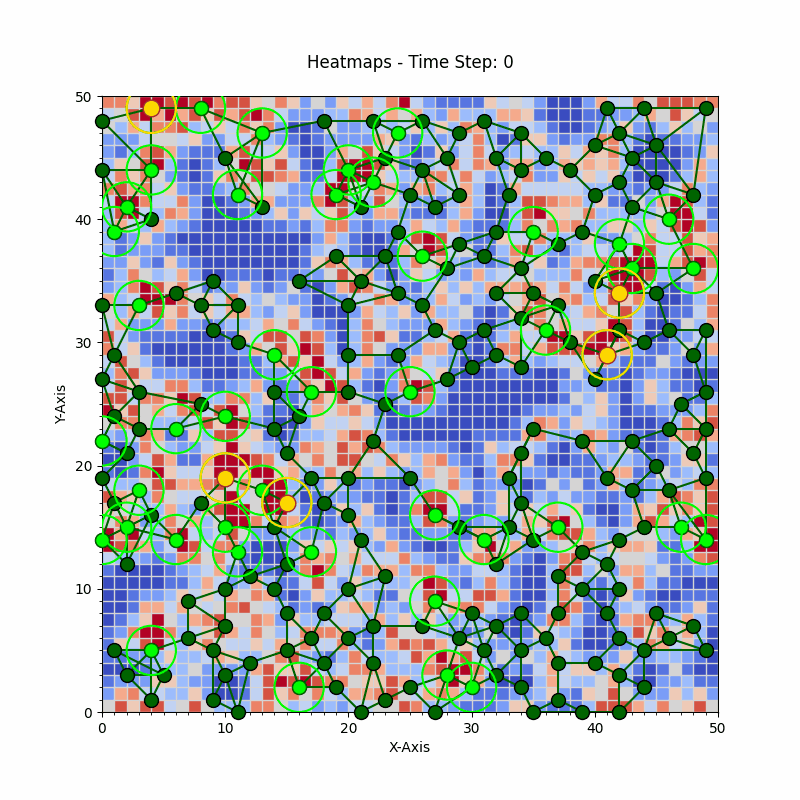}

    \vspace{-1.25cm}

    \includegraphics[width=0.7\textwidth]{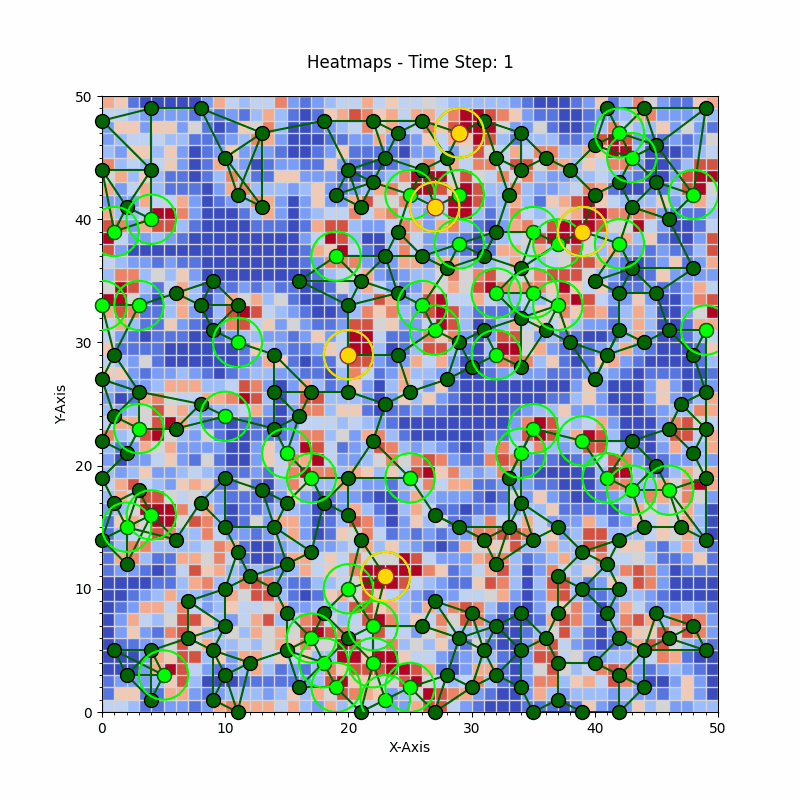} 
    \vspace{-0.6cm}
    \caption{TED Predictor results over two timesteps. Top: time 0 showing best nodes (gold) and active nodes (lime) with perfect knowledge. Bottom: time 1 showing best nodes (gold) and active nodes (lime) with perfect knowledge.}
    \label{fig:ted-predictors}
\end{figure}

\paragraph{Top Panel (Timestep 0):}
The Top panel \ref{fig:ted-predictors} displays the initial state of the waypoint network. It shows the distribution of active and inactive waypoints at timestep 0, along with any highly active waypoints in proximity to the top 5 most significant events at that timestep.

\paragraph{Bottom Panel (Timestep 1):}
The bottom panel \ref{fig:ted-predictors} illustrates the evolution of waypoint activity between timestep 0 and timestep 1. The colors of the waypoints reflect their activity levels at timestep 1, again indicating inactive, active, and highly active waypoints based on their proximity to the top 5 most significant events at that timestep.

By comparing the two panels, we can observe the changes in waypoint activity over time and identify how the network adapts to the evolving distribution of significant events.

\clearpage

\section{Phase 3: WAITR Planner (Perfect Knowledge) }

\subsubsection{Overview}
The WAITR (Weighted Aggregate Inter-Temporal Reward) Planner is the final phase in the ROBUST Network for mobile sensors. It focuses on calculating optimal sensor paths and positions based on event dynamics and network constraints.

\vspace{8mm}

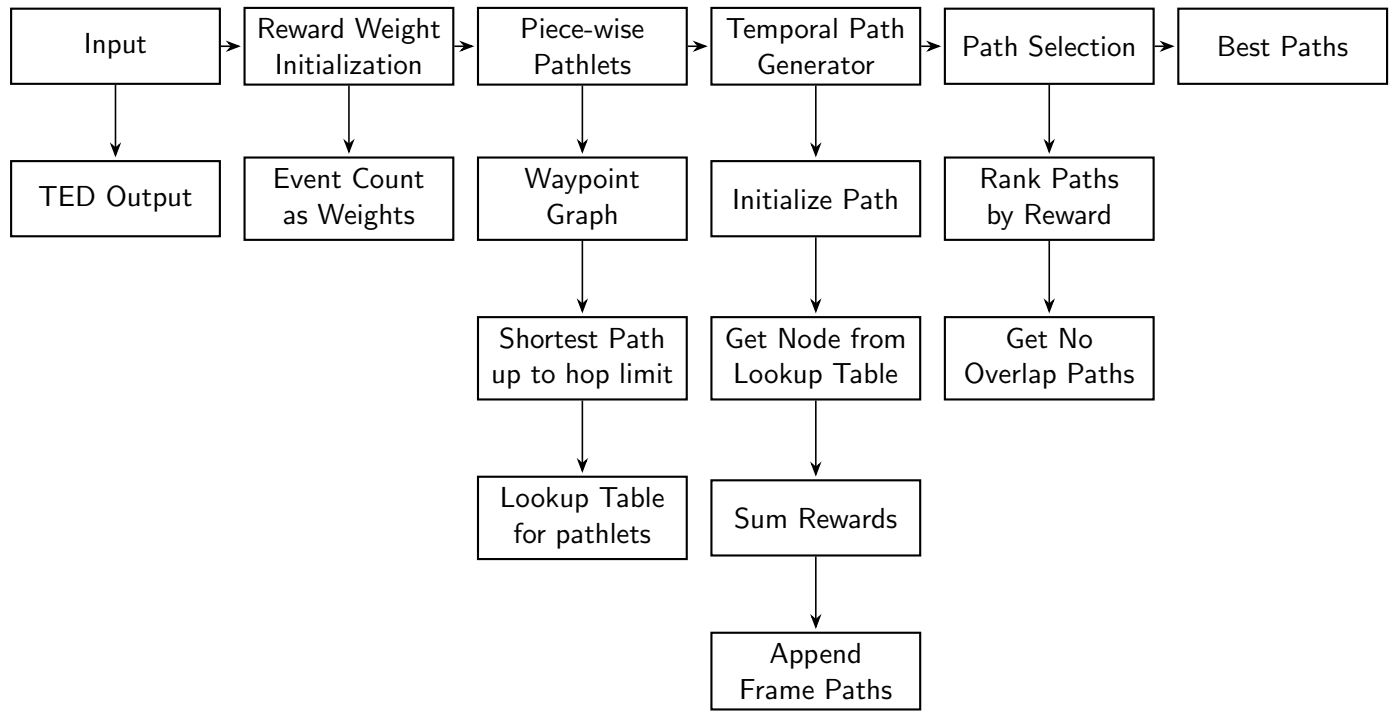
\begin{figure}[htbp]
    \centering
    \begin{tikzpicture}[
        node distance=1cm and 0.3cm,
        auto,
        thick,
        block/.style={rectangle, draw, text width=2.5cm, minimum height=1cm, align=center, font=\sffamily\small},
        line/.style={-Stealth, shorten >=1pt, semithick, font=\sffamily\footnotesize}
    ]

        \node[block] (input) {Input};
        \node[block, right=of input] (rewardWeight) {Reward Weight Initialization};
        \node[block, right=of rewardWeight] (pathlet) {Piece-wise Pathlets};
        \node[block, right=of pathlet] (timepaths) {Temporal Path Generator};
        \node[block, right=of timepaths] (pathselect) {Path Selection};
        \node[block, right=of pathselect] (toppaths) {Best Paths};
        
        \node[block, below=of input] (tedOutput) {TED Output};
        \node[block, right=of tedOutput] (eventCount) {Event Count as Weights};
        \node[block, right=of eventCount] (waypointGraph) {Waypoint Graph};
        \node[block, right=of waypointGraph] (initPath) {Initialize Path};
        \node[block, right=of initPath] (rankPath) {Rank Paths by Reward};

        \node[block, below=of waypointGraph] (shortPath) {Shortest Path up to hop limit};
        \node[block, right=of shortPath] (getNode) {Get Node from Lookup Table};
        \node[block, right=of getNode] (selecttop) {Get No Overlap Paths};

        \node[block, below=of shortPath] (lookupTable) {Lookup Table for pathlets};
        \node[block, right=of lookupTable] (sumRewards) {Sum Rewards};

        \node[block, below=of sumRewards] (appendPaths) {Append Frame Paths};

        \draw[line] (input) -- (rewardWeight);
        \draw[line] (rewardWeight) -- (pathlet);
        \draw[line] (pathlet) -- (timepaths);
        \draw[line] (timepaths) -- (pathselect);
        \draw[line] (pathselect) -- (toppaths);

        \draw[line] (input) -- (tedOutput);
        \draw[line] (rewardWeight) -- (eventCount);
        \draw[line] (pathlet) -- (waypointGraph);
        \draw[line] (timepaths) -- (initPath);
        \draw[line] (pathselect) -- (rankPath);

        \draw[line] (waypointGraph) -- (shortPath);
        \draw[line] (initPath) -- (getNode);
        \draw[line] (rankPath) -- (selecttop);

        \draw[line] (shortPath) -- (lookupTable);
        \draw[line] (getNode) -- (sumRewards);

        \draw[line] (sumRewards) -- (appendPaths);

    \end{tikzpicture}
    \caption{Flow chart representation of the process from input to best paths selection.}
    \label{fig:waitr-process-flow}
\end{figure}

\subsection{Weight Initialization}

In the weight initialization process, we assign weights to waypoint nodes based on the event counts. The weights are indicative of the number of events covered by each waypoint node at different timeframes.

\subsubsection{Mathematical Representation}
Let's define the following notations:
\begin{itemize}
    \item \( W \): A matrix representing the weights of waypoint nodes across different timeframes.
    \item \( E \): A matrix representing the event counts for each waypoint node across different timeframes, with missing counts represented as null or a specific placeholder (like NaN).
    \item \( N \): The number of waypoint nodes.
    \item \( T \): The number of timeframes.
\end{itemize}

The initialization of node weights can be mathematically expressed as:

\[
W_{t, n} = 
\begin{cases}
E_{t, n} & \text{if an event count is available for node } n \text{ at timeframe } t,\\
0 & \text{otherwise (if event count is missing or null).}
\end{cases}
\]

for each \( t \) from 1 to \( T \) and each \( n \) from 1 to \( N \).

\vspace{12pt}
In this expression, \( W_{t, n} \) denotes the weight assigned to the \( n \)-th waypoint node at the \( t \)-th timeframe. If \( E_{t, n} \) is available (i.e., not null or missing), \( W_{t, n} \) is set to \( E_{t, n} \). Otherwise, \( W_{t, n} \) is set to zero. This ensures that all waypoint nodes have a defined weight, either based on actual event counts or defaulting to zero where data is unavailable.

\subsection{Generate Piecewise Pathlets}

The generation of piecewise pathlets involves constructing a graph-based network of potential routes or 'pathlets' between nodes, taking into account a maximum number of hops. This methodology is integral to efficient sensor movement planning in the WAITR Planner.

\subsubsection{Mathematical Representation using Graph Theory}
Let's define the following notations:
\begin{itemize}
    \item \( G(V, E) \): A graph where \( V \) represents the set of nodes (waypoint nodes) and \( E \) represents the set of edges (unique links between nodes).
    \item \( H_{\text{max}} \): The maximum number of hops allowed for paths between nodes.
    \item \( P \): A lookup table for storing the shortest pathlets between nodes in \( G \).
\end{itemize}

\paragraph{}
The process of generating piecewise pathlets can be described as follows:

\begin{enumerate}
    \item \textbf{Graph Construction}: Formulate a graph \( G(V, E) \) using the waypoints as nodes \( V \) and the unique links as edges \( E \).

    \item \textbf{Pathlet Calculation}: For each node \( v \in V \), calculate the shortest pathlets to other nodes within \( H_{\text{max}} \) hops using an algorithm like Dijkstra's.

    \item \textbf{Lookup Table for Pathlets}: Populate the lookup table \( P \) with these pathlets, where \( P[v] \) contains all the pathlets originating from node \( v \) within the hop limit \( H_{\text{max}} \).
    
    \[
        P[v] = \{ \text{pathlets from node } v \text{ to other nodes within } H_{\text{max}} \text{ hops in } G \}
    \]

\end{enumerate}

\subsection{Temporal Path Generation}

The Temporal Path Generation process in the WAITR Planner is crucial for extending paths across different timeframes while accumulating rewards. This subsection will mathematically detail this process.

\paragraph{Notations:}
\begin{itemize}
    \item $G(V, E)$: Graph representing the network of waypoint nodes.
    \item $P$: Lookup table for storing paths between nodes in $G$.
    \item $W$: Matrix of node weights (rewards) across different timeframes.
    \item $T$: Number of timeframes.
    \item $R$: Reward value associated with each node.
    \item $NP$: New paths generated across timeframes.
    \item $OP$: Optimal paths selected considering the number of sensors.
\end{itemize}

\subsubsection{Temporal Path Append}
\begin{enumerate}
    \item \textbf{Initialize Current Paths}: At the start, for each node $v$ at timeframe $t_0$, create initial paths:
       \[ NP[v, 0] = [v] \]
    \item \textbf{Extend Paths Across Timeframes}: For each subsequent timeframe $t$, for each node $v$ in $G$:
       \begin{itemize}
           \item Retrieve possible next nodes from $P[v]$.
           \item Calculate the new accumulated weight (reward) for reaching each next node.
           \item Update $NP$ by appending these next nodes and their corresponding accumulated rewards.
       \end{itemize}
       Represented as:
       \[ NP[v, t] = \bigcup\{ (next\_v, R[next\_v] + NP[v, t-1]) : \text{for } next\_v \in P[v] \} \]
\end{enumerate}

\subsection{Path Selection}

The Path Selection process in the WAITR Planner is critical for choosing the most rewarding paths. This subsection will mathematically detail this process.

\paragraph{Notations:}
\begin{itemize}
    \item $TS$: Temporal scores, a set of tuples (weight, path) representing the reward and corresponding path.
    \item $N_S$: Number of sensors.
    \item $OP$: Optimal paths selected based on rewards and constraints.
    \item $UN$: Set of used nodes to avoid overlaps in path selection.
\end{itemize}

\subsubsection{Optimal Path Selection}
\begin{enumerate}
    \item \textbf{Flatten Temporal Scores}: Convert $TS$ into a list of tuples (weight, path) and sort them by weight in descending order.
        \[ \text{All\_Paths} = \text{sort}(\{(w, p) : (w, p) \in TS\}, \text{by } w, \text{descending}) \]
    \item \textbf{Select Paths}: Iterate through the sorted paths and select paths avoiding overlaps:
        \begin{itemize}
            \item Initialize $OP$ as an empty set and $UN$ as an empty set.
            \item For each path $p$ in $\text{All\_Paths}$:
                \begin{enumerate}
                    \item Check for overlap: if any node in $p$ is already in $UN$, skip to the next path.
                    \item Otherwise, add $p$ to $OP$ and add its nodes to $UN$.
                    \item If the size of $OP$ equals $N_S$, break the loop.
                \end{enumerate}
        \end{itemize}
        Expressed as:
        \[ OP = \{ p : p \in \text{All\_Paths}, \text{no overlap with } UN, \text{up to } N_S \text{ paths} \} \]
\end{enumerate}

\subsubsection{WAITR Visualization}
\begin{figure}[htbp]
    \centering
    \vspace{-0.5cm} 
    
    \includegraphics[width=0.7\textwidth]{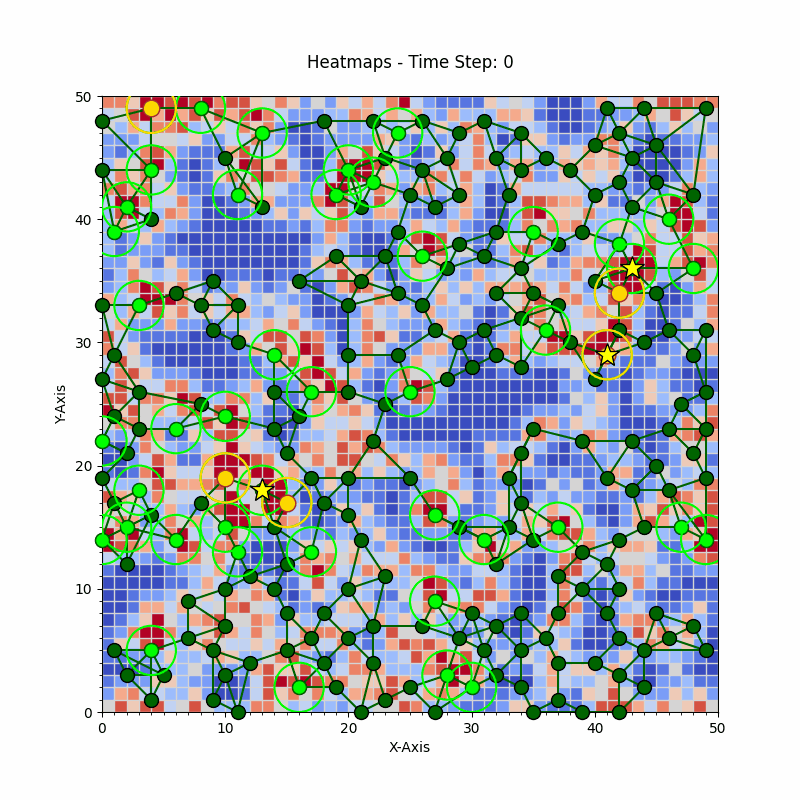}
    \label{fig:waitr-planner-t0}
    
    \vspace{-1.25cm} 
    
    \includegraphics[width=0.7\textwidth]{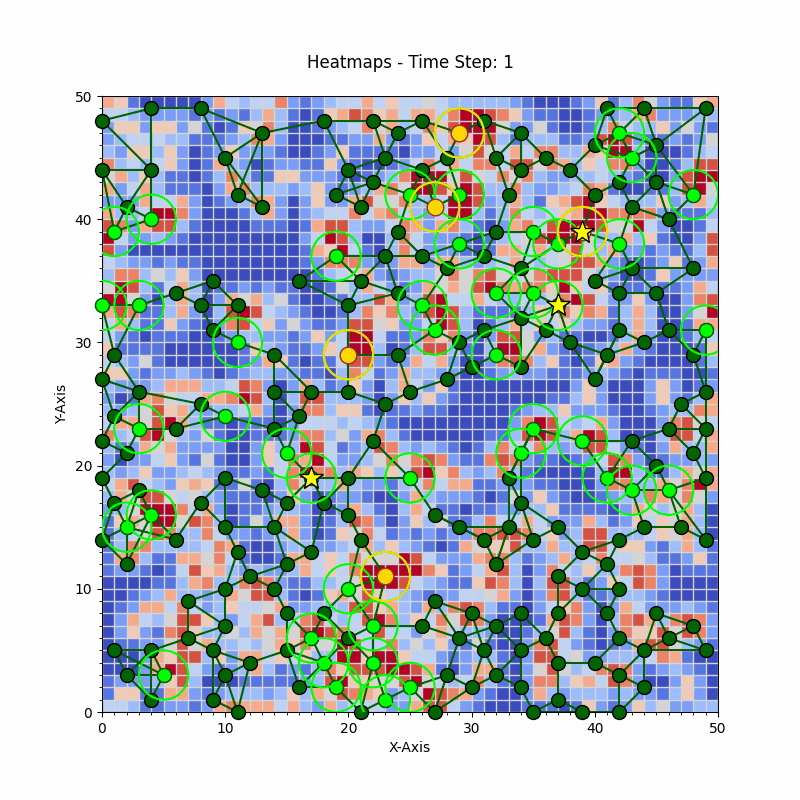}
    \label{fig:waitr-planner-t1}
    \vspace{-0.5cm} 
    \caption{Sensor placements in WAITR Planner over two frames. Top: time 0. Bottom: time 1. Yellow stars are sensors with 2-hop movements. Best nodes in gold, Active nodes in lime, Inactive in gray.}
    \label{fig:waitr-timesteps}
\end{figure}

This figure consists of two panels, side-by-side, that illustrate the placement of mobile sensors using WAITR at two distinct timesteps. The top panel (Figure~\ref{fig:waitr-timesteps}) depicts the initial sensor configuration at timestep 0, while the bottom panel (Figure~\ref{fig:waitr-timesteps}) shows the sensor arrangement at timestep 1.

\paragraph{Sensor Representation:}
\begin{itemize}
  \item \textbf{Yellow Stars:} These yellow stars represent the mobile sensors deployed in the simulation environment. Their placement adheres to specific constraints, such as a maximum movement range of two hops (denoting the number of neighboring waypoints a sensor can move to in a single step).
\end{itemize}

\paragraph{Node Colors:}
\begin{itemize}
    \item \textbf{Gray:} Inactive waypoints at the current timestep.
    \item \textbf{Lime Green:} Active waypoints at the current timestep.
    \item \textbf{Gold:} Highly active waypoints, in range of the most significant event at the current timestep (Top 5).
\end{itemize}

\paragraph{Timestep Differentiation:}

The key takeaway from this visualization is the difference in sensor placement between the two timesteps. By comparing the left and right panels, we can observe how the WAITR system strategically repositions its mobile sensors across the environment to potentially gather more valuable information or adapt to changing conditions.

Overall, this figure serves as a visual representation of the dynamic sensor planning process within the WAITR simulation.

\clearpage

\section{Phase 3: WAITR Planner (Imperfect Knowledge)}

\paragraph{Overview}
The WAITR (Weighted Aggregate Inter-Temporal Reward) Planner under Imperfect Knowledge involves complex decision-making processes, incorporating estimations and predictions about future events. This phase adapts to the inherent uncertainty of dynamic environments, aiming to optimize sensor paths and positions in the face of incomplete or uncertain data.

\subsection{Conceptualizing Reward in Observer Node Dynamics}
\begin{itemize}
    \item Under Imperfect Knowledge, the concept of 'reward' in observer node dynamics takes on a probabilistic aspect, factoring in the likelihood of event occurrences and their potential impact.
    \item This involves estimating the probability of events at various nodes and adjusting reward calculations accordingly.
\end{itemize}

\subsection{Conceptualizing Regret in Observer Node Dynamics}
\begin{itemize}
    \item 'Regret' in this context refers to the measure of missed opportunities or suboptimal decisions due to imperfect predictions.
    \item This involves assessing the difference between the outcomes of decisions made under uncertainty and those that would have been made with perfect information.
\end{itemize}

\subsection{Toward Spatiotemporal Regret Assessment}
\begin{itemize}
    \item The assessment of regret in spatiotemporal dynamics includes evaluating the temporal and spatial aspects of decision-making under uncertainty.
    \item This involves mapping regret over time and space, to understand where and when suboptimal decisions are likely to occur.
\end{itemize}

\subsection{Risk Aversion in Observer Node Strategy}
\begin{itemize}
    \item Strategies under Imperfect Knowledge often incorporate elements of risk aversion, prioritizing decisions that minimize potential negative outcomes.
    \item This involves weighting decisions based on their associated risks and potential rewards, considering the uncertainty of predictions.
\end{itemize}

\subsection{Cost-Benefit Analysis}
\begin{itemize}
    \item Cost-benefit analysis in this phase weighs the potential benefits of certain sensor placements against their costs, including the risk of inaccurate predictions.
    \item This involves a thorough assessment of potential outcomes and their respective probabilities.
\end{itemize}

\subsection{Imperfect Knowledge Visualization}
This subsection presents a visualization from the benchmarking dataset that highlights the challenges associated with imperfect knowledge, data acquisition, and information decay in a scenario with mobile sensors. The image depicts a grid-like environment with twenty frames, representing different timesteps.

\subsubsection*{Imperfect Knowledge:}
\begin{itemize}
    \item \textbf{Limited View:} Three mobile sensors are depicted, each with a limited field of view represented by the colorful cells surrounding them. This limited view restricts the sensors' ability to perceive the entire environment at once, forcing them to explore and gather information incrementally. The white cells surrounding the colored areas represent the unknown portions of the environment that the sensors haven't explored yet.
    \item \textbf{Data Acquisition:} As the sensors move through the environment, they discover new information and update their understanding of the world. This process of uncovering the unknown (white cells) and populating them with colors signifies data acquisition.
\end{itemize}

\subsubsection*{Information Decay (Outdated Information):}
\begin{itemize}
    \item \textbf{Faded Cells:} The visualization also includes faded colored cells, which can be interpreted as outdated information. Since the environment might be dynamic, these faded cells could represent previously observed states that are no longer accurate. This highlights the challenge of information decay, where previously acquired knowledge may become irrelevant over time.
\end{itemize}

Overall, this visualization effectively showcases how imperfect knowledge, data acquisition, and information decay can impact the performance of mobile sensors in an unknown environment.

\subsubsection*{Key Points:}
\begin{itemize}
    \item The sensors have a limited view of the environment.
    \item Sensors actively acquire data by exploring and uncovering the unknown.
    \item Faded colors represent potentially outdated information.
\end{itemize}

\captionsetup[figure]{margin=0pt} 

\begin{adjustwidth}{-2cm}{-2cm}
  \centering
  \foreach \n in {0,...,18}{
    \begin{minipage}[t]{0.6\textwidth}
      \centering
      \includegraphics[width=\linewidth]{images/imperfect-knowledge/imperfect-knowledge-\n.png}
      \captionof*{figure}{Imperfect Knowledge - Frame \n}
    \end{minipage}
    \ifodd\n \par\bigskip\else\hspace*{\fill}\fi
  }
\end{adjustwidth}

\clearpage

\section{Scalability of the ROBUST Network}

\subsection{Introduction to Scalability in the ROBUST Network}

The design and operational efficiency of the ROBUST Network are critically hinged on its scalability. Scalability, in the context of this network, refers to the ability to efficiently adapt and perform under varying scales of operation, ranging from small, localized events to large-scale, complex scenarios. This chapter delves into the multifaceted aspects of scalability within the ROBUST Network, examining how its architecture and algorithms cater to diverse and expanding operational needs.

\subsection{Matrix and Vector-Based Operations}

A core aspect of the ROBUST Network's scalability lies in its extensive use of matrix and vector-based operations. These mathematical structures provide a robust framework for representing complex data and relationships in a compact and computationally efficient manner. By leveraging linear algebraic constructs, the network achieves a high degree of flexibility and efficiency in processing large datasets, which is essential for scalability.

\subsubsection{Efficient Data Handling}
Matrix and vector operations allow for efficient handling and manipulation of large volumes of data. This efficiency is pivotal in scenarios where the network has to process extensive spatiotemporal data, such as large-scale event heatmaps or intricate sensor network layouts. The ability to perform bulk operations on matrices and vectors significantly reduces computational overhead, enabling the network to scale up operations without a proportional increase in processing time or resource utilization.

\subsubsection{Algorithmic Scalability}
The algorithms employed in the ROBUST Network, including those for waypoint identification, path generation, and sensor placement, are inherently scalable due to their matrix and vector-based nature. These algorithms can seamlessly adapt to varying sizes of input data, maintaining their effectiveness and efficiency. Whether it's processing a small set of nodes or accommodating an extensive network of sensors and waypoints, the underlying matrix and vector operations ensure consistent performance.

\subsection{Dynamic Adaptation to Network Size}

The ROBUST Network's architecture is designed to dynamically adapt to the size of the operational environment. This adaptability is crucial in ensuring scalability, as it allows the network to maintain optimal performance irrespective of the scale of deployment.

\subsubsection{Modular Design}
The ROBUST Network's modular design is exemplified in its distinct phases: the Mapper, Predictor, and Planner. Each of these phases represents a core module with specific operational responsibilities, tailored to address different aspects of network functionality. 

- The \textbf{Mapper Module} is responsible for generating and updating the network's spatial layout, identifying key waypoints and establishing links between them based on real-time data.
- The \textbf{Predictor Module}, as the name suggests, focuses on forecasting and determining active waypoints, using both Perfect and Imperfect Knowledge scenarios to enhance decision-making accuracy.
- The \textbf{Planner Module} is tasked with the strategic allocation of resources and sensors based on the insights provided by the Mapper and Predictor modules.

This division into distinct yet interdependent modules allows each part of the network to be scaled and adapted independently, according to the evolving needs and challenges of the operational environment. Such a modular approach not only augments the network's flexibility but also ensures that resources are allocated efficiently, optimizing overall performance across various scales of operation.

\subsection{Conclusion}

In conclusion, the scalability of the ROBUST Network is a fundamental attribute that underpins its effectiveness across diverse operational contexts. The network's design, leveraging matrix and vector-based operations, equips it with dynamic adaptation capabilities and an efficient approach to managing large-scale scenarios. This scalability is especially critical in temporal-based problems, where the sequential nature of time can lead to a combinatorial explosion in possible outcomes.

\section{ROBUST and Continuous Sampling}

\subsection{Introduction}
This chapter delves into the adaptation of the ROBUST Network to accommodate continuous dynamic sensor placements. In contrast to previous models that focused on the weight from the last node, this approach takes into account the cumulative weight of entire path segments. This is particularly relevant in scenarios where sensors have the capacity to traverse more than one position per timeframe.

\subsection{Phase 3: WAITR Planner (Perfect Knowledge)}

\paragraph{Overview}
Under the paradigm of Perfect Knowledge, the WAITR (Weighted Aggregate Inter-Temporal Reward) Planner plays a critical role in the network. Its goal is to compute optimal sensor paths and positions based on the total reward of entire path segments.

\begin{enumerate}
    \item Reward Weights Initialization
    \item Generation of Piecewise Pathlets
    \item Temporal Path Generation with Path Segment Weights
    \item Selection of Optimal Paths Considering Path Segment Rewards
\end{enumerate}

\subsection{Weight Initialization}
This process involves assigning weights to waypoint nodes based on the event counts at different timeframes, similar to the previous method.

\subsubsection{Generation of Piecewise Pathlets}
Constructing a graph-based network of `pathlets' between nodes, this step considers the maximum number of hops possible, which is essential for planning sensor movements.

\subsubsection{Temporal Path Generation with Path Segment Weights}
This section deviates from the previous approach by extending paths across timeframes and accumulating the rewards for entire path segments.
\subsubsection{Temporal Path Append with Path Segment Weights}
In this approach, paths are extended across timeframes by considering the total rewards of path segments, rather than just focusing on the last node.

\subsection{Path Selections Considering Path Segment Rewards}
During the Path selection stage this adaptation is consistent with the previous approach as it gives priority to paths based on their accumulated rewards across all nodes traversed, ensuring efficient sensor coverage and utility.

 \chapter{Case Study I: Oceanographic Monitoring}
\label{chap:oceanography}

\section{Introduction}
The Gulf of Mexico (GoM) is environmentally and economically vital to the US. Its coastline extends across five U.S. states: Texas, Louisiana, Mississippi, Alabama, and Florida. It hosts multiple major ports and transportation waterways which provide the US with many critical resources: oil, gas, wind, waves, and seafood \cite{b2}. A diverse group of commercial, academic, federal, and local organizations jointly support operations in the GoM to observe, measure, and study the region. However, the GoM is vast, with over 17,000 miles of shoreline, and its basin encompasses 600,000 square miles.  Despite the concerted effort between organizations to build a shared sensor array, the current number of sensors only observes a sparse fraction of the GoM \cite{b1}. It is critical to supply these institutions with guidance on where optimal new sensor placements may go to best contribute to the sensor array. When considering that the GoM continuously changes states, this problem becomes even more challenging.  Unlike land terrain which remains relatively stable, water bodies are dynamic systems \cite{b5}. Network models and analysis provide key insights into where to place new GCOOS sensors.   

\section{Background}
Various ongoing initiatives are engaged in monitoring and reporting both historical and real-time states of the GoM. This paper focuses on two: The Gulf of Mexico Coastal Ocean Observing System (GCOOS) and HYbrid Coordinate Ocean Model (HYCOM). 

\subsection{GCOOS}
The GCOOS is the Gulf of Mexico regional component of the U.S. Integrated Ocean Observing System (IOOS). It is the only certified comprehensive data collection and dissemination center for coastal and ocean data in the Gulf. GCOOS collects data from 1,655 sensors located at 163 non-federal and 159 federal stations \cite{b1}.

\subsection{HYCOM}
HYCOM is a real-time three-dimensional grid mesh ocean model with 1/25° horizontal resolution that provides eddy-resolving hindcast, nowcast, and forecast as numerical states of the GoM.  HYCOM assimilates data from various sensors, including satellites, buoys, ARGO floats, and autonomous underwater gliders (AUGs).  The forecast system is the Navy Coupled Ocean Data Assimilation (NCODA); a multivariate optimal interpolation scheme that assimilates surface observations. By combining these observations via data assimilation and using the dynamical interpolation skill of the model, a three-dimensional ocean state can be more accurately nowcast and forecast \cite{b4}.

\section{Objective}
This research aims to construct a ROBUST network using HYCOM and the observational sensor data from GCOOS. This ROBUST network aims to identify regions of interest within the HYCOM model to recommend how best to utilize the sensor array of GCOOS and provide guidance on where to expand it.

\section{Motivation}
Localized regions of temporal variability within HYCOM hinder the accuracy of its nowcast/forecast. A region of temporal variability occurs where significant changes in a numerical property exist within the same coordinate between two consecutive temporal frames\cite{Holmberg2014}. HYCOM produces nowcasts and forecasts by combining its real-time observations and prior historical data.   The forecasting error rate generally increases as the values between snapshots differ \cite{Holmberg2014}. The best approach to mitigate such regions of temporal variability is to acquire new observations to feed into HYCOM \cite{Holmberg2014}.  The next set of nowcasts and forecasts will then use the most up-to-date measures and ensure the error rate is as minimal as possible. By placing instruments into the regions of interest (RoI), GCOOS can get the data needed to maximize the accuracy rate in the HYCOM nowcasting and forecasting model.  However, the number of sensors is limited and proper planning should maximize their effectiveness in improving the nowcasting and forecasting model. 

\section{Approach}
By modeling a ROBUST network composed of a set of observable nodes representing the RoI within the GoM and a set of observer nodes representing the GCOOS sensors. An RoI is identified by taking a set of temporal snapshots from HYCOM and computing the residuals over time, where the residual is the magnitude difference between snapshots. The nodes representing the GCOOS sensors have attributes consistent with that particular instrument, such as its operational status, geo-coordinates, current data readings, mobility speed, and institutional membership. The ROBUST network establishes a link between the sensor nodes to all potential nearby RoI nodes. It facilitates the decision-making to assess which location to recommend planning for new installations for sensors, for relocation, or when to perform maintenance. Other decisions involving the sensor array might be when and where to grow the sensor network. Monte Carlo simulations identify optimal sensor placements by attempting to add a GCOOS node to the ROBUST network randomly, and its effectiveness is then subsequently evaluated.

\section{Methods}
This is a stochastic problem, therefore there is no deterministic solution. Consequently, it is best to rely on random samplings to construct temporal graph representations by evaluating potential outcomes between GCOOS sensor placements and RoI positions. Graph analysis identifies and selects the optimal positions to maximize the GCOOS coverage and its coverage robustness.   

\subsection{Temporal Graph Representation}
ROBUST network is a type of temporal graph. A temporal network is an ordered set of static graphs. The ordering is the static network's temporal occurrence or "snapshot" at a particular timestamp.
\begin{equation}
TG = \Big( G_{t0}, \; G_{t1}, \; ..., \;  G_{tn} \Big) \label{tg-eq}
\end{equation}

where a Graph is a set of Nodes and a set of Edges.

\begin{equation}
G = \Big( N, \; E \Big) \label{g-eq}
\end{equation}

\subsection{ROBUST Network Model}
A geo-spatiotemporal network comprises a set of geo-spatiotemporal nodes and a set of geo-spatiotemporal edges. A geo-spatiotemporal node has a geographical longitude/latitude coordinate and may occur at select times or be persistent across all times. It may also move over time or remain stationary. A geo-spatiotemporal edge connects two nodes and may occur just once, at multiple times, or across all times. Geo-spatiotemporal edges have a numerical weight representing the geodesic distance between the linked nodes. This graph model assumes a bipartite network structure, ideal for mapping relations between two sets of nodes \cite{b9}.\\

\subsubsection{ROBUST Nodes} 
\hfill \break
Since this is a bipartite network, there are two types of nodes, observers and observables. Both types of nodes in this ROBUST network represent geospatial coordinate points within the GoM, but they differ in the following ways, as outlined in this section.  \\

\paragraph{GCOOS Sensors (static)} 
\hfill \break
GCOOS sensor nodes are observer types. GCOOS sensor nodes are modeled as static or stationary, which means their geolocation is persistent across all time frames. The properties of each GCOOS node are in Table 1.  

\begin{table}[htbp]
\caption{(GCOOS) Observer Node Properties} 
\begin{center}

\begin{tabular}{c l} 
 \hline\hline
 \textbf{Label} & \textbf{Description} \\ [0.5ex] 
 \hline\hline
 id & Unique identifier number for each node \\ 
 \hline
 membership & Federal asset or local data node (ldn) asset \\
 \hline
 data source & Institution that operates the GCOOS sensor \\
 \hline
 platform & Name of observatory platform \\
 \hline
 mobility & Stationary or mobile \\  
 \hline
  geolocation & Latitude and Longitude of platform \\  
 \hline
 operational status & Active or Inactive  \\  
 \hline
 observations$^{\mathrm{a}}$ & Types of measures sampled by this platform  \\ 
 \hline
\end{tabular}
\end{center}
\footnotesize{$^{\mathrm{a}}$Observations were limited to those used by the HYCOM forecasting model, which are temperature, salinity, and ocean current velocities. }
\end{table}

\paragraph{HYCOM RoI Events (temporal)}
\hfill \break
HYCOM RoI nodes are observable types. HYCOM RoI nodes represent locations between consecutive snapshots where a significant change occurred in an observation. The properties for each RoI node are in Table II and Table III.  The residual formula quantifies the significance of the change between snapshots.   

\begin{table}[htbp]
\caption{(HYCOM RoI) Event Node Properties } 
\begin{center}

\begin{tabular}{c l} 
 \hline\hline
 \textbf{Label} & \textbf{Description} \\ [0.5ex] 
 \hline\hline
id & Unique identifier number for each node \\ 
 \hline
 geolocation & Latitude and Longitude of platform \\ 
 \hline
 RoI snapshots & Temporal dictionary with corresponding ROI data \\ & \textit{(see: Table III)}
 \\ 
 \hline
\end{tabular}
\end{center}
\end{table}

\begin{table}[htbp]
\caption{ RoI Snapshot Properties } 
\begin{center}

\begin{tabular}{c l} 
 \hline\hline
 \textbf{Label} & \textbf{Description} \\ [0.5ex] 
 \hline\hline
 snapshots & Nested Dictionary,
 outer keys are timestamps, \\ & inner keys are HYCOM observations.
\\ & The value is the residual score (real number) 
 \\ 
 \hline
\end{tabular}
\end{center}
\end{table}

\hspace{10mm} \textit{i) Residual formula} \\
In this paper, a residual is the squared difference between a given value from the same geospatial coordinate at two different times. Squaring the difference serves dual purposes. The first purpose is to ensure that the residual is always positive between the two times. The second purpose is to boost or diminish the residual based on the magnitude of its difference. If the difference is less than 1, it is diminished; if it is greater than 1, it is boosted. See Fig. 1 for a visualization of the residuals between two frames.  

\begin{equation}
residual = \big( value_{t_{n+1}} - value_{t_n} \big)^2
\end{equation}

\begin{figure}[H]
\centering
\includegraphics[width=0.4\textwidth]{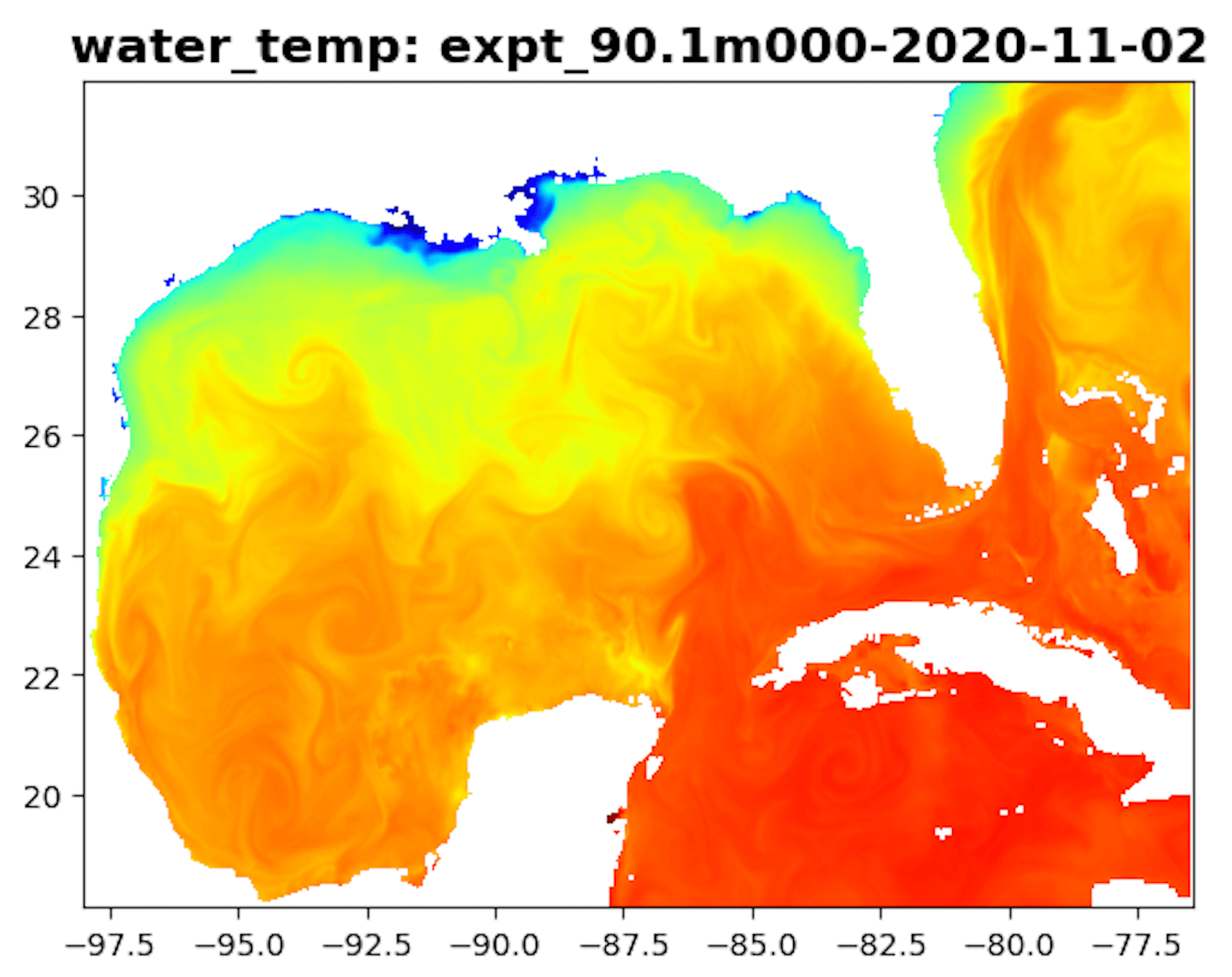}
\includegraphics[width=0.4\textwidth]{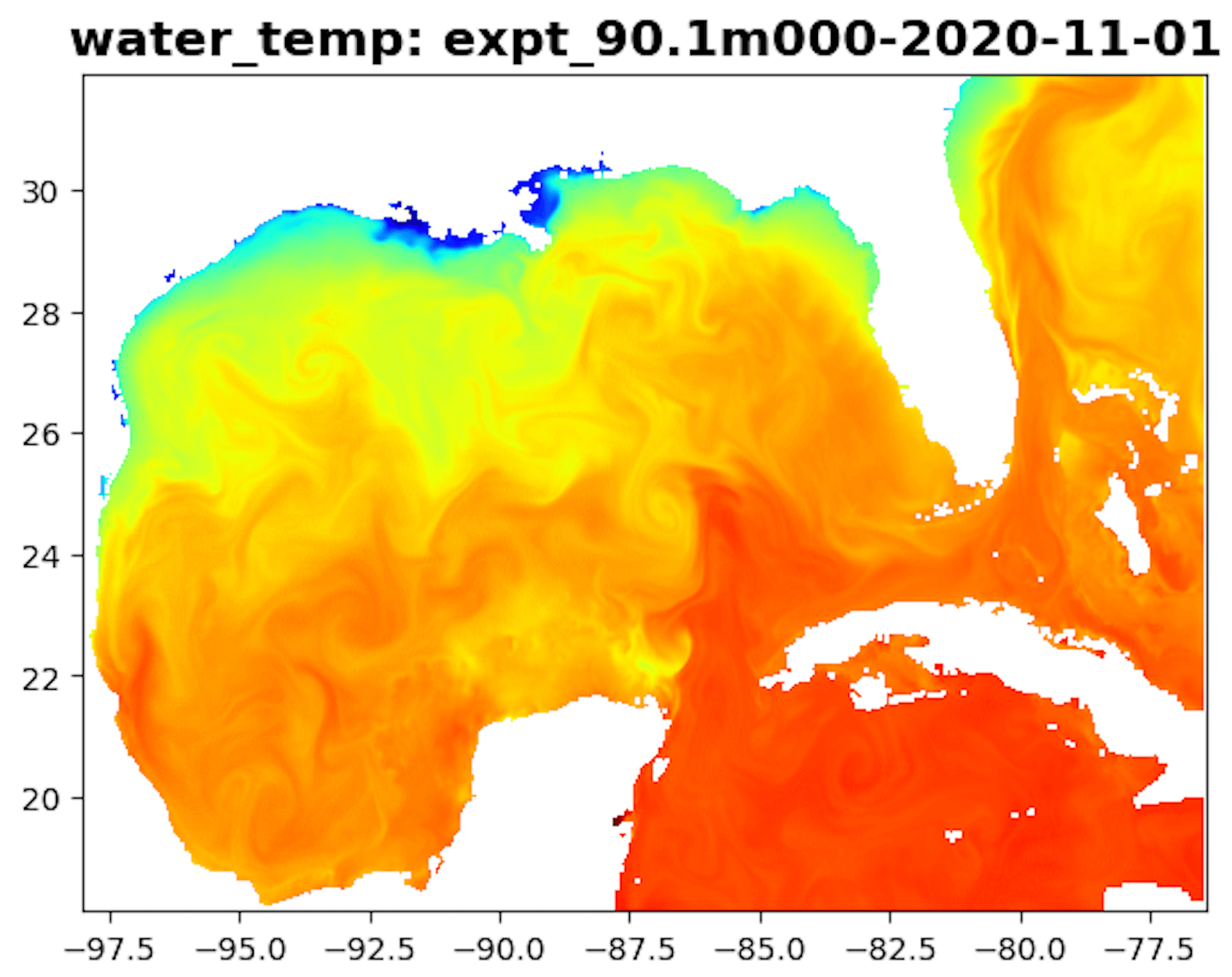}
\includegraphics[width=0.4\textwidth]{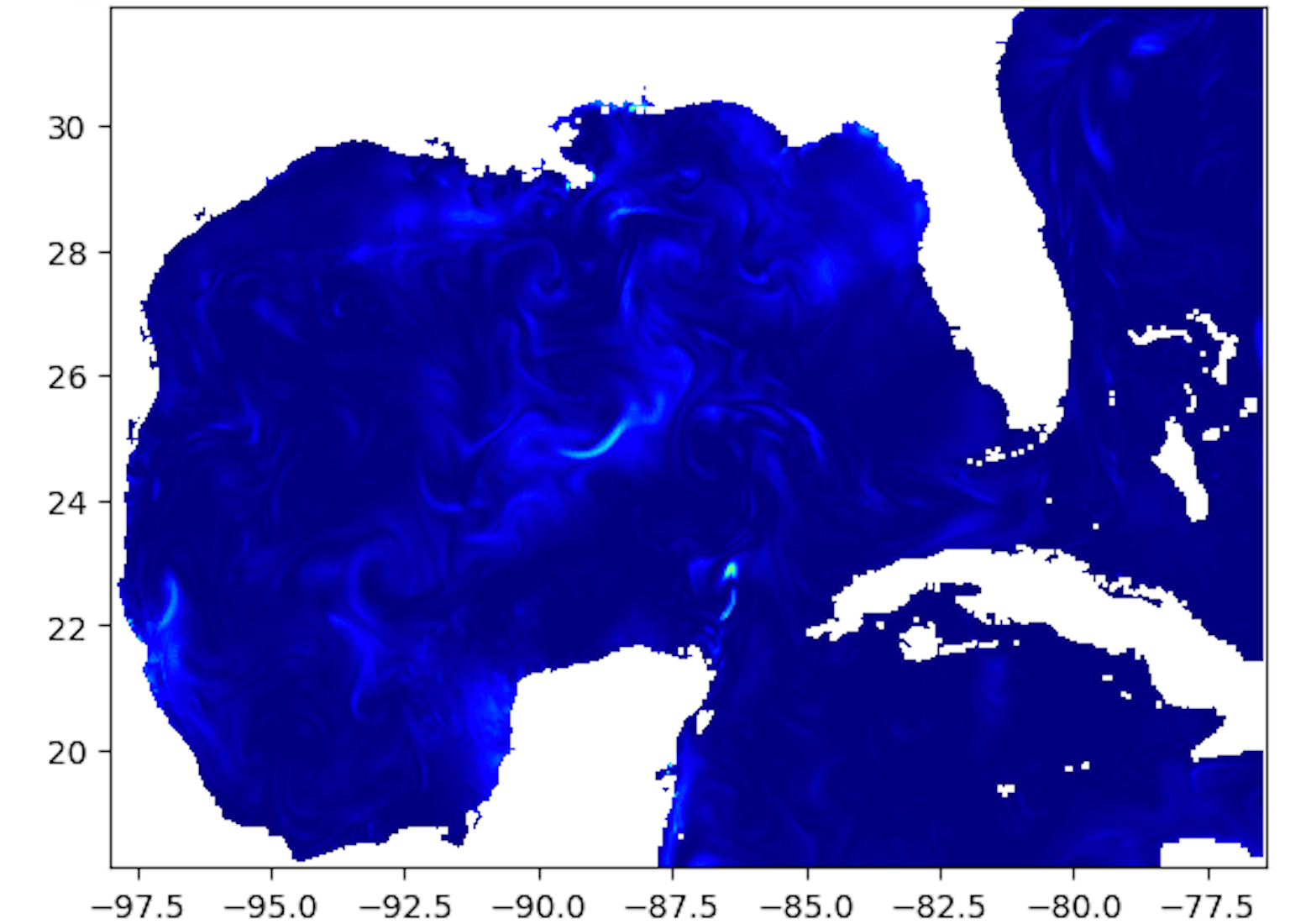}
\caption{Computing residuals between time frames}
\label{fig:residuals}
\end{figure}

In Fig. \ref{fig:residuals}, the top and middle images are colormaps of the water temperature from the HYCOM model separated by a 24-hour period.  The bottom image is a colormap illustrating the residual difference between the two frames. These brighter locations within the bottom image depict the regions of interest for water temperature. \\

\hspace{10mm} \textit{ii) RoI formula} \\
The RoI is computed as the sum of residual values across all observations between time t\textsubscript{n} and t\textsubscript{n+1}. The resulting value is then compared to a thresholding value to determine if it is an RoI or ignored. The threshold value in Fig. 2 and Fig. 3 is 0.5.  

\begin{equation}
RoI= \sum_{v \in \text{ observations}} \left\{ \begin{array}{l} residual(v) \geq   \text{threshold} \\ \text{otherwise }  0 
\end{array}
\right\} 
\end{equation}
\\
\subsubsection{ROBUST Edges}
\hfill \break
ROBUST edges link GCOOS nodes and HYCOM nodes. Edge generation starts with an RoI Node and pairs with a GCOOS Node based on the shortest geodesic distance from the RoI node to the closest GCOOS node. The geodesic distance, in this case, is the spherical distance between two points, otherwise known as the "great circle distance" or “haversine” distance.

\small{
\begin{equation}
\begin{array}{l}
dx = \sin\theta_{\text{LAT1}} \cdot \sin\theta_{\text{LAT2}} \\
dy = \cos\theta_{\text{LAT1}} \cdot \cos\theta_{\text{LAT2}} \cdot \cos(\theta_{\text{LON1}} - \theta_{\text{LON2}}) \\
distance = \arccos(dx + dy) \cdot R 
\end{array}
\end{equation}
}

Note:\\
\(distance\) = distance between two coordinates.\\
\(R\) = radius of Earth (approximately 6371.0090667KM)\\
\(\theta_{\text{LAT1}}\) = Latitude of the first coordinate in radians\\
\(\theta_{\text{LAT2}}\)= Latitude of the second coordinate in radians\\
\(\theta_{\text{LON1}}\)= Longitude of the first coordinate in radians\\
\(\theta_{\text{LON2}}\) = Longitude of the second coordinate in radians\\

\subsubsection{ROBUST Realizations}
\begin{figure}[H]
\centering
\includegraphics[width=0.4\textwidth]{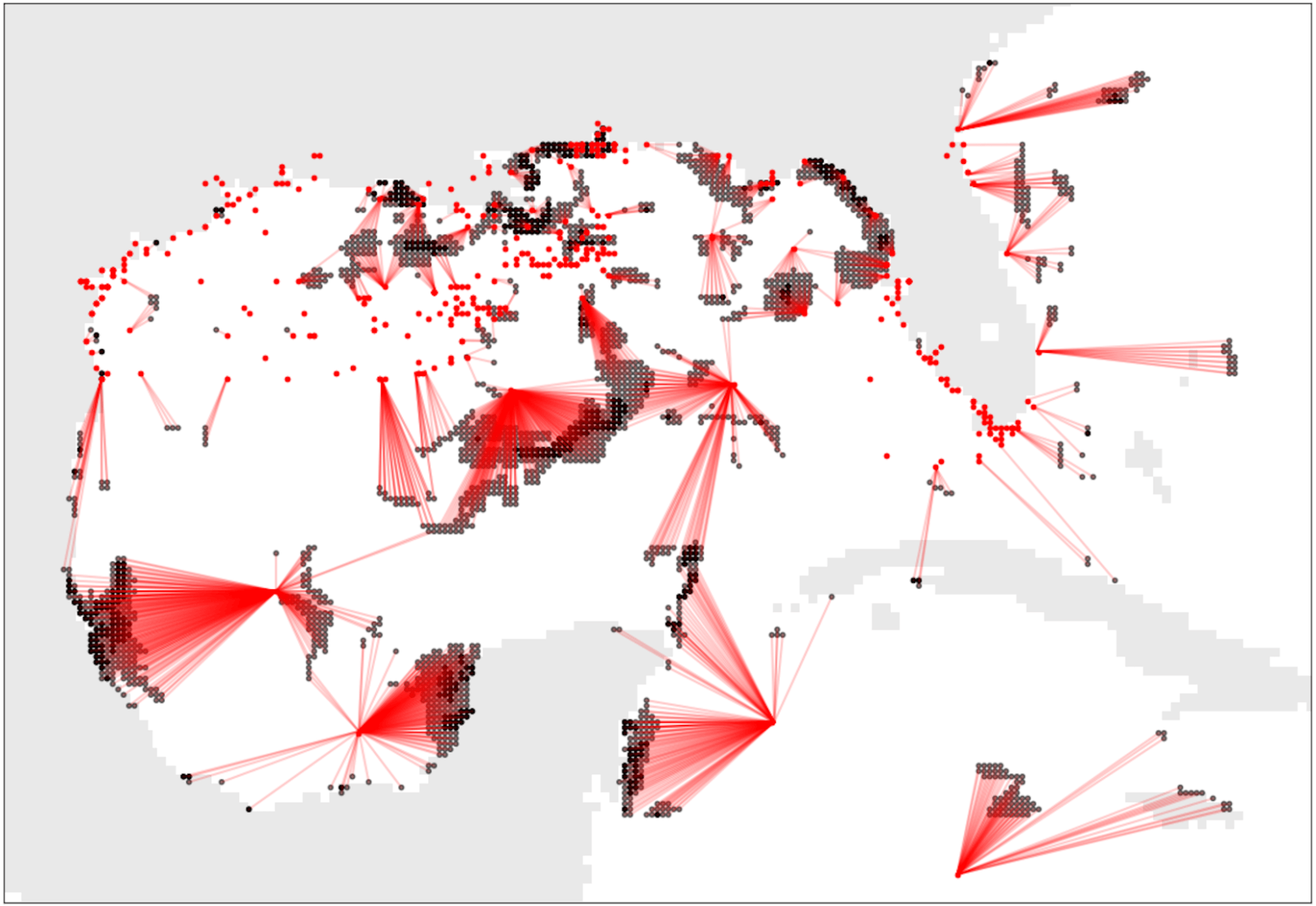}
\caption{ A snapshot of ROBUST Network at time \(t_0\). (Global view)}
\label{fig:global-view}
\end{figure}

\begin{figure}[H]
\centering
\includegraphics[width=0.4\textwidth]{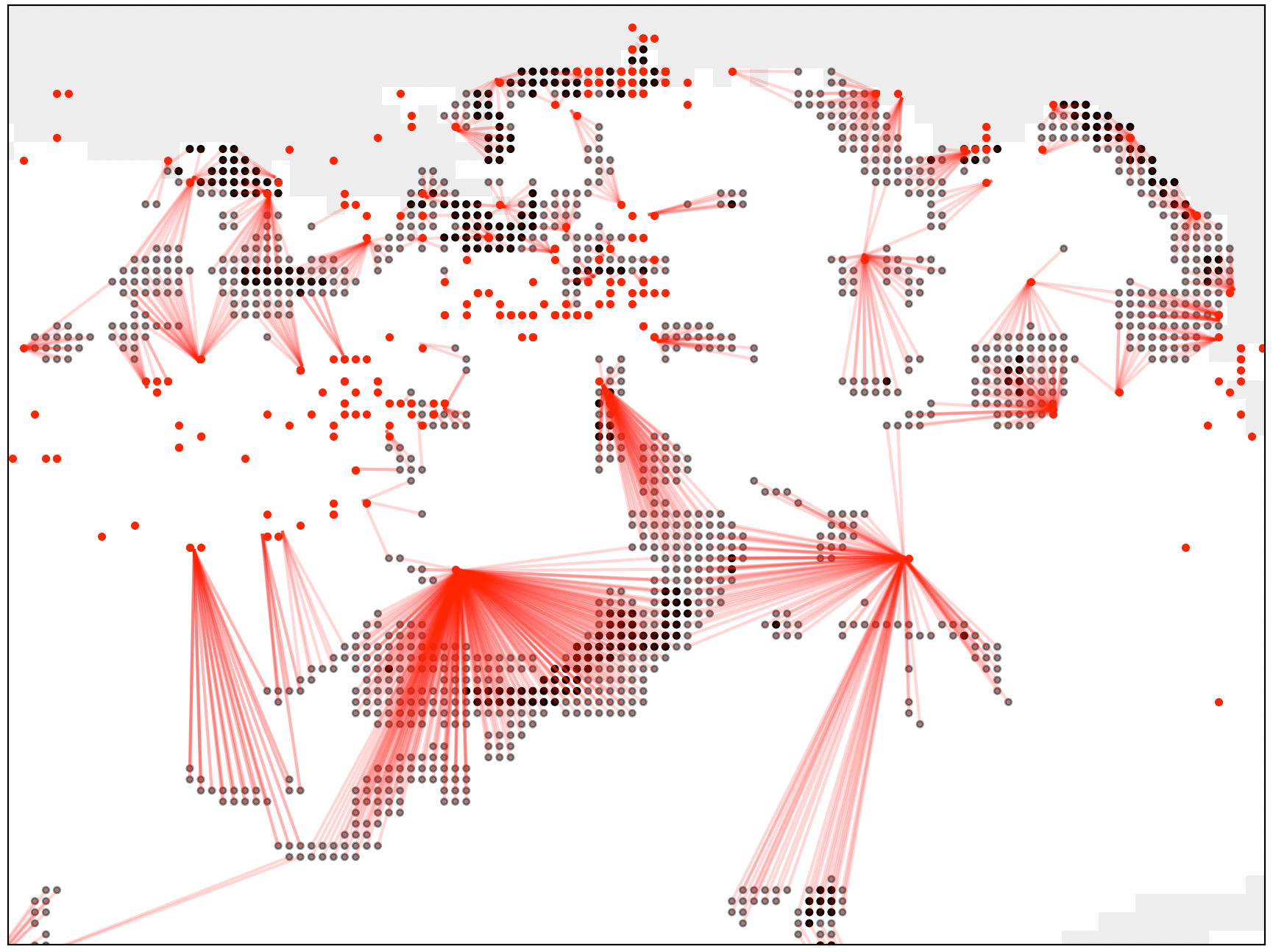}
\includegraphics[width=0.4\textwidth]{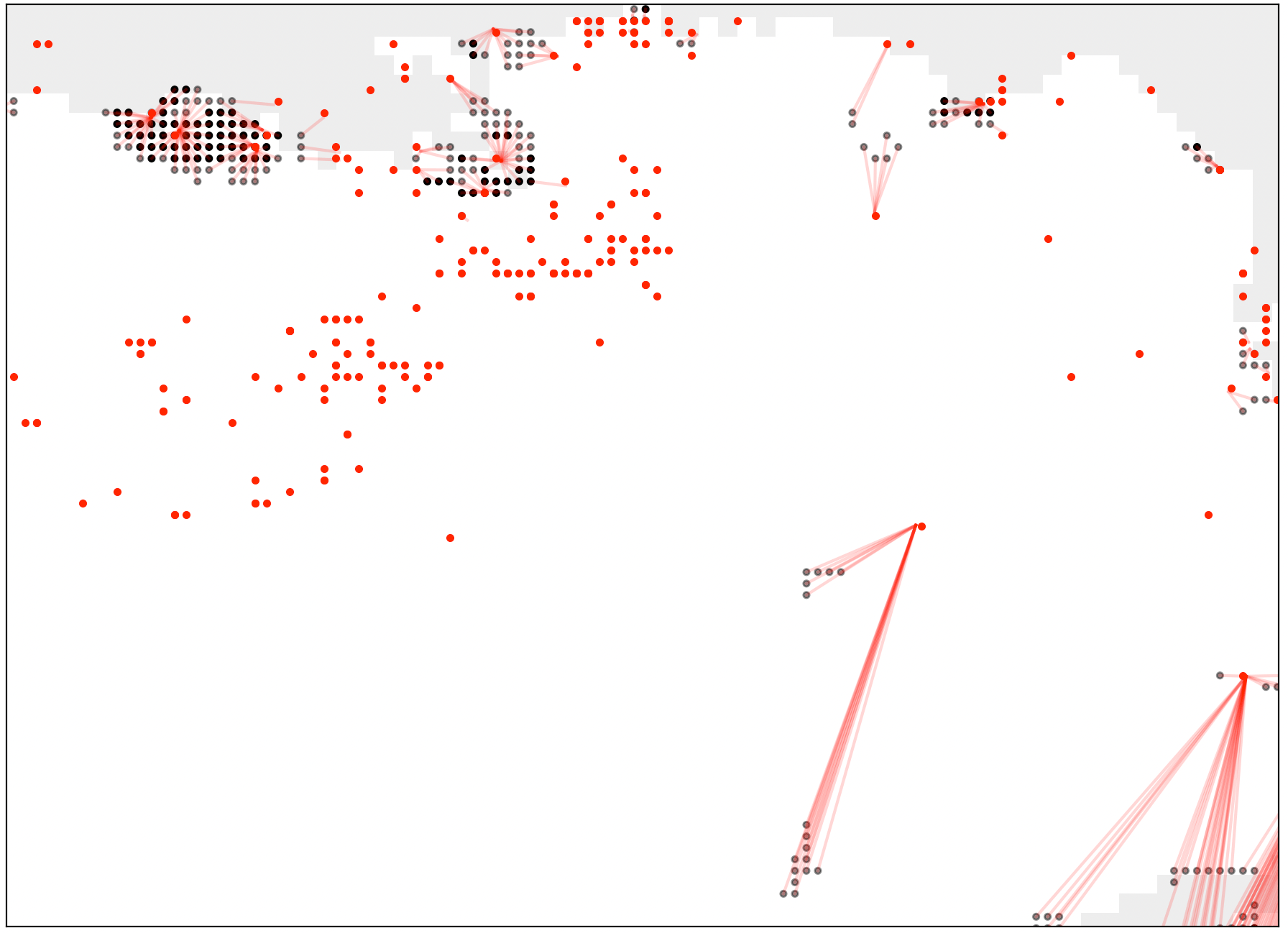}
\caption{Two temporal snapshots of the ROBUST network at times \(t_0\) and \(t_2\) from same region. (Zoom view)}
\label{fig:zoom-view}
\end{figure}

\hfill \break
In Fig. \ref{fig:global-view}, the GCOOS sensor nodes are colored red and remain stationary. In contrast, the HYCOM RoI nodes are grayscale with coloring relative to the strength of the residual and dynamic between frames.  In Fig. \ref{fig:zoom-view}, the zoom window is the same position but each snapshot differs as the set of RoI nodes varies, illustrating the inherent stochastic problem of selecting optimal placements for new sensors.

\section{Analysis}
The goal of the ROBUST network is to measure sensor placement strategies. Coverage and Coverage Robustness are the primary and secondary considerations used to measure the effectiveness of the spatial configuration of the ROBUST network sensor nodes. These measures are defined in the subsections below. 

\subsection{Coverage measure}
This measure quantifies the coverage of the GCOOS sensor nodes to the HYCOM RoI nodes across all snapshots in the ROBUST network. A maximal coverage would result in a GCOOS sensor node typically being close to an HYCOM RoI node. A suboptimal coverage could result in GCOOS sensor nodes being out of range from the HYCOM RoI nodes. \\

\subsubsection{Static Coverage }
\hfill \break
The static network coverage may be numerically computed as the sum of edge weights across all edges for a single discrete time step. This ROBUST Network encodes the geospatial distance between the closest GCOOS sensor node and an HYCOM RoI node as the edge weight. Given a set of weights, the formula below provides the coverage score for that timestep.  

\begin{equation}
coverage =  \sum edge \text{ } weights
\end{equation}

\vspace{2mm}

\subsubsection{Temporal Coverage }
\hfill \break
There are two approaches for measuring temporal network coverage by expanding the above definition for the static coverage measure: \\

\paragraph{Sum of Static Coverage Scores}
\hfill \break
The total temporal coverage is the sum of the static coverages across all timesteps within the ROBUST Network. 

\begin{equation}
total \text{ } temporal \text{ }  coverage = \sum_{t \in Timesteps} coverage(t)
\end{equation}

This approach may be unduly influenced by the presence of a bimodal distribution of coverage scores across the ROBUST Network timesteps.  One poor performance in a timestep heavily penalizes the total temporal coverage, or one positive performance greatly benefits it.   \\

\paragraph{Average of Static Coverage Scores}
\hfill \break
The average temporal coverage score better represents the expected coverage for any given time step within the ROBUST Network by using the quotient between the total temporal coverage and the total number of time steps. The Average Temporal Coverage is the primary measure used to rank the performance of the ROBUST Network configurations in this paper.

\small {
\begin{equation} 
average \text{ } temporal \text{ } coverage =  \frac {total \text{ } temporal \text{ } coverage} {number \text{ } of \text{ } timesteps} 
\end{equation}
}

\vspace{2mm}

\subsection{Coverage Robustness measure}
This measure quantifies the robustness of the GCOOS sensor nodes in its ability to cover the HYCOM RoI nodes adequately.  Maximal robustness results in a network configuration whereby the coverage would be minimally affected by removing a critical sensor node. With minimal robustness, removing a sensor node may substantially penalize the coverage score. Nodal centrality is a useful measure for determining the most critical sensor positions. To maximize robustness, the distribution of degree centrality should spread across multiple nodes instead of residing in only a few select critical nodes.  \\

\subsubsection{Static Degree Centrality }
\hfill \break
The static network degree centrality computes the distribution of edges across the GCOOS sensor nodes within the network. The distribution is the count of edges per node against the degree frequency across all nodes. \\

\subsubsection{Temporal Degree Centrality }
\hfill \break
There are two approaches to measuring the temporal network degree centrality.

\paragraph{Overall Centrality}
\hfill \break
The sum of all connections each node has through time over the entire temporal sequence \cite{b7}.

\paragraph{Per-Timestep Centrality}
\hfill \break
The sum of all connections each node has through time per time point
 \cite{b7}.   \\

Via  simulations, the network robustness is evaluated by removing the nodes with the highest nodal degree and recomputing the new network edges based on the revised distances to derive  the  new coverage score. If the coverage score increases significantly, the network is fragile to the loss of sensor nodes. If the coverage score remains relatively stable, the network is robust to the loss of sensor nodes.

\section{Optimizing Placements of New Sensor Nodes}
Identifying the optimal placements for new sensor nodes must start from the initial GCOOS sensor configuration. Given the stochastic nature of the HYCOM RoI nodes, a Monte Carlo simulation strategy determines the best locations for new GCOOS sensor nodes. The goal is to identify a nodal configuration that both distributes centrality and minimizes edge distances.

\subsection{Monte Carlo Simulation}
A Monte Carlo simulation is helpful to identify the probability of different outcomes in a non-deterministic environment due to the intervention of random variables \cite{b8}. In the case of this ROBUST network, it is the HYCOM RoI node placements per timestep that are random. 

The Monte Carlo simulation begins by selecting a random coordinate within the spatial domain of the ROBUST network. That coordinate is then used to insert a new GCOOS sensor node into the ROBUST network and recompute all of the edges. The updated edge list produces a new average temporal coverage score. That new coverage score compares against the current optimal placement's coverage score. If the new score is less than the current optimal score, then that random coordinate is saved at its optimal position. Repeating this process a suitably high number of times to exhaustively search the spatial space identifies the best location, which maximizes coverage across all timesteps in the testing dataset. 

An advantage of the Monte Carlo approach is that it is a distributed process at its core since each simulation is independent of the others. To identify optimal positions for multiple sensors within the ROBUST network, perform this process sequentially, one node at a time. 

\section{Case Study Results}
All results use the Average Total Coverage measure. The score for the initial GCOOS sensor configuration is below. 

\subsection{Initial Sensors}
\begin{quote}
coverage score: 180222.806856 \\
\end{quote}

\begin{figure}[H]
\centering
\includegraphics[width=0.38\textwidth]{images-paper1/gstbn-init-t0.png}
\includegraphics[width=0.38\textwidth]{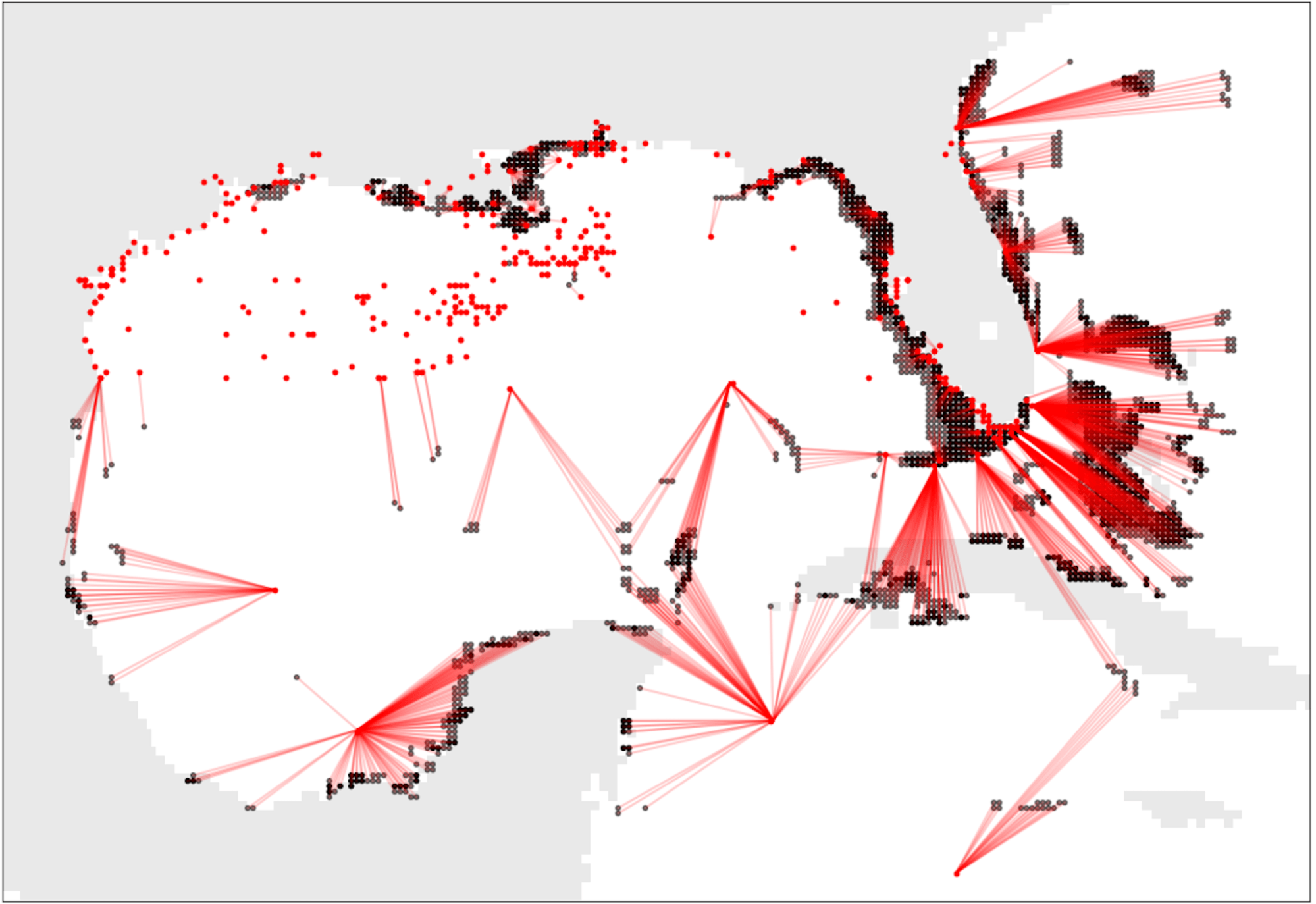}
\includegraphics[width=0.38\textwidth]{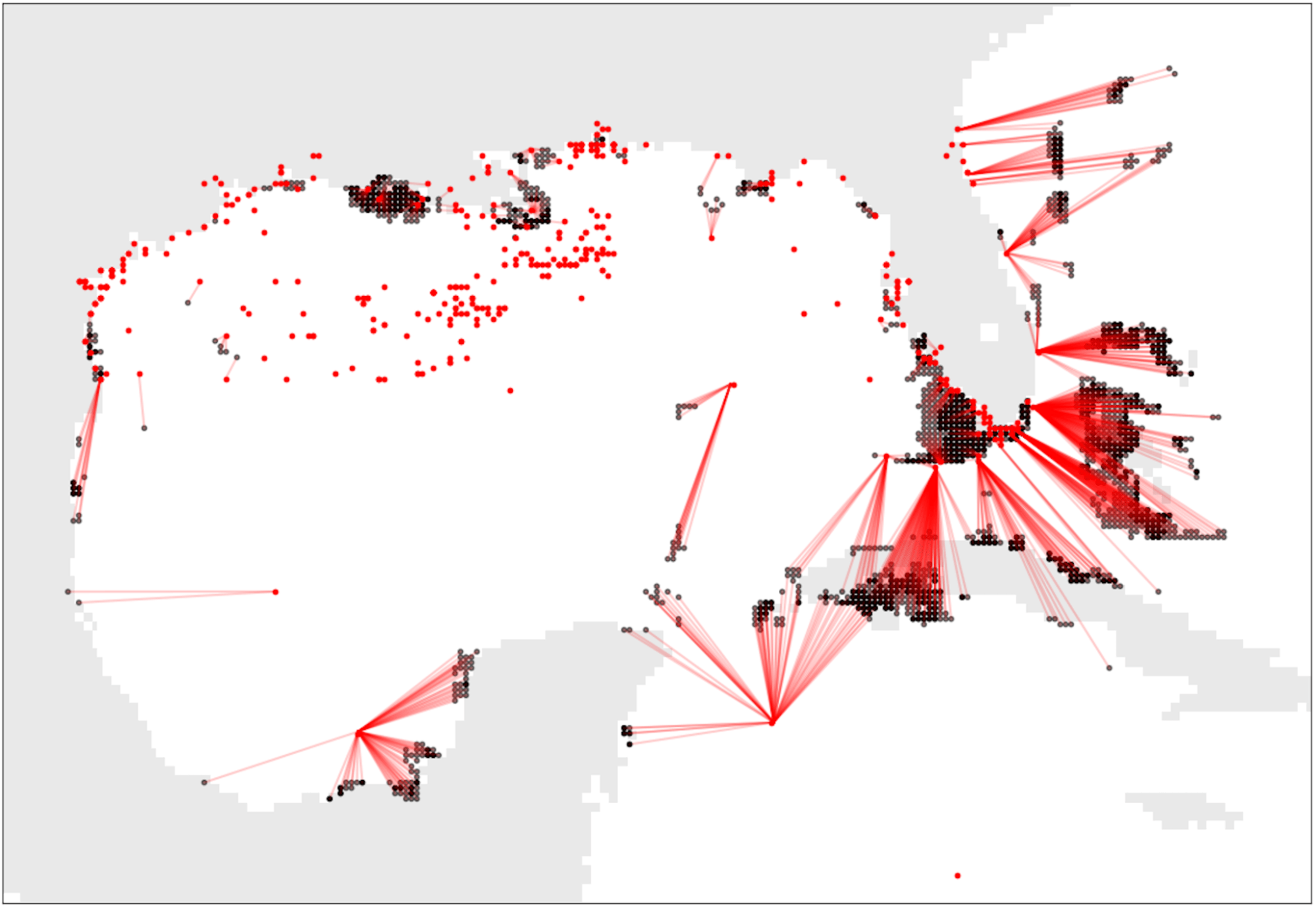}
\includegraphics[width=0.38\textwidth]{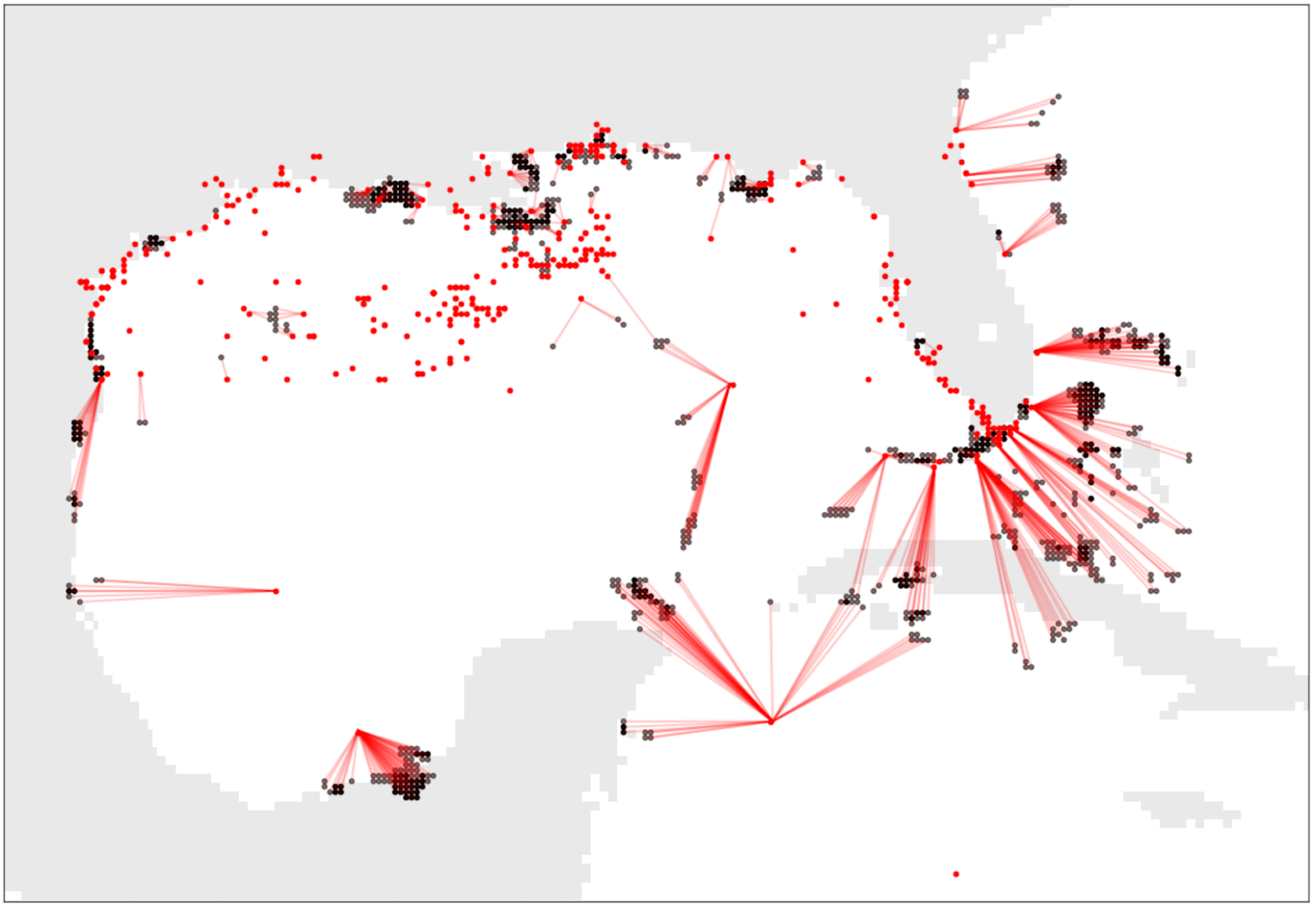}
\caption{Sequence of Temporal Snapshots of the ROBUST network with the initial GCOOS sensor configuration}
\label{fig:init-gcoos-net}
\end{figure}

\subsection{First New Sensor}
The Monte Carlo simulation  with 1,000 trials.
\begin{quote}
longitude: -78.7403109976445 \\ 
latitude: 24.385624429875215 \\
coverage score: 160873.88100 \\
\end{quote}

\begin{figure}[H]
\centering
\includegraphics[width=0.38\textwidth]{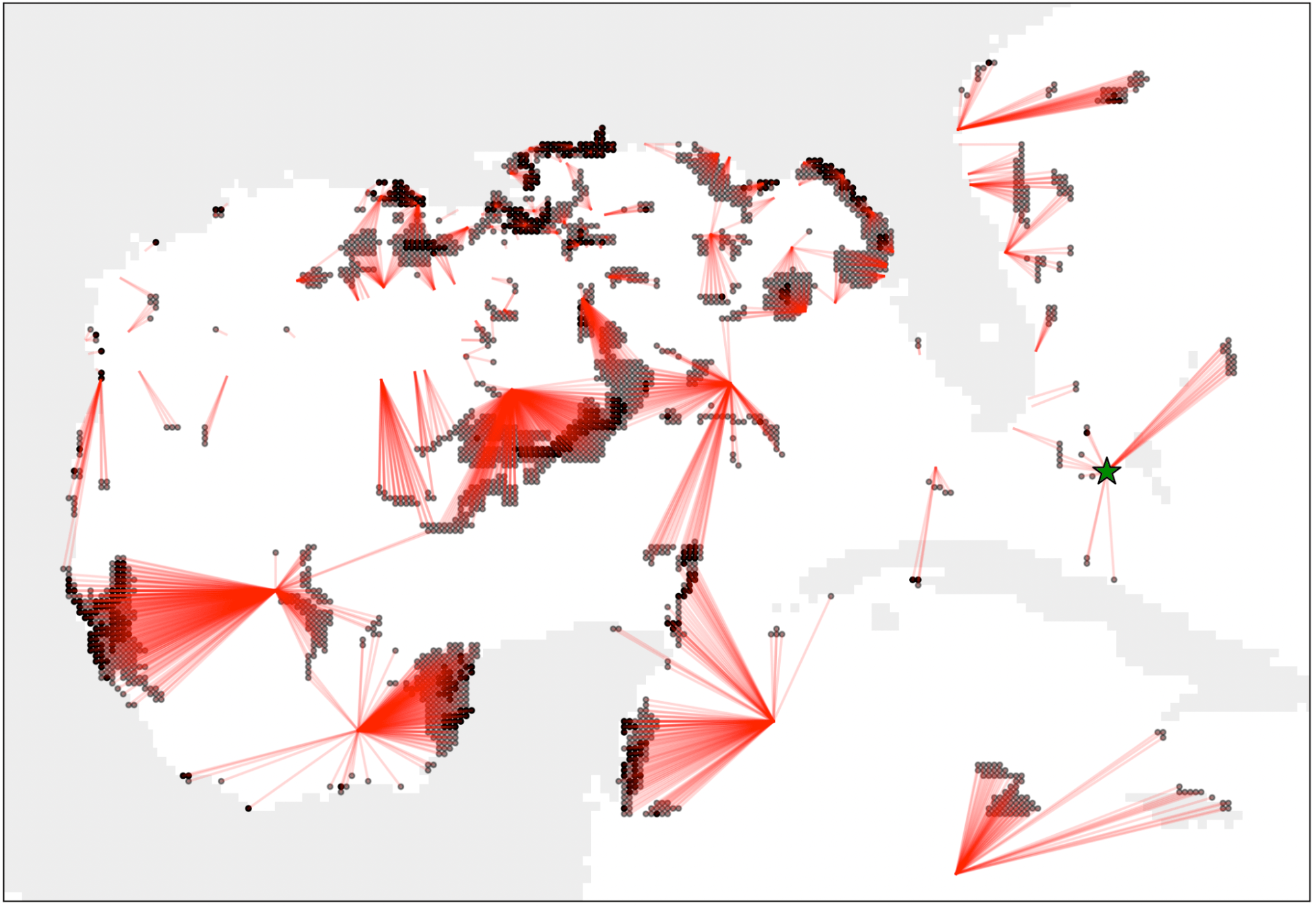}
\includegraphics[width=0.38\textwidth]{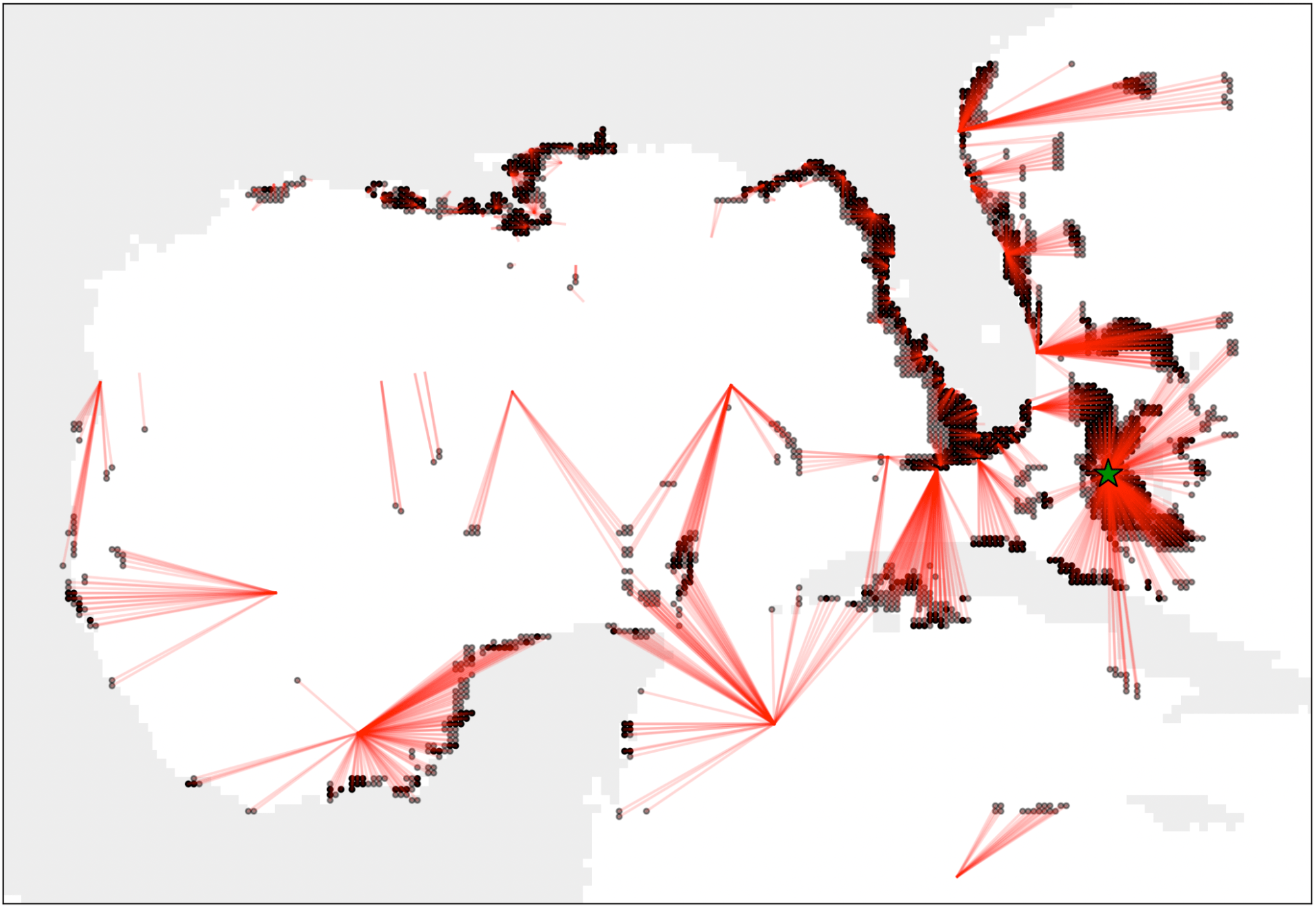}
\includegraphics[width=0.38\textwidth]{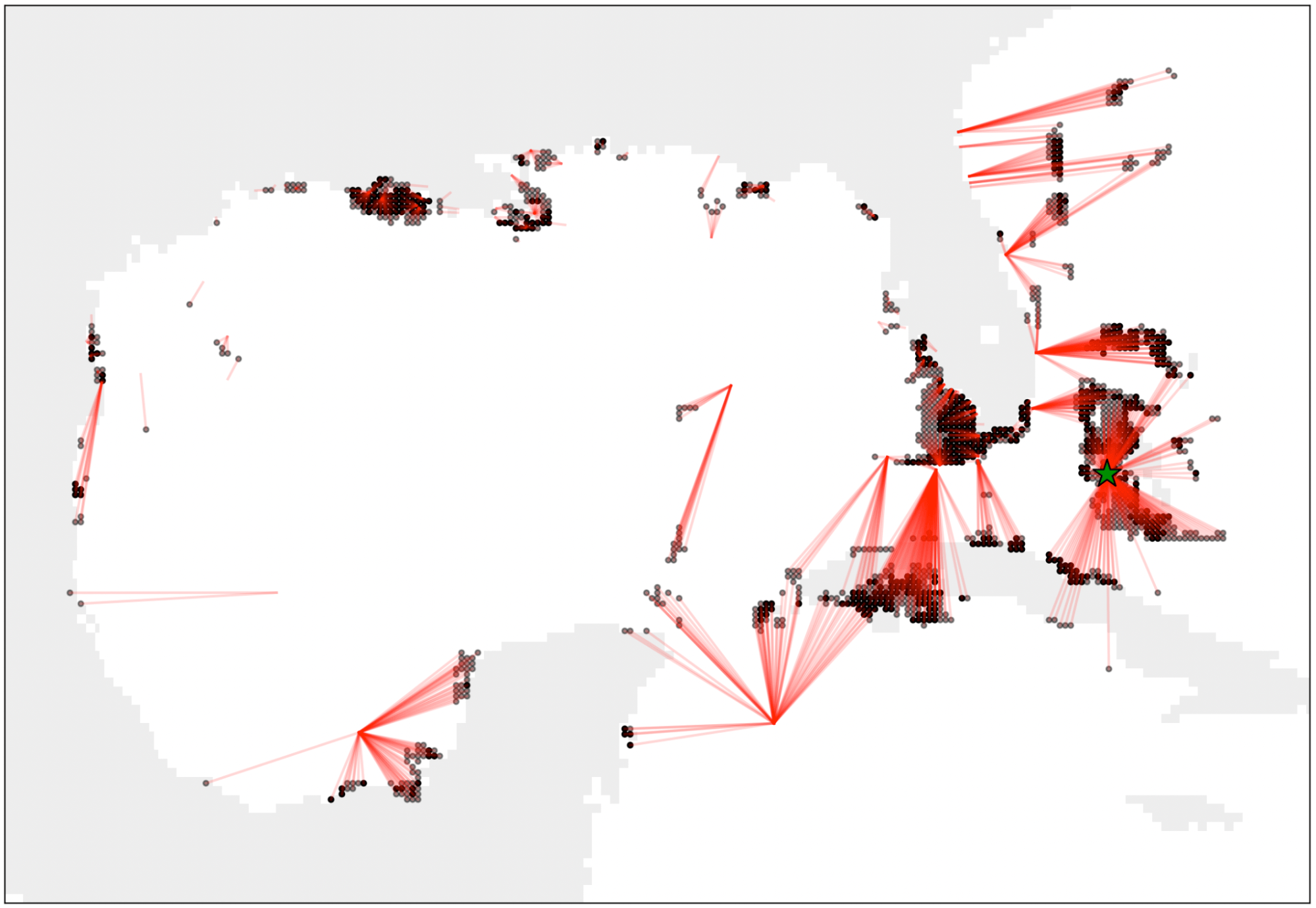}
\includegraphics[width=0.38\textwidth]{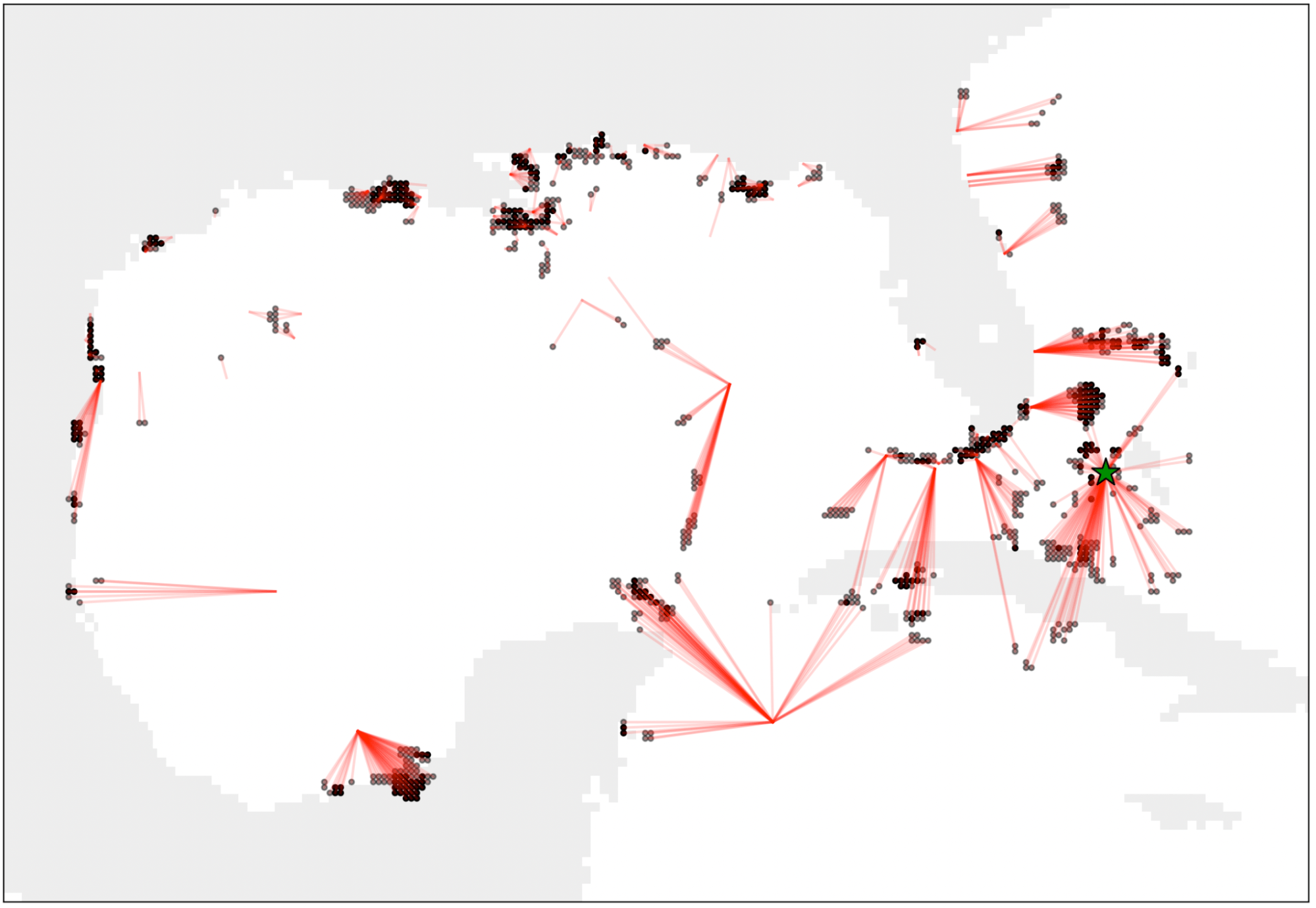}
\caption{Sequence of Temporal Snapshots of the ROBUST network with the suggested position for a new sensor represented as a green star}
\label{fig:gcoos-net-plus1}
\end{figure}

\subsection{Second New Sensor}
The Monte Carlo simulation  with 1,000 trials. 

\begin{quote}
longitude: -85.81532374804107 \\
latitude: 22.561994782989117 \\
coverage score: 147411.742470\\
\end{quote}

\begin{figure}[H]
\centering
\includegraphics[width=0.38\textwidth]{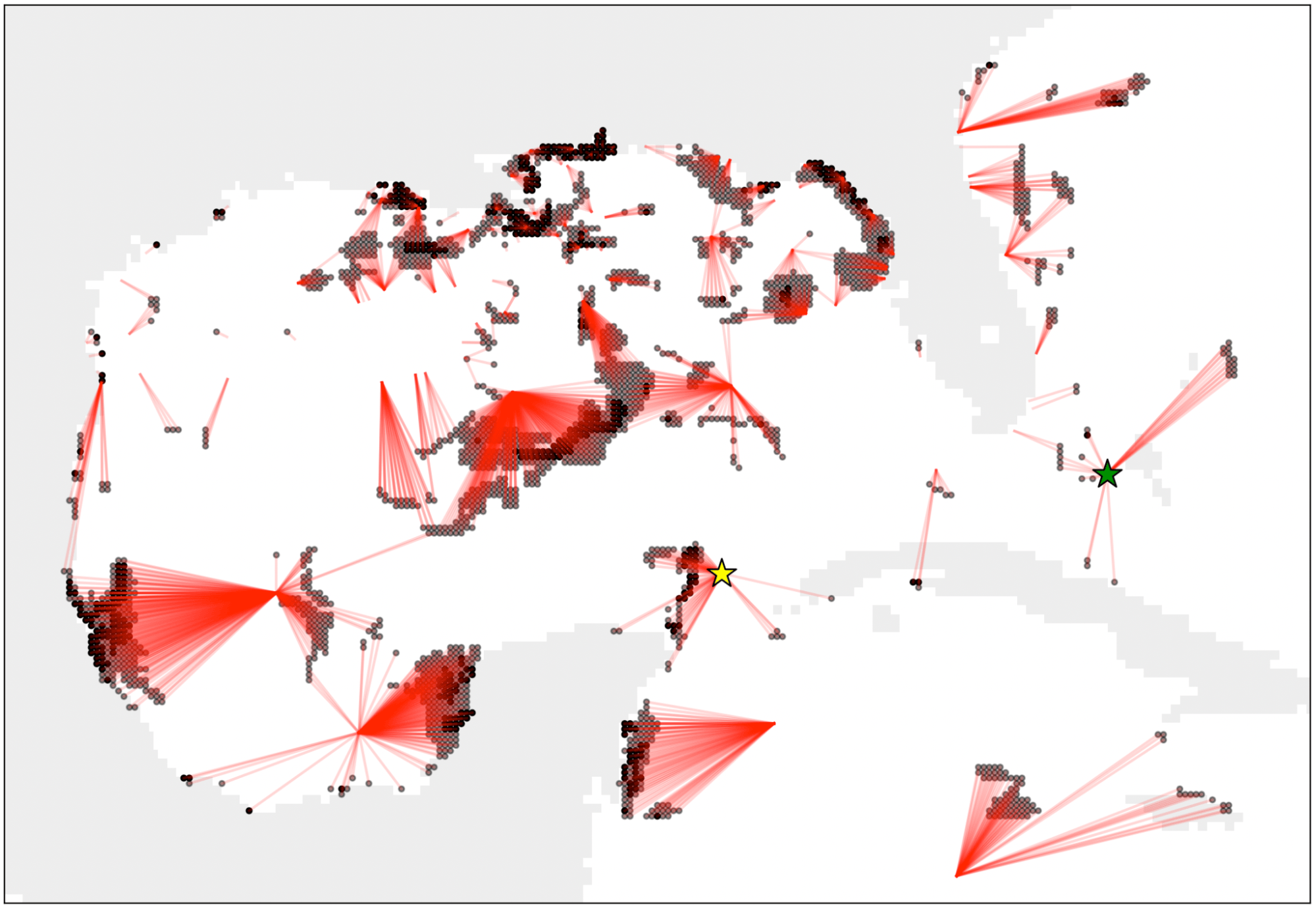}
\includegraphics[width=0.38\textwidth]{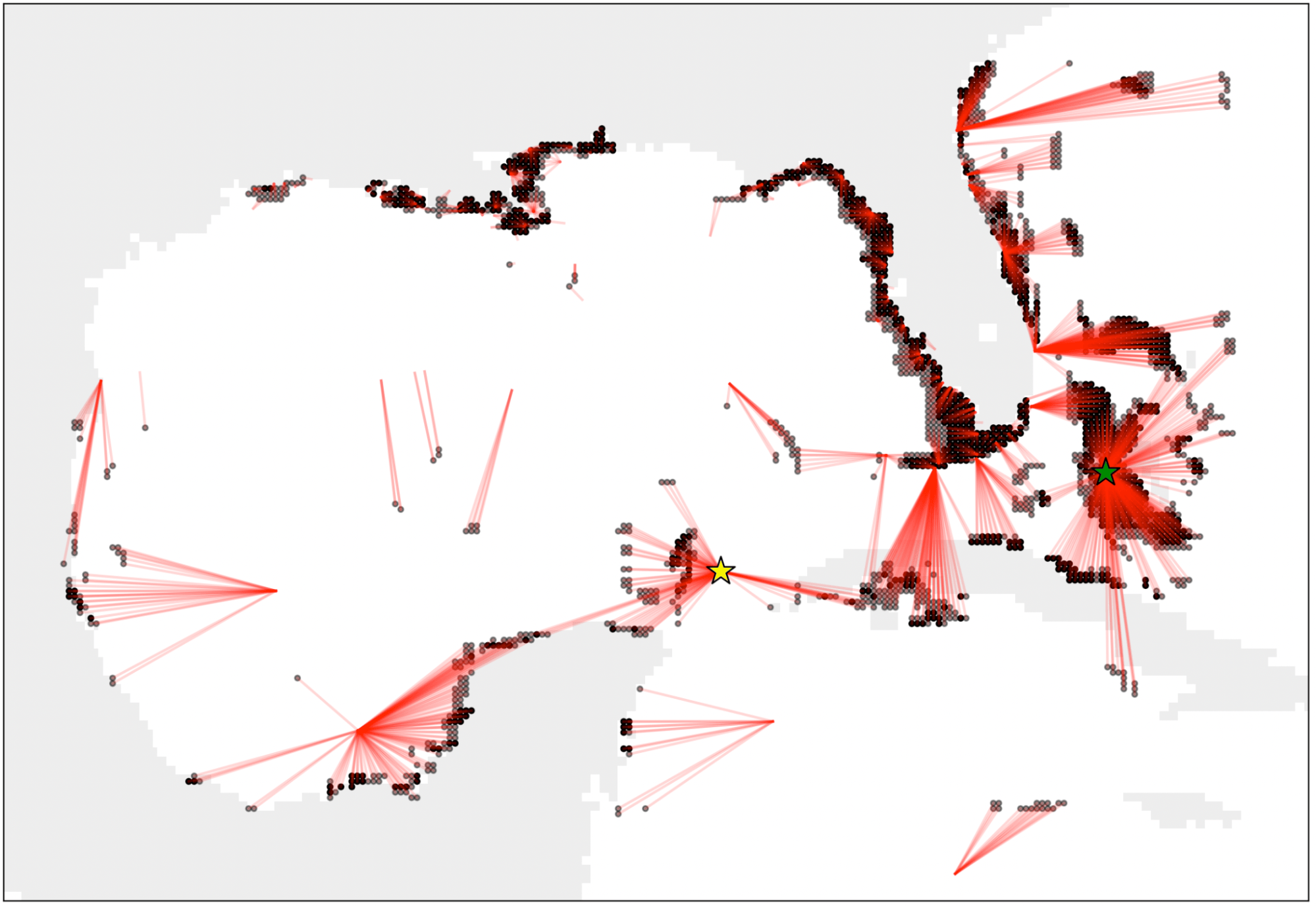}
\includegraphics[width=0.38\textwidth]{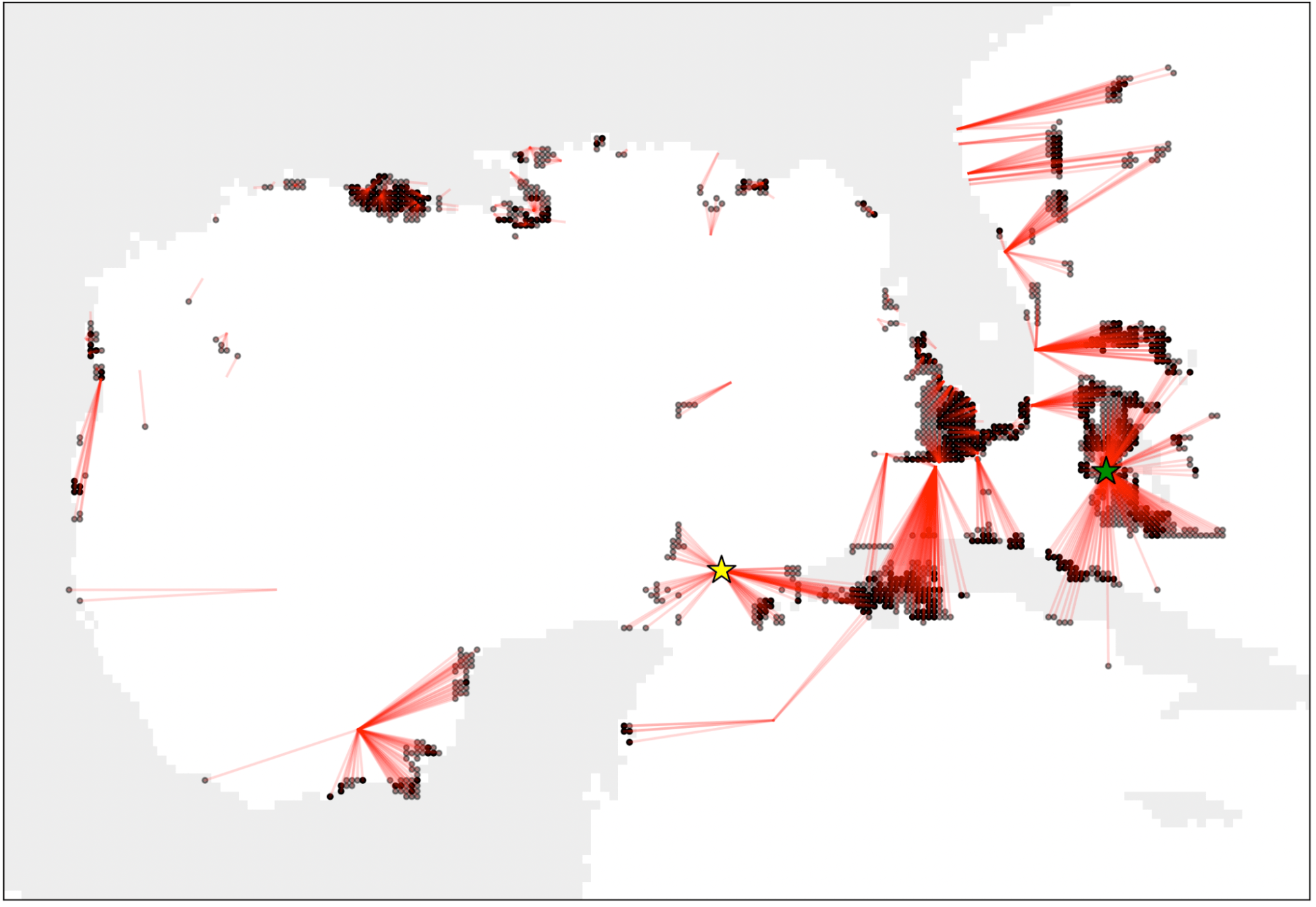}
\includegraphics[width=0.38\textwidth]{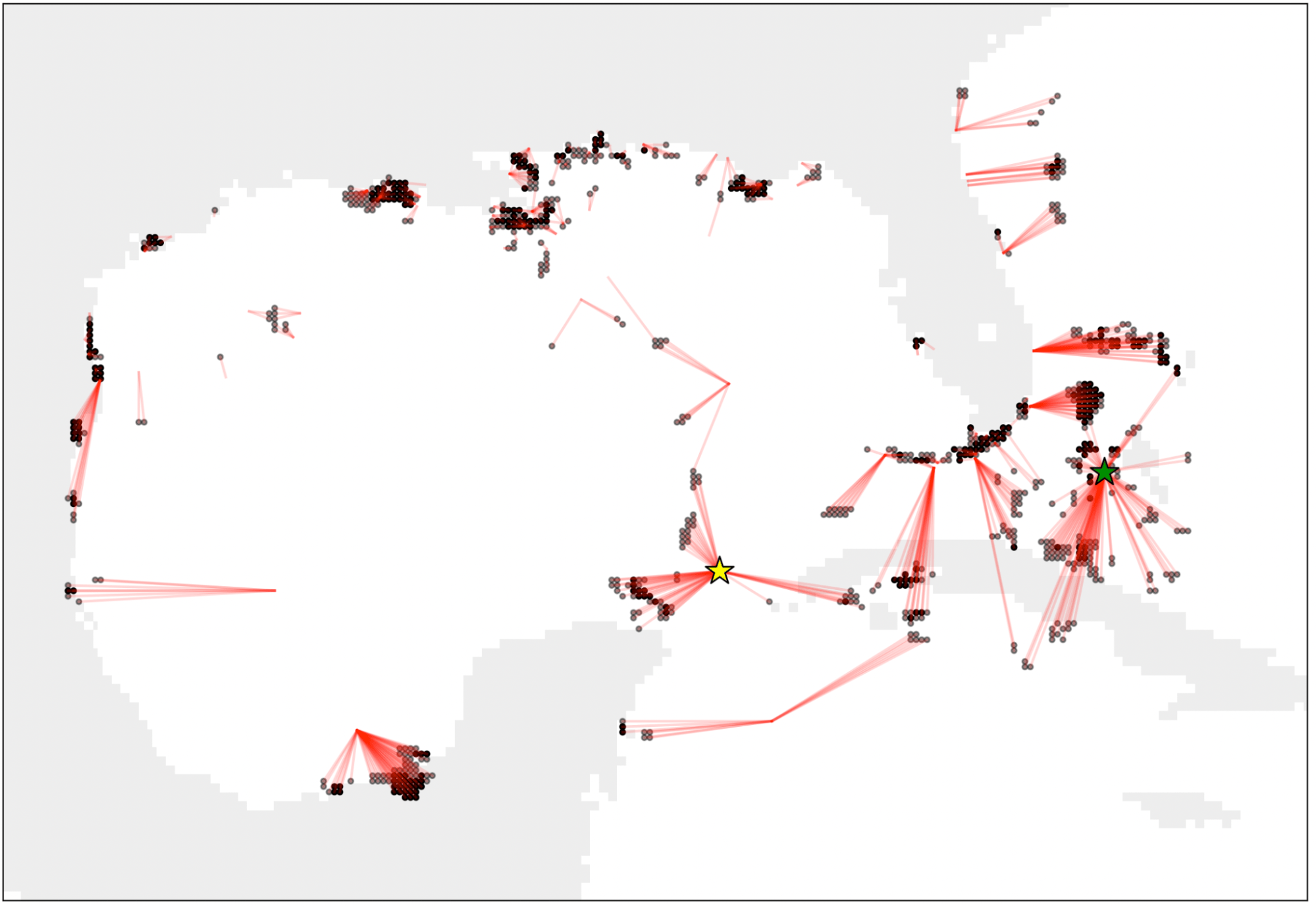}
\caption{Sequence of Temporal Snapshots of the ROBUST network with the suggested position for a 2nd sensor represented as a yellow star}
\label{fig:gcoos-net-plus2}
\end{figure}
\clearpage

\section{Discussion}
The results illustrate how this approach uses a bipartite network model to capture the coverage performance between persistent observers and observable stochastic events within a spatial environment and successfully identify near-optimal locations to expand the observer array. 

As seen in the Results section, the coverage score significantly decreases with the insertion of both the  first and second new nodes. Notice that the placement of each new node maximizes its coverage across all times.  

The results identified through this process would then support any management decisions from the platform/station operators/owners when they intend to add new observational equipment. 

We can expand the guidance provided with even more information by performing community clustering on the observer or the observable nodes. For example, we can cluster the GCOOS platform nodes into different communities based on their platform ownership status. Alternatively, the HYCOM RoI nodes can be clustered together in communities based on their observation type. The Robust network can model such constraints; however, these levels of fine detail and consideration are omitted from the scope of this paper.


\chapter{Case Study II: Urban Safety}
\label{chap:crime}
\section{Introduction}
In response to escalating crime and a diminished police presence \cite{SimermanAdelson2022}, New Orleans—besieged by over 250,000 Calls for Service (CFS) in 2022, of which over 13,000 were violent—has adopted Real-Time Crime Camera (RTCC) systems as a strategic countermeasure\cite{CityOfNewOrleans2022}. The RTCC, with 965 cameras saved over 2,000 hours of investigative work in its inaugural year\cite{Genetec2019}. However, a significant challenge persists: optimizing these systems to maximize their efficacy and coverage amidst the city's staggering crime rate.

This research focuses on the mathematical model of this issue through the lens of graph theory, exploring ROBUST networks. While the immediate application is crime surveillance, the principles and methodologies developed herein apply to various observer-type networks, including environmental monitoring systems.

The objectives are threefold: identify the most influential nodes, evaluate network efficacy, and enhance network performance through targeted node insertions. Employing a centrality measure, the methodology optimizes observer node placements in a spatial bipartite network. Through modeling relationships between service calls, violent events, and crime camera locations in New Orleans using knowledge graphs, the research contrasts our Spatiotemporal Ranged Observer-Observable clustering technique with alternative approaches, such as Kmeans\cite{{wu2021kmeans}}, DBSCAN \cite{schubert2017dbscan}, and time series analysis via statistical mode \cite{hastie2001}. Additionally, our approach, focusing on GPU-accelerated computation, ensures scalable and efficient data processing across large datasets \cite{rapids2023}.

\section{Background}
This section outlines the key datasets and mathematical concepts foundational to our exploration of spatiotemporal clustering in crime and surveillance data.

\subsection{Dataset: Crime Dynamics and Surveillance in NOLA}
This subsection introduces two datasets used to construct the ROBUST network.

\subsubsection{Real-Time Crime Camera (RTCC) Systems}
The RTCC system, set up by the New Orleans Office of Homeland Security in 2017, uses 965 cameras to monitor, respond to, and investigate criminal activity in the city. Of these cameras, 555 are owned by the city and 420 are privately owned \cite{Stein2021}.

\subsubsection{Calls for Service (CFS)}
The New Orleans Police Department’s (NOPD) Computer-Aided Dispatch (CAD) system logs Calls for Service (CFS). These include all requests for NOPD services and cover both calls initiated by citizens and officers \cite{CityOfNewOrleans2022}.

\subsection{Mathematical Background}
Utilizing the datasets introduced, our analysis hinges upon several mathematical and network principles, each playing a crucial role in interpreting the spatial and temporal dynamics of crime and surveillance in NOLA.

\vspace{6pt}
\subsubsection{Bipartite Networks}
Bipartite networks consist of two discrete sets of nodes with edges connecting nodes from separate sets. They effectively portray observer-observable relationships by defining the relationship between the two types.
Mathematically, a bipartite graph (or network) \( G = (U, V, E) \) is defined by two disjoint sets of vertices \( U \) and \( V \), and a set of edges \( E \) such that each edge connects a vertex in \( U \) with a vertex in \( V \). Formally, if \( e = (u, v) \) is an edge in \( E \), then \( u \in U \) and \( v \in V \). This ensures that nodes within the same set are not adjacent and that edges only connect vertices from different sets \cite{asratian1998}.

\vspace{6pt}
\subsubsection{Centrality}
Degree Centrality measures a node's influence based on its edge count and is crucial for identifying critical nodes within a network \cite{Nicosia_2013}. In bipartite networks, such as those modeling interactions between surveillance cameras (observers) and crime incidents (observables), centrality is vital for evaluating and optimizing camera positioning to ensure effective incident monitoring. Consequently, the degree centrality of the observer nodes is of particular concern.

\vspace{6pt}
\subsubsection{Spatiotemporal Networks}
Spatiotemporal networks (STNs) model entities and interactions that are both spatially and temporally situated. These networks can effectively represent geolocated events and observers by establishing connections based on spatiotemporal conditions. 

A Spatiotemporal Network (STN) is defined as a sequence of graphs representing spatial relationships among objects at discrete time points:
\begin{equation}
STN = \Big( G_{t_1}, \; G_{t_2}, \; ..., \;  G_{t_n} \Big) 
\end{equation}
where each graph \( G \) is defined as:
\begin{equation}
G_{t_i} = \Big( N_{t_i}, \; E_{t_i} \Big)
\end{equation}
and represents the network at time \( t_i \) with \( N_{t_i} \) and \( E_{t_i} \) denoting the set of nodes and edges at that time, respectively. \cite{Holme2012} 

In the context of this research, STNs are utilized to model relationships between crime cameras (observer nodes) and violent events (observable nodes) across New Orleans. Nodes represent objects, whereas edges depict spatiotemporal relationships, connecting observer nodes to observable nodes based on spatial proximity and temporal occurrence. The analysis of STNs allows for the extraction of insightful patterns, aiding in understanding and potentially mitigating the progression and spread of events throughout space and time. 

\vspace{6pt}
\subsubsection{Spatiotemporal Clustering}
Spatiotemporal clustering groups spatially and temporally proximate nodes to identify regions and periods of significant activity within a network. In the context of spatiotemporal networks, clusters might reveal hotspots of activity or periods of unusual event concentration. \cite{Holmberg2022}  Techniques for spatiotemporal clustering must consider both spatial and temporal proximity, ensuring that nodes are similar in both their location and their time of occurrence.

\section{Related Work}
Building upon the mathematical principles introduced, this section evaluates existing clustering techniques, highlighting their limitations when applied to bipartite networks and, particularly, in the context of our application. This exploration accentuates the challenges faced in utilizing conventional methods for optimizing bipartite networks and underscores the potential and necessity of the novel methodology proposed in this research. In the results section, we will present a comprehensive comparative analysis of the aforementioned methods against our proposed methodology, providing a robust evaluation and benchmarking in various scenarios relevant to bipartite spatiotemporal networks.

\subsection{K-means Clustering}
K-means clustering partitions datasets into \( k \) distinct clusters by minimizing intra-cluster variances.\cite{wu2021kmeans}  Despite its popularity, K-means exhibits several limitations when applied to spatiotemporal data. Firstly, it mandates the pre-specification of \( k \), which can be non-intuitive in real-world scenarios. Secondly, K-means insists on assigning every data point to a cluster, potentially obscuring subtle yet crucial spatial-temporal patterns in the data. This rigidity in assignment can lead to misrepresentations of the true underlying structures within the observer-observable networks, thereby limiting the utility and accuracy of the clusters for strategic node insertions. Critically, K-means does not accommodate the observational range of entities within the network, such as surveillance cameras, due to its inability to define a maximum radius or diameter for clusters \cite{dorabiala2022spatiotemporal}. This shortcoming hampers its applicability in scenarios where the spatial reach of items is a vital factor, prompting the need for alternative clustering methodologies that inherently incorporate spatial constraints.

\subsection{DBSCAN}
DBSCAN, which stands for Density-Based Spatial Clustering of Applications with Noise, identifies clusters by assessing density proximity. It begins from an arbitrary point and iteratively explores neighboring points to ascertain all directly or indirectly reachable points that fulfill a predetermined density condition. \cite{schubert2017dbscan, birant2007stdbscan} While DBSCAN offers certain advantages, such as the ability to form clusters without pre-specifying their number and to exclude points from clustering, its application to bipartite spatiotemporal networks exposes several limitations. Notably, its tendency to merge adjacent clusters may create larger, potentially less dense clusters, which could misrepresent genuine spatial-temporal patterns in the data. This merging approach also implies that DBSCAN does not inherently consider the observational range of entities in the network, a crucial factor in scenarios where the spatial influence of entities is vital for modeling and analyzing spatial networks. Furthermore, the absence of generated centroid points for the clusters formed by DBSCAN hampers its applicability for strategizing node insertions, especially in spatial-temporal contexts where centroids can serve as logical and efficient insertion candidates.

\subsection{Frequency Clustering}
The statistical frequency discerns the most frequent value within a dataset, offering a spotlight on singular, recurrent instances and providing a non-spatial measure of data density \cite{hastie2001}. While it excels in identifying dominant occurrences in a dataset, its glaring limitation is the neglect of spatial relationships among data points. In the context of optimizing bipartite spatiotemporal networks, this neglect poses a significant challenge. A technique that solely hinges on frequency overlooks the critical spatial dimension of the data, potentially bypassing spatial clusters that, while perhaps not housing the most frequent events, may nevertheless represent crucial hubs or hotspots in the network. This could lead to suboptimal strategies for node insertions, failing to fully harness the spatial-temporal patterns within the data.

\section{Approach}

\subsection{Problem Definition}

Let matrices representing observer nodes \(O\) and observable events \(E\) be defined, with their respective longitudinal \((O_{\text{lon}}, E_{\text{lon}})\) and latitudinal \((O_{\text{lat}}, E_{\text{lat}})\) coordinates. The objective is to formulate a framework that:
\begin{itemize}
    \item Computes distances between observers and events.
    \item Assesses the centrality of observer nodes.
    \item Classifies events according to their observability.
    \item Clusters unobserved points utilizing spatial proximity.
    \item Adds new observers to improve network performance.
\end{itemize}

\subsection{Objectives}

\begin{enumerate}
\item \textbf{Maximize Current Network Observability:} Optimize the placement or utilization of existing observer nodes to ensure a maximum number of events are observed.
\item \textbf{Identify and Target Key Unobserved Clusters:} Analyze unobserved events to identify significant clusters and understand their characteristics to inform future observer node placements.
\item \textbf{Strategize Future Observer Node Placement:} Develop strategies for placing new observer nodes to address unobserved clusters and prevent similar clusters from forming in the future.
\end{enumerate}

\subsection{Rationale}

The limitations identified within existing methodologies, as discussed in the Related Work section, underscore the need for an innovative approach to optimizing bipartite networks, particularly in the context of spatiotemporal data. Traditional clustering methodologies, such as K-means and DBSCAN, present challenges in terms of accommodating spatial constraints, managing computational complexity, and providing actionable insights for node insertions. Whereas non-spatial methods like statistical mode lack the capacity to truly harness the spatial-temporal patterns within the data, often leading to suboptimal strategies for node insertions.

Therefore, our approach hinges on creating a bipartite distance matrix to systematically evaluate the spatial relationships between observer nodes and events. Following this, clustering algorithms are implemented to group disconnected points, providing an understanding of the spatial dimensions of our data. This methodology evaluates the existing observer network and delivers strategic, data-driven insights to enhance future network configurations.

\begin{itemize}
    \item \textbf{Distance Matrix Calculation:} Utilizing geographical data to calculate distances between observer nodes and events, with particular attention to ensuring all possible combinations of nodes and events are considered, is paramount to understanding spatial relationships within the network.
    \item \textbf{Effectiveness Evaluation:} The centrality and effectiveness of observer nodes are crucial metrics that inform us about the current status of the network in terms of its observational capabilities.
    \item \textbf{Event Classification:} Classifying events into observed and unobserved categories helps in understanding the coverage of the observer network and identifying potential areas of improvement.
    \item \textbf{Clustering of Unobserved Points:} Identifying clusters amongst unobserved points guides the strategic placement of new observer nodes, ensuring they are positioned where they can maximize their observational impact.
\end{itemize}

\subsection{Challenges}

\begin{itemize}
    \item \textbf{Computational Complexity:} Given the potentially large number of observer nodes and events, computational complexity is a pertinent challenge, especially when calculating the bipartite distance matrix and implementing clustering algorithms.
    \item \textbf{Spatial Constraints:} Ensuring that the placement of new observer nodes adheres to geographical and logistical constraints while still maximizing their observational impact. The range of each observer node is a major consideration for any event clustering.
    \item \textbf{Temporal Dynamics:} Accounting for the temporal dynamics within the data, ensuring that the models and algorithms are robust enough to handle variations and fluctuations in the event occurrences over time.
\end{itemize}

\section{Methods}
This section provides a detailed account of the approaches and algorithms used to establish spatiotemporal relationships between observer nodes and observable events. The ensuing subsections systematically unfold the mathematical and algorithmic strategies employed in various processes: calculating distances using the Haversine formula, determining centrality and generating links, constructing bipartite and unipartite distance matrices, classifying events, initializing ROBUST network, and clustering disconnected events. Each subsection introduces relevant notations, explains the method through mathematical expressions, and outlines the procedural steps of the respective algorithm.

~\\
\rule{\textwidth}{0.4pt} 

\section*{Bipartite Distance Matrix}

Constructing a bipartite distance matrix is pivotal in capturing the spatial relationships between observer nodes and events. 

\textbf{Notations:}
\begin{itemize}
    \item \( O \): Matrix representing observer nodes with coordinates as its elements, where \( O_{ij} \) denotes the \( j \)-th coordinate of the \( i \)-th observer. Example: 
    \[
    O = \begin{bmatrix} x_1 & y_1 \\ x_2 & y_2 \\ \vdots & \vdots \end{bmatrix}
    \]
    \item \( E \): Matrix representing observable events with coordinates as its elements, where \( E_{ij} \) denotes the \( j \)-th coordinate of the \( i \)-th event. Example: 
    \[
    E = \begin{bmatrix} x'_1 & y'_1 \\ x'_2 & y'_2 \\ \vdots & \vdots \end{bmatrix}
    \]
    \item \( DM \): Distance Matrix, where \( DM_{ij} \) represents the distance between the \(i\)-th observer and the \(j\)-th event. Assuming there are \(m\) observers and \(n\) events, \( DM \) will be an \(m \times n\) matrix. Example: 
    \[
    DM = \begin{bmatrix} 
    d_{11} & d_{12} & \dots  & d_{1n} \\
    d_{21} & d_{22} & \dots  & d_{2n} \\
    \vdots & \vdots & \ddots & \vdots \\
    d_{m1} & d_{m2} & \dots  & d_{mn} \\
    \end{bmatrix}
    \]
    where each element \(d_{ij}\) represents the distance between the coordinates of the \(i\)-th observer and the \(j\)-th event.

\end{itemize}

\textbf{Methodology:}
\begin{enumerate}
    \item \textit{Haversine Distance Calculation:}
    Compute the Haversine distances for all combinations of observer-event pairs, populating the distance matrix \(DM\). Specifically, 
    \[
    DM_{ij} = \text{haversine}(O_i, E_j)
    \]
    where \( O_i \) and \( E_j \) are the coordinates of the \( i \)-th observer and \( j \)-th event. The Haversine formula calculates the shortest distance between two points on the surface of a sphere, given their latitude and longitude. The resulting distances in \(DM\) are given in kilometers, assuming the Earth's mean radius to be 6371 km.

    \vspace{6pt}
    \item \textit{Centrality Calculation:}
    The centrality of each observer is calculated as the count of events within a specified radius \( r \). This is mathematically expressed as:
    \[
    \text{Centrality}(o) = \sum_{e \in E} \mathbb{I}(DM_{oe} \leq r)
    \]
    In this equation, \( \mathbb{I} \) represents the indicator function, which is defined as:
    \[
    \mathbb{I}(P) = 
    \begin{cases} 
    1 & \text{if } P \text{ is true} \\
    0 & \text{if } P \text{ is false}
    \end{cases}
    \]
    Hence, \( \mathbb{I}(DM_{oe} \leq r) \) is 1 if the distance between observer \( o \) and event \( e \), denoted as \( DM_{oe} \), is less than or equal to \( r \), and 0 otherwise. The centrality thus provides a count of events that are within the radius \( r \) of each observer.

    \vspace{6pt}
    \item \textit{Link Generation:}
    Identify and create links between observers and observables that are within radius \( r \) of each other, forming a set \( L \) of pairs \((o, e)\) such that 
    \[
    L = \{(o, e) \,|\, DM_{oe} \leq r\}
    \]
    where each pair represents a link connecting observer \( o \) and event \( e \) in the bipartite graph, constrained by the specified radius.

\end{enumerate}

\textbf{Algorithm:}
\begin{enumerate}
    \item \textit{Step 1: Compute Haversine Distances}
    Calculate the Haversine distances between all pairs of observer-event coordinates, and store them in the distance matrix \( DM \). Utilize the coordinates from matrices \( O \) and \( E \), and apply the Haversine formula to calculate distances, ensuring they are represented in kilometers.

    \item \textit{Step 2: Determine Observer Centrality}
    Calculate the centrality of each observer, which is defined as the count of events that are within a specified radius \( r \). Use the indicator function to identify and sum the events that satisfy the distance criterion for each observer.

    \item \textit{Step 3: Generate Links}
    Identify and create links between observers and events that are within radius \( r \) of each other. Form a set \( L \) of pairs \((o, e)\) satisfying the distance constraint, representing the edges in the bipartite graph.
\end{enumerate}

\vspace{6pt}
\rule{\textwidth}{0.4pt} 
\section*{Event Classifier}

\textbf{Notations:}
\begin{itemize}
    \item \( E \): Matrix representing events, as defined in the "Bipartite Distance Matrix" section.
    \item \( DM \): Distance Matrix, as defined in the "Bipartite Distance Matrix" section.
    \item \( r \): Radius threshold.
    \item \( OE \) and \( UE \): Sets of indices of observed and unobserved events.
\end{itemize}

\textbf{Methodology:}
\begin{enumerate}
    \item \textit{Number of Observations Calculation:}
    Calculate the number of observations for each event.
    \[
    num_{observations}(e) = \sum_{i=1}^{|O|} \mathbb{I}(DM_{ie} \leq r) \, \forall e \in E
    \]
    where \( \mathbb{I} \) is the indicator function that is 1 if the condition inside is true and 0 otherwise, and \( |O| \) is the number of observers.
    
    \item \textit{Determine Observed and Unobserved Points:}
    Classify the events into observed and unobserved based on the number of observations.
    \[
    \begin{aligned}
        OE &= \{ e \,|\, num_{observations}(e) > 0, \, e \in E \} \\
        UE &= \{ e \,|\, num_{observations}(e) = 0, \, e \in E \}
    \end{aligned}
    \]
\end{enumerate}

\textbf{Algorithm:}
\begin{enumerate}
    \item \textit{Step 1: Calculate Observations -}
    Compute the number of observations for each event by summing the applicable entries within the radius \( r \) from the distance matrix \( DM \).

    \item \textit{Step 2: Classify Events} -
    Categorize the events into observed and unobserved by utilizing the number of observations for each event, creating sets \( OE \) and \( UE \) which denote the indices of observed and unobserved events, respectively.
\end{enumerate}

\vspace{6pt}
\rule{\textwidth}{0.4pt} 
\section*{Initialize ROBUST net}

\textbf{Notations:}
\begin{itemize}
    \item \( O \) and \( E \): Sets of RTCC (observers) and CFS (events).
    \item \( L \): Set of Links between observers and events.
    \item \( r \): Radius threshold, consistent with prior sections.
    \item \( OE \) and \( UE \): Sets of indices of observed and unobserved events, maintaining consistency with the "Event Classifier" section.
\end{itemize}

\textbf{Methodology:}
\begin{enumerate}
    \item \textit{Bipartite Distance Calculation:}
    Compute the distances among RTCC nodes and CFS nodes.
    \[
    DM, L  = \text{Bipartite\_Distance\_Matrix}(O, E, r)
    \]
    
    \item \textit{Determine Observed and Unobserved Points:}
    Identify which events are observed and which are not.
    \[
    OE, UE = \text{Event\_Classifier}(E, DM, r)
    \]
    
\end{enumerate}

\textbf{Algorithm:}
\begin{enumerate}
    \item \textit{Step 1: Calculate Distances}
    Compute the effectiveness and distances among RTCC nodes and CFS nodes, utilizing \( O \), \( E \), and \( r \) as inputs.
    
    \item \textit{Step 2: Classify Events}
    Determine which events are observed and unobserved, using \( E \), \( DM \), and \( r \) for the calculation.

\end{enumerate}

\vspace{6pt}
\rule{\textwidth}{0.4pt} 
\section*{Unipartite Distance Matrix}

\textbf{Notations:}
\begin{itemize}
    \item \( UE = \{e_1, e_2, \ldots, e_m\} \) be a set of unobserved event nodes.
    \item \( DM \) is a distance matrix, where \( DM_{ij} \) represents the distance between nodes \( e_i \) and \( e_j \).
    \item \( r \) is a radius threshold.
\end{itemize}

\textbf{Methodology:}
\begin{enumerate}
    \item \textit{Point Extraction:}
    Extract the longitudinal (\( x \)) and latitudinal (\( y \)) coordinates of unobserved nodes.
    \[
    P_x = \{x_1, x_2, \ldots, x_m\}, \quad P_y = \{y_1, y_2, \ldots, y_m\}
    \]

    \item \textit{Combination of Nodes:}
    Compute all possible combinations of nodes from \( UE \).
    \[
    C = \{(e_i, e_j) : e_i, e_j \in UE, i \neq j\}
    \]

    \item \textit{Haversine Distance Calculation:}
    Compute the haversine distance for each combination of nodes in \( C \) and construct the distance matrix \( DM \).
    \[
    DM_{ij} = \text{haversine}(e_i, e_j) \quad \forall (e_i, e_j) \in C
    \]
    where \(\text{haversine}(e_i, e_j)\) calculates the haversine distance between nodes \( e_i \) and \( e_j \).
    
    \item \textit{Distance Matrix Construction:}
    Create a matrix, \( DM \), of size \( m \times m \), where each element, \( DM_{ij} \), represents the haversine distance between nodes \( e_i \) and \( e_j \).
\end{enumerate}

\textbf{Algorithm:}
\begin{enumerate}
    \item \textit{Step 1:} Extract the \( x \) and \( y \) coordinates of unobserved nodes.
    \item \textit{Step 2:} Compute all possible combinations of nodes.
    \item \textit{Step 3:} Calculate the haversine distance between all combinations of nodes.
    \item \textit{Step 4:} Construct a distance matrix that contains the distances between all pairs of nodes.
\end{enumerate}

\vspace{6pt}
\rule{\textwidth}{0.4pt} 
\section*{Cluster Disconnected Events}

\textbf{Notations:}
\begin{itemize}
    \item \( UE = \{e_1, e_2, \ldots, e_m\} \) be a set of unobserved event nodes.
    \item \( DM \) is a distance matrix, where \( DM_{ij} \) represents the distance between nodes \( e_i \) and \( e_j \).
    \item \( r \) is a radius threshold.
    \item \( n \) is the desired number of densest clusters to return.
\end{itemize}

\textbf{Methodology:}
\begin{enumerate}
    \item \textit{Identifying Nodes within Radius:}
    Determine the nodes within radius \( r \) of each other using the distance matrix \( DM \).
    \[
    N(i) = \{j : DM_{ij} \leq r\}
    \]

    \item \textit{Creating Clusters:}
    Form clusters, \( C \), based on the proximity defined above.
    \[
    C_i = \{e_j : j \in N(i)\}
    \]

    \item \textit{Sorting Clusters by Density:}
    Sort clusters based on their size (density) in ascending order.
    \[
    C_{\text{sorted}} = \text{sort}(C, \text{key} = |C_i|)
    \]

    \item \textit{Identifying Densest Clusters:}
    Identify the densest clusters, ensuring that a denser cluster does not contain the centroid of a less dense cluster.
    Let \( D_C \) be the set of densest clusters:
    \[
    D_C = \left\{C_i \in C_{\text{sorted}} : C_i \text{ is maximal dense}\right\}
    \]
    where "maximal dense" means that there is no denser cluster that contains the centroid of \( C_i \).

    \item \textit{Selecting Top \( n \) Clusters:}
    Select the top \( n \) clusters from \( D_C \).
\end{enumerate}

\textbf{Algorithm:}
\begin{enumerate}
    \item \textit{Step 1:} Identify nodes within a specified radius (\( r \)) of each other using the precomputed distance matrix.
    \item \textit{Step 2:} Form clusters by grouping nodes that are within radius \( r \) of a given node.
    \item \textit{Step 3:} Sort the clusters based on their density (number of nodes).
    \item \textit{Step 4:} Identify the densest clusters ensuring a denser cluster does not contain the centroid of a less dense cluster.
    \item \textit{Step 5:} Select the top \( n \) densest clusters.
\end{enumerate}

\vspace{6pt}
\rule{\textwidth}{0.4pt} 
\section*{Add New Observers}

\textbf{Notations:}
\begin{itemize}
    \item \( UE \): Set of unobserved event nodes.
    \item \( r \): Radius threshold, consistent with prior sections.
    \item \( n \): Number of clusters to identify.
    \item \( DM \): Distance matrix among unobserved nodes.
    \item \( C \): Set representing the densest clusters.
\end{itemize}

\textbf{Methodology:}
\begin{enumerate}
    \item \textit{Unobserved Distances:}
    Compute the pairwise distances among all unobserved event nodes within a given radius \( r \) and represent them in a distance matrix \( DM \).
    \[
    DM = \text{unipartite\_distance\_matrix}(UE, r)
    \]
    where \(\text{unipartite\_distance\_matrix}(\cdot, \cdot)\) is a function that returns a distance matrix, calculating the pairwise distances between all points in set \( UE \) within radius \( r \).

    \vspace{6pt}
    \item \textit{Retrieve Densest Clusters:}
    Identify the \( n \) densest clusters among the unobserved event nodes \( UE \), utilizing the precomputed distance matrix \( DM \) and within the radius \( r \).
    \[
    C = \text{cluster\_disconnected}(DM, UE, r, n)
    \]
    where \(\text{cluster\_disconnected}(\cdot, \cdot, \cdot, \cdot)\) is a function that returns the \( n \) densest clusters from the unobserved events nodes \( UE \), using the precomputed distance matrix \( DM \) and within radius \( r \).
\end{enumerate}

\textbf{Algorithm:}
\begin{enumerate}
    \item \textit{Step 1: Compute Unobserved Event Distances}
   - Calculate the distance matrix \( DM \) for all unobserved event nodes within a specified radius \( r \).

    \item \textit{Step 2: Identify Clusters}
   - Determine the \( n \) clusters using the computed distance matrix \( DM \) and consider nodes within the specified radius \( r \).
\end{enumerate}

\renewcommand*\descriptionlabel[1]{\hspace\labelsep\normalfont #1:}

\section{Datasets}
This research revolves around constructing a ROBUST network to model relationships between events from CFS that are violent, and the location of crime cameras in New Orleans. This ROBUST network comprises observer nodes (symbolizing crime cameras) and observable nodes (representing violent event locations). Edges in the ROBUST network are created by pairing each CFS event node with the closest crime camera node based on spatial range. 

\subsection{Observer Node Set}
The observer node set comprises RTCC camera nodes utilized to monitor events in New Orleans. These nodes are either stationary or discretely mobile, meaning their location might instantaneously change between snapshots. Properties characterizing each observer node include:

\begin{table}[htbp]
\caption{ (RTCC) Observer Node Properties }
\begin{center}

\begin{tabular}{c l} 
 \hline\hline
 \textbf{Label} & \textbf{Description} \\ [0.5ex] 
 \hline\hline
 id & Unique identifier number for each node \\
 \hline
 membership & City-owned asset or Private-owned asset \\
 \hline
data source & Entity that operates the RTCC camera  \\
 \hline
  mobility & Stationary or mobile \\  
 \hline
 address &  Street Address of RTCC Camera \\
 \hline
  geolocation & Latitude and Longitude of platform \\  
 \hline
 x,y & Location in EPSG: 3452  \\  
 \hline
 
\end{tabular}
\end{center}
\end{table}

\begin{figure}[H]
\centering
\includegraphics[width=0.8\columnwidth]{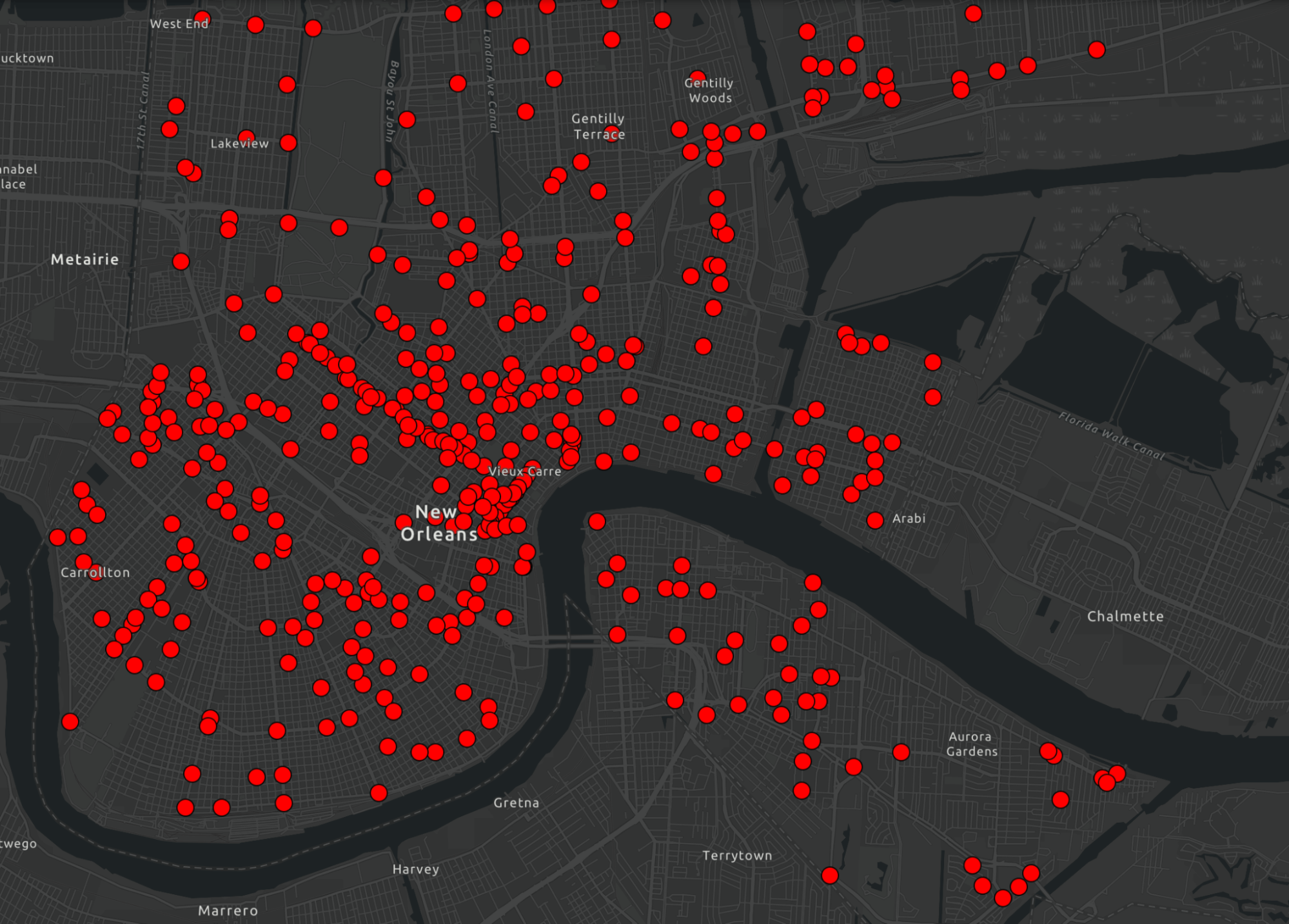}
\caption{RTCC Cameras. Locations of the RTCC cameras}
\label{fig:init-rtcc}
\end{figure}

\subsection{Observable Node Set}
CFS events serve as observable nodes in the ROBUST network, symbolizing violent event locations reported to the NOPD. Properties of each CFS event node include:

\begin{table}[htbp]
\caption{(CFS) Event Node Properties }
\begin{center}

\begin{tabular}{c l} 
 \hline\hline
 \textbf{Label} & \textbf{Description} \\ [0.5ex] 
 \hline\hline
 NOPD Item & Unique item number for the incident. \\ 
 \hline
 Type & The type associated with the CFS. \\ 
 \hline
 Type Text & TypeText associated with the CFS.
 \\ 
 \hline
 Priority & The priority associated with the CFS. \\  
& 3-highest, 2-emergency, 1-nonemergency, 0-none \\
\hline
InitialType &  InitialType associated with the CFS. \\
 \hline
 InitialTypeText & The description associated with the CFS. \\
 \hline
 Initial Priority & The initial priority associated with the CFS. \\
 \hline
 MapX & The X coordinate for CFS  in state plane. \\
 \hline
MapY & The Y coordinate for CFS  in state plane. \\
\hline
Time Create &  Creation time of the incident in the CAD. \\
\hline
Time Dispatch & Dispatch time by NOPD to the incident. \\
\hline
Time Arrive & Arrival time by NOPD to the incident. \\
\hline
Time Closed & Time the incident was closed in the CAD. \\
\hline
Disposition & The disposition associated with the CFS. \\
\hline
DispositionText & Text associated with the CFS. \\
\hline
Self-Initiated & If the Officer generates the CFS. \\
\hline
Beat & The area in Parish where occurred. \\
& The first number is District, the letter is the zone, \\
& and the numbers are the subzone.\\
\hline
Block Address & The BLOCK unique address number for CFS \\
\hline
Zip & The NOPD Zip associated with the CFS. \\
\hline
Police District &  The NOPD Police District associated with CFS. \\
\hline
Geolocation &  Latitude and Longitude \\
\hline

\end{tabular}
\end{center}
\end{table}

To pinpoint pertinent observable events, CFS events undergo filtering based on type and location. Only specific types within a designated police zone are taken into account. This filtering ensures that observable events of interest (violent events) occur within the targeted region.

\subsubsection{Violent Types}
The violent events considered in this research encompass homicides, non-fatal shootings, armed robberies, and aggravated assaults. By integrating both type and location filters, the resulting locations for observable nodes are derived.

\begin{figure}[H]
\centering

\includegraphics[width=0.75\columnwidth]{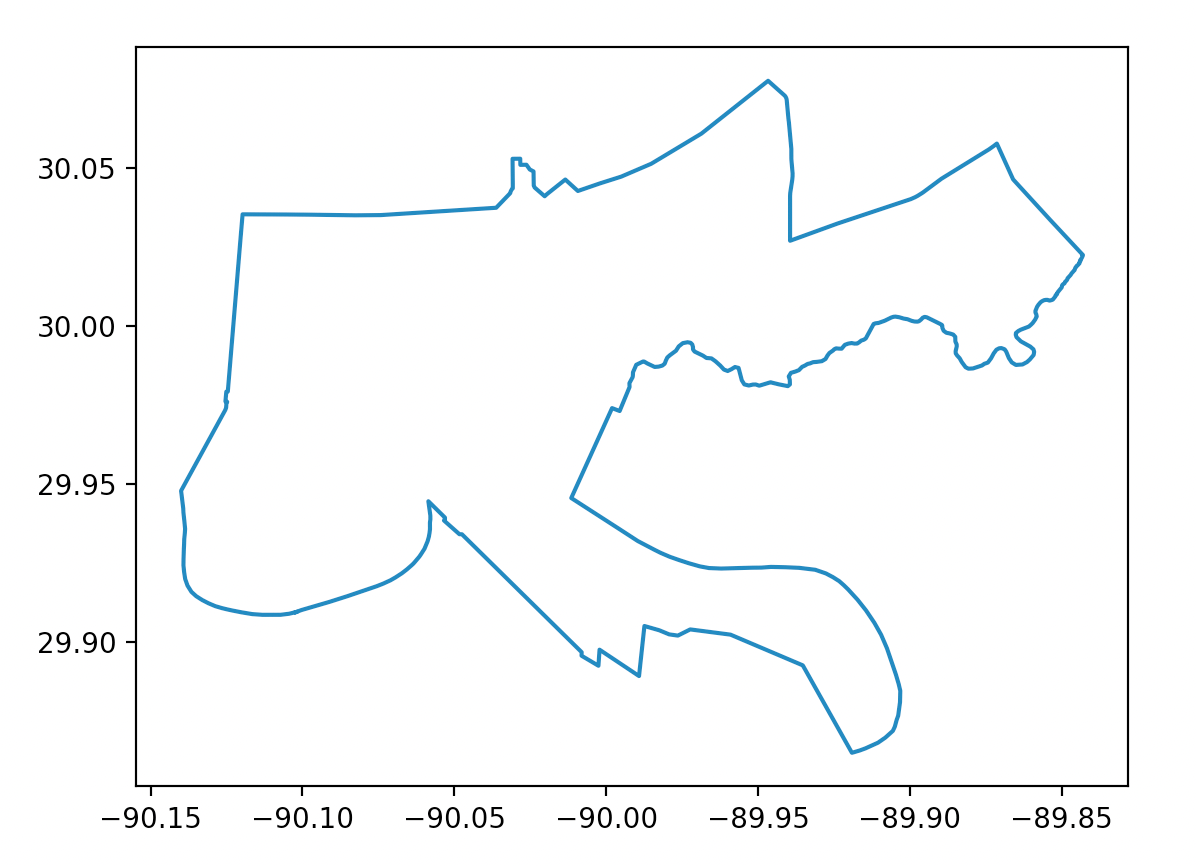}
\caption{Orleans Parish Inclusion Zone Polygon}
\label{fig:nola-inclusion-zone}
\end{figure}

\begin{table}[htbp]
\caption{ CFS Violent Types } 
\begin{center}

\begin{tabular}{c l} 
 \hline\hline
 \textbf{Violent Type} & \textbf{Description} \\ [0.5ex] 
 \hline\hline
 30, 30-C, 30-S  & Homicide - (C)Cutting, (S)Shooting \\
 \hline
 34, 34-C, 34-S & Aggravated Battery - (C)Cutting, (S)Shooting \\
  \hline
 35 & Simple Battery \\
  \hline
 37 & Aggravated Assault \\
  \hline
 38 & Simple Assault \\
  \hline
 55 & Aggravated criminal damage to property \\
  \hline
 60 & Aggravated burglary \\
  \hline
 64, 64-G, 64-J, 64-K & Armed robbery - (G)un, car(J)Jacking, (K)nife \\
  \hline
 65, 65-J, 65-P & Simple robbery \\
  \hline
 ASLT, ASLTI, ASLTWP & Simple Assault, (I) Injury, (WP) Weapon \\
  \hline
 MURDERST & Murder by Shooting \\
  \hline
 ROB, ROBCJ, ROBW & Robbery, (CJ) Carjacking, (W) Weapon \\
  \hline
 SHOTP & Assault with Weapon - Shooting \\
  \hline
 STABP & Assault with Weapon - Stabbing \\
  \hline
 STFIRED & Shots Fired \\ 
 \hline

\end{tabular}
\end{center}
\end{table}

\section{Experimental Setup}

Reproducibility and accessibility are paramount in our experimental approach. Thus, experiments were conducted on Google Colaboratory, chosen for its combination of reproducibility—via easily shareable, browser-based Python notebooks—and computational robustness through GPU access, particularly employing a Tesla T4 GPU to leverage CUDA-compliant libraries. The software environment, anchored in Ubuntu 20.04 LTS and Python 3.10, utilized modules such as cuDF, Dask cuDF, cuML, cuGraph, cuSpatial, and CuPy to ensure computations were not only GPU-optimized but also efficient. The availability of code and results is ensured through the platform, promoting transparent and repeatable research practices.

\section{Results}
The insights from the research, visualized through graphs, heatmaps, and network diagrams, delineate the foundational attributes and effectiveness of the ROBUSt network, providing a baseline for analyses and comparison with state-of-the-art approaches.

\subsection{Initial ROBUST network}
The initial state of the ROBUST network, before any methods were applied, is characterized using various visualization techniques as described below.

\begin{figure}[h!]
    \centering
    \includegraphics[width=0.95\columnwidth]{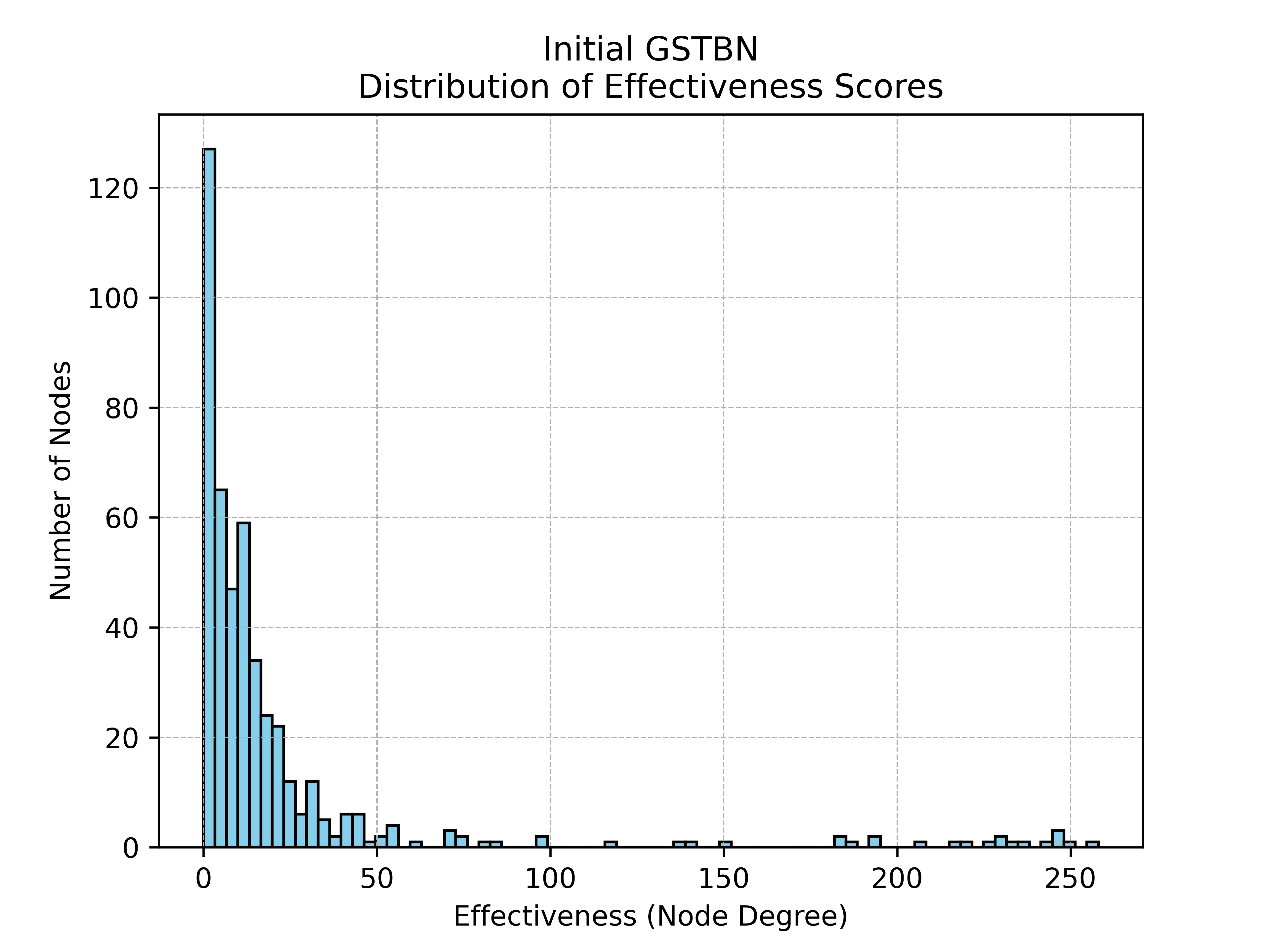}
    \caption{Histogram of effectiveness scores (node degrees) for observer nodes in the initial ROBUTS network, demonstrating a power-law distribution. The x-axis represents the degree count, signifying the ability of an observer node to detect events within a specified radius, while the y-axis shows the count of nodes for each degree.}
    \label{fig:degree_distro}
\end{figure}

\begin{figure}[h!]
    \centering
    \includegraphics[width=0.95\columnwidth]{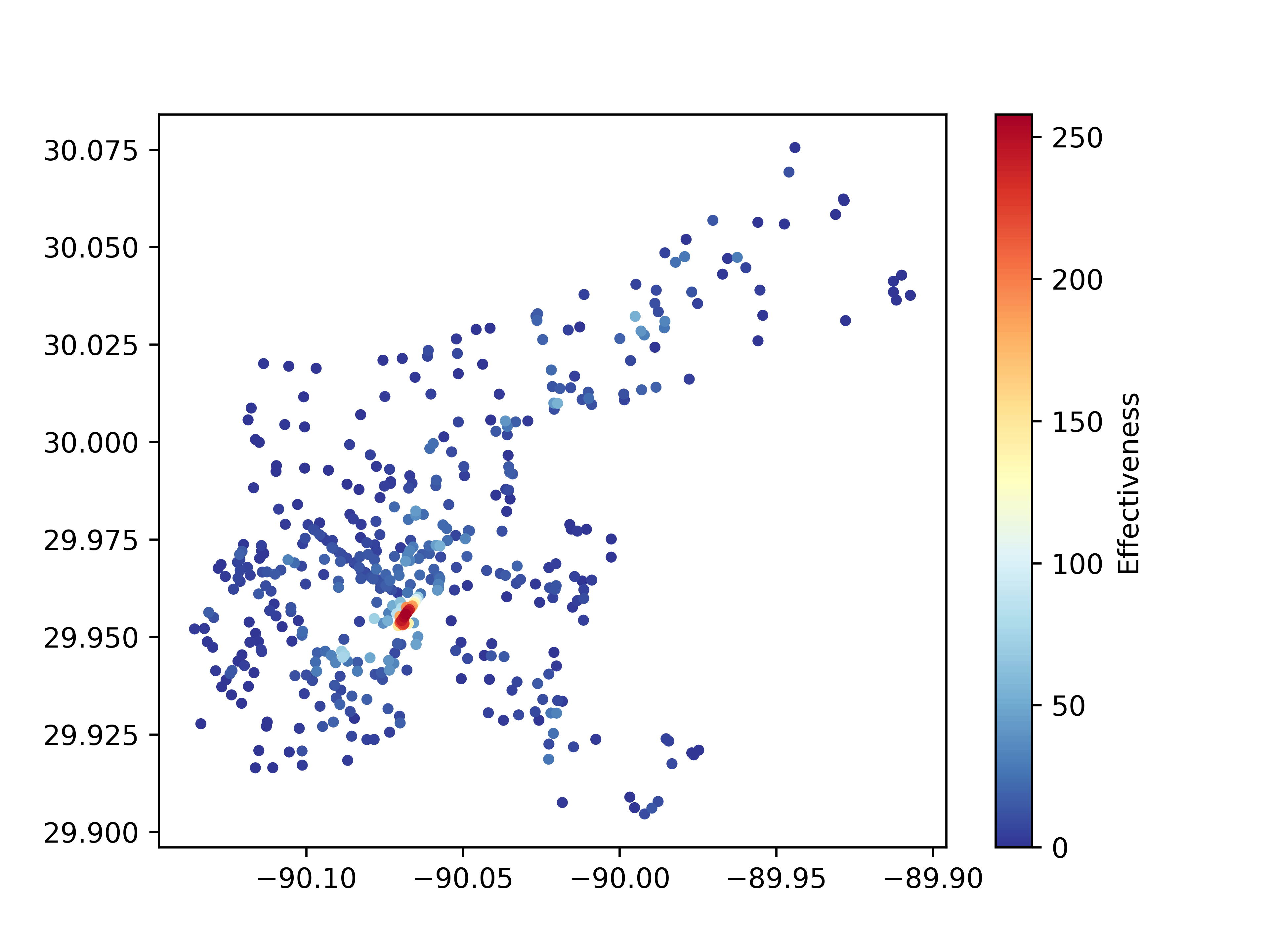}
    \caption{Heatmap of observer node effectiveness scores in the initial ROBUST network. Node spatial locations are plotted with color intensity indicating degree centrality and a gradient from blue (low effectiveness) to red (high effectiveness).}
    \label{fig:effectiveness_heatmap}
\end{figure}

\begin{figure}[h!]
    \centering
    \includegraphics[width=\columnwidth]{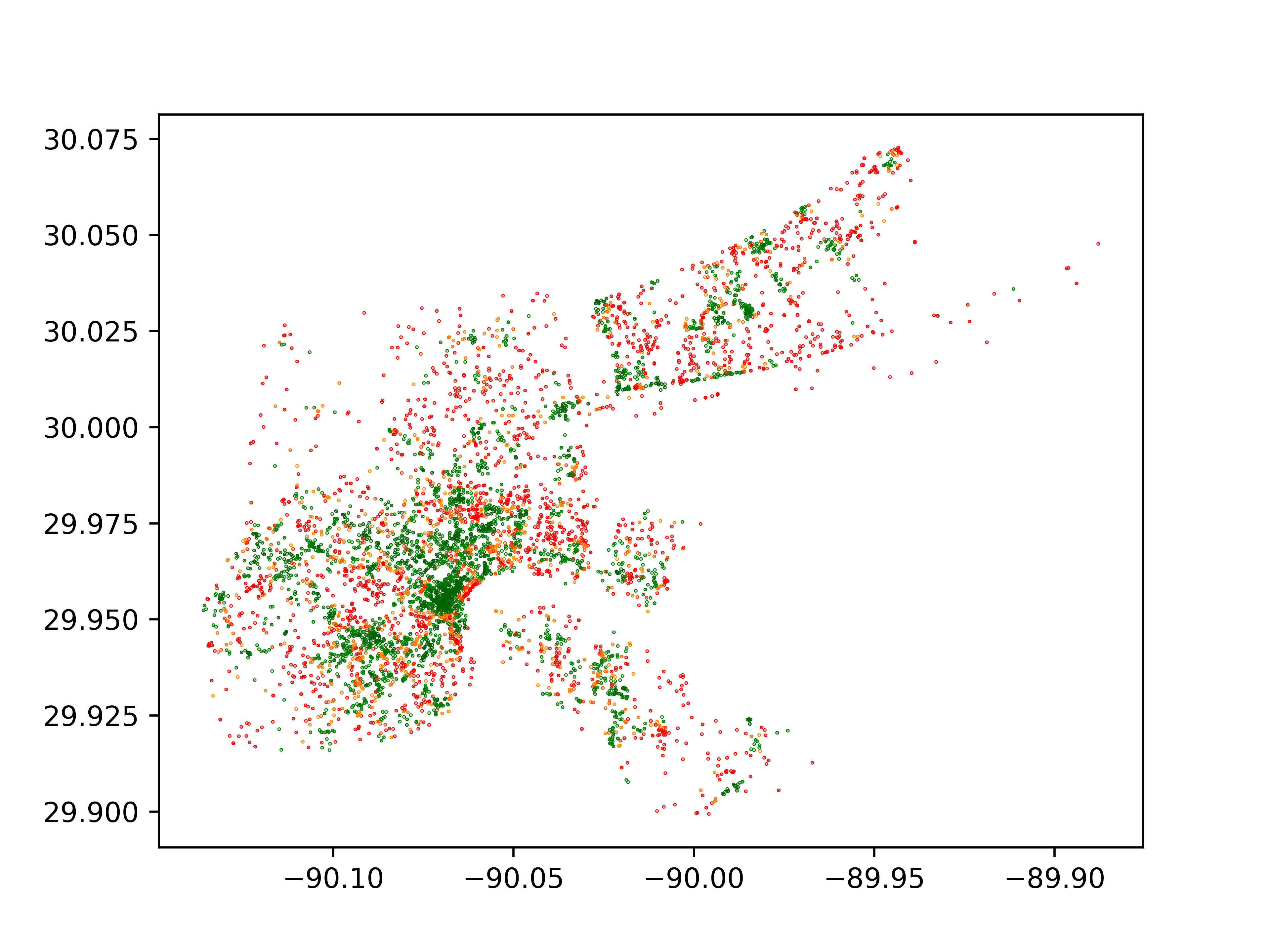}
    \caption{Spatial distribution and observational coverage of events in the initial ROBUST network. Event locations are color-coded to indicate observational coverage: deep green for multiple observers, green for a single observer, varied colors for near-observers, and red for unobserved events.}
    \label{fig:initial_gstbn_event_coverage}
\end{figure}

\FloatBarrier  

\subsection{Proposed Method Results: Proximal Recurrence}
The integration of 100 new nodes into the ROBUST network using the proximal recurrence strategy (with a specified radius of 0.2) produced both visual and quantitative outcomes, which are depicted and summarized in the following figures.

\begin{figure}[h!]
\centering
\includegraphics[width=\columnwidth]{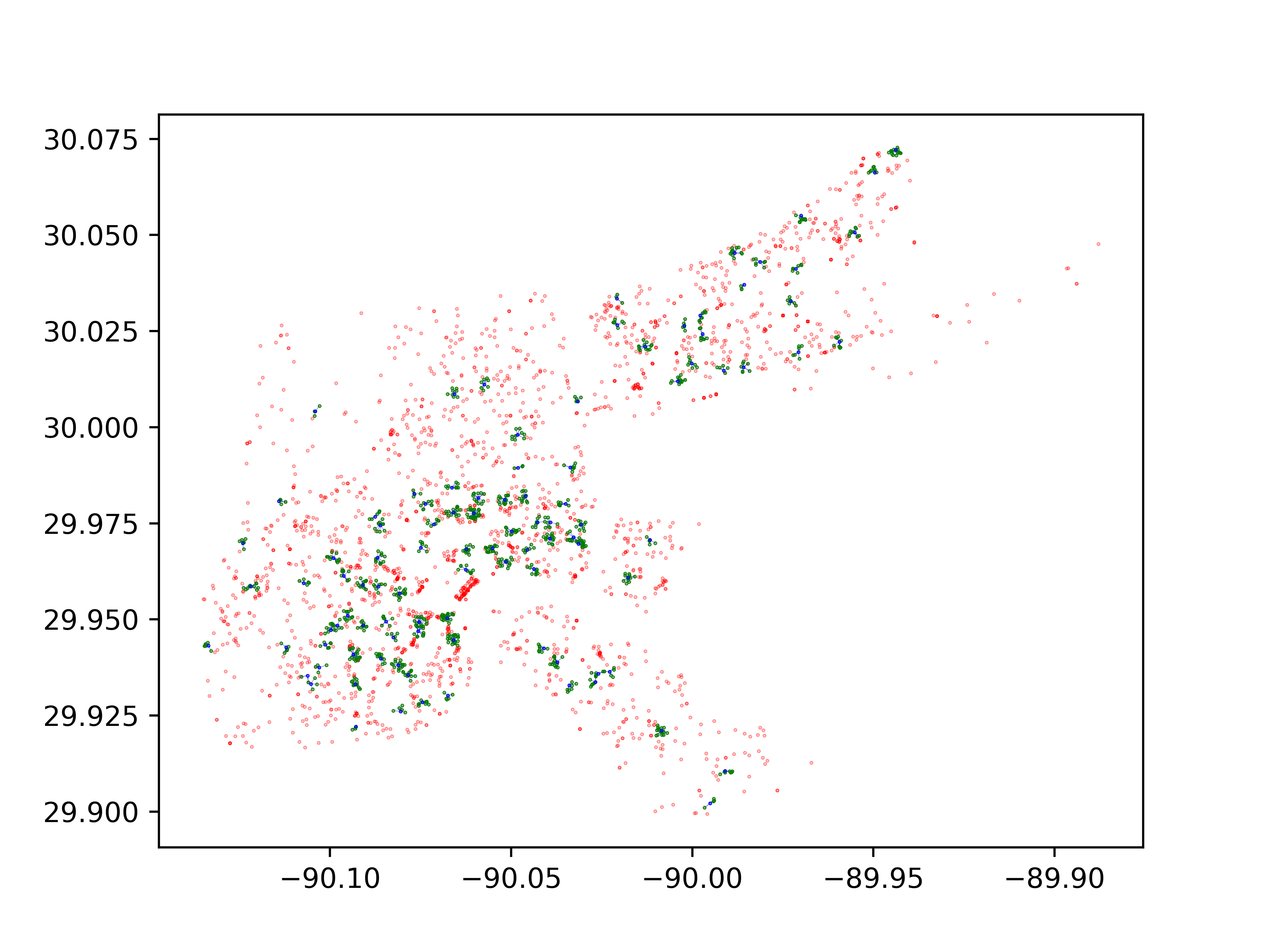}
\caption{ROBUST network visualization post proximal recurrence integration. New nodes are represented in blue, directly observed nodes in green, and unobserved nodes in red, with edges indicating observational relationships.}
\label{fig:gstbn_proximal_recurrence_subnet_with_red}
\end{figure}

\begin{figure}[h!]
\centering
\includegraphics[width=0.75\columnwidth]{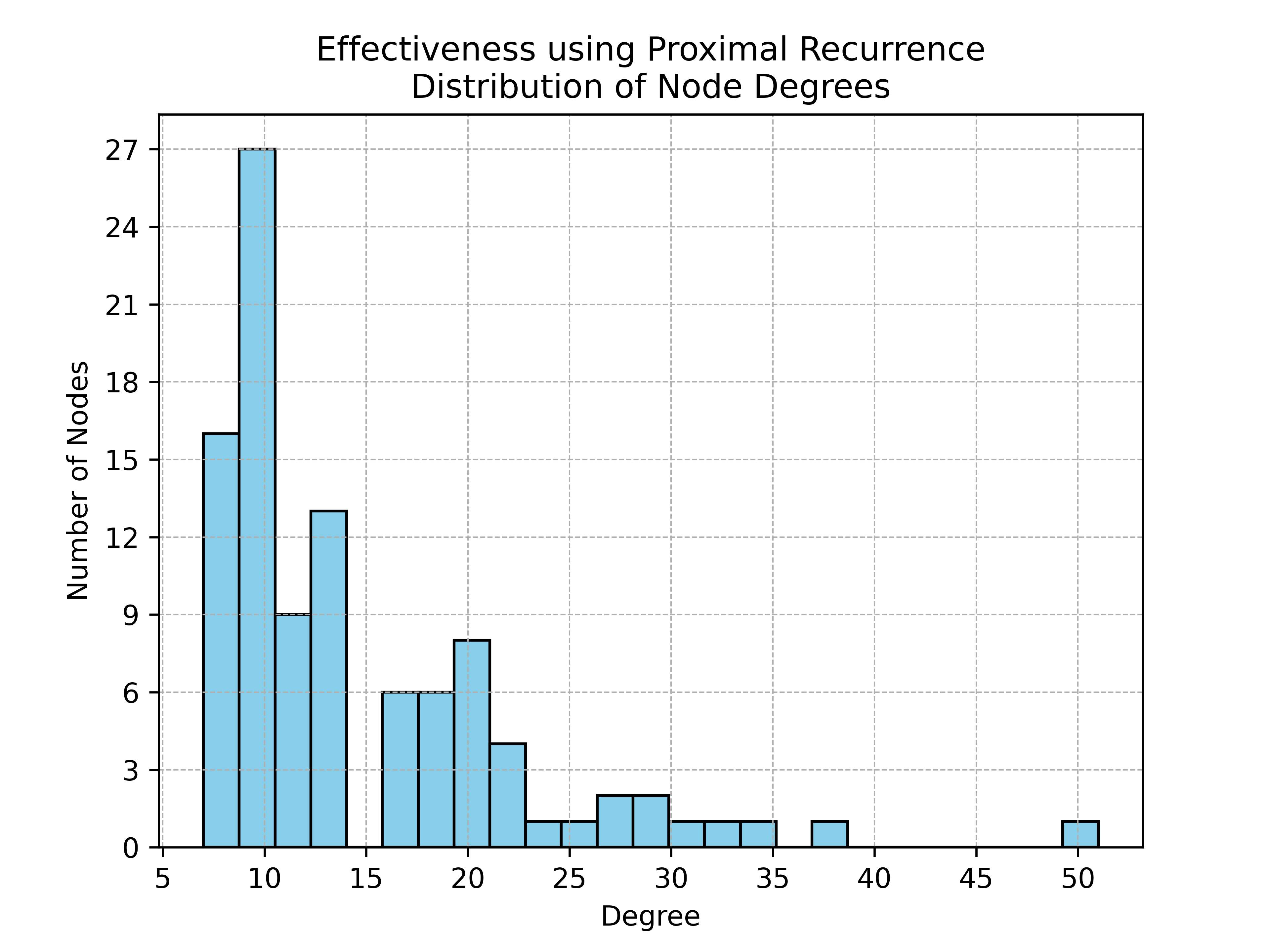}
\caption{Histogram of node degree distribution for the new nodes introduced through the proximal recurrence strategy. Axes represent degree and node count, respectively.}
\label{fig:effectiveness_proximal_recurrence}
\end{figure}

\begin{figure}[h!]
\centering
\includegraphics[width=0.75\columnwidth]{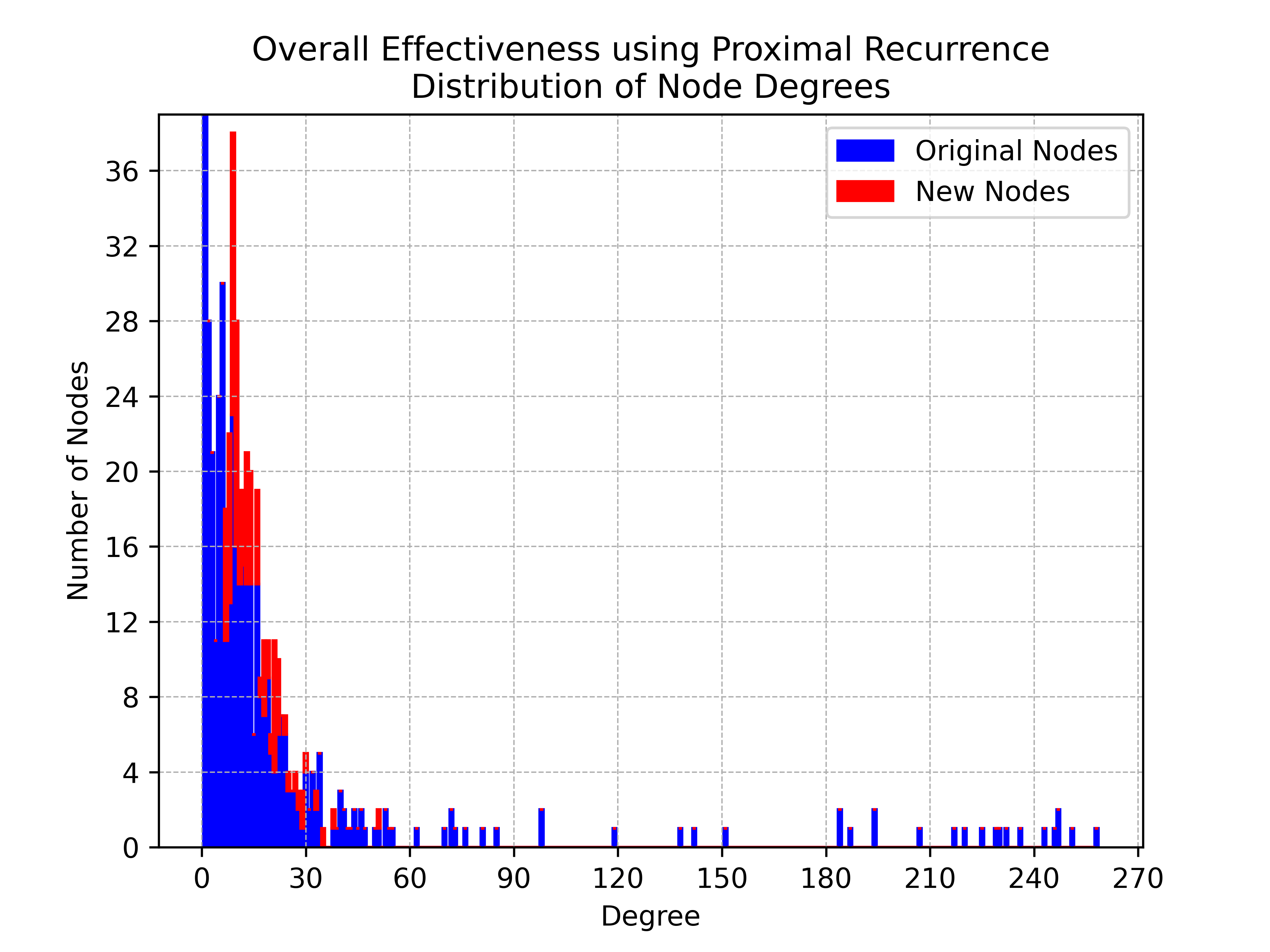}
\caption{Aggregate node degree distribution of the updated ROBUST network, with original nodes in blue and new nodes superimposed in red.}
\label{fig:overall_effectiveness_proximal_recurrence}
\end{figure}

\subsection{Comparison with Existing Methods}
This subsection presents the results of other techniques like k-means, mode, average, binning, and DBSCAN.
\\

\subsubsection{\textbf{DBScan}}
The DBScan clustering approach and its resultant effectiveness are visualized and analyzed below through various graphical representations and histograms.

\begin{figure}[h!]
    \centering
    \includegraphics[width=\columnwidth]{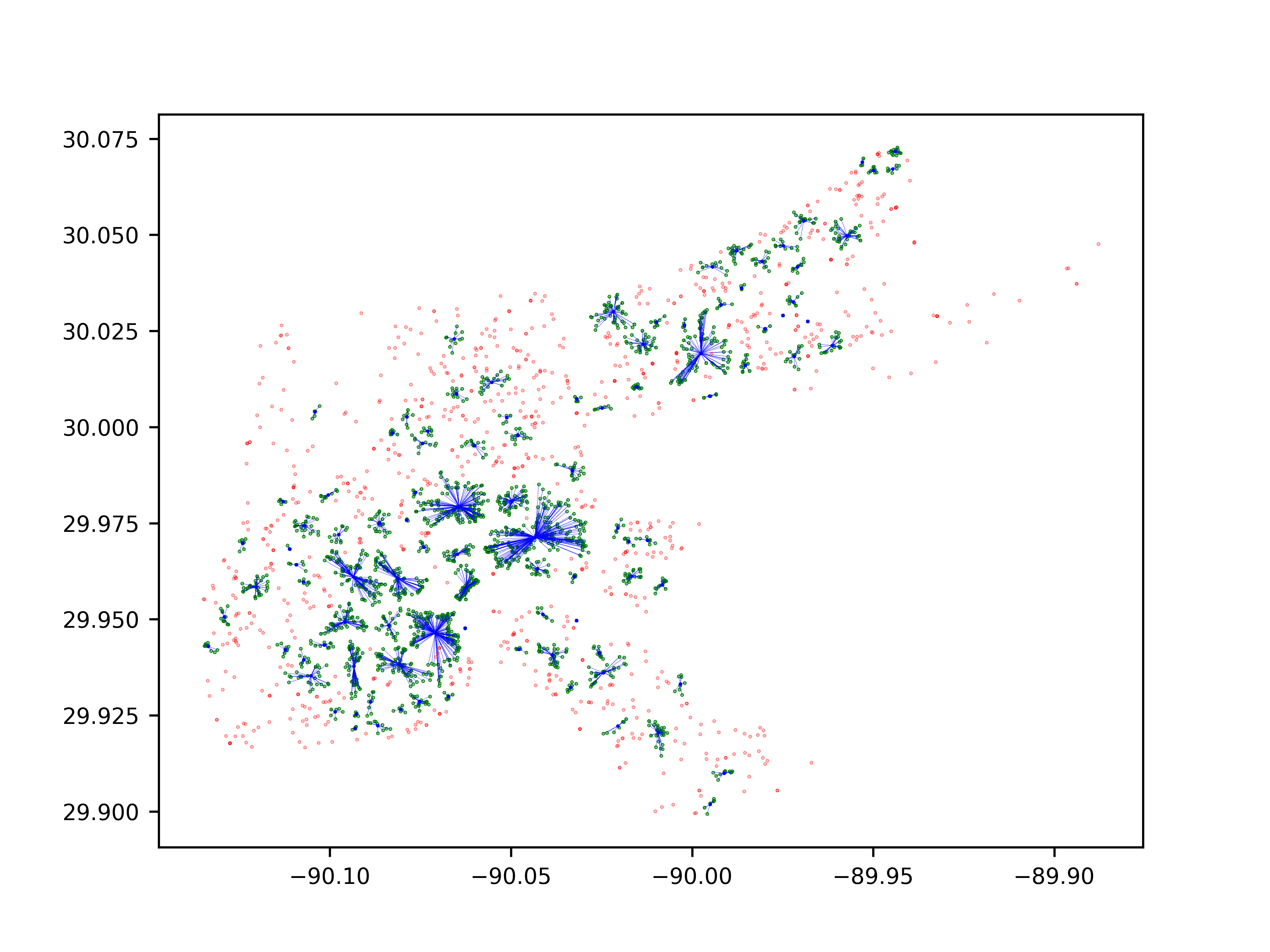}
    \caption{Centroids and associated events determined using DBSCAN clustering on unwitnessed events within the ROBUST network. Centroids (blue), clustered nodes (green), and unclustered nodes (red) are depicted, with edges connecting centroids and respective nodes.}
    \label{fig:dbscan_centroids_nonrange}
\end{figure}

\begin{figure}[h!]
    \centering
    \includegraphics[width=\columnwidth]{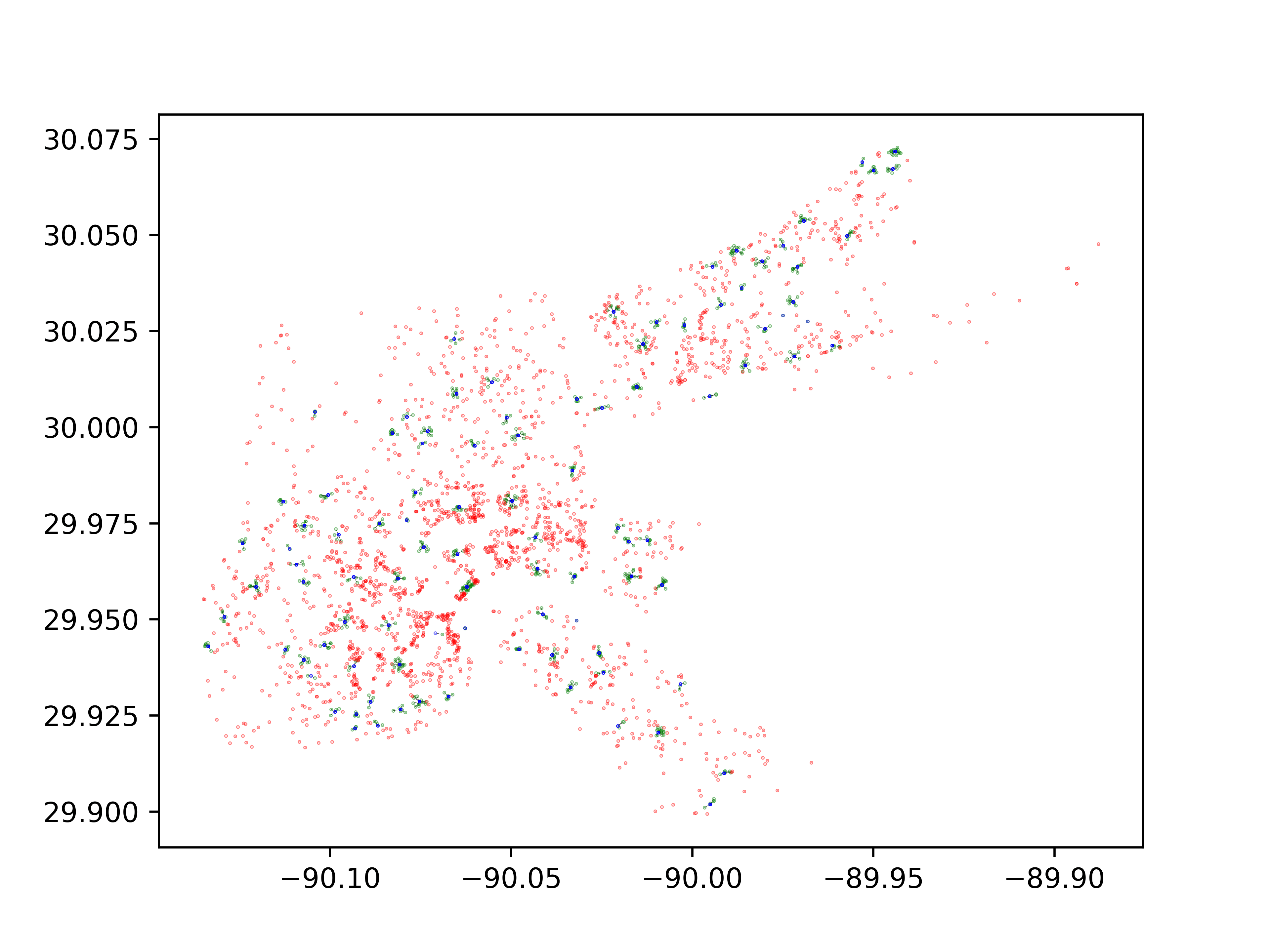}
    \caption{Node insertions based on DBSCAN clustering, emphasizing new DBSCAN nodes and filtering out events and nodes not within the centroid's spatial range.}
    \label{fig:dbscan_cluster_nonrange_gstbn}
\end{figure}

\begin{figure}[h!]
    \centering
    \includegraphics[width=0.75\columnwidth]{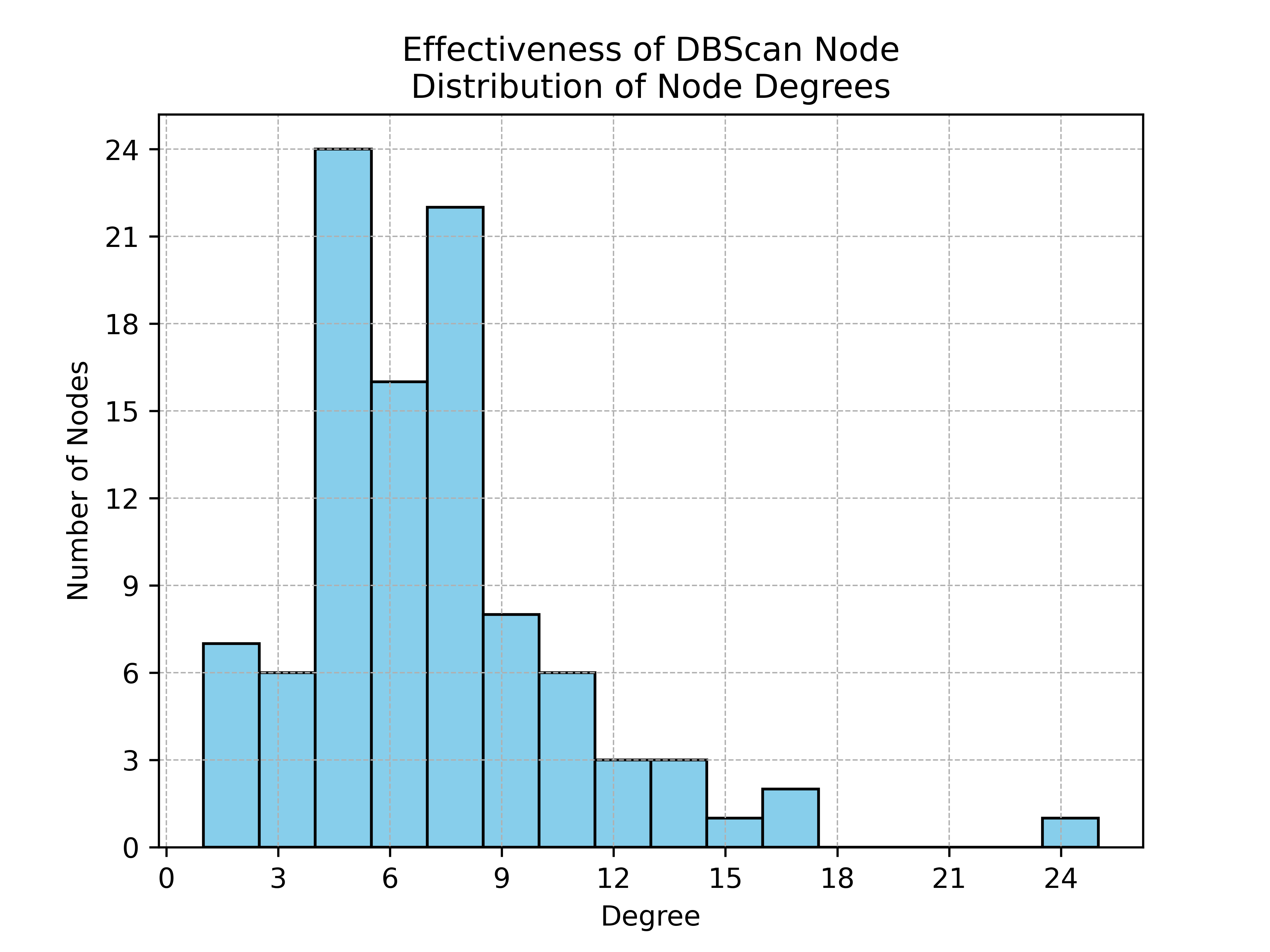}
    \caption{Distribution of node degrees for new nodes in the ROBUST network via the DBScan algorithm, with axes representing degree and node quantity.}
    \label{fig:effectiveness_dbscan_node}
\end{figure}

\begin{figure}[h!]
\centering
\includegraphics[width=0.75\columnwidth]{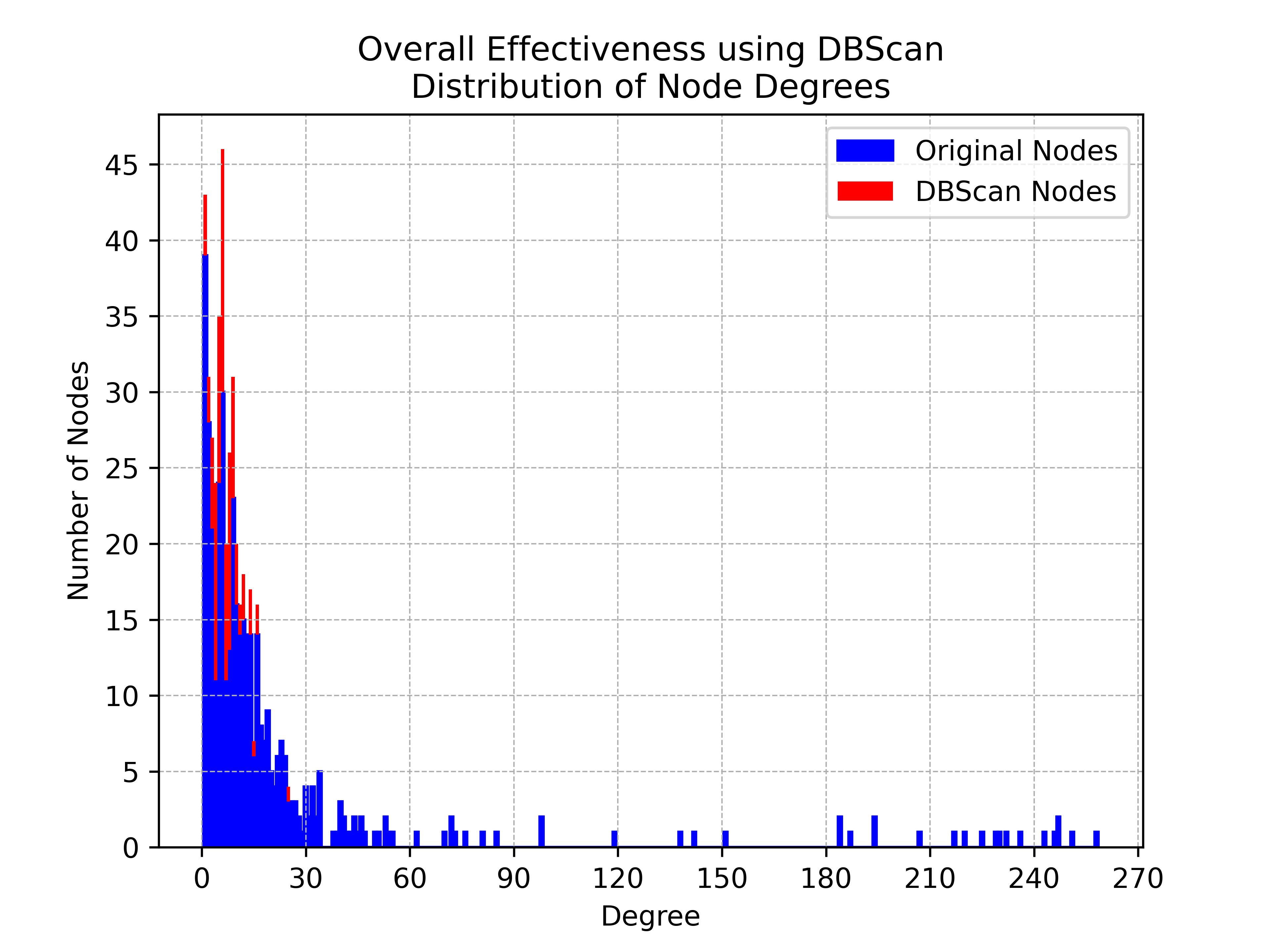}
\caption{Overall node degree distribution in the updated ROBUST network, showcasing the original distribution (blue) and the new DBScan nodes (red).}
\label{fig:overall_effectiveness_dbscan_nodes}
\end{figure}

\subsubsection{\textbf{K-means}}
The effectiveness of the K-means clustering method is illustrated through a series of visualizations and histograms.

\begin{figure}[h!]
    \centering
    \includegraphics[width=\columnwidth]{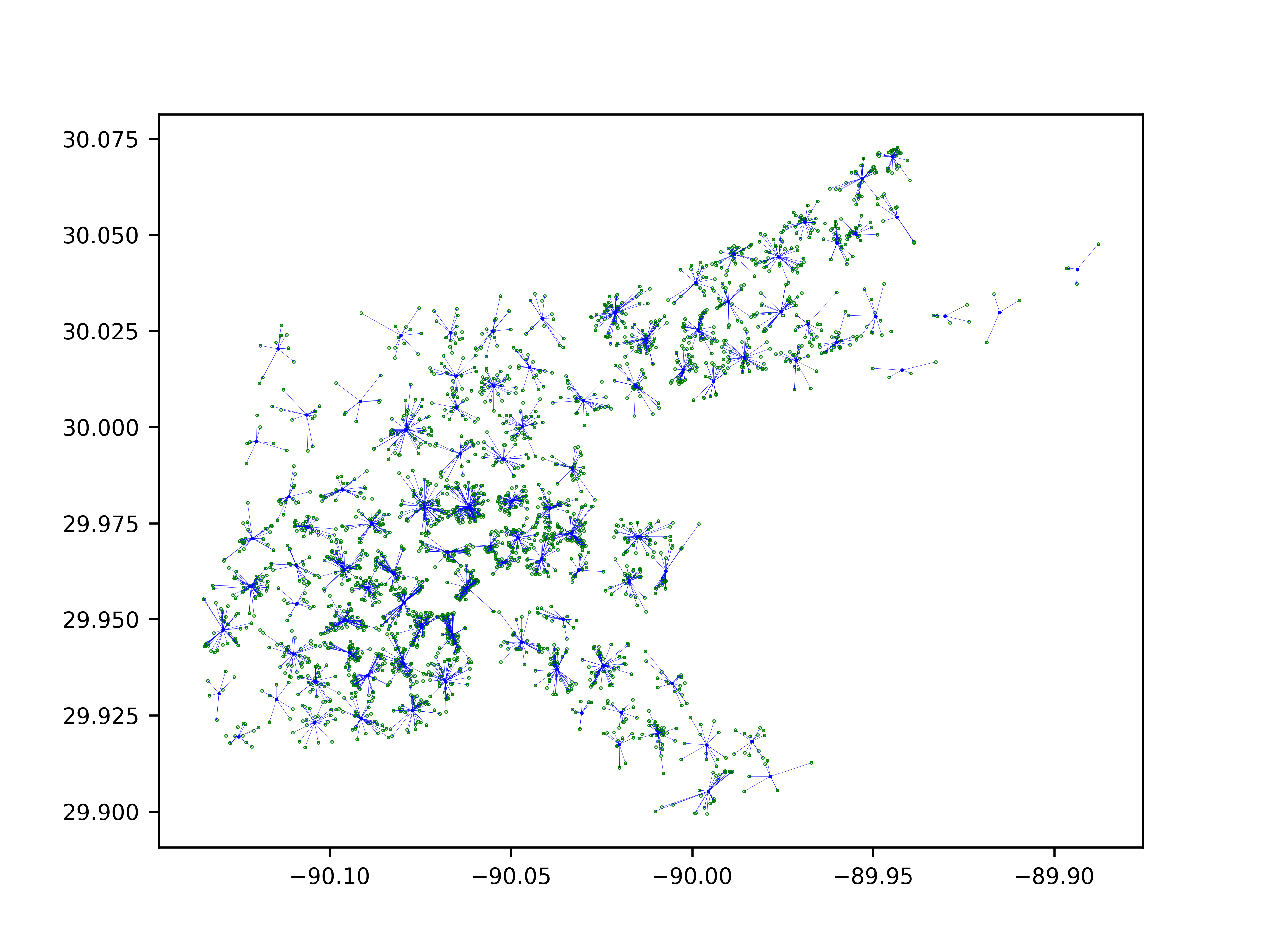}
    \caption{Visualization using K-means clustering on unwitnessed events within the ROBUST network. Centroids are in blue, clustered nodes in green, and they are interconnected with edges.}
    \label{fig:kmeans_centroids_nonrange}
\end{figure}

\begin{figure}[h!]
    \centering
    \includegraphics[width=\columnwidth]{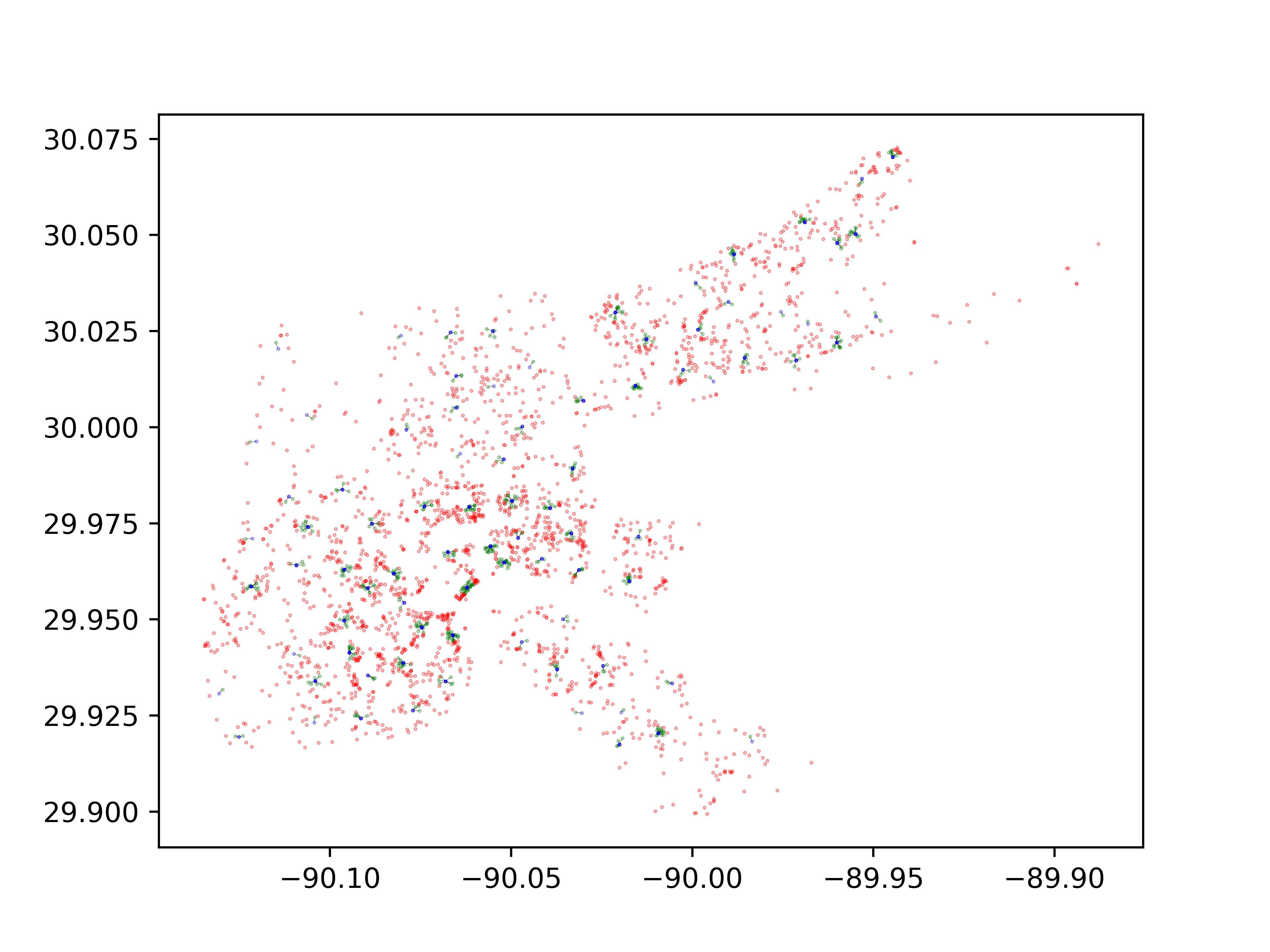}
    \caption{Centroid-based node insertions using K-means clustering, focusing on new nodes while filtering those out of the spatial range of the centroid.}
    \label{fig:kmean_cluster_nonrange_gstbn}
\end{figure}

\begin{figure}[h!]
    \centering
    \includegraphics[width=0.75\columnwidth]{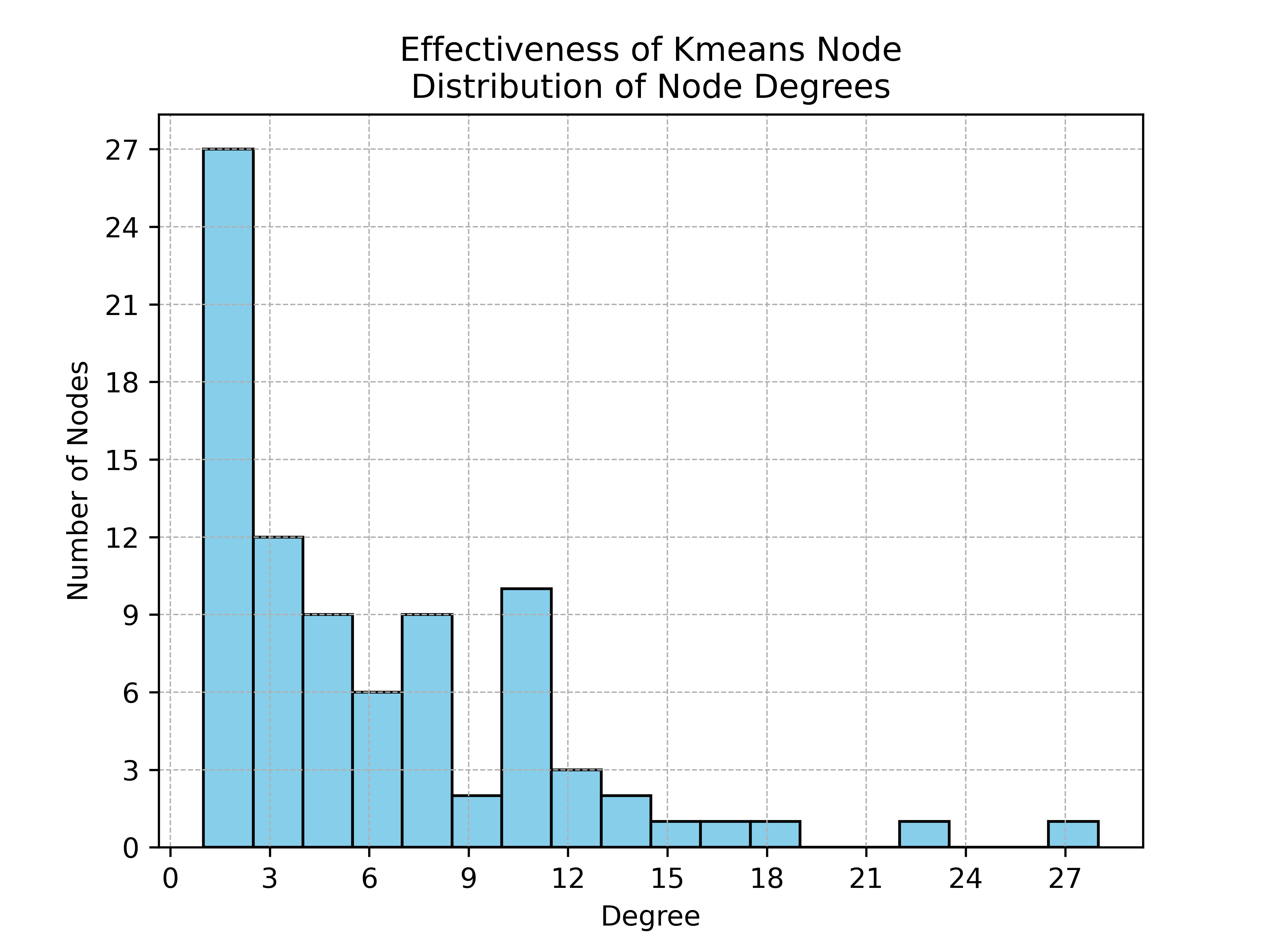}
    \caption{Histogram showing the distribution of node degrees for nodes added through the K-means method.}
    \label{fig:effectiveness_kmeans_node}
\end{figure}

\begin{figure}[h!]
\centering
\includegraphics[width=0.75\columnwidth]{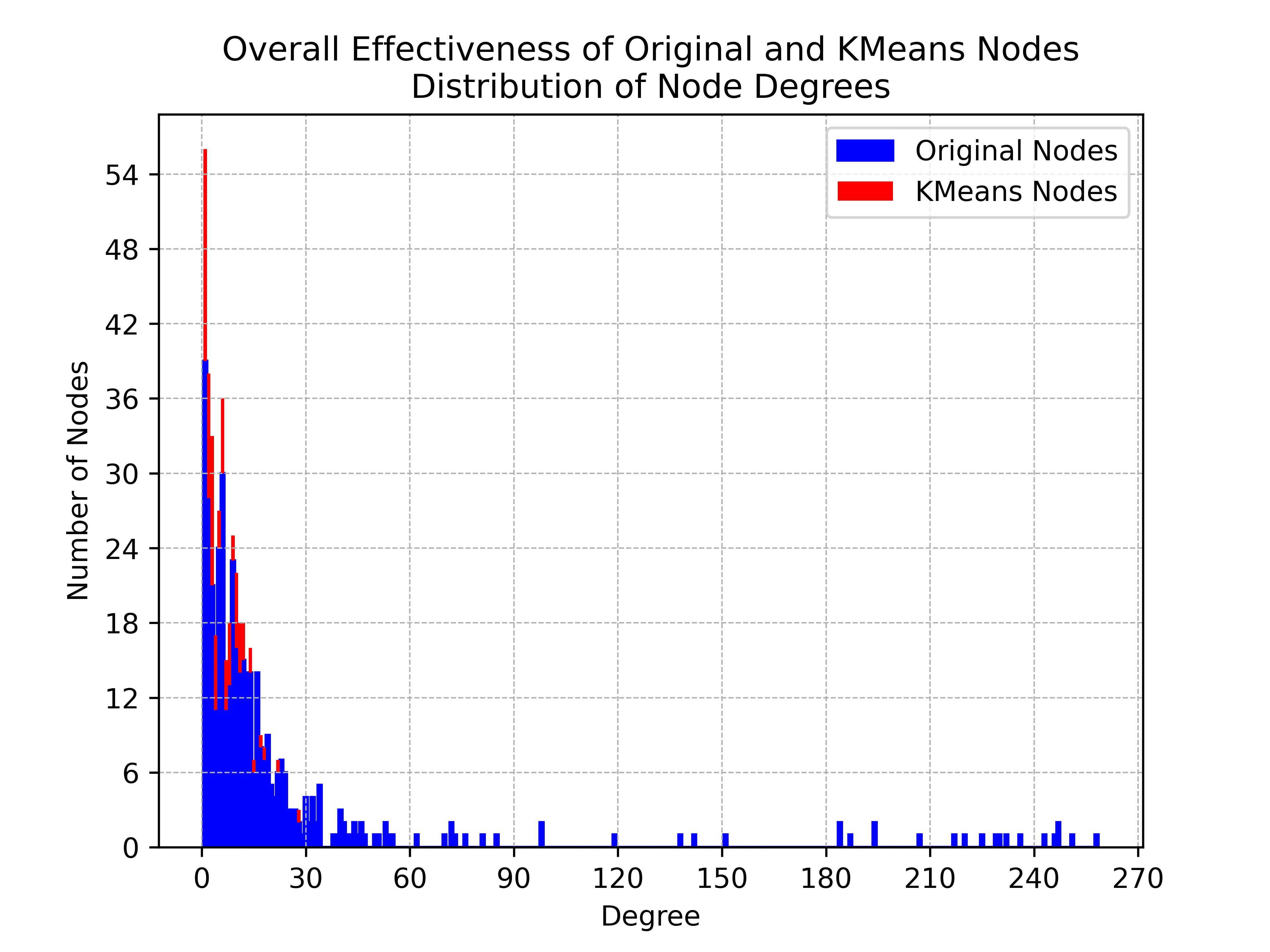}
\caption{Combined histogram indicating the node degree distribution in the updated ROBUST network, distinguishing between original (blue) and K-means added nodes (red).}
\label{fig:overall_effectiveness_kmeans_nodes}
\end{figure}

\subsubsection{\textbf{Mode Clustering}}

\begin{figure}[h!]
\centering
\includegraphics[width=\columnwidth]{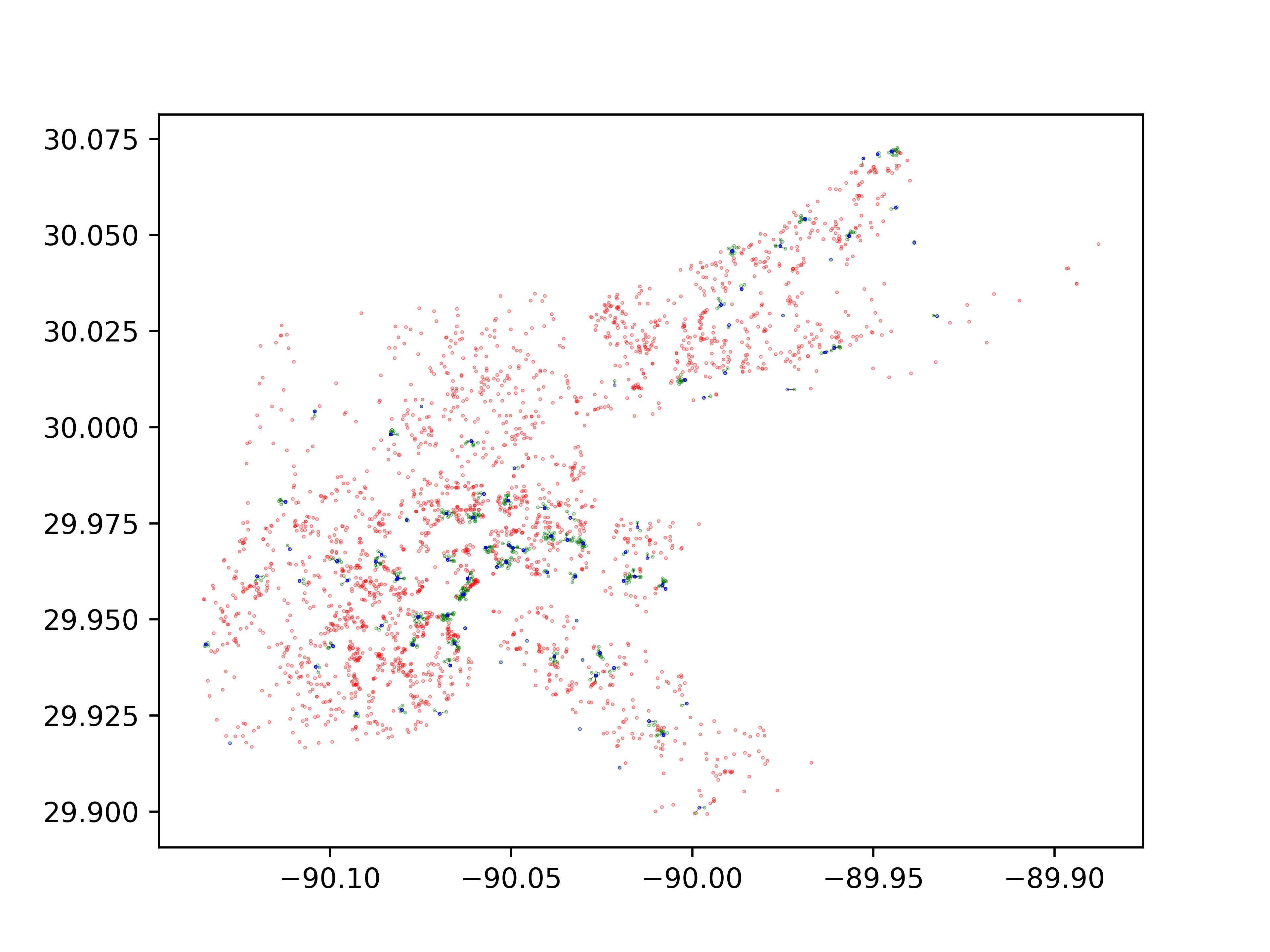}
\caption{ROBUST network utilizing the Mode-adjusted variant of K-means approach, featuring new nodes (blue), directly observed nodes (green), and remaining unobserved nodes (red).}
\label{fig:mode_standard_gstbn_unobserved}
\end{figure}

\begin{figure}[h!]
\centering
\includegraphics[width=0.75\columnwidth]{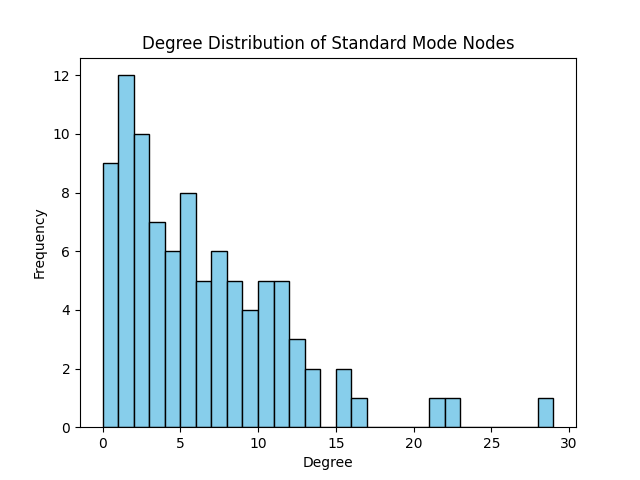}
\caption{Histogram indicating the distribution of node degrees for nodes added via the Standard Mode strategy.}
\label{fig:effectiveness_mode_standard_nodes}
\end{figure}

\begin{figure}[h!]
\centering
\includegraphics[width=0.75\columnwidth]{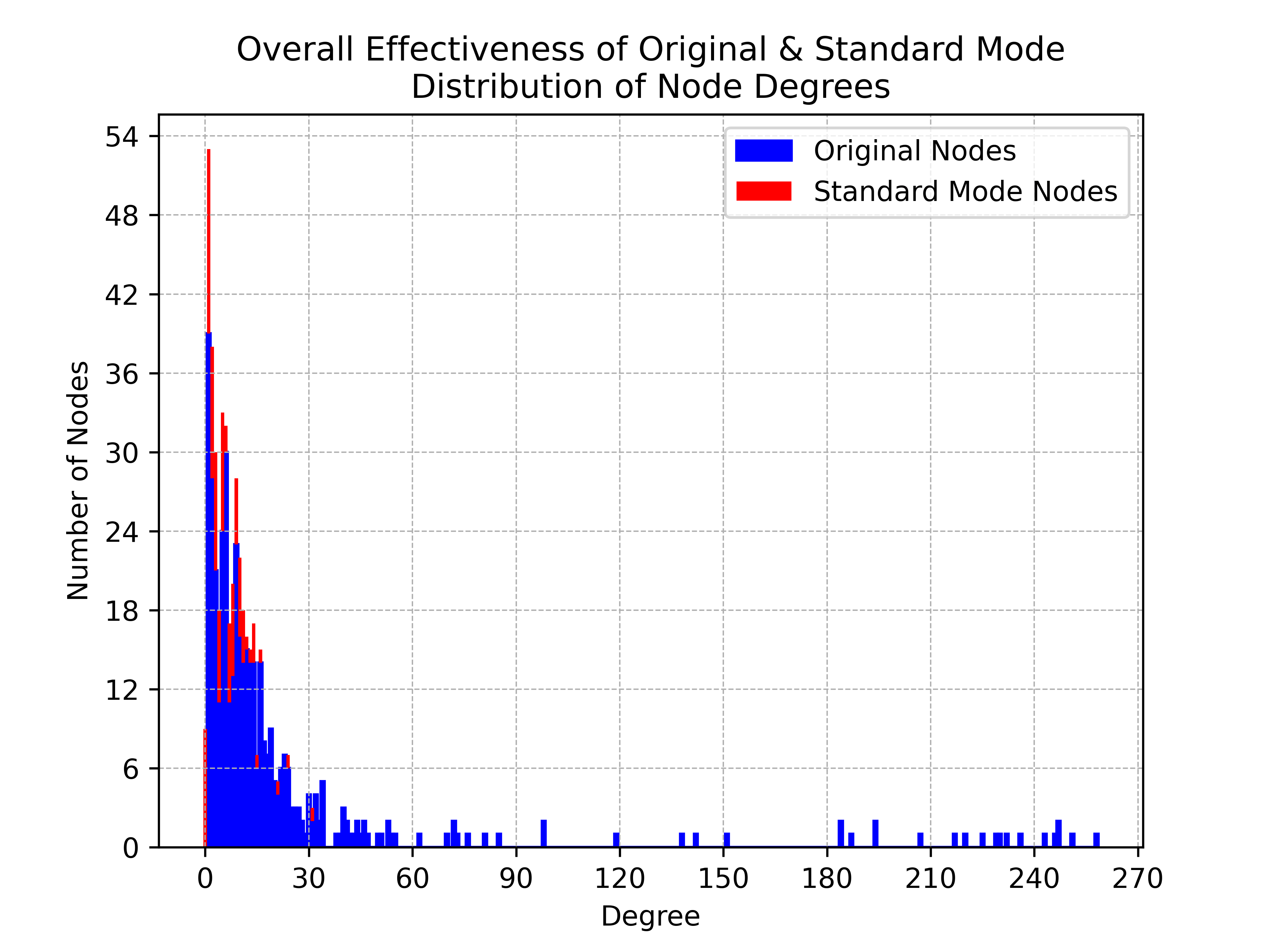}
\caption{Histogram of node degree distribution in the ROBUST network, illustrating both original (blue) and Standard Mode added nodes (red).}
\label{fig:overall_effectiveness_mode_kmeans_nodes}
\end{figure}

\section{Case Study Discussion}

\subsection{Evaluating Effectiveness}

\subsubsection{Degree Centrality}
Utilizing a bipartite behavior, the network connects observers to observable events through edges. The centrality of a node quantifies its efficacy in witnessing events. The distribution of node degrees provides insights into the network's capability to capture events through its observers.

\subsubsection{Desired Transformations in Degree Centrality Distributions}
The network initially exhibits a power-law distribution in degree centrality, where few nodes are highly effective and most are not. An objective is to shift this to a skewed Gaussian distribution by introducing new nodes, thereby redistributing node effectiveness and enhancing network performance. The combined histogram of original and new node centrality acts as a metric for evaluating this shift. A rightward histogram shift, visible in the results, indicates enhanced network performance due to the inclusion of new nodes with higher degree centrality. This transformation and the resulting outcomes set a benchmark for evaluating different clustering strategies in subsequent sections.

\subsection{Comparative Analysis of Clustering Approaches}

\subsubsection{Limitations of Conventional Clustering Strategies}
Conventional strategies like k-means and DBSCAN manifest limitations in controlling cluster diameter and centroid assignment, which can obscure genuine data patterns and hinder the transition from a power-law to a skewed Gaussian distribution upon the addition of new nodes. The proximal recurrence approach offers improved control and contextually relevant centroid assignment, addressing these issues.

\subsubsection{The Challenge of Centroid Averaging}
In domains such as crime analysis, centroid averaging, or calculating a point that represents the mean position of all cluster points, can misrepresent actual incident locations in spatially significant contexts. Traditional centroid assignment can produce misplaced centroids, leading to inaccurate analyses and strategies, such as incorrectly identifying crime hot spots. The proximal recurrence method ensures accurate data representation by managing centroid assignments and controlling cluster diameters, particularly in contexts requiring precise spatial data point accuracy.

\subsection{Inadequacies of Mode-Based Analysis}

\subsubsection{Neglect of Spatial Clustering}
Mode-based analysis effectively identifies high-recurrence points but neglects the spatial clustering of nearby values. This oversight can miss opportunities to place observer nodes in locations where a cluster of neighboring points within a specific range might yield higher effectiveness than a singular high-incidence point.

\subsubsection{Lack of Comprehensive Insights}
While mode-based analysis hones in on recurrence, it disregards insights from considering spatial and neighboring data, identifying locations with high incidents but failing to define a cluster shape. Consequently, it misses locations near points of interest, potentially omitting optimal node identification. The approach discussed counters some of these deficiencies by incorporating both recurrence and proximate incidents into its analysis, offering a more balanced view.

\subsection{Advantages of Proximal Recurrence Approach}

\subsubsection{Harmonizing Recurrence and Spatiality}
The proximal recurrence approach integrates incident recurrence and spatial analysis, ensuring a comprehensive data perspective. It does not only identify high-incident locations but also accounts for spatial contexts and neighboring events, safeguarding against isolated data interpretation and enhancing the analysis's comprehensiveness.

\subsubsection{A Middle Ground: Specificity vs. Generalization}
The approach balances specificity and generalization by identifying incident points and considering their spatial contexts. It prevents the dilution of specific data points in generalized clustering and avoids omitting relevant neighboring data in specific point analysis. It thereby ensures the analysis remains accurate and prevents omission of pivotal data points.

\subsubsection{Defined Constraints for Enhanced Performance}
The integration of defined constraints, such as a predetermined cluster diameter, optimizes performance and ensures contextually relevant insights. By adhering to predefined limits, the approach maintains a balanced analysis, ensuring insights are precise and applicable to real-world scenarios.

\subsection{Comparison with Histogram Approach}

\subsubsection{Hyperparameter Tuning and Large Datasets}
Histograms require tuning of the bin number, a task that can be particularly intricate for large datasets due to its impact on analysis and result quality. Conversely, the proximal recurrence approach demands no hyperparameter tuning, providing straightforward applicability without iterative refinement.

\subsubsection{Accuracy, Precision, and Flexibility}
Histograms may lose fine data structures by aggregating data into bins. In contrast, the proximal recurrence approach evaluates points on actual distances, capturing accurate clustering patterns and enabling identification of non-rectangular clusters, thereby providing a more realistic data pattern representation.

\subsubsection{Adaptability and Relationship Consideration}
While histograms use uniform bin sizes, potentially misrepresenting datasets with varied density regions, the proximal recurrence approach adapts to different densities by evaluating proximity, not a fixed grid, and utilizes explicit pairwise distances, ensuring direct measurement of point relationships.

\subsubsection{Customizability and Consistency}
While histograms offer bin size as the primary customizable parameter, the proximal recurrence approach provides parameters, such as distance thresholds, for tuning based on data and problem specifics, and ensures consistent results, unaffected by alignment considerations.

\section{Case Study Conclusion}

The exploration of ROBUST network, through various strategic node insertions and clustering methodologies, has yielded insights into the potential and limitations of different strategies in enhancing network performance, specifically in the realm of observer node effectiveness in detecting events. The initial state of the ROBUST network was characterized by a power-law distribution in node degree centrality, prompting a need for improved uniformity and elevated effectiveness in event detection across the network. 

\subsection{Key Findings}

\begin{enumerate}
    \item \textbf{Shift in Degree Centrality Distribution:}
    The integration of new nodes, utilizing various methodologies, influenced a shift towards a skewed Gaussian distribution in node degree centrality, reflecting an enhanced effectiveness in event detection across the ROBUST network. This is a pivotal step in optimizing the network’s observational capabilities.
    
    \item \textbf{Comparative Efficacy of Clustering Approaches:}
    While traditional clustering methods like k-means and DBSCAN revealed certain limitations in centroid assignment and cluster diameter control, the proximal recurrence approach demonstrated a noteworthy capability in managing these aspects, thereby enhancing context-relevant centroid assignment and optimizing cluster sizes.
    
    \item \textbf{Mode-Based Analysis:}
    The standard mode approach, despite its effectiveness in identifying high-recurrence points, revealed inadequacies in considering spatial clustering, often bypassing optimal node insertion points. 
    
    \item \textbf{Proximal Recurrence Approach:}
    Balancing specificity and generalization, the proximal recurrence approach not only identified high-incident points but also considered the spatial context, thus preventing data interpretation in isolation and enhancing the comprehensiveness and applicability of the analysis.
    
\end{enumerate}

\subsection{Implications}

The findings underscore the necessity of selecting an appropriate clustering strategy, one that not only identifies optimal node insertion points but also considers the spatial context and underlying distribution of data points in the network. Furthermore, these strategies should ensure accurate and contextually relevant data representation, especially in domains requiring stringent spatial data accuracy.

\subsection{Future Directions}

While the proximal recurrence approach shows promise, future research could explore:
\begin{itemize}
    \item \textbf{Algorithm Optimization:} Refining algorithms for varied use-cases and exploring the scalability of these strategies in larger, more complex networks.
    \item \textbf{Real-World Application:} Implementing these strategies in real-world scenarios, such as environmental monitoring systems, to validate and further refine them based on practical outcomes.
    \item \textbf{Additional Metrics:} Investigating additional metrics for network optimization and evaluating their impact on the network’s observational capabilities.
    \item \textbf{Integration with Machine Learning:} Exploring how machine learning algorithms could further optimize node placement and network effectiveness.
\end{itemize}

\subsection{Closing Remarks for Case Study}

The journey through varied clustering strategies and their respective impacts on ROBUST network's observational capabilities has shed light on not only the potential for optimization but also the intrinsic complexities and considerations pivotal to network enhancement. Balancing the need for specificity, spatial accuracy, and data representation accuracy remains paramount in optimizing such networks for enhanced, comprehensive event detection and observational capabilities.

\chapter{Case Study III: Multiagent Planning}
Multiagent Spatiotemporal Planning Strategies for AUVs with ROBUST Network. This research introduces an approach for optimizing a fleet of Autonomous Underwater Gliders (AUVs) using the principles of the ROBUST (Ranged Observer Bipartite-Unipartite SpatioTemporal) network, applied to the dynamic environment of the Gulf of Mexico. Leveraging the HYCOM oceanographic dataset, regions of high temporal variability in parameters such as temperature, water currents, and salinity are identified. A proximal recurrence clustering technique generates temporal waypoints that capture these variabilities. The uniqueness of the ROBUST network lies in its Bipartite-Unipartiteness grouping, allowing for an evaluation of current sensor placements. This facilitates the filtering and identification of unobserved regions, from which a waypoint network is built. The integration of these waypoints within the ROBUST network framework enables the multiagent system of AUVs to plan optimal paths, considering the spatiotemporal dynamics of the ocean and the coordination between multiple AUVs for maximal data collection efficiency.

\section{Introduction}
Effective deployment and coordination of Autonomous Underwater Gliders (AUVs) in the Gulf of Mexico's dynamic environment pose a complex spatiotemporal challenge. This research seeks to optimize path planning for AUV fleets, pioneering a long-term planning approach over traditional short-term, immediate-reward strategies. The primary objective is to maximize overall sampling efficiency across the entire mission, considering spatial constraints, such as observational range and movement speed, while minimizing redundant data collection. Although some placements might be sub-optimal at specific times, the strategy is designed for optimal overall mission outcomes. Strategic waypoint nodes are defined in a ROBUST network, focusing on areas of high temporal variability, with the goal of enhancing HYCOM model forecasts. This enhancement is achieved by utilizing the data collected by AUVs to improve the model's current state, leading to more precise predictions.  \cite{Holmberg2014}

\section{Background}
\subsection{HYCOM}
HYCOM (Hybrid Coordinate Ocean Model) is a three-dimensional, gridded marine model with hindcasting, nowcasting, and forecasting capabilities. It provides both a high-resolution global view and zoomed regional views such as the Gulf of Mexico. HYCOM assimilates data from various sources, including satellites, buoys, and AUVs, using the Navy Coupled Ocean Data Assimilation (NCODA) system \cite{Chassignet2007}. This data assimilation phase establishes a feedback loop with the sensor operators, to encourage strategic sensor relocation into better locations that improve the present state of HYCOM.  This research uses HYCOM's regions of high temporal variability to target sensor deployment to improve data collection for those points. \cite{Holmberg2022}

\subsection{Underwater Autonomous Gliders}
Gliders are inexpensive AUVs that navigate water using buoyancy changes and wings. Designed for extended missions, they can follow either pre-set or adjusted routes, making them well-suited for long-term data collection. They move slowly, at about 0.5 knots, and while their low power consumption and durability are advantageous for sustained operations, they do face challenges in strong currents. These gliders collect temperature, salinity, and water current measurements, transmitting the data via satellite for rapid ocean monitoring. \cite{Rudnick2015}

\section{Related Work}

\subsection{Proximal Recurrence Clustering}
Proximal Recurrence, a spatiotemporal clustering method, identifies clusters by analyzing both the frequency of occurrences and the spatial proximity of neighboring points, with parameters controlling clustering ranges. \cite{Holmberg2023}

This technique scans data points, considering their spatial coordinates and occurrence frequency. It identifies points of high recurrence and evaluates their spatial groupings within a defined range of influence. Points are then assigned to the densest cluster they belong to. A key feature to this method is that it allows points to remain unclustered, instead prioritizing consistency in cluster diameters.  

Its effectiveness compared to other clustering methods like k-means or DBSCAN in specific spatiotemporal scenarios suggests its potential use in this study for the purpose of identifying way-point locations. However, originally proposed for binary classification of observed and unobserved events, for use in crime camera efficiency analysis, adapting Proximal Recurrence to gradient-based data like the severity of temporal variability poses a unique challenge.

\subsection{STROOB Networks}
SpatioTemporal Ranged Observer-Observable Bipartite (STROOB) networks model the interaction between sensors (observers) and events (observables) in a network format. These networks apply graph theory to assess sensor effectiveness on specific events. This assessment guides the optimization of new sensor node placements within the network. \cite{Holmberg2023}

STROOB networks were initially developed for optimizing the placement of static sensors, offering a foundation for evaluating sensor effectiveness in designated areas. However, their original design did not encompass spatiotemporal path planning. This limitation presents a unique opportunity to extend STROOB network principles to the domain of multi-agent path planning, allowing for a more versatile application in varying operational contexts.

\subsection{Pathing Strategies}
\subsubsection{Dijkstra's Shortest Path}
Dijkstra's algorithm is a straightforward technique for finding the shortest path in graphs, and it serves as the initial framework for exploring pathing strategies in this research. Traditionally, the algorithm iteratively determines the shortest path from a starting node to all other nodes within a static graph. However, this approach lacks temporal considerations. \cite{Wu2014} To address this, the study adapts Dijkstra's method to incorporate temporal factors, laying the groundwork for developing a more sophisticated, temporally aware algorithm suitable for dynamic environments. 

\subsection{Temporal Greedy Strategy with Look-Ahead}
The temporal greedy approach with look-ahead represents a short-term planning strategy that seeks immediate rewards while incorporating a degree of future reward-seeking. This method aims to maximize gains in the current timeframe, yet it differs from traditional greedy algorithms by considering imminent future events. \cite{Rigas2020}

\paragraph{Algorithm: }
The key steps in the algorithm are as follows:

\begin{footnotesize}
\begin{enumerate}[label=\arabic*., noitemsep, topsep=0pt, partopsep=0pt]
\item For each timestep, iterate through each sensor.
\item Check if the current timestep offers an optimal point.
    \begin{enumerate}[label*=\alph*., noitemsep]
    \item If no optimal point is present, look ahead to the next timestep.
    \item Plan a move towards a point of interest in the next frame.
    \end{enumerate}
\item If an optimal point is present, select the next point with the maximum immediate reward.
\item Update the path and covered points for the sensor.
\end{enumerate}
\end{footnotesize}

\section{Approach}
This research adopts a comprehensive strategy to optimize the paths of AUVs in the GoM. The focus is on dynamically identifying and responding to significant temporal events, through a network system. The strategy involves:

\begin{enumerate}
    \item \textbf{Event Identification}: Leveraging temporal variability in HYCOM to pinpoint areas of significant change.
    \item \textbf{Waypoint Network Development}: Crafting a network that adapts to these identified events, into waypoints nodes, constrained by the AUV movement and viewing properties.
    \item \textbf{SpatioTemporal Path Planning}: Integrating predictive models and planning tools to ensure efficient, responsive navigation of multiple AUVs.
\end{enumerate}

Each phase is designed to build upon the other, resulting in a framework for adaptive multi-agent path planning in complex environments.

\section{Methods}

\subsection{Weighted Proximal Recurrence Clustering}
The Weighted Proximal Recurrence (WPR) method in this implementation extends the proximal recurrence concept to continuous data evaluation. It calculates the squared temporal differences within a specified radius around each data point, which magnifies the significant differences and diminishes the insignificant ones.

\paragraph{Algorithm}
The key steps in the algorithm are as follows:

\begin{footnotesize}
\begin{enumerate}[label=\arabic*., noitemsep, topsep=0pt, partopsep=0pt]
\item Calculate the squared temporal difference between values from two consecutive time frames.
\item Create a land mask and apply a buffer to exclude coastal areas.
\item Apply the land mask to the squared difference data.
\item Filter insignificant differences based on a threshold.
\item Create a fine grid for circle placement.
\item Compute the overlap grid, which represents the weighted sum of temperature differences within each circle's radius.
\item Parallelize the circle placement operation using GPU methods, as each GPU thread can work on placing a circle independently.
\item After all circle placements are complete, sort the results based on a specified criteria, such as density score or cluster size.
\end{enumerate}
\end{footnotesize}

\begin{figure}[ht]
\centering
\includegraphics[width=0.75\columnwidth]{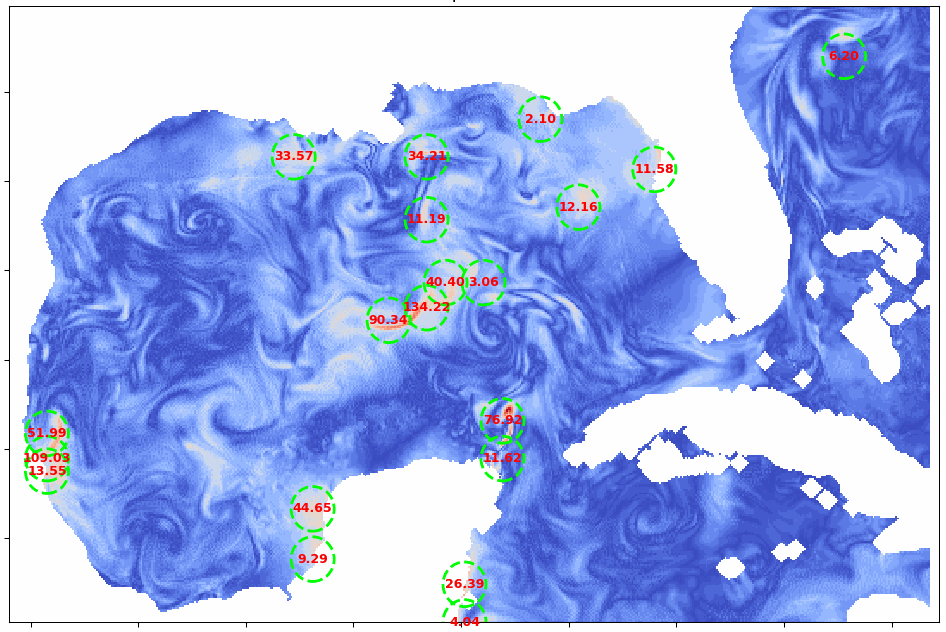} 
\caption{Shows densest cluster identifications for frame 1}
\label{fig:wpr} 
\end{figure}

\subsection{ROBUST Network Construction}
The construction process for the Ranged Observer Bipartite-Unipartite SpatioTemporal (ROBUST) Network involves two key steps: node generation and edge creation within the aggregated spatiotemporal network.

\subsubsection{Node Generation}
Utilizing the WPR methodology, the ROBUST Network identifies optimal locations for waypoints. WPR clusters spatiotemporal data to pinpoint areas of high activity, marking them as waypoints. These waypoints serve as potential sensor positions or nodes in the network, chosen for their significance in the temporal domain.

\subsubsection{Edge Generation}
Once the network nodes have been established, the next step involves generating edges that represent potential paths between waypoints. These edges are determined based on spatial proximity and environmental constraints, which collectively create a connected framework for dynamic sensor movement planning. It's important to note that these edges span across different time steps and do not connect waypoints that exist within the same timestep.

\begin{figure}[ht]
\centering
\includegraphics[width=0.75\columnwidth]{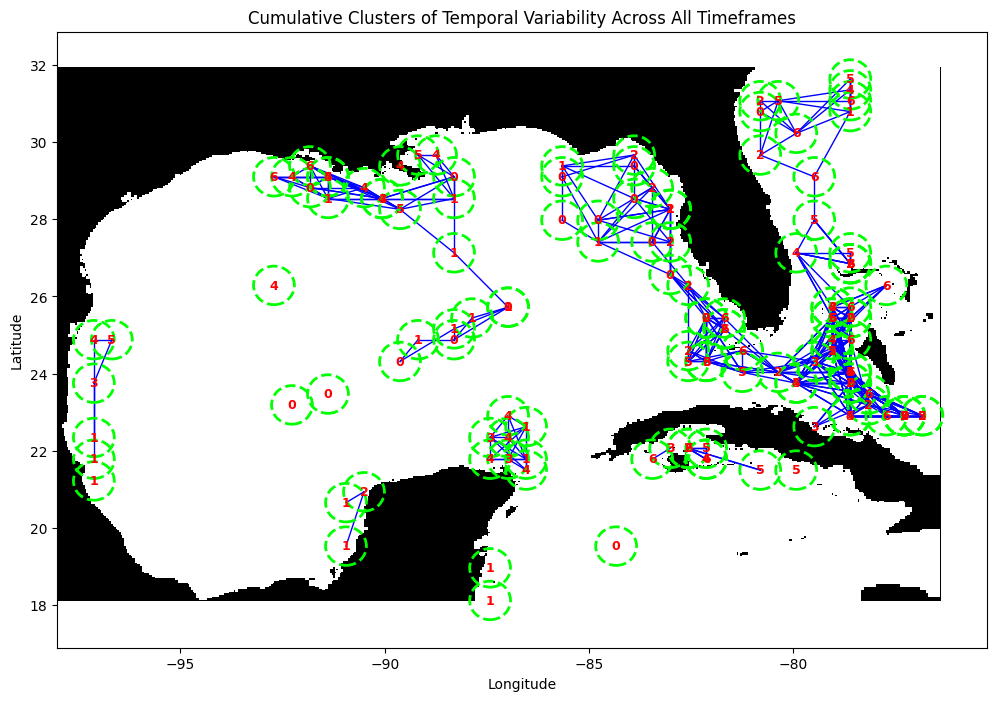} 
\caption{ROBUSTnet, Aggregated Spatiotemporal Graph}
\label{fig:robust-net} 
\end{figure}

\subsection{TED Predictor}
The TED (Temporal Event Dynamics) Predictor is a critical addition to the ROBUST Network in support of temporal path planning by focusing on the dynamic aspects of the environment. This predictor analyzes the Aggregated SpatioTemporal Network of waypoints, identifying active nodes for each timeframe. It assesses the influence of each node based on the quantity and significance of events it covers. This research uses the differences in the HYCOM forecasting for that period to predict points of interests.

\paragraph{Node Activation and Influence}
For each timeframe, the TED Predictor determines which nodes are active and evaluates their influence. This is achieved by analyzing event intensities and distributions, thus enabling the predictor to ascertain the most impactful nodes in the network for that time.

\begin{figure}[ht]
\centering
\begin{tikzpicture}
    \node[inner sep=0pt] (image) at (0,0) {\includegraphics[width=0.75\columnwidth]{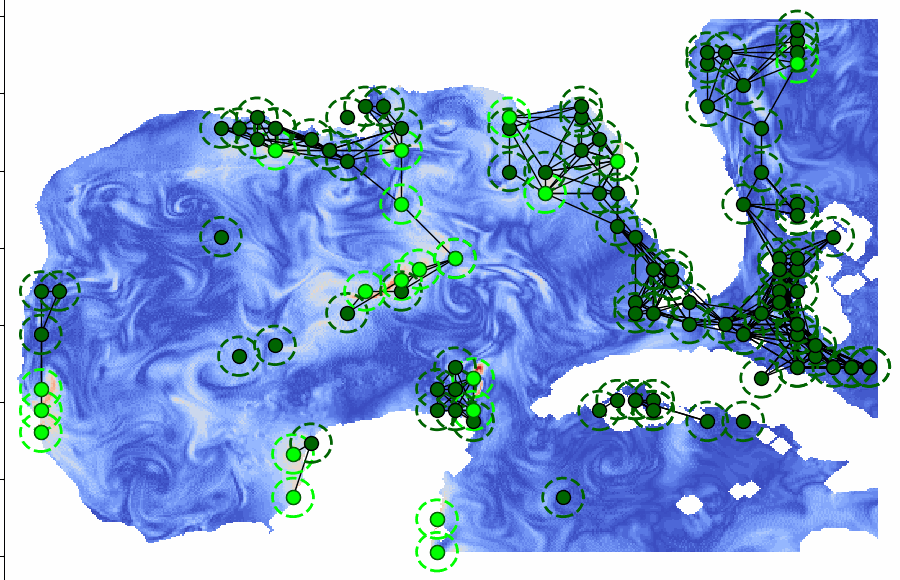}};
    \draw[black] (image.south west) rectangle (image.north east); 
\end{tikzpicture}
\caption{TED predictor shows node activations for frame 1}
\label{fig:ted-predictor}
\end{figure}

\subsection{WAITR Planner}
The Weighted Aggregate Inter-Temporal Reward (WAITR) Planner is the multi-agent path planning framework. Its design focuses on optimizing sensor paths and positions by taking into account the dynamic nature of events and constraints within the network. 

\paragraph{Process Outline}
The WAITR Planner consists of several key phases:
\begin{enumerate}
    \item \textit{Weight Initialization}: Assigning weights to waypoint nodes based on event counts to reflect their importance over different timeframes.
    \item \textit{Piecewise Pathlets Generation}: Constructing a network of potential routes between nodes, considering a maximum number of hops for efficient sensor movement.
    \item \textit{Temporal Path Generation}: Extending paths across different timeframes to accumulate rewards, adapting to the changing environmental conditions.
    \item \textit{Path Selection}: Choosing the most rewarding paths, factoring in the number of sensors and avoiding overlaps in their trajectories.
\end{enumerate}

\paragraph{Technical Implementation}
The WAITR Planner initializes node weights based on event data, generates graph-based pathlets using Dijkstra's algorithm, and extends these pathlets temporally to optimize for cumulative rewards. The final step involves a sorted selection of paths that maximize the density coverage, while adhering to operational constraints of the AUV fleet.

\begin{figure}[H]
\centering
\begin{subfigure}{.6\linewidth}
  \centering
  \includegraphics[width=\linewidth]{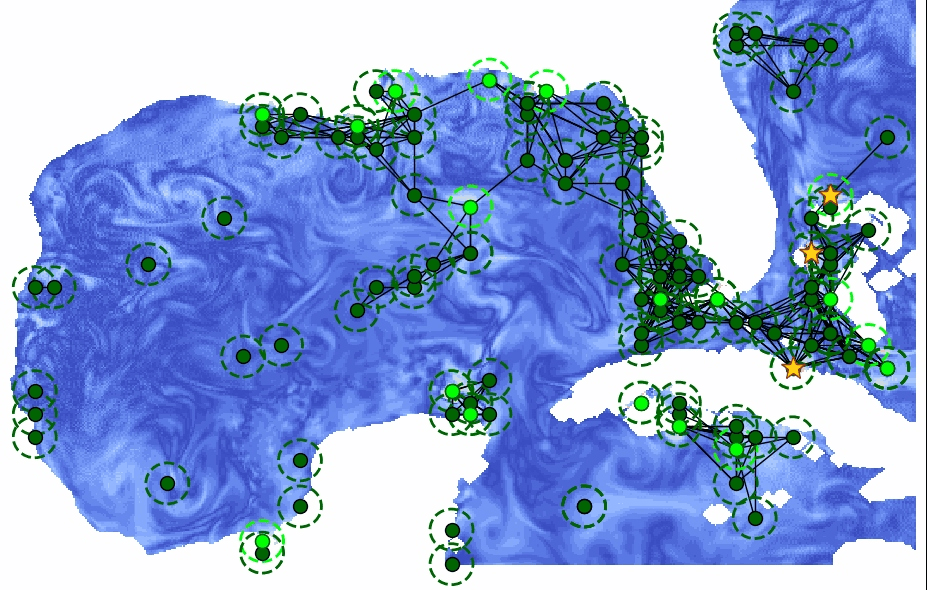}
  \caption{Frame 0}
\end{subfigure}

\begin{subfigure}{.6\linewidth}
  \centering
  \includegraphics[width=\linewidth]{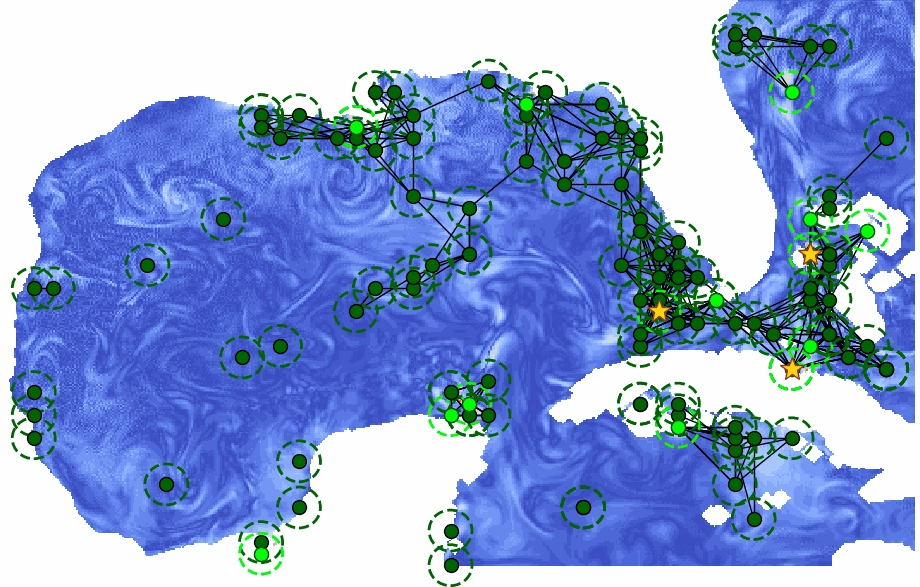}
  \caption{Frame 1}
\end{subfigure}

\begin{subfigure}{.6\linewidth}
  \centering
  \includegraphics[width=\linewidth]{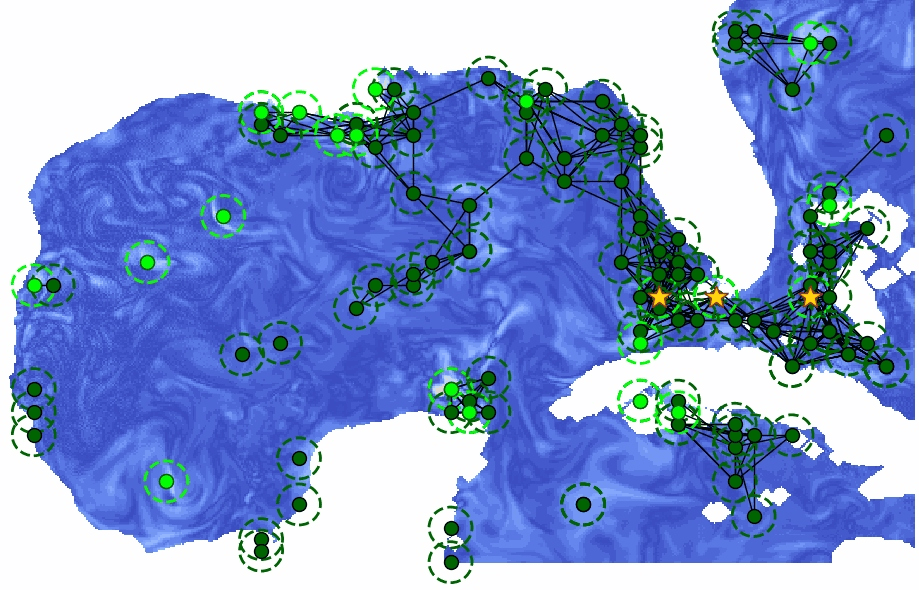}
  \caption{Frame 2}
\end{subfigure}

\caption{Sequential visualization of ROBUST-network path planning for three AUVs across Frames 0, 1, and 2.}
\label{fig:robust-sequence}
\end{figure}

\paragraph{Visualisation}
Figure \ref{fig:robust-sequence} displays heatmaps illustrating the temporal variability for temperature in the Gulf of Mexico. Dark green nodes represent aggregated waypoints, calculated by the WPR clustering method. Lime green nodes highlight the active waypoints selected based on their temporal significance and sensor coverage using the TED predictor. Golden stars represent optimal sensor locations determined by the WAITR strategy for efficient data collection. This visual summary illustrates the ROBUST network's ability to enhance AUV navigational planning for maximizing coverage.

\section{Analysis}
This section introduces metrics developed for assessing the effectiveness of the ROBUST network for the generated multi-agent spatiotemporal paths. These metrics aim to quantify the efficiency and coverage of the planned paths.

\subsection{Event Coverage Ratio (ECR)}
The Event Coverage Ratio (ECR) measures the proportion of significant environmental events that are within the range of the AUVs' sensors. It is calculated as:
\[
ECR = \frac{\text{Number of Covered Events}}{\text{Total Number of Significant Events}}
\]

\section{Results}

\begin{table}[H]
\centering
\resizebox{\columnwidth}{!}{%
\begin{tabular}{lcccc}
\hline
\textbf{Cluster Approach} & \textbf{Top 1} & \textbf{Top 5} & \textbf{Top 10} & \textbf{Top 20} \\ \hline
Aggregated WPR Counts                 & 1256           & 4726           & 7308            & 10378           \\
WPR\% (total=14273)        & 8.8\%          & 33.11\%        & 51.2\%          & 72.7\%          \\
\end{tabular}
}
\caption{Coverage comparisons using WPR clustering with a range=0.5.}
\label{tab:cluster_comparison}
\end{table}

\begin{table}[H]
\centering
\resizebox{\columnwidth}{!}{%
\begin{tabular}{lcc}
\hline
\textbf{Timestep} & \textbf{WAITR Planner (\%)} & \textbf{Greedy Planner (\%)} \\ \hline
Frame 0                   & \textcolor{red}{476}                        & \textcolor{green!60!black}{1463}                         \\
Frame 1                  & \textcolor{red}{18}                      & \textcolor{green!60!black}{116}                        \\
Frame 2                  & \textcolor{green!60!black}{1236}                       & \textcolor{red}{674}                        \\
Frame 3                 & \textcolor{green!60!black}{14}                      & \textcolor{red}{5}                         \\ 
Frame 4                 & \textcolor{green!60!black}{305}                       & \textcolor{red}{0}                         \\ 
Frame 5                 & \textcolor{green!60!black}{430}                       & \textcolor{red}{0}                         \\ 
Frame 6                 & \textcolor{green!60!black}{332}                       & \textcolor{red}{187}                         \\ \hline
\textbf{Total (10378)}                 & \textbf{2811 (27.1\%)}                       & \textbf{ 2445 (23.56\%)}                         \\ \hline
\end{tabular}
}
\caption{Efficiency comparison of WAITR and Greedy planning strategies for different numbers of gliders.}
\label{tab:glider_efficiency}
\end{table}

\section{Discussion}

\subsection{Coverage of WPR Clustering Techniques}
Table \ref{tab:cluster_comparison} highlights the WPR method's efficiency in identifying high temporal variability regions. Coverage significantly increases with more waypoints: from 8.8\% with the top waypoint to 72.7\% using the top 20. This ascending trend indicates WPR's potential for extensive area coverage and establishes an upper limit on achievable coverage, accessible for multi-agent path planning strategy development.

\subsection{Efficiency of Multiagent Path Planning}
Table \ref{tab:glider_efficiency} shows that the WAITR planner consistently outperforms the Greedy planner in terms of overall coverage, achieving a total of 27.1\% coverage compared to 23.56\% by the Greedy planner. Notably, the WAITR planner demonstrates a broader spread of samples across all timesteps, indicating its effectiveness in adapting to the dynamic changes in the environment. In contrast, the Greedy planner's performance is less consistent, with a significant drop in coverage in later frames.

\section{Conclusion}
This research lays the foundation for the effective optimization of AUVs plans by adapting the ROBUST network for path-planning tasks. The WPR clustering technique proves effective in identifying areas of high temporal variability, thus enhancing the potential effectiveness of AUV deployments. The development of the WAITR Planner and TED Predictor for the ROBUST network is key in planning efficient spatiotemporal paths.

\subsection{Future Work}
Future efforts will aim to integrate u-v vector fields into our models to account for water currents, enhancing the accuracy of AUV trajectory planning and better addressing the challenges posed by underwater currents. Future developments will explore more advanced pathfinding algorithms, like D*, to enhance navigation in marine environments. Such improvements are anticipated to significantly boost the efficiency and flexibility of AUV path planning, thus contributing to more effective and reliable oceanographic data collection.

\chapter{Benchmarking Approach }
\label{chap:benchmark_approach}

\section{Introduction}
This chapter presents benchmarking approach obtained using a synthetic spatiotemporal dataset designed to evaluate the approaches introduced in Chapter \ref{chap:lit_review}: \textit{Literature Review} and Chapter \ref{chap:robust_network}: \textit{Robust Network Theory}. The dataset, along with the benchmarking framework, are detailed in Chapter \ref{chap:bench_env}: \textit{Benchmarking Environment}. This framework provides a standardized platform for comparing and analyzing the performance of these approaches.

Our benchmarking focuses on two distinct challenges within spatiotemporal environments:

\begin{enumerate}
  \item \textbf{Static placements:} This presents a clustering challenge.

  \item \textbf{Mobile placements:} This constitutes a planning problem. 
\end{enumerate}

\section{Perfect Knowledge}

In this benchmarking process, we intentionally isolate the clustering and planning challenges from the complexities of prediction. To achieve this, we assume perfect knowledge for all approaches. In our context, perfect knowledge implies complete and accurate information about the spatiotemporal environment, including future conditions and potential events.

\subsection{Rationale}

Employing the assumption of perfect knowledge serves several purposes:

\begin{itemize}
    \item \textbf{Performance Baseline:} It helps establish a theoretical baseline against which to measure the efficiency and potential performance of our clustering and planning methods under optimal conditions.
    \item \textbf{Methodological Focus:}  This assumption allows us to identify the inherent strengths and limitations of the approaches themselves, disentangled from the complexities of imperfect information.
\end{itemize}

\subsection{Limitations}

While the assumption of perfect knowledge is a valuable theoretical tool, it's important to acknowledge its limitations. In real-world scenarios, perfect knowledge is rarely attainable. This simplification helps us establish a benchmark for ideal performance.

\subsection{Future Work}
A crucial area for further exploration is examining the impact of imperfect knowledge on the robustness and adaptability of these methods. Understanding how they perform under uncertain or incomplete information is essential for their real-world application.

\section{Static Sensor Placement:}

\subsection{A Clustering Challenge}
In spatiotemporal environments, optimal placement of static sensors is the goal. Clustering techniques can be used to identify groups of locations with similar patterns or temporal behaviors. This information aids in sensor placement strategies that maximize coverage, target areas of interest, or balance redundancy for resilience. 

The following clustering approaches, introduced in detail in Chapter \ref{chap:lit_review}: \textit{Literature Review} and Chapter \ref{chap:robust_static}:\textit{ROBUST and Static Observers}, will be benchmarked in this section:

\begin{enumerate}[noitemsep]
\item \textbf{Frequency Clustering:} Forms clusters solely based on the frequency of point occurrence.

\item \textbf{K-means Clustering:} Groups points by their proximity to cluster centroids, minimizing distances within each cluster.

\item \textbf{Improved K-means Clustering (Frequency-biased):} Adapts K-means by filtering out infrequent points, emphasizing recurring patterns.

\item \textbf{DBSCAN Clustering:} Identifies dense regions of points, separating them from less dense areas (noise).

\item \textbf{Improved DBSCAN Clustering (Frequency-biased):} Extends DBSCAN by filtering out infrequent points, focusing on recurring spatial patterns.

\item \textbf{Mixed Integer Linear Programming (MILP) Clustering:} Formulates clustering as an optimization problem, subject to user-defined constraints.

\item \textbf{ROBUST with PREP Clustering:} Identifies clusters based on the recurrence of point proximities over time, using the PREP algorithm. 
\end{enumerate}

\section{Mobile Sensor Placement}

\subsection{A Planning Challenge}

Mobile sensor placement in spatiotemporal environments presents a complex, multi-agent planning problem. Due to mobility constraints, such as limited speed and sensor range, these systems require strategies that optimize observation coverage across space and time.  Effective planning must anticipate dynamic environmental changes and coordinate the movement of multiple sensors. This section benchmarks three distinct planning approaches:

\begin{enumerate}
    \item  \textbf{Greedy Planning:} Prioritizes immediate rewards within a limited time horizon, potentially sacrificing long-term gains. (Chapter \ref{chap:lit_review})\ 
    \item  \textbf{Graph Signal Sampling:} Employs online learning for adaptive, long-term planning. Leverages graph theory to optimize sensor placement based on signal dynamics and network structure.\ (Chapter \ref{chap:lit_review}) 
    \item  \textbf{ROBUST with WAITR:} Employs the WAITR Planner to compute temporally optimized sensor paths, maximizing observation coverage. Leverages knowledge about predicted event dynamics and network structure. (Chapter \ref{chap:robust_mobile}) 
\end{enumerate}

\section{Benchmarking Objectives}

This benchmarking study aims to evaluate the coverage of points of interest (POIs) across all time frames within a spatiotemporal environment. Specific objectives include:

\begin{itemize}
    \item Assessing the ability of different sensor placement strategies to maximize POI observation throughout the monitoring period. 

    \item Evaluating the trade-off between maximizing the number of POIs observed at any given time and ensuring consistent observation coverage across all time frames.

    \item Benchmarking the robustness of sensor placement strategies to environmental changes or unexpected POI behavior by analyzing the consistency of coverage across all time frames.
\end{itemize}

\subsection{Metrics for Evaluation:}

\begin{itemize}
    \item \textbf{Average POI Coverage:} Measures the average number of POIs observed across all time frames.
    \item \textbf{Minimum POI Coverage:} Identifies the time frame with the lowest POI observation count, highlighting potential gaps.
    \item \textbf{Temporal Continuity of Coverage:} Evaluates the consistency of POI tracking over time by analyzing observation gaps.
    \item \textbf{Information Gain:} Measures the amount of valuable information collected about each POI, especially when POIs have varying significance. 
\end{itemize}

\section{Synthetic Dynamic Heatmap}

Chapter \ref{chap:bench_env}: \textit{Benchmarking Environment} describes the rationale for using synthetic data to evaluate and compare algorithm performance in a controlled setting. This section outlines the features of the selected synthetic dynamic heatmap dataset:

\begin{itemize}
    \item \textbf{Spatiotemporal Dynamics:} The dataset simulates various spatiotemporal behaviors, mirroring the dynamics observed in real-world environments. These behaviors could include evolutionary patterns, stochastic changes, or event-driven shifts in hot zones. 

    \item \textbf{Customizable Parameters:} The dataset generation process allows for tailoring parameters such as grid size, intensity metrics, time steps, and the frequency of hot zone events. This enables the creation of diverse scenarios for algorithm testing.

    \item \textbf{Benchmarking Focus:} The dataset is designed specifically for benchmarking the performance of algorithms in identifying, tracking, and responding to evolving hot zones within spatiotemporal environments. 
\end{itemize}

\subsection{Dimensionality}
The dynamic heatmap captures both spatial and temporal properties:

\begin{itemize}
    \item \textbf{Spatial Dimensions:} A 50x50 grid (50 rows, 50 columns).
    \item \textbf{Temporal Dimension:} 20 uniform time frames.
    \item \textbf{Dataset Shape:} (20, 50, 50)
\end{itemize}

\subsection{Points of Interest}
The dynamic heatmap is an abstract representation of spatiotemporal behaviors. Values within the heatmap are normalized between 0 and 1:

\begin{itemize}
    \item \textbf{Hot Cells:} Values of 1 represent the locations of interest for benchmarking.
    \item \textbf{Graded Intensity:} Values below 1 indicate varying levels of warmth, coolness, or decreased intensity.
\end{itemize}

\subsection{Evolutionary Behavior}
The dataset exhibits evolutionary dynamics, where each state is directly influenced by prior states. This evolutionary focus emphasizes temporal variability and how hot cells change locations over time.

\subsection{Additional Considerations:}
\begin{itemize}
    \item \textbf{Flexibility:} While abstract, it's implied that this dataset could model behaviors relevant to diverse domains. 
    \item \textbf{Algorithmic Focus:} The dataset is specifically tailored for benchmarking algorithms designed to locate, track, and potentially respond to evolving hot zones.
\end{itemize}

\subsection{Visualization: Initial Heatmap }
This subsection presents a visualization for all twenty frames of the benchmarking dataset.

\captionsetup[figure]{margin=0pt} 

\begin{adjustwidth}{-2cm}{-2cm}
  \centering
  \foreach \n in {0,...,19}{
    \begin{minipage}[t]{0.6\textwidth}
      \centering
      \includegraphics[width=\linewidth]{images/results/heatmaps/heatmaps-image-\n.png}
      \captionof*{figure}{Heatmap - Frame \n}
    \end{minipage}
    \ifodd\n \par\bigskip\else\hspace*{\fill}\fi
  }
\end{adjustwidth}

\clearpage

\chapter{Benchmarking Results }
\label{chap:benchmark_results}

This chapter presents the detailed results of the benchmarking process outlined in Chapter \ref{chap:benchmark_approach}: \textit{Benchmarking Approach}. Quantitative performance metrics and visualizations are used to evaluate and compare the effectiveness of different static and mobile sensor placement strategies within a simulated spatiotemporal environment.

\section{Static Placements}
This section details benchmarking results for static sensor placement strategies within a spatiotemporal environment that, while dynamic, interacts with non-mobile sensors. The experimental parameters set forth for all static placement strategies are as follows:
\begin{itemize}
    \item \textbf{Number of Static Sensors:} Ten (10)
    \item \textbf{Placement:}
        \begin{itemize}
            \item \textbf{Static Positioning:} Sensors are positioned without mobility, remaining at fixed locations throughout the observation period.
            \item \textbf{Sensor Range:} Each sensor has a view radius of three (in cell units), determined by the Euclidean distance formula, indicating the extent of the direct observation field. 
        \end{itemize}
    \item \textbf{Environment:} The spatiotemporal heatmap dataset described in Chapter \ref{chap:benchmark_approach}: \textit{Benchmarking Approach}, which is used to provide a consistent and repeatable context for evaluation.
    \item \textbf{Performance Metrics:} Employing the benchmarks introduced in Chapter \ref{chap:benchmark_approach}: \textit{Benchmarking Approach}
        \begin{itemize}
            \item \textbf{Overall POI Coverage}: The cumulative number of Points of Interest (POI) observed by all sensors throughout the benchmarking period.
            \item \textbf{Overall POI Coverage Ratio}: The proportion of POI observed relative to the total available POI during the entire set of time steps.
        \end{itemize}
\end{itemize}

\subsection{ Frequency Clustering}
Benchmarking results for Frequency Clustering indicate its performance in identifying clusters based solely on point frequency

\subsubsection{Visualization:}
This visualization presents a snapshot of a sensor network within a dynamic spatiotemporal environment. The heatmap is color-coded, with blue tones indicating cooler regions and red tones representing hotter regions. Areas of particular interest, which are the focus for sensor observation, are highlighted by dark red cells.

\begin{itemize}
    \item \textbf{Sensor Locations:} The green dots mark the current locations of the sensors within the environment. Inscribed within each dot is a number, indicating the total number of events that have occurred at that specific location.
\end{itemize}

\captionsetup[figure]{margin=0pt} 

\begin{adjustwidth}{-2cm}{-2cm}
  \centering
  \foreach \n in {0,...,19}{
    \begin{minipage}[t]{0.6\textwidth}
      \centering
      \includegraphics[width=\linewidth]{images/results/frequency-clustering/frequency-clustering-\n.png}
      \captionof*{figure}{Frequency clustering - Frame \n}
    \end{minipage}
    \ifodd\n \par\bigskip\else\hspace*{\fill}\fi
  }
\end{adjustwidth}

\clearpage

\subsection{ K-means Clustering}
Benchmarking results for K-means Clustering demonstrate its ability to group points based on spatial proximity.

\subsubsection{Visualization:}
 The color-coded heatmap background ranges from cool blue to warm red tones, with the most critical regions of interest highlighted in dark red.

\begin{itemize}
    \item \textbf{Event Aggregation:} The green dots distributed across the map represent the aggregated points of events, signifying the distribution of occurrences across all time steps within the environment.

    \item \textbf{Sensor Positions:} The red dots denote the current locations of sensors, which have been strategically placed based on the centroid placements derived from k-means clustering. 
\end{itemize}

\captionsetup[figure]{margin=0pt} 

\begin{adjustwidth}{-2cm}{-2cm}
  \centering
  \foreach \n in {0,...,19}{
    \begin{minipage}[t]{0.6\textwidth}
      \centering
      \includegraphics[width=\linewidth]{images/results/kmeans-clustering/kmeans-clustering-\n.png}
      \captionof*{figure}{K-means clustering - Frame \n}
    \end{minipage}
    \ifodd\n \par\bigskip\else\hspace*{\fill}\fi
  }
\end{adjustwidth}

\clearpage

\subsection{ Improved K-means Clustering}
Benchmarking results for Improved K-means Clustering highlight the impact of frequency-based filtering on cluster formation. This approach filters any non-recurring events from the k-means clustering.

\subsubsection{Visualization:}
This visualization showcases the results of an improved k-means clustering approach for sensor placement across a cumulative time frame, encompassing all 20 time steps. The heatmap in the background transitions from cool blue to warm red hues, with the most crucial areas of interest marked by dark red.

\begin{itemize}
    \item \textbf{Event Aggregation:} The green dots across the map indicate the locations where events have repeatedly occurred, highlighting the frequency of activity within specific regions over the entire period of observation. This aggregation helps identify zones of persistent environmental interest.

    \item \textbf{Sensor Positions:} The red dots illustrate the optimal sensor locations as determined by an advanced k-means clustering algorithm. This method has been refined to disregard non-recurring events, focusing the placement of sensors on areas with consistent activity in an attempt to enhance monitoring efficiency.
\end{itemize}

\captionsetup[figure]{margin=0pt} 

\begin{adjustwidth}{-2cm}{-2cm}
  \centering
  \foreach \n in {0,...,19}{
    \begin{minipage}[t]{0.6\textwidth}
      \centering
      \includegraphics[width=\linewidth]{images/results/improved-kmeans/improved-kmeans-\n.png}
      \captionof*{figure}{Improved K-means clustering - Frame \n}
    \end{minipage}
    \ifodd\n \par\bigskip\else\hspace*{\fill}\fi
  }
\end{adjustwidth}

\clearpage

\subsection{ DBSCAN Clustering}
Benchmarking results for DBSCAN Clustering reveal its effectiveness in detecting dense regions within the spatiotemporal data.

\subsubsection{Visualization:}
This visualization presents the outcome of a DBSCAN clustering method for sensor placement, integrating data across all 20 time steps. The background heatmap displays a spectrum of temperatures, shifting from cooler blue to warmer red areas, with the most significant regions of interest denoted in dark red.

\begin{itemize}
    \item \textbf{Event Aggregation:} The green dots scattered across the heatmap represent the accumulative points where events have occurred, reflecting the overall pattern of activity within the environment through the entire observation span. This comprehensive event aggregation underscores regions with a high incidence of environmental phenomena.

    \item \textbf{Sensor Positions:} The red dots pinpoint the finalized sensor placements, which are the result of the DBSCAN clustering process. This technique accounts for all events, ensuring that the sensor locations correspond to areas of frequent and significant environmental activity, thus providing a thorough and effective monitoring network.
\end{itemize}

\captionsetup[figure]{margin=0pt} 

\begin{adjustwidth}{-2cm}{-2cm}
  \centering
  \foreach \n in {0,...,19}{
    \begin{minipage}[t]{0.6\textwidth}
      \centering
      \includegraphics[width=\linewidth]{images/results/dbscan/dbscan-\n.png}
      \captionof*{figure}{DBSCAN clustering - Frame \n}
    \end{minipage}
    \ifodd\n \par\bigskip\else\hspace*{\fill}\fi
  }
\end{adjustwidth}

\clearpage

\subsection{ Improved DBSCAN Clustering}
Benchmarking results for Improved DBSCAN Clustering illustrate the effect of frequency-based filtering on its density-based cluster identification.

\subsubsection{Visualization:}
This visualization demonstrates the results of an enhanced DBSCAN clustering approach for sensor placement, consolidating data from all 20 time steps. The heatmap background gradates from cool blue to warm red hues, with the most pertinent areas of interest highlighted in dark red.

\begin{itemize}
    \item \textbf{Event Aggregation:} The green dots distributed across the map indicate the locations of recurring events, showing the frequency and consistency of activity within certain regions over the total duration of the observation. This selective event aggregation serves to identify zones of sustained environmental importance.

    \item \textbf{Sensor Positions:} The red dots mark the optimized locations of sensors, determined by a refined DBSCAN clustering algorithm that filters out non-recurring events. This approach focuses sensor placement on areas exhibiting persistent activity, aiming to optimize the efficiency and effectiveness of the environmental monitoring framework.
\end{itemize}

\captionsetup[figure]{margin=0pt} 

\begin{adjustwidth}{-2cm}{-2cm}
  \centering
  \foreach \n in {0,...,19}{
    \begin{minipage}[t]{0.6\textwidth}
      \centering
      \includegraphics[width=\linewidth]{images/results/improved-dbscan/improved-dbscan-\n.png}
      \captionof*{figure}{Improved DBSCAN clustering - Frame \n}
    \end{minipage}
    \ifodd\n \par\bigskip\else\hspace*{\fill}\fi
  }
\end{adjustwidth}

\clearpage

\subsection{ Mixed Integer Linear Programming (MILP)}
Benchmarking results for MILP Clustering showcase its performance in optimizing cluster formation subject to user-defined constraints.

\subsubsection{Visualization:}
This visualization portrays the application of a Mixed Integer Linear Programming (MILP) method for strategic sensor placement, utilizing data accumulated over all 20 time steps. The heatmap serves as the foundation, showcasing a temperature scale that transitions from a calm blue to a warmer red, with especially critical areas accentuated in dark red.

\begin{itemize}
    \item \textbf{Event Aggregation:} The black dots scattered throughout the heatmap represent all aggregated hot cells—zones of high temperature—across the entire observational period. This comprehensive data reflects the intensity and distribution of thermal events in the environment.
    
    \item \textbf{Sensor Influence:} The green dots depict the aggregated hot cells that fall within the sensors' field of view, illustrating the effective coverage area of the deployed sensors and the frequency of detected thermal events within these zones.
    
    \item \textbf{Sensor Locations:} The red dots indicate the final sensor placements as determined by the MILP optimization process. These locations have been computed to maximize the monitoring potential.
\end{itemize}

\captionsetup[figure]{margin=0pt} 

\begin{adjustwidth}{-2cm}{-2cm}
  \centering
  \foreach \n in {0,...,19}{
    \begin{minipage}[t]{0.6\textwidth}
      \centering
      \includegraphics[width=\linewidth]{images/results/milp/milp-\n.png}
      \captionof*{figure}{MILP clustering - Frame \n}
    \end{minipage}
    \ifodd\n \par\bigskip\else\hspace*{\fill}\fi
  }
\end{adjustwidth}

\clearpage

\subsection{ROBUST with PREP Clustering}
Benchmarking results for ROBUST with PREP Clustering evaluate its ability to identify clusters based on recurring spatiotemporal proximities.

\subsubsection{Visualization:}
This visualization showcases the PREP clustering method for sensor placement as defined in Chapter \ref{chap:robust_static}: \textit{ROBUST and Static Observers}. The underlying heatmap gradients from cool blues to warm reds, indicating temperature variations, with dark red spots marking regions of high importance or activity.

\begin{itemize}
    \item \textbf{Sensor Locations:} The green dots are strategically placed sensors within the environment. These locations were chosen based on the PREP clustering algorithm.
    
    \item \textbf{Sensor View Range:} Each sensor is encircled by a lime green line defining its view range, which visually represents the area that each sensor can effectively monitor.
\end{itemize}

\captionsetup[figure]{margin=0pt} 

\begin{adjustwidth}{-2cm}{-2cm}
  \centering
  \foreach \n in {0,...,19}{
    \begin{minipage}[t]{0.6\textwidth}
      \centering
      \includegraphics[width=\linewidth]{images/results/prep/prep-\n.png}
      \captionof*{figure}{PREP clustering - Frame \n}
    \end{minipage}
    \ifodd\n \par\bigskip\else\hspace*{\fill}\fi
  }
\end{adjustwidth}

\clearpage

\section{Mobile Placements }
This section presents benchmarking results for mobile sensor placement strategies within a dynamic spatiotemporal environment. The following experimental parameters were used across all evaluated approaches:
\begin{itemize}
    \item \textbf{Number of Mobile Sensors} Three (3)
    \item \textbf{Mobility:}
        \begin{itemize}
            \item \textbf{Movement Speed:} Up to 2 hops per time step (hop-based).
            \item \textbf{Sensor Range:} A view radius of two (in cell units), calculated using the Euclidean distance formula, defines each sensor's immediate observation area. 
        \end{itemize}
    \item \textbf{Environment:} The dynamic heatmap dataset introduced in Chapter \ref{chap:benchmark_approach}: \textit{Benchmarking Approach}, providing a controlled and replicable testing ground.
    \item \textbf{Performance Metrics:} As outlined in Chapter \ref{chap:benchmark_approach}: \textit{Benchmarking Approach} 
        \begin{itemize}
            \item \textbf{Overall POI Coverage}: sum of POI observed across all time steps for each agent.
            \item \textbf{Overall POI Coverage Ratio}: ratio of all observed points compared to total POI across all time steps.
        \end{itemize}
\end{itemize}

\subsection{ Greedy Planning}
Benchmarking results for Greedy Planning assess its effectiveness in achieving short-term gains and its potential trade-offs in long-term optimization.

\subsubsection{Visualization:}
These visualizations depict the snapshots of a mobile sensor network within a dynamic spatiotemporal environment. The underlying heatmap displays a color gradient, with blue tones indicating cooler areas and red tones signifying hotter areas. The dark red cells represent areas of interest within the environment.

\begin{itemize}
    \item \textbf{Network:} The network of circles and lines illustrates the waypoint network and their connections. This network is generated using the PREP clustering algorithm from chapter \ref{chap:robust_mobile},: \textit{ROBUST and Mobile Observers}.
    \item \textbf{Sensor Activity:} The dots are color-coded to indicate sensor activity.
        \begin{itemize}
            \item \textbf{Lime green nodes:} These represent active sensors that are currently within range of a hot cell (red area) in the heatmap.
            \item \textbf{Dark gray nodes:} These represent inactive sensors that are currently outside the range of any hot cell.
        \end{itemize}
    \item \textbf{Sensor Range:} The light green circles surrounding the active lime green nodes depict the sensor's possible view radius. This indicates the area that each sensor could observe from that waypoint.
    \item \textbf{Sensor Locations:} The yellow stars pinpoint the current locations of the mobile sensors within the environment.
\end{itemize}

\captionsetup[figure]{margin=0pt} 

\begin{adjustwidth}{-2cm}{-2cm}
  \centering
  \foreach \n in {0,...,19}{
    \begin{minipage}[t]{0.6\textwidth}
      \centering
      \includegraphics[width=\linewidth]{images/results/greedy/greedy-\n.png}
      \captionof*{figure}{Greedy Planner - Frame \n}
    \end{minipage}
    \ifodd\n \par\bigskip\else\hspace*{\fill}\fi
  }
\end{adjustwidth}

\clearpage

\subsection{ Graph Signal Sampling Planner}
Benchmarking results for Graph Signal Sampling analyze its adaptive capabilities in sensor placement optimization, leveraging graph theory and online learning.

\subsubsection{Visualization:}
The underlying heatmap displays a continuum of temperature, with cool zones represented by blue tones and warmer zones denoted by red tones. The most intense red cells highlight the most significant areas within the environment, which represent the points of interest.

\begin{itemize}
    \item \textbf{Network:} This waypoint network is generated using the Graph Signal Sampling approach, as detailed in chapter \ref{chap:lit_review}, titled \textit{Literature Review}.
    
    \item \textbf{Waypoint Nodes:} All nodes in this visualization are colored lime green regardless of activity status. 

    \item \textbf{Sensor Locations:} The yellow stars mark the current positions of the mobile sensors within the network.
\end{itemize}

\captionsetup[figure]{margin=0pt} 

\begin{adjustwidth}{-2cm}{-2cm}
  \centering
  \foreach \n in {0,...,19}{
    \begin{minipage}[t]{0.6\textwidth}
      \centering
      \includegraphics[width=\linewidth]{images/results/graph-signal/graph-signal-\n.png}
      \captionof*{figure}{Graph Signal Sampling - Frame \n}
    \end{minipage}
    \ifodd\n \par\bigskip\else\hspace*{\fill}\fi
  }
\end{adjustwidth}

\clearpage

\subsection { ROBUST with WAITR Planner }
Benchmarking results for ROBUST with WAITR examine its performance in maximizing observation coverage using predicted event dynamics and the WAITR Planner.

\subsubsection{Visualization:}
These visualizations depict the snapshots of a mobile sensor network within a dynamic spatiotemporal environment. The underlying heatmap displays a color gradient, with blue tones indicating cooler areas and red tones signifying hotter areas. The dark red cells represent areas of interest within the environment.

\begin{itemize}
    \item \textbf{Network:} The network of circles and lines illustrates the waypoint network and their connections. This network is generated using the PREP clustering algorithm from chapter \ref{chap:robust_mobile},: \textit{ROBUST and Mobile Observers}.
    \item \textbf{Sensor Activity:} The dots are color-coded to indicate sensor activity.
        \begin{itemize}
            \item \textbf{Lime green nodes:} These represent active sensors that are currently within range of a hot cell (red area) in the heatmap.
            \item \textbf{Dark gray nodes:} These represent inactive sensors that are currently outside the range of any hot cell.
        \end{itemize}
    \item \textbf{Sensor Range:} The light green circles surrounding the active lime green nodes depict the sensor's possible view radius. This indicates the area that each sensor could observe from that waypoint.
    \item \textbf{Sensor Locations:} The yellow stars pinpoint the current locations of the mobile sensors within the environment.
\end{itemize}

\captionsetup[figure]{margin=0pt} 

\begin{adjustwidth}{-2cm}{-2cm}
  \centering
  \foreach \n in {0,...,19}{
    \begin{minipage}[t]{0.6\textwidth}
      \centering
      \includegraphics[width=\linewidth]{images/results/waitr/waitr-\n.png}
      \captionof*{figure}{ROBUST with WAITR - Frame \n}
    \end{minipage}
    \ifodd\n \par\bigskip\else\hspace*{\fill}\fi
  }
\end{adjustwidth}

\clearpage

\chapter{Benchmarking Analysis }

\section{Static Placements}

\subsection{Introduction}
This section analyzes the performance of static sensor placement strategies based on the metrics outlined in Chapter \ref{chap:benchmark_results}: Benchmarking Results.  Table \ref{tab:sensor_placement} summarizes the results, comparing the total number of potential hot points present in the environment to the number of hot points successfully observed by each method.

\begin{table}[ht]
\centering
\begin{tabular}{lccc}
\hline
\textbf{Method} & \textbf{Total Hot Points} & \textbf{Observed Hot Points} & \textbf{Coverage Score} \\ \hline
ROBUST with PREP            & 1456                      & 406                         & 0.2788                  \\
MILP            & 1456                      & 396                         & 0.2720                  \\
Improved DBSCAN & 1456                      & 155                         & 0.1065                  \\
DBSCAN          & 1456                      & 118                         & 0.0810                  \\
Improved Kmeans & 1456                      & 270                         & 0.1854                  \\
Kmeans          & 1456                      & 178                         & 0.1223                  \\
Frequency       & 1456                      & 282                         & 0.1937                  \\ \hline
\end{tabular}
\caption{Comparison of Sensor Placement Strategies}
\label{tab:sensor_placement}
\end{table}

\subsection{Summary of Table \ref{tab:sensor_placement} }
This table compares the effectiveness of different clustering techniques for identifying potential sensor placement locations in a dynamic heatmap. Each row represents a clustering method.  All techniques used the same dynamic heatmap for consistent comparison. Columns include the total number of Points of Interest (POI), the number of POI successfully observed with sensors placed based on clustering results (using perfect knowledge), and the overall `Coverage Score'. The `Coverage Score' represents the ratio of observed POI to total POI, with higher scores indicating better sensor placement effectiveness.

\subsection{Comparative Analysis:}
PREP and MILP demonstrate the highest Coverage Scores, indicating their superior ability to locate sensors near hot points. However PREP with its vectorized approach scales much better than MILP. Frequency, K-means, and their improved variants perform moderately well, while standard DBSCAN exhibits the lowest coverage. The results highlight how critical recurrence (temporal clustering) is in addition to spatial clustering algorithms in observing the maximum number of hot spots. 

\subsubsection{Filtering Impact:} 
Filtering out noisy events improves K-means and DBSCAN performance, suggesting that infrequent points can distort cluster formation in these algorithms.

\subsubsection{Cluster Diameter Limitations:} 
Traditional K-means and DBSCAN focus on spatial proximity or density but may not prioritize clusters with restricted diameters. This mismatch, especially when the cluster diameter should ideally align with the sensor's range, limits the direct correlation between optimal cluster centroids and ideal sensor locations.

\subsubsection{Frequency's Strength:}  The surprisingly good performance of simple frequency clustering highlights that in this scenario, areas with frequent events naturally tend to be denser, making them easier to identify.

\subsubsection{Synthetic Datasets}
The flexibility of the synthetic heatmap generator provides an opportunity to systematically test how these sensor placement strategies perform under a variety of dataset conditions. By manipulating parameters such as the frequency, intensity, and evolutionary behavior of hot zones, we could gain even deeper insights into the strengths and limitations of each approach.

\section{Mobile Placements}

\subsection{Introduction}
This section evaluates mobile sensor placement strategies in a dynamic environment. Table \ref{tab:mobile_sensor_coverage} presents the coverage achieved by each sensor individually and their overall combined Coverage Score.

\begin{table}[H]
\centering
\caption{Sensor Coverage Analysis}
\label{tab:mobile_sensor_coverage}
\begin{tabular}{|l|c|c|c|c|c|}
\hline
\textbf{Method} & \textbf{Sensor 0} & \textbf{Sensor 1} & \textbf{Sensor 2} & \textbf{Sum} & \textbf{Score} \\ \hline
ROBUST with WAITR           & 64                & 60                & 49                & 173          & 0.11882        \\ \hline
Greedy          & 34                & 41                & 40                & 115          & 0.07898        \\ \hline
Graph Signal    & 55                & 50                & 50                & 155          & 0.10646        \\ \hline
\end{tabular}
\end{table}

\subsection{Summary of Table \ref{tab:mobile_sensor_coverage} }
This table analyzes the performance of different multi-agent planning strategies in a dynamic environment with a shared heatmap. Each row represents a strategy's outcomes with three mobile sensors (agent0, agent1, agent2). The columns depict individual sensor coverage (`Sensor 0', `Sensor 1', `Sensor 2'), the total team coverage (`Sum'), and the overall `Coverage Score'. The `Coverage Score' represents the ratio of successfully observed Points of Interest (POI) to the total number of POI that occurred during the mission, with higher scores indicating better performance.

\subsection{Comparative Analysis:}  
The WAITR planner demonstrates the highest overall coverage, suggesting its effectiveness in optimizing sensor paths over time. Graph Signal Sampling performs comparably to WAITR, indicating its ability to adapt to changing conditions. Greedy planning achieves the lowest coverage, highlighting potential limitations when prioritizing short-term gains in dynamic scenarios.

\subsection{Challenges of Imperfect Knowledge}
WAITR's reliance on Weighted Aggregate Inter-Temporal Rewards (WAITR) provides flexibility in how predictions are incorporated. While the perfect knowledge scenario used a uniform weighting scheme, real-world applications can leverage confidence levels and risk assessments to influence path calculation.

\subsubsection{WAITR's Risk-Reward Flexibility}  
In contrast to the uniform weighting under perfect knowledge, real-world scenarios necessitate a risk-aware approach where Sensors could prioritize routes with higher confidence hot cell predictions, demonstrating a risk-averse approach. Alternatively, a risk-tolerant strategy might prioritize paths even with greater uncertainty, focusing on areas with high potential rewards (e.g., denser clusters of hot cells). This trade-off between certainty and potential payoff highlights the importance of tailoring WAITR's weighting strategy to the specific application's risk tolerance. The graph-theoretic representation within the ROBUST network could facilitate the modeling of uncertainty alongside spatial relationships, allowing for richer risk analysis in WAITR's path calculation.

\subsubsection{Regret Scoring:}  
Incorporating risk and confidence-based weighting enables the calculation of 'regret scores', providing valuable insights into the trade-offs between risk-averse and risk-seeking sensor paths. These scores could inform dynamic adjustments to WAITR's behavior, potentially triggering replanning or switching between risk profiles based on real-time performance evaluations.

\subsubsection{Incorporating Real-Time Feedback into WAITR:}
In practical use, reliance on deep forecasting may introduce vulnerability if the dynamics of the environment become highly unpredictable. This could be mitigated with the following considerations: 

\begin{itemize}
    \item \textbf{Replanning Triggers:}  Suggest incorporating a mechanism to trigger recalculation of the sensor path under specific conditions. These could be:

    \item \textbf{Significant Deviation from Predictions:}  Measure the divergence between the predicted hot spot locations and the actual observations. If they exceed a certain threshold, initiate replanning.

    \item \textbf{Time-based Intervals:}  Replan paths at regular intervals to account for potential environmental changes, balancing proactivity with computational cost.
\end{itemize}

\subsection{WAITR's Role in Benchmarking}

\subsubsection{Theoretical Benchmark:} WAITR represents the theoretical upper bound on achievable coverage under the perfect knowledge assumption. This highlights its value in understanding the limits of performance in this specific scenario. Due to its reliance on viewing and movement constraints, as well as limited resources, WAITR provides a maximin outcome for multi-agent temporal path planning under ideal conditions while avoiding redundant sampling.

\subsubsection{Comparative Tool:} WAITR's optimal path serves as a powerful baseline for evaluating other sensor placement approaches. By comparing results to WAITR, we gain valuable insights into the impact of real-world constraints like imperfect predictions, limited sensor resources, or unpredictable environments. This comparison highlights the gap between theoretical optimality and the necessary adaptations required for practical implementation.

\chapter{Conclusions and Future Work}

\section{Summary of Findings}
This dissertation introduced novel spatial metrics that advance the field of ROBUST network analysis. These metrics outperformed traditional methods, providing greater precision and enabling more actionable insights, thereby validating Hypothesis 1. The foundation of this innovation lies in the novel graph-based measures introduced and explored within this study. 

\subsection{Novel Graph-Based Measures}
This study has introduced and explored novel graph-based measures for ROBUST network analysis, emphasizing the integration of spatial metrics and structural insights to unravel the complex interplay between nodes within a network.

\subsection{Demonstrates ROBUST Under Perfect Knowledge}
Our methodological approach assumes perfect knowledge of the network. This allows us to clearly demonstrate the capabilities of ROBUST networks for analyzing dynamic behaviors, evaluating network configurations, and identifying optimization strategies. By establishing a baseline under ideal conditions, we gain valuable insights into the potential of the ROBUST framework.

\subsection{Versatility Across Domains}
Particularly, our analysis has proven adept at examining observer-observable patterns, thereby facilitating the optimization of stationary nodes and planning paths in multi-agent systems. This success in optimizing resource allocation highlights the potential for ROBUST networks to outperform alternative models, as proposed in Hypothesis 2.

\section{Contributions to the Field}
\subsection{Extending Spatiotemporal Measures with ROBUST}
Our work makes contributions to the field of spatiotemporal network analysis. First, by developing a suite of spatial metrics tailored for robust network analysis, we offer a fresh lens through which the resilience, connectivity, and spatial properties of networks can be examined. The integration of these measures within the ROBUST framework directly contributed to the optimization improvements observed across various case studies.  This demonstrates the value of this methodological innovation, validating Hypothesis 3.

\subsection{Extending Spatiotemporal Dynamics with ROBUST}
The ROBUST framework offers a methodological innovation in network analysis, enhancing our understanding of spatiotemporal dynamics and the intricacies of bipartite interactions. It offers a novel perspective on the interaction of entities within networks, illuminating factors that influence network efficiency and resilience. ROBUST integrates an understanding of the roles and relationships inherent in observer-observable bipartite structures. This allows for the definition of optimality within observational and exploratory networks. Case studies validate the framework's effectiveness and adaptability across various real-world scenarios.

\section{Directions for Future Research}
\subsection{Incorporating Imperfect Knowledge}
Firstly, acknowledging the limitations of assuming perfect knowledge, there is a compelling need to explore models that incorporate imperfect knowledge. By integrating probabilistic modeling to represent imperfect knowledge, we can evaluate regret scores associated with decision-making, thereby enhancing the robustness of networks against unforeseen changes and uncertainties.

\subsection{Refinement of Temporal Pathing Methodologies}
Secondly, we intend to refine our methodologies for temporal pathing. Although the Dijkstra's approach has served well, the potential for optimization through alternative algorithms, such as A* pathfinding, warrants investigation. Such advancements could significantly improve the real-time adaptability of our models, especially in continuous sampling scenarios.

\subsection{Expanding the Scope of Case Studies}
Moreover, expanding the scope of our case studies to include a wider array of bipartite dynamics—such as collaborative versus competitive interactions—will be crucial. By modeling subcategories like invaders and defenders or hiders and seekers, we can gain deeper insights into the nuanced relationships that govern network behavior and efficiency.

\subsection{Concluding Remarks}
This dissertation not only validates key hypotheses that underpin the ROBUST framework but also expands on the traditional boundaries in network analysis through the introduction of novel spatial metrics. Our findings pave the way for a deeper understanding of observational network dynamics, offering methodological and theoretical advancements that promise significant implications for both academic research and practical applications. The journey embarked upon in this study, while laying a solid foundation, unveils a spectrum of opportunities for future exploration. The outlined directions for future research are not mere continuations but strategic extensions that seek to address the complexities and uncertainties inherent in real-world networks. As we advance, the ROBUST framework stands as an innovative approach with the potential to enhance network resilience and functionality across diverse domains.

\clearpage 
\addcontentsline{toc}{chapter}{Appendices} 

\begin{appendix}

\chapter{Theoretical Foundations }
\label{chap:theoretical_foundations}

\section{Transition to Graph-Based Approaches}
Although the approaches presented in Chapter~\ref{chap:lit_review} offer comprehensive solutions for sensor placement, complexity and scalability challenges require more efficient methodologies in handling intricate network relationships and spatial constraints. This paves the way for exploring graph-based approaches. Graph theory, with its ability to model complex interactions and constraints in a networked environment, offers promising alternatives. These approaches can efficiently manage the spatial and temporal dimensions inherent in sensor networks, making them suitable for extending the principles of combinatorial optimization via vectorization to more scalable applications. This chapter focuses on establishing the underlying theoretical foundations for defining and analyzing Spatiotemporal Networks.

\section{Static Networks}
\subsection{Network Basics in Graph Theory}
Networks in graph theory are abstract constructs used to model complex systems. They are defined as a collection of nodes and a collection of edges, where the spatial arrangement is not the primary concern. The abstract nature of graph theory allows for the analysis of the relational structure of networks, independent of their visual representation. \cite{DiestelGraphTheory2017}

A network can be mathematically represented as a graph \( G = (V, E) \), where:
\begin{itemize}
    \item \( V \) is the set of nodes (or vertices), \( V = \{v_1, v_2, \ldots, v_n\} \), representing the individual entities within the network.
    \item \( E \) is the set of edges, \( E \subseteq \{ (v_i, v_j) \mid v_i, v_j \in V, i \neq j \} \), representing the connections or interactions between pairs of nodes.
\end{itemize}

\begin{figure}[H]
    \centering
    \begin{tikzpicture}
        \node[shape=circle,draw=black] (v1) at (0,0) {$v_1$};
        \node[shape=circle,draw=black] (v2) at (2,0) {$v_2$};
        \node[shape=circle,draw=black] (v3) at (4,0) {$v_3$};
        \node[shape=circle,draw=black] (v4) at (2,-2) {$v_4$};
        \node[shape=circle,draw=black] (v5) at (4,-2) {$v_5$};

        \path [-] (v1) edge node[above] {$e_1$} (v2);
        \path [-] (v2) edge node[right] {$e_3$} (v4);
        \path [-] (v1) edge node[left] {$e_2$} (v4);
        \path [-] (v3) edge node[above] {$e_4$} (v4);
        \path [-] (v3) edge node[right] {$e_5$} (v5);
    \end{tikzpicture}
    \caption{Sample network representation in graph theory}
    \label{fig:sample-network}
\end{figure}
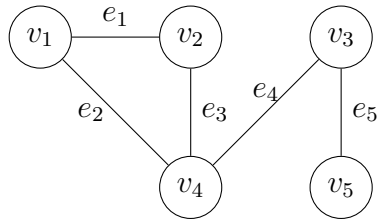

\subsubsection{Nodes (Vertices)}
Conceptually, nodes in a network represent the fundamental units, such as individuals in a social network, computers in a network, or stations in a transportation system. In mathematical terms, a node \( v_i \) is an element of the set \( V \) in the graph \( G \). \cite{DiestelGraphTheory2017}.

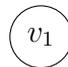
\begin{figure}[H]
    \centering
    \begin{tikzpicture}
        \node[shape=circle,draw=black] (v1) at (0,0) {$v_1$};
    \end{tikzpicture}
    \caption{Example of a node in a network}
    \label{fig:node}
\end{figure}

\subsubsection{Edges (Links)}
Edges in a network signify the relationships or interactions between nodes. An edge is conceptually understood as the connection between two nodes. Mathematically, an edge \( (v_i, v_j) \) is an ordered or unordered pair of nodes, indicating a connection between nodes \( v_i \) and \( v_j \) in the set \( E \). \cite{DiestelGraphTheory2017}.

\begin{figure}[H]
    \centering
    \begin{tikzpicture}
        \node[shape=circle,draw=black] (v1) at (0,0) {$v_1$};
        \node[shape=circle,draw=black] (v2) at (2,0) {$v_2$};
        \path [-] (v1) edge node[above] {$e_1$} (v2);
    \end{tikzpicture}
    \caption{Example of an edge connecting two nodes}
    \label{fig:edge}
\end{figure}
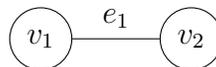

\subsubsection{Comparing Networks}
In graph theory, the spatial positioning of nodes and edges is abstracted away. Two graphs may be considered isomorphic (similar) based on their sets of nodes and edges, regardless of their visual layout or geometric arrangement. This abstraction focuses on the relational and structural properties of the network, rather than its physical representation. \cite{DiestelGraphTheory2017}

\begin{figure}[H]
    \centering
    \begin{tikzpicture}[scale=0.8, transform shape]
    \begin{scope}[xshift=-5cm]
        \node[shape=circle,draw=black] (a1) at (0,0) {$a_1$};
        \node[shape=circle,draw=black] (a2) at (0,2) {$a_2$};
        \node[shape=circle,draw=black] (a3) at (1,3) {$a_3$};
        \node[shape=circle,draw=black] (a4) at (2,2) {$a_4$};
        \node[shape=circle,draw=black] (a5) at (2,0) {$a_5$};

        \path [-] (a1) edge node[left] {} (a2);
        \path [-] (a2) edge node[above] {} (a3);
        \path [-] (a3) edge node[right] {} (a4);
        \path [-] (a4) edge node[below] {} (a5);
        \path [-] (a5) edge node[below] {} (a1);
    \end{scope}
    
    \begin{scope}[xshift=3cm]
        \node[shape=circle,draw=black] (b1) at (0,1) {$b_1$};
        \node[shape=circle,draw=black] (b2) at (1,3) {$b_2$};
        \node[shape=circle,draw=black] (b3) at (2,0) {$b_3$};
        \node[shape=circle,draw=black] (b4) at (3,3) {$b_4$};
        \node[shape=circle,draw=black] (b5) at (4,1) {$b_5$};

        \path [-] (b1) edge (b4);
        \path [-] (b2) edge (b3);
        \path [-] (b3) edge (b4);
        \path [-] (b2) edge (b5);
        \path [-] (b5) edge (b1);
    \end{scope}
    \end{tikzpicture}
    \caption{Two isomorphic networks with different spatial arrangements}
    \label{fig:isomorphic-networks}
\end{figure}
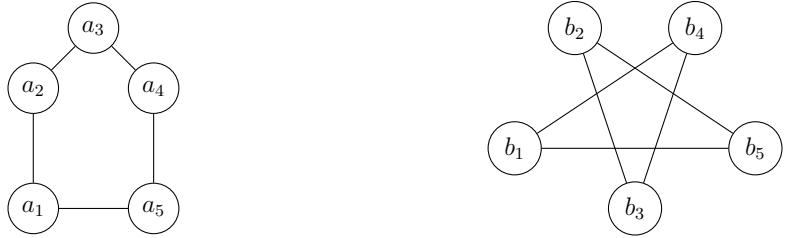

\subsection{Types of Networks}

\subsubsection{Undirected vs. Directed Networks}
Networks can be categorized based on the nature of the relationships they represent. These relationships can be undirected or directed.

\paragraph{Undirected Networks:}
In undirected networks, the edges represent bidirectional relationships. Mathematically, an undirected edge is an unordered pair of nodes. If \( v_i \) and \( v_j \) are two nodes in an undirected network, the edge connecting them is represented as \( \{v_i, v_j\} \) and is inherently bidirectional. \cite{grinberg2023introduction}

\begin{figure}[H]
    \centering
    \begin{tikzpicture}
        \node[shape=circle,draw=black] (v1) at (0,0) {$v_1$};
        \node[shape=circle,draw=black] (v2) at (2,0) {$v_2$};
        \node[shape=circle,draw=black] (v3) at (1,1.73) {$v_3$};

        \path [-] (v1) edge (v2);
        \path [-] (v2) edge (v3);
        \path [-] (v3) edge (v1);
    \end{tikzpicture}
    \caption{An example of an undirected network.}
    \label{fig:undirected-network}
\end{figure}

\paragraph{Directed Networks:}
Conversely, in directed networks, the edges represent unidirectional relationships. These are mathematically represented as ordered pairs of nodes. An edge in a directed network from node \( v_i \) to node \( v_j \) is denoted as \( (v_i, v_j) \), indicating a direction from \( v_i \) to \( v_j \). This directionality is crucial in systems where the relationship is asymmetrical, such as in a follower-following relationship on social media platforms or cause-effect relationships for temporal modeling. \cite{grinberg2023introduction}

\begin{figure}[H]
    \centering
    \begin{tikzpicture}
        \node[shape=circle,draw=black] (v1) at (0,0) {$v_1$};
        \node[shape=circle,draw=black] (v2) at (2,0) {$v_2$};
        \node[shape=circle,draw=black] (v3) at (1,1.73) {$v_3$};

        \path [->,>=stealth] (v1) edge (v2);
        \path [->,>=stealth] (v2) edge (v3);
        \path [->,>=stealth] (v3) edge (v1);
    \end{tikzpicture}
    \caption{An example of a directed network.}
    \label{fig:directed-network}
\end{figure}
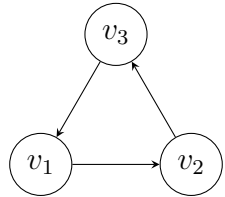

\subsubsection{Weighted vs. Unweighted Networks}
Networks can also be differentiated based on whether their edges carry weights.

\paragraph{Unweighted Networks:}
In unweighted networks, edges do not have any weights associated with them. Relationships are simply represented as connections without any indication of their strength or capacity. Mathematically, an unweighted edge is just a pair of nodes without any additional information. \cite{van2010graph}

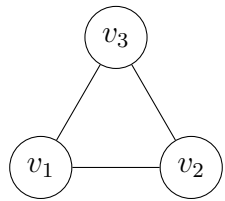
\begin{figure}[H]
    \centering
    \begin{tikzpicture}
        \node[shape=circle,draw=black] (v1) at (0,0) {$v_1$};
        \node[shape=circle,draw=black] (v2) at (2,0) {$v_2$};
        \node[shape=circle,draw=black] (v3) at (1,1.73) {$v_3$};

        \path [-] (v1) edge (v2);
        \path [-] (v2) edge (v3);
        \path [-] (v3) edge (v1);
    \end{tikzpicture}
    \caption{An example of an unweighted network.}
    \label{fig:unweighted-network}
\end{figure}

\paragraph{Weighted Networks:}
In contrast, weighted networks assign a numerical value (weight) to each edge, representing the strength or capacity of the connection. This is expressed as a triplet \( (v_i, v_j, w_{ij}) \), where \( v_i \) and \( v_j \) are the nodes connected by the edge, and \( w_{ij} \) is the weight of the edge. Weighted networks are essential in scenarios where the intensity or capacity of connections varies, such as in networks representing costs, where the edge weights could represent either the distance or time taken to travel between edges. \cite{van2010graph}

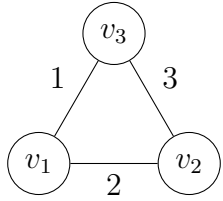
\begin{figure}[H]
    \centering
    \begin{tikzpicture}
        \node[shape=circle,draw=black] (v1) at (0,0) {$v_1$};
        \node[shape=circle,draw=black] (v2) at (2,0) {$v_2$};
        \node[shape=circle,draw=black] (v3) at (1,1.73) {$v_3$};

        \path [-] (v1) edge node[below] {$2$} (v2);
        \path [-] (v2) edge node[above right] {$3$} (v3);
        \path [-] (v3) edge node[above left] {$1$} (v1);
    \end{tikzpicture}
    \caption{An example of a weighted network with weights on edges.}
    \label{fig:weighted-network}
\end{figure}

\subsection{Network Topology and Structure}

\subsubsection{Degree Distribution}
Degree distribution in a network is a fundamental characteristic that describes how the number of connections (degree) per node varies across the network.

\paragraph{Degree of a Node:}
In graph theory, the degree of a node is defined by the number of connections it has. For a node \( v_i \), its degree is denoted as \( \deg(v_i) \). In an undirected network, \( \deg(v_i) \) is the number of edges incident to \( v_i \), while in a directed network, we may distinguish between in-degree and out-degree, depending on the direction of the edges. \cite{grinberg2023introduction}

\begin{figure}[H]
    \centering
    \begin{tikzpicture}[scale=0.8, transform shape]
        \node[shape=circle,draw=black] (v1) at (0,0) {$v_i$};
        
        \path [-] (v1) edge (2,1);
        \path [-] (v1) edge (-2,0);
        \path [-] (v1) edge (2,-1);

        \draw (0,-1) node[left] {$\deg(v_i) = 3$};

        \node[shape=circle,draw=black] (v2) at (7,0) {$v_i$};

        \path [->,>=stealth] (6,1) edge (v2);
        \path [->,>=stealth] (6,0) edge (v2);
        \path [->,>=stealth] (6,-1) edge (v2);
        \path [->,>=stealth] (v2) edge (8,1);
        \path [->,>=stealth] (v2) edge (8,0);

        \draw (6,1) node[left] {In-degree};
        \draw (6,0) node[left] {In-degree};
        \draw (6,-1) node[left] {In-degree};
        \draw (8,1) node[right] {Out-degree};
        \draw (8,0) node[right] {Out-degree};
        
    \end{tikzpicture}
    \caption{Illustration of node degree: undirected (left) with degree 3, and directed (right) with in-degree 3 and out-degree 2.}
    \label{fig:node-degree}
\end{figure}
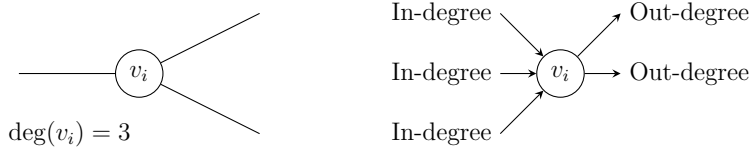

\paragraph{Degree Distribution:}
The degree distribution of a network is the probability distribution of these degrees over the whole network. It is denoted as \( P(k) \), where \( P(k) \) is the probability that a randomly selected node has degree \( k \). This distribution is crucial for understanding the network's structure and dynamics. For instance, in scale-free networks, the degree distribution follows a power law, indicating the presence of a few highly connected nodes (hubs) as well as many nodes with few connections. \cite{Newman2018, barabasi2003linked}

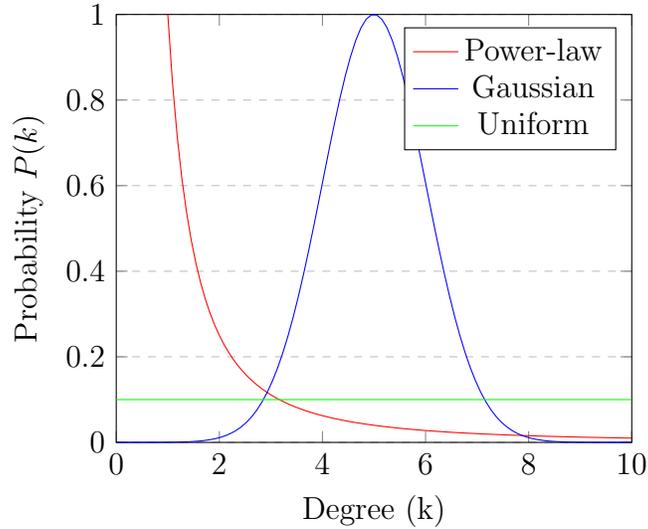
\begin{figure}[H]
    \centering
    \begin{tikzpicture}
    \begin{axis}[
        xlabel={Degree (k)},
        ylabel={Probability \( P(k) \)},
        xmin=0, xmax=10,
        ymin=0, ymax=1,
        legend pos=north east,
        ymajorgrids=true,
        grid style=dashed,
    ]
    
    \addplot[domain=1:10, samples=100, red] {1/x^2};
    \addlegendentry{Power-law}

    \addplot[domain=0:10, samples=100, blue] {exp(-((x-5)^2)/2)};
    \addlegendentry{Gaussian}
    
    \addplot[domain=0:10, samples=100, green, const plot] {0.1};
    \addlegendentry{Uniform}

    \end{axis}
    \end{tikzpicture}
    \caption{Comparison of different degree distributions.}
    \label{fig:degree-distributions}
\end{figure}

\subsubsection{Clustering Coefficient}
The clustering coefficient of a network provides insight into the tendency of nodes to cluster together, forming tightly knit groups within the network.

\paragraph{Local Clustering Coefficient:}
The local clustering coefficient of a node measures how close its neighbors are to being a complete graph (or clique). For a node \( v_i \), the local clustering coefficient \( C_i \) is defined as:

\[
C_i = \frac{2 |E_{v_i}|}{\deg(v_i)(\deg(v_i) - 1)}
\]

where \( |E_{v_i}| \) is the number of edges among the neighbors of \( v_i \). This coefficient ranges from 0 (no connections among the neighbors) to 1 (every neighbor is connected to every other neighbor). \cite{Newman2018}

\begin{figure}[H]
\centering
\begin{tikzpicture}
  [scale=1,auto=left,every node/.style={circle,fill=blue!20}]
  \node (n6) at (1,10) {v6};
  \node (n4) at (4,8)  {v4};
  \node (n5) at (8,9)  {v5};
  \node[fill=red!60] (n1) at (11,8) {v1};
  \node (n2) at (9,6)  {v2};
  \node (n3) at (5,5)  {v3};

  \foreach \from/\to in {n6/n4,n5/n1,n1/n2,n2/n3,n3/n4,n4/n5,n5/n3,n6/n5}
    \draw (\from) -- (\to);

  \foreach \from/\to in {n1/n2,n2/n3,n3/n1}
    \draw [line width=2pt,red] (\from) -- (\to);

\end{tikzpicture}
\caption{Example of a network with a highlighted local cluster around node \( v_1 \). The thick red edges represent \( |E_{v_i}| \).}
\label{fig:cluster}
\end{figure}
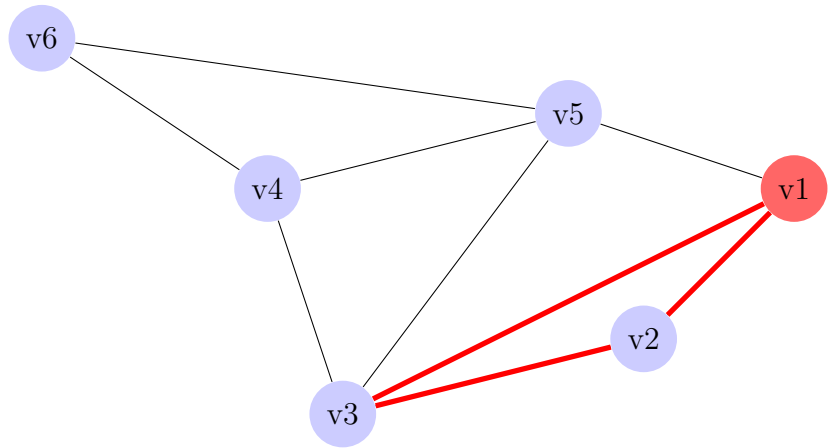

\paragraph{Global Clustering Coefficient:}
The global clustering coefficient is the average of the local clustering coefficients over all the nodes in the network. Alternatively, it can be defined by the ratio of the number of closed triplets (triangles) to the total number of triplets (closed and open) in the network.  \cite{Newman2018}

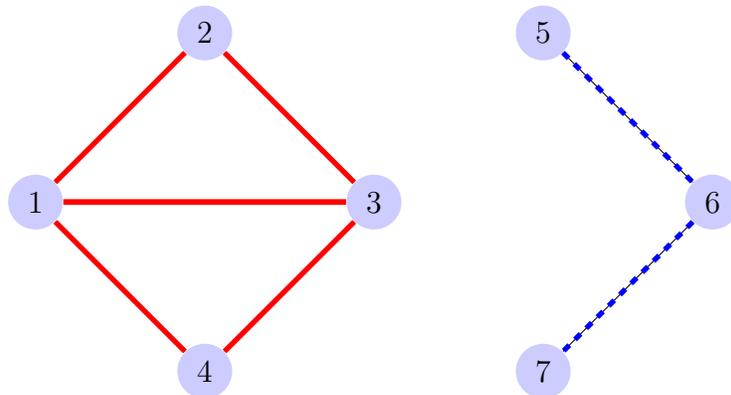
\begin{figure}[H]
\centering
\begin{tikzpicture}
  [scale=0.75,auto=left,every node/.style={circle,fill=blue!20}]
  \node (n1) at (1,4) {1};
  \node (n2) at (4,7) {2};
  \node (n3) at (7,4) {3};
  \node (n4) at (4,1) {4};
  \node (n5) at (10,7) {5};
  \node (n6) at (13,4) {6};
  \node (n7) at (10,1) {7};

  \foreach \from/\to in {n1/n2,n2/n3,n3/n1,n3/n4,n4/n1,n5/n6,n6/n7}
    \draw (\from) -- (\to);

  \foreach \from/\to in {n1/n2,n2/n3,n3/n1}
    \draw [line width=2pt,red] (\from) -- (\to);
  \foreach \from/\to in {n3/n4,n4/n1,n1/n3}
    \draw [line width=2pt,red] (\from) -- (\to);

  \foreach \from/\to in {n5/n6,n6/n7}
    \draw [line width=2pt,blue,dashed] (\from) -- (\to);

\end{tikzpicture}
\caption{Example of a network showing closed triplets (in red) and an open triplet (in blue). The global clustering coefficient is calculated based on these.}
\label{fig:global-clustering}
\end{figure}

\subsection{Path Analysis in Networks}

\subsubsection{Shortest Paths and Network Distance}
Understanding the concept of shortest paths and network distance is crucial in analyzing the efficiency and connectivity of a network.

\paragraph{Shortest Path:}
The shortest path between two nodes in a network is the path that has the minimum length (or distance) among all possible paths connecting them. The length of a path is typically defined as the number of edges it contains (in unweighted networks) or the sum of the weights of its edges (in weighted networks). \cite{Newman2018}

For any two nodes \( v_i \) and \( v_j \) in an unweighted network, the shortest path is the smallest number of edges that must be traversed to travel from \( v_i \) to \( v_j \).

\begin{figure}[H]
\centering

\begin{minipage}{0.48\textwidth}
    \centering
    \begin{tikzpicture}
      [scale=0.75,auto=left]
      \node[shape=circle,fill=blue!20] (n1) at (1,10) {1};
      \node[shape=circle,fill=blue!20] (n2) at (4,8)  {2};
      \node[shape=circle,fill=blue!20] (n3) at (1,6)  {3};
      \node[shape=circle,fill=red!60] (n4) at (4,4)  {4};
      \node[shape=circle,fill=red!60] (n5) at (7,6)  {5};
      \node[shape=circle,fill=red!60] (n6) at (7,10) {6};

      \draw (n1) -- (n2);
      \draw (n2) -- (n5);
      \draw (n5) -- (n6);
      \draw (n4) -- (n5);
      \draw (n4) -- (n3);
      \draw (n3) -- (n1);
      \draw (n2) -- (n4);
      \draw (n1) -- (n6);

      \draw [line width=2pt,red] (n4) -- (n5);
      \draw [line width=2pt,red] (n5) -- (n6);
    \end{tikzpicture}
    \caption{Unweighted network with the shortest path based on hops.}
    \label{fig:unweighted-shortest-path}
\end{minipage}
\hfill 
\begin{minipage}{0.48\textwidth}
    \centering
    \begin{tikzpicture}
      [scale=0.75,auto=left]
      \node[shape=circle,fill=blue!20] (n1) at (1,10) {1};
      \node[shape=circle,fill=red!60] (n2) at (4,8)  {2};
      \node[shape=circle,fill=blue!20] (n3) at (1,6)  {3};
      \node[shape=circle,fill=red!60] (n4) at (4,4)  {4};
      \node[shape=circle,fill=red!60] (n5) at (7,6)  {5};
      \node[shape=circle,fill=red!60] (n6) at (7,10) {6};

      \draw (n1) -- (n2) node[midway, fill=white, sloped] {3};
      \draw (n2) -- (n5) node[midway, fill=white, sloped] {2};
      \draw (n5) -- (n6) node[midway, fill=white, sloped] {1};
      \draw (n4) -- (n5) node[midway, fill=white, sloped] {4};
      \draw (n4) -- (n3) node[midway, fill=white, sloped] {5};
      \draw (n3) -- (n1) node[midway, fill=white, sloped] {6};
      \draw (n2) -- (n4) node[midway, fill=white] {1};
      \draw (n1) -- (n6) node[midway, fill=white] {2};

      \draw [line width=2pt,red] (n5) -- (n6);
      \draw [line width=2pt,red] (n5) -- (n2);
      \draw [line width=2pt,red] (n2) -- (n4);
    \end{tikzpicture}
    \caption{Weighted network with the shortest path based on edge weights.}
    \label{fig:weighted-shortest-path}
\end{minipage}

\end{figure}

\paragraph{Network Distance:}
Network distance measures the efficiency of information or resource transfer in the network. It is often quantified using the average shortest path length between all pairs of nodes.

\subsubsection{Betweenness Centrality}
Betweenness centrality quantifies the importance of nodes based on their role as intermediaries along the shortest paths between other nodes in the network.

\paragraph{Definition:}
For a node \( v \), its betweenness centrality \( C_B(v) \) is calculated as:

\[
C_B(v) = \sum_{s \neq v \neq t} \frac{\sigma_{st}(v)}{\sigma_{st}}
\]

where \( \sigma_{st} \) is the total number of shortest paths from node \( s \) to node \( t \) and \( \sigma_{st}(v) \) is the number of those paths that pass through \( v \). This measure reflects how often a node acts as a bridge along the shortest paths between other nodes. \cite{Newman2018}

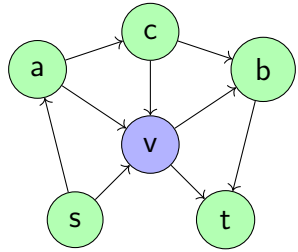
\begin{figure}[htbp]
  \centering
  \begin{tikzpicture}
    \tikzset{
      basic/.style  = {circle, draw, inner sep=5pt, font=\sffamily, fill=green!30},
      root/.style   = {basic, fill=blue!30}, 
      level-2/.style = {basic} 
    }
    
    \node[root] (v) at (0,0) {v};
    \node[level-2] (s) at (-1,-1) {s}; 
    \node[level-2] (t) at (1,-1) {t};  
    \node[level-2] (a) at (-1.5,1) {a};  
    \node[level-2] (b) at (1.5,1) {b};   
    \node[level-2] (c) at (0,1.5) {c}; 

    \draw[->] (s) -- (v);
    \draw[->] (v) -- (t);
    \draw[->] (a) -- (v);
    \draw[->] (v) -- (b);
    \draw[->] (c) -- (v);
    \draw[->] (s) -- (a);
    \draw[->] (a) -- (c);
    \draw[->] (c) -- (b);
    \draw[->] (b) -- (t);
    
  \end{tikzpicture}
  \caption{Visualization of Betweenness Centrality}
  \label{fig:betweenness-centrality}
\end{figure}

Figure~\ref{fig:betweenness-centrality} illustrates the concept of betweenness centrality. The central node \( v \) is a key connector, significantly influencing the shortest paths between the other nodes (\( s, t, a, b, c \)), highlighted by its unique position and connectivity. The computed betweenness centrality for node \( v \) is \( C_B(v) = 3 \).

\subsubsection{Community Structure}
Community structure in networks refers to the division of the network into clusters, or communities, where nodes within the same community are more densely connected to each other than to nodes in other communities. \cite{Fortunato_2010}

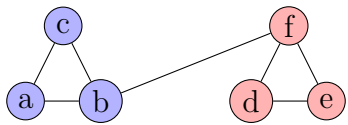
\begin{figure}[htbp]
  \centering
  \begin{tikzpicture}
    \tikzset{
      node_style/.style={circle, draw, inner sep=2pt, minimum size=0.5cm},
      community1/.style={node_style, fill=blue!30},
      community2/.style={node_style, fill=red!30}
    }
    
    \node[community1] (a) at (0,0) {a};
    \node[community1] (b) at (1,0) {b};
    \node[community1] (c) at (0.5,1) {c};

    \node[community2] (d) at (3,0) {d};
    \node[community2] (e) at (4,0) {e};
    \node[community2] (f) at (3.5,1) {f};

    \draw (a) -- (b);
    \draw (a) -- (c);
    \draw (b) -- (c);

    \draw (d) -- (e);
    \draw (d) -- (f);
    \draw (e) -- (f);

    \draw (b) -- (f);

  \end{tikzpicture}
  \caption{Illustration of Community Structure in a Network}
  \label{fig:community-structure}
\end{figure}

Figure~\ref{fig:community-structure} illustrates a network with two distinct communities. Community 1, in blue, includes nodes \(a\), \(b\), and \(c\), while Community 2, in red, consists of nodes \(d\), \(e\), and \(f\). Each community exhibits dense connections among its nodes, indicative of strong intra-community ties. The edge between node \(b\) of Community 1 and node \(f\) of Community 2 highlights the sparser inter-community interactions, emphasizing the network's modular structure.

\paragraph{ \textit{Modularity:} }
A common measure of community structure is modularity, which quantifies the strength of the division of a network into communities. High modularity indicates a strong division into communities with dense connections internally and sparser connections between communities. Identifying the community structure in networks reveals how sub-groups within the network are interconnected.

Mathematically, modularity \( Q \) is defined as:

\[
Q = \frac{1}{2m} \sum_{ij} \left[ A_{ij} - \frac{k_i k_j}{2m} \right] \delta(c_i, c_j)
\]

where \( A_{ij} \) represents the edge weight between nodes \( i \) and \( j \), \( k_i \) and \( k_j \) are the degrees of nodes \( i \) and \( j \), \( m \) is the total number of edges in the network, and \( \delta(c_i, c_j) \) is 1 if nodes \( i \) and \( j \) are in the same community and 0 otherwise. \cite{Fortunato_2010}

\paragraph{ \textit{Modularity Example:} }
The application of the modularity formula to the network shown in Figure~\ref{fig:community-structure} yields a modularity score \( Q \) of approximately \( 0.357 \). This value suggests a moderate level of community structure within the network. It is composed of two distinct communities, however, the modularity score is reduced due to the existence of an edge linking nodes \( b \) and \( f \) from different communities. This inter-community edge indicates a degree of interaction across the community boundaries, thus impacting the overall modularity.

\subsection{Graph Theory Algorithms}

\subsubsection{Algorithmic Analysis}
Various algorithms in graph theory are pivotal for network analysis,by addressing tasks such as finding shortest paths and identifying communities.

\paragraph{ \textit{Shortest Path Algorithms:}}
Dijkstra's algorithm is used to find the shortest path in a network and is defined as follows:

Given a graph \( G = (V, E) \) with source node \( s \), Dijkstra's algorithm computes the shortest path distance from \( s \) to every other node in the graph. The algorithm maintains a set \( S \) of nodes whose shortest distance from the source is known and repeatedly selects the node \( u \notin S \) with the smallest distance estimate, adding \( u \) to \( S \) and relaxing all edges leaving \( u \). \cite{Dijkstra1959}

\vspace{12pt}

\begin{algorithm}[H]
\setstretch{1}
\SetAlgoLined
\LinesNumbered
\KwIn{Graph $G$ with vertices $V$ and edges $E$, source vertex $s$}
\KwOut{Shortest path length from $s$ to all other vertices}
\Begin{
    Initialize a set $Q$ to contain all vertices in $V$\;
    Set distance to $s$ as 0 and to all other vertices as infinity\;
    \While{$Q$ is not empty}{
        $u \leftarrow$ vertex in $Q$ with the smallest distance\;
        Remove $u$ from $Q$\;
        \ForEach{neighbor $v$ of $u$}{
            \If{$v$ is in $Q$}{
                alt $\leftarrow$ distance[$u$] + length($u$, $v$)\;
                \If{alt $<$ distance[$v$]}{
                    distance[$v$] $\leftarrow$ alt\;
                    previous[$v$] $\leftarrow$ $u$\;
                }
            }
        }
    }
    \KwRet{distance[], previous[]}\;
}
\caption{Dijkstra's Algorithm for Shortest Path}
\end{algorithm}

\paragraph{ \textit{Community Detection Algorithms:}}
The Girvan-Newman method is used to identify communities within networks. The Girvan-Newman algorithm repeatedly removes edges from the network that have the highest betweenness centrality, which separates the network into communities. \cite{GirvanNewman2002}

\vspace{12pt}

\begin{algorithm}[H]
\setstretch{1}
\SetAlgoLined
\LinesNumbered
\KwIn{Graph $G$ with vertices $V$ and edges $E$}
\KwOut{Division of $G$ into communities}
\Begin{
    Calculate the betweenness centrality for all edges in $E$\;
    \While{Graph $G$ is not divided into communities}{
        Identify the edge $e$ with the highest betweenness centrality\;
        Remove edge $e$ from $G$\;
        Recalculate betweenness centralities for all edges affected by the removal of $e$\;
        \If{removal of $e$ results in $G$ being divided into more components}{
            Label each component as a separate community\;
        }
    }
    \KwRet{Communities in $G$}\;
}
\caption{Girvan-Newman Algorithm for Community Detection}
\end{algorithm}

\subsubsection{Computational Complexity}
The analysis of large networks poses significant computational challenges, particularly concerning time complexity and memory requirements. 

\paragraph{ \textit{Scalability Issues:} }
As the size of the network grows, the time and memory needed to process the network using standard algorithms can increase exponentially. For instance, Dijkstra's algorithm has a time complexity of \( O(V^2) \) in its basic form, where \( V \) is the number of vertices, making it less efficient for very large graphs. \cite{Cormen2009}

\paragraph{ \textit{Approximation and Heuristics:} }
To address these challenges, approximation algorithms and heuristic methods are often used. These methods provide solutions that are close to optimal and require significantly less computational time. For example, in community detection, heuristic methods like Louvain's algorithm are popular for their ability to handle large networks more efficiently. \cite{Blondel_2008}

\paragraph{ \textit{Distributed and Parallel Computing:} }
Another approach to mitigating computational challenges in large networks is the use of distributed and parallel computing techniques. These techniques involve dividing the computational tasks among multiple processors or machines, significantly reducing the processing time. \cite{Harish2007}

\subsection{Graph Representations}

For computational analysis and efficient handling of networks, graph representations like adjacency matrices and lists are widely used. These data structures provide different ways to model the relationships and structures within a network.

\subsubsection{Adjacency Matrix}
The adjacency matrix is a common way to represent a graph in a matrix format, where each cell indicates whether a pair of nodes is connected.

\paragraph{ \textit{Definition:} }
Given a graph \( G = (V, E) \) with \( n \) nodes, the adjacency matrix \( A \) is an \( n \times n \) matrix where each element \( A_{ij} \) represents the presence or absence of an edge between nodes \( i \) and \( j \). For an unweighted graph, \( A_{ij} \) is 1 if there is an edge between \( i \) and \( j \), and 0 otherwise. In a weighted graph, \( A_{ij} \) equals the weight of the edge. \cite{Newman2018}

\paragraph{ \textit{Mathematical Representation:} }
\[
A_{ij} = 
\begin{cases}
1 & \text{if } (i, j) \in E \text{ (for unweighted graphs)} \\
w_{ij} & \text{if } (i, j) \in E \text{ (for weighted graphs)} \\
0 & \text{if } (i, j) \notin E
\end{cases}
\]

\paragraph{ \textit{Usage:} }
The adjacency matrix is particularly useful in algorithms that require frequent checking of whether an edge exists between two nodes. However, it can be memory-intensive for large networks since its size is proportional to the square of the number of nodes.

\begin{figure}[htbp]
  \begin{minipage}[t]{0.49\textwidth}
    \vspace{0mm}
    \centering
    \begin{tikzpicture}[scale=1]
      \tikzset{
        node_style/.style={circle, draw, inner sep=0pt, minimum size=0.75cm},
      }
      
      \node[node_style] (a) at (0,0) {a};
      \node[node_style] (b) at (1,1) {b};
      \node[node_style] (c) at (2,0) {c};
      \node[node_style] (d) at (1,-1) {d};
      \node[node_style] (e) at (3,1) {e};

      \draw (a) -- (b);
      \draw (b) -- (c);
      \draw (c) -- (d);
      \draw (b) -- (d);
      \draw (c) -- (e);
    \end{tikzpicture}
  \end{minipage}
  \begin{minipage}[t]{0.49\textwidth}
    \vspace{0mm}
    \begin{tabular}{c|ccccc}
      & a & b & c & d & e \\
      \hline
      a & 0 & 1 & 0 & 0 & 0 \\
      b & 1 & 0 & 1 & 1 & 0 \\
      c & 0 & 1 & 0 & 1 & 1 \\
      d & 0 & 1 & 1 & 0 & 0 \\
      e & 0 & 0 & 1 & 0 & 0 \\
    \end{tabular}
  \end{minipage}
  \caption{Graph and its Adjacency Matrix}
  \label{fig:graph-and-adj-matrix}
\end{figure}
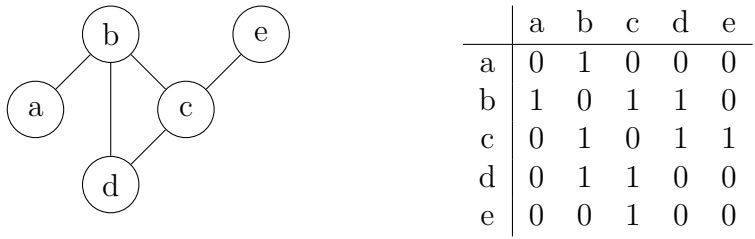

\subsubsection{Adjacency List}
An adjacency list is another way to represent a graph, and is particularly useful for sparse graphs where the number of edges is much less than the square of the number of nodes.

\paragraph{ \textit{Definition:} }
For each node \( i \) in the graph \( G = (V, E) \), the adjacency list maintains a list of all the nodes that are adjacent to \( i \). This can be implemented as an array or a list of lists, where each sublist corresponds to a node and contains the nodes to which it is connected. \cite{Newman2018}

\paragraph{ \textit{Mathematical Representation:} }
For a graph \( G = (V, E) \), the adjacency list \( L \) can be represented as:
\[
L_i = \{j \mid (i, j) \in E\} \quad \text{for each node } i \in V
\]

\paragraph{ \textit{Usage:} }
Adjacency lists are efficient in terms of space, especially for large networks with relatively few edges, as they only store actual connections. They are preferred for algorithms that need to iterate over the neighbors of a node.

\begin{figure}[htbp]
  \begin{minipage}[t]{0.49\textwidth}
    \vspace{0mm}
    \centering
    \begin{tikzpicture}[scale=1]
      \tikzset{
        node_style/.style={circle, draw, inner sep=0pt, minimum size=0.75cm},
      }
      
      \node[node_style] (a) at (0,0) {a};
      \node[node_style] (b) at (1,1) {b};
      \node[node_style] (c) at (2,0) {c};
      \node[node_style] (d) at (1,-1) {d};
      \node[node_style] (e) at (3,1) {e};

      \draw (a) -- (b);
      \draw (b) -- (c);
      \draw (c) -- (d);
      \draw (b) -- (d);
      \draw (c) -- (e);
    \end{tikzpicture}
  \end{minipage}
  \hfill
  \begin{minipage}[t]{0.49\textwidth}
    \vspace{0mm}
    \centering
    \begin{tabular}{l|l}
      Node & Adjacent Nodes \\
      \hline
      a & b \\
      b & a, c, d \\
      c & b, d, e \\
      d & b, c \\
      e & c \\
    \end{tabular}
  \end{minipage}
  \caption{Graph and its Adjacency List}
  \label{fig:graph-and-adj-list}
\end{figure}

\subsubsection{Usage Considerations: Storage versus Time }
The choice between an adjacency matrix and an adjacency list for graph representation depends on the network's size, density, and the operations to be performed.

\subsection{Network Resilience and Robustness}

This section examines the response of a network to failures and assesses its resilience and robustness. The focus is on how the structural properties of a network determine its capacity to endure failures.

\subsubsection{Resilience to Failures}
Network resilience refers to the ability of a network to maintain its overall structure and functionality despite the failure of one or more of its components (nodes or edges). 

\paragraph{ \textit{Node Failure:} }
The failure of a node in a network can be represented by the removal of the node and its associated edges. Mathematically, if node \( v_i \) fails in a graph \( G = (V, E) \), the resulting graph \( G' \) is defined as \( G' = (V - \{v_i\}, E - \{(v_i, v_j) \mid v_j \in V\}) \). The impact of this failure depends on the node's role and position within the network. For instance, the failure of a highly connected node (hub) in a scale-free network can have a more significant impact than the failure of a less connected node. \cite{Albert2002, Newman2018, Ellens2013, Goldschmidt1994}

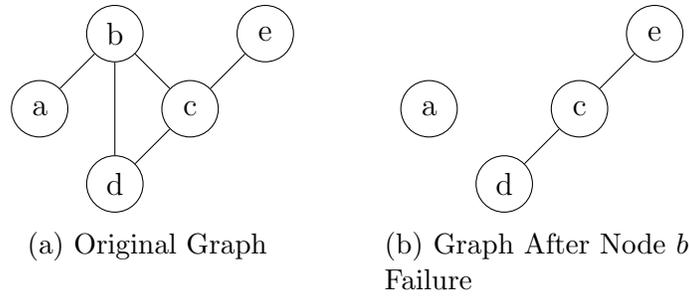
\begin{figure}[htbp]
\centering
\subfloat[Original Graph]{
\begin{tikzpicture}[scale=1, auto, swap]
    \tikzset{
      node_style/.style={circle, draw, inner sep=0pt, minimum size=0.75cm},
      edge_style/.style={draw=black},
    }
    \node[node_style] (a) at (0,0) {a};
    \node[node_style] (b) at (1,1) {b};
    \node[node_style] (c) at (2,0) {c};
    \node[node_style] (d) at (1,-1) {d};
    \node[node_style] (e) at (3,1) {e};
    \draw[edge_style] (a) -- (b);
    \draw[edge_style] (b) -- (c);
    \draw[edge_style] (c) -- (d);
    \draw[edge_style] (b) -- (d);
    \draw[edge_style] (c) -- (e);
\end{tikzpicture}
}
\hspace{1cm} 
\subfloat[Graph After Node \( b \) Failure]{
\begin{tikzpicture}[scale=1, auto, swap]
    \tikzset{
      node_style/.style={circle, draw, inner sep=0pt, minimum size=0.75cm},
      edge_style/.style={draw=black},
      disabled_node_style/.style={circle, draw=gray, dashed, inner sep=0pt, minimum size=0.75cm},
      disabled_edge_style/.style={draw=gray, dashed},
    }
    \node[node_style] (a) at (0,0) {a};
    \node[node_style] (c) at (2,0) {c};
    \node[node_style] (d) at (1,-1) {d};
    \node[node_style] (e) at (3,1) {e};
    \draw[edge_style] (c) -- (d);
    \draw[edge_style] (c) -- (e);
\end{tikzpicture}
}
\caption{Illustration of Node Failure in a Graph}
\label{fig:node_failure}
\end{figure}

\paragraph{ \textit{Edge Failure:} }
Similarly, edge failure involves the removal of an edge from the graph. If edge \( (v_i, v_j) \) fails, the resulting graph \( G' \) is \( G' = (V, E - \{(v_i, v_j)\}) \). The network's ability to withstand edge failures depends on its redundancy and alternative paths between nodes. \cite{Albert2002, Newman2018, Ellens2013}

\begin{figure}[htbp]
\centering
\subfloat[Original Graph]{
\begin{tikzpicture}[scale=1, auto, swap]
    \tikzset{
      node_style/.style={circle, draw, inner sep=0pt, minimum size=0.75cm},
      edge_style/.style={draw=black},
    }
    \node[node_style] (a) at (0,0) {a};
    \node[node_style] (b) at (1,1) {b};
    \node[node_style] (c) at (2,0) {c};
    \node[node_style] (d) at (1,-1) {d};
    \node[node_style] (e) at (3,1) {e};
    \draw[edge_style] (a) -- (b);
    \draw[edge_style] (b) -- (c);
    \draw[edge_style] (c) -- (d);
    \draw[edge_style] (b) -- (d);
    \draw[edge_style] (c) -- (e);
\end{tikzpicture}
}
\hspace{1cm} 
\subfloat[Graph After Edge \( (b, c) \) Failure]{
\begin{tikzpicture}[scale=1, auto, swap]
    \tikzset{
      node_style/.style={circle, draw, inner sep=0pt, minimum size=0.75cm},
      edge_style/.style={draw=black},
      disabled_edge_style/.style={draw=gray, dashed},
    }
    \node[node_style] (a) at (0,0) {a};
    \node[node_style] (b) at (1,1) {b};
    \node[node_style] (c) at (2,0) {c};
    \node[node_style] (d) at (1,-1) {d};
    \node[node_style] (e) at (3,1) {e};
    \draw[edge_style] (a) -- (b);
    \draw[edge_style] (c) -- (d);
    \draw[edge_style] (b) -- (d);
    \draw[edge_style] (c) -- (e);
\end{tikzpicture}
}
\caption{Illustration of Edge Failure in a Graph}
\label{fig:edge_failure}
\end{figure}
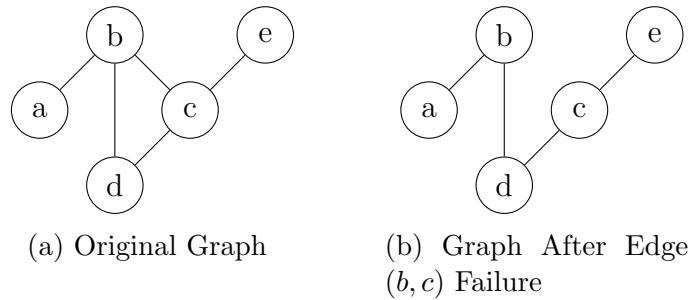

\paragraph{\textit{Network Connectivity:}}
The resilience of a network is often measured by its connectivity, or the ability to find alternative paths between nodes after failures. A key metric is the size of the Largest Connected Component (LCC) or subgraph in the network after removing nodes or edges. Mathematically, this can be represented as follows:

\[
LCC(G') = \max \{|C| : C \text{ is a connected component of } G'\}
\]

Where,
\begin{itemize}
    \item \( G' = (V', E') \) represents the resulting graph after failures.
    \item \( V' \subseteq V \) and \( E' \subseteq E \) indicate that the sets of vertices and edges in \( G' \) are subsets of those in the original graph \( G \).
    \item \( \max \{|C|\} \) denotes the cardinality (size) of the largest connected component in \( G' \).
\end{itemize}

 in the example shown in Figure \ref{fig:node_failure}, after the failure of node \( b \), the size of the Largest Connected Component (LCC) in the resulting graph is 3, because the largest group of interconnected nodes consists of nodes \( c \), \( d \), and \( e \).

\subsubsection{Impact of Network Topology}
Different network topologies exhibit varying degrees of resilience:

\paragraph{Random Networks:}
In random networks, failures are less likely to disrupt the network significantly due to the uniform distribution of connections. These networks are typically more resilient to random node failures.

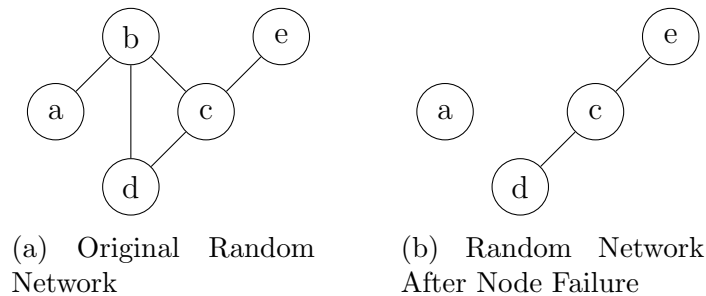
\begin{figure}[htbp]
\centering
\subfloat[Original Random Network]{
\begin{tikzpicture}[scale=1, auto, swap]
    \tikzset{
      node_style/.style={circle, draw, inner sep=0pt, minimum size=0.75cm},
      edge_style/.style={draw=black},
    }
    \node[node_style] (a) at (0,0) {a};
    \node[node_style] (b) at (1,1) {b};
    \node[node_style] (c) at (2,0) {c};
    \node[node_style] (d) at (1,-1) {d};
    \node[node_style] (e) at (3,1) {e};
    \draw[edge_style] (a) -- (b);
    \draw[edge_style] (c) -- (d);
    \draw[edge_style] (b) -- (d);
    \draw[edge_style] (c) -- (e);
    \draw[edge_style] (b) -- (c);
\end{tikzpicture}
}
\hspace{1cm} 
\subfloat[Random Network After Node Failure]{
\begin{tikzpicture}[scale=1, auto, swap]
    \tikzset{
      node_style/.style={circle, draw, inner sep=0pt, minimum size=0.75cm},
      edge_style/.style={draw=black},
    }
    \node[node_style] (a) at (0,0) {a};
    \node[node_style] (c) at (2,0) {c};
    \node[node_style] (d) at (1,-1) {d};
    \node[node_style] (e) at (3,1) {e};
    \draw[edge_style] (c) -- (e);
    \draw[edge_style] (c) -- (d);
\end{tikzpicture}
}
\caption{Impact of Node Failure on a Random Network}
\label{fig:random_network_failure}
\end{figure}

\paragraph{Scale-Free Networks:}
Scale-free networks, characterized by a few highly connected hubs and many nodes with fewer connections, are more vulnerable to targeted attacks on their hubs. However, they are relatively resilient to random failures.

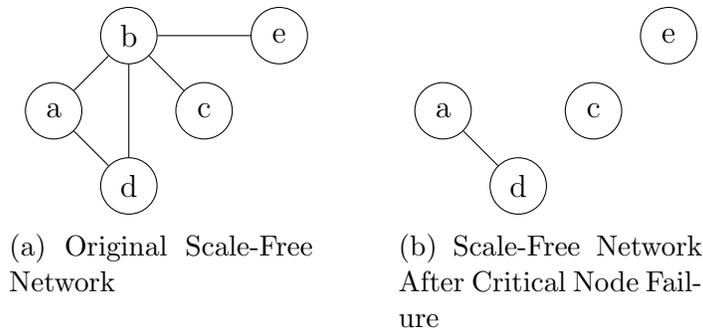
\begin{figure}[htbp]
\centering
\subfloat[Original Scale-Free Network]{
\begin{tikzpicture}[scale=1, auto, swap]
    \tikzset{
      node_style/.style={circle, draw, inner sep=0pt, minimum size=0.75cm},
      edge_style/.style={draw=black},
    }
    \node[node_style] (a) at (0,0) {a};
    \node[node_style] (b) at (1,1) {b}; 
    \node[node_style] (c) at (2,0) {c};
    \node[node_style] (d) at (1,-1) {d};
    \node[node_style] (e) at (3,1) {e};
    \draw[edge_style] (a) -- (b);
    \draw[edge_style] (b) -- (c);
    \draw[edge_style] (b) -- (d);
    \draw[edge_style] (b) -- (e);
    \draw[edge_style] (d) -- (a);
\end{tikzpicture}
}
\hspace{1cm} 
\subfloat[Scale-Free Network After Critical Node Failure]{
\begin{tikzpicture}[scale=1, auto, swap]
    \tikzset{
      node_style/.style={circle, draw, inner sep=0pt, minimum size=0.75cm},
      edge_style/.style={draw=black},
    }
    \node[node_style] (a) at (0,0) {a};
    \node[node_style] (c) at (2,0) {c};
    \node[node_style] (d) at (1,-1) {d};
    \node[node_style] (e) at (3,1) {e};
\draw[edge_style] (d) -- (a);
\end{tikzpicture}
}
\caption{Impact of Critical Node Failure on a Scale-Free Network}
\label{fig:scale_free_network_failure}
\end{figure}

\paragraph{Small-World Networks:}
Small-world networks are known for their short path lengths and high clustering. They can maintain their overall structure even after the failure of several nodes due to their high clustering coefficient and redundant pathways.

\begin{figure}[htbp]
\centering
\subfloat[Original Small World Network]{
\begin{tikzpicture}[scale=1, auto, swap]
    \tikzset{
      node_style/.style={circle, draw, inner sep=0pt, minimum size=0.75cm},
      edge_style/.style={draw=black},
    }
    \node[node_style] (a) at (0,0) {a};
    \node[node_style] (b) at (1,1) {b};
    \node[node_style] (c) at (2,0) {c};
    \node[node_style] (d) at (1,-1) {d};
    \node[node_style] (e) at (3,1) {e};
    \draw[edge_style] (a) -- (b);
    \draw[edge_style] (b) -- (c);
    \draw[edge_style] (c) -- (d);
    \draw[edge_style] (d) -- (a);
    \draw[edge_style] (b) -- (d);
    \draw[edge_style] (a) -- (e);
    \draw[edge_style] (c) -- (e);
\end{tikzpicture}
}
\hspace{1cm} 
\subfloat[Small World Network After Node Failure]{
\begin{tikzpicture}[scale=1, auto, swap]
    \tikzset{
      node_style/.style={circle, draw, inner sep=0pt, minimum size=0.75cm},
      edge_style/.style={draw=black},
    }
    \node[node_style] (a) at (0,0) {a};
    \node[node_style] (b) at (1,1) {b};
    \node[node_style] (c) at (2,0) {c};
    \node[node_style] (d) at (1,-1) {d};
    \draw[edge_style] (a) -- (b);
    \draw[edge_style] (b) -- (c);
    \draw[edge_style] (c) -- (d);
    \draw[edge_style] (d) -- (a);
    \draw[edge_style] (b) -- (d);
    \end{tikzpicture}
}
\caption{Impact of Node Failure on a Small World Network}
\label{fig:small_world_network_failure}
\end{figure}
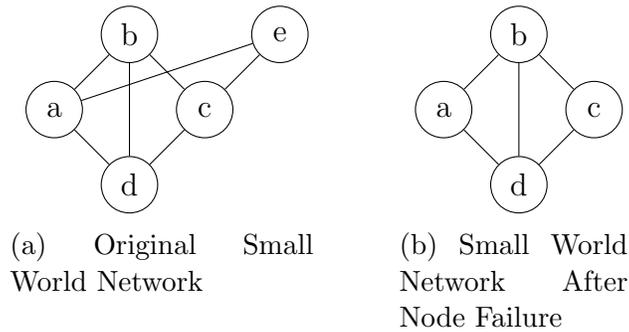

\paragraph{Topology Considerations}
The resilience and robustness of a network to failures are closely tied to its structure and topology. Understanding how different network models respond to such disruptions is key to designing and maintaining resilient systems in various domains.

\subsection{Unipartite vs. Bipartite Networks}
This section explores the fundamental differences between unipartite networks, where all nodes are of a single type, and bipartite networks, which consist of two distinct sets of nodes with connections only between these different sets.

\begin{figure}[H]
\centering
\begin{minipage}[b]{0.45\textwidth}
    \centering
    \begin{tikzpicture}
      [scale=1,auto=left,every node/.style={circle,fill=blue!20}]
      \node (n1) at (1,1) {N1};
      \node (n2) at (0,3)  {N2};
      \node (n3) at (2,3)  {N3};

      \draw (n1) -- (n2);
      \draw (n2) -- (n3);
      \draw (n3) -- (n1);
    \end{tikzpicture}
    \caption{Example of a Unipartite Network}
    \label{fig:unipartite-network}
\end{minipage}
\hfill
\begin{minipage}[b]{0.45\textwidth}
    \centering
    \begin{tikzpicture}
      [scale=1,auto=left,every node/.style={circle}]
      \node[fill=red!20] (a1) at (1,3) {A1};
      \node[fill=red!20] (a2) at (3,3) {A2};
      \node[fill=blue!20] (b1) at (0,1) {B1};
      \node[fill=blue!20] (b2) at (2,1) {B2};
      \node[fill=blue!20] (b3) at (4,1) {B3};

      \draw (a1) -- (b1);
      \draw (a1) -- (b2);
      \draw (a2) -- (b2);
      \draw (a2) -- (b3);
    \end{tikzpicture}
    \caption{Example of a Bipartite Network}
    \label{fig:bipartite-network}
\end{minipage}
\end{figure}

\subsubsection{Unipartite Networks}
Unipartite networks represent a fundamental type of graph where all nodes belong to a single type and can potentially connect with any other node.

\paragraph{ \textit{Definition:} }
Mathematically, a unipartite graph \( G = (V, E) \) consists of a set of nodes \( V \) and a set of edges \( E \) where each edge \( (v_i, v_j) \) connects nodes \( v_i \) and \( v_j \) that belong to the same set \( V \). \cite{DiestelGraphTheory2017}

\paragraph{ \textit{Characteristics:} }
In these networks, edges symbolize interactions or relationships between entities of a similar type.

\subsubsection{Bipartite Networks}
Bipartite networks, or two-mode networks, are characterized by having two distinct sets of nodes, with edges connecting only nodes from different sets.

\paragraph{ \textit{Definition:} }
A bipartite graph \( G = (V_1, V_2, E) \) consists of two distinct sets of nodes, \( V_1 \) and \( V_2 \), and a set of edges \( E \), where each edge \( (v_i, v_j) \) connects a node \( v_i \) from set \( V_1 \) to a node \( v_j \) from set \( V_2 \). \cite{DiestelGraphTheory2017}

\paragraph{ \textit{Characteristics:} }
This structure is definitive in scenarios involving two different types of entities.

\paragraph{ \textit{Analytical Implications:} }
The analysis of bipartite networks differs significantly from unipartite networks. Key aspects such as degree distribution, centrality measures, and clustering coefficients have to be redefined or adapted to account for the two distinct sets of nodes.

\subsection{Edge-based Measures}

\subsubsection{Edge Density}
Edge density is a fundamental measure in graph theory that quantifies how close a network is to being fully connected. It is defined as the ratio of the number of actual edges in the network to the maximum possible number of edges. For a network with \( n \) nodes, the maximum number of edges in an undirected graph is \( \frac{n(n-1)}{2} \). The edge density formula is expressed as:

\begin{equation}
\text{Edge Density} = \frac{\text{Number of Actual Edges}}{\frac{n(n-1)}{2}}
\end{equation}

This measure provides insight into the overall connectivity of the network, with higher values indicating a more densely connected graph. \cite{Easley2010, DiestelGraphTheory2017}

\begin{figure}[htbp]
    \centering
    \begin{minipage}[b]{0.45\textwidth}
        \centering
        \begin{tikzpicture}
            \node[shape=circle,draw,fill=blue!20] (A) at (-0.5,1) {A};
            \node[shape=circle,draw, fill=blue!20] (B) at (1,2) {B};
            \node[shape=circle,draw, fill=blue!20] (C) at (2.5,1) {C};
            \node[shape=circle,draw, fill=blue!20] (D) at (0,-0.5) {D};
            \node[shape=circle,draw, fill=blue!20] (E) at (2,-0.5) {E};

            \draw[ultra thick] (A) -- (B);
            \draw[ultra thick] (A) -- (C);
            \draw[ultra thick] (A) -- (D);
            \draw[ultra thick] (A) -- (E);
            \draw[ultra thick] (B) -- (C);
            \draw[ultra thick] (B) -- (D);
            \draw[ultra thick] (B) -- (E);
            \draw[ultra thick] (C) -- (D);
            \draw[ultra thick] (C) -- (E);
            \draw[ultra thick] (D) -- (E);
        \end{tikzpicture}
        \caption*{Maximally Connected Network}
    \end{minipage}
    \hfill
    \begin{minipage}[b]{0.45\textwidth}
        \centering
        \begin{tikzpicture}
            \node[shape=circle,draw,fill=blue!20] (A) at (-0.5,1) {A};
            \node[shape=circle,draw, fill=blue!20] (B) at (1,2) {B};
            \node[shape=circle,draw, fill=blue!20] (C) at (2.5,1) {C};
            \node[shape=circle,draw, fill=blue!20] (D) at (0,-0.5) {D};
            \node[shape=circle,draw, fill=blue!20] (E) at (2,-0.5) {E};

            \draw[ultra thick] (A) -- (B);
            \draw[ultra thick] (A) -- (D);
            \draw[ultra thick] (B) -- (C);
            \draw[ultra thick] (C) -- (E);
            \draw[ultra thick] (D) -- (E);
        \end{tikzpicture}
        \caption*{Less Connected Network}
    \end{minipage}
    \caption{Comparison of Edge Density}
    \label{fig:edge-density}
\end{figure}

Figure \ref{fig:edge-density} illustrates a comparison of edge density between two network configurations. On the left, the Maximally Connected Network represents a fully connected graph with each node linked to every other node, indicating the maximum edge density achievable. On the right, the Less Connected Network shows a sparser connectivity with fewer edges, signifying a lower edge density. The edge density is computed as the ratio of the number of actual edges to the number of possible edges, which for the Less Connected Network is (5 actual edges / 10 possible edges), resulting in an edge density of 0.5.

\subsubsection{Edge Betweenness Centrality}
Edge centrality measures the importance of edges in a network. It is often used to identify the most crucial connections within the network that, if removed, would significantly impact the network's structure and efficiency. One common form of edge centrality is betweenness centrality, which is calculated as the number of shortest paths that pass through an edge, normalized by the total number of shortest paths in the network. The formula for edge betweenness centrality for an edge \( e \) is:

\begin{equation}
\text{Edge Betweenness Centrality (of edge \( e \))} = \sum_{s,t \in V} \frac{\sigma_{st}(e)}{\sigma_{st}}
\end{equation}

Here, \( \sigma_{st} \) is the total number of shortest paths from node \( s \) to node \( t \), and \( \sigma_{st}(e) \) is the number of those paths passing through edge \( e \). Edge centrality highlights the role of individual edges in facilitating communication and connectivity within the network. \cite{Freeman1977, Newman2018}

\begin{figure}[htbp]
    \centering
    \begin{tikzpicture}[auto,node distance=2cm,
        thick,main node/.style={circle,draw,fill=blue!20,font=\sffamily\Large\bfseries}]

        \node[main node] (1) {A};
        \node[main node] (2) [below left of=1] {B};
        \node[main node] (3) [below right of=1] {C};

        \draw (1) -- (2) node[midway, above left] {12};
        \draw[line width=2pt, red] (2) -- (3) node[midway, below] {1};
        \draw (3) -- (1) node[midway, above right] {10};
    \end{tikzpicture}
    \caption{Example for Edge Centrality where B-C is influential}
    \label{fig:edge-centrality}
\end{figure}
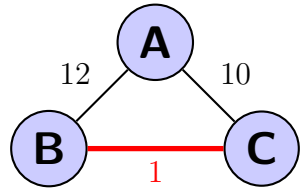

Figure \ref{fig:edge-centrality} illustrates an example of edge centrality in a simple network. The edge centrality score for B-C can be computed based on the number of shortest paths that pass through it.

Considering the shortest paths:
\begin{itemize}
    \item From A to C: The shortest path is A-C, and it does not pass through edge B-C.
    \item From A to B: The shortest path is A-C-B, which passes through edge B-C.
    \item From B to C: The shortest path is B-C, which is the edge itself.
\end{itemize}

Thus, the edge betweenness centrality for edge B-C is calculated as follows:

\begin{equation}
\text{Edge Betweenness Centrality (of edge B-C)} = \frac{1}{3} (0 + 1 + 1) = \frac{2}{3}
\end{equation}

This value of \(\frac{2}{3}\) reflects the high importance of edge B-C in the network's connectivity, as it is involved in two out of the three shortest paths.

\begin{figure}[htbp]
    \centering
    \begin{tikzpicture}[auto,node distance=2cm,
        thick,main node/.style={circle,draw,fill=blue!20,font=\sffamily\Large\bfseries}]

        \node[main node] (A) {A};
        \node[main node] (C) [below right of=A] {C};
        \node[main node] (B) [below left of=C] {B};
        \node[main node] (D) [right of=C] {D};
        \node[main node] (E) [above right of=D] {E};
        \node[main node] (F) [below right of=D] {F};

        \draw (A) -- (C);
        \draw (B) -- (C);
        \draw (C) -- (D);
        \draw (D) -- (E);
        \draw (D) -- (F);

        \draw[line width=2pt,red] (C) -- (D) node[midway, above] {};
    \end{tikzpicture}
    \caption{Example of Edge Betweenness Centrality with High Centrality Score for Edge C-D}
    \label{fig:edge-betweenness-centrality}
\end{figure}
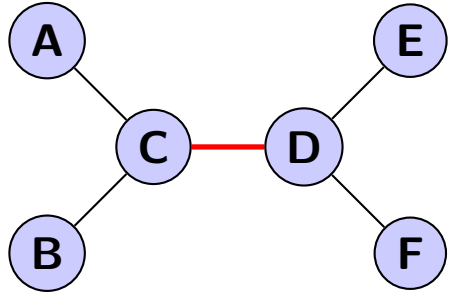

Figure \ref{fig:edge-betweenness-centrality} displays an example of edge betweenness centrality in a unweighted network comprising nodes A, B, C, D, E, and F. The edge C-D is highlighted, indicating its high centrality score due to its critical role in connecting multiple node pairs.

Considering the shortest paths:
\begin{itemize}
    \item From A to D, E, and F: The shortest paths pass through edge C-D.
    \item From B to D, E, and F: The shortest paths pass through edge C-D.
    \item From E to A, B, and C: The shortest paths pass through edge C-D.
    \item From F to A, B, and C: The shortest paths pass through edge C-D.
\end{itemize}

Given that there are 6 nodes in the network, there are a total of \(\frac{6 \times (6 - 1)}{2} = 15\) pairs of nodes. Considering the shortest paths listed above, edge C-D is involved in the shortest path of 12 out of these 15 pairs (excluding pairs A-B, B-A, and C-D).

Therefore, the edge betweenness centrality for edge C-D is:

\begin{equation}
\text{Edge Betweenness Centrality (of edge C-D)} = \frac{12}{15} = 0.8
\end{equation}

This value of 0.8 reflects the high importance of edge C-D in the network's connectivity, as it is a part of the shortest paths for most pairs of nodes in the network.

\paragraph{Edge Closeness Centrality}
Edge closeness centrality could be defined as a measure reflecting the collective closeness of the two nodes an edge connects, in relation to the rest of the network. This measure assesses the edge's importance based on the premise that edges connecting nodes with high closeness centrality are strategically significant for efficient network communication.

For an edge \( e \) connecting nodes \( u \) and \( v \), the edge closeness centrality \( C_{C}(e) \) could be formulated as:

\begin{equation}
C_{C}(e) = \frac{C_{C}(u) + C_{C}(v)}{2}
\end{equation}

where \( C_{C}(u) \) and \( C_{C}(v) \) are the closeness centralities of nodes \( u \) and \( v \) respectively. This formula averages the closeness centralities of the two nodes connected by the edge, thereby capturing the edge's role in connecting parts of the network that are, on average, closer to all other nodes. \cite{Freeman1977, Newman2018}

This conceptualization of edge closeness centrality provides a unique perspective on the significance of an edge, not just as a link between two nodes, but as a connection between two critical points in the network that collectively enhance the overall network connectivity.

\begin{figure}[htbp]
    \centering
    \begin{tikzpicture}[auto,node distance=2cm,
        thick,main node/.style={circle,draw,font=\sffamily\Large\bfseries}]

        \node[main node, fill=blue!20] (U) {U};
        \node[main node, fill=blue!20] (V) [right of=U] {V};

        \node[main node] (A) [above of=U] {A};
        \node[main node] (B) [below of=U] {B};
        \node[main node] (C) [above of=V] {C};
        \node[main node] (D) [below of=V] {D};

        \draw[red, ultra thick] (U) -- (V);
        \draw (U) -- (A);
        \draw (U) -- (B);
        \draw (V) -- (C);
        \draw (V) -- (D);
        \draw (A) -- (C);
        \draw (B) -- (D);
    \end{tikzpicture}
    \caption{Network with Central Nodes U and V Connected by an Edge and Surrounded by Nodes A, B, C, D}
    \label{fig:network-uv}
\end{figure}
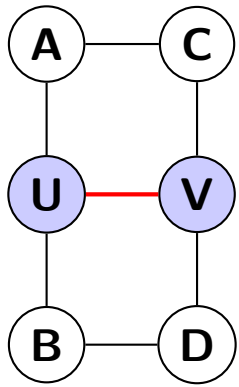

\section{Static Spatial Networks}
This section delves into graphs that include space in their representation, examining the impacts on graph structure, node attributes, edge properties, isomorphic mappings, clustering, and network resilience.

\subsection{Definition of Spatial Networks}

\subsubsection{Contrast with Traditional Graphs}
In traditional graphs, space is an abstract concept without quantifiable dimensions, and there is no definitive spatial orientation. The concept of neighboring nodes in these graphs is based solely on the connections specified by the set of edges. In contrast, spatial graphs introduce a well-defined spatial dimension, making the physical placement of nodes as integral to the network's structure as the nodes and edges themselves. This spatial aspect allows for an additional layer in evaluating relationships between nodes, extending beyond mere link-based connectivity. \cite{Barthelemy2022} 

\begin{figure}[H]
\centering
\begin{tikzpicture}
    \draw[->] (0,0) -- (6,0) node[anchor=north] {X};
    \draw[->] (0,0) -- (0,6) node[anchor=east] {Y};

    \foreach \x in {1,2,...,5}
        \draw (\x,1pt) -- (\x,-3pt) node[anchor=north] {$\x$};
    \foreach \y in {1,2,...,5}
        \draw (1pt,\y) -- (-3pt,\y) node[anchor=east] {$\y$};

    \node[shape=circle,fill=blue!20] (n1) at (1,5) {1};
    \node[shape=circle,fill=blue!20] (n2) at (4,4)  {2};
    \node[shape=circle,fill=blue!20] (n3) at (2,2)  {3};
    \node[shape=circle,fill=blue!20] (n4) at (5,1)  {4};

    \draw (n1) -- (n2);
    \draw (n2) -- (n4);
    \draw (n1) -- (n3);
    \draw (n3) -- (n4);
\end{tikzpicture}
\caption{Example of a Static Spatial Network. Nodes are embedded in a spatial coordinate system.}
\label{fig:basic-spatial-network}
\end{figure}
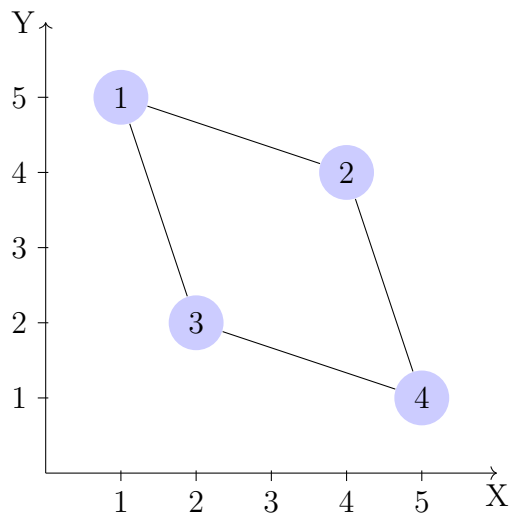

\subsubsection{Mathematical Definition}
Spatial networks are characterized by the integration of a spatial dimension into network topology. Mathematically, a spatial network can be represented as: \cite{Barthelemy2022} 

\begin{equation}
    G_s = (V, E, P) 
\end{equation}

where:
\begin{itemize}
    \item \( V \) represents the set of nodes.
    \item \( E \) denotes the set of edges.
    \item \( P \) represents the set of positions for each node.
\end{itemize}

\subsection{Spatial Properties and Nodes}
In spatial networks, nodes are not just abstract entities but are defined by their specific locations in space, which significantly influences their interaction and functionality within the network.

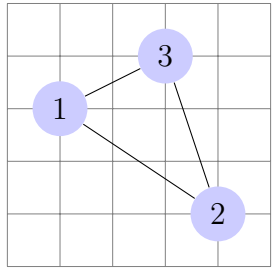
\begin{figure}[htbp]
    \centering
    \begin{tikzpicture}[scale=0.7,auto=left]
      \draw[step=1cm,gray,very thin] (0,0) grid (5,5);

      \node[shape=circle,fill=blue!20] (n1) at (1,3) {1};
      \node[shape=circle,fill=blue!20] (n2) at (4,1)  {2};
      \node[shape=circle,fill=blue!20] (n3) at (3,4)  {3};

      \draw (n1) -- (n2);
      \draw (n2) -- (n3);
      \draw (n1) -- (n3);
    \end{tikzpicture}
    \caption{Spatial network with meaningful node placement. Nodes are positioned according to spatial coordinates.}
    \label{fig:spatial-network-nodes}
\end{figure}

\subsubsection{Spatial Embedding of Nodes}

Spatial coordinates provide a physical or geometric dimension to nodes in a network. This can be mathematically represented as:

\begin{equation}
P: V \rightarrow \mathbb{R}^n
\end{equation}

where \( P(v) \) denotes the spatial position of node \( v \) in an \( n \)-dimensional space, and \( V \) is the set of nodes. This spatial embedding is crucial for understanding the network's structure and dynamics. \cite{Barthelemy2022}

\begin{table}[htbp]
\centering
\caption{Sample Spatial Embedding of Nodes in a Network}
\label{tab:spatial-embedding}
\begin{tabular}{|c|c|}
\hline
\textbf{Node (V)} & \textbf{Spatial Position (P)} \\
\hline
A & ($x_1$, $y_1$, $z_1$) \\
\hline
B & ($x_2$, $y_2$, $z_2$) \\
\hline
C & ($x_3$, $y_3$, $z_3$) \\
\hline
... & ... \\
\hline
\end{tabular}
\end{table}

The table above (Table \ref{tab:spatial-embedding}) provides an example of how nodes (V) in a spatial network are associated with specific coordinates (P) in a given space. Each node is assigned a unique set of coordinates, representing its position in a potentially multidimensional space.

\subsubsection{Curse of Dimensionality}
The dimensionality of the spatial embedding in networks, whether 2D, 3D, or higher dimensions, significantly influences the representation and analysis of nodes. For example, nodes in a 3D space can exhibit complex layers of interaction and connectivity, offering a stark contrast to the more straightforward interactions in 2D spaces. Each added dimension in spatial embedding can exponentially increase the computational cost for algorithms used in clustering, pathfinding, and other forms of network analysis. This scenario is commonly referred to as the ``curse of dimensionality."  \cite{peng2024interpreting, little2021balancing, Steinbach2004}

\begin{figure}[htbp]
    \centering
    \includegraphics[width=0.8\textwidth]{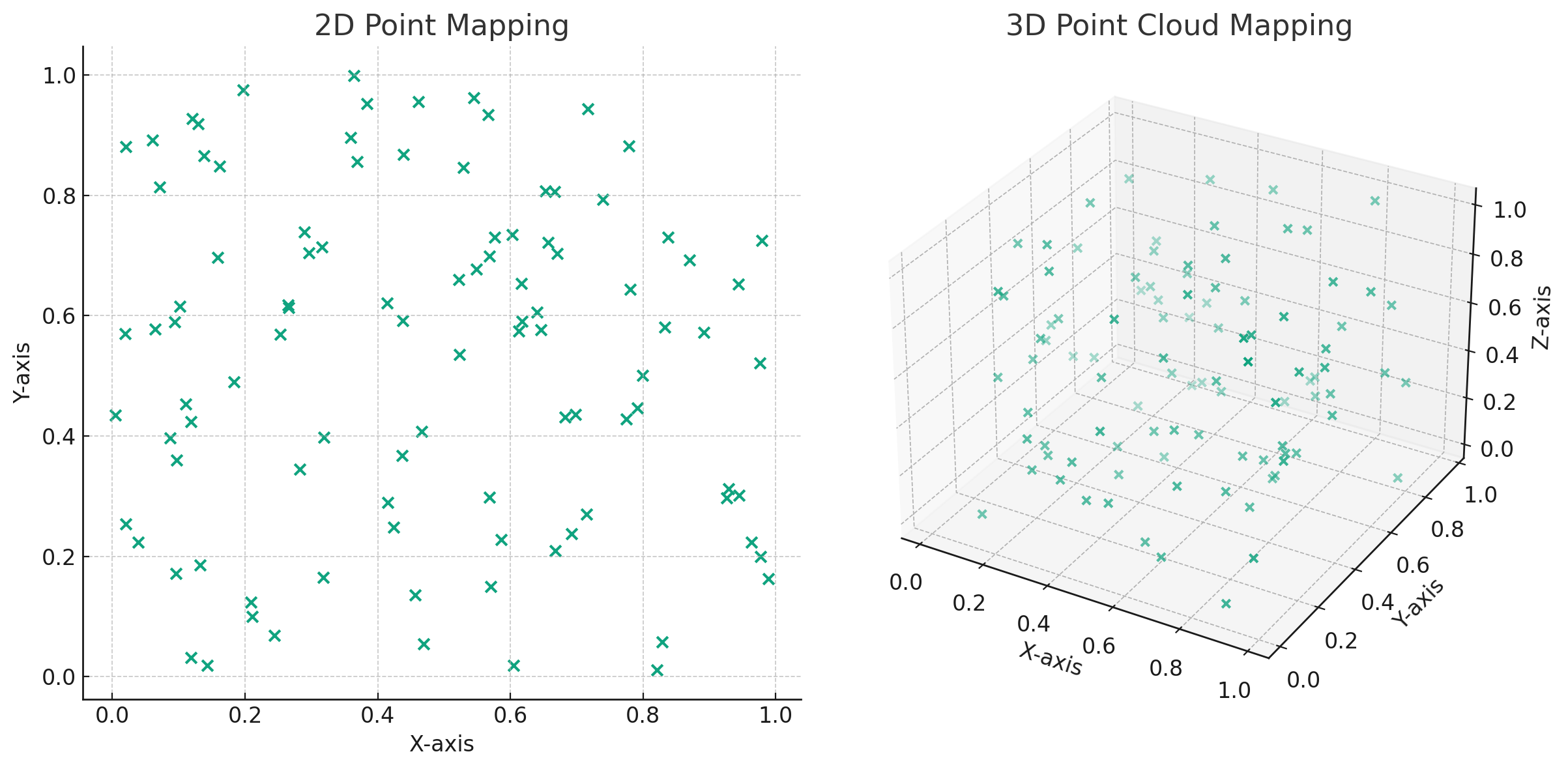}
    \caption{Comparison of 2D and 3D point mappings.}
    \label{fig:2Dvs3D}
\end{figure}

In Figure \ref{fig:2Dvs3D} The left side shows a 2D mapping of points on an x-y axis, while the right side depicts a 3D point cloud on x-y-z axes, highlighting the increased complexity in higher dimensions.

\subsubsection{Spatial Heterogeneity}

 Spatial heterogeneity in networks refers to the variation in spatial attributes among nodes, such as location, distance to other nodes, and distribution within the network space. The positions of nodes in a spatial network influence not only their physical locations but also their roles and interactions within the network.

For example, nodes in densely clustered areas may exhibit different connectivity patterns compared to those in more sparsely populated regions. Nodes positioned at strategic locations, like network centers, might assume critical roles in processes like information dissemination, resource distribution, or contributing to network resilience. Understanding spatial attributes helps in identifying key nodes, predicting interaction patterns, and assessing the network's structural properties and resilience in the face of disruptions or changes in spatial configurations. \cite{Barthelemy2022, Mhatre2004}

\begin{figure}[htbp]
\centering
\includegraphics[width=0.8\textwidth]{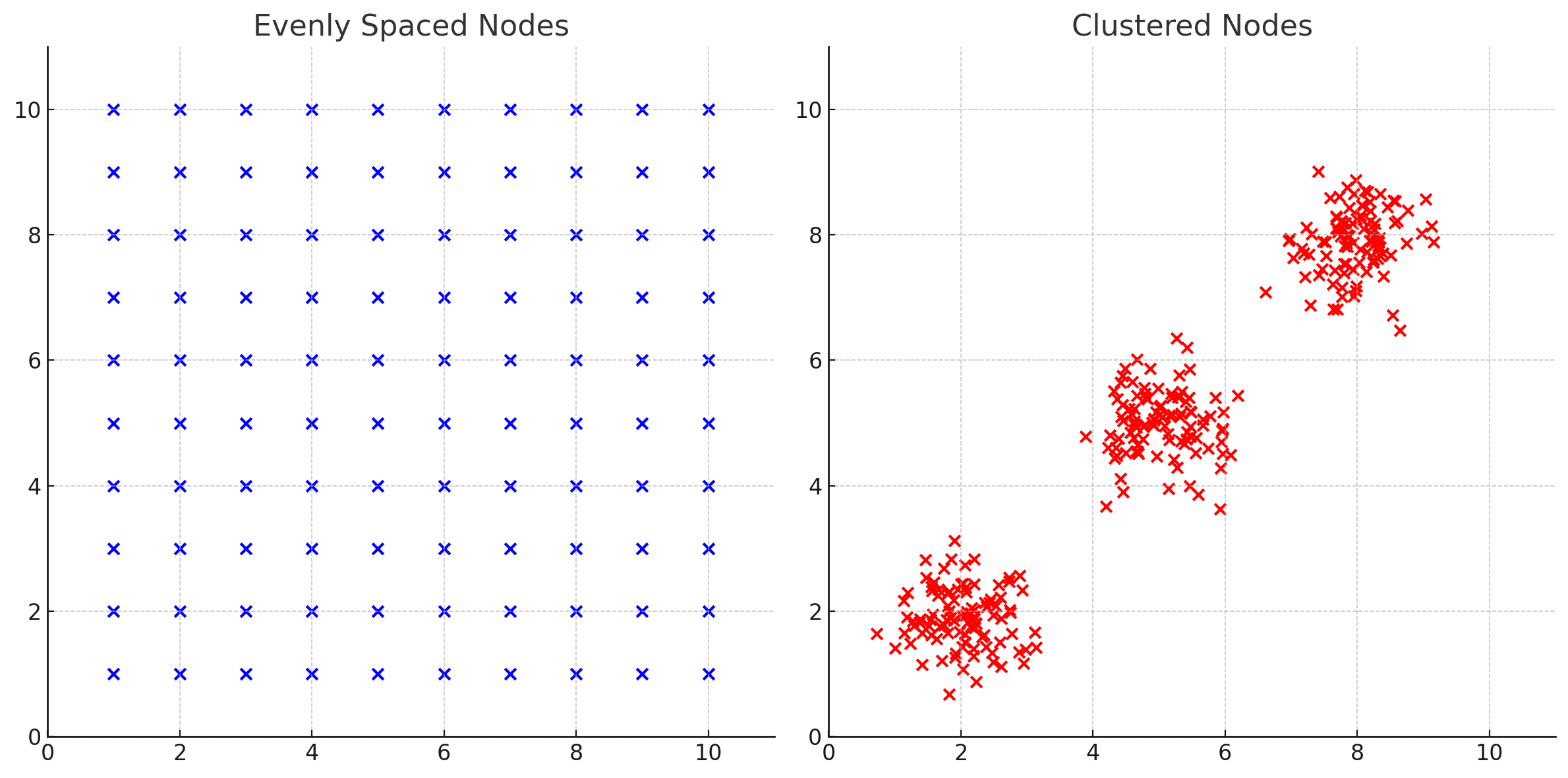}
\caption{Comparison of Spatial Node Distributions. Left: Evenly Spaced Nodes. Right: Clustered Nodes.}
\label{fig:spatial_node_distributions}
\end{figure}

Figure \ref{fig:spatial_node_distributions} depicts two different spatial distributions. The left column shows evenly spaced nodes, while the right column displays nodes that are clustered into groups.

\subsection{Spatial Properties and Edges}

In spatial networks, edges are deeply influenced by the spatial properties of the nodes they connect. This subsection explores the theoretical aspects and dynamics of edge formation and properties in the context of spatial networks. 

\subsubsection{Role of Spatial Proximity in Edge Formation}
Spatial proximity is a key factor in determining edge formation between nodes in spatial networks. The probability of an edge existing between two nodes typically decreases as the spatial distance between them increases. Mathematically, this relationship can be expressed as:

\begin{equation}
P((v_i, v_j) \in E) = f(d(P(v_i), P(v_j)))
\end{equation}

where \( E \) is the set of edges, \( d \) represents a distance metric, and \( f \) is a function that models the relationship between distance and edge formation. \cite{Barthelemy2022, Barthélemy_2003}

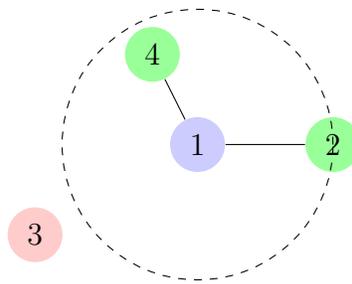
\begin{figure}[htbp]
    \centering
    \begin{tikzpicture}[scale=1.2,auto=left]
      \node[shape=circle,fill=blue!20] (c1) at (2,2) {1};
      \node[shape=circle,fill=green!40] (n1) at (3.5,2) {2};
      \node[shape=circle,fill=red!20] (n2) at (0.2,1) {3};
      \node[shape=circle,fill=green!40] (n3) at (1.5,3) {4};

      \draw (c1) -- (n1);
      \draw (c1) -- (n3);
      \draw[dashed] (2,2) circle (1.5cm);
    \end{tikzpicture}
    \caption{Edge formation influenced by spatial relationships.}
    \label{fig:edge-formation}
\end{figure}

Figure \ref{fig:edge-formation} illustrates edge formation in a spatial network based on proximity. Edges are formed between nodes that are within a certain range, indicated by the dashed circle. Nodes connected by green edges fall within this range, signifying established connections due to their proximity. Conversely, potential connections to nodes outside this range, beyond the dashed circle, are not established and are represented by red nodes. 

\subsubsection{Edges and Spatial Distances}
The spatial characteristics of edges are primarily shaped by the physical or geometric distance between the nodes they connect. This distance is a central factor in determining the nature of the edges, particularly in how they are represented in weighted and unweighted networks.

In weighted spatial networks, the physical distance between nodes is often directly translated into edge weights. This approach acknowledges that longer distances might entail greater resource investment, energy consumption, or time duration for traversal or communication. \cite{Gastner_2006}

On the other hand, unweighted spatial networks tend to abstract away the physical distances, focusing instead on the mere existence of connections. In these networks, edges are typically conceptualized as hops, emphasizing the presence of a connection over its spatial characteristics. This model is particularly relevant in scenarios where the exact distances are less critical to the network's function or where connections are uniform in nature. \cite{Gastner_2006}

\begin{figure}[htbp]
    \centering
    \begin{minipage}[b]{0.45\textwidth}
        \centering
        \begin{tikzpicture}[scale=.80,auto=left]
            \draw[thick,->] (-1,-1) -- (5,-1) node[anchor=north] {X};
            \draw[thick,->] (-1,-1) -- (-1,5) node[anchor=east] {Y};
    
            \node[shape=circle,fill=blue!20] (a) at (0,0) {a};
            \node[shape=circle,fill=blue!20] (b) at (2,0) {b};
            \node[shape=circle,fill=blue!20] (c) at (5,1) {c};
            \node[shape=circle,fill=blue!20] (d) at (0.5,4) {d};
            \node[shape=circle,fill=blue!20] (e) at (2,2) {e};
            \node[shape=circle,fill=blue!20] (f) at (3.5,3) {f};
    
            \draw (a) -- (b) node[midway, below] {1};
            \draw (b) -- (e) node[midway, right] {1};
            \draw (d) -- (e) node[midway, left] {2};
            \draw (c) -- (f) node[midway, above, xshift=5pt] {2};
            \draw (f) -- (e) node[midway] {1};
            \draw (b) -- (c) node[midway, above] {3};
            \draw (e) -- (c) node[midway, above right] {3};
        \end{tikzpicture}
        \caption*{Weighted Network}
    \end{minipage}
    \hfill
    \begin{minipage}[b]{0.45\textwidth}
        \centering
        \begin{tikzpicture}[scale=.80,auto=left]
            \draw[thick,->] (-1,-1) -- (5,-1) node[anchor=north] {X};
            \draw[thick,->] (-1,-1) -- (-1,5) node[anchor=east] {Y};
    
            \node[shape=circle,fill=red!20] (a) at (0,0) {a};
            \node[shape=circle,fill=red!20] (b) at (2,0) {b};
            \node[shape=circle,fill=red!20] (c) at (4,0) {c};
            \node[shape=circle,fill=red!20] (d) at (0,2) {d};
            \node[shape=circle,fill=red!20] (e) at (2,2) {e};
            \node[shape=circle,fill=red!20] (f) at (4,2) {f};
            \node[shape=circle,fill=red!20] (g) at (0,4) {g};
            \node[shape=circle,fill=red!20] (h) at (2,4) {h};
            \node[shape=circle,fill=red!20] (i) at (4,4) {i};
    
            \draw (a) -- (b) -- (c) -- (f) -- (i) -- (h) -- (g) -- (d) -- (a);
            \draw (b) -- (e) -- (h);
            \draw (d) -- (e) -- (f);
        \end{tikzpicture}
        \caption*{Unweighted Network}
    \end{minipage}

    \caption{Comparison of Weighted and Unweighted Spatial Networks}
    \label{fig:weighted-unweighted-networks}
\end{figure}

Figure \ref{fig:weighted-unweighted-networks} contrasts weighted and unweighted spatial networks. The left side shows a weighted network where edge weights reflect physical distances between nodes. The right side illustrates an unweighted network with uniform node distances supporting single spatial hops.

\subsubsection{ Edges and Spatial Constraints}
Spatial constraints play a critical role in shaping both the formation of edges and the overall connectivity of a network. These constraints encompass a variety of factors, including physical barriers, limitations in distance, or specific rules for connectivity that are unique to certain regions. In the context of spatial networks, terrain features such as mountains, valleys, bodies of water, or urban structures can significantly influence edge properties. For example, a mountain range may act as a barrier, preventing edge formation, or it might increase the 'cost' associated with an edge in a weighted network, reflecting the additional effort or resources needed to traverse such terrain. 

Mathematically, the adjacency matrix of a network, denoted by \( A \), can be adapted to reflect these spatial constraints:

\begin{equation}
A_{ij} = 
\begin{cases}
1, & \text{if } d(P(v_i), P(v_j)) \leq \theta \; \text{and other constraints, such as terrain, are met} \\
0, & \text{otherwise}
\end{cases}
\end{equation}

This equation represents how spatial constraints, including terrain, influence the likelihood of edge existence and characteristics within the network.

\begin{figure}[htbp]
    \centering
    \begin{minipage}{0.45\linewidth}
        \centering
        \begin{tikzpicture}[scale=2,auto=left]
            \node[shape=circle,fill=blue!20] (node1) at (0,0) {A};
            \node[shape=circle,fill=blue!20] (node2) at (1,0) {B};
            \node[shape=circle,fill=blue!20] (node3) at (0,1) {C};
            \node[shape=circle,fill=blue!20] (node4) at (1,1) {D};
            \node[shape=circle,fill=blue!20] (node5) at (0.5,1.8) {E};
            
            \draw[very thick] (node1) -- (node2);
            \draw[very thick] (node1) -- (node4);
            \draw[very thick] (node2) -- (node3);
            \draw[very thick] (node2) -- (node5);
            \draw[very thick] (node3) -- (node4);
            \draw[very thick] (node4) -- (node5);
            \draw[fill=red] (-0.2,0.4) rectangle (0.2,0.6); 
            \draw[fill=red] (0.95,0.4) rectangle (1.35,0.6); 
            \draw[fill=red] (-0.2,1.4) rectangle (0.2,1.6); 

    
        \end{tikzpicture}
        \caption*{Realized Network}
    \end{minipage}
    \hfill
    \begin{minipage}{0.45\linewidth}
        \centering
        \caption*{Adjacency Matrix of Spatial Constraints}
        \resizebox{\textwidth}{!}{ 
            \begin{tabular}{|c|c|c|c|c|c|}
                \hline
                \textbf{Nodes} & \textbf{A} & \textbf{B} & \textbf{C} & \textbf{D} & \textbf{E} \\
                \hline
                A & 0 & 1 & 0 & 1 & 0 \\
                \hline
                B & 1 & 0 & 1 & 0 & 1 \\
                \hline
                C & 0 & 1 & 0 & 1 & 0 \\
                \hline
                D & 1 & 0 & 1 & 0 & 1 \\
                \hline
                E & 0 & 1 & 0 & 1 & 0 \\
                \hline
            \end{tabular}
        }
    \end{minipage}
     \caption{Example of Edge-based Spatial Constraints}
    \label{tab:spatial-constraints}
\end{figure}

Figure \ref{tab:spatial-constraints} illustrates a realized network with five nodes labeled A to E, connected by edges. Red rectangles are placed between certain nodes to signify spatial constraints or obstacles, indicating the absence of direct edges between these node pairs due to these constraints. The accompanying table shows the corresponding adjacency matrix, reflecting these spatially constrained relationships.

\subsubsection{Planarity in Spatial Networks}
Planarity is a concept that relates to the network's ability to be drawn on a plane without edge crossings. In spatial networks, particularly waypoint networks or pathways, planarity of edges is important consideration for avoiding potential collision points, as edges represent movement flow. A planar layout is therefore key for safe, efficient navigation, minimizing the risks and complexities of intersecting paths between nodes. \cite{Barthelemy2022}

A network is considered planar if:

\begin{equation}
\forall e_1, e_2 \in E, \; e_1 \cap e_2 = \emptyset \; \text{or} \; e_1 \cap e_2 = \{v\}, \; v \in V
\end{equation}

This equation asserts that for every pair of edges \(e_1\) and \(e_2\) in the edge set \(E\), the edges either do not intersect at all (\(e_1 \cap e_2 = \emptyset\)) or only intersect at a common vertex (\(e_1 \cap e_2 = \{v\}\), where \(v\) is a node in the vertex set \(V\)).

\begin{figure}[htbp]
    \centering
    \begin{minipage}[b]{0.45\textwidth}
        \centering
        \begin{tikzpicture}[scale=.80,auto=left]
            \draw[thick,->] (-1,-1) -- (5,-1) node[anchor=north] {X};
            \draw[thick,->] (-1,-1) -- (-1,5) node[anchor=east] {Y};
    
            \node[shape=circle,fill=blue!20] (a) at (0,0) {a};
            \node[shape=circle,fill=blue!20] (b) at (2,0) {b};
            \node[shape=circle,fill=blue!20] (c) at (4,0) {c};
            \node[shape=circle,fill=blue!20] (d) at (0,2) {d};
            \node[shape=circle,fill=blue!20] (e) at (2,2) {e};
            \node[shape=circle,fill=blue!20] (f) at (4,2) {f};
            \node[shape=circle,fill=blue!20] (g) at (0,4) {g};
            \node[shape=circle,fill=blue!20] (h) at (2,4) {h};
            \node[shape=circle,fill=blue!20] (i) at (4,4) {i};
    
            \draw (a) -- (b) -- (c) -- (f) -- (i) -- (h) -- (g) -- (d) -- (a);
            \draw (b) -- (e) -- (h);
            \draw (d) -- (e) -- (f);
        \end{tikzpicture}
        \caption*{Planar Network}
    \end{minipage}
    \hfill
    \begin{minipage}[b]{0.45\textwidth}
        \centering
        \begin{tikzpicture}[scale=.80,auto=left]
            \draw[thick,->] (-1,-1) -- (5,-1) node[anchor=north] {X};
            \draw[thick,->] (-1,-1) -- (-1,5) node[anchor=east] {Y};
    
            \node[shape=circle,fill=red!20] (a) at (0,0) {a};
            \node[shape=circle,fill=red!20] (b) at (2,0) {b};
            \node[shape=circle,fill=red!20] (c) at (4,0) {c};
            \node[shape=circle,fill=red!20] (d) at (0,2) {d};
            \node[shape=circle,fill=red!20] (e) at (2,2) {e};
            \node[shape=circle,fill=red!20] (f) at (4,2) {f};
            \node[shape=circle,fill=red!20] (g) at (0,4) {g};
            \node[shape=circle,fill=red!20] (h) at (2,4) {h};
            \node[shape=circle,fill=red!20] (i) at (4,4) {i};
    
            \draw (g) -- (e) -- (c);
            \draw (i) -- (e) -- (a);
            \draw (h) -- (f);
            \draw (d) -- (b);
            \draw (f) -- (b);
            \draw (d) -- (h);
            
        \end{tikzpicture}
        \caption*{Non-Planar Network}
    \end{minipage}

    \caption{Comparison of Planarity in Spatial Networks}
    \label{fig:planarity-networks}
\end{figure}

Figure \ref{fig:planarity-networks} presents a side-by-side comparison of planar and non-planar networks. On the left, the Planar Network showcases nodes connected without any edge crossings, illustrating a planar graph layout. On the right, the Non-Planar Network displays overlapping edges among nodes, representing a non-planar structure.

\subsection{Spatial Properties and Network Structures}
This subsection examines the impact of spatial characteristics on the structure of networks. It focuses on how the physical positioning of nodes and the distances between them shape the network's topology, leading to unique emergent patterns and dynamics.

\subsubsection{Spatial Segmentation: Clustering Space in Networks}

Spatial segmentation in networks applies clustering analysis to the spatial domain, focusing on the division of space around nodes based on their spatial attributes and proximities. This results in a unique cluster space for each node.

Voronoi Tessellation is utilized for spatial segmentation. It mathematically allocates regions based on the proximity to the nearest node, formalized as:

\begin{equation}
\text{Voronoi}(v) = \{ x \in \mathbb{R}^n \mid \forall u \in V \setminus \{v\}, d(x, P(v)) < d(x, P(u)) \}
\end{equation}

Here, \(\text{Voronoi}(v)\) signifies the spatial region closest to node \(v\), where \(P(v)\) indicates \(v\)'s location and \(d\) is the distance function. This segmentation is crucial for understanding the spatial influence of nodes within the network.

\begin{figure}[htbp]
    \centering
    \includegraphics[width=0.6\textwidth]{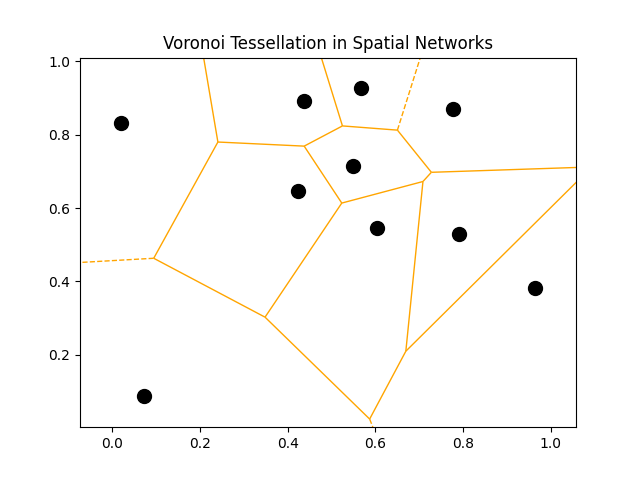}
    \caption{Voronoi Tessellation in Spatial Networks}
    \label{fig:voronoi-tessellation}
\end{figure}

Figure \ref{fig:voronoi-tessellation} demonstrates Voronoi tessellation in a spatial network. Ten nodes (black dots) are segmented into distinct regions (orange outlines), each closest to a single node, showcasing spatial segmentation.

\subsubsection{Correlation Analysis in Spatial Networks}

The Pearson correlation coefficient, a widely accepted statistical measure, quantifies the linear relationships between spatial and network metrics.  

Given two variables, \(X\) and \(Y\), which could represent any spatial or network metric, the Pearson correlation coefficient is defined as:

\begin{equation}
\rho_{X,Y} = \frac{\text{cov}(X,Y)}{\sigma_X \sigma_Y}
\end{equation}

Here, \(\text{cov}(X,Y)\) represents the covariance between the variables \(X\) and \(Y\), while \(\sigma_X\) and \(\sigma_Y\) denote the standard deviations of \(X\) and \(Y\), respectively. This coefficient ranges from -1 to 1, where 1 indicates a perfect positive linear relationship, -1 indicates a perfect negative linear relationship, and 0 signifies no linear correlation. \cite{Newman2018}

In the context of spatial networks, \(X\) and \(Y\) could represent various aspects, such as the physical distance between nodes (a spatial metric) and the number of connections (a network metric). By applying this correlation analysis, researchers can uncover significant patterns and dependencies. For instance, one might investigate whether nodes that are closer in space tend to have more connections or whether longer distances correlate with specific network behaviors.

\begin{figure}[htbp]
    \centering
    \includegraphics[width=\textwidth]{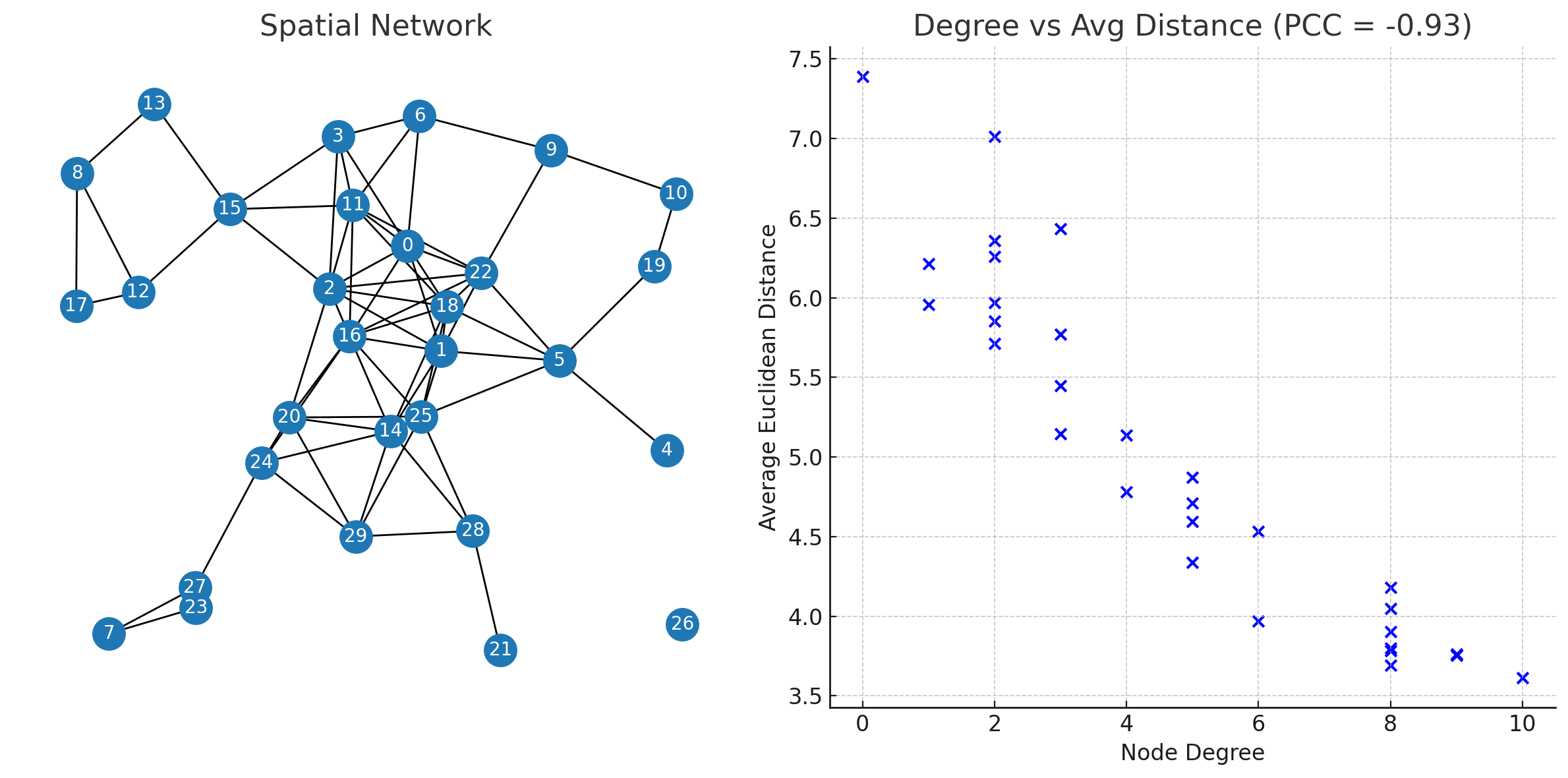}
    \caption{Illustration of Pearson Correlation in Spatial Networks}
    \label{fig:pearson-correlation}
\end{figure}

Figure \ref{fig:pearson-correlation} shows a spatial network where nodes are connected based on proximity, visualized with larger nodes and white labels for clarity. The accompanying scatter plot demonstrates a negative Pearson Correlation Coefficient between node degrees and their average distances, suggesting that highly connected nodes tend to be closer to each other. This pattern highlights the role of spatial proximity in network connectivity and has significant implications for the analysis and design of spatial networks in practical fields.

\subsection{Spatial Properties and Isomorphism}

The concept of isomorphism in spatial networks introduces unique considerations compared to traditional graph theory. Spatial isomorphism considers not only the structure of the network but also the spatial positioning of nodes.

\subsubsection{Theoretical Definition of Spatial Isomorphism}

Spatial isomorphism between two spatial networks \( G_{s1} = (V_1, E_1, P_1) \) and \( G_{s2} = (V_2, E_2, P_2) \) occurs when there is a bijection \( f: V_1 \rightarrow V_2 \) such that:

\begin{equation}
(v_i, v_j) \in E_1 \Leftrightarrow (f(v_i), f(v_j)) \in E_2 \quad \text{and} \quad P_1(v_i) = P_2(f(v_i)) \quad \forall v_i \in V_1
\end{equation}

This definition implies that corresponding nodes and edges in the two networks are not only connected in the same way but also have the same spatial coordinates.

\begin{figure}[H]
\centering

\begin{minipage}[b]{0.25\textwidth}
    \centering
    \begin{tikzpicture}[scale=0.8,auto=left]
      \draw[step=1cm,gray,very thin] (0,0) grid (4,4);
      \node[shape=circle,fill=blue!20] (n1) at (1,1) {1};
      \node[shape=circle,fill=blue!20] (n2) at (2,3)  {2};
      \node[shape=circle,fill=blue!20] (n3) at (3,1)  {3};
      \draw[line width=2pt] (n1) -- (n2);
      \draw[line width=2pt] (n2) -- (n3);
      \draw[line width=2pt] (n1) -- (n3);
    \end{tikzpicture}
    \caption{ \( G_{1} \) is a Spatial network}
    \label{fig:spatial-network-1}
\end{minipage}
\raisebox{3cm}{\Huge \( = \)}
\begin{minipage}[b]{0.25\textwidth}
    \centering
    \begin{tikzpicture}[scale=0.8,auto=left]
      \draw[step=1cm,gray,very thin] (0,0) grid (4,4);
      \node[shape=circle,fill=red!20] (n1) at (1,1) {A};
      \node[shape=circle,fill=red!20] (n2) at (2,3)  {B};
      \node[shape=circle,fill=red!20] (n3) at (3,1)  {C};
      \draw[line width=2pt] (n1) -- (n2);
      \draw[line width=2pt] (n2) -- (n3);
      \draw[line width=2pt] (n1) -- (n3);
    \end{tikzpicture}
    \caption{\( G_{2} \) \& \( G_{1} \) are isomorphic }
    \label{fig:spatial-network-2}
\end{minipage}
\raisebox{3cm}{\Huge \( \neq \)}
\begin{minipage}[b]{0.25\textwidth}
    \centering
    \begin{tikzpicture}[scale=0.8,auto=left]
      \draw[step=1cm,gray,very thin] (0,0) grid (4,4);
      \node[shape=circle,fill=green!20] (n1) at (1,3) {1};
      \node[shape=circle,fill=green!20] (n2) at (2,1)  {2};
      \node[shape=circle,fill=green!20] (n3) at (3,3)  {3};
      \draw[line width=2pt] (n1) -- (n2);
      \draw[line width=2pt] (n2) -- (n3);
      \draw[line width=2pt] (n1) -- (n3);
    \end{tikzpicture}
    \caption{\( G_{3} \) is not isomorphic to \( G_{1} \) \& \( G_{2} \)}
    \label{fig:non-spatial-network}
\end{minipage}

\end{figure}

\subsubsection{Distinctions Between Spatial and Traditional Graph Isomorphism}

Unlike traditional graph isomorphism, which focuses solely on the preservation of adjacency relationships between nodes, spatial isomorphism requires the preservation of node locations. This additional spatial constraint means that spatially isomorphic networks are more strictly defined than their non-spatial counterparts.

\subsubsection{Theoretical Implications of Spatial Isomorphism}

Spatial isomorphism has several theoretical implications:
\begin{itemize}
    \item \textbf{Network Equivalence}: Spatial isomorphism implies a stronger form of equivalence between networks, considering both their structure and spatial arrangement.
    \item \textbf{Pattern Recognition}: It is crucial in the analysis of spatial patterns and structures within networks, as it helps identify similar spatial configurations.
\end{itemize}

\section{Dynamic Networks}

\subsection{Static vs. Dynamic Networks}

Dynamic networks represent a significant shift from static network models, accommodating changes in structure or attributes. This section explores the foundational concepts of dynamic networks, contrasting them with static networks

\paragraph{Static Networks:}
Defined as a graph \( G = (V, E) \), a static network's topology, consisting of nodes \( V \) and edges \( E \), remains constant. Such networks are ideal for systems with stable interactions or scenarios where the evolution of network connections is not a focus.

\paragraph{Dynamic Networks:}
Dynamic networks are distinguished by their flexibility, showing variations in their architecture and attributes as they respond to different conditions or contexts. Mathematically, these networks can be denoted as:
\[
G_i = (V_i, E_i), \quad i = 1, 2, \ldots, N
\]

where \( V_i \) and \( E_i \) denote the nodes and edges in the \( i \)-th configuration, and \( N \) represents the total number of unique configurations of the network. In essence, a single graph can take on multiple configurations, each adapting to a set of parameters that may be influenced by time, specific features, or particular situational variables. \cite{Mesbahi2010}

\begin{figure}[H]
\centering

\begin{minipage}{0.45\textwidth}
    \centering
    \begin{tikzpicture}[scale=0.9,auto=left,every node/.style={circle}]
      \node[fill=red!20] (a1) at (1,3) {\( v_{1i} \)};
      \node[fill=red!20] (a2) at (3,3) {\( v_{2i} \)};
      \node[fill=red!20] (a3) at (2,1) {\( v_{3i} \)};
      \draw (a1) -- node[above] {1} (a2);
      \draw (a2) -- node[right] {2} (a3);
      \draw (a3) -- node[left]  {3} (a1);
    \end{tikzpicture}
    \caption{State \( i \) of dynamic network}
    \label{fig:dynamic-network-state-i}
\end{minipage}
\hfill 
\begin{minipage}{0.45\textwidth}
    \centering
    \begin{tikzpicture}[scale=0.9,auto=left,every node/.style={circle}]
      \node[fill=blue!20] (b1) at (1,3) {\( v_{1j} \)};
      \node[fill=blue!20] (b2) at (3,3) {\( v_{2j} \)};
      \node[fill=blue!20] (b3) at (2,1) {\( v_{3j} \)};
      \draw (b1) -- node[above] {2} (b2);
      \draw[dashed] (b2) -- node[right] {\(\varnothing\)} (b3);
      \draw (b3) -- node[left]  {5} (b1);
    \end{tikzpicture}
    \caption{State \( j \) of dynamic network}
    \label{fig:dynamic-network-state-j}
\end{minipage}

\end{figure}

\subsection{Variability in Dynamic Networks:}
Dynamic networks may occur due to changes influenced by a wide array of factors not confined to temporal shifts. For instance, alterations in network topology could be driven by feature-based changes in nodes or edges, decision-making processes, interaction dynamics, or external environmental conditions. This aspect highlights the versatility of dynamic networks in representing a single underlying graph through multiple realizations, each tailored to specific conditions or influences. \cite{Mesbahi2010}

\subsection{Transition from Static to Dynamic Models}
The progression from static to dynamic network models represents a shift towards a more holistic perspective, where networks are not just snapshots of interactions but are capable of reflecting a spectrum of changes. This broader view is crucial in understanding systems that exhibit varied behaviors or adaptations in different scenarios, making dynamic networks a powerful tool in capturing the complexity of real-world systems. \cite{Mikko2014}

\subsection{Example Dynamic Network: Adaptive Pathing Algorithm}
To illustrate the concept of a dynamic network, consider an adaptive pathing algorithm used for navigation using a set of waypoint nodes. This example showcases the network's adaptability not only in terms of the paths (edges) available based on various factors but also in the mobility and adjustability of the waypoints (nodes) themselves.

\paragraph{Dynamic Edges:}
In this network, the edges represent potential paths between waypoints, and their characteristics can vary depending on the agent's requirements and environmental factors, such as:
\begin{itemize}
    \item \textbf{Risk-Aversion}: Paths can be safer or riskier, depending on the agent's tolerance for risk.
    \item \textbf{Movement Mode}: Different modes of movement (walking, driving, flying) necessitate varying path options and constraints.
    \item \textbf{Fuel-Cost}: Paths may be optimized for fuel efficiency.
    \item \textbf{Environmental Constraints}: The paths chosen can vary based on weather conditions or terrain.
\end{itemize}

\begin{table}[htbp]
\centering
\resizebox{\textwidth}{!}{%
\begin{tabular}{|c|c|c|c|c|c|c|}
\hline
\textbf{Edge} & \textbf{Risk Level} & \textbf{Fuel Cost} & \textbf{Speed} & \textbf{Active (Min Risk)} & \textbf{Active (Min Fuel)} & \textbf{Active (Max Speed)} \\
\hline
A-B & Low & High & Fast & Yes & No & Yes \\
\hline
A-C & High & Low & Slow & No & Yes & No \\
\hline
B-C & Medium & Medium & Medium & Yes & Yes & Maybe \\
\hline
C-D & Low & High & Slow & Yes & No & No \\
\hline
D-A & High & Low & Fast & No & Yes & Yes \\
\hline
\end{tabular}
}
\caption{Edge Characteristics in the Adaptive Pathing Algorithm}
\label{tab:edge-characteristics}
\end{table}

\begin{figure}[htbp]
\centering
\begin{minipage}{.32\linewidth}
\centering
\begin{tikzpicture}[scale=0.6, every node/.style={circle, fill=blue!20}]
    \node (a) at (0,0) {A};
    \node (b) at (2,2) {B};
    \node (c) at (4,0) {C};
    \node (d) at (2,-2) {D};
    \draw[thick] (a) -- (b); 
    \draw[thick, dashed] (b) -- (c); 
    \draw[thick] (c) -- (d); 
\end{tikzpicture}
\caption*{Minimizing Risk}
\end{minipage}%
\begin{tikzpicture}[remember picture, overlay]
    \draw[<->,thick] (-0.5,-1.75) -- (0.5,-1.75);
\end{tikzpicture}
\begin{minipage}{.32\linewidth}
\centering
\begin{tikzpicture}[scale=0.6, every node/.style={circle, fill=blue!20}]
    \node (a) at (0,0) {A};
    \node (b) at (2,2) {B};
    \node (c) at (4,0) {C};
    \node (d) at (2,-2) {D};
    \draw[thick] (a) -- (c); 
    \draw[thick, dashed] (b) -- (c); 
    \draw[thick] (d) -- (a); 
\end{tikzpicture}
\caption*{Minimizing Fuel}
\end{minipage}%
\begin{tikzpicture}[remember picture, overlay]
    \draw[<->,thick] (-0.5,-1.75) -- (0.5,-1.75);
\end{tikzpicture}
\begin{minipage}{.32\linewidth}
\centering
\begin{tikzpicture}[scale=0.6, every node/.style={circle, fill=blue!20}]
\node (a) at (0,0) {A};
\node (b) at (2,2) {B};
\node (c) at (4,0) {C};
\node (d) at (2,-2) {D};
\draw[thick] (a) -- (b); 
\draw[thick, dashed] (b) -- (c); 
\draw[thick] (d) -- (a); 
\end{tikzpicture}
\caption*{Maximizing Speed}
\end{minipage}%

\caption{Dynamic Edges in Adaptive Pathing Network under Different Conditions}
\label{fig:dynamic-edges-conditions}
\end{figure}

\paragraph{Dynamic Nodes:}
Moreover, the waypoints themselves are dynamic. Their position or availability can change based on factors like the speed of the agent or power usage. This dynamic aspect includes:
\begin{itemize}
    \item In high-speed scenarios, waypoints might be placed farther apart.
    \item For energy-efficient navigation, waypoints could be positioned to minimize energy consumption.
    \item terrain or environmental factors may influence how waypoints may be placed, such as ground-based vs air-based movement. 
\end{itemize}

\begin{table}[htbp]
\centering
\resizebox{\textwidth}{!}{%
\begin{tabular}{|c|c|c|c|c|c|}
\hline
\textbf{Node} & \textbf{Standard Position} & \textbf{High Speed} & \textbf{Energy Efficient} & \textbf{Ground Access} & \textbf{Air Access} \\
\hline
A & (0,0) & (0,0) & (0,0) & Yes & Yes \\
\hline
B & (2,2) & (4,4) & (1,1) & No & Yes \\
\hline
C & (4,0) & (8,0) & (2,0) & Yes & Yes \\
\hline
D & (2,-2) & (4,-4) & (1,-1) & Yes & No \\
\hline
\end{tabular}
}
\caption{Dynamic Node Positions in Different Scenarios}
\label{tab:node-characteristics}
\end{table}

\begin{figure}[htbp]
\centering
\begin{minipage}[b]{.4\linewidth}
\centering
\caption*{Ground Access Node Positions}
\begin{tikzpicture}[scale=0.5]
    \node[circle, fill=blue!20] (a_std) at (-2,0) {A};
    \node[circle, fill=red!30] (a_hs) at (-4,0) {A};
    \node[circle, fill=green!30] (a_ee) at (0,0) {A};
    \node[circle, fill=blue!20] (b_std) at (1,-4) {D};
    \node[circle, fill=red!30] (b_hs) at (1,-6) {D};
    \node[circle, fill=green!30] (b_ee) at (1,-2) {D};
    \node[circle, fill=blue!20] (c_std) at (4,0) {C};
    \node[circle, fill=red!30] (c_hs) at (6,0) {C};
    \node[circle, fill=green!30] (c_ee) at (2,0) {C};
\end{tikzpicture}
\end{minipage}%
\begin{minipage}[b]{.4\linewidth}
\centering
\caption*{Air Access Node Positions}
\begin{tikzpicture}[scale=0.5]
    \node[circle, fill=blue!20] (a_std) at (-2,0) {A};
    \node[circle, fill=red!30] (a_hs) at (-4,0) {A};
    \node[circle, fill=green!30] (a_ee) at (0,0) {A};
    \node[circle, fill=blue!20] (b_std) at (1,4) {B};
    \node[circle, fill=red!30] (b_hs) at (1,6) {B};
    \node[circle, fill=green!30] (b_ee) at (1,2) {B};
    \node[circle, fill=blue!20] (c_std) at (4,0) {C};
    \node[circle, fill=red!30] (c_hs) at (6,0) {C};
    \node[circle, fill=green!30] (c_ee) at (2,0) {C};
\end{tikzpicture}
\end{minipage}%
\begin{minipage}[b]{.19\linewidth}
\centering
\begin{tikzpicture}[scale=0.5]
\node[circle, fill=blue!20, label=right:Standard] at (0,3) {};
\node[circle, fill=red!30, label=right:High Speed] at (0,2) {};
\node[circle, fill=green!30, label=right:Energy Efficient] at (0,1) {};
\node[circle, fill=white] at (0,-4) {};
\end{tikzpicture}
\end{minipage}

\caption{Comparative Dynamic Nodes in Adaptive Pathing Network}
\label{fig:dynamic-nodes-comparative}
\end{figure}

\paragraph{Adaptive Network Characteristics:}
This pathing algorithm exemplifies a dynamic network where both the nodes (waypoints) and edges (paths) are variable. The network adapts to the agent's specific characteristics and environmental conditions, offering a range of possible paths and waypoint configurations. This adaptability makes the network ideal for applications in fields like robotics, autonomous vehicle navigation, and adaptive logistics systems.

\section{Temporal Networks}

\subsection{Temporal Networks are Dynamic}
Temporal networks represent a specialized category within the broader spectrum of dynamic networks, distinguished primarily by their explicit incorporation of the temporal dimension into network structure and analysis. This section introduces temporal networks, differentiates them from general dynamic networks, and underscores the significance of the temporal aspect in network studies. 

\paragraph{Definition and Differentiation}
Temporal networks are defined by the connections between nodes that are not only defined by their existence but also by the time at which these connections occur. The explicit consideration of the timing and order of interactions distinguishes temporal networks from other types of dynamic networks, which may involve changes in network structure or attributes that are not inherently time-ordered. \cite{Holme_2012}

\begin{figure}[H]
\centering
\tikzset{
  node_style/.style={circle, fill=black, inner sep=0pt, minimum size=0.25cm},
  new_node/.style={circle, fill=red, inner sep=0pt, minimum size=0.25cm},
  new_edge/.style={draw=red, thick},
  delete_edge/.style={draw=red, dashed, thick},
  time_frame/.style={draw=blue, very thick},
  time_label/.style={rectangle, fill=none, draw=none, font=\Large} 
}

\setlength{\tabcolsep}{-5pt} 
\begin{tabular}{ccccc}

\begin{minipage}{0.2\textwidth}
\begin{tikzpicture}[scale=0.5, every node/.style={node_style}]
  \node (n1) at (1, 2) {};
  \node (n2) at (0,3.5) {};
  \node (n3) at (1.75,4) {};
  \node (n4) at (2,1) {};
  \node (n5) at (0,0.8)  {};
  \node (n6) at (1.25,0)  {};

  \draw (n1) -- (n2);
  \draw (n1) -- (n3);
  \draw (n1) -- (n4);
  \draw (n1) -- (n5);
  \draw (n1) -- (n6);

    \draw[time_frame] (-1,-2) -- (3,0) -- (3,6) -- (-1,4) -- cycle;

   \node[time_label] (t0) at (2,-2) {t=0}; 
\end{tikzpicture}
\end{minipage}

\begin{minipage}{0.2\textwidth}
\begin{tikzpicture}[scale=0.5, every node/.style={node_style}]
  \node (n1) at (1, 2) {};
  \node (n2) at (0,3.5) {};
  \node (n3) at (1.75,4) {};
  \node (n4) at (2,1) {};
  \node (n5) at (0,0.8)  {};
  \node (n6) at (1.25,0)  {};

  \draw (n1) -- (n2);
  \draw (n1) -- (n3);
  \draw (n1) -- (n4);
  \draw (n1) -- (n5);
  \draw (n1) -- (n6);
  \draw[new_edge] (n4) -- (n3);

   \draw[time_frame] (-1,-2) -- (3,0) -- (3,6) -- (-1,4) -- cycle;

   \node[time_label] (t0) at (2,-2) {t=1}; 
\end{tikzpicture}
\end{minipage}

\begin{minipage}{0.2\textwidth}
\begin{tikzpicture}[scale=0.5, every node/.style={node_style}]
  \node (n1) at (1, 2) {};
  \node (n2) at (0,3.5) {};
  \node (n3) at (1.75,4) {};
  \node (n4) at (2,1) {};
  \node (n5) at (0,0.8)  {};
  \node (n6) at (1.25,0)  {};

  \draw (n1) -- (n2);
  \draw[delete_edge] (n1) -- (n3);
  \draw (n1) -- (n4);
  \draw (n1) -- (n5);
  \draw (n1) -- (n6);
   \draw(n4) -- (n3);

  \draw[time_frame] (-1,-2) -- (3,0) -- (3,6) -- (-1,4) -- cycle;

   \node[time_label] (t0) at (2,-2) {t=2}; 
\end{tikzpicture}
\end{minipage}

\begin{minipage}{0.2\textwidth}
\begin{tikzpicture}[scale=0.5, every node/.style={node_style}]
  \node (n1) at (1, 2) {};
  \node (n2) at (0,3.5) {};
  \node (n3) at (1.75,4) {};
  \node (n4) at (2,1) {};
  \node (n5) at (0,0.8)  {};
  \node (n6) at (1.25,0)  {};
  \node[new_node] (n7) at (-0.3, 2) {};

  \draw (n1) -- (n2);
  \draw (n1) -- (n4);
  \draw (n1) -- (n5);
  \draw (n1) -- (n6);
  \draw (n4) -- (n3);
  \draw[new_edge] (n1) -- (n7);

   \draw[time_frame] (-1,-2) -- (3,0) -- (3,6) -- (-1,4) -- cycle;

   \node[time_label] (t0) at (2,-2) {t=3}; 
\end{tikzpicture}
\end{minipage}

\end{tabular}

\caption{Temporal evolution of the network}
\label{fig:temporal-network}
\end{figure}
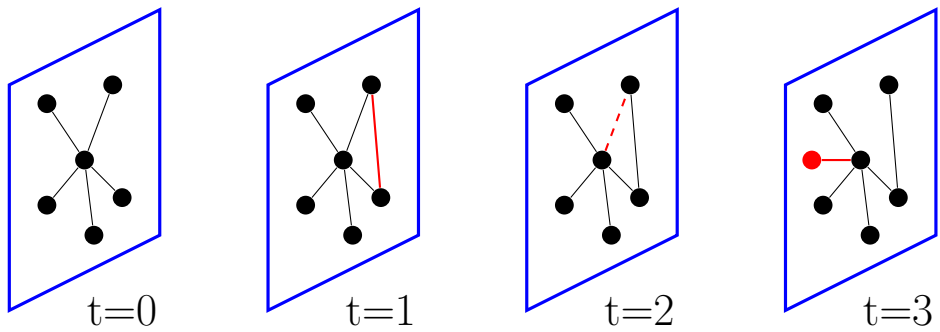

\subsection{Importance of Temporal Dimension}
The temporal dimension introduces a critical layer of complexity to network analysis. In temporal networks, the direction and timing of edge formations and dissolutions significantly influence network properties such as connectivity and path lengths. This aspect allows for a detailed representation of real-world systems where interactions are time-dependent. 

Unlike undirected dynamic networks, temporal networks often necessitate treating the sequence of state changes as an ordered set. \cite{Holme_2012}

\paragraph{Directed Dynamic Networks}
Temporal networks encapsulate the concept of causality, wherein time is perceived as flowing from previous states to subsequent states. This causality underpins the representation of a temporal network as a series of temporal states, each directed from one to the next in an irreversible relationship. Such a characteristic distinctly positions temporal networks as a special case of directed dynamic networks, emphasizing the sequential nature of time-dependent interactions. \cite{Nicosia_2013, michail2015introduction}

\begin{figure}[ht!]
\centering
\tikzset{
  node_style/.style={circle, fill=black, inner sep=0pt, minimum size=0.25cm},
  active_edge/.style={draw=red, thick},
  transition_arrow/.style={->, very thick, shorten >=1pt}
}

\begin{tikzpicture}

\begin{scope}[xshift=0cm]
  \node[node_style] (A1) at (0, 0) {};
  \node[node_style] (B1) at (1, 1.5) {};
  \node[node_style] (C1) at (2, 0) {};

  \draw[active_edge] (A1) -- (B1);
  \draw[active_edge] (B1) -- (C1);
\end{scope}

\draw[transition_arrow] (2.5, 0.75) -- (3.5, 0.75);

\begin{scope}[xshift=4cm]
  \node[node_style] (A2) at (0, 0) {};
  \node[node_style] (B2) at (1, 1.5) {};
  \node[node_style] (C2) at (2, 0) {};

  \draw[active_edge] (C2) -- (A2);
  \draw[active_edge] (B2) -- (A2);
\end{scope}

\draw[transition_arrow] (6.5, 0.75) -- (7.5, 0.75);

\begin{scope}[xshift=8cm]
  \node[node_style] (A3) at (0, 0) {};
  \node[node_style] (B3) at (1, 1.5) {};
  \node[node_style] (C3) at (2, 0) {};

  \draw[active_edge] (A3) -- (C3);
  \draw[active_edge] (C3) -- (B3);
\end{scope}

\end{tikzpicture}
\caption{Illustration of Directed Dynamic Networks showcasing the progression and directedness of network changes. Each graph represents a different state with varying edge activations.}
\label{fig:directed-dynamic-networks}
\end{figure}
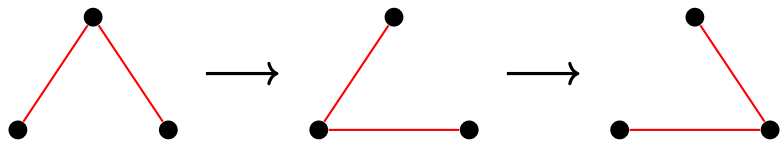

\subsection{Mathematical Representations of Temporal Networks}

Two common approaches to representing these networks are event-based representation and snapshot representation.

\subsubsection{Event-Based Representation}
In an event-based representation, a temporal network is described by a series of events, each representing an interaction between a pair of nodes at a specific time. This can be mathematically formulated as a set of triples:

\[
E_{\text{event}} = \{(u, v, t) | u, v \in V, t \in \mathbb{T}\}
\]

where \( (u, v) \) represents an edge between nodes \( u \) and \( v \) at time \( t \), and \( \mathbb{T} \) is the set of all time points at which events occur. This representation is particularly useful for capturing the precise timing of interactions. \cite{Masuda2016}

\begin{table}[htbp]
  \centering
  \caption{Example: Event-based graph log with activation and deactivation times}
  \label{tab:event-based-log}
  \begin{tabular}{ccc}
    \toprule
    Edge & Activation Time & Deactivation Time \\
    \midrule
    (A, B) & t=1 & t=2 \\
    (B, C) & t=2 & t=3 \\
    (C, D) & t=3 & t=5 \\
    (A, D) & t=4 & t=6 \\
    (B, D) & t=5 & t=7 \\
    (C, A) & t=6 & t=8 \\
    (A, B) & t=9 & t=10 \\ 
    \bottomrule
  \end{tabular}
\end{table}

Table~\ref{tab:event-based-log} demonstrates an example of an event-based graph, where the table records the time intervals during which each edge is active, capturing the dynamic nature of temporal network interactions, including instances of reactivation as seen with the edge (A, B).

\subsubsection{Snapshot Representation}
Alternatively, in snapshot representation, the temporal network is divided into a sequence of static graphs (snapshots), each representing the network at a particular time interval. This can be expressed as:

\[
G(t) = (V(t), E(t)), \quad t = 1, 2, \ldots, T
\]

where:
\begin{itemize}
    \item \( V(t) \) is the set of nodes at time \( t \),
    \item \( E(t) \) is the set of edges present during the time interval \( t \),
    \item \( T \) is the total number of discrete time intervals.
\end{itemize}

\begin{table}[htbp]
  \centering
  \caption{Example: Snapshot representation of a temporal network}
  \label{tab:snapshot-representation}
  \begin{tabular}{cc}
    \toprule
    Time Interval & Active Edges \\
    \midrule
    t=1 & (A, B) \\
    t=2 & (B, C) \\
    t=3 & (C, D) \\
    t=4 & (A, D), (C, D) \\
    t=5 & (A, D), (B, D), (C, D) \\
    t=6 & (B, D), (C, A), (C, D) \\
    t=7 & (B, D), (C, A) \\
    t=8 & (C, A) \\
    t=9 & (A, B) \\
    t=10 &  \\
    \bottomrule
  \end{tabular}
\end{table}

In contrast to the event-based representation, the snapshot representation provides a sequential view of the network's state at discrete time intervals. This approach is advantageous for understanding the network's structure at specific moments in time \cite{Masuda2016}. Table~\ref{tab:snapshot-representation} illustrates the snapshot representation of the same temporal network, where each row corresponds to a different time interval, indicating which edges are active at that specific time.

\subsubsection{Choice of Representation}
The choice between event-based and snapshot representations depends on the nature of the temporal data and the specific analysis goals. Event-based representation is more granular and suitable for data with precise time stamps, while snapshot representation simplifies the network into discrete intervals, which can be advantageous for certain types of analysis. \cite{Blonder2012}

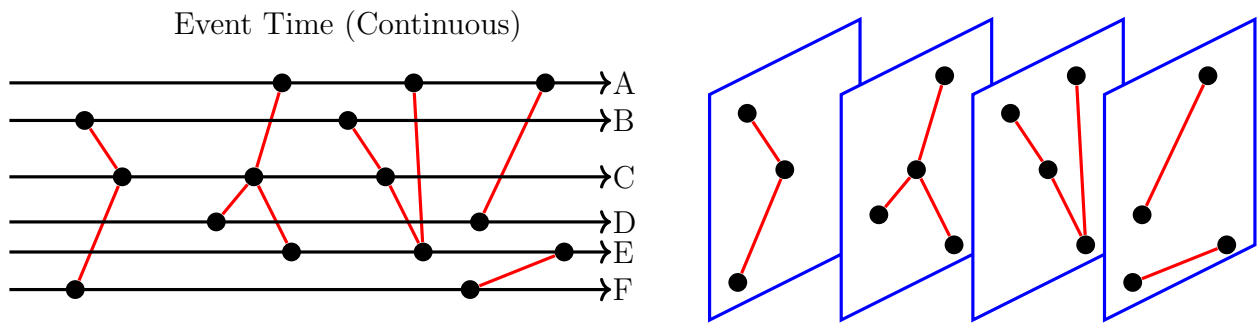
\begin{figure}[ht!]
\centering
\tikzset{
  node_style/.style={circle, fill=black, inner sep=0pt, minimum size=0.25cm},
  new_node/.style={circle, fill=red, inner sep=0pt, minimum size=0.25cm},
  new_edge/.style={draw=red, thick},
  delete_edge/.style={draw=red, dashed, thick},
  time_frame/.style={draw=blue, very thick, fill=white},
  time_label/.style={rectangle, fill=none, draw=none} 
}

\begin{minipage}{\textwidth}
  \setlength{\tabcolsep}{-5pt} 
  \begin{tabular}{cc}
\centering
   \begin{minipage}{0.3\textwidth}
    \begin{tikzpicture}[scale=0.5, every node/.style={node_style}]
      
      \node (n1t0) at (1, 2) {};
      \node (n2t0) at (0,3.5) {};
      \node (n6t0) at (-0.25,-1)  {};
    
      \draw[new_edge, very thick] (n1t0) -- (n2t0);
      \draw[new_edge, very thick] (n1t0) -- (n6t0);

      \node (n1t1) at (4.5, 2) {};
      \node (n3t1) at (5.25,4.5) {};
      \node (n4t1) at (5.5,0) {};
      \node (n5t1) at (3.5,0.8)  {};
    
      \draw[new_edge, very thick] (n1t1) -- (n3t1);
      \draw[new_edge, very thick] (n1t1) -- (n5t1);
      \draw[new_edge, very thick] (n4t1) -- (n1t1);


      \node (n1t2) at (8, 2) {};
      \node (n2t2) at (7,3.5) {};
      \node (n3t2) at (8.75,4.5) {};
      \node (n4t2) at (9,0) {};
    
      \draw[new_edge, very thick] (n1t2) -- (n2t2);
      \draw[new_edge, very thick] (n1t2) -- (n4t2);
      \draw[new_edge, very thick] (n4t2) -- (n3t2);


          \node (n3t3) at (12.25,4.5) {};
          \node (n4t3) at (12.75,0) {};
          \node (n5t3) at (10.5,0.8)  {};
          \node (n6t3) at (10.25,-1)  {};
        
          \draw[new_edge, very thick] (n3t3) -- (n5t3);
          \draw[new_edge ,very thick] (n4t3) -- (n6t3);

        \draw[->, very thick] (-2,2) -- (14,2) node[time_label, anchor=west] {C};
        \draw[->, very thick] (-2,3.5) -- (14,3.5) node[time_label, anchor=west] {B};
        \draw[->, very thick] (-2,4.5) -- (14,4.5) node[time_label, anchor=west] {A};
         \draw[->, very thick] (-2,-1) -- (14,-1) node[time_label, anchor=west] {F};
         \draw[->, very thick] (-2,0) -- (14,0) node[time_label, anchor=west] {E};
         \draw[->, very thick] (-2,0.8) -- (14,0.8) node[time_label, anchor=west] {D};
       \node[time_label] (t0) at (7,-2.75) {}; 
       \node[time_label] (t0) at (7,6) {Event Time (Continuous)};

        \end{tikzpicture}
    \end{minipage}
    
    &
    \hspace{130pt}
    \begin{minipage}{0.3\textwidth}
    \begin{tikzpicture}[scale=0.5, every node/.style={node_style}]
      \draw[time_frame] (-1,-2) -- (3,0) -- (3,6) -- (-1,4) -- cycle;

      \node (n1t0) at (1, 2) {};
      \node (n2t0) at (0,3.5) {};
      \node (n6t0) at (-0.25,-1)  {};
    
      \draw[new_edge, very thick] (n1t0) -- (n2t0);
      \draw[new_edge, very thick] (n1t0) -- (n6t0);
    
       \node[time_label] (t0) at (-0.5,-2.75) {}; 

       \draw[time_frame] (2.5,-2) -- (6.5,0) -- (6.5,6) -- (2.5,4) -- cycle;
      \node (n1t1) at (4.5, 2) {};
      \node (n3t1) at (5.25,4.5) {};
      \node (n4t1) at (5.5,0) {};
      \node (n5t1) at (3.5,0.8)  {};
    
      \draw[new_edge, very thick] (n1t1) -- (n3t1);
      \draw[new_edge, very thick] (n1t1) -- (n5t1);
      \draw[new_edge, very thick] (n4t1) -- (n1t1);
    
       \node[time_label] (t1) at (3,-2.75) {}; 

      \draw[time_frame] (6,-2) -- (10,0) -- (10,6) -- (6,4) -- cycle;

      \node (n1t2) at (8, 2) {};
      \node (n2t2) at (7,3.5) {};
      \node (n3t2) at (8.75,4.5) {};
      \node (n4t2) at (9,0) {};
    
      \draw[new_edge, very thick] (n1t2) -- (n2t2);
      \draw[new_edge, very thick] (n1t2) -- (n4t2);
      \draw[new_edge, very thick] (n4t2) -- (n3t2);

       \node[time_label] (t2) at (6.5,-2.75) {}; 

        \draw[time_frame] (9.5,-2) -- (13.5,0) -- (13.5,6) -- (9.5,4) -- cycle;

          \node (n3t3) at (12.25,4.5) {};
          \node (n4t3) at (12.75,0) {};
          \node (n5t3) at (10.5,0.8)  {};
          \node (n6t3) at (10.25,-1)  {};
        
          \draw[new_edge, very thick] (n3t3) -- (n5t3);
          \draw[new_edge ,very thick] (n4t3) -- (n6t3);
        
           \node[time_label] (t3) at (10.25,-2.75) {}; 
       
        \end{tikzpicture}
    \end{minipage}

  \end{tabular}
  \end{minipage}

\caption{Event-based versus Snapshot-based}
\label{fig:events-vs-frames}  
\end{figure}

Figure~\ref{fig:events-vs-frames} illustrates the two principal methods of representing temporal networks: event-based and snapshot-based. On the left, the event-based approach is depicted, where interactions are represented as a continuous stream, with edges drawn at the exact times of occurrence. Each horizontal line corresponds to a node, and the directed edges represent interactions occurring over time, marked by their temporal order. On the right, the snapshot-based method is visualized by segmenting the timeline into discrete frames, each encapsulating the network's state within a specific time interval.

\subsection{General Dynamics in Temporal Networks}

This section delves into the intricate dynamics of temporal networks, focusing on the distinct characteristics and implications of Markovian random, Bayesian random, and Uniform random behaviors. \cite{porter2015dynamical}

\subsubsection{Evolutionary Dynamics}

\paragraph{ \textit{Definition:}}
Markovian random dynamics in temporal networks refer to situations where the current state of the network is influenced by its previous states. This type of dynamics is characterized by a dependency on the history of the network. \cite{Sheng2023, Galdeman2023}

\paragraph{ \textit{Mathematical Representation:} }
Markovian random dynamics can be mathematically represented by incorporating historical dependency into the network model. For instance:

\[
G(t) = f(G(t-1), G(t-2), \ldots, G(t-k))
\]

where \( G(t) \) is the state of the network at time \( t \), and \( f \) is a function that determines the current state based on \( k \) previous states.

\paragraph{ \textit{Significance:} }
In Markovian random temporal networks, patterns and structures evolve in a more predictable manner, following the network's historical trends. This predictability is essential in scenarios where past interactions influence future connections.

\begin{table}[htbp]
  \centering
  \caption{Markvoian Random Dynamics in a Temporal Network}
  \label{tab:evolutionary-dynamics}
  \begin{tabular}{cc}
    \toprule
    Time & Active Edges \\
    \midrule
    t=1 & A-B \\
    t=2 & B-C \\
    t=3 & C-D \\
    t=4 & D-E \\
    t=5 & E-F \\
    t=6 & F-A \\
    \bottomrule
  \end{tabular}
\end{table}

  Table~\ref{tab:evolutionary-dynamics} illustrates a network that evolves through a sequence of edge activations over time. At each time step, a new edge forms, creating a chain of connections that demonstrate the influence of prior states on subsequent ones. This pattern of edge activations showcases an evolutionary-like nature of the network, where each state is a progression from the previous one, leading to a cyclical connection pattern that completes at time \( t=6 \) with the edge F-A.

\subsubsection{Bayesian Random Dynamics}
\paragraph{ \textit{Definition:} }
Bayesian Random dynamics in temporal networks describe scenarios where the current state of the network is estimated by the aggregate of prior states in a probabilistic manner. While there is no direct dependency, as seen in Markovian Random networks, past states inform the likelihood or probability of various future events or states occurring. \cite{Newman2014}

\paragraph{ \textit{Mathematical Representation:} }
The stochastic nature of such networks can be mathematically represented as:

\[
G(t) = f(P(G(t-1)), P(G(t-2)), \ldots, P(G(t-k)))
\]

Here, \( G(t) \) denotes the state of the network at time \( t \), and \( f \) is a function that integrates the probabilities \( P \) associated with \( k \) previous states to determine the likely state at \( t \).

\paragraph{ \textit{Significance:} }
Bayesian dynamics model systems where prior states increase or decrease the likelihood of future occurrences without dictating them. This approach is especially useful in fields where historical trends impact, but do not fully determine, future outcomes. \cite{Grimmett2001}

\begin{table}[htbp]
  \centering
  \caption{Bayesian Stochastic Dynamics in a Temporal Network}
  \label{tab:stochastic-dynamics}
  \begin{tabular}{cc}
    \toprule
    Time & Active Edges \\
    \midrule
    t=1 & A-B \\
    t=2 & A-B \\
    t=3 & C-E \\
    t=4 & A-B \\
    t=5 & F-D \\
    t=6 & A-B \\
    \bottomrule
  \end{tabular}
\end{table}

Table~\ref{tab:stochastic-dynamics} presents an example of Bayesian dynamics in a temporal network. The table illustrates a sequence of active edges over different time intervals. While some edges, like A-B, recur at multiple time steps, others, such as C-E and F-D, appear sporadically. This pattern reflects the probabilistic nature of stochastic dynamics, where past states influence the likelihood of future states but do not determine them with certainty. The randomness embedded in the occurrence of edges like C-E and F-D contrasts with the more predictable reoccurrence of the A-B edge, showcasing the Bayesian behavior of the network.

\subsubsection{Uniform Random Dynamics}
\paragraph{ \textit{Definition:} }
Uniform Random dynamics in temporal networks imply that the state of the network at any given time is independent of its past states. These dynamics are often modeled as stochastic processes.  \cite{newman2006structure}

\paragraph{ \textit{Mathematical Representation:} }
A uniform random dynamic temporal network can be represented as:

\[
G(t) = f(\epsilon_t)
\]

where \( \epsilon_t \) is a random variable or a stochastic process that influences the network state at time \( t \) independently of previous states.

\paragraph{ \textit{Significance:} }
Uniform Random dynamics model systems where interactions are sporadic and unpredictable. The independence from historical data makes these models suitable for scenarios where future states are not influenced by the past. \cite{newman2006structure}

\begin{table}[htbp]
  \centering
  \caption{Uniform Random Dynamics in a Temporal Network}
  \label{tab:random-dynamics}
  \begin{tabular}{cc}
    \toprule
    Time & Active Edges \\
    \midrule
    t=1 & C-D \\
    t=2 & E-F \\
    t=3 & A-C \\
    t=4 & B-D \\
    t=5 & E-A \\
    t=6 & F-B \\
    \bottomrule
  \end{tabular}
\end{table}

As illustrated in Table~\ref{tab:random-dynamics}, random dynamics within temporal networks are characterized by the lack of dependency between the network's states over time. Each time interval reveals a set of active edges that do not follow a predictable pattern or historical influence, emphasizing the network's non-deterministic and independent nature at each time step.

\subsection{Analytical Challenges in Temporal Networks}

Temporal networks pose unique challenges due to their time-ordered nature and the computational complexities involved. This subsection addresses these challenges and introduces methods for effective analysis.

\subsubsection{Challenges with Time-Ordered Data}
\paragraph{Complexity of Time-Ordered Interactions:}
The primary challenge in analyzing temporal networks lies in the need to account for the time ordering of interactions. This temporal aspect can significantly influence network dynamics and requires careful handling to accurately capture the network's evolution. \cite{michail2015introduction}

\paragraph{Computational Complexity:}
Temporal networks often involve large datasets with fine-grained time stamps, leading to high computational complexity. Efficient data structures and algorithms are needed to manage and process this time-ordered data effectively. \cite{michail2015introduction}

\subsubsection{Methods for Managing Time-ordered Data}
\paragraph{Time-Aggregation:}
One approach to simplifying the analysis of temporal networks is time-aggregation, where interactions over a specified time period are aggregated into a single network snapshot. Mathematically, this can be represented as:

\[
G_{\text{agg}}(T) = \bigcup_{t=1}^{T} G(t)
\]

where \( G_{\text{agg}}(T) \) represents the aggregated network over the time interval \( T \), and \( G(t) \) is the network state at each time step \( t \). This method reduces the complexity of the network but can also lead to the loss of fine-grained temporal information. \cite{holme2021temporal, George2008, Nicosia_2012}

\begin{figure}[H]
\centering
\tikzset{
  node_style/.style={circle, fill=black, inner sep=0pt, minimum size=0.25cm},
  new_node/.style={circle, fill=red, inner sep=0pt, minimum size=0.25cm},
  new_edge/.style={draw=red, thick},
  delete_edge/.style={draw=red, dashed, thick},
  time_frame/.style={draw=blue, very thick, fill=white},
  time_label/.style={rectangle, fill=none, draw=none} 
}

\begin{minipage}{\textwidth}
  \setlength{\tabcolsep}{-5pt} 
  \begin{tabular}{cc}
    \begin{minipage}{0.48\textwidth}
    \begin{tikzpicture}[scale=0.5, every node/.style={node_style}]
      \draw[time_frame] (-1,-2) -- (3,0) -- (3,6) -- (-1,4) -- cycle;
      \node (n1t0) at (1, 2) {};
      \node (n2t0) at (0,3.5) {};
      \node (n3t0) at (1.75,4) {};
      \node (n4t0) at (2,1) {};
      \node (n5t0) at (0,0.8)  {};
      \node (n6t0) at (1.25,0)  {};
    
      \draw (n1t0) -- (n2t0);
      \draw (n1t0) -- (n3t0);
      \draw (n1t0) -- (n4t0);
      \draw (n1t0) -- (n5t0);
      \draw (n1t0) -- (n6t0);
    
       \node[time_label] (t0) at (1.25,-2) {$\Delta t_0$}; 

       \draw[time_frame] (2.5,-2) -- (6.5,0) -- (6.5,6) -- (2.5,4) -- cycle;
      \node (n1t1) at (4.5, 2) {};
      \node (n2t1) at (3.5,3.5) {};
      \node (n3t1) at (5.25,4) {};
      \node (n4t1) at (5.5,1) {};
      \node (n5t1) at (3.5,0.8)  {};
      \node (n6t1) at (4.75,0)  {};
    
      \draw (n1t1) -- (n2t1);
      \draw (n1t1) -- (n3t1);
      \draw (n1t1) -- (n4t1);
      \draw (n1t1) -- (n5t1);
      \draw (n1t1) -- (n6t1);
      \draw[new_edge] (n4t1) -- (n3t1);
    
       \node[time_label] (t1) at (4.5,-2) {$\Delta t_1$}; 

      \draw[time_frame] (6,-2) -- (10,0) -- (10,6) -- (6,4) -- cycle;
      
      \node (n1t2) at (8, 2) {};
      \node (n2t2) at (7,3.5) {};
      \node (n3t2) at (8.75,4) {};
      \node (n4t2) at (9,1) {};
      \node (n5t2) at (7,0.8)  {};
      \node (n6t2) at (8.25,0)  {};
    
      \draw (n1t2) -- (n2t2);
      \draw[delete_edge] (n1t2) -- (n3t2);
      \draw (n1t2) -- (n4t2);
      \draw (n1t2) -- (n5t2);
      \draw (n1t2) -- (n6t2);
      \draw(n4t2) -- (n3t2);

       \node[time_label] (t2) at (8,-2) {$\Delta t_2$}; 

        \draw[time_frame] (9.5,-2) -- (13.5,0) -- (13.5,6) -- (9.5,4) -- cycle;
        
          \node (n1t3) at (11.5, 2) {};
          \node (n2t3) at (10.5,3.5) {};
          \node (n3t3) at (12.25,4) {};
          \node (n4t3) at (12.5,1) {};
          \node (n5t3) at (10.5,0.8)  {};
          \node (n6t3) at (11.75,0)  {};
          \node[new_node] (n7t3) at (10.2, 2) {};
        
          \draw (n1t3) -- (n2t3);
          \draw (n1t3) -- (n4t3);
          \draw (n1t3) -- (n5t3);
          \draw (n1t3) -- (n6t3);
          \draw (n4t3) -- (n3t3);
          \draw[new_edge] (n1t3) -- (n7t3);
        
           \node[time_label] (t3) at (11.75,-2) {$\Delta t_3$};

        \end{tikzpicture}
        \caption{Time-varying Graph}
    \end{minipage}
    
    &
    \begin{minipage}{0.48\textwidth}
    \centering
    \begin{tikzpicture}[scale=1, every node/.style={node_style}]
      \node (n1) at (1, 2) {};
      \node (n2) at (0,3.5) {};
      \node (n3) at (1.75,4) {};
      \node (n4) at (2,1) {};
      \node (n5) at (0,0.8)  {};
      \node (n6) at (1.25,0)  {};
      \node (n7) at (-0.3, 2) {};
    
      \draw[very thick] (n1) -- (n2);
      \draw[very thick] (n1) -- (n3);
      \draw[very thick] (n1) -- (n4);
      \draw[very thick] (n1) -- (n5);
      \draw[very thick] (n1) -- (n6);
      \draw[very thick] (n1) -- (n7);
      \draw[very thick] (n3) -- (n4);
    \end{tikzpicture}
    \caption{Time-Aggregated Graph}
    \end{minipage}
  \end{tabular}
\end{minipage}

\label{fig:temporal-network-aggregate}
\end{figure}

\paragraph{Window-Based Analysis:}
 Another method is window-based analysis, where the network is analyzed over time using specific windows. Each window captures a subset of the temporal data. \cite{Masuda2019} Depending on the analysis objectives, these windows can be:

\begin{itemize}
  \item \textbf{Overlapping Windows (Rolling Windows):} These provide an evolutionary view of network dynamics, ensuring continuity in the network's edges across consecutive windows. This approach facilitates the observation of gradual changes and trends over time. \cite{Klobas2023Sliding} Additionally, this approach is prime for ARIMA (AutoRegressive Integrated Moving Average) time series forecasting \cite{Sikdar_2016}. 
  \item \textbf{Discrete Windows:} They offer distinct and clear segmentation of the network's state at specific time intervals. This approach is useful for analyzing changes that occur between separate, non-overlapping periods. \cite{Batagelj_2016}
\end{itemize}

\textbf{Mathematical Representation:} \\
The general approach of window-based analysis can be mathematically expressed as:

\[
G_{\text{window}}(t, w) = \bigcup_{\tau=t}^{t+w} G(\tau)
\]

where \( G_{\text{window}}(t, w) \) represents the network within the window starting at time \( t \) and spanning \( w \) time units. This method offers a versatile approach to capturing temporal dynamics while managing computational feasibility. \cite{Batagelj_2016}

\vspace{12pt}

\begin{figure}[ht!]
\centering
\tikzset{
  node_style/.style={circle, fill=black, inner sep=0pt, minimum size=0.25cm},
  new_node/.style={circle, fill=red, inner sep=0pt, minimum size=0.25cm},
  new_edge/.style={draw=red, thick},
  delete_edge/.style={draw=red, dashed, thick},
  time_frame/.style={draw=blue, very thick, fill=white},
  time_label/.style={rectangle, fill=none, draw=none} 
}

\begin{minipage}{\textwidth}
  \setlength{\tabcolsep}{-5pt} 
  \begin{tabular}{cc}
\centering
   \begin{minipage}{0.3\textwidth}
    \begin{tikzpicture}[scale=0.5, every node/.style={node_style}]
      
      \node (n1t0) at (1, 2) {};
      \node (n2t0) at (0,3.5) {};
      \node (n6t0) at (-0.25,-1)  {};
    
      \draw[new_edge, very thick] (n1t0) -- (n2t0);
      \draw[new_edge, very thick] (n1t0) -- (n6t0);

      \node (n1t1) at (4.5, 2) {};
      \node (n3t1) at (5.25,4.5) {};
      \node (n4t1) at (5.5,0) {};
      \node (n5t1) at (3.5,0.8)  {};
    
      \draw[new_edge, very thick] (n1t1) -- (n3t1);
      \draw[new_edge, very thick] (n1t1) -- (n5t1);
      \draw[new_edge, very thick] (n4t1) -- (n1t1);


      \node (n1t2) at (8, 2) {};
      \node (n2t2) at (7,3.5) {};
      \node (n3t2) at (8.75,4.5) {};
      \node (n4t2) at (9,0) {};
    
      \draw[new_edge, very thick] (n1t2) -- (n2t2);
      \draw[new_edge, very thick] (n1t2) -- (n4t2);
      \draw[new_edge, very thick] (n4t2) -- (n3t2);


          \node (n3t3) at (12.25,4.5) {};
          \node (n4t3) at (12.75,0) {};
          \node (n5t3) at (10.5,0.8)  {};
          \node (n6t3) at (10.25,-1)  {};
        
          \draw[new_edge, very thick] (n3t3) -- (n5t3);
          \draw[new_edge ,very thick] (n4t3) -- (n6t3);

        \draw[->, very thick] (-2,2) -- (14,2) node[time_label, anchor=west] {3};
        \draw[->, very thick] (-2,3.5) -- (14,3.5) node[time_label, anchor=west] {2};
        \draw[->, very thick] (-2,4.5) -- (14,4.5) node[time_label, anchor=west] {1};
         \draw[->, very thick] (-2,-1) -- (14,-1) node[time_label, anchor=west] {6};
         \draw[->, very thick] (-2,0) -- (14,0) node[time_label, anchor=west] {5};
         \draw[->, very thick] (-2,0.8) -- (14,0.8) node[time_label, anchor=west] {4};
       \node[time_label] (t0) at (7,-2.75) {Continuous Time}; 
       \node[time_label] (t0) at (7,6) {Stream of Interactions};

        \end{tikzpicture}
    \end{minipage}
    
    &
    \hspace{130pt}
    \begin{minipage}{0.3\textwidth}
    \begin{tikzpicture}[scale=0.5, every node/.style={node_style}]
      \draw[time_frame] (-1,-2) -- (3,0) -- (3,6) -- (-1,4) -- cycle;
      \draw[time_frame] (3,6) -- (6.5,6);
      \draw[time_frame] (-1,4) -- (2.5, 4);
      \draw[time_frame] (-1,-2) -- (2.5, -2);
      \node (n1t0) at (1, 2) {};
      \node (n2t0) at (0,3.5) {};
      \node (n6t0) at (-0.25,-1)  {};
    
      \draw[new_edge, very thick] (n1t0) -- (n2t0);
      \draw[new_edge, very thick] (n1t0) -- (n6t0);
    
       \node[time_label] (t0) at (-0.5,-2.75) {start}; 

       \draw[time_frame] (2.5,-2) -- (6.5,0) -- (6.5,6) -- (2.5,4) -- cycle;
      \node (n1t1) at (4.5, 2) {};
      \node (n3t1) at (5.25,4.5) {};
      \node (n4t1) at (5.5,0) {};
      \node (n5t1) at (3.5,0.8)  {};
    
      \draw[new_edge, very thick] (n1t1) -- (n3t1);
      \draw[new_edge, very thick] (n1t1) -- (n5t1);
      \draw[new_edge, very thick] (n4t1) -- (n1t1);
    
       \node[time_label] (t1) at (3,-2.75) {end}; 

      \draw[time_frame] (6,-2) -- (10,0) -- (10,6) -- (6,4) -- cycle;

      \draw[time_frame] (10,6) -- (13.5,6);
      \draw[time_frame] (6,4) -- (9.5, 4);
      \draw[time_frame] (6,-2) -- (9.5, -2);
      
      \node (n1t2) at (8, 2) {};
      \node (n2t2) at (7,3.5) {};
      \node (n3t2) at (8.75,4.5) {};
      \node (n4t2) at (9,0) {};
    
      \draw[new_edge, very thick] (n1t2) -- (n2t2);
      \draw[new_edge, very thick] (n1t2) -- (n4t2);
      \draw[new_edge, very thick] (n4t2) -- (n3t2);

       \node[time_label] (t2) at (6.5,-2.75) {start}; 

        \draw[time_frame] (9.5,-2) -- (13.5,0) -- (13.5,6) -- (9.5,4) -- cycle;

          \node (n3t3) at (12.25,4.5) {};
          \node (n4t3) at (12.75,0) {};
          \node (n5t3) at (10.5,0.8)  {};
          \node (n6t3) at (10.25,-1)  {};
        
          \draw[new_edge, very thick] (n3t3) -- (n5t3);
          \draw[new_edge ,very thick] (n4t3) -- (n6t3);
        
           \node[time_label] (t3) at (10.25,-2.75) {end}; 

        \draw[->, thick, blue, dashed] (-2,2) -- (15,2) node[time_label, anchor=east] {};
        \draw[->, thick, blue, dashed] (-2,3.5) -- (15,3.5) node[time_label, anchor=east] {};
        \draw[->, thick, blue, dashed] (-2,4.5) -- (15,4.5) node[time_label, anchor=east] {};
         \draw[->, thick, blue, dashed] (-2,-1) -- (15,-1) node[time_label, anchor=east] {};
         \draw[->, thick, blue, dashed] (-2,0) -- (15,0) node[time_label, anchor=east] {};
         \draw[->, thick, blue, dashed] (-2,0.8) -- (15,0.8) node[time_label, anchor=east] {};

        \end{tikzpicture}
    \end{minipage}

  \end{tabular}
  \end{minipage}
\begin{minipage}{\textwidth}
\vspace{14pt}
\centering
\setlength{\tabcolsep}{0pt}
\begin{tabular}{cc}
 
\begin{minipage}{0.5\textwidth}
\centering
\begin{tikzpicture}[scale=0.75, every node/.style={node_style}]
  \draw[blue, thick, dashed] (-1.5, -3) rectangle (11.5, 5.5);
  \node (n1) at (0.5, 2) {};
  \node (n2) at (-0.5, 3.5) {};
  \node (n3) at (1.25, 4.5) {};
  \node (n4) at (1.5, 0) {};
  \node (n5) at (-0.5, 0.8) {};
  \node (n6) at (-0.75, -1) {};

  \draw[new_edge, very thick] (n1) -- (n2);
  \draw[new_edge, very thick] (n1) -- (n6);

  \draw[blue, thick] (-1, -2) rectangle (2, 5);
  \node[time_label] at (0.5, -2.5) {$\Delta t_1$};

  \node (n1) at (3.5, 2) {};
  \node (n2) at (2.5, 3.5) {};
  \node (n3) at (4.25, 4.5) {};
  \node (n4) at (4.5, 0) {};
  \node (n5) at (2.5, 0.8) {};
  \node (n6) at (2.25, -1) {};

  \draw[very thick] (n1) -- (n2);
  \draw[very thick] (n1) -- (n6);
  \draw[new_edge, very thick] (n1) -- (n4);
  \draw[new_edge, very thick] (n1) -- (n3);
  \draw[new_edge, very thick] (n1) -- (n5);

  \draw[blue, thick] (2, -2) rectangle (5, 5);
  \node[time_label] at (3.5, -2.5) {$\Delta t_2$};

  \node (n1) at (6.5, 2) {};
  \node (n2) at (5.5, 3.5) {};
  \node (n3) at (7.25, 4.5) {};
  \node (n4) at (7.5, 0) {};
  \node (n5) at (5.5, 0.8) {};
  \node (n6) at (5.25, -1) {};

  \draw[very thick] (n1) -- (n4);
  \draw[very thick] (n1) -- (n3);
  \draw[very thick] (n1) -- (n5);
  \draw[new_edge, very thick] (n1) -- (n2);
  \draw[new_edge, very thick] (n1) -- (n4);
  \draw[new_edge, very thick] (n4) -- (n3);

  \draw[blue, thick] (5, -2) rectangle (8, 5);
  \node[time_label] at (6.5, -2.5) {$\Delta t_3$};

  \node (n1) at (9.5, 2) {};
  \node (n2) at (8.5, 3.5) {};
  \node (n3) at (10.25, 4.5) {};
  \node (n4) at (10.5, 0) {};
  \node (n5) at (8.5, 0.8) {};
  \node (n6) at (8.25, -1) {};

  \draw[very thick] (n1) -- (n2);
  \draw[very thick] (n1) -- (n4);
  \draw[very thick] (n4) -- (n3);
  \draw[new_edge, very thick] (n3) -- (n5);
  \draw[new_edge, very thick] (n4) -- (n6);

  \draw[blue, thick] (8, -2) rectangle (11, 5);
  \node[time_label] at (9.5, -2.5) {$\Delta t_4$};

\end{tikzpicture}
\end{minipage}
  
  &
  \hspace{30pt}
\begin{minipage}{0.5\textwidth}
\centering
\begin{tikzpicture}[scale=0.75, every node/.style={node_style}]
  \draw[blue, thick, dashed] (-1.5, -3) rectangle (5.5, 5.5);
  \node (n1) at (0.5, 2) {};
  \node (n2) at (-0.5, 3.5) {};
  \node (n3) at (1.25, 4.5) {};
  \node (n4) at (1.5, 0) {};
  \node (n5) at (-0.5, 0.8) {};
  \node (n6) at (-0.75, -1) {};

  \draw[new_edge, very thick] (n1) -- (n2);
  \draw[new_edge, very thick] (n1) -- (n4);
  \draw[new_edge, very thick] (n1) -- (n6);
  \draw[new_edge, very thick] (n1) -- (n3);
  \draw[new_edge, very thick] (n1) -- (n5);

  \draw[blue, thick] (-1, -2) rectangle (2, 5);
  \node[time_label] at (0.5, -2.5) {$\Delta t_1$};

  \node (n1) at (3.5, 2) {};
  \node (n2) at (2.5, 3.5) {};
  \node (n3) at (4.25, 4.5) {};
  \node (n4) at (4.5, 0) {};
  \node (n5) at (2.5, 0.8) {};
  \node (n6) at (2.25, -1) {};

  \draw[new_edge, very thick] (n1) -- (n2);
  \draw[new_edge, very thick] (n1) -- (n4);
  \draw[new_edge, very thick] (n4) -- (n3);
  \draw[new_edge, very thick] (n3) -- (n5);
  \draw[new_edge, very thick] (n4) -- (n6);

  \draw[blue, thick] (2, -2) rectangle (5, 5);
  \node[time_label] at (3.5, -2.5) {$\Delta t_2$};
\end{tikzpicture}
\end{minipage}

\end{tabular}
\end{minipage}

\caption{Time Windowed Graphs}
\label{fig:temporal-windowing-demo}
\end{figure}

Figure \ref{fig:temporal-windowing-demo} illustrates the concept of window-based analysis applied to a stream of network interactions over continuous time. The set of four blue frames on the left demonstrates overlapping windows (\(\Delta t_1\) to \(\Delta t_4\)), where each window captures a segment of the network that partially overlaps with the next, providing a rolling analysis of network changes. The two blue frames on the right represent discrete windows (\(\Delta t_1\) and \(\Delta t_2\)), which showcase non-overlapping segments for a more segmented analysis. This visualization aids in comprehending the evolution of network connectivity as it changes over time within both types of windowed segments.

\subsubsection{Implications for Temporal Network Analysis}
The choice of method—time-aggregation or window-based analysis—depends on the specific goals of the analysis and the nature of the temporal data. While time-aggregation is suitable for understanding overall trends and patterns, window-based analysis provides insights into the evolution of network dynamics over time. \cite{jiang2023exploring}

\subsection{Key Concepts in Temporal Network Analysis}

Temporal network analysis examines the dynamics of time-ordered structures and this subsection introduces its fundamental concepts and their relevance.

\subsubsection{Temporal Distances and Time-Ordered Paths}
\paragraph{ \textit{Definition:} }
A temporal path in a network is a sequence of edges where each edge’s activation time is after the previous edge in the sequence. The temporal distance between two nodes is the minimum time required to travel from one node to the other following temporal paths. \cite{Delling2009}

\paragraph{ \textit{Mathematical Representation:} }
If \( P \) represents a path in a temporal network, and \( t(P) \) represents the time taken to traverse this path, the temporal distance \( D(u, v) \) between nodes \( u \) and \( v \) is given by:

\[
D(u, v) = \min \{ t(P) | P \text{ is a path from } u \text{ to } v \}
\]

\paragraph{ \textit{Significance:} }
Temporal paths and distances are fundamental for understanding the efficiency and speed of information or process flow in a network, which can be drastically different from their static counterparts. \cite{Scholtes_2016, Betsy2007}

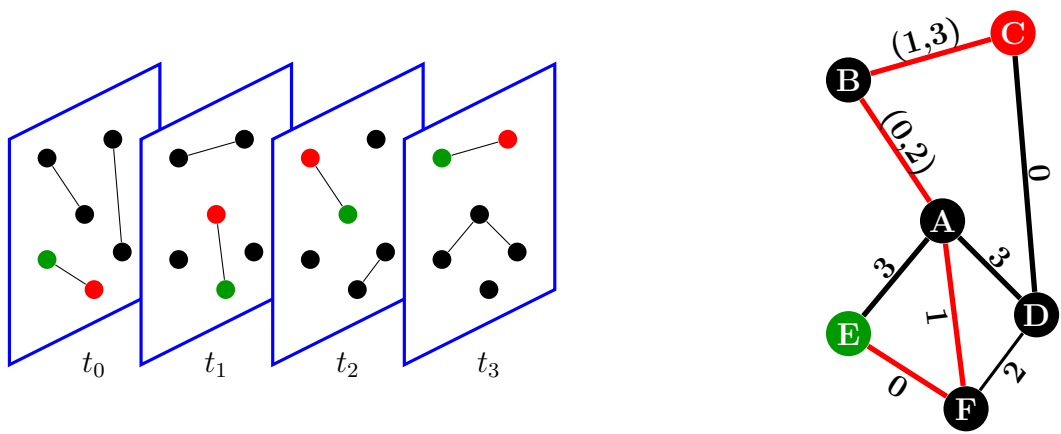
\begin{figure}[H]
    \centering
    \tikzset{ 
        node_style/.style={
            circle, fill=black, inner sep=0pt,       minimum size=0.25cm
        },
        new_node/.style={
            circle, fill=red, inner sep=0pt,           minimum size=0.25cm
        },
        new_edge/.style={
            draw=red, thick
        },
        delete_edge/.style={
            draw=red, dashed, thick
        },
        time_frame/.style={
            draw=blue, very thick, fill=white
        },
        time_label/.style={
            rectangle, fill=none, draw=none
        },
        edge_label/.style={
            font=\small, 
            font=\bfseries,
            fill=none,
            draw=none,
            rectangle,
            inner sep=0pt
        }
    }
    
\begin{minipage}{\textwidth}
    \setlength{\tabcolsep}{-5pt} 
    \begin{tabular}{cc}
    \begin{minipage}{0.48\textwidth}
    \begin{tikzpicture}[scale=0.5, every node/.style={node_style}]
      \draw[time_frame] (-1,-2) -- (3,0) -- (3,6) -- (-1,4) -- cycle;
      \node (n1t0) at (1, 2) {};
      \node (n2t0) at (0,3.5) {};
      \node (n3t0) at (1.75,4) {};
      \node (n4t0) at (2,1) {};
      \node[green!60!black] (n5t0) at (0,0.8)  {};
      \node[new_node] (n6t0) at (1.25,0)  {};
    
      \draw (n1t0) -- (n2t0);
      \draw (n3t0) -- (n4t0);
      \draw (n5t0) -- (n6t0);
    
       \node[time_label] (t0) at (1.25,-2) {$t_0$}; 

       \draw[time_frame] (2.5,-2) -- (6.5,0) -- (6.5,6) -- (2.5,4) -- cycle;
      \node[new_node] (n1t1) at (4.5, 2) {};
      \node (n2t1) at (3.5,3.5) {};
      \node (n3t1) at (5.25,4) {};
      \node (n4t1) at (5.5,1) {};
      \node (n5t1) at (3.5,0.8)  {};
      \node[green!60!black] (n6t1) at (4.75,0)  {};
    
      \draw (n2t1) -- (n3t1);
      \draw (n1t1) -- (n6t1);
    
       \node[time_label] (t1) at (4.5,-2) {$t_1$}; 

      \draw[time_frame] (6,-2) -- (10,0) -- (10,6) -- (6,4) -- cycle;
      
      \node[green!60!black] (n1t2) at (8, 2) {};
      \node[new_node] (n2t2) at (7,3.5) {};
      \node (n3t2) at (8.75,4) {};
      \node (n4t2) at (9,1) {};
      \node (n5t2) at (7,0.8)  {};
      \node (n6t2) at (8.25,0)  {};
    
      \draw (n1t2) -- (n2t2);
      \draw (n4t2) -- (n6t2);

       \node[time_label] (t2) at (8,-2) {$t_2$}; 

        \draw[time_frame] (9.5,-2) -- (13.5,0) -- (13.5,6) -- (9.5,4) -- cycle;
        
          \node (n1t3) at (11.5, 2) {};
          \node[green!60!black]  (n2t3) at (10.5,3.5) {};
          \node[new_node] (n3t3) at (12.25,4) {};
          \node (n4t3) at (12.5,1) {};
          \node (n5t3) at (10.5,0.8)  {};
          \node (n6t3) at (11.75,0)  {};
        
          \draw (n2t3) -- (n3t3);
          \draw (n1t3) -- (n5t3);
          \draw (n1t3) -- (n4t3);
        
           \node[time_label] (t3) at (11.75,-2) {$t_3$};

    \end{tikzpicture}
    \end{minipage}
    
    &
    \begin{minipage}{0.60\textwidth} 
    \centering
  \begin{tikzpicture}[scale=1.25, every node/.style={node_style, font=\bfseries, minimum size=0.6cm}] 
      \node[text=white] (n1) at (1, 2) {A};
      \node[text=white] (n2) at (0,3.5) {B};
      \node[new_node, minimum size=0.6cm, text=white] (n3) at (1.75,4) {C};
      \node[text=white] (n4) at (2,1) {D};
      \node[green!60!black, text=white] (n5) at (0,0.8) {E};
      \node[text=white] (n6) at (1.25,0) {F};
      
      \draw[new_edge, line width=2pt] (n1) -- (n2) node[edge_label, above, midway, sloped, yshift=-4pt] {(0,2)};
      \draw[line width=2pt] (n3) -- (n4) node[edge_label, midway, above, sloped, yshift=-4pt] {0};
      \draw[new_edge, line width=2pt] (n5) -- (n6) node[edge_label, midway, below, sloped, yshift=3pt] {0};
      \draw[new_edge, line width=2pt] (n2) -- (n3) node[edge_label, midway, above, sloped, yshift=-3pt] {(1,3)};
      \draw[new_edge, line width=2pt] (n1) -- (n6) node[edge_label, midway, below, sloped, yshift=3pt] {1};
      \draw[very thick] (n4) -- (n6) node[edge_label, midway, below, sloped, yshift=3pt] {2};
      \draw[line width=2pt] (n1) -- (n5) node[edge_label, midway, above, sloped, yshift=-4pt] {3};
      \draw[line width=2pt] (n1) -- (n4) node[edge_label, midway, above, sloped, yshift=-4pt ] {3};
    \end{tikzpicture}
    \label{fig:time-aggregated-path}
    \end{minipage}

  \end{tabular}
\end{minipage}
\caption{Time-varying Path \& Time-Aggregated Path}
\label{fig:temporal-network-paths}
\end{figure}

Figure \ref{fig:temporal-network-paths} illustrates the concept of temporal paths within a network and presents an aggregated view of the network over time. On the left, a sequence of frames represents the network at different time steps ($t_0$ to $t_3$), with each frame showing the active edges for that time slice. The green node signifies the starting node for each timeframe, while the red node marks the end of the active timeframe. On the right, the Time-Aggregated Path compiles the temporal activity into a single static graph. This graph depicts the potential paths from the start node (E) to the end node (C) across the entire observation period. The red edges on the right-hand side graph represent the sequence of edges taken from node E to node C, with the edge labels indicating the time frames when each edge is available. This visual representation emphasizes the temporal distance concept, which accounts for the sequence and timing of edge activations necessary to traverse from node E to node C.

\subsubsection{Temporal Components}
\paragraph{Dynamic Connected Components:}
Connected components in a temporal network can change over time as edges appear and disappear. Analyzing these components helps in understanding the evolving connectivity of the network. 

\paragraph{Mathematical Representation:}
A temporal component at time \( t \) can be represented as a subgraph \( C(t) \subseteq G(t) \), where each node in \( C(t) \) can reach every other node following temporal paths within \( C(t) \). \cite{casteigts2012timevarying}

\subsubsection{Centrality Measures}
\paragraph{Time-Independent vs. Time-Dependent Centrality:}
In temporal networks, centrality measures can be time-independent (measuring centrality over the entire network) or time-dependent (varying with time). \cite{Holme_2012}

\paragraph{Example of Time-Dependent Centrality:}
A simple time-dependent centrality measure could be the count of a node’s active edges at each time step: \cite{Pan_2011}

\[
C_t(v) = |\{e \in E(t) | v \in e\}|
\]

\subsubsection{Temporal Motifs and Correlation}
\paragraph{Temporal Motifs:}
Motifs in temporal networks are recurring patterns of interactions over time. They are critical for understanding repeated structures and behaviors in the network. \cite{Kovanen_2011}

\paragraph{Temporal Correlation Analysis:}
This involves studying the dependencies and correlations between events in the network, crucial for understanding the causality and influence patterns. \cite{Kovanen_2011}

\subsubsection{Community Detection and Dynamics}
\paragraph{Evolving Communities:}
Communities in temporal networks are groups of nodes that interact more frequently with each other than with the rest of the network, and these communities can evolve over time. \cite{aynaud2010}

\paragraph{Detection Methods:}
Methods for detecting communities in temporal networks often involve adapting static community detection algorithms to consider the temporal dimension, allowing for the tracking of community evolution. \cite{aynaud2010}

\subsection{Detecting Edge Patterns in Temporal Networks}

Understanding the patterns and dynamics of edges in temporal networks is crucial for analyzing how interactions evolve over time. This subsection focuses on the statistical analysis of event times and the methodologies for link prediction.

\subsubsection{Statistical Analysis of Event Times}
\paragraph{Inter-Event Times:}
In temporal networks, the time intervals between consecutive events (inter-event times) reveal critical information about the network's dynamics. 

\paragraph{Mathematical Analysis:}
The inter-event time \( \Delta t_i \) for the \( i \)-th event is mathematically defined as:
\[
\Delta t_i = t_{i+1} - t_i
\]
where \( t_i \) denotes the time of the \( i \)-th event. Analyzing the set \( \{\Delta t_i\} \) helps in identifying patterns such as periodicity, burstiness, or irregularities in interactions, providing insights into underlying processes. \cite{Holme_2012}

\subsubsection{Link Prediction}
\paragraph{Predicting Future Connections:}
Link prediction in temporal networks involves forecasting which connections (edges) are likely to occur in the future based on historical data.

\paragraph{Mathematical Approaches:}
Various algorithms and models are employed for link prediction. One common approach is to calculate a probability score for the formation of an edge between two nodes, based on factors like previous interactions, node similarity, and network topology. For instance, the probability of an edge forming between nodes \( u \) and \( v \) at time \( t \) can be represented as:
\[
P_{uv}(t) = f(\text{historical data}, \text{node features}, \text{network structure})
\]
where \( P_{uv}(t) \) is the probability of an edge forming between \( u \) and \( v \) at time \( t \), and \( f \) is a function that incorporates various predictive factors.

By analyzing edge patterns through statistical methods and predictive modeling, researchers can gain a deeper understanding of the temporal dynamics in networks. This insight is vital for applications ranging from social network analysis to predictive maintenance in engineering systems. \cite{Lu_2011}

\subsection{Detecting Node Patterns in Temporal Networks}

The analysis of node patterns in temporal networks is key to understanding individual behaviors and roles within the network's evolving structure. This subsection focuses on network embedding techniques and anomaly analysis to detect and interpret unusual node behaviors and patterns.

\subsubsection{Network Embedding}
\paragraph{Purpose and Techniques:}
Network embedding in temporal networks aims to map nodes into a low-dimensional space while preserving temporal dynamics and relationships. This technique facilitates tasks like visualization, clustering, and machine learning on network data. \cite{hamilton2018representation}

\paragraph{Mathematical Representation:}
An embedding can be mathematically represented by a function \( \phi \) that maps each node \( v \) to a vector in a \( d \)-dimensional space:
\[
\phi(v) \in \mathbb{R}^d
\]
The goal is to ensure that the embedding \( \phi(v) \) captures the temporal dynamics and relationships of node \( v \) in the network. \cite{Goyal_2020}

\subsubsection{Anomaly Analysis}
\paragraph{Detecting Unusual Node Behaviors:}
Anomaly analysis in temporal networks involves identifying nodes that exhibit behaviors deviating significantly from the norm. This can indicate potential errors, fraud, or significant events. \cite{Ranshous2015}

\paragraph{Mathematical Techniques:}
Statistical tests or machine learning algorithms are used to detect anomalies. For example, a node's behavior can be modeled as a time series, and anomalies can be detected based on deviations from predicted patterns:
\[
\text{Anomaly}_v(t) = \text{Detect}(\{v(t') | t' < t\})
\]
where \( \text{Anomaly}_v(t) \) indicates whether node \( v \) exhibits anomalous behavior at time \( t \), based on its past behaviors. \cite{Ranshous2015}

\subsection{Detecting Structural Patterns in Temporal Networks}

Analyzing structural patterns in temporal networks involves understanding how the overall network structure changes over time. This subsection focuses on temporal correlation and hypothesis testing, as well as change point detection, to identify significant structural changes in the network. \cite{Holme_2012}

\subsubsection{Temporal Correlation and Hypothesis Testing}
\paragraph{Analyzing Correlation Patterns:}
Temporal correlation analysis involves examining the relationships between different events or interactions over time, helping to identify underlying patterns and dependencies in the network structure. \cite{Nicosia_2013}

\paragraph{Null Models for Hypothesis Testing:}
Null models are used to test hypotheses about network structures by comparing observed patterns to those expected under random conditions.

\paragraph{Mathematical Representation of Null Models:}
A null model in a temporal network can be represented as a randomized version of the network, denoted as \( G_{\text{null}} \). This model is used to assess the significance of observed structural patterns:
\[
G_{\text{null}} = \text{Randomize}(G_{\text{observed}})
\]
where \( G_{\text{observed}} \) is the observed temporal network, and \( G_{\text{null}} \) is the randomized version for comparison. 

\subsubsection{Change Point Detection}
\paragraph{Identifying Structural Changes:}
Change point detection in temporal networks is the process of identifying times at which the network structure undergoes significant changes. This is important for understanding transitions in network dynamics and for detecting events that lead to these changes. \cite{sulem2022graph}

\paragraph{Mathematical Techniques for Change Point Detection:}
Change point detection often involves statistical methods that identify times when the network's structural properties, such as connectivity or centrality measures, change significantly. This can be represented as:
\[
\text{ChangePoint}(t) = \text{Detect}(\{G(t') | t' \leq t\})
\]
where \( \text{ChangePoint}(t) \) indicates a significant structural change at time \( t \) based on the history of the network up to that point. \cite{sulem2022graph}

\section{Spatiotemporal Networks}

\subsection{Introduction to Spatiotemporal Networks}

Spatiotemporal networks represent a sophisticated integration of spatial and temporal dimensions in network theory. They are characterized by their ability to model complex systems where both spatial positioning and temporal evolution play critical roles.

\begin{figure}[H]
\centering
\tikzset{
  node_style/.style={circle, fill=black, inner sep=0pt, minimum size=0.25cm},
  new_node/.style={circle, fill=red, inner sep=0pt, minimum size=0.25cm},
  new_edge/.style={draw=red, thick},
  delete_edge/.style={draw=red, dashed, thick},
  time_frame/.style={draw=blue, very thick},
  time_label/.style={rectangle, fill=none, draw=none, font=\Large} 
}

\setlength{\tabcolsep}{-5pt} 
\begin{tabular}{ccccc}

\begin{minipage}{0.2\textwidth}
\begin{tikzpicture}[scale=0.5, every node/.style={node_style}]

    \draw[time_frame] (-1,-2) -- (3,0) -- (3,6) -- (-1,4) -- cycle;

  \def\dx{0.5} 
  \def\dy{0.5} 

  \foreach \y in {-2,-1.5,...,4} {
    \draw[gray] (-1,\y) -- ++(4,2);
  }

  \foreach \x in {-1,-0.5,...,3} {
    \pgfmathsetmacro{\ystart}{-1.5 + (\x *0.5 )}
    \pgfmathsetmacro{\yend}{4.5 + (\x *0.5)}
    \draw[gray] (\x,\ystart) -- (\x,\yend);
  }

  \node (n1) at (1, 2) {};
  \node (n2) at (0,3.5) {};
  \node (n3) at (1.75,4) {};
  \node (n4) at (2,1) {};
  \node (n5) at (0,0.8)  {};
  \node (n6) at (1.25,0)  {};

  \draw[very thick] (n1) -- (n2);
  \draw[very thick] (n1) -- (n3);
  \draw[very thick] (n1) -- (n4);
  \draw[very thick] (n1) -- (n5);
  \draw[very thick] (n1) -- (n6);

   \node[time_label] (t0) at (2,-2) {$\Delta t_0$}; 
   
\end{tikzpicture}
\end{minipage}

\begin{minipage}{0.2\textwidth}
\begin{tikzpicture}[scale=0.5, every node/.style={node_style}]

   \draw[time_frame] (-1,-2) -- (3,0) -- (3,6) -- (-1,4) -- cycle;

  \def\dx{0.5} 
  \def\dy{0.5} 

  \foreach \y in {-2,-1.5,...,4} {
    \draw[gray] (-1,\y) -- ++(4,2);
  }

  \foreach \x in {-1,-0.5,...,3} {
    \pgfmathsetmacro{\ystart}{-1.5 + (\x *0.5 )}
    \pgfmathsetmacro{\yend}{4.5 + (\x *0.5)}
    \draw[gray] (\x,\ystart) -- (\x,\yend);
  }

  \node (n1) at (1, 2) {};
  \node (n2) at (0,3.5) {};
  \node (n3) at (1.75,4) {};
  \node (n4) at (2,1) {};
  \node (n5) at (0,0.8)  {};
  \node (n6) at (1.25,0)  {};

  \draw[very thick] (n1) -- (n2);
  \draw[very thick] (n1) -- (n3);
  \draw[very thick] (n1) -- (n4);
  \draw[very thick] (n1) -- (n5);
  \draw[very thick] (n1) -- (n6);
  \draw[new_edge] (n4) -- (n3);

   \node[time_label] (t0) at (2,-2) {$\Delta t_1$}; 
\end{tikzpicture}
\end{minipage}

\begin{minipage}{0.2\textwidth}
\begin{tikzpicture}[scale=0.5, every node/.style={node_style}]

  \draw[time_frame] (-1,-2) -- (3,0) -- (3,6) -- (-1,4) -- cycle;

  \def\dx{0.5} 
  \def\dy{0.5} 

  \foreach \y in {-2,-1.5,...,4} {
    \draw[gray] (-1,\y) -- ++(4,2);
  }

  \foreach \x in {-1,-0.5,...,3} {
    \pgfmathsetmacro{\ystart}{-1.5 + (\x *0.5 )}
    \pgfmathsetmacro{\yend}{4.5 + (\x *0.5)}
    \draw[gray] (\x,\ystart) -- (\x,\yend);
  }

  \node (n1) at (1, 2) {};
  \node (n2) at (0,3.5) {};
  \node (n3) at (1.75,4) {};
  \node (n4) at (2,1) {};
  \node (n5) at (0,0.8)  {};
  \node (n6) at (1.25,0)  {};

  \draw[very thick] (n1) -- (n2);
  \draw[delete_edge, very thick] (n1) -- (n3);
  \draw[very thick] (n1) -- (n4);
  \draw[very thick] (n1) -- (n5);
  \draw[very thick] (n1) -- (n6);
   \draw[very thick] (n4) -- (n3);

   \node[time_label] (t0) at (2,-2) {$\Delta t_2$}; 
\end{tikzpicture}
\end{minipage}

\begin{minipage}{0.2\textwidth}
\begin{tikzpicture}[scale=0.5, every node/.style={node_style}]

   \draw[time_frame] (-1,-2) -- (3,0) -- (3,6) -- (-1,4) -- cycle;

  \def\dx{0.5} 
  \def\dy{0.5} 

  \foreach \y in {-2,-1.5,...,4} {
    \draw[gray] (-1,\y) -- ++(4,2);
  }

  \foreach \x in {-1,-0.5,...,3} {
    \pgfmathsetmacro{\ystart}{-1.5 + (\x *0.5 )}
    \pgfmathsetmacro{\yend}{4.5 + (\x *0.5)}
    \draw[gray] (\x,\ystart) -- (\x,\yend);
  }

  \node (n1) at (1, 2) {};
  \node (n2) at (0,3.5) {};
  \node (n3) at (1.75,4) {};
  \node (n4) at (2,1) {};
  \node (n5) at (0,0.8)  {};
  \node (n6) at (1.25,0)  {};
  \node[new_node] (n7) at (-0.3, 2) {};

  \draw[very thick] (n1) -- (n2);
  \draw[very thick] (n1) -- (n4);
  \draw[very thick] (n1) -- (n5);
  \draw[very thick] (n1) -- (n6);
  \draw[very thick] (n4) -- (n3);
  \draw[new_edge, very thick] (n1) -- (n7);

   \node[time_label] (t0) at (2,-2) {$\Delta t_3$}; 
\end{tikzpicture}
\end{minipage}

\end{tabular}

\caption{SpatioTemporal evolution of the network}
\label{fig:spatiotemporal-network}
\end{figure}
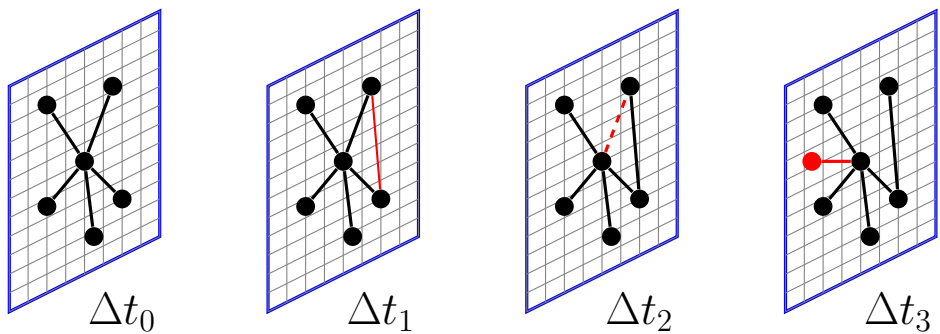

\subsubsection{Defining Spatiotemporal Networks}

A spatiotemporal network can be defined as a network where both the spatial and temporal aspects are essential to its structure and function. This type of network is denoted as:

\begin{equation}
G_{st} = (V, E, P, T)
\end{equation}

where:
\begin{itemize}
    \item \( V \) is the set of all nodes that may appear at any time in \( T \).
    \item \( E \) represents the set of all edges that appear between nodes in \( V \) at any time in \( T \).
    \item \( P: V \times T \rightarrow \mathbb{R}^n \) is a function defining the spatial position of nodes over time.
    \item \( T \) is the temporal domain over which the network is defined.
    \item \( V(t) \subseteq V \) is a function that returns the set of nodes active at time \( t \).
    \item \( E(t) \subseteq E \) is a function that returns the set of edges active between nodes in \( V(t) \) at time \( t \).
\end{itemize}

This definition encapsulates the dual nature of spatiotemporal networks, where both the spatial location of nodes and their temporal dynamics are fundamental. \cite{George2012SpatioTemporal, Batagelj2014}

\subsubsection{Modeling Systems with Spatial and Temporal Dimensions}

Spatiotemporal networks are particularly adept at modeling systems where changes in spatial relationships occur over time. The significance of these networks lies in their ability to capture dynamic interactions and transformations within the network, offering a more comprehensive understanding of complex systems. 

\subsubsection{Complexity and Richness of Spatiotemporal Networks}

The integration of spatial and temporal dynamics in spatiotemporal networks introduces a level of complexity and richness not present in networks considering only one of these dimensions. This complexity is manifest in the network's evolving topology, the changing spatial relationships among nodes, and the temporal variation in network properties.

\subsection{Mathematical Representation of Spatiotemporal Networks}

While the foundational model of spatiotemporal networks has been introduced, a deeper exploration into the mathematical representation reveals the dynamic nature of these networks. This section focuses on the representation of nodes and edges, highlighting their temporal and spatial variability.

\subsubsection{Temporal Evolution of Node Attributes}

In spatiotemporal networks, nodes can have attributes that change over time. These changes can be represented as functions of time:

\begin{equation}
A_V: V \times T \rightarrow \mathbb{A}
\end{equation}

where \( A_V \) represents a function assigning attributes from a set \( \mathbb{A} \) to each node in \( V \) at different times in \( T \). This can include changes in node status, capacity, or other relevant characteristics. \cite{Holme_2012}

\subsubsection{Dynamic Edges in Spatiotemporal Context}

Edges in spatiotemporal networks may also have attributes that vary over time, reflecting the evolving nature of the relationships between nodes:

\begin{equation}
A_E: E \times T \rightarrow \mathbb{B}
\end{equation}

where \( A_E \) is a function that maps each edge in \( E \) to a set of attributes \( \mathbb{B} \) over the temporal domain \( T \). This allows the representation of temporal variations in edge properties, such as connectivity strength or transmission capacity.   \cite{Holme_2012}

\subsubsection{Modeling Spatiotemporal Dynamics}

The spatiotemporal dynamics of the network can be represented by combining the spatial positioning function with the temporal attribute functions:

\begin{equation}
G_{st} = (V, E, P, A_V, A_E, T)
\end{equation}

In this expanded model, \( A_V \) and \( A_E \) add layers of temporal dynamics to the spatial framework defined by \( P \), providing a more comprehensive understanding of the network's evolution over time. 

\subsubsection{Handling Complexity in Representation}

While this mathematical representation provides a detailed view of spatiotemporal networks, it also underscores the complexity involved in modeling such networks. Each element \( v \) in \( V \) and each edge \( e \) in \( E \) can have a multitude of states or properties that evolve, making the analysis and computation more challenging.

\subsection{Dynamics of Spatiotemporal Networks}

Spatiotemporal networks are characterized by their dynamic nature, where both spatial and temporal dimensions play a crucial role in shaping the network's evolution and behavior. This section explores the key aspects of these dynamics.

\subsubsection{Evolution of Network Topology Over Time}

In spatiotemporal networks, the topology is not static but evolves over time. This evolution can be represented by a time-dependent adjacency matrix:

\begin{equation}
A(t) = [a_{ij}(t)] \quad \text{where} \; a_{ij}(t) = 
\begin{cases} 
1, & \text{if there is an edge between } v_i \text{ and } v_j \text{ at time } t \\
0, & \text{otherwise}
\end{cases}
\end{equation}

The temporal dimension allows for the modeling of dynamic processes such as the formation and dissolution of edges and the emergence or disappearance of nodes. 

\subsubsection{Temporal Changes in Spatial Relationships}

The spatial relationships among nodes in spatiotemporal networks can change over time, affecting how nodes interact and connect. These changes can be quantified by a time-dependent spatial positioning function:

\begin{equation}
P(v_i, t) \quad \text{denoting the position of node } v_i \text{ at time } t
\end{equation}

This function allows for the analysis of how the movement or changing spatial proximity of nodes influences the network structure.

\subsubsection{Impact on Network Analysis}

The interplay of spatial and temporal dimensions significantly impacts network analysis:
\begin{itemize}
    \item \textbf{Time-Dependent Network Measures}: Traditional network metrics like centrality, clustering coefficient, and path lengths become time-dependent, requiring temporal data for accurate analysis.
    \item \textbf{Modeling Network Dynamics}: Understanding the dynamics of spatiotemporal networks is crucial for predicting future states and behaviors, particularly in systems where spatial and temporal factors are closely intertwined.
\end{itemize}

\subsection{Computational and Analytical Complexity}

The integration of both spatial and temporal dimensions in spatiotemporal networks introduces significant computational and analytical complexities. This section discusses the nature of these challenges and the methodologies employed to address them.

\subsubsection{Increased Computational Complexity}

The complexity of spatiotemporal networks arises from the dynamic nature of their components. Each node and edge can change over time, leading to a constantly evolving network structure. This dynamic nature can be represented as:

\begin{equation}
G_{st}(t) = (V(t), E(t), P(t), T)
\end{equation}

where \( V(t) \) and \( E(t) \) represent the sets of nodes and edges at time \( t \), and \( P(t) \) denotes the spatial positions of nodes at time \( t \). This time-dependent representation implies that traditional static network analysis techniques are insufficient, as they must be adapted to handle temporal changes.

\subsubsection{Challenges in Analyzing Spatiotemporal Networks}

The primary challenges in analyzing spatiotemporal networks include:
\begin{itemize}
    \item \textbf{Data Volume and Variety}: The volume of data in spatiotemporal networks is typically large, especially when considering fine-grained temporal changes.
    \item \textbf{Dynamic Topology}: The evolving nature of the network topology requires algorithms that can adapt to changes over time.
    \item \textbf{Spatial-Temporal Interdependencies}: The interplay between spatial and temporal factors complicates the analysis, as changes in one dimension can significantly impact the other.
\end{itemize}

\subsubsection{Strategies for Addressing Computational Challenges}

To address these challenges, several strategies and methodologies are employed:
\begin{itemize}
    \item \textbf{Time-Slicing Techniques}: Analyzing the network at discrete time intervals can simplify the dynamics, allowing for the application of static network analysis techniques at each time slice.
    \item \textbf{Parallel Computing}: Leveraging parallel computing architectures to handle the computational load of processing large spatiotemporal datasets.
    \item \textbf{Algorithmic Adaptations}: Modifying existing algorithms or developing new ones to account for the temporal dynamics and spatial dependencies in the network.
\end{itemize}

\subsection{Conclusion and Transition to ROBUST Networks}

This section concludes our exploration of spatiotemporal networks and sets the stage for introducing the concept of ROBUST networks. The key points of spatiotemporal networks and their relevance are summarized, highlighting the complexity and sophistication that the integration of spatial and temporal dimensions brings to network analysis.

\subsubsection{Summary of Spatiotemporal Networks}

Spatiotemporal networks, as a fusion of spatial and temporal dynamics, offer a comprehensive framework for understanding systems that evolve over time and space. Key points include:
\begin{itemize}
    \item \textbf{Hybrid Nature}: The integration of spatial layouts with temporal changes, encapsulating the dynamics of real-world systems.
    \item \textbf{Mathematical Complexity}: The intricate representation that combines spatial positioning with temporal attributes, reflecting the evolving nature of the networks.
    \item \textbf{Computational Challenges}: Addressing the increased complexity in analyzing these networks, due to the dual aspects of space and time.
\end{itemize}

\subsubsection{Importance in Understanding ROBUST Networks}

Understanding the principles of spatiotemporal networks is crucial for delving into the study of ROBUST networks:
\begin{itemize}
    \item \textbf{Foundation for Advanced Concepts}: The concepts and methodologies developed in the context of spatiotemporal networks lay the groundwork for understanding the more nuanced and specialized ROBUST networks.
    \item \textbf{Complexity and Dynamics}: The insights gained from spatiotemporal networks about handling complexity and dynamic changes are directly applicable to the analysis of ROBUST networks.
\end{itemize}

\subsubsection{Transition to ROBUST Networks}

As we transition to the discussion of ROBUST networks, it is evident that the thorough understanding of spatiotemporal networks provides a solid foundation for appreciating the intricacies involved in ROBUST networks. The subsequent sections will delve into the unique characteristics of ROBUST networks, building upon the theoretical framework established in the context of spatiotemporal networks.

\chapter{ROBUST Dynamics}
\label{chap:robust-dynamics}

In the ROBUST Network, entities are grouped into two distinct sets, embodying the classic observer-observable pattern.  However, upon deeper inspection, these asymmetrical relationships between node sets may vary significantly,leading to the identification of several subcategories. Each subcategory is defined by its specific bipartite dynamics, which impacts the node interactions, link generation, and node placement. This section aims to dissect these subcategories in depth, assessing their influence on the network’s design and operational strategies, thereby laying the foundation for a thorough exploration of differing observer-observable analytics and considerations.

\begin{figure}[H]
  \centering
  \begin{tikzpicture}[->,>=Stealth,auto,node distance=2cm,semithick, scale=0.9, every node/.style={scale=0.9}]
    \node[draw, thick, rectangle, minimum height=2.5cm, minimum width=2cm, label=above:Observables] (observableSet) at (0, 0) {};
    \node[align=left, anchor=north west] at ([xshift=5pt, yshift=-5pt]observableSet.north west) {event 1\\event 2\\$\vdots$\\event M};

    \node[draw, thick, rectangle, minimum height=2.5cm, minimum width=2cm, label=above:Observers, right=of observableSet] (observerSet) {};
    \node[align=left, anchor=north west] at ([xshift=5pt, yshift=-5pt]observerSet.north west) {sensor 1\\sensor 2\\$\vdots$\\sensor N};

    \draw[<->] (observableSet) -- node[above] {\Huge ?} (observerSet);
  \end{tikzpicture}
  \caption{Identifying the Various Bipartite Relationships}
  \label{fig:bipartite-relationships}
\end{figure}
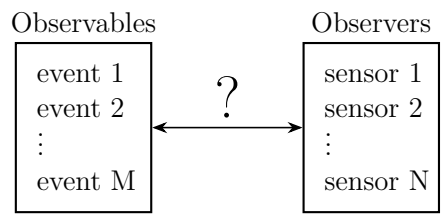

\section{Observer-Observable Dynamics}
The observer-observable relationship forms the foundational dynamic of the ROBUST Network. In this interaction, observer nodes are tasked with the crucial role of monitoring and collecting data from the observable events. Key elements of this relationship include:

\begin{itemize}
    \item \textbf{Spatial and Temporal Proximity:} The effectiveness of observation is heavily influenced by the spatial and temporal proximity of the observers to the events. Observers must be strategically placed and timed to capture relevant data accurately.

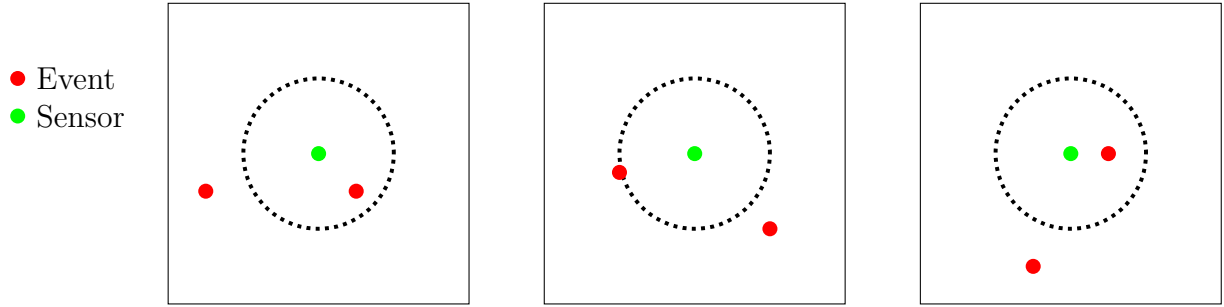
\begin{figure}[H]
  \centering
  \begin{tikzpicture}
    \node at (-2, 2.5) [circle,fill=green,inner sep=2pt,label=right:Sensor] {};
    \node at (-2, 3) [circle,fill=red,inner sep=2pt,label=right:Event] {};
    
    \begin{scope}[xshift=0cm]
      \draw (0,0) -- (0,4) -- (4,4) -- (4,0) -- cycle;
      \fill[green] (2,2) circle (0.1cm); 
      \draw[dotted, line width=1.5pt] (2,2) circle (1cm);
      \fill[red] (0.5,1.5) circle (0.1cm); 
      \fill[red] (2.5,1.5) circle (0.1cm); 
    \end{scope}
    
    \begin{scope}[xshift=5cm]
      \draw (0,0) -- (0,4) -- (4,4) -- (4,0) -- cycle;
      \fill[green] (2,2) circle (0.1cm); 
      \draw[dotted, line width=1.5pt] (2,2) circle (1cm);
      \fill[red] (1,1.75) circle (0.1cm); 
      \fill[red] (3,1) circle (0.1cm); 
    \end{scope}
    
    \begin{scope}[xshift=10cm]
      \draw (0,0) -- (0,4) -- (4,4) -- (4,0) -- cycle;
      \fill[green] (2,2) circle (0.1cm); 
      \draw[dotted, line width=1.5pt] (2,2) circle (1cm);
      \fill[red] (1.5,0.5) circle (0.1cm); 
      \fill[red] (2.5,2) circle (0.1cm); 
    \end{scope}
  \end{tikzpicture}
  \caption{Sensor and events in and out of range across three timesteps.}
\end{figure}

    \item \textbf{Adaptability and Continuous Observation:} Observers are required to continuously monitor their environment. Additionally, they must be adaptable and capable of adjusting in response to changes in the observable landscape. This ongoing observation is to track dynamic events or changes over time. This ongoing observation is key to identifying long-term patterns and trends and the adaptability for maximizing the relevance of the data collected.

\begin{figure}[H]
  \centering
  \begin{tikzpicture}
    \node at (-2, 2.5) [circle,fill=green,inner sep=2pt,label=right:Sensor] {};
    \node at (-2, 3) [circle,fill=red,inner sep=2pt,label=right:Event] {};

    \begin{scope}[xshift=0cm]
      \draw (0,0) -- (0,4) -- (4,4) -- (4,0) -- cycle;
      \fill[green] (2,2) circle (0.1cm); 
      \draw[dotted, line width=1.5pt] (2,2) circle (1cm);
      \fill[red] (2.5,1.5) circle (0.1cm); 
    \end{scope}

    \begin{scope}[xshift=5cm]
      \draw (0,0) -- (0,4) -- (4,4) -- (4,0) -- cycle;
      \fill[green] (2,2) circle (0.1cm); 
      \draw[dotted, line width=1.5pt] (2,2) circle (1cm);
      \fill[red] (0.5,1.5) circle (0.1cm); 
      \fill[red] (2.5,1.5) circle (0.1cm); 
      \fill[red] (1.5,0.5) circle (0.1cm); 
    \end{scope}
    
    \begin{scope}[xshift=10cm]
       \draw (0,0) -- (0,4) -- (4,4) -- (4,0) -- cycle;
      \fill[green] (1.5,1.25) circle (0.1cm); 
      \draw[dotted, line width=1.5pt] (1.5,1.25) circle (1cm);
      \fill[red] (0.5,1.5) circle (0.1cm); 
      \fill[red] (2.5,1.5) circle (0.1cm); 
      \fill[red] (1.5,0.5) circle (0.1cm); 
    \end{scope}
    
  \end{tikzpicture}
  \caption{Sensor adapts position for events across three frames.}
\end{figure}
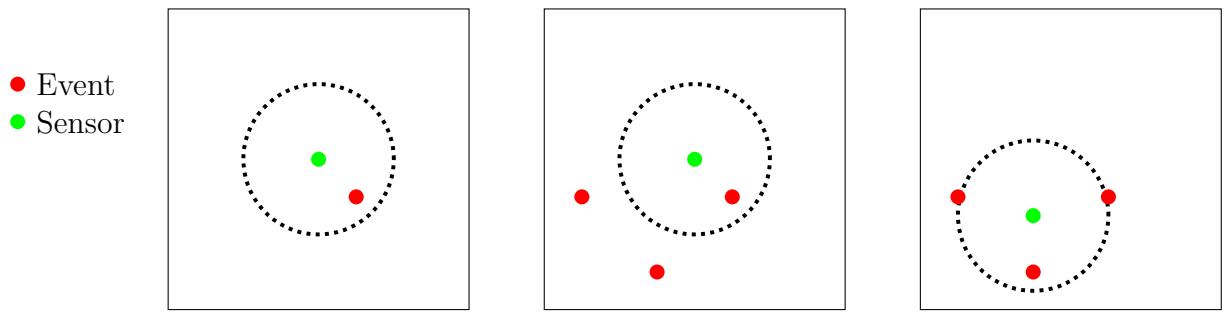

    \item \textbf{Data Processing and Analysis:} Beyond mere observation, the observer nodes are also responsible for the initial processing and analysis of the collected data. This step is vital for transforming raw observational data into actionable insights.

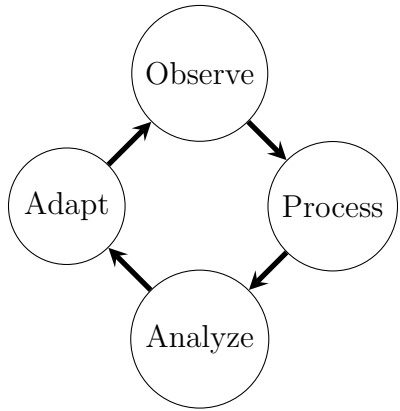
\begin{figure}[H]
     \centering
    \begin{tikzpicture}[auto,node distance=2.5cm]
      \tikzstyle{every state}=[fill=white,draw=black,text=black]
      \tikzset{every edge/.append style={->, >=stealth, line width=2pt}} 
    
      \node[state]         (A)                    {Observe};
      \node[state]         (B) [below right of=A] {Process};
      \node[state]         (C) [below left of=B] {Analyze};
      \node[state]         (D) [above left of=C]  {Adapt};
    
      \path (A) edge[->]             node {} (B)
            (B) edge[->]             node {} (C)
            (C) edge[->]              node {} (D)
            (D) edge[->]              node {} (A);
    \end{tikzpicture}
  \caption{Finite State Machine for the Responsive Observer Cycle.}
\end{figure}

\end{itemize}

The observer-observable relationship  drives the network's functionality and forms the basis of its various applications.  These are common properties that are shared among all subcategories.

\section{Identifying Subcategories by Behavior}

The ROBUST network framework uniquely addresses the complexities of multiagent systems by focusing on the intricate dynamics between observers and observables. This chapter explores the nuances of agent interactions—ranging from the selfish to collaborative behaviors within unipartite dynamics, and in bipartite dynamics,   the interplay of  between different nodes ranging from the competitive to the cooperative.  By examining the dynamics between nodes, we can apply ROBUST to a variety of domains, from biological to physical to virtual.

\section{Unipartite Relationships}
Unipartite relationships within the ROBUST network framework delve into the interactions among similar nodes—either observers or observables—within the same partition. These relationships are pivotal in shaping the network's dynamics, influencing its stability, efficiency, and overall behavior. By understanding the inherent motivations and behaviors of nodes, architects can design and manage networks more effectively, preventing destructive patterns and fostering constructive interactions.

\subsection{Selfish Dynamics}
In scenarios where individual agents (nodes) act in their self-interest, the network faces a unique set of challenges. Selfish dynamics emerge within partitions where nodes of the same type prioritize their own utility above the collective well-being of the network. This behavior is particularly prevalent in competitive settings or when resources are limited, and it directly impacts how nodes within the same group interact.

\subsubsection{Example: Clients in a Restaurant}
Clients are inherently selfish, prioritizing their own dining experience. They expect prompt service, desirable tables, and may exhibit impatience or demanding behavior.  This focus on individual needs can lead them to overlook the impact their actions (like lingering after a meal) have on wait times for others and the overall strain on waiters.

\subsubsection{Selfish dynamics often manifest as:}

\begin{itemize}
    \item \textbf{Resource Hoarding:} Nodes may accumulate or over-consume shared resources, depriving others within their partition.
    \item \textbf{Strategic Non-cooperation:} Nodes might withhold information, support, or participation, hindering collective efforts and overall network performance.
\end{itemize}

\subsubsection{Consequences} 
Unchecked selfish dynamics can destabilize the network, leading to inefficiencies, conflicts, and a decline in the system's overall health.

\subsubsection{ROBUST Design  }
Understanding these dynamics is crucial for network architects. By recognizing the potential for selfish behavior within node partitions,  systems can be designed with mechanisms to mitigate negative impacts and promote healthier long-term outcomes

\subsection{Collaborative Dynamics}
Collaborative dynamics emerge within partitions where agents (nodes) of the same type work together towards shared goals, often leading to greater success than if they acted individually. This behavior fosters a supportive network environment, enhancing efficiency, resilience, and the system's overall well-being.  Collaborative dynamics are particularly common when success hinges on collective efforts or where resources are best utilized through sharing.

\subsubsection{Example: Waiters in a Restaurant}
Waiters often exhibit collaborative dynamics to ensure a smooth dining experience for all customers. They may:

\begin{itemize}
    \item \textbf{Share Workload:} Communicate effectively, cover for each other during busy periods, and balance table assignments fairly.
    \item \textbf{Pool Resources:} Share information about table availability, customer preferences, or potential issues that could affect service.
\end{itemize}

\paragraph{} Collaborative dynamics often manifest as:

\begin{itemize}
    \item \textbf{Resource Sharing:} Nodes pool resources, knowledge, or skills, benefiting the entire group and expanding what's individually achievable.
    \item \textbf{Joint Problem-Solving:} Nodes work together to overcome complex challenges, often leading to more innovative and effective solutions than any individual could achieve.
\end{itemize}

\subsubsection{Benefits}  
Collaborative dynamics promote efficiency, fairness, and positive experiences within the network. In the restaurant context, this translates to happy customers, smoother operations, and greater overall success for the establishment.

\subsubsection{ROBUST Design}  
Network architects can encourage collaborative dynamics. This is particularly important when selfish tendencies may be present, emphasizing the long-term benefits outweigh potential short-term sacrifices. Strategies include:

\begin{itemize}
    \item \textbf{Communication Protocols:} Facilitate information sharing and coordination among nodes, reducing misunderstandings and building trust.
    \item \textbf{Incentive Structures:} Reward collective achievements or tie individual success to the well-being of the network, fostering an ``all for one" mentality.
\end{itemize}

\section{Bipartite Relationships}

In the ROBUST network framework, bipartite relationships form the core of interactions between distinct node partitions—observers and observables. These relationships are pivotal, shaping how the network functions, evolves, and adapts to challenges. Within this context, bipartite dynamics can be broadly classified into neutral, cooperative, or competitive, each influencing the network's structure and behavior in unique ways. This section provides an overview of cooperative and competitive dynamics, setting the stage for a deeper exploration in subsequent chapters.

\subsection{Cooperative Dynamics}

Cooperative dynamics in bipartite relationships occur when agents from different groups work together towards shared goals, leading to mutually beneficial outcomes. This form of interaction is characterized by a symbiotic relationship where the success of one agent directly contributes to the success of the other. Cooperation can manifest in various forms, including resource sharing, joint problem-solving, and collaborative innovation.

In cooperative dynamics, observers and observables engage in a positive-sum game where the combined effort results in greater achievements than what could be accomplished individually. For instance, in a client-server model, the server (observer) provides services or data that the client (observable) needs, enhancing the user experience and creating value for both parties.

\subsubsection{Example: Waiter - Client Relationship}
In a restaurant setting, while clients may exhibit selfish behavior towards other clients by prioritizing their own dining experience, they establish a cooperative relationship with waiters that significantly enhances the dining process for all parties involved. Clients demonstrate patience by waiting to be seated and served, adhere to the restaurant's operational protocols such as making reservations, and communicate their needs clearly to facilitate efficient service. Waiters are cooperative with clients by being attentive and responsive, tailoring their service to meet the specific needs and preferences of each client. This reciprocal understanding between clients and waiters leads to a mutual coordination where clients respect the waiters' efforts contributing to a more efficient and pleasant dining experience overall.

\subsection{Competitive Dynamics}

Competitive dynamics describe scenarios where the relationship between observers and observables is characterized by rivalry, opposition, or conflict. In these dynamics, the parties may compete for resources, influence, or dominance within the network. Unlike cooperative dynamics, competitive interactions often resemble a zero-sum game, where the gain of one party equates to the loss of another.

Competition can drive innovation, efficiency, and evolution within the network as agents strive to outperform their rivals. However, unchecked competition can also lead to conflicts, inefficiencies, and the potential for destructive behaviors. Recognizing and managing competitive dynamics is crucial for maintaining balance and ensuring the network's long-term viability and growth.

In appendix chapters, we will explore competitive dynamics in depth, examining how they manifest across different scenarios and strategies for mitigating negative impacts while harnessing competition's positive aspects for network improvement.

\subsubsection{Example: Predator-Prey Relationship}
In a natural ecosystem, predators (observers) and prey (observables) exhibit a fundamentally adversarial relationship driven by survival.  Prey animals prioritize their survival, employing tactics to evade. Predators, focused on obtaining food, actively seek and pursue prey, refining their hunting strategies to increase their chances of success.  This dynamic is inherently competitive, as the survival of one type of agent (predator) directly depends on the failure of the other (prey). 

\subsection{Conclusion}
Through understanding both cooperative and competitive bipartite dynamics, the ROBUST network framework is flexible to design and steer multiagent systems toward desired outcomes. See Appendix for more bipartite dynamics.

\chapter{Collaborative Bipartite Dynamics}
This chapter explores the collaborative dynamics exhibited by more specialized subcategories of the base observer-observable model and highlights how varied bipartite relationships manifest within the ROBUST network. This chapter details interactions between roles: Client-Server, Waiter-Client, Manager-Worker and Guard-Citizen relationships.

\section{Server-Client Dynamics}
This subsection explores the server-client model within the ROBUST Network, advancing from the basic observer-observable relationship.  The client-initiated request-response cycle reshapes network configuration and optimization, fundamentally altering network analysis methods. Finally, identifying server-client interactions influences the broader network properties.

\paragraph{Definition of a Server}
A server is an entity designed to respond to requests from clients. It operates as an observer by waiting to process and execute incoming requests, and subsequently deliver services or responses. Servers function within operational constraints, such as handling a maximum number of concurrent clients. Their effectiveness and efficiency in service delivery are significantly influenced by their strategic spatiotemporal placement within the network.

\paragraph{Definition of a Client}
A client acts as an initiator in the system, generating requests directed towards servers. Reflecting the characteristics of observables, clients articulate their service or information needs to trigger the servers' response mechanisms.

\subsection{Examples of Server-Client Relationships}

To illustrate the server-client dynamics in various contexts, this subsection presents real-world and digital examples.

\paragraph{Grocery Store Example}
In a grocery store, employees act as servers, and  shoppers act as clients. Shoppers initiate requests for assistance, information, or services, triggering employees to respond accordingly. This interaction exemplifies the server-client dynamic within a physical space. Importantly, certain sections of the store, like the deli, bakery, or checkout lines, often experience higher demand for service. Moreover, demand varies significantly throughout the day, with peak times such as evenings or weekends requiring more staff to manage the influx of shoppers efficiently. This uneven distribution of requests and temporal variation in shopper activity highlight the critical need for strategic server placement and the dynamic scheduling of resources (employees) to ensure service efficiency and shopper satisfaction at all times.

\paragraph{Object-Oriented Programming Example}
In object-oriented programming, the concept of servers and clients is mirrored through objects interacting within a software system. Here, `servers' are objects that offer functionalities or services, and `clients' are objects that request these services. This interaction creates a server-client dynamic that is fundamental to the system's architecture, similar to relationships seen in broader network models.

Spatial considerations emerge from how these objects are interdependent, influencing the ease with which a client object can utilize a server's services. The term "distance" metaphorically describes the number of steps or the complexity a client object must overcome to reach a server object, directly affecting the system's design for modularity, efficiency, and maintainability.
To minimize the need for clients to understand complex dependencies — effectively bringing servers ``closer" to clients — software engineers often utilize design principles such as the Factory pattern. This approach strategically positions a new 'server' within the architecture to simplify access to services, analogous to optimizing the layout of a network for better service reachability. This method underlines the importance of spatial and temporal planning even in abstract, non-physical environments, ensuring that systems remain adaptable and maintainable while meeting clients' needs efficiently.

\subsection{Node Interactions: Request-Response Cycle}
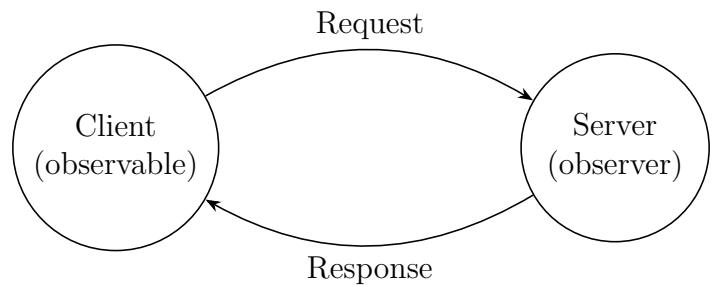
\begin{figure}[H]
     \centering
    \begin{tikzpicture}[>=Stealth, auto, semithick]
        \tikzstyle{state}=[fill=white,draw=black,text=black, circle, minimum size=2.5cm, align=center]
    
        \node[state] (client) {Client \\ (observable)};
        \node[state] (server) [right=4cm of client] {Server \\ (observer)}; 
    
        \draw[->] (client) edge[bend left] node {Request} (server);
        \draw[->] (server) edge[bend left] node {Response} (client);
        
        \node[fit=(client) (server),inner sep=0cm] {}; 
    
    \end{tikzpicture}
    \caption{Finite State Machine showing the request-response cycle.}
    \label{fig:server-client}
\end{figure}

The client-driven request-response cycle, as shown in Figure \ref{fig:server-client}, highlights the server-client node interaction, requiring network configurations that enable quick and dependable service exchanges. In such a relationship, servers should be positioned in locations to maximize their engagement with clients at the times of requests. This strategic positioning directly impacts the efficiency of service delivery and the network's overall performance.

The cycle is initiated with a request from the client (observables), with the server (observer) processing and satisfying the request, reflecting the observer's role in responding to observables. This server-client exchange introduces a bottleneck regarding the number of clients a server can handle at any given time. It underscores the importance of network optimization strategies that include load balancing, prioritization of requests, and dynamic resource allocation to mitigate the effects of this bottleneck. Efficient handling of this cycle, characterized by minimal response times and high-quality responses, directly contributes to enhanced service delivery, client satisfaction, and network efficiency.

\subsection{Link Generation and Constraints}
In the server-client model, link generation between servers and clients is influenced by two critical factors: the server's maximum degree (\textit{max\_degree}) and the temporal nature of links. The \textit{max\_degree} of a server node represents the maximum number of concurrent client connections it can handle, a constraint that directly impacts network topology and service delivery strategies.

\paragraph{Server Max Degree}
The concept of \textit{max\_degree} introduces a capacity limitation for servers, necessitating network optimization techniques that ensure an equitable and efficient distribution of service requests across the network. This constraint requires:

\begin{itemize}
    \item \textbf{Load Balancing:} Distributing client requests evenly across servers to prevent overloading and ensure timely responses.
    \item \textbf{Priority Queuing:} Implementing prioritization mechanisms for handling requests based on criteria such as urgency, client importance, or request complexity.
\end{itemize}

\paragraph{Temporal Nature of Links}
Links in the server-client model are inherently temporal, existing for the duration required to complete the request-response cycle. This transient characteristic of links emphasizes the network's dynamic adaptability, with connections being formed and dissolved based on real-time service demands. Implications include:

\begin{itemize}
    \item \textbf{Dynamic Network Topology:} The network's structure continuously evolves, reflecting the changing patterns of client-server interactions over time.
    \item \textbf{Resource Allocation Strategies:} Adaptive resource allocation becomes essential, with servers dynamically adjusting their available capacity to accommodate incoming requests efficiently.
\end{itemize}

\subsection{Implications for Network Optimization and Analysis}
The \textit{max\_degree} limitation and the temporal nature of links necessitate sophisticated network optimization and analysis approaches. Strategies must account for these dynamic and constrained aspects of the network, focusing on:

\begin{itemize}
    \item \textbf{Optimizing Server Utilization:} Ensuring servers operate within their capacity limits while minimizing response times and maximizing service quality.
    \item \textbf{Analyzing Temporal Patterns:} Understanding the temporal dynamics of client requests to forecast demand and adjust server configurations preemptively.
\end{itemize}

Incorporating considerations of server \textit{max\_degree} and the temporal characteristics of links enriches the network's design and operational framework. These factors underscore the complexity of managing server-client interactions within the ROBUST Network, highlighting the need for adaptive and responsive network strategies that can accommodate the dynamic nature of service requests and delivery.

\subsection{Network Optimization}
Optimization focuses on enhancing request-response efficiency, leveraging spatiotemporal considerations to ensure optimal server positioning and timing to meet client demands. Strategies include strategic server placement and adaptive service provision, aiming to minimize response times and adapt to client request patterns.

\subsection{Spatiotemporal Considerations}
The strategic placement and scheduling of servers are crucial for optimizing service delivery, guided by the following considerations:
\begin{itemize}
    \item \textbf{Spatial Domain:} Ensuring service interactions occur within minimized distances or latencies to enhance service immediacy and accessibility for clients.
    \item \textbf{Temporal Blocks:} Allocating service provisions into defined time blocks, allowing for the agile reallocation of server resources in response to fluctuating demand and availability.
\end{itemize}

\subsection{Efficiency Metrics}
Service delivery efficiency is quantitatively assessed through metrics designed to capture the network's performance:
\begin{equation}
    \text{Efficiency} = f\left(\text{Coverage Ratio}, \text{Average Response Time}, \text{Load Distribution}\right)
\end{equation}
Key metrics include:
\begin{itemize}
    \item \textbf{Coverage Ratio:} The fraction of clients that are within optimal reach of server services.
    \item \textbf{Average Response Time:} The mean duration from client request initiation to server response.
    \item \textbf{Load Distribution:} The equitable distribution of service demands across the available server resources.
\end{itemize}

\subsection{Server Placement Strategies}
In optimizing the server-client dynamic within the network, traditional considerations of client density are expanded to include the activity levels of clients and the potential influence of certain regions on generating service requests. This nuanced approach to server placement emphasizes not just physical proximity to a large number of clients but strategic proximity to highly active or influential client clusters.

\paragraph{Identifying High-Activity and Influential Regions}
The network's optimization strategies incorporate advanced analytics to:
\begin{itemize}
    \item Identify clients or regions with high service request rates, indicating areas where servers can be most impactful in terms of response efficiency and satisfaction.
    \item Analyze patterns of service requests to determine regions that, due to various factors, might influence an increase in service requests, allowing for proactive server placement to manage anticipated demand.
\end{itemize}

\section{Waiter-Client Dynamics}
This subsection introduces the waiter-client model. In this relationship, the waiter initiates interactions in a prompt-response cycle. Such a dynamic impacts the waiter nodes' ability to interact with clients, which must be accounted for when determining optimal spatiotemporal placements in a ROBUST network.

\paragraph{Definition of a Waiter}
A waiter represents a specialized observer-type who initiates interactions with a recipient to offer services, recommendations, or assistance. This role is defined by ongoing engagement, where the provider checks on the needs of the assigned recipient. The effectiveness of a waiter is measured by their expediency in providing for recipients and managing multiple interactions across multiple clients efficiently. Their ability to reduce lag or delay in service is dependent on spatiotemporal planning. 

\paragraph{Definition of a Client}
A client represents a specialized observable-type who waits for the waiter to prompt them before making any requests. Clients are decision-making agents based on options the waiter provides and have some terminating goal or task that leads to their ending the interaction.

\subsection{Examples of Waiter-Client Relationships}

\paragraph{Restaurant Example}
In a restaurant, waiters are assigned to specific zones or tables, mirroring the spatial partitioning essential for ROBUST modeling. Clients enter this environment seeking service. The relationship initiates with the waiter, once the client enters by prompting them where to sit and maintains this engagement throughout the dining experience. The continuous interaction of the waiter-client dynamic distinguishes it from the transactional server-client model, with the waiter's capacity to serve effectively constrained by the number of clients they can attentively monitor. This introduces a critical bottleneck, emphasizing the need for strategic waiter allocation and scheduling based on client flow and dining peak times. Additionally, for large groups, multiple waiters may need to be flexed onto a single table to ensure efficient service delivery without compromising quality. Optimizing this distribution involves both spatial (zoning) and temporal (scheduling) considerations to ensure each waiter maximizes their effectiveness without compromising service quality.

\subsection{Node Interactions: Prompt-Response Cycle}
The prompt-response cycle defines the waiter-client relationship, with waiters (observers) initiating interactions with clients (observables) through prompting. This cycle begins with the waiter's initial engagement and is sustained throughout the client's responses, influencing the sequence of service actions.

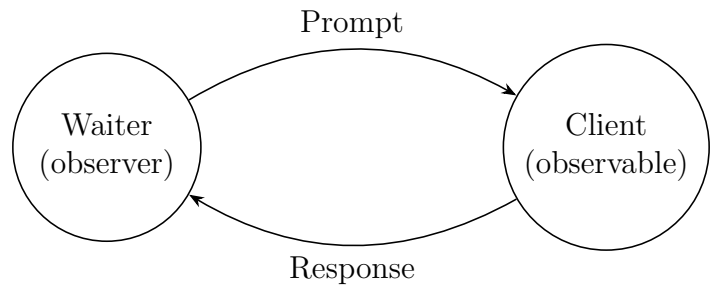
\begin{figure}[H]
     \centering
    \begin{tikzpicture}[>=Stealth, auto, semithick]
        \tikzstyle{state}=[fill=white,draw=black,text=black, circle, minimum size=2.5cm, align=center]
    
        \node[state] (waiter) {Waiter \\ (observer)};
        \node[state] (client) [right=4cm of waiter] {Client \\ (observable)}; 
    
        \draw[->] (waiter) edge[bend left] node {Prompt} (client);
        \draw[->] (client) edge[bend left] node {Response} (waiter);
        
        \node[fit=(waiter) (client),inner sep=0cm] {}; 
    
    \end{tikzpicture}
    \caption{Finite State Machine showing the prompt-response cycle.}
    \label{fig:waiter-client}
\end{figure}

This interaction model introduces a capacity constraint on the number of clients a waiter can effectively manage, creating a bottleneck that impacts service delivery efficiency and client satisfaction. Addressing this bottleneck requires strategic placement and scheduling of waiters to balance service demand and ensure high-quality, personalized service. This optimization emphasizes the need for careful consideration of spatial (waiter zoning) and temporal (scheduling) elements, aligning with ROBUST modeling principles to enhance both service quality and operational efficiency.

\subsection{Link Generation and Constraints}
In the waiter-client model, link generation between waiters and clients is influenced by two critical factors: the waiter's service capacity (\textit{service\_capacity}) and the temporal nature of service interactions. The \textit{service\_capacity} of a waiter represents the maximum number of clients they can effectively serve at once, a constraint that directly impacts service delivery and client satisfaction strategies.

\paragraph{Waiter Service Capacity}
The concept of \textit{service\_capacity} introduces a capacity limitation for waiters, necessitating optimization techniques to ensure a balanced and efficient distribution of client interactions across the service area. This constraint necessitates:

\begin{itemize}
    \item \textbf{Service Balancing:} Distributing client interactions evenly across waiters to prevent service delays and ensure minimal lag times.
    \item \textbf{Priority Service:} Implementing prioritization mechanisms for managing service requests based on criteria such as service complexity, client needs, or special requests.
\end{itemize}

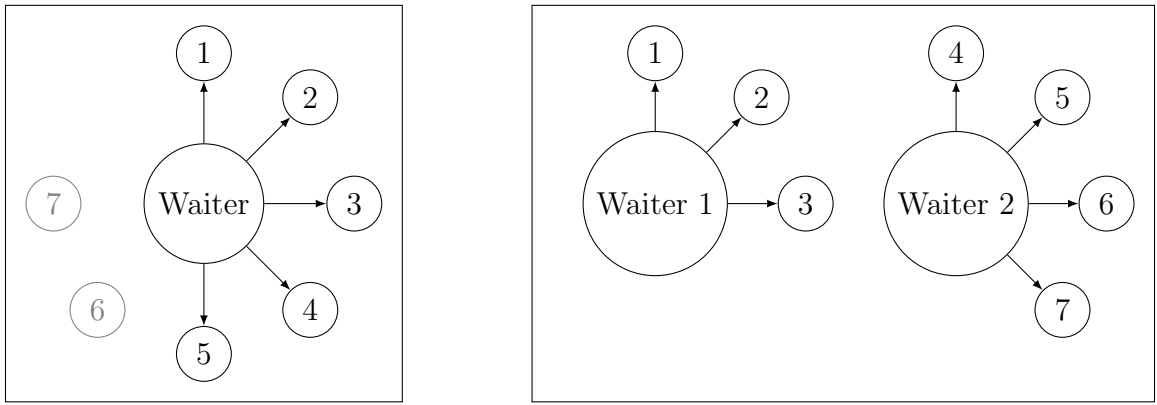
\begin{figure}[H]
     \centering
    \begin{tikzpicture}[>=latex]
        \coordinate (NW1) at (-2.5,2.5);
        \coordinate (SE1) at (2.5,-2.5);
    
        \node[circle, draw, minimum size=1cm] (waiter) at (0,0) {Waiter};
        
        \foreach \angle/\label in {90/1, 45/2, 0/3, 315/4, 270/5, 225/6, 180/7} {
            \ifthenelse{\angle=225 \OR \angle=180}{
                \node[circle, draw, minimum size=0.5cm, gray] (client\label) at (\angle:2cm) {\label};
            }{
                \node[circle, draw, minimum size=0.5cm] (client\label) at (\angle:2cm) {\label};
                \draw[->] (waiter) -- (client\label);
            }
        }
        
        \node[draw, rectangle, fit=(NW1) (SE1)] (box1) {};
        
        \begin{scope}[xshift=6cm]
            \coordinate (NW2) at (-1.5,2.5);
            \coordinate (SE2) at (6.5,-2.5);
        
            \node[circle, draw, minimum size=1cm] (waiter1) at (0,0) {Waiter 1};
            
            \node[circle, draw, minimum size=1cm] (waiter2) at (4,0) {Waiter 2};
            
            \foreach \angle/\label in {90/1, 45/2, 0/3} {
                \node[circle, draw, minimum size=0.5cm] (client\label) at ([shift=(\angle:2cm)]waiter1) {\label};
                \draw[->] (waiter1) -- (client\label);
            }
            
            \foreach \angle/\label in {90/4, 45/5, 0/6, 315/7} {
                \node[circle, draw, minimum size=0.5cm] (client\label) at ([shift=(\angle:2cm)]waiter2) {\label};
                \draw[->] (waiter2) -- (client\label);
            }
            
            \node[draw, rectangle, fit=(NW2) (SE2)] (box2) {};
        \end{scope}
    \end{tikzpicture}
    \caption{Left: A waiter node with a max\_degree of 5 can only connect to 5 clients, leaving clients 6 and 7 disconnected. Right: Load balancing clients across two waiters, ensuring all clients are connected.}
    \label{fig:waiter-load-balancing-combined}
\end{figure}

\paragraph{Temporal Nature of Links}
Links in the waiter-client model are inherently temporal, existing for the duration required to complete the service cycle. This characteristic emphasizes the adaptability of service delivery, with interactions being initiated and concluded based on real-time service demands. Implications include:

\begin{itemize}
    \item \textbf{Dynamic Service Structure:} The service structure continuously evolves, reflecting the changing patterns of waiter-client interactions over time.
    \item \textbf{Adaptive Service Strategies:} Adaptive service strategies become essential, with waiters dynamically adjusting their service approach to efficiently accommodate client needs.
\end{itemize}

\subsubsection{Implications for Network Optimization and Analysis}
The \textit{service\_capacity} limitation and the temporal nature of interactions necessitate sophisticated service optimization and analysis approaches. Strategies must account for these dynamic and constrained aspects of the service model, focusing on:

\begin{itemize}
    \item \textbf{Maximizing Waiter Utilization:} Ensuring waiters operate within their capacity limits while minimizing wait times and maximizing service quality.
    \item \textbf{Analyzing Service Patterns:} Understanding the temporal dynamics of client interactions to forecast demand and adjust waiter assignments preemptively.
\end{itemize}

Incorporating considerations of waiter \textit{service\_capacity} and the temporal characteristics of service interactions enriches the service model's design and operational framework. These factors underscore the complexity of managing waiter-client interactions within the service area, highlighting the need for adaptive and responsive service strategies that can accommodate the dynamic nature of client needs and expectations.

\subsection{Network Optimization}
Optimization focuses on enhancing the efficiency of the prompt-response cycle, leveraging spatiotemporal considerations to ensure optimal waiter placement and timing to meet client needs. Strategies include strategic waiter placement and adaptive service provision, aiming to minimize wait times and adapt to client interaction patterns.

\subsection{Spatiotemporal Considerations}
The strategic placement and scheduling of waiters are crucial for optimizing service delivery,and are guided by the following considerations:
\begin{itemize}
    \item \textbf{Spatial Domain:} Ensuring service interactions occur within minimized distances or latencies enhances service immediacy and personalization for clients.
    \item \textbf{Temporal Blocks:} Allocating service provisions into defined time blocks allows for the agile reallocation of waiter resources in response to fluctuating demand and availability.
\end{itemize}

\subsection{Efficiency Metrics}
Service delivery efficiency is quantitatively assessed through metrics designed to capture the service model's performance:
\begin{equation}
    \text{Service Efficiency} = f\left(\text{Coverage Ratio}, \text{Average Service Time}, \text{Client Satisfaction}\right)
\end{equation}
Key metrics include:
\begin{itemize}
    \item \textbf{Coverage Ratio:} The fraction of clients that receive timely and personalized service.
    \item \textbf{Average Service Time:} The mean duration from service initiation to completion.
    \item \textbf{Client Satisfaction:} The quality of service perceived by clients, which is influenced by service speed, personalization, and waiter responsiveness.
\end{itemize}

\subsection{Waiter Placement Strategies}
Optimizing the waiter-client dynamic within the service area involves considerations not just of client density but also of the activity levels of clients and the potential influence of specific zones on service demand. This approach to waiter placement emphasizes not only physical proximity to a large number of clients but strategic proximity to areas with high activity or special service needs.

\paragraph{Identifying High-Activity and Service Demand Zones}
Service optimization strategies incorporate analytics to:
\begin{itemize}
    \item Identify zones or times with high service demand, indicating areas where waiters can be most impactful in terms of service efficiency and client satisfaction.
    \item Analyze patterns of service interactions to determine zones that might experience a surge in service demand, allowing for proactive waiter placement to manage anticipated needs.
\end{itemize}

\section{Manager-Worker Dynamics}

This section delves into the Manager-Worker model within the ROBUST Network, building upon the foundational observer-observable dynamic. It focuses on optimizing the allocation and execution of tasks within a structured hierarchy, aiming to maximize network efficiency and performance.

\paragraph{Definition of a Manager}
A manager operates as a specialized observer within the network, and is responsible for overseeing worker nodes, allocating tasks, and ensuring optimal performance. The manager's capacity to effectively manage and engage workers is constrained by a finite bandwidth, necessitating strategic planning and delegation to maximize productivity and worker engagement. The efficiency of a manager is significantly influenced by their ability to dynamically allocate tasks, manage worker capacities, and adjust strategies based on real-time feedback.

\paragraph{Definition of a Worker}
A worker functions as an observable within the network, executing tasks allocated by the manager. Workers initiate communication with managers to request resources, report progress, or seek guidance, triggering managerial responses. The worker's role is characterized by their responsiveness to managerial directives and their contribution to the network's collective objectives.

\subsection{Examples of Manager-Worker Dynamics}

\paragraph{Corporate Setting Example}
In a corporate environment, managers and workers engage in a continuous cycle of task delegation and execution. Managers assess project needs, worker capabilities, and resource availability to distribute tasks effectively. Workers, in turn, undertake these tasks, providing updates and feedback that inform subsequent managerial decisions. This dynamic highlights the necessity for effective communication, trust, and a shared commitment to achieving organizational goals.

\paragraph{Manufacturing Example}
In a manufacturing context, managers allocate production tasks to workers based on production goals, worker expertise, and machinery availability. Workers execute these tasks, with the efficiency of production lines heavily dependent on the managers' ability to balance workloads and respond to production challenges. This setting exemplifies the manager-worker dynamic in a tangible operational environment.

\subsection{Node Interactions: Task Allocation and Execution Cycle}
The Manager-Worker relationship is characterized by a task allocation and execution cycle, where managers delegate tasks and workers execute them. This cycle necessitates a network configuration that supports efficient task distribution and execution, with a focus on minimizing idle times and maximizing worker engagement.

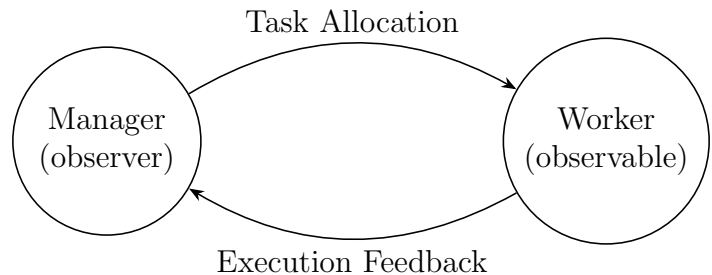
\begin{figure}[H]
     \centering
    \begin{tikzpicture}[>=Stealth, auto, semithick]
        \tikzstyle{state}=[fill=white,draw=black,text=black, circle, minimum size=2.5cm, align=center]
    
        \node[state] (manager) {Manager \\ (observer)};
        \node[state] (worker) [right=4cm of manager] {Worker \\ (observable)}; 
    
        \draw[->] (manager) edge[bend left] node {Task Allocation} (worker);
        \draw[->] (worker) edge[bend left] node {Execution Feedback} (manager);
        
        \node[fit=(manager) (worker),inner sep=0cm] {}; 
    
    \end{tikzpicture}
    \caption{Finite State Machine showing the task allocation and execution cycle.}
    \label{fig:manager-worker}
\end{figure}

This interaction model introduces a capacity constraint on the number of workers a manager can effectively oversee, creating a bottleneck that impacts task execution efficiency and worker satisfaction. Addressing this bottleneck requires strategic manager placement and task scheduling to ensure a balanced workload and effective worker engagement.

\subsection{Optimization Strategies for Manager-Worker Interactions}
Optimization strategies for the Manager-Worker dynamic focus on enhancing the task allocation and execution cycle's efficiency, aiming to maximize worker productivity while ensuring managers effectively communicate expectations and resources.

\begin{itemize}
    \item \textbf{Dynamic Task Allocation:}  Managers dynamically adjust task assignments based on worker feedback, task complexity, and completion rates to balance workloads and maintain high engagement.
    \item \textbf{Worker Activation Strategies:} Techniques are implemented to quickly engage idle or at-risk workers, such as diversified task assignments and motivational incentives.
    \item \textbf{Efficiency Metrics:} Metrics to assess the effectiveness of the task allocation and execution cycle are developed, focusing on task completion rates, worker engagement levels, and active versus idle worker ratios.
    \item \textbf{Collaborative Management:} Managers collaborate in planning and configuring task allocations, sharing best practices and optimizing strategies network-wide to ensure consistent worker activity and reduce execution bottlenecks.
\end{itemize}

\subsection{Network Optimization}
Optimization of the Manager-Worker dynamic emphasizes maximizing task execution efficiency while ensuring managers can effectively oversee and guide workers. Key to this optimization is the strategic scheduling of tasks and the allocation of managerial resources to align with worker capabilities and availability.

\subsection{Spatiotemporal Considerations in Task Allocation}
The strategic scheduling and distribution of tasks are guided by the need to minimize idle times and optimize worker engagement, with a focus on:

\begin{itemize}
    \item \textbf{Minimizing Response Times:} Ensuring tasks are allocated to reduce the time between assignment and execution, enhancing task completion efficiency.
    \item \textbf{Optimizing Task Scheduling:} Aligning task assignments with worker availability and peak productivity periods to facilitate efficient task execution.
\end{itemize}

\subsection{Conclusion}
The Manager-Worker dynamic within the ROBUST network underscores the importance of strategic task allocation, effective managerial oversight, and the implementation of optimization strategies to maximize worker engagement and productivity. Through careful planning, dynamic task allocation, and fostering a collaborative environment, managers can guide their teams towards achieving strategic objectives, thereby enhancing the overall performance and efficiency of the network.

\section{Guard-Citizen Relationship}
The Guard-Citizen Relationship within a ROBUST network offers a unique perspective on the dynamics of protection and freedom within a defined spatial domain. This relationship can be distinguished from the Hide-Seek and Predator-Prey dynamics by its focus on communal safety and individual freedom, as well as the proactive and reactive roles of guards (observers) and citizens (observables). Here's how this relationship can be modeled:

\subsection{Operational Framework}
In the Guard-Citizen dynamic, guards are tasked with overseeing the safety and freedom of movement of citizens within a spatial domain, such as a community or digital network. Their primary goals include maximizing the citizens' ability to roam freely while identifying and neutralizing threats from rogue agents.

\subsubsection{Definition of a Guard:} Guards are observers within the network, responsible for maintaining order, ensuring the safety of citizens, and responding to incidents. They employ surveillance, pattern recognition, and intervention strategies to protect citizens from potential harm.
  
\subsubsection{Definition of a Citizen:} Citizens are observables that engage in regular activities within the network's domain. While most citizens contribute positively to the community, a subset may act as rogue agents, posing threats to others.

\subsection{Node Interactions: Surveillance and Intervention Cycle}
The Guard-Citizen dynamic is characterized by a cycle of surveillance and intervention, where guards continuously monitor citizen activities to detect abnormal or threatening behaviors. Upon identifying a potential threat, guards intervene to isolate the rogue citizen and mitigate any harm, ensuring the continued freedom and safety of the community.

\subsection{Strategies for Maximizing Freedom and Safety}
To effectively manage the Guard-Citizen relationship, strategies must be developed to balance the dual objectives of maximizing citizen freedom and ensuring communal safety:

\begin{itemize}
    \item \textbf{Surveillance Optimization:} Implements sophisticated surveillance systems that respect citizens' privacy while effectively identifying potential threats.
  
    \item \textbf{Rapid Response Protocols}: Establishes protocols for quick intervention when rogue behavior is detected, minimizing harm to the community.
  
    \item \textbf{Community Engagement:} Encourages` citizen participation in safety measures, such as reporting suspicious activities or participating in community watch programs.
\end{itemize}

\subsection{Spatiotemporal Considerations}
The efficiency of the Guard-Citizen relationship is influenced by both spatial and temporal factors:
\begin{itemize}
    \item \textbf{Spatial Coverage:} Ensuring comprehensive surveillance and protection coverage across the entire domain to prevent areas of vulnerability.
  
    \item \textbf{Temporal Responsiveness:} The speed at which guards can respond to incidents is crucial for effective threat mitigation and maintaining public trust.
\end{itemize}

\subsection{Efficiency Metrics}
Efficiency in the Guard-Citizen dynamic can be measured through metrics such as:

\subsubsection{Incident Response Time:} 
The average time taken by guards to respond to and resolve incidents.
  
\subsubsection{Citizen Satisfaction:}
Measures of citizen satisfaction with safety measures and freedom of movement can provide insight into the relationship's effectiveness.

\subsubsection{Threat Neutralization Rate:} 
The percentage of identified threats that are successfully neutralized by the guards.

\chapter{Competitive Bipartite Dynamics}
This chapter explores the competitive dynamics, which stand in stark contrast to the relationships in the previous chapter, but similarly builds upon the base observer-observable model.  This chapter highlights how different competitive bipartite relationships may be within the ROBUST network.  This chapter details interactions between roles: Invader-Defender, Predator-Prey, Hide-Seek, Host-Parasite relationships.

\section{Defender-Invader Dynamics}

In the context of protecting spatial areas and assets, the Defender-Invader relationship is central to strategies that guard against unauthorized access or harm. This section dives into the dynamics of this relationship, exploring operational frameworks, defense optimization strategies, and the influence of spatial and temporal factors on defense efficacy.

\begin{figure}[H]
  \centering
  \includegraphics[width=0.6\linewidth]{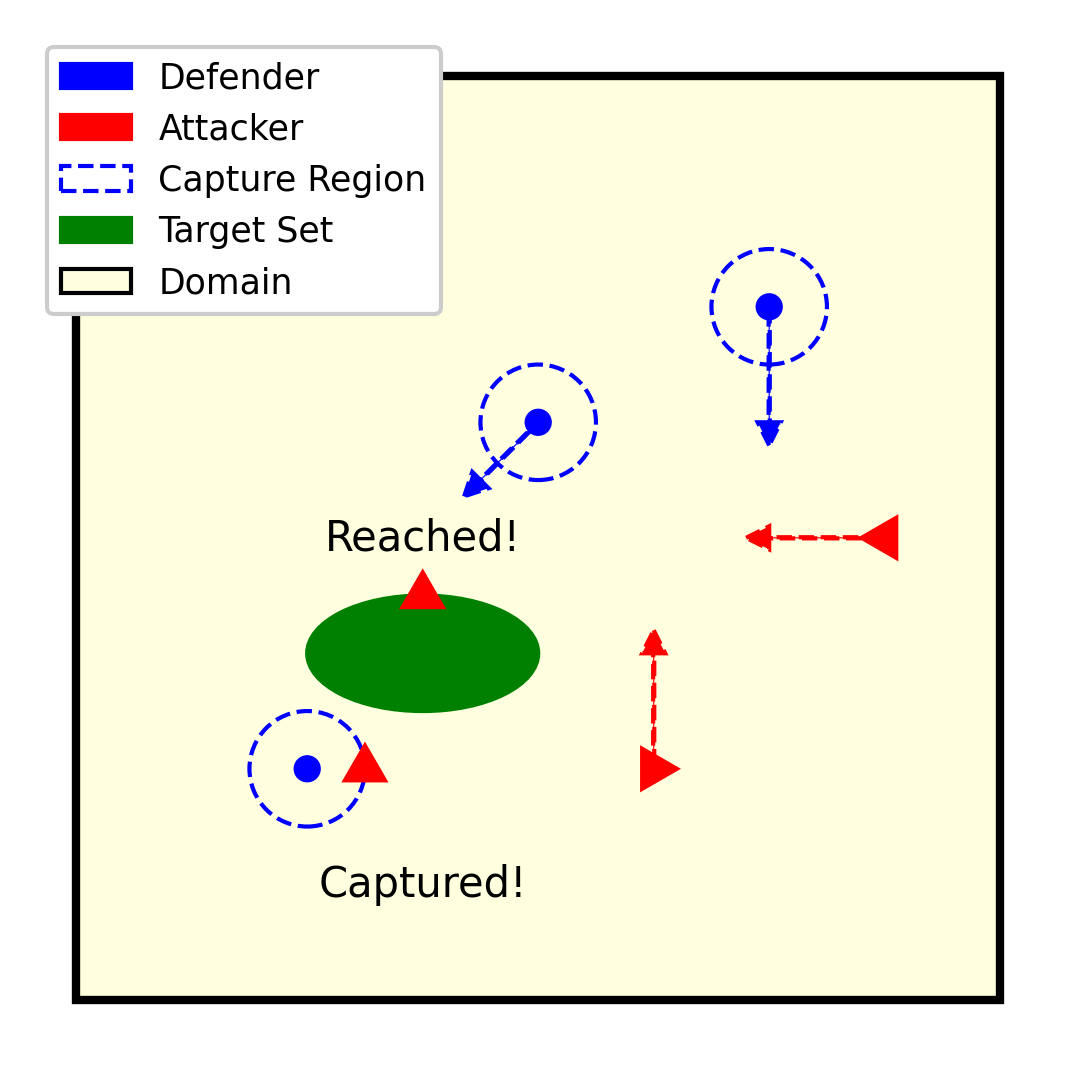} 
  \caption{Invader-Defender relationship}
  \label{fig:invader-defender-demo} 
\end{figure}

\subsection{Operational Framework}

The Defender-Invader dynamic is underscored by defenders' proactive strategies to safeguard specific regions or assets from invaders aiming to compromise, breach, or exploit them. Defenders, acting as specialized observers, employ detection, interception, and neutralization tactics against invaders.

\paragraph{Definition of a Defender}
A defender is tasked with the protection of a spatial area, utilizing a dual-layer strategy that encompasses both the anticipation of invader actions and their subsequent interception. Their effectiveness is measured by their ability to predict potential breaches and implement effective countermeasures to prevent invader success.

\paragraph{Definition of an Invader}
An invader seeks to penetrate defined protective barriers with the goal of disruption or unauthorized access. The unpredictable and complex nature of invader strategies demands a defense approach that is both dynamic and adaptable.

\subsection{Node Interactions: Prediction and Interception Cycle}

At the heart of the Defender-Invader relationship is a continuous cycle of prediction and interception. Defenders leverage predictive analytics to forecast invader movements and strategize the allocation of defensive resources for effective interception.

\subsection{Strategic Defense Optimization}

Optimization strategies for defense involve developing and executing algorithms that enhance defenders' capability to efficiently allocate resources, ensuring broad coverage and high rates of interception. This process includes real-time online convex programming and scalable hierarchical iterative methods.

\subsection{Spatiotemporal Considerations}

Effective defense against invaders hinges on strategic considerations of both space and time:

Spatial Optimization: Identifies the optimal deployment of defenders to cover potential invasion paths and safeguard critical assets.
Temporal Adaptability: Involves the real-time adjustment of defense strategies in response to evolving invader tactics and behaviors.

\subsection{Efficiency Metrics}

The efficiency of defense strategies in the Defender-Invader dynamic is crucial for maintaining the security and functionality of the protected spatial area. This efficiency is quantified through several key metrics:

\paragraph{Interception Rate:} This metric measures the percentage of successfully intercepted invaders out of the total number of attempted breaches. A higher interception rate indicates a more effective defense strategy.

   \[
   \text{Interception Rate} = \frac{\text{Number of Interceptions}}{\text{Total Number of Invasion Attempts}}
   \]

\paragraph{Time to Neutralize Threats:} This metric assesses the average time taken by defenders to detect, engage, and neutralize invaders from the moment of detection. Shorter times suggest a more responsive and efficient defense system.

   \[
   \text{Time to Neutralize Threats} = \frac{\sum \text{Time Taken to Neutralize Each Invader}}{\text{Number of Neutralized Invaders}}
   \]

\paragraph{Integrity of the Protected Area Post-Interception:} Evaluates the condition of the protected spatial area after defense actions have been taken. This includes assessing any damages or breaches that occurred despite interception efforts. High integrity levels post-interception signify minimal damage and effective containment of threats.

   \[
   \text{Integrity} = \frac{\text{Uncompromised Area}}{\text{Total Protected Area}}
   \]

\paragraph{Resource Efficiency:} Analyzes how effectively the defenders utilize their available resources, including personnel, technology, and energy, to maximize interception rates and minimize response times.

   \[
   \text{Resource Efficiency} = \frac{\text{Number of Successful Interceptions}}{\text{Resources Expended}}
   \]

\paragraph{Adaptability Score:} Reflects the defense system's ability to adapt to new or evolving threats over time. This includes incorporating lessons learned from past invasions to improve future defense responses.

   \[
   \text{Adaptability Score} = \frac{\text{Number of Adaptations to New Threats}}{\text{Total Encountered Threats}}
   \]

These metrics together provide a comprehensive overview of a defense strategy's effectiveness, allowing for targeted improvements and adjustments to enhance overall security against invader threats.

\subsection{Conclusion}

The Defender-Invader relationship is pivotal in the context of spatial protection, embodying the ongoing struggle between safeguarding assets and the challenge of invasive threats. Through strategic optimization, predictive analysis, and adaptable defensive mechanisms, defenders strive to thwart invaders, ensuring the security and resilience of the protected spaces. This dynamic underscores the necessity for anticipatory defense strategies and the continuous adaptation to new and evolving threats, highlighting the complexity of maintaining security in variable spatial domains.

\section{Predator-Prey Dynamics}

In contrast to the adversarial Defender-Invader relationship, the Predator-Prey dynamics present a different interaction model within spatial domains. This section explores the subtleties of Predator-Prey dynamics, focusing on the roles of seekers and hiders, the strategies they employ, and the implications of spatial and temporal considerations on the game's outcomes.

\subsection{Operational Framework}

The essence of Predator-Prey lies in the interplay between seekers, who aim to locate hidden participants within a defined area, and prey, who strive to remain undetected and evade. Unlike the Defender-Invader dynamic, the Predator-Prey game is less about safeguarding and more about the challenge of detection and evasion.

\paragraph{Definition of a Predator}
A predator, analogous to the observer in other dynamics, actively searches for hiders within the game's spatial domain. The seeker's role is to discover all hiders using various search strategies, without necessarily protecting a specific asset or domain from harm.

\paragraph{Definition of a Prey}
Prey, acting as the observables, employ strategies of concealment and evasion to remain undetected by the predator. Their objective may be to invade or disrupt or hide within the environment until the predator threat leaves.

\subsection{Node Interactions: Search and Evasion Cycle}

At the core of Hide-Seek dynamics is the cycle of search and evasion, where seekers employ various methods to locate hiders. This cycle is continuous and evolves as the game progresses, with seekers adjusting their search patterns and hiders potentially moving to maintain concealment.

\subsection{Strategic Evasion and Search Optimization}

Optimization strategies in Predator-Prey focus on maximizing the efficacy of search patterns for Predator and enhancing hiding strategies for hiders. Predators may use probabilistic methods to predict likely hiding spots, while prey analyze the predator's behavior to choose the most effective hiding places.

\subsection{Spatiotemporal Considerations}

The game's outcome heavily depends on strategic considerations of both space and time:

\begin{itemize}
    \item \textbf{Spatial Optimization for Prey:} Involves selecting hiding spots that offer the best chance of remaining undetected, considering the seeker's search patterns and the environment's features.
    \item  \textbf{Temporal Adaptability for Predator:} Predators may need to adjust their search strategy over time, responding to the elapsed time without finding hiders or to cues indicating a hider's recent movement.
\end{itemize}

\subsection{Efficiency Metrics}

Efficiency in Predator-Prey dynamics is measured differently, focusing on the duration of concealment for hiders and the time required for seekers to locate all hiders:

\paragraph{Evasion Duration:} Measures the average time hiders remain undetected. Longer durations indicate more effective hiding strategies.

\[
\text{Evasion Duration} = \frac{\sum \text{Time Each Hider Remains Undetected}}{\text{Number of Prey}}
\]

\paragraph{Search Efficiency:} Assesses how quickly and effectively a seeker locates all hiders. Greater efficiency is indicated by shorter search times.

\[
\text{Search Efficiency} = \frac{\text{Number of Prey Found}}{\text{Time Spent Searching}}
\]

\section{ Hide-Seek Dynamics}

Exploring an inverse interpretation of Predator-Prey dynamics unveils a scenario where hiders assume the role of observers, actively monitoring the seeker's movements to adapt and evade detection. This inversion of traditional roles introduces a nuanced layer of strategy and awareness to the game.

\subsection{Operational Framework}

The revised Hide-Seek dynamics pivot around the concept of hiders as active participants, leveraging observational strategies to avoid being found by seekers. This shift portrays seekers as persistent threats within the environment, whose presence and behavior must be continuously monitored and outmaneuvered by hiders.

\paragraph{Definition of a Hider}
In this alternative dynamic, hiders are reimagined as proactive observers, employing tactical awareness and environmental knowledge to evade seekers. Their success is predicated on their ability to predict the seeker's search patterns and dynamically adjust their hiding strategies accordingly.

\paragraph{Definition of a Seeker}
Seekers, now cast as the observables, embody the constant challenge hiders must navigate. Their movements and search strategies, while methodical, become the basis upon which hiders strategize their evasion tactics.

\subsection{Node Interactions: Surveillance and Evasion Cycle}

Central to this reinterpretation is a surveillance and evasion cycle, where hiders actively observe and react to the seeker's movements. This cycle emphasizes the cognitive and strategic elements of evasion, highlighting the hider's agency in dictating the pace and direction of the game.

\subsection{Strategic Evasion and Surveillance Optimization}

The optimization of hiding strategies now involves a sophisticated blend of predictive analytics, environmental manipulation, and psychological tactics to mislead or divert the seeker. Conversely, seekers must enhance their search methods to counteract the hiders' evasive maneuvers effectively.

\subsection{Spatiotemporal Considerations}

\begin{itemize}
    \item \textbf{Spatial Awareness for Hiders:} Achieving undetectability hinges on the hider's ability to interpret spatial cues and anticipate the seeker's movements, requiring a nuanced understanding of the environment.
    \item \textbf{Temporal Dynamics of Evasion:} The timing of movements and the decision when to shift hiding spots are critical, with hiders needing to balance the risk of movement against the potential benefits of repositioning.
\end{itemize}

\subsection{Efficiency Metrics}

\paragraph{Evasion Success Rate:} Quantifies the effectiveness of hiders in remaining undetected, factoring in the duration of evasion and the adaptability of hiding strategies.

\paragraph{Seeker Detection Time:} Measures the time taken for seekers to locate hiders, offering insights into the complexity and cunning of hider tactics.

\section{Host-Parasite Dynamics}

The Host-Parasite dynamics introduce a complex interplay within ROBUST Networks, focusing on the relationship between hosts (observers) and parasites (observables). This dynamic is characterized by the cyclical nature of infection, response, and adaptation, underscoring the spatial and temporal challenges inherent in managing parasitic interactions.

\subsection{Operational Framework}

Within ROBUST Networks, the Host-Parasite relationship is defined by the interaction between the host nodes, which aim to maintain the network's integrity and functionality, and the parasite nodes, which seek to exploit the network's resources for their proliferation. This dynamic framework introduces unique challenges, including infection spread, host resistance, and network resilience.

\paragraph{Definition of a Host}
A host in the ROBUST Network is an entity that provides necessary resources or conditions for parasites to live and reproduce. As observers, hosts are tasked with detecting and responding to parasitic invasions, employing strategies to mitigate damage and prevent further spread.

\paragraph{Definition of a Parasite}
A parasite within this context is an observable entity that relies on the host for survival, often at the host's expense. Parasites aim to exploit the network's resources, leading to potential degradation of network performance or integrity.

\subsection{Node Interactions: Infection and Response Cycle}

The Host-Parasite dynamic revolves around the infection and response cycle. Parasites attempt to infect host nodes, exploiting their resources for reproduction and spread. Hosts, in turn, activate defense mechanisms to detect, isolate, or eliminate the parasitic influence, often involving recovery processes to restore affected nodes.

\subsection{Strategies for Infection Control and Host Resilience}

Effective management of Host-Parasite dynamics within ROBUST Networks involves strategic planning and optimization to control infection rates and enhance host resilience. This includes:

\begin{itemize}
    \item \textbf{Infection Rate Management:} Employing strategies to limit the rate at which parasites can infect host nodes, potentially through network segmentation or enhanced detection mechanisms.
    \item \textbf{Lag Time Reduction:}  Decreasing the time between infection detection and response, thereby minimizing the impact of each parasitic invasion.
    \item \textbf{Recovery Time Optimization:}  Enhancing the network's ability to recover from parasitic infections, ensuring quick restoration of affected nodes to full functionality.
    \item \textbf{Host-to-Host Interaction for Infection Limitation:}  Facilitating communication between host nodes to share information about detected parasites, enabling preemptive defenses and coordinated responses.
\end{itemize}

\subsection{Spatiotemporal Considerations}

The dynamics of Host-Parasite interactions within ROBUST Networks are significantly influenced by spatial and temporal factors:

\begin{itemize}
    \item \textbf{Spatial Distribution and Connectivity:} The spatial arrangement and connectivity of host nodes can impact the ease and speed with which parasites spread throughout the network.
    \item \textbf{Temporal Dynamics of Infection and Response:} The timing of infection detection, response initiation, and recovery completion plays a critical role in the network's overall resilience to parasitic threats.
\end{itemize}

\subsection{Efficiency Metrics}

The effectiveness of strategies against Host-Parasite dynamics is quantified through several key metrics:

\paragraph{Infection Rate:}
\[
\text{Infection Rate} = \frac{\text{Number of New Infections}}{\text{Time Interval}}
\]

\paragraph{Recovery Time:}
\[
\text{Recovery Time} = \frac{\sum \text{Time Taken for Hosts to Recover}}{\text{Number of Recovered Hosts}}
\]

\paragraph{Network Integrity Post-Recovery:}
\[
\text{Network Integrity} = \frac{\text{Number of Fully Functional Hosts Post-Recovery}}{\text{Total Number of Hosts}}
\]

\subsection{Conclusion}

The Host-Parasite dynamics within ROBUST Networks highlight a critical aspect of network management, focusing on the balance between resource exploitation by parasites and the resilience mechanisms of hosts. Through strategic spatial and temporal planning, alongside effective infection control strategies, networks can enhance their resilience against parasitic threats, ensuring sustained integrity and functionality. This dynamic underscores the importance of continuous monitoring, rapid response, and adaptive strategies in maintaining network health and performance amidst ongoing parasitic challenges.

\chapter{Potential Cases for ROBUST network} 

\section{ Introduction}
This chapter explores the diverse practical applications of the ROBUST framework developed in this research, illustrating its adaptability and relevance across multiple domains. Each example demonstrates how the theoretical constructs and methodologies from earlier chapters can be applied to solve real-world problems.

\section{ Software Development:}
\paragraph{Code Coverage:} In software testing, analysis tools (observers) strategically examine the spatial layout of the codebase, focusing on complex modules or functions and their interdependencies (spatial attributes). By tracking the evolution and execution sequence of these specific sections over time (temporal attributes), these tools measure the execution percentage. The optimization challenge lies in determining the minimal set of tests needed to achieve comprehensive coverage, allowing developers to pinpoint untested or frequently modified areas, thus ensuring the software's robustness and reliability.

\section{ Environmental Modeling:}
\paragraph{Oceanic Forecasting:} Buoys and underwater gliders (observers) are deployed to monitor specific regions of the ocean (observables) that are vital for forecasting due to their spatial or temporal variability. The data collected, such as temperature and salinity, refines marine models, enhancing their forecasting capabilities. Strategically positioning these observers ensures we gather the most crucial data, thereby significantly impacting climate models.

\section{ Urban Planning:}
\paragraph{Crime and Safety Monitoring:}  In urban environments, the spatial arrangement of security cameras (observers) is crucial. These cameras are strategically placed to monitor areas with high foot traffic or historically high crime rates (spatial attributes). Over time (temporal attributes), patterns of criminal activities (observables) emerge, and this data is vital. Optimizing the placement and responsiveness of these cameras can deter potential criminals, lead to rapid police responses, and, ultimately, create safer urban spaces.

\section{ Traffic Flow Optimization:} Urban traffic has both spatial and temporal dynamics. Traffic cameras (observers) are positioned at critical junctions, monitoring areas that historically experience congestion during rush hours (spatial and temporal attributes). By observing vehicle flow and congestion patterns (observables) in real-time, data-driven decisions can be made. This might involve adjusting traffic light timings or providing real-time traffic updates to commuters. Strategically optimizing these observation points can lead to smoother traffic flow, reduced congestion, and more efficient urban transportation systems.

\section{ Disaster Planning:}
\paragraph{Emergency Response:}  In the aftermath of natural disasters, the temporal progression of the disaster's impact becomes a critical focus. Drones or first responders (observers) are deployed to scan affected regions, prioritizing areas that report the highest number of distress calls or historically vulnerable zones (spatial attributes). By tracking real-time developments (temporal attributes), these observers identify stranded or injured individuals (observables). Strategic deployment of these observation tools ensures timely rescue operations and maximizes the efficacy of relief efforts.

\section{ Online Platforms:}
\paragraph{Content Moderation:} Digital platforms like Wikipedia or online forums are vast and continually evolving. Here, the challenge lies in the spatial distribution of content across various topics and the temporal dynamics of user interactions. Editors or automated systems (observers) are calibrated to focus on high-traffic pages or trending topics (spatial attributes) and to keep an eye on rapidly changing or frequently updated content (temporal attributes). By monitoring and reviewing public entries or comments (observables), these systems ensure the quality, accuracy, and trustworthiness of the platform's content. Optimizing the efficiency of these observers is crucial to maintaining a platform's credibility and user trust.

\section{ Retail and Consumer Behavior:}
\paragraph{In-Store Shopping Dynamics:} Brick-and-mortar stores are intricately laid out to maximize client engagement and sales. In-store cameras or sensors (observers) may purposefully be positioned to capture areas of high client interaction or prolonged stay (spatial attributes). As shopping patterns evolve (temporal attributes), these tools detect notable shifts in client behavior and movement (observables). Insights drawn from this data empower retailers to tweak store designs, adjust product displays, and modify promotional zones. The primary aim is to elevate the in-store experience, boost sales, and nurture client loyalty.

\section{ Online Shopping Patterns:} Digital storefronts are dynamic, often tailored to individual user behaviors. Web analytics tools (observers) are set to monitor pages with high engagement or track lateral movements from one product page to another (spatial attributes). Over successive user visits (temporal attributes), these tools identify patterns in page visits, product views, and purchase behaviors (observables). This data is pivotal for refining recommendation systems, personalizing user experiences, and strategically placing promotional content. The overarching objective is to enhance user engagement, optimize conversions, and foster repeat visits.

\section{ Biology and Medicine:}
\paragraph{Epidemiology:} In the realm of public health, spatial distribution and temporal evolution of diseases are crucial. Health agencies and researchers (observers) are designed to monitor specific regions or communities that might be more susceptible to outbreaks (spatial attributes). They also keep a keen eye on the progression and spread of diseases over time (temporal attributes). By observing the breakout patterns of viruses or other diseases (observables), informed strategies can be devised to curtail their spread, ensuring a healthier populace and reduced strain on healthcare resources.

\paragraph{ Pharmacovigilance:} Post-market drug safety is both a spatial and temporal challenge. Professionals in pharmacovigilance (observers) are tasked with monitoring specific demographics or regions that might show varied drug responses (spatial attributes). They also track the onset and progression of adverse effects over time (temporal attributes). By keeping a vigilant watch on drug-related issues (observables) post-market release, they ensure patient safety and can swiftly intervene when new drug risks emerge.

\section{ Education:}
\paragraph{Student Retention:} Maintaining high retention rates is a key objective for academic institutions. To achieve this, educators (observers) monitor cohort performance across courses and over time (spatial and temporal attributes). Through this observation, patterns and anomalies in engagement and performance (observables) emerge. Whether challenges arise from course material, instructional methods, resource constraints, or individual struggles, timely identification is crucial. The optimization challenge lies in swiftly pinpointing and responding to at-risk students or cohorts. By doing so, institutions can provide rapid interventions and support, maximizing the chances of student success and minimizing dropouts.

\paragraph{ Personalized Learning:} Today's digital learning environments are vast, encompassing various modules, topics, and resources. Learning management systems (observers) are calibrated to focus on the most accessed resources or challenging modules (spatial attributes). They also track students' learning patterns and feedback over the duration of courses (temporal attributes). By observing individual student's interactions and feedback (observables), these systems can tailor content and resources to cater to unique learning needs, fostering a more personalized and effective learning experience.

\section{ Economics and Finance:}
\paragraph{Market Analysis:} In the world of finance, financial analysts (observers) strategically focus on specific markets or sectors (spatial attributes) and track the evolution of market indicators and events over time (temporal attributes). By keenly observing these fluctuations (observables), analysts can make informed predictions, aiding investors and policymakers. The optimization challenge lies in forecasting economic trends with minimal error, ensuring stable economic planning and smart investment strategies.

\paragraph{ Credit Analysis and Community Trends:} Banking institutions (observers) analyze both individual transaction patterns and broader trends within specific communities or demographic groups. These trends have spatial attributes, such as behaviors in distinct neighborhoods, and temporal dimensions, tracking their evolution over time. By observing individual and group patterns (observables), banks compute accurate credit scores and identify broader lending trends. This optimization ensures both responsible individual lending and an understanding of broader financial shifts.
Possible Domains of Application: 

\section{ Astronomy:}
\paragraph{Celestial Monitoring:} The cosmos are vast, with countless celestial events occurring across different regions (spatial attributes) and evolving over millennia (temporal attributes). Telescopes or satellites (observers) are positioned to track specific celestial events or objects (observables), gathering invaluable data. The optimization challenge in astronomy is to determine the best observation points and times to expand our understanding of the universe, guiding future space missions and enhancing our grasp of cosmic phenomena.

\section{ Transportation and Logistics:}
\paragraph{Fleet Optimization:} In the constantly shifting landscape of logistics, the location and demand of delivery and pickup points (events) evolve rapidly. Trucks, monitored by GPS and sensors (observers), act as dynamic entities navigating this landscape (observables). By tracking their real-time positions and status (spatial attributes) against the backdrop of ever-changing delivery schedules and priorities (temporal attributes), logistics teams can dynamically reroute and optimize their fleets. The ultimate goal is to ensure timely pickups and deliveries, maximize truck utilization, and minimize operational costs.

\section{ Energy and Infrastructure:}
\paragraph{Grid Efficiency and Reliability:} Electrical grids, with their vast network of nodes, transmission lines, substations, and other components, act as the critical observers in the realm of power distribution. These grids have to cater to spatially diverse consumers, from hospitals and industrial units that are high-priority, to residential areas with varied demands. Each of these consumers represents an observable event, with unique power needs and consumption patterns. Given the spatial complexity of these demands, and the potential for faults or disruptions, especially during disasters like hurricanes, the grid continuously monitors and adjusts power distribution (spatial attributes). Over time (temporal attributes), patterns of power consumption, outages, and system stresses emerge. The optimization challenge for energy providers is twofold: ensuring that critical infrastructures like hospitals always have power, and when disruptions occur, restoring power in a manner that maximizes the number of consumers reconnected in the shortest time, all while ensuring the overall robustness and resilience of the grid.

\chapter{Benchmarking Environment}
\label{chap:bench_env}

\section{Study Design}
This chapter outlines the approach to dissecting and evaluating distinct spatiotemporal data behaviors as observed in two primary case studies: Oceanography Data and Crime Data. The objective is to extrapolate and understand varying spatiotemporal behaviors and their implications. This analysis sets the stage for the creation and utilization of synthetic datasets, designed to generalize these behaviors. The synthetic datasets enable testing and evaluation of different analytical methods under a spectrum of spatiotemporal scenarios, thereby facilitating a deeper understanding and optimization of various data analysis techniques.

\section{Data Collection}
This section delves into the detailed analysis of the datasets used in the primary case studies: Oceanography Data and Crime Data. Each dataset exhibits distinct spatiotemporal characteristics, providing valuable insights into the diverse behaviors of spatiotemporal data. 

\subsection{Oceanography Data Case Study}
The Oceanography Data Case Study exhibits evolutionary spatiotemporal behavior, where the influence of neighboring spaces and prior states plays a pivotal role in the dataset's dynamics, mirroring natural ecological systems. The focus on regions with significant spatial and temporal variability pinpoints areas of environmental interest. The evolutionary properties of this data are useful for predictive modeling, particularly in strategic sensor placement for ecological monitoring. The interconnected nature of these states illustrates the spatiotemporal dynamics that approaches must be able to plan for.  For a detailed description of the Oceanography Data, see Chapter~\ref{chap:oceanography}.

\subsection{Crime Data Case Study}
In contrast to the Oceanography Data's evolutionary dynamics, the Crime Data Case Study exhibits a distinctly different stochastic behavior. It operates on a probabilistic model, where incidents are marked as discrete events within shifting spatial clusters. These clusters signify areas with a heightened likelihood of crime occurrences, but unlike the Oceanography Data, events here are less influenced by historical patterns and more by current probabilistic trends. This distinction underscores the need for precise event detection, as it captures the unpredictable yet patterned nature of urban and social spatiotemporal dynamics.  For a detailed description of the Crime Data, see Chapter~\ref{chap:crime}.

\subsection{Spatiotemporal Insights}
These two complimentary case studies  provide a broad perspective on spatiotemporal data, underscoring the diversity of behaviors and the importance of selecting suitable analytical methods for different types of datasets. The contrasting nature of the datasets – one depicting environmental interdependence and the other illustrating random, isolated events – serves as a robust foundation for understanding the complexity and adaptability required in spatiotemporal data analysis.

\section{Synthetic Spatiotemporal Data}
This section explores the development and use of synthetic spatiotemporal datasets, drawing on behaviors identified in the Oceanography and Crime Data case studies. Such datasets are essential for simulating common spatiotemporal behaviors in a controlled environment, focusing on the comparative performance of algorithms rather than replicating specific real-world scenarios.

\subsection{Rationale for Synthetic Data}
\subsubsection{Bridging Real-World Complexity}
Synthetic datasets act as intermediaries, capturing key aspects of real-world data while providing a structured setting for analytical testing. They allow for the emulation of identified behaviors, aiding in the evaluation of various analytical methods.

\subsubsection{Customizing Spatiotemporal Behaviors}
These datasets facilitate customization by enabling the study of behaviors such as the evolutionary patterns in oceanography or the stochastic nature of crime data. This approach is instrumental in exploring a range of spatiotemporal dynamics and their analytical implications.

\subsection{Advantages of Synthetic Spatiotemporal Data}
\subsubsection{Controlled Experimentation}
Synthetic data offers controlled conditions for experimentation, providing a platform for precise hypothesis testing and model evaluation.

\subsubsection{Flexibility and Scalability}
The flexibility and scalability of synthetic datasets make them adaptable to various research requirements, supporting comprehensive testing across different scenarios.

\subsection{Disadvantages of Synthetic Spatiotemporal Data}
\subsubsection{Limitations in Emulating Real-World Complexity}
 Valuable synthetic data may not capture the full unpredictability and intricacies of natural datasets, since the focus is on common behaviors rather than specific real-world scenarios.

\subsubsection{Bias and Generalizability Concerns}
The creation process may introduce biases, thereby impacting the generalizability of results. However, the primary goal is to compare algorithm performances under standardized conditions, rather than to tightly coupling the data with specific real-world use cases.

\section{Dynamic Heatmaps}

\subsection{Dynamic Heat Map Abstraction}
A dynamic heat map abstraction may be employed to capture the temporal hot zones as they appear and propagate across the search space. In such an abstraction, the values are typically normalized between 0 and 1 to standardize the intensity measures across different scales.

Let \( H_t \) denote the heat map at time \( t \). The heat map \( H_t \) is configured as a grid of size \( n \times m \), with each cell \( H_{i,j,t} \) indicating the intensity of the hot zone at location \( (i, j) \) and time \( t \).
Intensity is defined by a suitable metric, which may include factors such as crisis zone, temporal variability, or other task-specific points of interest.

\begin{figure}[htbp]
  \begin{adjustwidth}{-0cm}{-0cm}
    \centering
    \begin{minipage}[b]{0.48\textwidth}
      \centering
      \includegraphics[width=\textwidth]{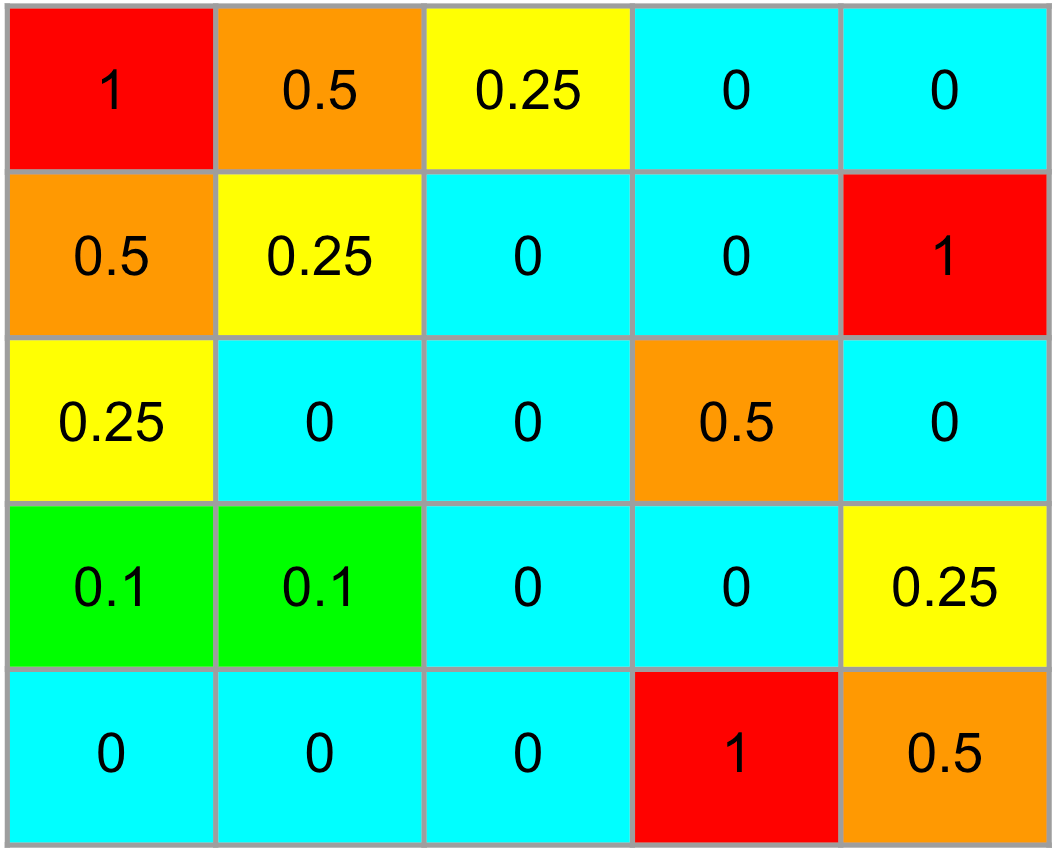}
      \caption{Heat map \( H_t \) at initial time}
      \label{fig:abstract-heatmap-t0}
    \end{minipage}
    \hfill
    \begin{minipage}[b]{0.48\textwidth}
      \centering
      \includegraphics[width=\textwidth]{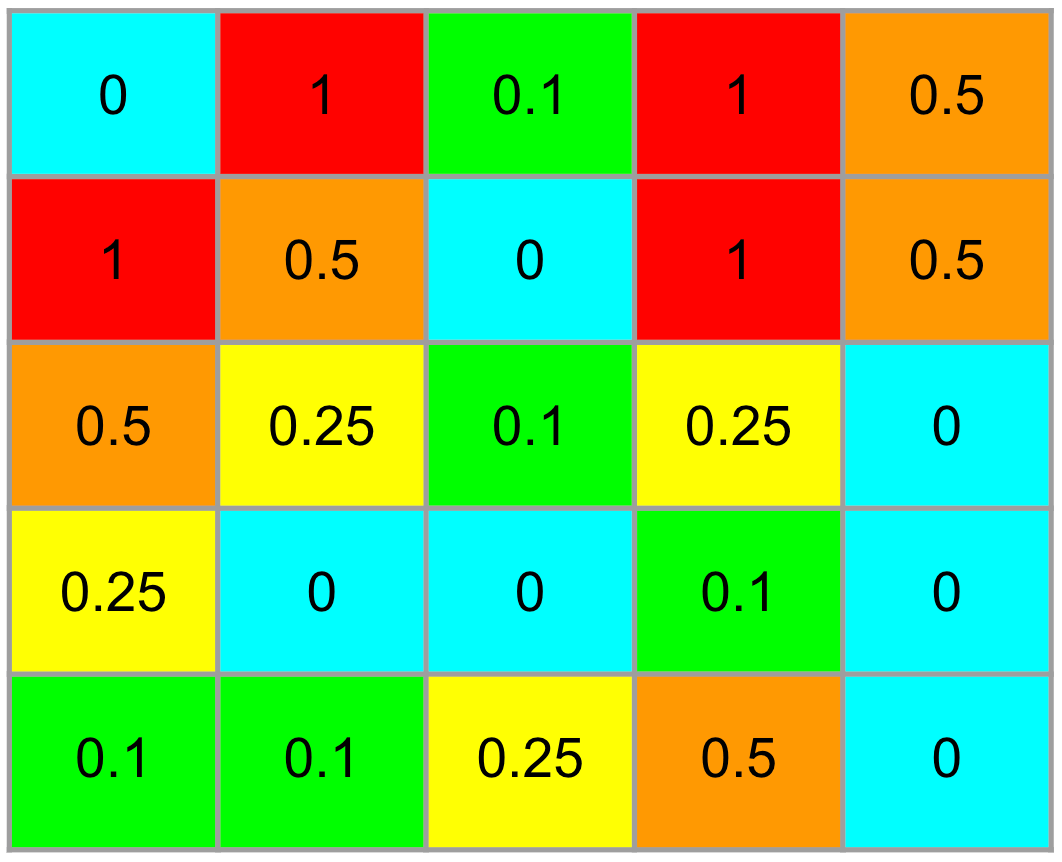}
      \caption{Heat map \( H_{t+1} \) at a later time}
      \label{fig:abstract-heatmap-t1}
    \end{minipage}
  \end{adjustwidth}
\end{figure}

In the accompanying illustrations (Figures \ref{fig:abstract-heatmap-t0} and \ref{fig:abstract-heatmap-t1}), the numerical values are mapped to a color scale where 0 represents no intensity and 1 represents the maximum intensity. This color-coding provides a clear and immediate visual representation of the distribution and evolution of hot zones over time, which is particularly beneficial for evaluators when comparing the effectiveness of different analytical approaches.

\subsection{Mathematical Formulation of Dynamic Heatmaps}

The mathematical foundation of dynamic heatmaps is critical for spatial and temporal data analysis. This section outlines the mathematical models defining dynamic heatmaps.

\subsubsection{Heatmap Representation}
The dynamic heatmap is defined as a vector \( H \):
\[
H = (H_0, H_1, \dots, H_T)
\]
where each element \( H_t \) in the vector represents the state of the heatmap at a specific time step \( t \). Each \( H_t \) is a grid of size \( n \times m \), where \( n \) and \( m \) are the dimensions of the grid. The intensity at each cell within this grid is denoted by \( h_{ij}^t \).

The matrix representation of the heatmap at time \( t \) is given by:
\[
H_t = \begin{bmatrix}
h_{11}^t & h_{12}^t & \dots  & h_{1m}^t \\
h_{21}^t & h_{22}^t & \dots  & h_{2m}^t \\
\vdots   & \vdots   & \ddots & \vdots   \\
h_{n1}^t & h_{n2}^t & \dots  & h_{nm}^t
\end{bmatrix}
\]

Here, \( h_{ij}^t \) represents the intensity value at the cell located in the \( i \)-th row and \( j \)-th column at time \( t \).

\subsection{Dynamic Changes in Heatmaps}
Heatmaps serve as a synthesis platform to visualize outcomes from diverse predictive methodologies. Whether adopting a frequentist approach, which may consider the frequency of events, or a Markovian approach, which relies on the state-dependent probabilities, the resulting heatmap effectively abstracts temporal dynamics into discernible hot and cold zones. This abstraction is achieved by representing changes attributed to environmental factors or external events, and it showcases the versatility of heatmaps in depicting various scenarios, from natural phenomena to complex societal dynamics. The unified visual representation provided by the heatmap enables quick and easy comparative analysis of predictions, regardless of the underlying predictive model employed.

\subsubsection{Random Heatmap}
A random heatmap represents a spatial distribution of values that are assigned completely at random, without influence from previous states or any specific probabilistic structure.

\begin{figure}[htbp]
  \begin{adjustwidth}{-0cm}{-1cm}
    \centering
    \begin{minipage}[b]{0.33\textwidth}
      \centering
      \includegraphics[width=\textwidth]{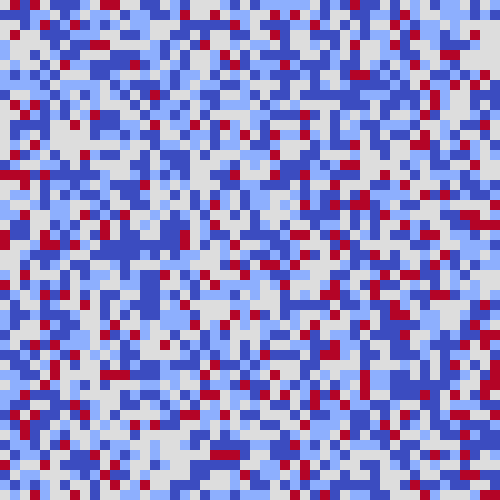}
      \caption{Random \( H_1 \)}
      \label{fig:rnd-heatmap-t0}
    \end{minipage}
    \hfill
    \begin{minipage}[b]{0.33\textwidth}
      \centering
      \includegraphics[width=\textwidth]{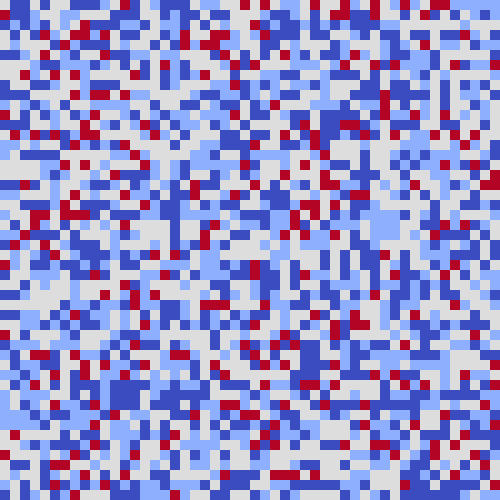}
      \caption{Random \( H_2 \)}
      \label{fig:rnd-heatmap-t1}
    \end{minipage}
    \hfill
    \begin{minipage}[b]{0.33\textwidth}
      \centering
      \includegraphics[width=\textwidth]{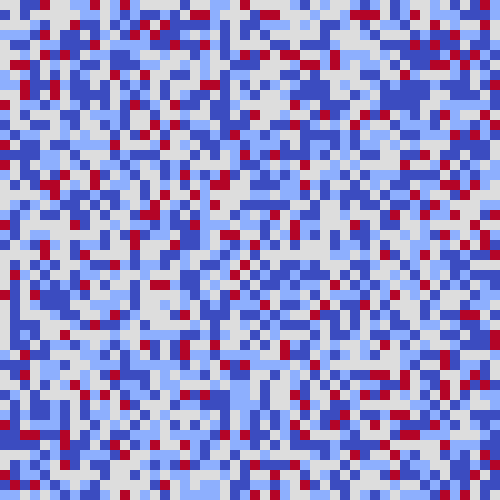}
      \caption{Random \( H_3 \)}
      \label{fig:rnd-heatmap-t2}
    \end{minipage}
  \end{adjustwidth}
\end{figure}

Figures \ref{fig:rnd-heatmap-t0}, \ref{fig:rnd-heatmap-t1}, and \ref{fig:rnd-heatmap-t2} display instances of a random heatmap at three consecutive time steps. Each figure visualizes a unique and independent random distribution of intensities across the grid, showcasing the lack of temporal or spatial correlation.

\subsubsection{Stochastic Heatmap}
A stochastic heatmap represents a spatial distribution where values are assigned according to a non-uniform probability distribution. This can model systems where some regions have a higher likelihood of events, informed by historical patterns or data, while maintaining independence between successive states.

\begin{figure}[htbp]
  \begin{adjustwidth}{-0cm}{-1cm}
    \centering
    \begin{minipage}[b]{0.33\textwidth}
      \centering
      \includegraphics[width=\textwidth]{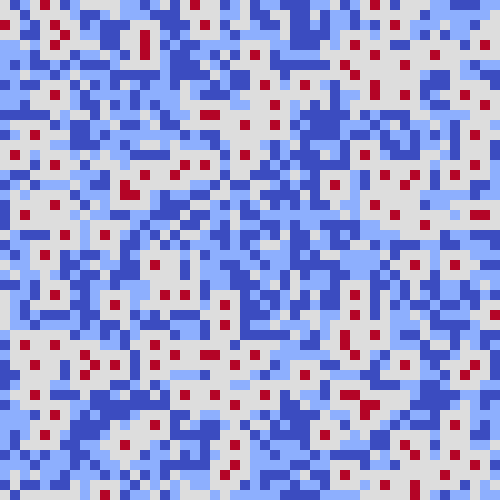}
      \caption{Stochastic \( H_1 \)}
      \label{fig:sto-heatmap-t0}
    \end{minipage}
    \hfill
    \begin{minipage}[b]{0.33\textwidth}
      \centering
      \includegraphics[width=\textwidth]{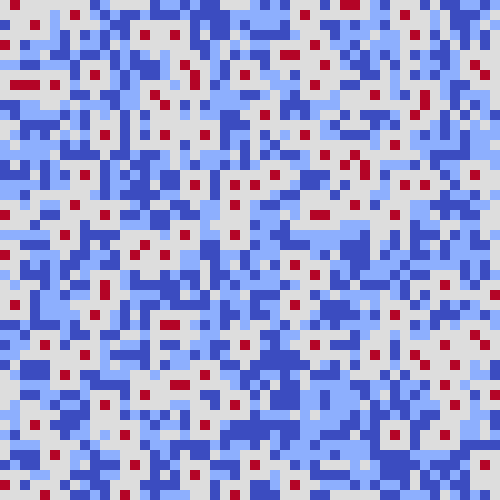}
      \caption{Stochastic \( H_2 \)}
      \label{fig:sto-heatmap-t1}
    \end{minipage}
    \hfill
    \begin{minipage}[b]{0.33\textwidth}
      \centering
      \includegraphics[width=\textwidth]{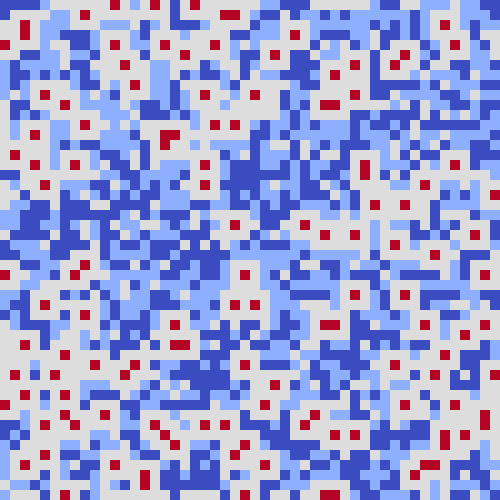}
      \caption{Stochastic \( H_3 \)}
      \label{fig:sto-heatmap-t2}
    \end{minipage}
  \end{adjustwidth}
\end{figure}

\subsubsection{Evolutionary Heatmap}
An evolutionary heatmap is derived or heavily influenced from previous states, capturing the ongoing dynamics of a system. Each state is an evolution of the last, reflecting changes within the system due to environmental factors, internal dynamics, or event-driven modifications.

\begin{figure}[htbp]
  \begin{adjustwidth}{-0cm}{-1cm}
    \centering
    \begin{minipage}[b]{0.33\textwidth}
      \centering
      \includegraphics[width=\textwidth]{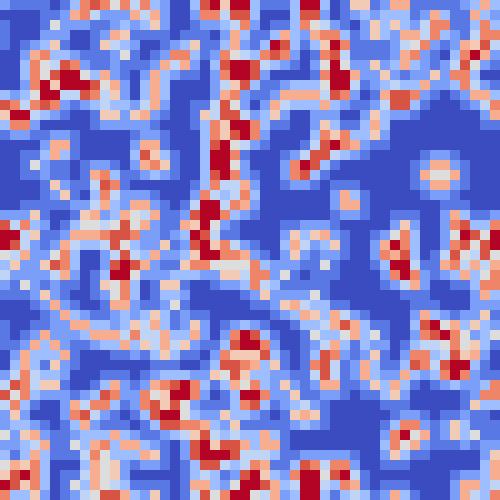}
      \caption{Evolutionary \( H_1 \)}
      \label{fig:evo-heatmap-t0}
    \end{minipage}
    \hfill
    \begin{minipage}[b]{0.33\textwidth}
      \centering
      \includegraphics[width=\textwidth]{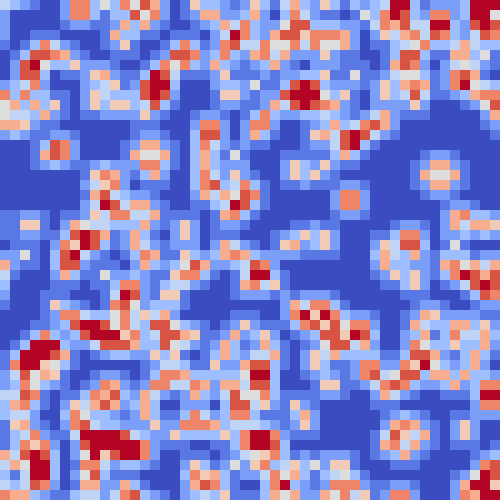}
      \caption{Evolutionary \( H_2 \)}
      \label{fig:evo-heatmap-t1}
    \end{minipage}
    \hfill
    \begin{minipage}[b]{0.33\textwidth}
      \centering
      \includegraphics[width=\textwidth]{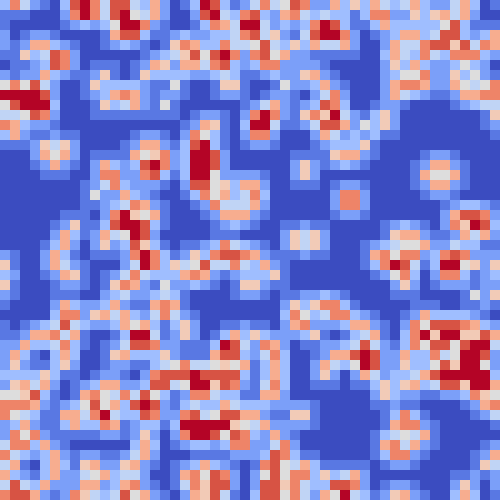}
      \caption{Evolutionary \( H_3 \)}
      \label{fig:evo-heatmap-t2}
    \end{minipage}
  \end{adjustwidth}
\end{figure}

Figure \ref{fig:evo-heatmap-t0}, \ref{fig:evo-heatmap-t1}, and \ref{fig:evo-heatmap-t2} illustrate successive states of an evolutionary heatmap. The transition from \( H_1 \) to \( H_3 \) depicts the system's response over time to various factors. Each figure represents a unique time step in the system’s evolution, with color intensity indicating the level of activity or the presence of specific events at each grid cell.

\subsection{Applications and Implications}
Dynamic heatmaps go beyond visual representation, capturing and generalizing spatiotemporal behaviors and patterns. This approach aids in transforming abstract data behaviors into tangible analysis, enhancing the development of analytical methods and bridging the gap between theory and practical application.

\section{Dynamic Heatmap Generator}

\subsection{Motivation and Objectives}
The Dynamic Heatmap Generator is an innovative web application specifically designed for generating synthetic spatiotemporal datasets. Its primary purpose is to provide a scalable and configurable platform for simulating various spatiotemporal behaviors and patterns. This simulation is crucial for testing and evaluating algorithms and approaches that are intended to operate within spatiotemporal environments.

Key objectives of the Dynamic Heatmap Generator include:

\begin{itemize}
    \item \textbf{Synthetic Dataset Generation:} To generate scalable datasets that mimic real-world spatiotemporal dynamics, allowing for the effective testing of algorithms under various conditions.

    \item \textbf{Customization for Diverse Test Scenarios:} Enabling users to tailor spatiotemporal behaviors and patterns, thus creating diverse scenarios for comprehensive algorithm testing.

    \item \textbf{Interactive Visualization for Data Exploration:} Incorporating interactive elements such as sliders for time navigation and play/pause functions, to visualize and understand the synthetic data as it evolves over time and space.

    \item \textbf{Enabling Scalability and Complexity Adjustments:} Allowing researchers to scale the complexity of datasets up or down, thus providing insights into which conditions algorithms might perform optimally.
\end{itemize}

The heatmap generator stands out not just as a visualization tool but as a sophisticated platform for synthesizing complex spatiotemporal patterns, primarily aimed at assessing the effectiveness of various adaptive strategies and algorithms in dynamic environments.

\subsection{Tool Overview and Comprehensive Capabilities}

The Dynamic Heatmap Generator is a web-based application designed with a focus on advanced functionality and user accessibility. Developed using modern web technologies like HTML5 and JavaScript, it provides a user-friendly interface that makes the complex task of creating and analyzing synthetic spatiotemporal datasets intuitive and efficient.

\subsubsection{Advanced Functionality and User Interface}
The tool's interface is meticulously crafted to cater to both novice users and seasoned data analysts. Key features include:

\begin{itemize}
    \item \textbf{User-Friendly Interface:} The application boasts a straightforward and clean interface, ensuring that users of all skill levels can navigate and utilize the tool with ease.

    \item \textbf{Browser Development:} Leveraging the latest web technologies, the heatmap generator offers a seamless and responsive experience across various devices and platforms.

    \item \textbf{Handling Diverse Dynamical Behaviors:} It is equipped to handle a wide array of spatiotemporal behaviors and patterns, making it versatile for different kinds of data analysis tasks.

    \item \textbf{Customization Features:} The tool offers extensive customization options, allowing users to adjust parameters like grid size, intensity metrics, and time steps. This level of customization ensures that the heatmaps can be tailored to fit specific research questions or hypothesis testing.

    \item \textbf{Interactive Visualization:} The application includes interactive elements like sliders for time navigation and play/pause functionality for animations. These features not only make the tool more engaging but also aid in a deeper understanding of the data.

    \item \textbf{Data Export Capabilities:} Users have the option to download the generated heatmap data in a JSON format, enabling further analysis outside the tool.

    \item \textbf{Scalability:} The generator is designed to scale, both in terms of the size and complexity of the datasets it can handle, ensuring its applicability in various research contexts.

    \item \textbf{Application Suitability:} The heatmap generator is suitable for a wide range of experimental applications, from environmental studies to mock configurable spatiotemporal behaviors and patterns.
\end{itemize}

\subsection{Interactive Elements and User Engagement}

The Dynamic Heatmap Generator integrates a range of interactive elements designed to enhance user engagement and facilitate a comprehensive understanding of the heatmap data. These elements, crafted through advanced web development practices, contribute to an intuitive and dynamic user experience.

\subsubsection{User Interface and Control Panel}
The application features a control panel that allows users to interact with the heatmap generator through various input options and controls. Key aspects include:

\begin{itemize}
    \item \textbf{Heatmap Type Selection:} A dropdown menu (\texttt{heatmap-type}) enables users to choose from different heatmap types, such as Evolutionary Heatmap with Spatial or Temporal Rules and Randomly Weighted Heatmap with or without Proximity Constraints. This selection sets the basis for the type of heatmap to be generated.

    \item \textbf{General Options Form:} Users can input general parameters like the length, width, cell size, and number of snapshots through the \texttt{GeneralOptionsForm}. These parameters define the basic structure and scale of the heatmap.

    \item \textbf{Specific Options Form:} Depending on the chosen heatmap type, the \texttt{SpecificOptionsForm} presents additional customizable options. For instance, in evolutionary heatmaps, users can adjust the probability of a cell being alive (\texttt{pAlive}), while in random weighted heatmaps, thresholds for hot, warm, and cool cells are settable.

    \item \textbf{Generation and Visualization Controls:} After setting the parameters, users can generate the heatmap using the "Generate" button. The heatmap is then rendered on a canvas (\texttt{heatmapCanvas}), allowing for visual data analysis.
\end{itemize}

\begin{figure}[htbp]
  \begin{adjustwidth}{-0cm}{-1cm}
    \centering
    \begin{minipage}[b]{0.48\textwidth}
      \centering
      \frame{\includegraphics[width=\textwidth]{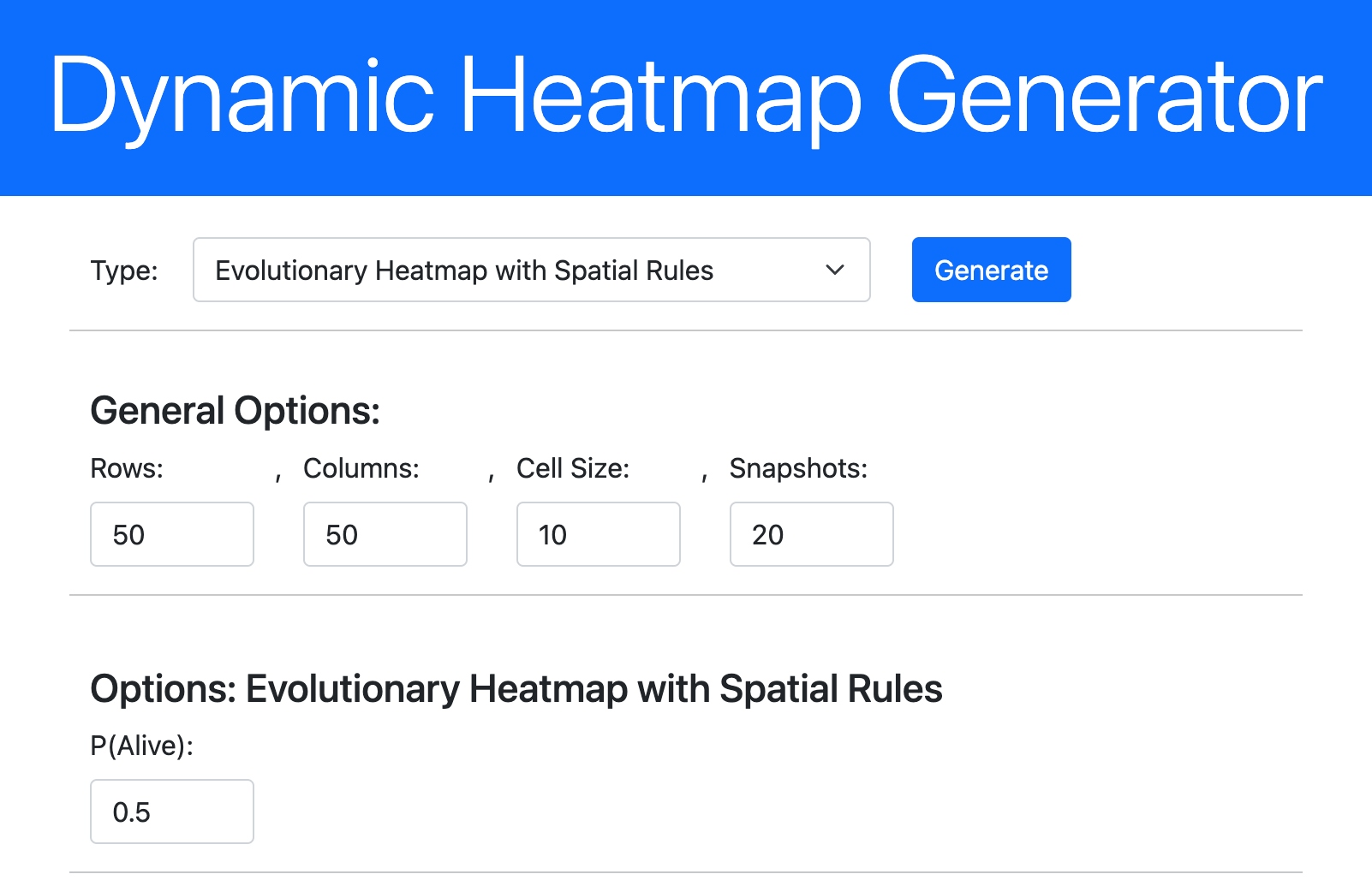}}
      \caption{\( H_1 \) Options}
      \label{fig:heatmap-gui-1}
    \end{minipage}
    \hfill
    \begin{minipage}[b]{0.48\textwidth}
      \centering
      \frame{\includegraphics[width=\textwidth]{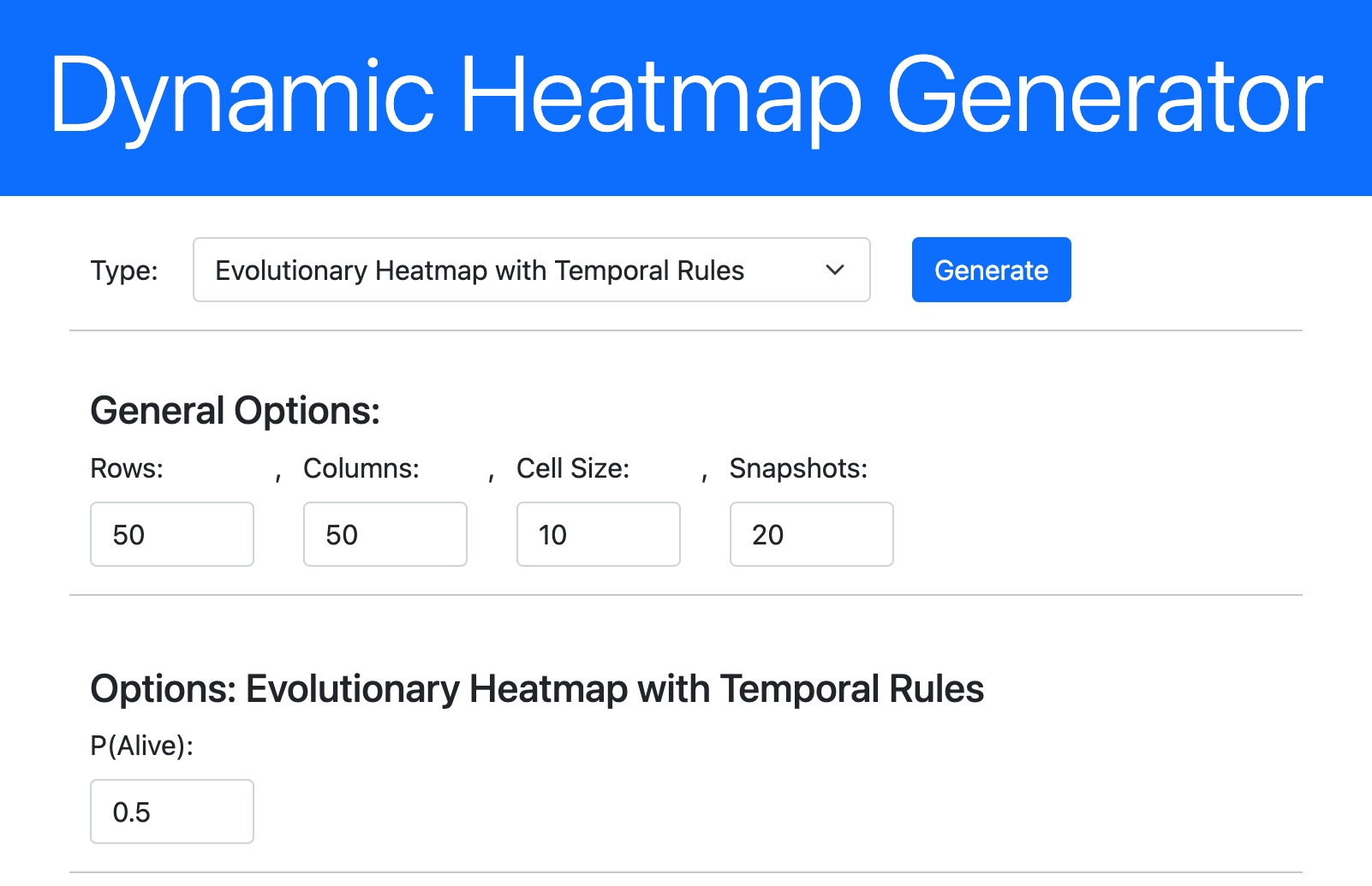}}
      \caption{\( H_2 \) Options}
      \label{fig:heatmap-gui-2}
    \end{minipage}
    \hfill
    \begin{minipage}[b]{0.48\textwidth}
      \centering
      \frame{\includegraphics[width=\textwidth]{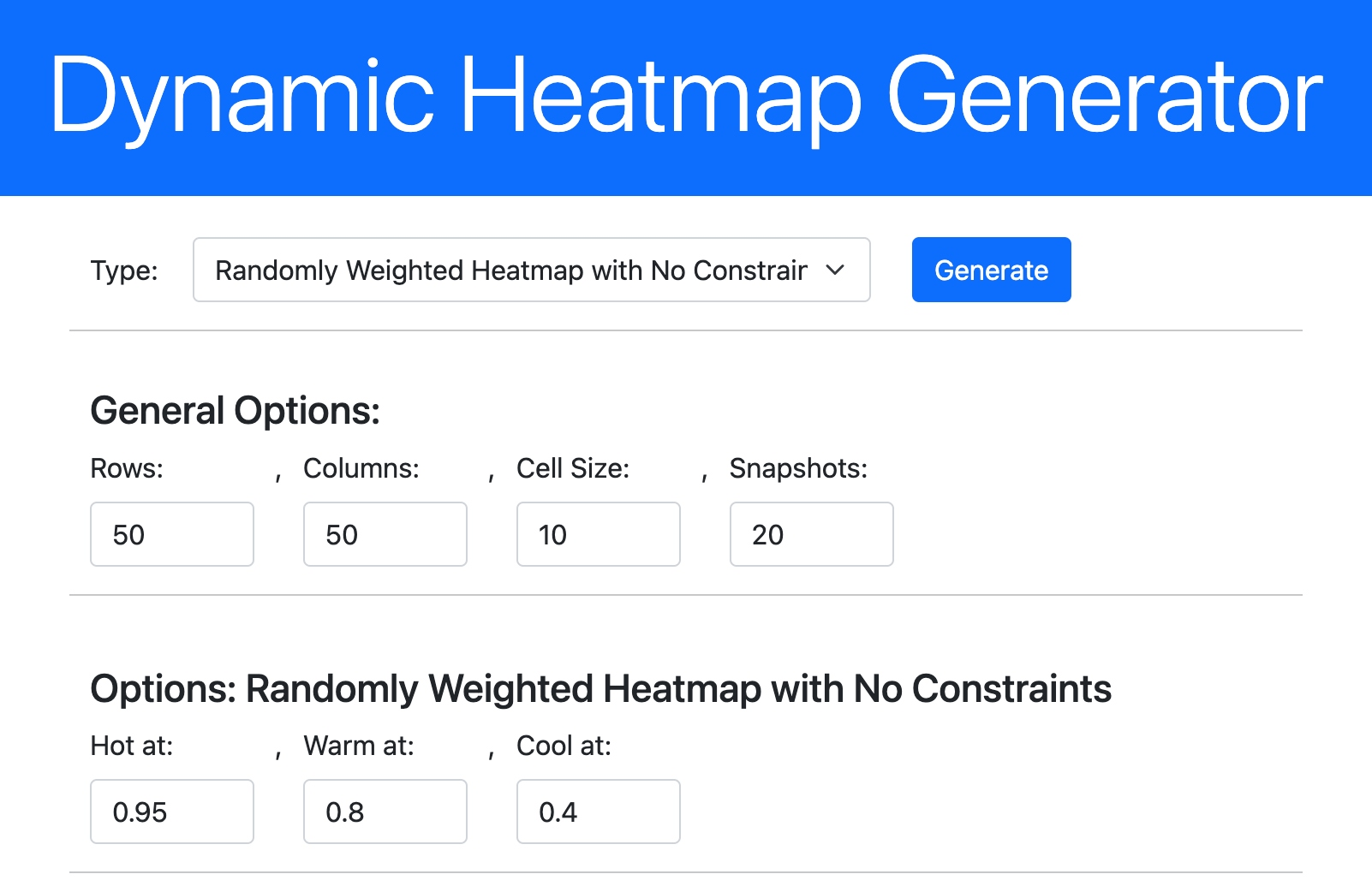}}
      \caption{\( H_3 \) Options}
      \label{fig:heatmap-gui-3}
    \end{minipage}
    \hfill
    \begin{minipage}[b]{0.48\textwidth}
      \centering
      \frame{\includegraphics[width=\textwidth]{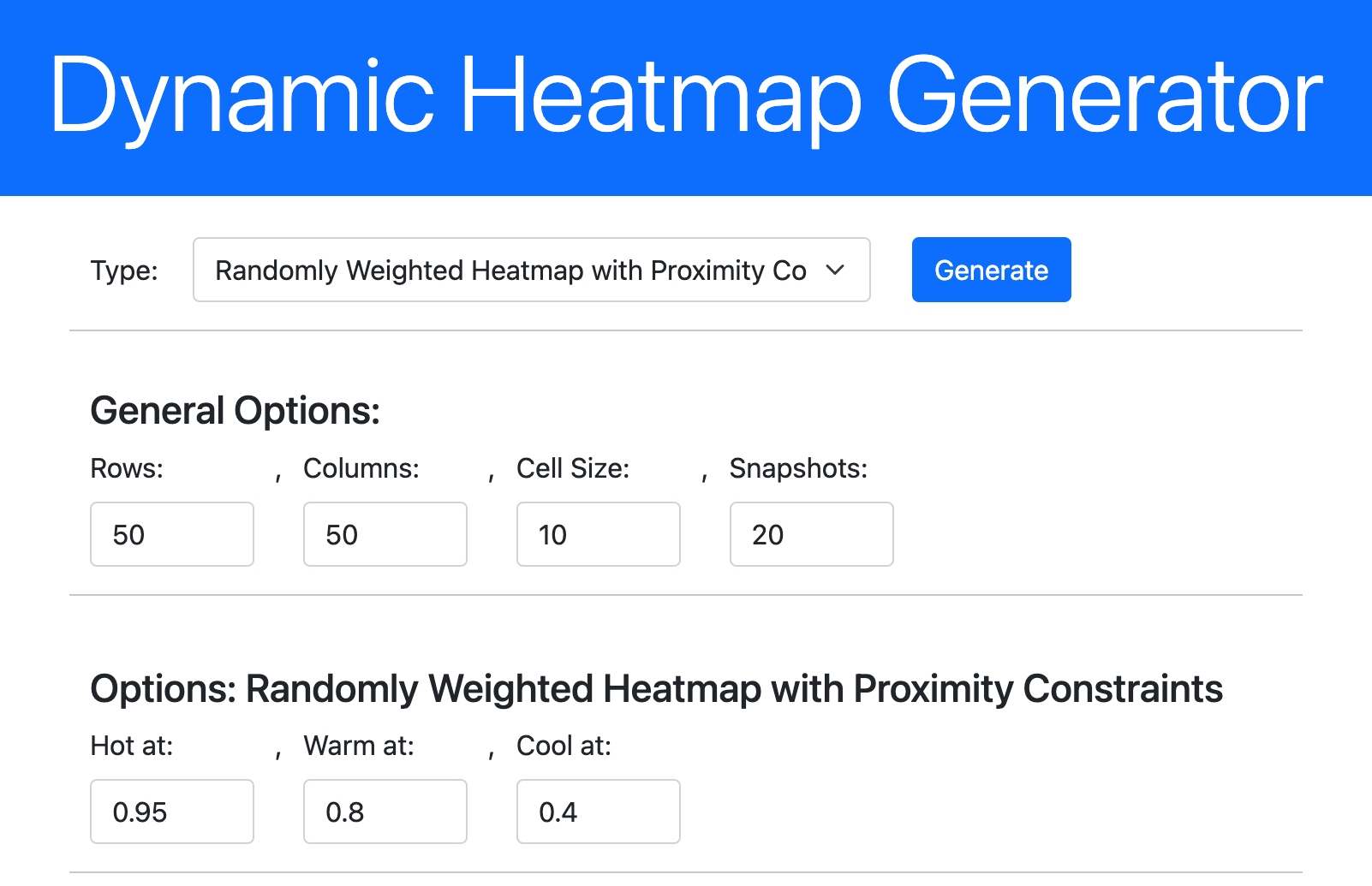}}
      \caption{\( H_4 \) Options}
      \label{fig:heatmap-gui-4}
    \end{minipage}
  \end{adjustwidth}
\end{figure}

\subsubsection{Dynamic Visualization and Data Exploration}
The application's visualization capabilities are enhanced by several interactive features:

\begin{itemize}
    \item \textbf{Time Slider:} The \texttt{timeSlider} allows users to navigate through different time steps of the heatmap, facilitating the exploration of temporal changes in the data.

    \item \textbf{Timestep Input and Control:} An input field (\texttt{timestepInput}) provides precise control over the visualization, enabling users to jump to specific time steps directly.

    \item \textbf{Play/Pause Animation:} The play/pause button (\texttt{playPause}) adds an animation feature, allowing users to observe the evolution of the heatmap over time in a continuous sequence.

    \item \textbf{Data Export Functionality:} A key feature is the ability to download the generated heatmap data as a JSON file (\texttt{download}), which can be used for further analysis or documentation purposes.
\end{itemize}

\subsubsection{Implementation and User Experience Design}
The interactive elements are implemented using a combination of HTML, CSS, and JavaScript, ensuring a seamless and responsive user experience. The layout and design are thoughtfully structured to guide the user naturally through the process of generating and analyzing heatmaps. The choice of Bootstrap for styling and layout further enhances the application's aesthetic and functional appeal.

\subsubsection{Description and Help Texts}
Additionally, the application includes descriptive texts and help sections (\texttt{DescriptionView}), providing users with contextual information about the different heatmap types and their potential applications. This feature aids in decision-making and enhances the user's understanding of the tool's capabilities.

\subsection{Application Architecture and Deployment}

The Dynamic Heatmap Generator is architected as a client-centric, serverless web application, ensuring streamlined efficiency and user-focused interactivity. This architecture supports the application's core functionality to generate and visualize heatmaps within a web browser, leveraging the computational power of the client's machine.

\subsubsection{Client-Side Architecture}
The entire processing workload, from rendering heatmaps to handling user interactions, is managed client-side. This serverless architecture model enables the application to run directly in the user's browser without backend dependencies, leading to rapid responsiveness and a reduction in server-side load and reliable availability.

\begin{itemize}
    \item \textbf{Modular Script Organization:} The application's scripts are organized into modules, separating concerns and enhancing maintainability. The 'views' module contains scripts that manage the user interface components, ensuring a clean and organized presentation layer.

    \item \textbf{Responsive Design:} Bootstrap, a front-end framework, is integrated to provide a consistent and aesthetically pleasing design, ensuring that the application is not only functional but also visually engaging.
\end{itemize}

\subsubsection{Public Deployment on GitHub Pages}
Deployment of the Dynamic Heatmap Generator employs GitHub Pages, providing a serverless hosting solution that aligns with the application's client-side processing philosophy.

\begin{itemize}
    \item \textbf{Global Accessibility:} Hosted on GitHub Pages, the application is universally accessible, allowing for immediate, global reach without the need for users to install or configure software.

    \item \textbf{Continuous Integration:} The GitHub repository facilitates continuous integration, allowing for seamless updates and iteration of the application without service interruption.
\end{itemize}

\subsubsection{Enhanced User Experience}
The Dynamic Heatmap Generator's deployment strategy is designed with user experience as a top priority, ensuring the tool is as user-friendly as it is powerful.

\begin{itemize}
    \item \textbf{Ease of Use:} The application's user interface is intuitive, allowing users to interact with the heatmap generator without the need for in-depth technical knowledge or extensive documentation.

    \item \textbf{Self-Sufficient Application:} By leveraging client-side processing, the application operates independently of server-side constraints, providing a robust and self-sufficient tool for users.

    \item \textbf{Open Source Collaboration:} The open-source nature of the project encourages collaboration, contribution, and community support, fostering an environment of shared knowledge and continuous improvement.
\end{itemize}

\subsection{API Documentation}

\subsubsection*{class \texttt{\textbf{HeatmapRandomBase}}}
\noindent 
{\small 
\setstretch{1} 
\begin{longtable}{|p{6cm}|p{10cm}|}

\hline
\multicolumn{2}{|c|}{\textbf{Fields}} \\
\hline
\rowcolor{lightgray}\texttt{canvasId} & ID of the canvas element where the heatmap is rendered. \\
\rowcolor{white}\texttt{length} & Number of rows in the heatmap grid. \\
\rowcolor{lightgray}\texttt{width} & Number of columns in the heatmap grid. \\
\rowcolor{white}\texttt{cellSize} & Size of each cell in the heatmap. \\
\rowcolor{lightgray}\texttt{snapshots} & Number of heatmap snapshots to be generated. \\
\rowcolor{white}\texttt{hotspotDistribution} & Distribution thresholds for hot, warm, and cool spots. \\
\hline
\multicolumn{2}{|c|}{\textbf{Methods}} \\
\hline
\rowcolor{lightgray}\texttt{\textbf{constructor}} & Initializes an object with specified properties for canvas, grid dimensions, cell size, number of snapshots, and distribution of hotspots. \\
\hline
\rowcolor{white}\texttt{\textbf{generateRandomGrid()}} & Creates a grid with random values [0,1] for each cell. \\
\hline
\rowcolor{lightgray}\texttt{\textbf{applyHotspotDistribution(grid)}} & Applies the hotspot distribution thresholds to the grid, categorizing cells into different heat levels. \\
\hline
\rowcolor{white}\texttt{\textbf{drawHeatMap(heatMap)}} & Renders the heatmap on the canvas with color coding based on cell heat levels. \\
\hline
\rowcolor{lightgray}\texttt{\textbf{setHeatmaps()}} & Generates a series of heatmaps based on the random distribution and stores them. \\
\hline
\end{longtable}
} 
\addcontentsline{lot}{table}{\protect\numberline{\thetable} class HeatmapRandomBase}

\subsubsection*{class \texttt{\textbf{HeatmapRandomWeightedWithConstraints}}}
\noindent 
{\small 
\setstretch{1} 
\begin{longtable}{|p{6cm}|p{10cm}|}

\hline
\multicolumn{2}{|c|}{\textbf{Fields}} \\
\hline
\rowcolor{lightgray}\texttt{canvasId} & ID of the canvas element where the heatmap is rendered. \\
\rowcolor{white}\texttt{length} & Number of rows in the heatmap grid. \\
\rowcolor{lightgray}\texttt{width} & Number of columns in the heatmap grid. \\
\rowcolor{white}\texttt{cellSize} & Size of each cell in the heatmap. \\
\rowcolor{lightgray}\texttt{snapshots} & Number of heatmap snapshots to be generated. \\
\rowcolor{white}\texttt{hotspotDistribution} & Distribution thresholds for hot, warm, and cool spots. \\
\hline
\multicolumn{2}{|c|}{\textbf{Methods}} \\
\hline
\rowcolor{lightgray}\texttt{\textbf{constructor}} & Initializes an object with specified properties for canvas, grid dimensions, cell size, number of snapshots, and hotspot distribution. \\
\hline
\rowcolor{white}\texttt{\textbf{generateRandomGrid()}} & Creates a grid with random values [0,1] for each cell. \\
\hline
\rowcolor{lightgray}\texttt{\textbf{applyHotspotDistribution(grid)}} & Applies the hotspot distribution thresholds to the grid, categorizing cells into different heat levels. \\
\hline
\rowcolor{white}\texttt{\textbf{enforceNeighborhood(grid)}} & Adjusts the values of cells in the grid to enforce spatial constraints, ensuring that hotspots are surrounded by warmer cells. \\
\hline
\rowcolor{lightgray}\texttt{\textbf{setHeatmaps()}} & Prepares and stores a series of heatmaps for visualization based on random distribution with spatial constraints. \\
\hline
\rowcolor{white}\texttt{\textbf{getHeatmaps()}} & Returns the prepared series of heatmaps. If not already set, it triggers the generation of heatmaps. \\
\hline
\rowcolor{lightgray}\texttt{\textbf{run()}} & Initiates the process of generating and visualizing constrained heatmaps over time. \\

\hline
\end{longtable}
} 
\addcontentsline{lot}{table}{\protect\numberline{\thetable} class HeatmapRandomWeightedWithConstraints}

\subsubsection*{class \texttt{\textbf{HeatmapRandomWeighted}}}
\noindent 
{\small 
\setstretch{1} 
\begin{longtable}{|p{6cm}|p{10cm}|}

\hline
\multicolumn{2}{|c|}{\textbf{Fields}} \\
\hline
\rowcolor{lightgray}\texttt{canvasId} & ID of the canvas element where the heatmap is rendered. \\
\rowcolor{white}\texttt{length} & Number of rows in the heatmap grid. \\
\rowcolor{lightgray}\texttt{width} & Number of columns in the heatmap grid. \\
\rowcolor{white}\texttt{cellSize} & Size of each cell in the heatmap. \\
\rowcolor{lightgray}\texttt{snapshots} & Number of heatmap snapshots to be generated. \\
\rowcolor{white}\texttt{hotspotDistribution} & Distribution thresholds for hot, warm, and cool spots. \\
\hline
\multicolumn{2}{|c|}{\textbf{Methods}} \\
\hline
\rowcolor{lightgray}\texttt{\textbf{constructor}} & Initializes an object with specified properties for canvas, grid dimensions, cell size, number of snapshots, and hotspot distribution. \\
\hline
\rowcolor{white}\texttt{\textbf{generateRandomGrid()}} & Creates a grid with random values [0,1] for each cell. \\
\hline
\rowcolor{lightgray}\texttt{\textbf{applyHotspotDistribution(grid)}} & Applies the hotspot distribution thresholds to the grid, categorizing cells into different heat levels. \\
\hline
\rowcolor{white}\texttt{\textbf{drawHeatMap(heatMap)}} & Renders the heatmap on the canvas with color coding based on cell heat levels. \\
\hline
\rowcolor{lightgray}\texttt{\textbf{generateHeatMap()}} & Generates a series of heatmaps with random weighted distributions. \\
\hline
\rowcolor{white}\texttt{\textbf{setHeatmaps()}} & Prepares and stores a series of heatmaps for visualization. \\
\hline
\rowcolor{lightgray}\texttt{\textbf{getHeatmaps()}} & Returns the prepared series of heatmaps. If not already set, it triggers the generation of heatmaps. \\
\hline
\rowcolor{white}\texttt{\textbf{run()}} & Initiates the process of generating and visualizing random weighted heatmaps over time. \\

\hline
\end{longtable}
} 
\addcontentsline{lot}{table}{\protect\numberline{\thetable} class HeatmapRandomWeighted}

\subsubsection*{class \texttt{\textbf{HeatmapConwayGoL}}}
\noindent 
{\small 
\setstretch{1} 
\begin{longtable}{|p{6cm}|p{10cm}|}

\hline
\multicolumn{2}{|c|}{\textbf{Fields}} \\
\hline
\rowcolor{lightgray}\texttt{canvasId} & ID of the canvas element where the heatmap is rendered. \\
\rowcolor{white}\texttt{length} & Number of rows in the heatmap grid. \\
\rowcolor{lightgray}\texttt{width} & Number of columns in the heatmap grid. \\
\rowcolor{white}\texttt{cellSize} & Size of each cell in the heatmap. \\
\rowcolor{lightgray}\texttt{snapshots} & Number of heatmap snapshots to be generated. \\
\rowcolor{white}\texttt{pAlive} & Probability of a cell being alive at the start. \\
\hline
\multicolumn{2}{|c|}{\textbf{Methods}} \\
\hline
\rowcolor{lightgray}\texttt{\textbf{constructor}} & Initializes a HeatmapConwayGoL object with specified properties for canvas, grid dimensions, cell size, number of snapshots, and probability of life. \\
\hline
\rowcolor{white}\texttt{\textbf{randomChoice(arr, p)}} & Selects a random element from an array based on provided probabilities. \\
\hline
\rowcolor{lightgray}\texttt{\textbf{generateGrid()}} & Generates a grid with cells randomly set to alive (1) or dead (0) based on the probability of life. \\
\hline
\rowcolor{white}\texttt{\textbf{countAliveNeighbors(grid, row, col)}} & Counts the number of alive neighbors for a cell at the specified position in the grid. \\
\hline
\rowcolor{lightgray}\texttt{\textbf{evolveGrid(grid, rules)}} & Evolves the grid state based on Conway's rules or other specified rules. \\
\hline
\rowcolor{white}\texttt{\textbf{conwaysRules(cell, numAliveNeighbors)}} & Applies Conway's Game of Life rules to determine the next state of a cell. \\
\hline
\rowcolor{lightgray}\texttt{\textbf{drawHeatMap(heatMap)}} & Renders the heatmap on the canvas with color coding based on cell states. \\
\hline
\rowcolor{white}\texttt{\textbf{getHeatmaps()}} & Returns the generated heatmaps; if not set, it triggers the generation. \\
\hline
\rowcolor{lightgray}\texttt{\textbf{setHeatmaps()}} & Generates and stores a series of heatmaps representing the evolution of the game. \\
\hline
\end{longtable}
} 
\addcontentsline{lot}{table}{\protect\numberline{\thetable} class HeatmapConwayGoL}

\subsubsection*{class \texttt{\textbf{HeatmapConwayGoLWithSpatialRules}}}
\noindent 
{\small 
\setstretch{1} 
\begin{longtable}{|p{6cm}|p{10cm}|}

\hline
\multicolumn{2}{|c|}{\textbf{Fields}} \\
\hline
\rowcolor{lightgray}\texttt{canvasId} & ID of the canvas element where the heatmap is rendered. \\
\rowcolor{white}\texttt{length} & Number of rows in the heatmap grid. \\
\rowcolor{lightgray}\texttt{width} & Number of columns in the heatmap grid. \\
\rowcolor{white}\texttt{cellSize} & Size of each cell in the heatmap. \\
\rowcolor{lightgray}\texttt{snapshots} & Number of heatmap snapshots to be generated. \\
\rowcolor{white}\texttt{pAlive} & Probability of a cell being alive at the start. \\
\hline
\multicolumn{2}{|c|}{\textbf{Methods}} \\
\hline
\rowcolor{lightgray}\texttt{\textbf{constructor}} & Inherits the constructor from HeatmapConwayGoL and initializes a HeatmapConwayGoLWithSpatialRules object. \\
\hline
\rowcolor{white}\texttt{\textbf{gridToHeatMap(grid)}} & Transforms the grid into a heatmap, with cells having different heat levels based on the number of alive neighbors and their own state. \\
\hline
\rowcolor{lightgray}\texttt{\textbf{run()}} & Generates and visualizes heatmaps over time, evolving the grid according to Conway's Game of Life rules and applying spatial rules to the heatmap. \\
\hline
\end{longtable}
} 
\addcontentsline{lot}{table}{\protect\numberline{\thetable} class HeatmapConwayGoLWithSpatialRules}

\subsubsection*{class \texttt{\textbf{HeatmapConwayGoLWithTemporalRules}}}
\noindent 
{\small 
\setstretch{1} 
\begin{longtable}{|p{6cm}|p{10cm}|}

\hline
\multicolumn{2}{|c|}{\textbf{Fields}} \\
\hline
\rowcolor{lightgray}\texttt{canvasId} & ID of the canvas element where the heatmap is rendered. \\
\rowcolor{white}\texttt{length} & Number of rows in the heatmap grid. \\
\rowcolor{lightgray}\texttt{width} & Number of columns in the heatmap grid. \\
\rowcolor{white}\texttt{cellSize} & Size of each cell in the heatmap. \\
\rowcolor{lightgray}\texttt{snapshots} & Number of heatmap snapshots to be generated. \\
\rowcolor{white}\texttt{pAlive} & Probability of a cell being alive at the start. \\
\hline
\multicolumn{2}{|c|}{\textbf{Methods}} \\
\hline
\rowcolor{lightgray}\texttt{\textbf{constructor}} & Inherits the constructor from HeatmapConwayGoL and initializes a HeatmapConwayGoLWithTemporalRules object. \\
\hline
\rowcolor{white}\texttt{\textbf{generateHeatMaps(length, width, cellSize, snapshots, pAlive, rules)}} & Generates a series of heatmaps using Conway's rules or other specified rules, considering both the current and next grid states. \\
\hline
\rowcolor{lightgray}\texttt{\textbf{gridToHeatMap(grid, nextGrid)}} & Transforms the grid into a heatmap by comparing the current and next states of each cell, factoring in the number of alive neighbors. \\
\hline
\rowcolor{white}\texttt{\textbf{setHeatmaps()}} & Generates and stores a series of heatmaps representing the temporal evolution of the game. \\
\hline
\rowcolor{lightgray}\texttt{\textbf{run()}} & Runs the simulation, evolving the grid according to Conway's Game of Life rules and visualizing the changes over time. \\
\hline
\end{longtable}
} 
\addcontentsline{lot}{table}{\protect\numberline{\thetable} class HeatmapConwayGoLWithTemporalRules}

\subsection{Algorithmic Deep Dive and Mathematical Formulation}

\subsubsection{Evolutionary Algorithms}

\paragraph{Evolutionary Approach:}
A meticulous exposition of Conway's Game of Life is provided to underpin the generation of Dynamic Heatmaps. This classic cellular automaton was chosen for its renowned emergent spatiotemporal behaviors, offering a well-established framework conducive to analytical adaptation and synthesis.

\paragraph{Overview of Evolutionary Algorithms:}
\begin{itemize}
    \item \textbf{Random Choice:} An algorithmic method for selecting elements from a set with associated probabilities, foundational for introducing stochastic elements into the model.
    \item \textbf{Generate Grid:} The initial step in constructing a spatial framework, where the grid is populated according to specified parameters, serving as the base state for subsequent evolution.
    \item \textbf{Alive Neighbors:} A preparatory function that sets up a reference for the cells' states, critical for determining the rules' application in the grid's evolution.
    \item \textbf{Count Alive Neighbors:} This function computes the number of adjacent cells in a live state, which influences the transition of a cell's state based on Conway's rules.
    \item \textbf{Evolve Grid:} The core iterative process that updates the grid's state across discrete time steps, simulating the dynamism of the system.
    \item \textbf{Conway's Rules:} The set of prescribed rules that dictate the survival, death, or birth of cells, encapsulating the fundamental logic of the Game of Life.
    \item \textbf{Set Heatmaps:} A method that compiles a series of heatmap snapshots, encapsulating the evolutionary states of the grid over time.
    \item \textbf{Grid to Heatmap with Spatial-biased Evolution:} An algorithm that translates the grid states into a heatmap with an emphasis on spatial relationships and neighborhood influences.
    \item \textbf{Grid to Heatmap with Temporal-biased Evolution:} This variation focuses on the temporal evolution of the grid, highlighting changes over time in the heatmap visualization.
\end{itemize}

\subsubsection{Random Choice}
\begin{algorithm}[H]
\setstretch{1}
\SetAlgoLined
\LinesNumbered
\KwResult{Selects a random element from an array based on provided probabilities}
\Begin{
    \KwData{array arr, probability array p}
    last $\leftarrow$ (length of arr) - 1 \; 
    p\_sum $\leftarrow$ 0 \; 
    \For{each value in p}{
        p\_sum += value\;
    }
    rand $\leftarrow$ random in range [0, sum] \; 
    cumulative\_sum $\leftarrow$ 0 \; 
    \For{i from 0 to last}{
        cumulative\_sum += p[i] \; 
        \If{rand $\le$ cumulative\_sum}{
            \KwRet{arr[i]}\; 
        }
    }
    \KwRet{arr[last]}\; 
}
\caption{Random Choice method}
\end{algorithm}

\subsubsection{Generate Grid}
\begin{algorithm}[H]
\setstretch{1}
\SetAlgoLined
\LinesNumbered  
\KwResult{Generates a grid for the heatmap}
\Begin{
    \KwData{length, width, pAlive}
    grid $\leftarrow$ initialize empty grid of size: length $\times$ width\; 
    \For{j: each row in grid}{ 
        \For{i: each column in row}{
            grid[i,j] $\leftarrow$  value using randomChoice([0, 1], [1 - pAlive, pAlive])\;
        }
    }
    \KwRet{grid}\;
}
\caption{generateGrid method for Heatmap Generation}
\end{algorithm}

\subsubsection{Count Alive Neighbors}
\begin{algorithm}[H]
\setstretch{1}
\SetAlgoLined
\LinesNumbered
\KwResult{Counts the number of alive neighbors for a given cell}
\Begin{
    \KwData{grid, row index, column index}
    numRows $\leftarrow$ length of grid\; 
    numCols $\leftarrow$ length of grid[0]\; 
    aliveNeighbors $\leftarrow$ 0\; 
    \For{rowOffset from -1 to 1}{
        \For{colOffset from -1 to 1}{
            \If{rowOffset $\neq$ 0 and colOffset $\neq$ 0}{
                neighborRow $\leftarrow$ (row + rowOffset + numRows) mod numRows\; 
                neighborCol $\leftarrow$  (col + colOffset + numCols) mod numCols\; 
                \If{ \textnormal{grid[neighborRow][neighborCol]} is 1}{
                    aliveNeighbors += 1\;
                }
            }
        }
    }
    \KwRet{aliveNeighbors}\;
}
\caption{countAliveNeighbors method for Counting Neighbors}
\end{algorithm}

\subsubsection{Evolve Grid}
\begin{algorithm}[H]
\setstretch{1}
\SetAlgoLined
\LinesNumbered  
\KwResult{Evolves the state of the grid based on specified rules}
\Begin{
    \KwData{grid, rules function}
    numRows $\leftarrow$ length of grid\; 
    numCols $\leftarrow$ length of grid[0]\; 
    initialize newGrid as a deep copy of grid\; 
    \For{each row in newGrid}{
        \For{each column in row}{
            numAliveNeighbors $\leftarrow$ countAliveNeighbors(grid, row, col)\;
            newGrid[row][col] $\leftarrow$ rules(grid[row][col], numAliveNeighbors)\;
        }
    }
    \KwRet{newGrid}\;
}
\caption{evolveGrid method for Grid Evolution}
\end{algorithm}

\subsubsection{Conway's Rules}
\begin{algorithm}[H]
\setstretch{1}
\SetAlgoLined
\LinesNumbered  
\KwResult{Determines the next state of a cell based on Conway's Game of Life rules}
\Begin{
    \KwData{current cell state (cell), number of alive neighbors (numAliveNeighbors)}
    \uIf{ numAliveNeighbors is 3}{
        \KwRet{1}\;  
    }
    \uElseIf{cell is 1 \textnormal{and} numAliveNeighbors is 2}{
        \KwRet{1}\;  
    }
    \Else{
        \KwRet{0}\;  
    }
}
\caption{conwaysRules method for Conway's Game of Life}
\end{algorithm}

\subsubsection{Set Heatmaps}
\begin{algorithm}[H]
\setstretch{1}
\SetAlgoLined
\LinesNumbered  
\KwResult{Generates \& stores a series of heatmaps based on Conway's Game of Life}
\Begin{
    \KwData{number of snapshots (snapshots)}
    grid $\leftarrow$ generateGrid()\;
    heatmaps $\leftarrow$ new array size of snapshots\;
    \For{i from 0 to (snapshots - 1)}{
        grid $\leftarrow$ evolveGrid(grid, conwaysRules)\;
        heatmap $\leftarrow$ gridToHeatMap(grid)\;
        heatmaps.push(heatmap)\;
    }
}
\caption{setHeatmaps method for storing series of heatmaps}
\end{algorithm}

\subsubsection{Grid to HeatMap with Spatial-biased Evolution}
\begin{algorithm}[H]
\setstretch{1}
\SetAlgoLined
\LinesNumbered  
\KwResult{Converts a binary grid representation into a heatmap based on the number of alive neighbors}
\Begin{
    \KwData{grid}
    numRows $\leftarrow$ length of grid\;
    numCols $\leftarrow$ length of grid[0]\;
    heatMap $\leftarrow$ initialize empty grid of size numRows $\times$ numCols with all values set to 0\;
    \For{row from 0 to (numRows - 1)}{
        \For{col from 0 to (numCols - 1)}{
            numAliveNeighbors $\leftarrow$ countAliveNeighbors(grid, row, col)\;
            \eIf{grid[row][col] is 1}{
                heatMap[row][col] $\leftarrow$ min(0.5 + numAliveNeighbors $\times$ 0.1, 1)\;
            }{
                heatMap[row][col] $\leftarrow$ min(numAliveNeighbors $\times$ 0.1, 1)\;
            }
        }
    }
    \KwRet{heatMap}\;
}
\caption{gridToHeatMap method for converting grid to heatmap with spatial evolution}
\end{algorithm}

\subsubsection{Grid to HeatMap with Temporal-biased Evolution}
\begin{algorithm}[H]
\setstretch{1}
\SetAlgoLined
\LinesNumbered  
\KwResult{Converts a binary grid representation into a heatmap based on the current and next state, including the number of alive neighbors}
\Begin{
    \KwData{grid, nextGrid}
    numRows $\leftarrow$ length of grid\;
    numCols $\leftarrow$ length of grid[0]\;
    heatMap $\leftarrow$ new 2D array of size: [numRows] $\times$ [numCols]\;
    \For{row from 0 to (numRows - 1)}{
        \For{col from 0 to (numCols - 1)}{
            currentState $\leftarrow$ grid[row][col]\;
            nextState $\leftarrow$ nextGrid[row][col]\;
            numAliveNeighbors $\leftarrow$ countAliveNeighbors(grid, row, col)\;
            \eIf{currentState is 1 and nextState is 1}{
                heatMap[row][col] $\leftarrow$ 1\;  
            }{
                \If{currentState is 1}{
                    numAliveNeighbors += 1\;  
                }
                heatMap[row][col] $\leftarrow$ min(numAliveNeighbors / 8, 0.9)\;  
            }
        }
    }
    \KwRet{heatMap}\;
}
\caption{gridToHeatMap method for converting grid to heatmap with temporal evolution}
\end{algorithm}

\subsubsection{Stochastic Algorithms}

\paragraph{Stochastic Approach:}
This approach models stochastic spatiotemporal dynamics, characterized by independently occurring events that are not influenced by prior states. Spatial clustering is observed where certain regions have a higher likelihood of event occurrence. Additionally, sporadic, noisy events are distributed across the global space, introducing an element of unpredictability and randomness to the system.

\paragraph{Overview of Stochastic Algorithms}
\begin{itemize}
\item \textbf{Generate Random Grid}: Creates a grid where each cell's state is randomly determined, laying the foundation for stochastic spatial patterns.
\item \textbf{Apply Hotspot Distribution}: Assigns varying levels of 'heat' or intensity to cells based on predefined thresholds, introducing heterogeneity into the grid.
\item \textbf{Enforce Neighborhood}: Adjusts cell states based on the states of their neighboring cells, embedding a local spatial influence that can lead to emergent patterns.
\item \textbf{Generate HeatMap}: Compiles a series of heatmaps from the grids generated through stochastic processes, capturing the dynamic evolution of the system over time.
\end{itemize}

\subsubsection{Generate Random Grid}
\begin{algorithm}[H]
\setstretch{1}
\SetAlgoLined
\LinesNumbered  
\KwResult{Generates a grid filled with random values}
\Begin{
    \KwData{gridLength, gridWidth}
    grid $\leftarrow$ new 2d array of size: gridLength $\times$ gridWidth\;
    \For{j from 0 to (gridLength - 1) }{
        \For{i from 0 to (gridWidth - 1)}{
            grid[j][i] $\leftarrow$ random real value between [0,1]\;
        }
    }
    \KwRet{grid}\;
}
\caption{generateRandomGrid method for Heatmap Generation}
\end{algorithm}

\subsubsection{Apply Hotspot Distribution}
\begin{algorithm}[H]
\setstretch{1}
\SetAlgoLined
\LinesNumbered  
\KwResult{Applies hotspot distribution to a grid based on thresholds}
\Begin{
    \KwData{grid, hotThreshold, warmThreshold, coolThreshold}
    lastRow $\leftarrow$ length of grid -1\;
    lastCol $\leftarrow$ length of grid[0] - 1\;
    \For{row from 0 to lastRow}{
        \For{col from 0 to lastCol}{
            \uIf{grid[row][col] $>$ hotThreshold}{
                grid[row][col] $\leftarrow$ 1\;  
            }
            \uElseIf{grid[row][col] $>$ warmThreshold}{
                grid[row][col] $\leftarrow$ 0.5\;  
            }
            \uElseIf{grid[row][col] $>$ coolThreshold}{
                grid[row][col] $\leftarrow$ 0.25\;  
            }
            \Else{
                grid[row][col] $\leftarrow$ 0\;  
            }
        }
    }
    \KwRet{grid}\;
}
\caption{applyHotspotDistribution method for modifying grid based on heat levels}
\end{algorithm}

\subsubsection{Enforce Neighborhood}
\begin{algorithm}[H]
\setstretch{1}
\SetAlgoLined
\LinesNumbered  
\KwResult{Enforces neighborhood constraints in a grid based on warm thresholds}
\Begin{
    \KwData{grid, warmThreshold}
    newGrid $\leftarrow$ deep copy of grid\;  
    lenRow $\leftarrow$ length of grid \;
    lenCol $\leftarrow$ length of grid[0] \;
    \For{row from 0 to (lenRow - 1) }{
        \For{col from 0 to (lenCol -1)}{
            neighbors $\leftarrow$ empty list\;
            \For{rowOffset from -1 to 1}{
                \For{colOffset from -1 to 1}{
                    \If{rowOffset $\neq$ 0 or colOffset $\neq$ 0}{
                        neighborRow $\leftarrow$ (row + rowOffset + lenRow) mod lenRow\;
                        neighborCol $\leftarrow$ (col + colOffset + lenCol) mod lenCol\;
                        neighborCell $\leftarrow$ [neighborRow, neighborCol]\; 
                        neighbors.push(neighborCell)\;
                    }
                }
            }
            \If{grid[row][col] is 1}{  
                \For{each [r, c] in neighbors}{
                    \If{newGrid[r][c] $<$ warmThreshold}{
                        newGrid[r][c] $\leftarrow$ 0.5\;  
                    }
                }
            }
        }
    }
    \KwRet{newGrid}\;
}
\caption{enforceNeighborhood method for adjusting neighborhood values}
\end{algorithm}

\subsubsection{Generate HeatMap}
\begin{algorithm}[H]
\setstretch{1}
\SetAlgoLined
\LinesNumbered  
\KwResult{Generates a series of heatmaps based on random grid generation and rules application}
\Begin{
    \KwData{snapshots, isNeighborhoodRulesEnabled}
    heatMaps $\leftarrow$ empty list\;  
    \For{i from 0 to (snapshots - 1)}{
        grid $\leftarrow$ generateRandomGrid()\;  
        grid $\leftarrow$ applyHotspotDistribution(grid)\;  
        \If{isNeighborhoodRulesEnabled}{
            grid $\leftarrow$  enforceNeighborhood(grid)\;  
        }
        heatMaps.push(grid)\;  
    }
    \KwRet{heatMaps}\;  
}
\caption{generateHeatMap method for Heatmap Generation}
\end{algorithm}

\section{Synthetic Spatiotemporal Techniques}

In our research, the creation and analysis of synthetic datasets are pivotal for modeling the nuanced interplay of space and time. These datasets are crafted to mimic real-world complexities, capturing the essence of spatial interactions and temporal progressions. Spatiotemporal modeling techniques are employed to ensure each synthetic dataset accurately reflects the dynamics of the system it represents.

The spatial component is modeled using grid-based structures or networks, where each node or cell encapsulates spatial attributes and interactions. Temporal aspects are integrated through discrete time steps or continuous simulations, allowing for the observation of dynamic changes and the evolution of spatial relationships over time. 

\section{Synthetic Stochastic Dynamics}

Stochastic and evolutionary behaviors within our synthetic datasets are implemented through a careful balance of randomness and rule-based evolution. Spatial rules govern the local interactions between adjacent elements, dictating the conditions under which they influence each other. Temporal rules guide the system's evolution, determining how the current state transitions to the next in response to both stochastic factors and deterministic laws.

Random behaviors are introduced to simulate variability and unpredictability, key features of real-world systems. This randomness is counterbalanced by evolutionary algorithms that simulate natural selection and adaptation, steering the system towards emergent patterns and structures over successive evolutions.

\section{Computational Environment}

In the pursuit of maintaining the integrity and reproducibility of our experimental research, we have selected Google Colaboratory as our primary computational platform. Google Colab offers an amalgamation of benefits that cater to the core requirements of our research methodology. It stands out for its ability to provide a reproducible environment through shareable, web-based Python notebooks. Additionally, its infrastructure supports extensive computational tasks with the provision of GPU access, specifically incorporating the Tesla T4 GPU. This choice empowers us to utilize CUDA-compliant libraries, essential for our computational experiments.

The software environment is underpinned by Ubuntu 20.04 LTS and Python 3.10, which provides a stable and up-to-date foundation for our computational needs. We have harnessed a suite of modules—namely cuDF, Dask cuDF, cuML, cuGraph, cuSpatial, and CuPy—ensuring that our computations are not only GPU-accelerated but also executed with maximum efficiency. These modules have been integral to our ability to handle large datasets and complex computations that are typical in data-intensive research studies.

Google Colaboratory's framework enhances the accessibility of our work, allowing for the seamless sharing and replication of code and results. This approach aligns with our commitment to transparency and the promotion of open and repeatable scientific practices. By utilizing this platform, we affirm our dedication to contributing to a body of knowledge that is both reliable and accessible to the wider research community.

\end{appendix}

\newgeometry{left=1in,right=1in,top=0.5in,bottom=1in}
\chapter*{Vita}
\addcontentsline{toc}{chapter}{Vita}

\begin{spacing}{1} 

Ted Holmberg was born in New Orleans, Louisiana. He earned his Bachelor of Arts in Philosophy from the University of New Orleans. Subsequently, Ted pursued higher education at the same university, obtaining a Master of Science in Computer Science. His thesis was titled "Data Visualization to Evaluate and Facilitate Targeted Data Acquisitions in Support of a Real-time Ocean Forecasting System." Currently, Ted is a PhD candidate in Engineering \& Applied Science at the University of New Orleans.

\end{spacing}

\end{document}